\documentclass[a4paper,12pt]{article}
\pdfoutput=1				

\usepackage[margin=0.9in]{geometry}

\usepackage{ucs}			
\usepackage[utf8x]{inputenc}

\usepackage{bm}			
\usepackage{makecell}		
\usepackage{mathtools}		
\usepackage{multirow}		
\usepackage[compress,numbers,sort]{natbib}		
\usepackage[percent]{overpic}	
\usepackage{slashed}		
\usepackage{xcolor}

\usepackage{amssymb}		
\usepackage{dsfont}			
\usepackage{eucal}			
\usepackage{mathrsfs}		

\usepackage{tocloft}			
\usepackage{etoc}			

\allowdisplaybreaks[1]				
\numberwithin{equation}{section}		
\usepackage{hyperref}
\hypersetup{colorlinks, bookmarksopen, bookmarksnumbered, citecolor=verdeCP3, linkcolor=bluCP3, pdfstartview=FitH, urlcolor=rossoCP3}

\definecolor{Appe}{HTML}{24A6A4}
\definecolor{rossoCP3}{cmyk}{0, 0.88, 0.77, 0.40}
\definecolor{verdeCP3}{rgb}{0.09765625, 0.57421875, 0.1015625}
\definecolor{bluCP3}{rgb}{0, 0.23, 0.67}
\definecolor{arxcol}{HTML}{b31b1b}

\newcommand{\arx}[1]{\href{https://arxiv.org/abs/#1}{\textcolor{arxcol}{#1}}}
\newcommand{\doi}[1]{\href{https://doi.org/#1}{\textcolor{verdeCP3}{#1}}}

\newcommand{\website}{\href{https://sites.google.com/view/appendiciario/}{\textcolor{Appe}{website}}}
\newcommand{\figcode}{\vspace{1.4mm}\footnotesize{\textsf{The code to generate this figure is available on the \website~\cite{Appe-website}.}}}

\newcommand{\pnt}{\rule[-2mm]{0mm}{6mm}}

\newcommand{\eg}{e.g.~}
\newcommand{\ie}{i.e.~}

\newcommand{\Eq}[1]{Eq.~\eqref{#1}}
\newcommand{\Sec}[1]{Sec.~\ref{#1}}
\newcommand{\Fig}[1]{Fig.~\ref{#1}}
\newcommand{\Tab}[1]{Table~\ref{#1}}
\newcommand{\QnA}[1]{Q\&A~\ref{#1}}

\DeclareMathOperator{\Tr}{Tr}		
\newcommand{\tr}{\mathsf{T}}		
\DeclareMathOperator{\erf}{erf}
\DeclareMathOperator{\sgn}{sgn}

\newcommand{\beq}{\begin{equation}}
\newcommand{\eeq}{\end{equation}}
\newcommand{\ud}{\text{d}}

\newcommand{\DM}{\text{DM}}
\newcommand{\NR}{\text{NR}}

\newcommand{\fvec}[1]{\mathsf{#1}}			
\newcommand{\genp}[1]{\mathfrak{#1}}		
\newcommand{\amom}[1]{\CMcal{#1}}		
\newcommand{\mulop}[1]{\mathcal{#1}}		

\newcommand{\ket}[1]{| #1 \rangle}
\newcommand{\bra}[1]{\langle #1 |}
\newcommand{\braket}[2]{\langle #1 | #2 \rangle}
\newcommand{\matel}[3]{\langle #1 | #2 | #3 \rangle}
\newcommand{\redmatel}[3]{\langle #1 || #2 || #3 \rangle}
\newcommand{\Nmatel}[1]{\matel{N'}{#1}{N}}
\newcommand{\CG}[3]{\braket{#1 ; #2}{#3}}

\newcommand{\Ord}{\mathcal{O}}		
\newcommand{\Op}{\mathscr{O}}		
\newcommand{\Lag}{\mathscr{L}}		
\newcommand{\Mel}{\mathscr{M}}		

\newcommand{\bol}[1]{\bm{#1}}
\newcommand{\unop}{\mathds{1}}
\newcommand{\unom}{I}
\newcommand{\Cmat}{\CMcal{C}}
\newcommand{\NReq}{\overset{\text{\tiny{NR}}}{=}}
\newcommand{\qeq}{\overset{q}{=}}
\newcommand{\figmath}[1]{\emph{$#1$}}

\newcommand{\mDM}{m}				
\newcommand{\mN}{m_\text{N}}		

\newcommand{\ER}{E_\text{R}}			
\newcommand{\Ed}{E'}				
\newcommand{\Emin}{E_\text{min}}

\newcommand{\vmin}{v_\text{min}}
\newcommand{\vE}{v_\text{E}}
\newcommand{\vesc}{v_\text{esc}}
\newcommand{\vmax}{v_\text{max}}
\newcommand{\zmin}{z_\text{min}}
\newcommand{\zE}{z_\text{E}}
\newcommand{\zesc}{z_\text{esc}}

\newcommand{\fG}{f}
\newcommand{\fE}{f_\text{E}}

\newcommand{\aS}{\alpha_\text{s}}

\newcommand{\keV}{\text{keV}}
\newcommand{\MeV}{\text{MeV}}
\newcommand{\GeV}{\text{GeV}}
\newcommand{\TeV}{\text{TeV}}
\newcommand{\fm}{\text{fm}}
\newcommand{\cm}{\text{cm}}
\newcommand{\mt}{\text{m}}
\newcommand{\km}{\text{km}}
\newcommand{\pc}{\text{pc}}
\newcommand{\sd}{\text{s}}
\newcommand{\dd}{\text{day}}
\newcommand{\yr}{\text{yr}}
\newcommand{\gr}{\text{g}}
\newcommand{\kg}{\text{kg}}
\newcommand{\ton}{\text{t}}
\newcommand{\amu}{\text{u}}
\newcommand{\pb}{\text{pb}}
\newcommand{\cpd}{\text{cpd}}

\begin{document}

\begin{center}
{\Huge\textbf{\textcolor{Appe}{Appendiciario}}}
\\[4mm]
{\Large\emph{A hands-on manual on the theory \\[2mm] of direct Dark Matter detection}}

\medskip
\bigskip\color{black}\vspace{0.6cm}

{\large\textbf{Eugenio Del Nobile}}
\\[3mm]
\textit{\href{https://www.nottingham.ac.uk/physics/}{School of Physics \& Astronomy}, University of Nottingham, \\ University Park, Nottingham, NG7 2RD, UK}
\\[3mm]
\textit{\href{https://www.dfa.unipd.it/}{Dipartimento di Fisica e Astronomia ``G.~Galilei''}, \\ Universit\`a di Padova and INFN, Sezione di Padova, Via Marzolo 8, 35131 Padova, Italy}
\\[3mm]
\href{mailto:eugenio.delnobile@pd.infn.it}{eugenio.delnobile@pd.infn.it}
\end{center}

\bigskip

\centerline{\large\textbf{Abstract}}
\begin{quote}
\large
A manual for computations in direct Dark Matter detection phenomenology. Featuring self-contained sections on non-relativistic expansion, elastic and inelastic scattering kinematics, Dark Matter velocity distribution, hadronic matrix elements, nuclear form factors, cross sections, rate spectra and parameter-space constraints, as well as a handy two-page summary and Q\&A section for a quick reference. A pedagogical, yet general and model independent guide, with examples from standard and non-standard particle Dark Matter models.
\end{quote}

\newpage

\tableofcontents

\listoffigures

\listoftables

\newpage

\appendix

\section{Introduction}
From the point of view of particle physics, there are different levels of complexity in doing a computation for direct Dark Matter (DM) detection, also related to the fact that the DM particle interacts not with a free, fundamental particle but rather with a whole nucleus. In the simplest case, the spin-independent (SI) interaction with isosinglet couplings, one can get away with a prescription to derive the DM-nucleus scattering cross section from the DM-nucleon cross section scaled up by a factor $A^2$ ($A$ being the number of nucleons in the nucleus), times a form factor. For the rushed researcher, there is no need to understand deeply how all these factors come about. The relevant formulas can be easily found in the literature, ready to be `consumed', and conveniently parametrized in terms of a \emph{zero-momentum transfer cross section} $\sigma_0$. Computations for the spin-dependent (SD) interaction are a bit more involved, but again all relevant formulas can be easily retrieved (though not all details of the derivation). The SD interaction is often disregarded since the lack of the $A^2$ enhancement, according to the common lore, makes it negligible with respect to the SI interaction. This is however not always true and must be checked on a case by case basis (see \eg Refs.~\cite{Bednyakov:2008gv, Marcos:2015dza}).

Things start to get more involved when considering other interactions, for instance if the DM interacts with massless mediators like the photon or has non-relativistically (NR) suppressed interactions. In these cases it may not even be possible to define $\sigma_0$, since it may diverge or vanish, while in other cases it may be possible but not very useful. Moreover, the interaction does not need to be of a single type: the scattering cross section for a Dirac DM particle interacting through its anomalous magnetic dipole moment, for instance, has both a SI-like (charge-dipole) and a SD-like (dipole-dipole) part, with the latter dominating in certain regimes. In computing the unpolarized cross section, the gamma-matrix algebra cannot be used with impunity as done for scattering of fundamental fermions. In the NR limit, different space-time components of a nucleon fermion bilinear are associated to different nuclear properties, which can give wildly different contributions to the DM-nucleus scattering cross section. For this reason it is necessary to separate these components (opening up the belly of the Dirac matrices, if needed) and identify their contribution to the interaction. The need to do so is not immediately obvious if one is only interested in the SI and SD interactions, but it becomes crucial for other interactions.

In these notes we focus mostly on the particle-physics phenomenology of direct DM detection and on how to compute the DM-nucleus scattering rate. We restrict our attention to Weakly Interacting Massive Particles (WIMPs), as direct detection experiments are mostly concerned with this type of DM particles; for this reason, we use `DM' as a synonym for `WIMP'. We do not, however, enforce any strict definition of WIMPs, rather thinking of them in broad terms, as DM particles that can be detected on Earth through scattering off nuclei. In this sense, we will not delimit the range of WIMP mass a priori, but rather work out what masses experiments can be sensitive to. Likewise, we will not focus on any specific WIMP candidate, but rather try to be as general and model independent as possible. Some features of the less standard scenarios challenge our simplest intuition of the scattering process, thus giving us a chance to improve our understanding of the physics involved which could be valuable even for those who are only interested in the most standard DM-nucleon interactions.

An effort has been made to present the material of these notes in a form compatible with the different notations adopted in the literature, so that it is readily comparable with results found elsewhere. The discussion is kept as general as possible, however we restrict ourselves to elementary DM particles with spin $0$ and $1/2$ in our examples in Secs.~\ref{qg to N},~\ref{sigma},~\ref{Pheno} and in the treatment of NR operators and related form factors in Secs.~\ref{DM-N},~\ref{Form factors} (see \eg Refs.~\cite{Gondolo:2020wge, Gondolo:2021fqo} and references therein for DM with higher spins). Assumptions are spelled out systematically, and our notation is summarized in \Sec{Notation} for a quick reference.

These notes ideally follow the spirit of Ref.~\cite{Lewin:1995rx}, although without its convenient conciseness. Computations are worked out in all their crucial steps, and a number of examples are presented throughout to complement and illustrate the theoretical arguments. A code for generating most of the figures of these notes is also publicly available on the \website~\cite{Appe-website}, which already contains some of the machinery needed for a direct DM detection analysis and can be used as a playground or as a starting point for an actual analysis. The single sections are conceived as self-contained and as much as possible independent of one another, with \Sec{Rate} and \Sec{Pheno} working as a frame to the various parts. A two-page summary and a Q\&A section can be found in \Sec{Summary}, to the advantage of the reader seeking quick responses.

These notes are organized as follows. We first spell out our notation in \Sec{Notation}. We then discuss the general grounds of direct DM detection in \Sec{Rate}, where we write down the differential recoil rate $\ud R_T / \ud \ER$ and introduce the `ingredients' needed to compute it, which are then individually discussed in the subsequent sections. In \Sec{scattering kinematics} we explore the scattering kinematics, which contributes to the rate through the $\vmin(\ER)$ function, while a discussion of the DM velocity distribution $\fE(\bol{v}, t)$ is deferred to \Sec{velocity}. We then begin a journey into the DM interactions, which will take us to compute in \Sec{sigma} the DM-nucleus differential scattering cross section $\ud \sigma_T / \ud \ER$. We start from the most fundamental level in \Sec{qg to N}, with the DM-quark/gluon interaction operators $\Op_{q, \text{g}}$ and their hadronic matrix elements $\Nmatel{\Op_{q, \text{g}}}$. In \Sec{DM-N} we see how to compute the NR limit of the DM-nucleon scattering amplitude $\Mel_N$ and derive the corresponding NR interaction operator $\Op_\NR^N$. In \Sec{Form factors} we get a qualitative understanding of nuclear form factors and compute the NR DM-nucleus scattering amplitude $\Mel$. The various results are finally collected in \Sec{Pheno} (which together with \Sec{Rate} works as a frame to the various parts), where an example phenomenological analysis of a (pretend) experimental result is carried out to show how the different ingredients contribute to the rate. A two-page summary is then presented in \Sec{Summary}, which also features a handy Q\&A subsection.

For illustrative purposes and as a quick guide, going from microscopic to macroscopic, the above ingredients may be concisely connected as follows:
\begin{align*}
[\Op_{q, \text{g}} \to \Nmatel{\Op_{q, \text{g}}}]~\text{(\Sec{qg to N})}
\longrightarrow
[\Mel_N \to \Op_\NR^N]~\text{(\Sec{DM-N})}
\longrightarrow
\Mel~\text{(\Sec{Form factors})}
\longrightarrow
\frac{\ud \sigma_T}{\ud \ER}~\text{(\Sec{sigma})} \ ,
\\
\frac{\ud \sigma_T}{\ud \ER}~\text{(\Sec{sigma})}
+
\fE~\text{(\Sec{velocity})}
+
\vmin(\ER)~\text{(\Sec{scattering kinematics})}
+
\rho, \zeta_T, m_T~\text{(\Sec{Rate})}
\longrightarrow
\frac{\ud R_T}{\ud \ER}~\text{(\Sec{Rate})} \ ,
\end{align*}
where arrows mean that the quantities to the left are needed to compute those to the right. However, rather than starting right away with discussing the most fundamental bricks involved in direct DM detection, we first try to establish a connection with its experimental side, without giving for granted that experiments look at DM scattering off nuclei but rather asking (and trying to answer): what can we see with galactic DM scattering off a target, and why are nuclei effective targets? This starting point is then used as a motivation for the subsequent theoretical discussion and computations.

Reviews covering in some detail different aspects of the theory of direct DM detection include the old classics, Refs.~\cite{Engel:1992bf, Jungman:1995df, Lewin:1995rx}, and the most recent Refs.~\cite{Vergados:2006sy, Bednyakov:2008gv, Cerdeno:2010jj, Freese:2012xd, Strigari:2013iaa, Gelmini:2015zpa, Lisanti:2016jxe, Mambrini-Histories, Salati:2007zz}. Some references for topics that will not be touched upon here are: history of direct DM detection~\cite{Gelmini:2011xz, Bertone:2016nfn}, experimental techniques and ongoing and planned experiments~\cite{Gaitskell:2004gd, Primack:1988zm, Schnee:2011ef, Rau:2011zz, Saab:2012th, Strigari:2013iaa, Cooley:2014aya, Undagoitia:2015gya, Liu:2017drf, Schumann:2019eaa}, available codes and online resources~\cite{Gondolo:2004sc, Belanger:2008sj, Yellin:2008da, DelNobile:2013sia, Anand:2013yka, Shan:2014upa, Backovic:2015cra, DEramo:2016gos, Kavanagh:2016pyr, Workgroup:2017lvb, Bishara:2017nnn, Bringmann:2018lay, Kang:2018rad, Belanger:2020gnr, DMTools, WIMPLimitPlotter, DarkMatterPortal, DAMNED, DMonlinetools, DMHub}, directional detection~\cite{Mayet:2016zxu}, the neutrino floor~\cite{Monroe:2007xp, Vergados:2008jp, Strigari:2009bq, Gutlein:2010tq, Harnik:2012ni, Billard:2013qya, Gutlein:2014gma, Ruppin:2014bra, Davis:2014ama, Dent:2016iht, OHare:2016pjy, Dent:2016wor, Gelmini:2018ogy, Boehm:2018sux}, and direct detection with DM-electron scattering~\cite{Essig:2011nj, Roberts:2015lga, Essig:2017kqs, Catena:2019gfa}.

\section{Notation}
\label{Notation}
It is always a good idea to spend a word or two about the notation one is adopting. Natural units $\hbar = c = 1$ and the `mostly minus' Minkowski metric $g^{\mu \nu} = \text{diag}(+1, -1, -1, -1)$ are used throughout these notes. Useful unit identities are (see \eg Ref.~\cite{Zyla:2020zbs}):
\begin{align}
\label{units}
1 \approx 197~\MeV~\fm \ ,
&&
1~\GeV \approx 1.78 \times 10^{-27}~\kg \ ,
&&
1~\pb = 10^{-36}~\cm^2 \ .
\end{align}
The label $T$ indicates a target nucleus, and most precisely a nuclide, unless otherwise stated. $N = p, n$ indicates nucleon type, either proton or neutron. $\psi$ denotes a generic spin-$1/2$ field, while $\phi$, $\chi$ denote a spin-$0$ and a spin-$1/2$ DM field, respectively ($\chi$ is a Dirac field unless otherwise noted). $\unom_d$ indicates the $d$-dimensional unit matrix. We use the following acronyms:
\begin{itemize}
\item CM = center of momentum,
\item DM = Dark Matter,
\item EFT = effective field theory,
\item LSR = local standard of rest,
\item NR = non relativistic,
\item PLN = point-like nucleus,
\item QCD = Quantum Chromodynamics,
\item SHM = Standard Halo Model,
\item SI/SD = spin-independent/spin-dependent,
\item SM = Standard Model of particle physics.
\end{itemize}
Also, isospin always refers to strong isospin (as opposed to weak isospin).

$\mDM$ and $m_T$ denote the DM and nuclear mass, respectively. The nucleon mass is
\beq
\mN \equiv \frac{m_p + m_n}{2} \approx 939~\MeV
\eeq
(see below for an explanation of our use of the $\approx$ symbol). The symbols $m_p$ and $m_n$ for the proton and neutron mass, respectively, are only employed where required by certain definitions, as in \Eq{nuclear magneton} below and for some specific hadronic form factors in \Sec{Vector couplings}. $\mu_T$ and $\mu_\text{N}$ denote the DM-nucleus and DM-nucleon reduced mass, respectively. To avoid confusion, the \emph{nuclear magneton}
\beq\label{nuclear magneton}
\hat{\mu}_\text{N} \equiv \frac{e}{2 m_p} \approx 0.105 \, e~\fm \approx 0.16~\GeV^{-1} \ ,
\eeq
a unit of magnetic dipole moment, is indicated with a hat.

With few exceptions, the letters $p$ and $k$ always indicate the momenta of DM and target nucleus, respectively. An exception is that in Secs.~\ref{qg to N} and~\ref{DM-N} $k$ indicates the momenta of the interacting nucleon rather than that of the whole nucleus. We adopt different styles for the different types of momenta: $\fvec{Sans}$ $\fvec{serif}$ symbols (\eg $\fvec{p}$, $\fvec{k}$) denote four-vectors, $\bol{bold}$ symbols (\eg $\bol{p}$, $\bol{k}$) denote three-vectors, and $plain$ symbols (\eg $p \equiv |\bol{p}|$, $k \equiv |\bol{k}|$) denote absolute values of three-vectors. The same notation can also apply to other quantities, \eg $\fvec{x}$, $\bol{x}$ and $x \equiv |\bol{x}|$ could all be used to indicate the position of something. Scalar products between three-vectors and between four-vectors are both denoted with a dot, \eg $\bol{q} \cdot \bol{x}$ and $\fvec{p} \cdot \fvec{p}' = \fvec{p}^\mu \fvec{p}'_\mu$. Hats over bold symbols denote unit vectors, as in $\hat{\bol{q}} \equiv \bol{q} / q$ and $|\hat{\bol{n}}| = 1$. A prime usually indicates final-state quantities, \eg $\fvec{p}'$ and $\fvec{k}'$ indicate the final DM and nucleus/nucleon momenta, after the scattering has occurred; likewise, $\mDM' = \mDM + \delta$ indicates the final DM mass, with $\delta$ the DM mass splitting (however, $\Ed$ indicates the quantity actually measured by an experiment to infer the energy of a scattering event). To avoid writing twice formulas that apply equally to DM and nuclei/nucleons, we adopt $\genp{Fraktur}$ symbols to denote quantities that can refer to both the DM and the target, in the initial or in the final state. For example, $\genp{m}$ denotes the mass of a generic particle, while $\bol{\genp{p}}$ ($\bol{\genp{bold}}$) and $\genp{p} \equiv | \bol{\genp{p}} |$ ($\genp{plain}$) denote its three-momentum and the momentum absolute value. We can immediately use this style to define the energy of a generic particle with mass $\genp{m}$ and momentum $\bol{\genp{p}}$ as
\beq\label{particle energy}
E_\genp{p} \equiv \sqrt{\genp{m}^2 + \genp{p}^2} \ .
\eeq
It thus remains understood \eg that $E_{k'}$ indicates the final energy of the nucleus. The space-time components of a four-vector as \eg $\fvec{p}$ are indicated as $\fvec{p}^\mu = (E_p, \bol{p})^\tr$, with the superscript $\tr$ signalling this is a column vector that we wrote here as a row for simplicity.

\beq
\bol{v} \equiv \bol{v}_\DM - \bol{v}_{T, N}
\eeq
is the DM-nucleus or DM-nucleon (depending on the context) relative velocity, and $v \equiv \bol{v}$ is the relative speed. $\fvec{q}^\mu = (q^0, \bol{q})^\tr$ is the momentum transfer, defined as
\beq
\fvec{q} \equiv \fvec{p} - \fvec{p}' \ ,
\eeq
and
\beq
\ER = \frac{q^2}{2 m_T}
\eeq
is the nuclear recoil energy. \Fig{fig: basics} and \Fig{fig: momnot} below may be useful in quickly recalling some aspects of our notation.

As for other symbols, we use several different signs for equalities. A plain $=$ sign has no particular meaning attached to the equality, while $\equiv$ indicates a definition. We use $\NReq$ to indicate equalities that are only valid at some finite order of the NR expansion (see \eg \Eq{NR gamma} for an example of how this symbol is used). Analogously, the $\qeq$ sign means the equality is only valid at some finite order of an expansion in powers of $q / \mN$ (see below \Eq{<p'|A^a|p>} where this sign is introduced). In other cases where an approximation is to be stressed, in particular when the error is controlled by one or more parameters, we employ the $\simeq$ sign, although we remark that signalling all the approximations involved in the computations carried out here is out of the scope of these notes. For numerical equalities we use the $=$ sign whenever a relation is exact or otherwise it has an attached uncertainty, while we use $\approx$ otherwise: for instance, we may equally write $\pi = 3.14 \pm 0.01$ or $\pi = 3.14(1)$ or $\pi \approx 3.14$. The first two expressions have the same meaning and illustrate the two distinct notations we use to express uncertainties on numerical results. The latter expression is used whenever the uncertainty is so small it can be ignored for all our practical purposes. We also use the $=$ sign when a numerical value (without uncertainty) is assigned to a variable for the sake of definiteness in our computations and plots, \eg when setting $\rho = 0.3~\GeV / \cm^3$ (the values we adopt for the other astrophysical constants are summarized in \Eq{speeds}).

One-particle momentum eigenstates are normalized according to
\beq\label{statenorm}
\braket{\bol{\genp{p}}'}{\bol{\genp{p}}} = \rho(\genp{p}) \, (2 \pi)^3 \delta^{(3)}(\bol{\genp{p}} - \bol{\genp{p}}') \ .
\eeq
In NR Quantum Mechanics, this normalization together with $\braket{\bol{x}}{\bol{y}} = \delta^{(3)}(\bol{x} - \bol{y})$ yields
\beq\label{NR free wave function}
\psi_{\bol{\genp{p}}}(\bol{x}) \equiv \braket{\bol{x}}{\bol{\genp{p}}} = \sqrt{\rho(\genp{p})} \, e^{i \bol{\genp{p}} \cdot \bol{x}}
\eeq
for the wave function of a plane wave, and
\begin{align}
\label{unit op}
\int \ud^3 x \, \ket{\bol{x}} \bra{\bol{x}} \ ,
&&&
\int \frac{\ud^3 \genp{p}}{(2 \pi)^3 \rho(\genp{p})} \, \ket{\bol{\genp{p}}} \bra{\bol{\genp{p}}} \ ,
\end{align}
for the unit operator. Since $\braket{\bol{\genp{p}}'}{\bol{\genp{p}}} = \braket{\bol{\genp{p}}}{\bol{\genp{p}}'}^*$, $\rho(\genp{p})$ must be real and can be interpreted as the number of particles per unit volume, \ie the particle number density ($| \psi_{\bol{\genp{p}}}(\bol{x}) |^2 = \rho(\genp{p})$). The standard normalization in NR Quantum Mechanics in an infinite volume is $\rho(\genp{p}) = 1 / (2 \pi)^3$, so that $\braket{\bol{\genp{p}}}{\bol{\genp{p}}'} = \delta^{(3)}(\bol{\genp{p}} - \bol{\genp{p}}')$ and $\psi_{\bol{\genp{p}}}(\bol{x}) = e^{i \bol{\genp{p}} \cdot \bol{x}} / (2 \pi)^{3/2}$. In relativistic theories it is instead convenient to choose
\beq\label{rhohere}
\rho(\genp{p}) = 2 E_\genp{p} \ ,
\eeq
with $E_\genp{p}$ defined in \Eq{particle energy}, so that the state normalization
\beq\label{statenormhere}
\braket{\bol{\genp{p}}}{\bol{\genp{p}}'} = 2 E_\genp{p} \, (2 \pi)^3 \delta^{(3)}(\bol{\genp{p}} - \bol{\genp{p}}')
\eeq
is Lorentz invariant. This is the normalization adopted in these notes, though on some occasions we will make the dependence on $\rho(\genp{p})$ explicit to show how certain quantities depend on the adopted normalization. In the NR expansion, detailed in \Sec{kinematics notation}, \Eq{statenormhere} reads at leading order
\beq\label{statenormhere NR}
\braket{\bol{\genp{p}}}{\bol{\genp{p}}'} \NReq 2 \genp{m} \, (2 \pi)^3 \delta^{(3)}(\bol{\genp{p}} - \bol{\genp{p}}') \ .
\eeq
States of definite angular momentum are normalized as
\beq\label{amom norm}
\braket{J', M'}{J, M} = \delta_{J J'} \delta_{M M'} \ ,
\eeq
and our notation for Clebsch-Gordan coefficients is $\CG{J_1, M_1}{J_2, M_2}{J_3, M_3}$.

$\ket{\DM^{(\prime)}}$ and $\ket{N^{(\prime)}}$ are shorthand notation for $\ket{\DM^{(\prime)}(\bol{p}^{(\prime)}, s^{(\prime)})}$ and $\ket{N(\bol{k}^{(\prime)}, r^{(\prime)})}$, respectively, with $s$ and $r$ ($s'$ and $r'$) the spin index of the incoming (outgoing) DM particle and nucleon, respectively. Operator matrix elements are understood to be evaluated at the origin, unless the position is indicated explicitly; for instance, considering a generic operator $\Op(\fvec{x})$, in our sloppy notation $\Nmatel{\Op}$ actually means $\matel{N(\bol{k}', r')}{\Op(\fvec{0})}{N(\bol{k}, r)}$. Similarly as above, we use $u^{(\prime)}_\chi$, $u^{(\prime)}_N$ as shorthand for spin-$1/2$ DM and nucleon Dirac spinors $u^{(\prime)}_\chi(\bol{p}^{(\prime)}, s^{(\prime)})$, $u_N(\bol{k}^{(\prime)}, r^{(\prime)})$, respectively (notice that $u^{(\prime)}_\chi$ refers to a DM particle with mass $\mDM^{(\prime)}$). Our normalization for Dirac spinors is
\beq\label{spinor normalization}
\bar{u}(\bol{\genp{p}}, \genp{s}') \gamma^0 u(\bol{\genp{p}}, \genp{s}) = \bar{v}(\bol{\genp{p}}, \genp{s}') \gamma^0 v(\bol{\genp{p}}, \genp{s}) = 2 E_\genp{p} \, \delta_{\genp{s} \genp{s}'} \ ,
\eeq
which implies
\beq
\bar{u}(\bol{\genp{p}}, \genp{s}') u(\bol{\genp{p}}, \genp{s}) = - \bar{v}(\bol{\genp{p}}, \genp{s}') v(\bol{\genp{p}}, \genp{s}) = 2 \genp{m} \, \delta_{\genp{s} \genp{s}'} \ .
\eeq

\section{Rate}
\label{Rate}
We start this Section by establishing why Earth-borne nuclei are effective targets for scattering of galactic DM particles, and what recoil energies direct DM searches need being sensitive to for detection to occur. We then proceed by deriving the scattering rate and the detection rate, whose ingredients will be studied in greater detail in the next Sections. We will then resume our discussion in \Sec{Pheno} for an in-depth analysis of the properties of the rate.

\subsection{Basics}
\label{Rate preliminaries}
\begin{figure}[t]
\begin{center}
\includegraphics[width=.65\textwidth]{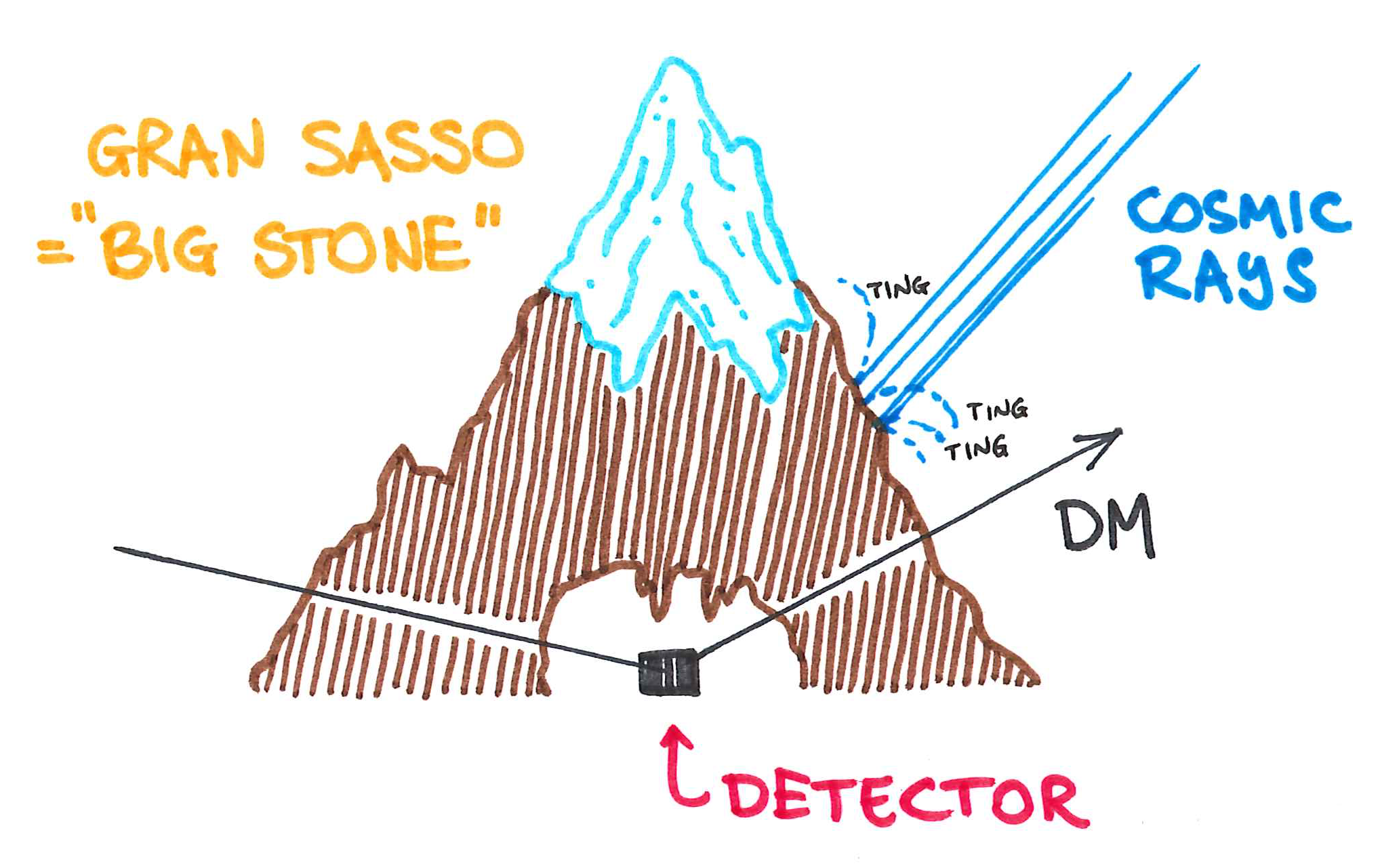}
\hfill
\raisebox{5mm}{
\begin{overpic}[width=.32\textwidth]{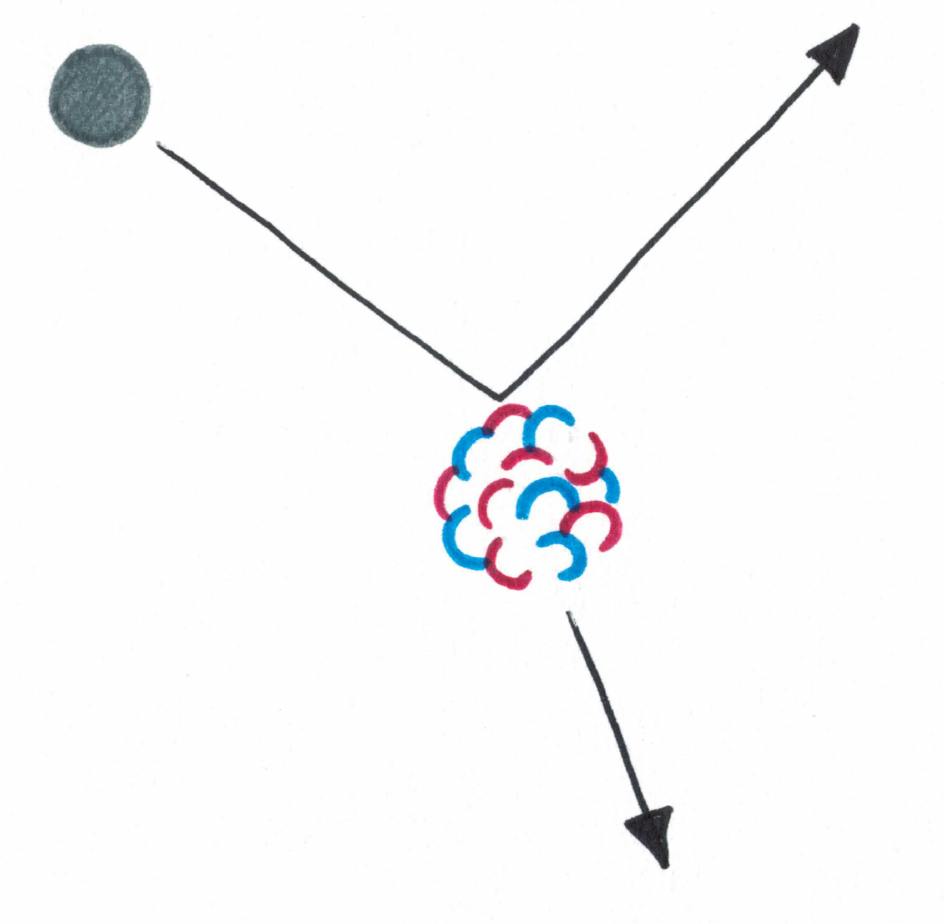}
\put (20, 85){DM}
\put (18, 43){nucleus}
\put (38, 70){$\bol{v}$}
\put (68, 25){$\ER$}
\end{overpic}
}
\caption[Basics of direct DM detection]{\label{fig: basics}\emph{Basics of direct DM detection. \textbf{Left:} on a \figmath{\sim \km} scale. \textbf{Right:} on a \figmath{\sim \fm} scale.}}
\end{center}
\end{figure}

Direct detection experiments attempt at measuring the energy released in the detector by DM particles scattering off detector nuclei. Brutally speaking, and from a theoretician viewpoint only, a detector may be thought of as a chunk of material covered with sensors. When a DM particle with velocity $\bol{v}$ reaches the detector, it may undergo scattering off a nucleus in the material. The nucleus is initially at rest, to a very good approximation, and it recoils with a recoil energy $\ER$ when it is struck by the DM particle. The detector is placed in a cave deep underground to be screened by cosmic rays, and further artificial shields are added to reduce the background due to natural radioactivity. \Fig{fig: basics} elucidates these concepts in the clearest possible manner.

This sort of DM searches, like others, relies on the assumption that the DM interacts with the standard matter other than just gravitationally. Unfortunately, we have no evidence that this is actually the case. We will just close our eyes here and assume that the DM couples to the standard matter beyond gravity. For large enough couplings, DM particles may scatter already in Earth's atmosphere or in the rock overburden before getting to the detector. This type of DM is called Strongly Interacting Massive Particle or SIMP, and was studied \eg in Ref.~\cite{Starkman:1990nj}; existing bounds on this candidate can be found summarized in Refs.~\cite{Taoso:2007qk, Davis:2017noy, Kavanagh:2017cru, Digman:2019wdm}. We will be interested instead in particles with weak-scale interactions, the so-called Weakly Interacting Massive Particles (WIMPs), with much smaller couplings to the standard matter. In this case the probability that a DM particle interacts multiple times inside the detector, or that it interacts other than gravitationally with the Sun or the planets (including Earth) in the Solar System before reaching the detector~\cite{Peter:2009mi, Peter:2009mm}, is negligible and can be safely ignored. In the same way we can neglect the small modifications that DM scattering within these bodies causes in the local (meaning at Earth's location) DM density and velocity distribution, although the gravitational effect of the Sun on the DM distribution can in principle be observable~\cite{Griest:1987vc, Sikivie:2002bj, Alenazi:2006wu, Patla:2013vza, Lee:2013wza, Bozorgnia:2014dqa} (see \Sec{velocity}).

How do we know that the DM interacts with a whole nucleus, and not just with part of it (or maybe with a whole atom)? As we will see in Secs.~\ref{Scattering amplitude},~\ref{nuc form factors}, the $2 \to 2$ scattering amplitude between a DM particle and a (generic) target $T$ has the form
\beq\label{support}
\Mel \sim \int \ud^3 x \, \varrho(\bol{x}) \, e^{i \bol{q} \cdot \bol{x}} \ ,
\eeq
with $\bol{q}$ the three-momentum transferred by the DM to the target during the scattering process. Since nuclei are initially at rest in the detector's rest frame, the final nuclear momentum equals $q$ and the nuclear recoil energy is $\ER = q^2 / 2 m_T$. The exponential originates from the product of the initial and final DM wave functions. The function $\varrho(\bol{x})$ is related to the internal structure of the target, and reflects its spatial extension. For a nuclear target, it may be related to a nucleon-specific nuclear density (\eg a number density, or a spin density, of either protons or neutrons), but it may also be a $\bol{q}$-dependent quantity if DM-nucleon interactions depend on momentum transfer. Denoting with $R$ the size of the target, we see that the integral in \Eq{support} gets suppressed for values of momentum transfer such that $q R \gg 1$, as the integrand gets averaged to zero by the rapid oscillations of the exponential. In other words, the interaction is coherent across distances of order $1 / q$ (or smaller). For $q = 0$ the coherence is complete, the scattering does not probe the internal structure of the target at all and the target behaves effectively as a point-like particle (in fact, taking $q = 0$ is indistinguishable from substituting $\varrho(\bol{x}) \propto \delta^{(3)}(\bol{x})$). To disentangle the $\bol{q}$ dependence of $\varrho(\bol{x})$ from that of $e^{i \bol{q} \cdot \bol{x}}$, we refer to the limit of point-like target when the exponential is neglected. The $q \to 0$ limit is known as \emph{long-wavelength limit} in the nuclear theory of electron-nucleus scattering, where $q$ is the momentum of the single particle (usually the photon) mediating the scattering in the tree-level approximation. Notice however that referring to $\sim 1 / q$ as a wavelength can only be meaningful in the context of a one-particle exchange approximation: in fact, no one intermediate particle is required to have momentum $q$ in a loop diagram.

\begin{figure}[t]
\begin{center}
\includegraphics[width=.49\textwidth]{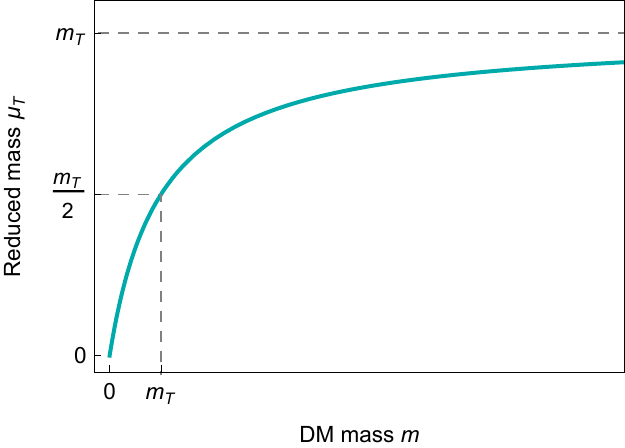}
\includegraphics[width=.49\textwidth]{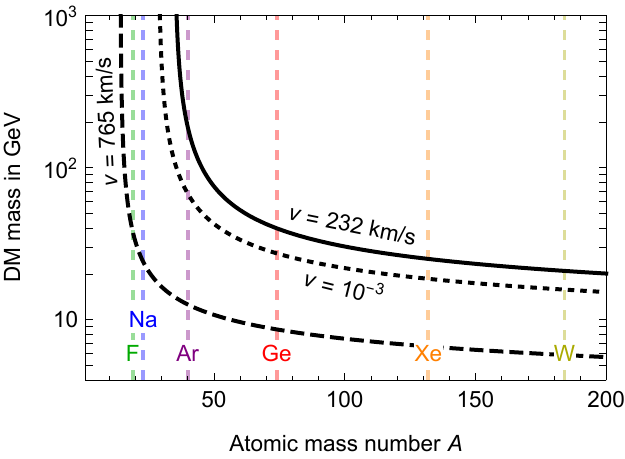}
\caption[Reduced mass. $\mDM$--$A$ relation derived from $q R = 1$]{\label{fig: reduced mass + qR=1}\emph{\textbf{Left:} linear plot of the reduced mass \figmath{\mu_T}, for fixed \figmath{m_T}, as a function of \figmath{\mDM}. \textbf{Right:} relation between the DM mass and the mass number \figmath{A} derived from \figmath{q R = 1} (see \Eq{Max coherent DM mass}), for three values of \figmath{v}: the indicative value \figmath{10^{-3}} in speed of light units, the typical DM speed \figmath{232~\km / \sd}, and the time-averaged maximum DM speed \figmath{765~\km / \sd} (see \Sec{velocity} and Eqs.~\eqref{speeds},~\eqref{vmax} below). For a given value of \figmath{A}, DM particles with masses well below the black line scatter coherently with the entire nucleus in the whole kinematically accessible range of momentum transfer. The mass number for a representative set of nuclides, \figmath{^{19}\text{F}}, \figmath{^{23}\text{Na}}, \figmath{^{40}\text{Ar}}, \figmath{^{74}\text{Ge}}, \figmath{^{132}\text{Xe}}, and \figmath{^{184}\text{W}} (the most abundant isotopes of the corresponding elements, see \Tab{tab: nuclides}), is indicated by the vertical dashed lines.}}
\figcode
\end{center}
\end{figure}

As we will see in \Sec{scattering kinematics} (see in particular \Eq{ERinterval}), the momentum transfer is $q \leqslant 2 \mu_T v$ for an elastic scattering, with $\mu_T$ the DM-target reduced mass and $v$ the initial DM speed. Complete coherence is thus ensured across the whole kinematically accessible range of momentum transfer if
\beq\label{mu_T max}
\mu_T \ll \frac{1}{2 v R} \ .
\eeq
For galactic DM, \ie DM particles that are gravitationally bound to the halo of our galaxy, the DM speed at the location of Earth in the detector's rest frame is expected to be a few hundred $\km / \sd$, $v \sim 10^{-3}$ in speed of light units (see \Sec{DM velocity distribution}). As a consequence, since $0.2~\GeV~\fm \approx 1$ in natural units (see~\Eq{units}), a target as large as a few $\fm$ and with a mass such that $\mu_T \ll 100~\GeV$ would guarantee that the scattering is at least partially coherent. For this reason atomic nuclei, which have masses no larger than few hundred $\GeV$ and sizes no larger than few $\fm$, are a good target to search for DM through the matrix element in \Eq{support} (notice that $\mu_T < \mDM, m_T$, as shown in the left panel of \Fig{fig: reduced mass + qR=1} below and discussed in more detail in \Sec{kinematics notation}). Crucially, experiments could be developed that are at least partially sensitive to the recoil energies produced by halo DM particles scattering off nuclear targets (see \eg the right panel of \Fig{fig: TypicalER} below). Nuclei are thus effective targets as their scattering with DM particles yields $\ER$ values that are both large enough for detection and small enough for the scattering to be at least partially coherent, so that the signal is not overly suppressed.

From now on we will exclusively consider nuclear targets. Notice that only nuclear elements or compounds satisfying certain technical requirements related to the experimental design can be employed in direct DM searches. Therefore, not all nuclei constitute good targets. A selection of nuclides of interest for direct DM detection experiments is reported in \Tab{tab: nuclides}, which also details some of their properties (atomic number, mass number, mass, isotopic abundance, spin, and magnetic dipole moment).

\begin{table}[t!]
\footnotesize
\begin{center}
\begin{tabular}{|>{\rule[-1mm]{0mm}{4.5mm}} c | c | c | c | c | c | c | c |}
\hline
Element & Symbol & $Z$ & $A$ & Mass ($\GeV$) & Abundance & Spin & Magnetic moment ($\hat{\mu}_\text{N}$)
\\
\hline
\raisebox{-0.4mm}{\multirow{2}{*}{carbon}} & \raisebox{-0.4mm}{\multirow{2}{*}{C}} & \raisebox{-0.4mm}{\multirow{2}{*}{$6$}} & $12$ & $11$ & $99 \%$ & $0$ & $0$
\\
& & & $13$ & $12$ & $1 \%$ & $1/2$ & $+0.70$
\\
\hline
oxygen & O & $8$ & $16$ & $15$ & $100 \%$ & $0$ & $0$
\\
\hline
fluorine & F & $9$ & $19$ & $18$ & $100 \%$ & $1/2$ & $+2.63$
\\
\hline
\raisebox{-0.4mm}{\multirow{2}{*}{neon}} & \raisebox{-0.4mm}{\multirow{2}{*}{Ne}} & \raisebox{-0.4mm}{\multirow{2}{*}{$10$}} & $20$ & $19$ & $90 \%$ & $0$ & $0$
\\
& & & $22$ & $20$ & $9.2 \%$ & $0$ & $0$
\\
\hline
sodium & Na & $11$ & $23$ & $21$ & $100 \%$ & $3/2$ & $+2.22$
\\
\hline
aluminium & Al & $13$ & $27$ & $25$ & $100 \%$ & $5/2$ & $+3.64$
\\
\hline
\raisebox{-0.4mm}{\multirow{3}{*}{silicon}} & \raisebox{-0.4mm}{\multirow{3}{*}{Si}} & \raisebox{-0.4mm}{\multirow{3}{*}{$14$}} & $28$ & $26$ & $92 \%$ & $0$ & $0$
\\
& & & $29$ & $27$ & $4.7 \%$ & $1/2$ & $-0.55$
\\
& & & $30$ & $28$ & $3.1 \%$ & $0$ & $0$
\\
\hline
argon & Ar & $18$ & $40$ & $37$ & $100 \%$ & $0$ & $0$
\\
\hline
\raisebox{-0.4mm}{\multirow{2}{*}{calcium}} & \raisebox{-0.4mm}{\multirow{2}{*}{Ca}} & \raisebox{-0.4mm}{\multirow{2}{*}{$20$}} & $40$ & $37$ & $97 \%$ & $0$ & $0$
\\
& & & $44$ & $41$ & $2.1 \%$ & $0$ & $0$
\\
\hline
\raisebox{-0.4mm}{\multirow{5}{*}{germanium}} & \raisebox{-0.4mm}{\multirow{5}{*}{Ge}} & \raisebox{-0.4mm}{\multirow{5}{*}{$32$}} & $70$ & $65$ & $20 \%$ & $0$ & $0$
\\
& & & $72$ & $67$ & $27 \%$ & $0$ & $0$
\\
& & & $73$ & $68$ & $7.8 \%$ & $9/2$ & $-0.88$
\\
& & & $74$ & $69$ & $37 \%$ & $0$ & $0$
\\
& & & $76$ & $71$ & $7.8 \%$ & $0$ & $0$
\\
\hline
iodine & I & $53$ & $127$ & $118$ & $100 \%$ & $5/2$ & $+2.81$
\\
\hline
\raisebox{-0.4mm}{\multirow{7}{*}{xenon}} & \raisebox{-0.4mm}{\multirow{7}{*}{Xe}} & \raisebox{-0.4mm}{\multirow{7}{*}{$54$}} & $128$ & $119$ & $1.9 \%$ & $0$ & $0$
\\
& & & $129$ & $120$ & $26 \%$ & $1/2$ & $-0.78$
\\
& & & $130$ & $121$ & $4.1 \%$ & $0$ & $0$
\\
& & & $131$ & $122$ & $21 \%$ & $3/2$ & $+0.69$
\\
& & & $132$ & $123$ & $27 \%$ & $0$ & $0$
\\
& & & $134$ & $125$ & $10 \%$ & $0$ & $0$
\\
& & & $136$ & $126$ & $8.9 \%$ & $0$ & $0$
\\
\hline
cesium & Cs & $55$ & $133$ & $124$ & $100 \%$ & $7/2$ & $+2.58$
\\
\hline
\raisebox{-0.4mm}{\multirow{4}{*}{tungsten}} & \raisebox{-0.4mm}{\multirow{4}{*}{W}} & \raisebox{-0.4mm}{\multirow{4}{*}{$74$}} & $182$ & $169$ & $27 \%$ & $0$ & $0$
\\
& & & $183$ & $170$ & $14 \%$ & $1/2$ & $+0.12$
\\
& & & $184$ & $171$ & $31 \%$ & $0$ & $0$
\\
& & & $186$ & $173$ & $28 \%$ & $0$ & $0$
\\
\hline
\end{tabular}
\caption[Significant stable nuclides and their properties]{\label{tab: nuclides}\emph{Properties of the main observationally stable isotopes of nuclear elements of interest to direct DM detection experiments: atomic number \figmath{Z}, mass number \figmath{A}, nuclear mass \figmath{m_T} in \figmath{\GeV}, natural isotopic fractional abundance \figmath{\tilde{\xi}_T}, spin \figmath{\amom{J}}, and magnetic dipole moment in units of the nuclear magneton \figmath{\hat{\mu}_\text{N}}, see \Eq{nuclear magneton}. Data from Refs.~\cite{Stone, Mathematica IsotopeData}; other references for nuclear properties are \eg Refs.~\cite{PDGnuclei, NIST, NNDC}.}}
\end{center}
\end{table}

One may be tempted to approximate $q \sim p$, with $p = \mDM v$ the initial DM momentum, so that $1 / q$ corresponds to the de Broglie wavelength of the incoming DM particle (divided by $2 \pi$). However, this can be a very poor approximation, even as an order of magnitude estimate. Let us consider for instance a DM particle with mass $\mDM = 10~\TeV$ scattering off a sodium nucleus, $m_T \approx 20~\GeV$ (see \Tab{tab: nuclides}). Then the minimum value of $1 / q$ is $\Ord(1)~\fm$ while $1 / p = \Ord(10^{-2})~\fm$. The scattering may well occur with the whole nucleus, contrary to what the result of approximating $q \sim p$ would suggest. A better approximation could be $q \sim \mu_T \bar{v}$, that is, taking the scale of $q$ to be that of its maximum value in an elastic scattering (see \Eq{ERinterval}), with $\bar{v} = \Ord(10^{-3})$ the typical DM speed. This estimate may stand if we already knew that the momentum transfer distribution (\ie the scattering rate, see \Sec{Scattering rate}) is approximately constant in the energy range probed by the experiment (if not peaked at $\mu_T \bar{v}$ for some reason), as it is \eg for the SI interaction for sufficiently heavy DM and sufficiently light target nuclei (see \Fig{fig: SI spectra} below). However, this may not be true in other situations, \eg if the experimental sensitivity reaches down to $q$ values much smaller than $\mu_T \bar{v}$, and at the same time the momentum transfer distribution (\ie the recoil spectrum) is peaked at low energies. This may happen with either a light enough DM particle, so that experiments can only probe the high-speed tail of the DM velocity distribution, or with a light or massless mediator exchanged in the $t$ channel, whose propagator makes the cross section increase significantly at low energies (see \Sec{Pheno} and in particular \Fig{fig: spectra}). In either case, the scattering events detected by an experiment would have $q$ values very close to the lowest end of its sensitivity region, whereas $\mu_T \bar{v}$ has nothing to do with the experimental sensitivity and thus it is not expected to necessarily provide a good approximation to $q$ in this case. To be more concrete, the sensitivity of a typical xenon detector currently extends down to recoil energies of $\Ord(1)~\keV$, leading to a typical value of $q = \sqrt{2 m_T \ER} = \Ord(10)~\MeV$ for a Xe nucleus (see \Tab{tab: nuclides}) if the spectrum decreases steeply. On the other hand, for $\mDM \gg m_T$, we get $\mu_T \bar{v} \simeq m_T \bar{v} = \Ord(100)~\MeV$, off by about one order of magnitude (\ie two orders of magnitude in $\ER$).

We can characterize the nuclear radius as $R = 1.2 A^{1/3}~\fm$~\cite{Krane:1987ky, Walecka, Povh} (with $A^{1/3}$ lying between $2$ and $6$ for the nuclei in \Tab{tab: nuclides}), whereas the harmonic oscillator parameter $b \equiv 1 / \sqrt{\mN \omega}$ with $\omega \equiv (45 A^{- 1/3} - 25 A^{- 2/3})~\MeV$ (see \eg Refs.~\cite{Blomqvist:1968zz, Kirson:2007rmw}) is also sometimes taken as a measure of the nuclear size. We can also approximate the nuclear mass as $m_T \simeq A~\amu$, with $\amu \approx 931~\MeV$ the unified atomic mass unit (numerically similar to the nucleon mass $\mN \approx 939~\MeV$). Using \Eq{inverse mu_T} below, we then get that the maximum DM mass saturating \Eq{mu_T max} (\ie yielding an equality) is
\begin{align}
\label{Max coherent DM mass}
\frac{m_T}{2 v R \, m_T - 1} \approx \frac{A}{12.18 \, v A^{4/3} - \GeV / \amu}~\GeV &&& \text{for}~A > \left( \frac{\GeV / \amu}{12.18 \, v} \right)^{3/4} \approx 28.8 \left( \frac{v}{10^{-3}} \right)^{- 3/4} \ ,
\end{align}
while no positive value of the DM mass saturates \Eq{mu_T max} for smaller mass numbers. This result, which is a refinement of that\footnote{In Ref.~\cite{Engel:1992bf} the approximations $\amu \approx 1~\GeV$, $0.2~\GeV~\fm \approx 1$, and $v \approx 10^{-3}$ are used, which imply a slightly different version of \Eq{Max coherent DM mass}, with the corresponding condition on $A$ reading $A \geqslant 28$. Notice also that the $+$ sign in the denominator in the formula in Ref.~\cite{Engel:1992bf} is a typo.} in Ref.~\cite{Engel:1992bf}, is illustrated in the right panel of \Fig{fig: reduced mass + qR=1} for three values of $v$: the indicative value $10^{-3}$ in speed of light units, the typical DM speed $232~\km / \sd$, and the time-averaged maximum DM speed $765~\km / \sd$ (see \Sec{velocity} and Eqs.~\eqref{speeds},~\eqref{vmax} below). For reference, the DM mass saturating \Eq{mu_T max} for $^{131}$Xe, one of the heaviest nuclides employed in direct detection, is about $19~\GeV$ for $v = 10^{-3}$.

The typical nuclear recoil energy induced by elastic scattering with a halo DM particle can be determined from the `average' $q^2$ value $2 \mu_T^2 v^2$, see \Eq{ERinterval} below (notice that this may not be representative of the values relevant to specific models, as discussed above). This is shown in \Fig{fig: TypicalER} as a function of the DM mass $\mDM$, for a representative sample of nuclides employed in direct detection experiments and a typical DM speed of $232~\km / \sd$ at Earth's location. This sets the sensitivity ballpark for experiments to be able to detect DM-nucleus scattering by looking at nuclear recoils. While the information on $\ER$ (right panel of \Fig{fig: TypicalER}) can be more easily compared with the sensitivity windows of the experiments, the information on $q$ (left panel) can be more immediately related to the target size and consequent form factors entering the scattering cross section, as well as to the mass of light interaction mediators whose propagator can endow the scattering cross section with a specific energy dependence (see Secs.~\ref{Vector-mediated interaction},~\ref{Scalar-mediated interaction},~\ref{Pheno} for some examples). The $\ER$ ($q$) curves in the right (left) panel of \Fig{fig: TypicalER} shift upwards as $v^2$ (as $v$) for DM speeds larger than $232~\km / \sd$, up to a factor of about $11$ (about $3.3$) for DM particles with speed $765~\km / \sd$ (see \Eq{vmax}). Notice however that the DM speed distribution is thought to drop quickly at these high speeds, and very few (or no) particles with speeds close to the maximum value are expected. If one considers the maximum value of $\ER$ ($q$) kinematically allowed for an elastic scattering, instead of the typical values displayed in \Fig{fig: TypicalER}, the curves shift upwards by another factor of $2$ ($\sqrt{2}$). All values of $\ER$ (and $q$) are kinematically allowed below these lines. As one can see, DM particles with mass above few $\GeV$ can be in principle detected, at least those with the highest speeds, if the experiments are sensitive to nuclear recoil energies of about $1~\keV$ (see also \Fig{fig: minDMmass} below). Heavier DM particles can yield larger $\ER$, while lighter DM can only be detected extending the experimental sensitivity to recoil energies below $1~\keV$. One can also notice that the $q$ dependence of the propagator in a tree-level $t$-channel scattering can be safely neglected if the interaction mediator is much heavier than few $\GeV$, while it may be needed taking into proper account otherwise, depending on the DM and target masses. The dashed horizontal lines in \Fig{fig: TypicalER} mark the $q$ and corresponding $\ER$ values where $q R = 1$, for each considered nuclide. Their position is indicative of the order of magnitude above which the DM-nucleus scattering amplitude gets highly suppressed by the loss of coherence.

\begin{figure}[t]
\begin{center}
\includegraphics[width=.49\textwidth]{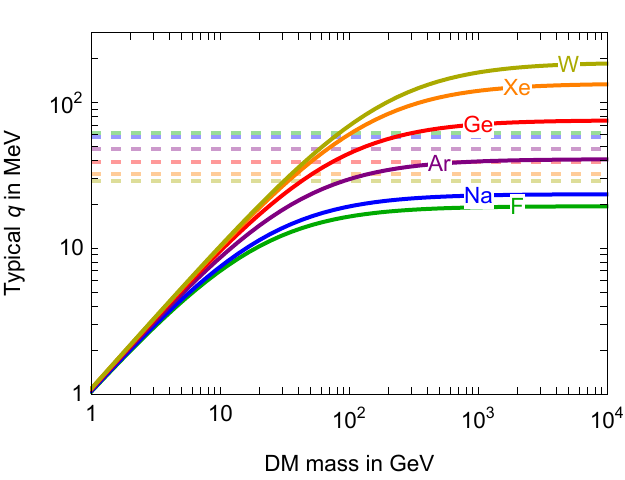}
\includegraphics[width=.49\textwidth]{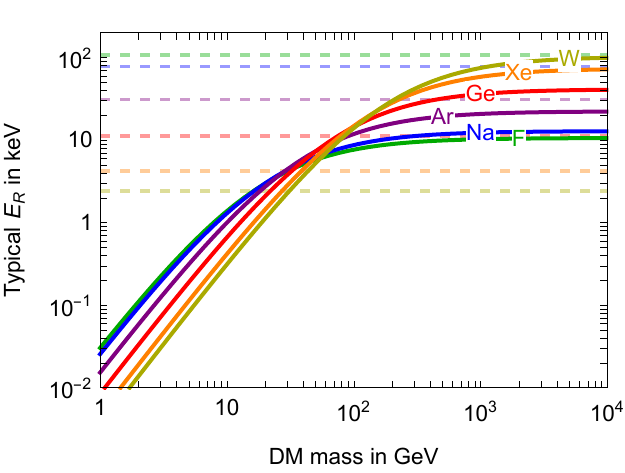}
\caption[Typical momentum transfer and corresponding recoil energy]{\label{fig: TypicalER}\emph{Typical momentum transfer \figmath{q = \sqrt{2} \mu_T v} (\textbf{left}) and corresponding recoil energy \figmath{\ER = q^2 / 2 m_T} (\textbf{right}) as functions of the DM mass for different target elements used in direct detection experiments. The DM speed has been fixed to the typical value \figmath{232~\km / \sd} (see \Sec{DM velocity distribution} and in particular \Eq{speeds}). The \figmath{q} (\figmath{\ER}) curves scale as \figmath{v} (\figmath{v^2}), thus shifting upwards by a factor of about \figmath{3.3} (about \figmath{11}) for halo DM particles with the time-averaged maximum DM speed \figmath{765~\km / \sd}, see \Eq{vmax}. The maximum kinematically allowed \figmath{q} (\figmath{\ER}) at given \figmath{v} in an elastic scattering, which is used in the right panel of \Fig{fig: reduced mass + qR=1}, is \figmath{\sqrt{2}} (\figmath{2}) times the typical \figmath{q} (\figmath{\ER}) shown by the curves. For each element, the most abundant isotope (same as in the right panel of \Fig{fig: reduced mass + qR=1}) has been chosen as representative (the spread in the curve due to considering other isotopes is much smaller than that due to taking into account the whole DM speed distribution). The dashed horizontal lines mark the value of momentum transfer for which \figmath{q R = 1} (\textbf{left}), and corresponding \figmath{\ER} (\textbf{right}), for each considered nuclide (this value decreases with increasing \figmath{A}).}}
\figcode
\end{center}
\end{figure}

\subsection{Scattering rate}
\label{Scattering rate}
The scattering cross section $\sigma_T$, for a DM particle traveling with velocity $\bol{v}$ in the target rest frame, is defined by
\beq\label{event number}
\frac{\ud N_T}{\ud V \ud t} \equiv n_\text{T} n \, v \, \sigma_T(\bol{v}) \ ,
\eeq
where $N_T$ is the number of scattering processes and $n_T$ and $n$ are the number densities of the target $T$ and the DM, respectively. Generally one assumes that the cross section only depends on the DM speed $v$ and not on the whole velocity vector $\bol{v}$. This is always true if both the DM flux and the nuclei in the detector are unpolarized, since $\sigma_T$ is invariant under rotations; were the DM polarized along a direction $\hat{\bol{n}}$, for instance, $\sigma_T$ could depend on both $v$ and $\bol{v} \cdot \hat{\bol{n}}$.

Dividing \Eq{event number} by the detector mass density $\ud M / \ud V$ we get the scattering rate per unit detector mass (henceforth simply `the rate')
\beq\label{rate}
R_T \equiv \frac{\ud N_T}{\ud t \, \ud M} = \frac{\ud \mathcal{N}_T}{\ud M} n \, v \, \sigma_T(\bol{v}) \ ,
\eeq
where $\mathcal{N}_T$ is the number of targets. If the detector is composed of nuclei of the same species with mass $m_T$ (\ie a single nuclide), the number of targets per unit detector mass is simply $\ud \mathcal{N}_T / \ud M = 1 / m_T$. Since $m_T \simeq A~\amu$, with $A$ the mass number and $\amu \equiv \frac{1}{12} m_{^{12}\text{C}} \approx 0.931~\GeV$ the unified atomic mass unit, related to the Avogadro constant $N_\text{A} \approx 6.022 \times 10^{23}~\text{mol}^{-1}$ by $\amu = M_\amu / N_\text{A}$ with $M_\amu = 1$~g/mol the molar mass constant, some authors approximate $1 / m_T \approx N_\text{A} / A M_\amu$. For a compound detector (different isotopes and/or different elements), denoting with $\xi_T$ the numerical abundance of a target nuclide $T$ in the detector, an amount of detector substance with mass $\sum_T \xi_T m_T$ contains $\xi_T$ nuclei of species $T$. We have therefore
\beq
\frac{\ud \mathcal{N}_T}{\ud M} = \frac{\xi_T}{\sum_{T'} \xi_{T'} m_{T'}} = \frac{\zeta_T}{m_T} \ ,
\eeq
where
\beq\label{zeta_T}
\zeta_T = \frac{\xi_T m_T}{\sum_{T'} \xi_{T'} m_{T'}}
\eeq
is the target mass fraction. $R_T$ in \Eq{rate} denotes now the rate of DM scattering with nuclei of the specific nuclear species $T$. Denoting isotopic abundances with $\tilde{\xi}_T$ (see \Tab{tab: nuclides} for a list), for a compound such as X$_x$Y$_y \dots$Z$_z$ (\eg C$_3$F$_8$, or CaWO$_4$) we have
\beq
\xi_{T_\text{X}} = \tilde{\xi}_{T_\text{X}} \frac{x}{x + y + \dots + z}
\eeq
for each isotope $T_\text{X}$ of X: for instance, $\xi_{^{186}\text{W}} \approx 4.7 \%$ and $\xi_{^{16}\text{O}} \approx 67 \%$ in a CaWO$_4$ detector. $\zeta_T / m_T$ can be converted from $\GeV^{-1}$ to $\kg^{-1}$ by using \Eq{units}.

Since the DM particles are not monochromatic in velocity, the DM density $n$ in \Eq{rate} must be substituted with its differential
\beq
\ud n = \bar{n} \fE(\bol{v}, t) \, \ud^3 v \ ,
\eeq
where $\fE(\bol{v}, t)$ is the DM velocity distribution at Earth's location in the detector's rest frame (see \Sec{velocity} for more details), normalized so that
\beq\label{f norm E}
\int \ud^3 v \, \fE(\bol{v}, t) = 1 \ ,
\eeq
and $\bar{n}$ is the DM number density at Earth's location. The rate for DM scattering off a specific target $T$ in the detector reads then
\beq\label{rate 2}
\ud R_T(\bol{v}, t) = \frac{\zeta_T}{m_T} \bar{n} \fE(\bol{v}, t) \, v \, \sigma_T(\bol{v}) \, \ud^3 v \ .
\eeq
The DM distribution is not expected to change significantly over the timescale of an experiment (years). The time dependence in $\fE(\bol{v}, t)$ is primarily due to Earth's revolution around the Sun, and causes the scattering rate to be annually modulated. A more thorough discussion on the DM velocity distribution and the rate annual modulation is postponed to \Sec{velocity}.

If the DM is composed by only one type of particle with mass $\mDM$, as we will assume throughout these notes, we can write $\bar{n} = \rho / \mDM$ with $\rho$ the local DM mass density, while if the DM has several components $\bar{n}$ must be scaled accordingly. $\rho$ is mainly determined from the study of the vertical kinematics of stars near the Sun, or is extrapolated from stellar rotation curves (see \eg Ref.~\cite{Read:2014qva} for a review). The recent determinations have best-fit values in the range $\rho = 0.2$ -- $0.6~\GeV / \cm^3$, with uncertainties lying indicatively in the $0.05$ -- $0.5~\GeV / \cm^3$ range (see \eg Refs.~\cite{Read:2014qva, Green:2017odb}).
\beq\label{rho}
\rho = 0.3~\GeV / \cm^3 \approx 8 \times 10^{-3}~M_\odot / \pc^3 \approx 5 \times 10^{-25}~\gr / \cm^3
\eeq
(see \Eq{units}) is historically the reference value adopted in the direct detection literature, although sometimes the value $\rho = 0.4~\GeV / \cm^3$ is preferred. Notice from \Eq{rate 2} that the DM density is completely degenerate with the overall size of the cross section, thus a precise measurement of $\rho$ is necessary in order to infer the actual value of the scattering cross section from data in case of detection. It is therefore important to keep in mind that astrophysical data only allow to determine an average value of $\rho$ over a few hundred parsecs. The presence of unresolved subhalos on smaller scales would imply that the actual DM density at Earth's position (\ie the quantity entering \Eq{rate 2}) may be significantly larger with respect to the average value, if we were sitting inside one of them. However the likelihood of this happening is very small, while there is a higher chance that we lie in the smooth component of the DM distribution, where the density is actually slightly smaller than the average value due to some of the DM being accumulated in substructures~\cite{Kamionkowski:2008vw}.

Fixing a value for $\rho$ implies that, assuming only one kind of DM particles, the larger the DM mass $\mDM$ the lower their numerical density close to Earth. Therefore, heavy enough DM particles may never cross a detector for the entire time of its operations. We can be more quantitative by considering the DM average differential flux, $(\rho / \mDM) \, \bar{v}$ with $\bar{v} \equiv \int v \fE \, \ud^3 v = \Ord(10^{-3}) = \Ord(300)~\km / \sd$ the average DM speed. Assuming $\rho = 0.3~\GeV / \cm^3$ we obtain for the DM flux $10^7 (\mDM / \GeV)^{-1}~\cm^{-2}~\sd^{-1}$. This means that considering a detector with linear size of order $10~\cm$ and a data-taking period of $10~\yr$ we can expect (on average) less than $1$ DM particle crossing the detector for DM heavier than roughly $10^{17}~\GeV$ (see \eg Refs.~\cite{Jacobs:2014yca, Kavanagh:2017cru} for analogous computations).

Apart from the rate of scattering events occurred in their detectors, direct detection experiments try to measure the energy $\ER$ of the recoiling nucleus in each event. For a fixed DM speed $v$ in the detector's rest frame, and therefore a fixed amount of kinetic energy in the DM-nucleus system, there is a maximum $\ER$ the scattering can yield, call it $\ER^\text{max}(v)$. As we will see more quantitatively in \Sec{scattering kinematics}, this maximum energy transfer occurs when the DM particle bounces backwards in the center of momentum (CM) frame of the system, \ie when the scattering angle is $\pi$. On the contrary, the minimum energy exchange occurs when the DM particle keeps traveling in the same directions it came from, with zero scattering angle. Intermediate energy exchanges occur at intermediate scattering angles. All DM particles with speed larger than our fixed value $v$ can therefore cause the nucleus to recoil with energy equal to $\ER^\text{max}(v)$. Inversely, for a fixed value of $\ER$, there is a minimum speed $\vmin(\ER)$ a DM particle must have in order to be able to transfer an energy $\ER$ to the nucleus. The actual form of the $\vmin(\ER)$ function depends on the scattering kinematics (we will compute it for elastic and inelastic $2 \to 2$ scattering in \Sec{scattering kinematics}). The differential scattering rate for target nuclei recoiling with energy $\ER$ is then
\beq\label{diffrate}
\frac{\ud R_T}{\ud \ER}(\ER, t) = \frac{\rho}{\mDM} \frac{\zeta_T}{m_T} \int_{v \geqslant \vmin(\ER)} \ud^3 v \, \fE(\bol{v}, t) \, v \, \frac{\ud \sigma_T}{\ud \ER}(\ER, \bol{v}) \ ,
\eeq
where $\ud \sigma_T / \ud \ER$ is the differential scattering cross section with respect to $\ER$. $\ud R_T / \ud \ER$ is usually expressed in $\cpd / (\kg~\keV) = 1 / (\dd~\kg~\keV)$, with `$\cpd$' short for `counts per day'.\footnote{Expressing $\rho / \mDM$ in $\cm^{-3}$, $\zeta_T / m_T$ in $\kg^{-1}$, and $\ud \sigma_T / \ud \ER$ in $\keV^{-3}$, the differential rate is automatically given numerically in $\cpd / (\kg~\keV)$ due to the approximate equality $1 / (\cm^3~\keV^2) \approx 1.01 / \dd$ (see \Eq{units}).}

In the standard assumption that both DM particles and target nuclei are unpolarized, the differential cross section only depends on $\bol{v}$ through its absolute value $v$. In the NR expansion of the scattering amplitude in powers of $\bol{v}$, to be discussed in \Sec{DM-N} below, the unpolarized differential cross section can be written as
\beq\label{Taylor cross section}
\frac{\ud \sigma_T}{\ud \ER}(\ER, \bol{v}) \NReq \frac{1}{v^2} \sum_{n = 0} g_n^T(\ER) \, v^{2 n} \ ,
\eeq
where the $v^{2 n}$ factors come from the expansion of the squared scattering amplitude, while the $1 / v^2$ factor comes about when deriving $\ud \sigma_T / \ud \ER$ from $\ud \sigma_T / \ud \cos\theta$, see \Sec{Differential cross section} (for the simple case of elastic scattering, $\ud \ER \propto v^2 \, \ud \cos\theta$). Only one term is often relevant in the sum, as for the SI and SD interactions discussed in \Sec{SI interaction} and \Sec{SD interaction}, but there are also cases where two terms contribute at the same order of the NR expansion, see \eg \Sec{Magnetic-dipole DM}, or even cases where truncating the expansion at leading order may not provide a good approximation (see discussion in \Sec{Non-relativistic expansion}). The lack of odd powers of $\bol{v}$ in the expansion of the spin-summed squared matrix element can be understood as a consequence of not having available three-vectors to form rotational invariants with $\bol{v}$, apart from $\bol{v}$ itself (as shown later on in \Sec{scattering kinematics}, NR kinematics entails that $\bol{v} \cdot \bol{q}$ actually does not depend on $\bol{v}$). We can then write the differential rate as
\beq\label{Taylor rate}
\frac{\ud R_T}{\ud \ER}(\ER, t) \NReq \frac{\rho}{\mDM} \frac{\zeta_T}{m_T} \sum_{n = 0} g_n^T(\ER) \, \eta_n(\vmin(\ER), t) \ ,
\eeq
where we defined the velocity integrals
\beq\label{eta}
\eta_n(\vmin, t) \equiv \int_{v \geqslant \vmin} \ud^3 v \, \frac{\fE(\bol{v}, t)}{v} \, v^{2 n} \ .
\eeq
As we will see in \Sec{velocity}, under certain (quite standard) circumstances the velocity integrals can be approximated as
\beq\label{eta bar tilde}
\eta_n(\vmin, t) \simeq \overline{\eta}_n(\vmin) + \widetilde{\eta}_n(\vmin) \cos \left[ 2 \pi \frac{t - t_0}{\yr} \right] ,
\eeq
with $t_0$ the time of maximum Earth's speed in the galactic frame, see \Eq{Taylor eta}. Consequently, the differential rate can be approximated as
\beq\label{Taylor R}
\frac{\ud R_T}{\ud \ER}(\ER, t) \simeq \frac{\ud \overline{R}_T}{\ud \ER}(\ER) + \frac{\ud \widetilde{R}_T}{\ud \ER}(\ER) \cos \left[ 2 \pi \frac{t - t_0}{\yr} \right] ,
\eeq
where $\ud \overline{R}_T / \ud \ER$ only involves the $\overline{\eta}_n$'s while $\ud \widetilde{R}_T / \ud \ER$ only involves the $\widetilde{\eta}_n$'s. The latter term describes the annual modulation of the signal due to the periodic variation of DM flux at Earth caused by the rotation around the Sun, see \Sec{Modulation}. This modulation has distinctive features that can help telling a putative DM signal from mismodeled or unaccounted for backgrounds, and can be studied with an appropriate analysis. For most purposes, however, it can often be neglected, see \Sec{velocity}.

Of all terms in the NR expansion of the scattering rate in \Eq{Taylor rate}, only one or two typically matter. The most common case is when, as it happens for the SI (\Sec{SI interaction}) and SD (\Sec{SD interaction}) interactions, the zeroth-order ($n = 0$) term dominates the NR expansion, so that effectively $\ud \sigma_T / \ud \ER \propto 1 / v^2$ and only $\eta_0$ contributes significantly to the rate. However, it may happen that the zeroth-order term is suppressed by an $\Ord(10^{-6})$ (or smaller) factor such as $q^2 / \mu_T^2$, so that the $n = 1$ term of the expansion also becomes relevant: this is the case, for instance, of a DM particle interacting electromagnetically with nuclei through its magnetic dipole or anapole moment (see discussion in Secs.~\ref{Electromagnetic interactions},~\ref{Magnetic-dipole DM},~\ref{Example models and setup}), where the rate depends on both $\eta_0$ and $\eta_1$. Notice that the $\eta_n$ integrals are mutually related, as noted \eg in Ref.~\cite{DelNobile:2015rmp} and discussed in \Sec{DM velocity distribution}. Also, they depend solely on the local astrophysical properties of the DM halo, and are therefore the same functions of $\vmin$ and $t$ for all experiments. The $\ER$ functions $\eta_n(\vmin(\ER), t)$ entering the rate are versions of these integrals mapped onto $\ER$ in an $\mDM$ and $m_T$ dependent way, see \eg discussion in \Sec{Pheno}.

\subsection{Detection rate}
To properly reproduce the event rate measured by the experiments, we need to take into account detector effects such as finite energy resolution, efficiency, quenching and so forth. Experiments do not measure $\ER$ directly, rather they measure a quantity $\Ed$ that is statistically related to it. Depending on the experimental setup, $\Ed$ is often an energy or a number of photoelectrons. If it is an energy, as we will assume in the following for definiteness, it is usually quoted in $\keV_\text{ee}$ for \emph{electron equivalent}, to distinguish it from the nuclear recoil energy $\ER$ which is quoted in $\keV_\text{nr}$. The scattering rate must be convolved with a (target-dependent) resolution function $\mathcal{K}_T(\ER, \Ed)$, indicating the probability that a recoil energy $\ER$ is measured as $\Ed$. In the simplest case, this can be approximated with a Gaussian distribution with possibly energy-dependent width. Because some of the scattering energy goes into unmeasured channels (quenching), $\mathcal{K}_T$ peaks at $\langle \Ed \rangle = Q_T(\ER) \ER$, with $0 \leqslant Q_T(\ER) \leqslant 1$ the quenching factor. We also need to include the detector's efficiency and cut acceptance $\epsilon(\Ed)$, and to sum over all nuclides $T$ employed in the detector. The differential detection rate as a function of the detected signal $\Ed$ is then
\beq\label{dRdEd}
\frac{\ud R}{\ud \Ed}(\Ed, t) = \sum_{T} \epsilon(\Ed) \int_0^\infty \ud \ER \, \mathcal{K}_T(\ER, \Ed) \frac{\ud R_T}{\ud \ER}(\ER, t) \ .
\eeq
Experimental data are usually analysed between a lower $\Ed$ value, the experimental threshold, and an upper $\Ed$ value. The detection rate within a $\Ed$ interval $[\Ed_1, \Ed_2]$ is
\beq
R_{[\Ed_1, \Ed_2]}(t) = \int_{\Ed_1}^{\Ed_2} \ud \Ed \, \frac{\ud R}{\ud \Ed}(\Ed, t) \ .
\eeq
For computational purposes it may be more convenient to perform the $\Ed$ integral first,
\beq
R_{[\Ed_1, \Ed_2]}(t) = \frac{\rho}{\mDM} \sum_T \frac{\zeta_T}{m_T} \int_0^\infty \ud \ER \int_{v \geqslant \vmin(\ER)} \ud^3 v \, \fE(\bol{v}, t) \, v \, \frac{\ud \sigma_T}{\ud \ER}(\ER, \bol{v}) \, \mathcal{D}^T_{[\Ed_1, \Ed_2]}(\ER) \ ,
\eeq
where the functions
\beq
\mathcal{D}^T_{[\Ed_1, \Ed_2]}(\ER) \equiv \int_{\Ed_1}^{\Ed_2} \ud\Ed \, \epsilon(\Ed) \, \mathcal{K}_T(\ER, \Ed)
\eeq
do not depend on the DM model and can be computed once and for all for each nuclide and each relevant energy interval. Following \Eq{Taylor R}, $R_{[\Ed_1, \Ed_2]}$ may be approximated as
\beq\label{Taylor R(Ed)}
R_{[\Ed_1, \Ed_2]}(t) \simeq \overline{R}_{[\Ed_1, \Ed_2]} + \widetilde{R}_{[\Ed_1, \Ed_2]} \cos \left[ 2 \pi \frac{t - t_0}{\yr} \right] ,
\eeq
where $\overline{R}_{[\Ed_1, \Ed_2]}$ (also denoted $S_0$ in the literature) only involves the annual-average $\ud \overline{R}_T / \ud \ER$ while $\widetilde{R}_{[\Ed_1, \Ed_2]}$ (also denoted $S_\text{m}$) only involves the annual-modulation $\ud \widetilde{R}_T / \ud \ER$. Finally, the number of events detected in $[\Ed_1, \Ed_2]$ within a time interval $[T_1, T_2]$ is
\beq\label{Nevents}
N_{[\Ed_1, \Ed_2]} = M \int_{T_1}^{T_2} R_{[\Ed_1, \Ed_2]}(t) \, \ud t \simeq w \, \overline{R}_{[\Ed_1, \Ed_2]} \ ,
\eeq
where $M$ is the mass of the detector material and in the second equality we neglected the last term in \Eq{Taylor R(Ed)}. The experimental exposure $w \equiv M (T_2 - T_1)$ is usually expressed in $\kg~\dd$, although some experiments have reached exposures in the $\ton~\yr$ ballpark.

\section{Scattering kinematics}
\label{scattering kinematics}
In this Section we work out the kinematics of a $2 \to 2$ scattering. We perform the computations using relativistic (four-vector) notation, and then take the NR limit.

\subsection{Preliminaries}
\label{kinematics notation}
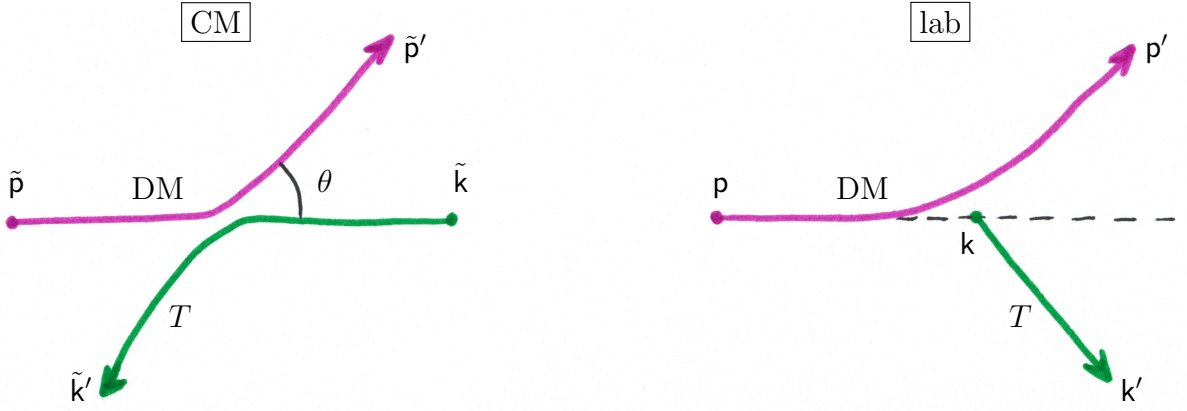
\begin{figure}[t]
\begin{center}
\begin{overpic}[width=.99\textwidth]{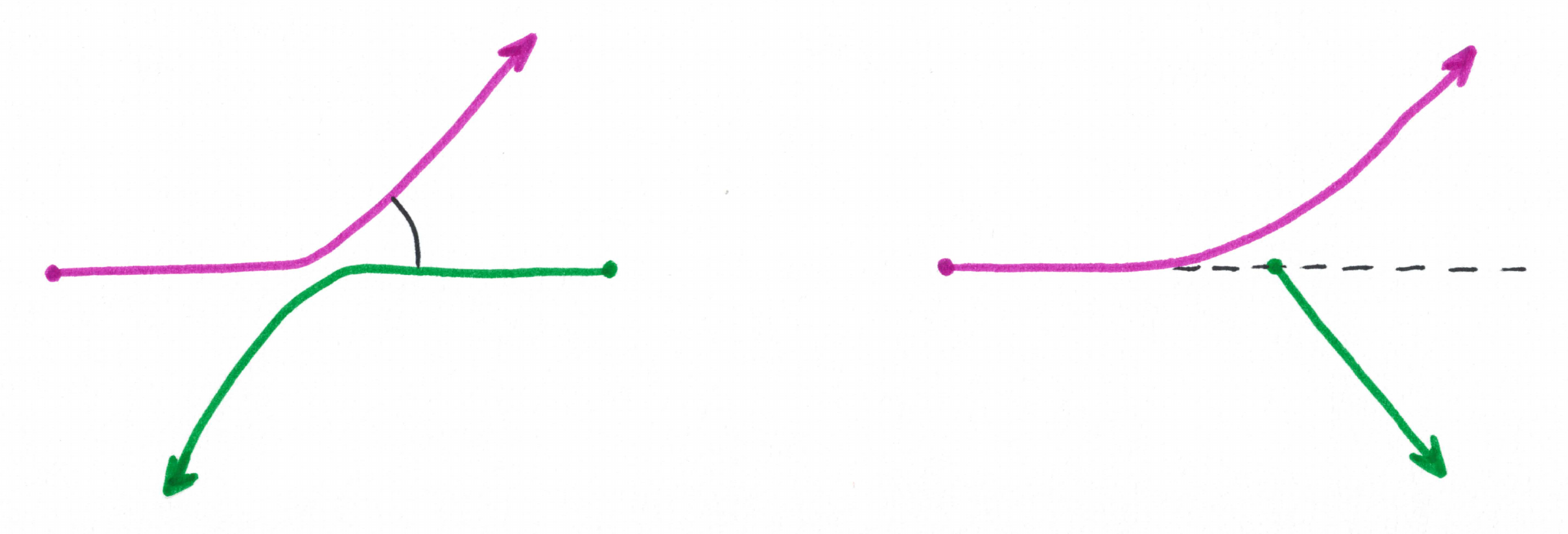}
\put (13, 18.5){DM}
\put (16, 8){$T$}
\put (3, 19){$\tilde{\fvec{p}}$}
\put (35, 30){$\tilde{\fvec{p}}'$}
\put (39, 19){$\tilde{\fvec{k}}$}
\put (8, 2){$\tilde{\fvec{k}}'$}
\put (28, 19){$\theta$}
\put (17, 32){\framebox{CM}}
\put (70, 18.5){DM}
\put (84, 8){$T$}
\put (60, 19){$\fvec{p}$}
\put (95, 30){$\fvec{p}'$}
\put (80, 14){$\fvec{k}$}
\put (93, 2){$\fvec{k}'$}
\put (76, 32){\framebox{lab}}
\put(45,20){\textcolor{white}{\rule{5mm}{5mm}}}
\end{overpic}
\caption[Kinematics notation]{\label{fig: momnot}\emph{Our notation for DM (purple line) and nucleus (green line) initial and final four-momenta. \textbf{Left:} in the CM frame. \figmath{\theta} denotes the CM-frame scattering angle. \textbf{Right:} in the lab frame.}}
\end{center}
\end{figure}

We start by fixing some notation we will use throughout this Section, see \Fig{fig: momnot} for a visual reference. A tilde denotes momenta and velocities in the center of momentum (CM) frame. All other momenta and velocities refer in this Section to the laboratory (lab) frame, where the target nucleus is at rest. Therefore we have for the (initial) DM and nucleus four-momenta in the CM frame
\begin{align}
\tilde{\fvec{p}}^\mu =
\begin{pmatrix}
E_{\tilde{p}}
\\
\tilde{\bol{p}}
\end{pmatrix}
,
&&&
\tilde{\fvec{k}}^\mu =
\begin{pmatrix}
E_{\tilde{k}}
\\
\tilde{\bol{k}} = - \tilde{\bol{p}}
\end{pmatrix}
,
\end{align}
while for the lab frame we have
\begin{align}
\fvec{p}^\mu =
\begin{pmatrix}
E_p
\\
\bol{p}
\end{pmatrix}
,
&&&
\fvec{k}^\mu =
\begin{pmatrix}
E_k
\\
\bol{k}
\end{pmatrix}
=
\begin{pmatrix}
m_T
\\
\bol{0}
\end{pmatrix}
.
\end{align}
We refer the reader to \Sec{Notation} for further information on our notation. We only recall here our definition of the momentum transfer four-vector,
\beq\label{momentum transfer}
\fvec{q} \equiv \fvec{p} - \fvec{p}' = \fvec{k}' - \fvec{k} \ ,
\eeq
where the last equality is due to energy-momentum conservation. Its coordinates are $\fvec{q}^\mu = (q^0, \bol{q})^\tr$, so that $\bol{k}' = \bol{q}$. The nuclear recoil energy is then defined as
\beq
\ER \equiv \frac{{k'}^2}{2 m_T} = \frac{q^2}{2 m_T} \ .
\eeq
We denote with $\theta$ the scattering angle in the CM frame, \ie $\cos \theta = \hat{\tilde{\bol{p}}} \cdot \hat{\tilde{\bol{p}}}' = \hat{\tilde{\bol{k}}} \cdot \hat{\tilde{\bol{k}}}'$.

It will be useful to bear in mind the following properties of the DM-nucleus reduced mass
\beq
\mu_T \equiv \frac{\mDM m_T}{\mDM + m_T} = \left( \frac{1}{\mDM} + \frac{1}{m_T} \right)^{-1}
\eeq
(see the left panel of \Fig{fig: reduced mass + qR=1}): it is symmetric under $\mDM \leftrightarrow m_T$ exchange; fixed one of the two masses, $\mu_T$ is an increasing function of the other; it is smaller than both the DM and nuclear mass, $\mu_T < \mDM, m_T$; it approaches the smallest mass in the limit in which one is much larger than the other, $\mu_T \xrightarrow{\mDM \gg m_T} m_T$ and $\mu_T \xrightarrow{m_T \gg \mDM} \mDM$. The reduced mass is therefore of the same order of magnitude of the smallest among $\mDM$ and $m_T$, which is useful remembering for quick qualitative estimates. Its inverse function is also sometimes useful,
\beq\label{inverse mu_T}
\mDM(\mu_T) = \frac{m_T \mu_T}{m_T - \mu_T} \ .
\eeq

As anticipated above, we will take the NR limit of fully relativistic results. The NR expansion consists in expanding four-momenta in powers of the particle speed. We will mostly truncate the expansion at leading order, although some of our results will be, when explicitly stated, at next-to-leading order (see also the discussion in \Sec{Non-relativistic expansion}). We will indicate with the symbol $\NReq$ equalities that are only valid at some finite order of the NR expansion. For instance we can write for the Lorentz factor of a generic particle with speed $\genp{u}$
\beq\label{NR gamma}
\gamma(\genp{u}) = 1 + \frac{\genp{u}^2}{2} + \Ord(\genp{u}^4) \NReq 1 + \frac{\genp{u}^2}{2} \ ,
\eeq
at next-to-leading order (the leading-order truncation being simply $\gamma(\genp{u}) \NReq 1$). \Eq{NR gamma} can be taken to define our NR expansion, in that it implies $\bol{\genp{p}} \NReq \genp{m} \bol{\genp{u}}$ at leading order and $E_\genp{p} \NReq \genp{m} + \frac{\genp{p}^2}{2 \genp{m}}$ at next-to-leading order. Notice that, with these approximations, one recovers the standard NR physics with its Galilean symmetry, which will apply to all our NR results. The only Galilean-invariant speed relevant to this problem being the (initial) DM-nucleus relative speed $v$, we can think of the NR expansion as an expansion in powers of $v$ (notice that the final DM-nucleus speed can also be expressed as a power series in $v$, as we will see). In this sense, momenta are of order $\Ord(v)$ while kinetic energies are of order $\Ord(v^2)$. In particular, $q / \mu_T \sim \Ord(v)$.

\subsection{Two-particle kinematics}
The internal dynamics of a $2 \to 2$ scattering is controlled by two parameters: the energy in the CM frame (associated with the masses and the relative motion of the particles), and the scattering angle in the CM frame (or alternatively the momentum transfer). The details of the scattering in any other reference frame can of course be determined by performing the appropriate Lorentz boost. Better yet, one can use Lorentz-invariant variables so that the result is automatically valid in any frame, \eg the Mandelstam variables $s$ and $t$. $s$ is related to the energy in the CM frame while $t$ is a measure of the momentum transfer. In terms of NR physics, it is convenient to use the DM-nucleus relative speed in place of $s$, and the nuclear recoil energy (or alternatively the momentum transfer) in place of $t$, both of which are Galilean invariant at the order we truncate the NR expansion.

To express the energy in the CM frame in terms of $s$ we can exploit the fact that $\tilde{\bol{p}} = - \tilde{\bol{k}}$ to write
\beq
s = (\tilde{\fvec{p}} + \tilde{\fvec{k}})^2 = (E_{\tilde{p}} + E_{\tilde{k}})^2 \ .
\eeq
The chain of equalities
\beq
E_{\tilde{p}}^2 - \mDM^2 = \tilde{p}^2 = \tilde{k}^2 = E_{\tilde{k}}^2 - m_T^2
\eeq
implies then
\beq
\sqrt{s} - E_{\tilde{p}} = E_{\tilde{k}} = \sqrt{E_{\tilde{p}}^2 - \mDM^2 + m_T^2} \ .
\eeq
Squaring, one yields the solution
\begin{align}
E_{\tilde{p}} &= \frac{s + \mDM^2 - m_T^2}{2 \sqrt{s}} \ ,
&&&
E_{\tilde{k}} &= \frac{s + m_T^2 - \mDM^2}{2 \sqrt{s}} \ .
\end{align}
Using these expressions, one obtains also
\beq\label{tildep^2}
\begin{split}
\tilde{p}^2 = E_{\tilde{p}}^2 - \mDM^2 &= \frac{s^2 + (\mDM^2 - m_T^2)^2 - 2 s (\mDM^2 + m_T^2)}{4 s}
\\
&= \frac{\left[ s - (\mDM + m_T)^2 \right] \left[ s - (\mDM - m_T)^2 \right]}{4 s}
\\
&= \frac{(\tilde{\fvec{p}} \cdot \tilde{\fvec{k}})^2 - \mDM^2 m_T^2}{s}
\\
&= \frac{\lambda(s, \mDM^2, m_T^2)}{4s} \ ,
\end{split}
\eeq
where
\beq
\lambda(x, y, z) \equiv x^2 + y^2 + z^2 - 2xy - 2 xz - 2yz
\eeq
is the K\"all\'en function. Notice that all the above formulas, while written explicitly in terms of initial-state quantities, also apply to final-state masses and momenta (just attach a $'$ to everything).

For NR particles in the CM frame, $\tilde{p} \NReq \mDM \tilde{v}_\DM \NReq m_T \tilde{v}_T$ at $\Ord(v^3)$ and therefore
\beq
\tilde{p} = \frac{\mDM \tilde{p} + m_T \tilde{p}}{\mDM + m_T} \NReq \frac{\mDM m_T}{\mDM + m_T} (\tilde{v}_\DM + \tilde{v}_T) = \mu_T | \tilde{\bol{v}}_\DM - \tilde{\bol{v}}_T | \ ,
\eeq
with $\tilde{\bol{v}}_\DM$, $\tilde{\bol{v}}_T$ the DM and target velocities in the CM frame, respectively. Exploiting the relative velocity $\bol{v} \equiv \bol{v}_\DM - \bol{v}_T$ being Galilean invariant at this order in the NR approximation, we get
\beq\label{tildep = mu v}
\tilde{\bol{p}} \NReq \mu_T \bol{v} \ .
\eeq
It is also convenient to compute the Mandelstam $s$ variable at next-to-leading order in the NR expansion:
\beq
\begin{split}
s &\NReq \left( \mDM + m_T + \frac{\tilde{p}^2}{2 \mu_T} \right)^2
\\
&\NReq (\mDM + m_T) \left( \mDM + m_T + \frac{\tilde{p}^2}{\mu_T} \right)
\\
&= (\mDM + m_T)^2 \left( 1 + \frac{\tilde{p}^2}{\mDM m_T} \right) .
\end{split}
\eeq
Notice that the $\tilde{p}^2 / 2 \mu_T \NReq \frac{1}{2} \mu_T v^2$ factor appearing in the parenthesis in the first line is the kinetic energy available in the CM frame.

The Mandelstam variable $t$ is
\beq
t = (\tilde{\fvec{p}} - \tilde{\fvec{p}}')^2 = (\tilde{\fvec{k}}' - \tilde{\fvec{k}})^2 \ ,
\eeq
which equals the squared momentum transfer four-vector (see \Eq{momentum transfer}). At $\Ord(v)$ in the NR expansion we have $\fvec{q}^\mu \NReq (0, \bol{q})^\tr$ (see also \Eq{q^0} below), therefore
\beq\label{t = - q^2}
t = \fvec{q}^2 \NReq - q^2 = - 2 m_T \ER \ .
\eeq
In the following we derive the relation between $q^2$ and $\theta$, the scattering angle in the CM frame, first for elastic scattering ($\delta = 0$) and then for a general value of the DM mass splitting $\delta$.

\subsection{Elastic scattering}
\label{Elastic scattering}
Elastic scattering occurs when the DM and nuclear masses remain unchanged in the interaction. From \Eq{tildep^2} one can see that $\tilde{p}$ (and $\tilde{p}'$) only depends on $s$ and the masses. Since $s$ is conserved, $\mDM$ and $m_T$ being the same before and after the scattering implies $\tilde{p} = \tilde{p}'$. Therefore we have
\beq\label{tildep tildep' tildek tildek'}
\tilde{p} = \tilde{p}' = \tilde{k} = \tilde{k}' \ ,
\eeq
and consequently
\begin{align}
E_{\tilde{p}} = E_{\tilde{p}'} \ ,
&&&
E_{\tilde{k}} = E_{\tilde{k}'} \ .
\end{align}
Being Lorentz invariant, the scalar product in the lab frame
\beq\label{tildek tildek'}
\fvec{k} \cdot \fvec{k}' = m_T E_{k'}
\eeq
equals that in the CM frame,
\beq
\tilde{\fvec{k}} \cdot \tilde{\fvec{k}}' = E_{\tilde{k}}^2 - \tilde{p}^2 \cos\theta \NReq m_T^2 + \tilde{p}^2 (1 - \cos \theta) \ ,
\eeq
thus we get in the NR limit
\begin{align}
\ER \NReq \frac{\mu_T^2 v^2}{m_T} (1 - \cos\theta) \ ,
&&&
q^2 \NReq 2 \mu_T^2 v^2 (1 - \cos\theta) \ .
\end{align}
The one-to-one correspondence between $\cos\theta$ and $q$ (or $\ER$) implies that these variables can be used interchangeably to describe the scattering process.

\begin{figure}[t]
\begin{center}
\includegraphics[width=.32\textwidth]{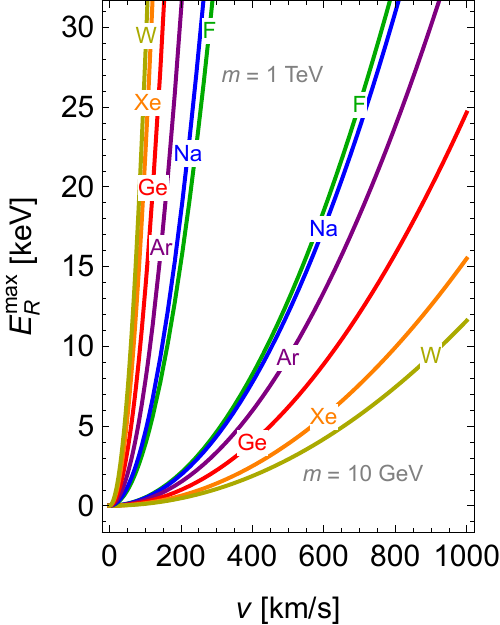}
\includegraphics[width=.32\textwidth]{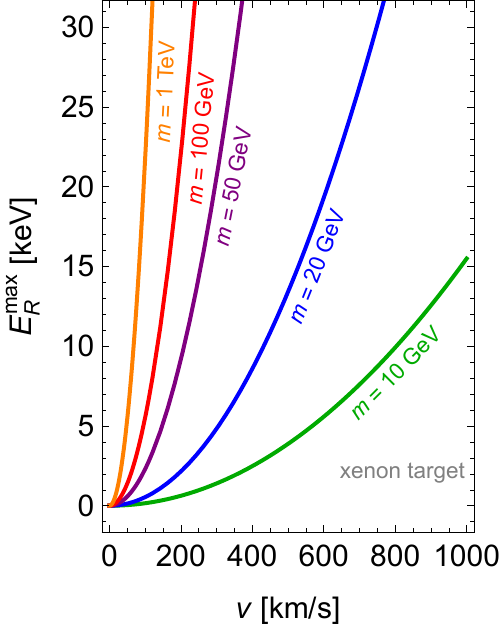}
\caption[$\ER^\text{max}(v)$, or alternatively $\vmin(\ER)$, for elastic scattering]{\label{fig: ElasticKinematics}\emph{\figmath{\ER^\text{max}} as a function of \figmath{v}, see \Eq{ERinterval}. Alternatively, the curves in the plots can be read as \figmath{\vmin}, on the horizontal axis, as a function of \figmath{\ER}, on the vertical axis, see \Eq{vmin}. The kinematically allowed values of \figmath{\ER} at fixed \figmath{v} (\figmath{\vmin} at fixed \figmath{\ER}) are those below (to the right of) the curve. For each element, the most abundant isotope (same as in the right panel of \Fig{fig: reduced mass + qR=1}) has been chosen as representative. \textbf{Left:} for different target nuclei at fixed \figmath{\mDM}, with \figmath{\mDM = 10~\GeV} (right) and \figmath{\mDM = 1~\TeV} (left). The opposite ordering of the targets for the two DM masses can be explained by noticing \eg that \figmath{\ER^\text{max} \sim 1 / m_T} for \figmath{\mDM \ll m_T} while \figmath{\ER^\text{max} \sim m_T} for \figmath{\mDM \gg m_T}, see text. \textbf{Right:} for different values of \figmath{\mDM} for a \figmath{\text{Xe}} target.}}
\figcode
\end{center}
\end{figure}

$\ER$ is maximum at maximum $q^2$, \ie when the DM particle bounces backwards, $\cos\theta = -1$, while it is minimum when the DM particle keeps traveling in the same direction after the scattering, $\cos\theta = 1$. Therefore, at fixed $v$, recoil energy and momentum transfer can take the values
\begin{align}
\label{ERinterval}
0 \leqslant \ER \leqslant \ER^\text{max}(v) \equiv \frac{2 \mu_T^2 v^2}{m_T} \ ,
&&&
0 \leqslant q \leqslant 2 \mu_T v \ .
\end{align}
The $\mu_T^2 / m_T$ dependence of $\ER^\text{max}$ implies that it can be approximated as $\ER^\text{max}(v) \simeq 2 \mDM^2 v^2 / m_T$ for $\mDM \ll m_T$, and that $\ER^\text{max}$ increases with $\mDM$ up to $\ER^\text{max}(v) \simeq 2 m_T v^2$. Therefore, the scattering is more kinematically favored for heavier DM, and for lighter (heavier) targets if the DM is sufficiently light (heavy). The non-trivial dependence of the scattering kinematics on $m_T$ may be better understood by analysing
\beq
\frac{\ud \ER^\text{max}(v)}{\ud m_T} = \frac{\ER^\text{max}(v)}{m_T} \frac{\mDM - m_T}{\mDM + m_T} \ ,
\eeq
which implies that $\ER^\text{max}$ at fixed $\mDM$ and $v$ has a single local maximum for $m_T = \mDM$, and thus increases (decreases) with $m_T$ for $m_T < \mDM$ ($m_T > \mDM$). The maximum value in \Eq{ERinterval} translates into a lower bound on the DM speed at fixed $\ER$ and $q$, $v \geqslant \vmin(\ER)$ with
\beq\label{vmin}
\vmin(\ER) \equiv \sqrt{\frac{m_T \ER}{2 \mu_T^2}} = \frac{q}{2 \mu_T} \ .
\eeq
$\vmin$ is the minimum speed a DM particle must have in Earth's rest frame to impart a recoil energy $\ER$ onto a target nucleus. The $\ER$ and velocity integrals in the rate can therefore be exchanged as
\beq
\int \ud^3 v \int_{0}^{\ER^\text{max}(v)} \ud \ER = \int_0^\infty \ud \ER \int_{v \geqslant \vmin(\ER)} \ud^3 v \ .
\eeq
$\vmin$ can be thought of as yet another variable, alternative to $\cos\theta$ and $q$ (or $\ER$), to describe the scattering process; it is the variable through which the velocity integral is most naturally defined, see Secs.~\ref{Scattering rate},~\ref{velocity}, and we will see in \Sec{Pheno} that it is sometimes the most convenient. The $v$ dependence of $\ER^\text{max}$, or alternatively the $\ER$ dependence of $\vmin$, is illustrated in \Fig{fig: ElasticKinematics} for different target nuclei and different values of the DM mass. If the recoil energy integral is performed first, one has to integrate below the curve, if instead the velocity integral is performed first one has to integrate to the right of the curve.

Denoting with $\bol{v}' \equiv \bol{v}'_\DM - \bol{v}'_T$ the final relative velocity, we have
\beq
\bol{v}'_\DM - \bol{v}'_T
\NReq
\frac{\bol{p}'}{\mDM} - \frac{\bol{k}'}{m_T}
=
\frac{\bol{p} - \bol{q}}{\mDM} - \frac{\bol{k} + \bol{q}}{m_T}
\NReq
\bol{v}_\DM - \bol{v}_T - \frac{\bol{q}}{\mu_T} \ ,
\eeq
and therefore
\beq\label{v v'}
\bol{v}' \NReq \bol{v} - \frac{\bol{q}}{\mu_T} \ .
\eeq
Exploiting Galilean invariance, \Eq{tildep tildep' tildek tildek'} implies
\beq
v' \NReq v \ .
\eeq
Squaring \Eq{v v'} we get therefore
\beq
\bol{v} \cdot \bol{q} \NReq \frac{q^2}{2 \mu_T} \ .
\eeq
As a consequence,
\beq\label{vperp}
\frac{1}{2} (\bol{v} + \bol{v}') \NReq \bol{v} - \frac{\bol{q}}{2 \mu_T} \NReq \bol{v} - (\bol{v} \cdot \hat{\bol{q}}) \, \hat{\bol{q}} \equiv \bol{v}^\perp_T
\eeq
is the component of $\bol{v}$ orthogonal to $\bol{q}$,
\begin{align}
\label{vperp.q}
\bol{v}^\perp_T \cdot \bol{q} = 0 \ ,
&&&
\bol{v} \cdot \hat{\bol{v}}^\perp_T \NReq v^\perp_T \ .
\end{align}
The DM-nucleon transverse velocity $\bol{v}^\perp_N$ can be defined analogously (see \Sec{NR operators}). We also have that
\beq\label{vperp^2}
{v^\perp_T}^2 \NReq v^2 - \frac{q^2}{4 \mu_T^2} = v^2 - \vmin^2 \ ,
\eeq
where $\vmin$, defined in \Eq{vmin}, is the minimum relative speed allowing the exchange of momentum $q$. In terms of DM and nucleus momenta, $\bol{v}^\perp_T$ can be written as
\beq\label{vperp pk}
2 \, \bol{v}^\perp_T \NReq \frac{\bol{P}}{\mDM} - \frac{\bol{K}}{m_T} \ ,
\eeq
where
\begin{align}
\bol{P} \equiv \bol{p} + \bol{p}' \ ,
&&&
\bol{K} \equiv \bol{k} + \bol{k}' \ .
\end{align}
$\bol{P}$ and $\bol{K}$ are sometimes defined with an extra factor of $\frac{1}{2}$ and called \emph{average momenta}. These vectors have the property that, in the CM frame,
\beq
\tilde{\bol{q}} \cdot \tilde{\bol{P}} = \tilde{\bol{q}} \cdot \tilde{\bol{K}} = 0 \ ,
\eeq
due to \Eq{tildep tildep' tildek tildek'}. The (Lorentz-invariant) four-vector version of this result is
\beq\label{q.P = q.K = 0}
\fvec{q} \cdot \fvec{P} = \fvec{q} \cdot \fvec{K} = 0 \ ,
\eeq
which can be derived directly from the definitions
\begin{align}
\fvec{P} \equiv \fvec{p} + \fvec{p}' \ ,
&&&
\fvec{K} \equiv \fvec{k} + \fvec{k}' \ .
\end{align}

\subsection{Inelastic scattering}
\label{Inelastic scattering}
Inelastic scattering occurs when a DM particle of mass $\mDM$ is excited to a state of mass $\mDM' = \mDM + \delta$ upon scattering off a nucleus. Models exist for both $\delta > 0$~\cite{TuckerSmith:2001hy} and $\delta < 0$~\cite{Graham:2010ca, Batell:2009vb}, with $\delta = 0$ corresponding to elastic scattering. In reactions with $\delta > 0$, part of the initial kinetic energy is absorbed from the final DM particle in the form of mass and the scattering is thus called \emph{endothermic}. On the contrary $\delta < 0$ implies that more kinetic energy is available in the final state than in the initial state, thus the reaction is \emph{exothermic}. Since the kinetic energy is small compared to the DM and nuclear masses, the scattering is kinematically allowed only if $\delta \ll \mDM$. This subsection generalizes the previous results to general values of $\delta$. The possibility that the nucleus undergoes a transition to an excited state has also been studied in the literature (see \eg Refs.~\cite{Ellis:1988nb, Engel:1999kv, Vergados:2003st, Vergados:2013raa, Baudis:2013bba, Vietze:2014vsa, McCabe:2015eia}), but we will not consider this possibility here.

From energy conservation in the CM frame we get, in the NR limit,
\beq
\frac{\tilde{p}^2}{2 \mDM} + \frac{\tilde{k}^2}{2 m_T} \NReq \delta + \frac{\tilde{p}'^2}{2 \mDM} + \frac{\tilde{k}'^2}{2 m_T} \ ,
\eeq
implying that the maximum possible value of $\delta$ for the scattering to occur is the kinetic energy initially available in the CM frame, $E_\text{kin} \equiv \tilde{p}^2 / 2 \mu_T \NReq \frac{1}{2} \mu_T v^2$. In the following we will treat $\delta / \mu_T$ as a parameter of order $\Ord(v^2)$ in the NR expansion for both endothermic and exothermic scattering. \Eq{tildep = mu v} and the above energy-conservation condition yield
\beq
\frac{1}{2} \mu_T v^2 \NReq \frac{1}{2} \mu'_T {v'}^2 + \delta \ ,
\eeq
with $\mu'_T$ the final DM-nucleus reduced mass, from which we obtain at leading order in $v^2$ and $\delta$
\beq\label{v v' delta}
{v'}^2 \NReq v^2 - 2 \frac{\delta}{\mu_T} \ .
\eeq
Following the steps of the above discussion on elastic scattering, we have
\beq
\tilde{\fvec{k}} \cdot \tilde{\fvec{k}}' = E_{\tilde{k}} E_{\tilde{k}'} - \tilde{p} \tilde{p}' \cos\theta \NReq m_T^2 + \mu_T^2 v^2 \left( 1 - \cos\theta \sqrt{1 - \frac{2 \delta}{\mu_T v^2}} \right) - \mu_T \delta \ ,
\eeq
where we used
\begin{align}
E_{\tilde{k}^{(\prime)}} &\NReq m_T + \frac{\tilde{p}^{(\prime) 2}}{2 m_T} \ ,
&
\label{tildep'^2}
\tilde{p}'^2 &\NReq {\mu_T'}^2 {v'}^2 \NReq \mu_T^2 v^2 - 2 \mu_T \delta \ .
\end{align}
Equating this to \Eq{tildek tildek'} we get
\beq\label{ERinel}
\ER \NReq \frac{\mu_T^2 v^2}{m_T} \left( 1 - \cos\theta \sqrt{1 - \frac{2 \delta}{\mu_T v^2}} \right) - \frac{\mu_T \delta}{m_T} \ .
\eeq
We can see from the square root that $\delta$ cannot be greater than the kinetic energy initially available in the CM frame, as already discussed above. We can also see that for a fixed DM speed there are a maximum and a minimum allowed recoil energy, $\ER^+$ and $\ER^-$, corresponding to $\cos\theta = \mp 1$:
\beq\label{ER^pm}
\ER^\pm(v) \equiv \frac{\mu_T^2 v^2}{2 m_T} \left( 1 \pm \sqrt{1 - \frac{2 \delta}{\mu_T v^2}} \right)^2 \ .
\eeq
For $\delta = 0$, the $\ER^+$ branch corresponds to $\ER^\text{max}$ defined in \Eq{ERinterval}, while $\ER^- = 0$. For $\delta \neq 0$, $\ER^+$ and $\ER^-$ have mutually inverse dependences on $v$, as one can see by noticing that
\begin{align}
\label{E_delta}
\ER^+ \ER^- = E_\delta^2
&&
\text{with}
&&
E_\delta \equiv |\delta| \frac{\mu_T}{m_T} \ .
\end{align}
$E_\delta$ is the value of $\ER^+$ and $\ER^-$ common to both, that is obtained by setting in \Eq{ER^pm} either $v = 0$ (which is only possible for $\delta \leqslant 0$) or $v = \sqrt{2 \delta / \mu_T}$ (which is only possible for $\delta > 0$), see \Eq{v_delta} below.

The kinematically allowed speed range for a DM particle to impart a target nucleus with a given nuclear recoil energy can be derived as follows. Since the scattering angle covers the whole round angle in the CM frame, $\tilde{\bol{q}} = \tilde{\bol{p}} - \tilde{\bol{p}}'$ implies
\beq
(\tilde{p} - \tilde{p}')^2 \leqslant \tilde{q}^2 \leqslant (\tilde{p} + \tilde{p}')^2 \ ,
\eeq
from which
\beq
\left| \tilde{q}^2 - \tilde{p}^2 - \tilde{p}'^2 \right| \leqslant 2 \tilde{p} \tilde{p}' \ .
\eeq
Squaring and using Eqs.~\eqref{tildep = mu v},~\eqref{tildep'^2} we then get
\beq
(\tilde{q}^2 + 2 \mu_T \delta)^2 \leqslant 4 \mu_T^2 v^2 \tilde{q}^2 \ .
\eeq
We now notice that $\bol{q}$ is proportional to a velocity difference at $\Ord(v)$, and as such it is Galilean invariant at this order of the NR expansion. This can be seen explicitly \eg by noticing that boosting $\bol{q} = \bol{p} - \bol{p}'$ into the CM frame yields $\tilde{\bol{p}} - \tilde{\bol{p}}'$, modulo a subdominant $\Ord(v^3)$ correction. Using then $\bol{q} \NReq \tilde{\bol{q}}$ we obtain
\beq
2 \mu_T v q \geqslant \left| q^2 + 2 \mu_T \delta \right| ,
\eeq
or in other words $v \geqslant \vmin(\ER)$ with
\beq\label{vmin_inelastic}
\vmin(\ER) \equiv \left| \frac{q}{2 \mu_T} + \frac{\delta}{q} \right| = \frac{1}{\sqrt{2 m_T \ER}} \left| \frac{m_T \ER}{\mu_T} + \delta \right| .
\eeq
Writing
\beq
\vmin(\ER) = \sqrt{\frac{m_T}{2 \mu_T^2 \ER}} \left| \ER + E_\delta \sgn\delta \right| = \sqrt{\frac{|\delta|}{2 \mu_T}} \left| h + \frac{\sgn\delta}{h} \right| = \sqrt{\frac{2 |\delta|}{\mu_T}} \times
\begin{cases}
\cosh y & \delta > 0,
\\
\sinh |y| & \delta < 0,
\end{cases}
\eeq
with
\begin{align}
h \equiv \sqrt{\frac{\ER}{E_\delta}} = \frac{q}{\sqrt{2 \mu_T |\delta|}} \ ,
&&&
y \equiv \log h \ ,
\end{align}
it can be noted that $\vmin$ is symmetric under $h \leftrightarrow h^{-1}$ exchange and in fact it is an even function of $y$, or in other words
\beq\label{vmin E_delta}
\vmin(\ER) = \vmin(E_\delta^2 / \ER) \ ,
\eeq
as could be already noted from \Eq{E_delta}. It can also be noted that $\vmin(\ER)$ has a minimum, call it $v_\delta$, at $\ER = E_\delta$, with value
\beq\label{v_delta}
v_\delta =
\begin{cases}
\sqrt{2 \delta / \mu_T} & \delta > 0,
\\
0 & \delta \leqslant 0.
\end{cases}
\eeq
This minimum occurs when $E_\text{kin}$ is minimum ($E_\text{kin} = \delta$ for $\delta > 0$, $E_\text{kin} = 0$ for $\delta \leqslant 0$), or equivalently when $\tilde{p}'^2$ is minimum ($\tilde{p}'^2 = 0$ for $\delta > 0$, $q^2 = \tilde{p}'^2 = 2 \mu_T |\delta|$ for $\delta \leqslant 0$, see \Eq{tildep'^2}). The energy and velocity integrals in the scattering rate can be exchanged as
\beq
\int_{v \geqslant v_\delta} \ud^3 v \int_{\ER^-(v)}^{\ER^+(v)} \ud \ER = \int_0^\infty \ud \ER \int_{v \geqslant \vmin(\ER)} \ud^3 v \ .
\eeq

\begin{figure}[t!]
\begin{center}
\includegraphics[width=.32\textwidth]{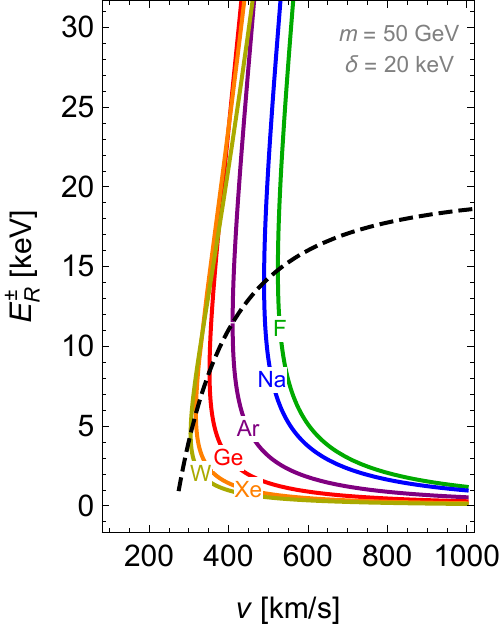}
\includegraphics[width=.32\textwidth]{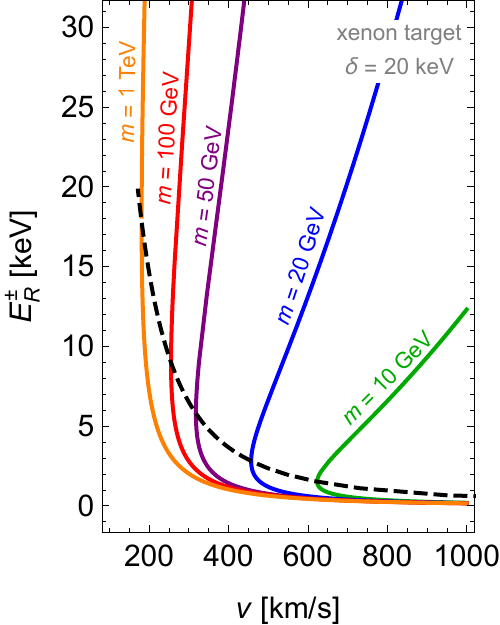}
\includegraphics[width=.32\textwidth]{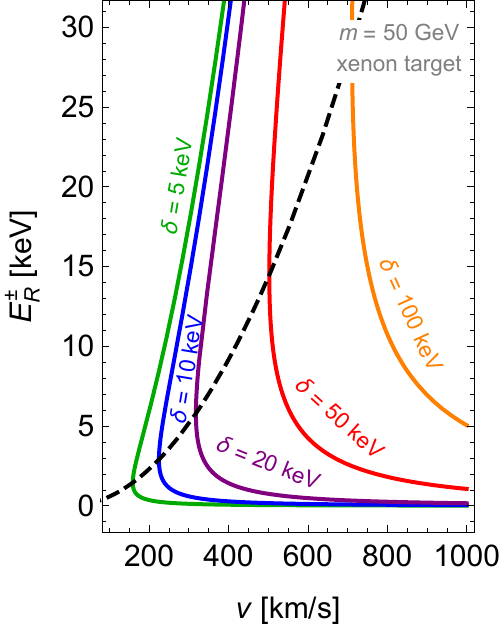}
\\[5mm]
\includegraphics[width=.32\textwidth]{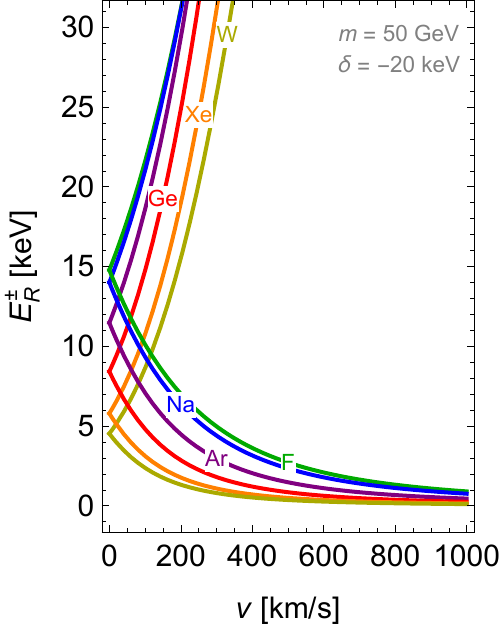}
\includegraphics[width=.32\textwidth]{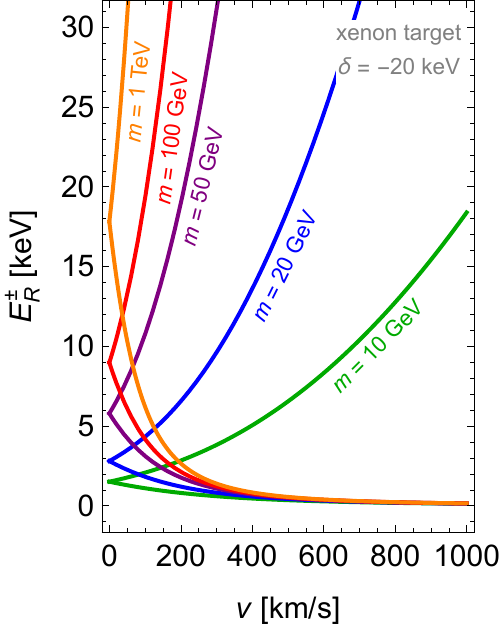}
\includegraphics[width=.32\textwidth]{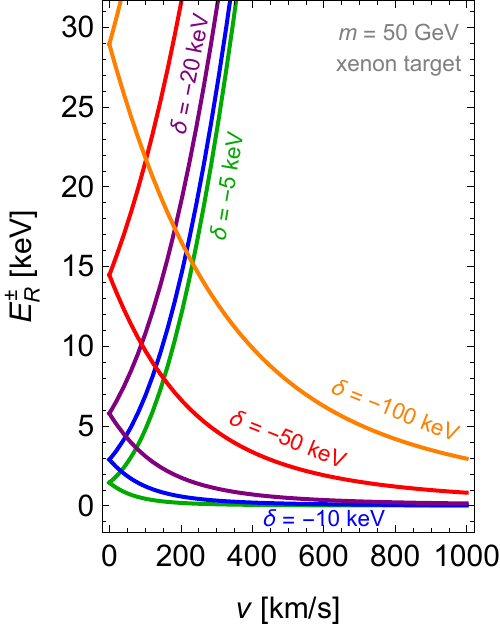}
\caption[$\ER^\pm(v)$, or alternatively $\vmin(\ER)$, for inelastic scattering]{\label{fig: InelasticKinematics}\emph{\figmath{\ER^\pm} as a function of \figmath{v}, see \Eq{ER^pm}. The kinematically allowed values of \figmath{\ER} at fixed \figmath{v} are those between the upper (\figmath{\ER^+}) and lower (\figmath{\ER^-}) branch of the curves. Alternatively, the curves in the plots can be read as \figmath{\vmin}, on the horizontal axis, as a function of \figmath{\ER}, on the vertical axis, see \Eq{vmin_inelastic}. The kinematically allowed values of \figmath{\vmin} at fixed \figmath{\ER} are those to the right of the curve. For each element, the most abundant isotope (same as in the right panel of \Fig{fig: reduced mass + qR=1}) has been chosen as representative. \textbf{Left:} for different target nuclei at fixed \figmath{\mDM} and \figmath{\delta}. \textbf{Center:} for different values of \figmath{\mDM} at fixed \figmath{\delta}, for a Xe target. \textbf{Right:} for different values of \figmath{\delta} at fixed \figmath{\mDM}, for a \figmath{\text{Xe}} target. \textbf{Top:} for \figmath{\delta > 0}. The dashed black line indicates the location of \figmath{v_\delta}, the minimum \figmath{\vmin} value allowed by kinematics (see \Eq{v_delta}). The \figmath{\ER^+} (\figmath{\ER^-}) branch is located above (below) it. \textbf{Bottom:} for \figmath{\delta < 0}. The \figmath{\ER^+} (\figmath{\ER^-}) branch is that above (below) the \figmath{v = 0} point of the curve.}}
\figcode
\end{center}
\end{figure}

The $v$ dependence of $\ER^\pm$, or alternatively the $\ER$ dependence of $\vmin$, is illustrated in \Fig{fig: InelasticKinematics} for different target nuclei and different values of the DM mass and the mass splitting. If the recoil-energy integral is performed first, one has to integrate between the upper and lower branch of the curves (\ie between $\ER^+$ and $\ER^-$); for $\delta > 0$ the two branches are separated by the dashed black line, representing the location of $v_\delta$. If instead the velocity integral is performed first, one has to integrate to the right of the curve. Notice that a single $\vmin$ value corresponds for $\delta \neq 0$ to two values of $\ER$, related as $\ER^+$ and $\ER^-$ in \Eq{E_delta} (see \Eq{vmin E_delta}), and therefore the velocity integral is the same at the two recoil energies. Notice also that, contrary to the case of elastic scattering, if $\delta \neq 0$ small nuclear recoil energies can only be obtained with sufficiently large DM speeds.

Inspection of \Eq{ER^pm} reveals that, for $\delta \geqslant 0$, $\ER^+$ decreases and $\ER^-$ increases with increasing $\delta$: this implies that elastic scattering is more kinematically favored than endothermic scattering, which itself is less and less kinematically favored for larger $\delta$. The same conclusion can be reached by noticing in \Eq{vmin_inelastic} that larger values of $\delta \geqslant 0$ cause $\vmin(\ER)$ to increase. We can also see that, for $\delta \leqslant 0$, $\ER^+$ and $\ER^-$ both decrease with increasing $\delta$ (\ie decreasing $|\delta|$), since the square root in \Eq{ER^pm} is larger than $1$. Therefore, for larger $\delta < 0$ the scattering is more (less) kinematically favored for sufficiently small (large) energies. More precisely, for two mass-splitting values $\delta_1 < \delta_2 \leqslant 0$, the scattering with $\delta_1$ is more (less) kinematically favored than for $\delta_2$ at recoil energies larger (smaller) than $(E_{\delta_1} + E_{\delta_2}) / 2$, as can be concluded by comparing \Eq{vmin_inelastic} for the two values of $\delta$ at fixed $\mDM$, $m_T$, $\ER$ with $E_{\delta_1} > \ER > E_{\delta_2}$. As a special case, then, exothermic scattering is more (less) kinematically favored than elastic scattering for $\ER$ larger (smaller) than $E_\delta / 2$, as can also be seen directly by comparing $\ER^-$ for $\delta < 0$ in \Eq{ER^pm} with $\ER^\text{max}$ in \Eq{ERinterval}. Moreover, comparing endothermic and exothermic scattering, it is clear from \Eq{vmin_inelastic} that $\delta < 0$ is always more kinematically favored than $\delta > 0$ at given $|\delta|$. We can perform a similar analysis to determine the effect of varying the DM mass on the scattering kinematics. From \Eq{vmin_inelastic} we can notice for instance that, for $\delta \geqslant 0$, $\vmin(\ER)$ decreases for larger $\mDM$ values, implying that $\ER^+$ increases and $\ER^-$ decreases with $\mDM$. This makes endothermic scattering (as elastic scattering) more kinematically favored for heavier DM. For $\delta < 0$, instead, $\vmin(\ER)$ decreases (increases) with increasing $\mDM$ for $\ER > E_\delta$ ($\ER < E_\delta$), meaning that both $\ER^+$ and $\ER^-$ increase with $\mDM$. Regarding the dependence of $\ER^+$ on $m_T$, for $\delta \neq 0$ one observes the opposite behavior for light and heavy DM that was already discussed for $\delta = 0$ after \Eq{ERinterval} and shown in the left panel of \Fig{fig: ElasticKinematics}: for sufficiently light (heavy) DM, $\ER^+$ decreases (increases) with $m_T$, as can be seen by taking the $\mDM \ll m_T$ ($\mDM \gg m_T$) limit of \Eq{ER^pm}. It also follows from Eqs.~\eqref{ER^pm},~\eqref{vmin_inelastic} that, for $\delta \neq 0$, $\ER^-$ decreases with $m_T$ in both the $\mDM \ll m_T$ and $\mDM \gg m_T$ regimes. We can be more precise by studying the sign of
\beq
\frac{\ud \vmin(\ER)}{\ud m_T} = \frac{s}{2 m_T \sqrt{2 m_T \ER}} \left( \frac{m_T - \mDM}{\mDM} \ER - \delta \right) = \frac{s |\delta|}{2 m_T \sqrt{2 m_T \ER}} \left( \frac{m_T - \mDM}{m_T + \mDM} \frac{\ER}{E_\delta} - \sgn\delta \right) ,
\eeq
with
\beq
s \equiv \sgn \left( \frac{m_T \ER}{\mu_T} + \delta \right) =
\begin{cases}
-1 & \delta < 0~\text{and}~\ER < E_\delta,
\\
+1 & \text{otherwise}.
\end{cases}
\eeq
It will be useful to note that $-1 < \frac{m_T - \mDM}{m_T + \mDM} < 1$. We can see that $\vmin(\ER)$ certainly decreases with increasing $m_T$ (thus making the scattering more kinematically favored for heavier targets) for $\ER < E_\delta$, corresponding to $\ER^-$ decreasing with $m_T$. For $\ER > E_\delta$, instead, $\vmin(\ER)$ certainly decreases (increases) with $m_T$ for $\delta \geqslant 0$ and $\mDM > m_T$ ($\delta < 0$ and $\mDM < m_T$), corresponding to $\ER^+$ increasing (decreasing) with $m_T$. In the remaining cases ($\ER > E_\delta$ with $\delta \geqslant 0$, $\mDM < m_T$ or with $\delta < 0$, $\mDM > m_T$), $\vmin(\ER)$ decreases with $m_T$ for $\ER$ values so that $\frac{m_T - \mDM}{\mDM} \ER < \delta$, and increases otherwise. For $\delta \geqslant 0$ we can conclude that the scattering is more kinematically favored for: larger $\mDM$; smaller $\delta$; larger $m_T$, as long as $\mDM > m_T$ or otherwise for recoil energies $\ER < \frac{\mDM}{m_T - \mDM} \delta$. For $\delta < 0$ instead we can summarize by saying that both $\ER^+$ and $\ER^-$ increase (decrease) with increasing $\mDM$ (increasing $\delta$, \ie decreasing $|\delta|$), thus `following' $E_\delta$. Likewise, $\ER^+$ and $\ER^-$ decrease with increasing $m_T$ at least for $\mDM < m_T$; in the opposite regime, $\ER^-$ keeps decreasing with $m_T$ while $\ER^+$ only decreases as long as $\ER^+(v) < \frac{\mDM}{\mDM - m_T} |\delta|$ (corresponding to $v < \sqrt{\frac{2 m_T |\delta|}{\mDM (\mDM - m_T)}}$, as can be checked through \Eq{vmin_inelastic}), and increases otherwise. This, in the $\mDM \gg m_T$ limit, results in the scattering being more kinematically favored for heavier targets, apart for only very small values of $v$.

As we will see in \Sec{Rate spectrum}, a particularly useful piece of information is the maximum $\ER$ value (thus $\ER^+$) attainable with light DM particles, as this is the parameter controlling the sensitivity loss of direct detection experiments caused by their finite threshold. The sensitivity reach of current experiments extends down to DM masses of few $\GeV$ or lower, thus we can safely assume $\mDM < m_T$ for the sake of this discussion (see \Tab{tab: nuclides}). For $\delta < 0$, from the above analysis we have that $\ER^+$ decreases with $m_T$, \ie the lighter the target the more kinematically favored light-DM scattering at recoil energies $\ER > E_\delta$, as for elastic scattering. For $\delta > 0$ the same is only true for $\ER > \frac{\mDM}{m_T - \mDM} \delta$, so that lighter targets are kinematically favored for sufficiently large recoil energies, which must be higher the larger $\delta$. We conclude that, regardless of the sign of $\delta$, the scattering of light enough DM particles at recoil energies $\ER > E_\delta$ is more kinematically favored for lighter targets, although for $\delta > 0$ this only happens for sufficiently large $\ER$ values (with larger mass splittings requiring larger recoil energies), heavier targets being otherwise favored.

The transverse velocity can be determined as follows. Squaring \Eq{v v'}, which holds at leading order for inelastic scattering, and using \Eq{v v' delta}, we get
\beq
\bol{v} \cdot \bol{q} \NReq \frac{q^2}{2 \mu_T} + \delta \ .
\eeq
As a consequence, the component of $\bol{v}$ orthogonal to $\bol{q}$, see \Eq{vperp}, is
\beq\label{vperpinel}
\bol{v}^\perp_{\text{inel}, T} \NReq \bol{v} - \bol{q} \left( \frac{1}{2 \mu_T} + \frac{\delta}{q^2} \right) ,
\eeq
so that \Eq{vperp.q} is obeyed with $\bol{v}^\perp_{\text{inel}, T}$ in place of $\bol{v}^\perp_T$, and we get
\beq
{v^\perp_{\text{inel}, T}}^2 \NReq v^2 - \left( \frac{q}{2 \mu_T} + \frac{\delta}{q} \right)^2 = v^2 - \vmin^2 \ .
\eeq
The generalization of \Eq{q.P = q.K = 0} is now
\begin{align}
\label{q.P neq 0, q.K = 0}
\fvec{q} \cdot \fvec{P} = - 2 \mDM \delta - \delta^2 \ ,
&&&
\fvec{q} \cdot \fvec{K} = 0 \ .
\end{align}

\section{From quarks and gluons to nucleons}
\label{qg to N}
We now abandon the realm of kinematics to venture into the dynamics of the DM-nucleus system. We do this in steps. In this Section we work out the hadronic matrix elements of quark and gluon operators, with which the DM-nucleon scattering amplitude $\Mel_N$ can be computed. We will then see in \Sec{DM-N} how to compute the NR expansion of $\Mel_N$ and how to match the result to a NR DM-nucleon interaction operator. From this operator the full DM-nucleus scattering amplitude $\Mel$ can be computed as described in \Sec{Form factors}. Finally, in \Sec{sigma} we write the differential cross section entering \Eq{diffrate} in terms of $\Mel$, and work out some specific examples.

\subsection{Hadronic matrix elements}
While we are ultimately interested in evaluating some operators between nuclear states to compute the DM-nucleus scattering amplitude, our interaction Lagrangian may involve quark and gluon degrees of freedom. We must learn therefore how to evaluate operators built out of quark and gluon fields between nuclear states. As discussed above, the first step is to compute the DM-nucleon scattering amplitude, which entails computing matrix elements of our relativistic quark and gluon operators between nucleon states.

As an example we can consider an effective operator, in a theory where the interaction mediators are very heavy and have been integrated out. In this limit the force between two particles does not propagate and the interaction region is point-like, hence the name \emph{contact interaction}. However our procedure can be used with trivial modifications to treat other types of operators, as in theories where the interaction mediators are light (see example below): in fact, the nucleon matrix element of quark and gluon operators, which is what we will be interested in in this Section, can be the same in both theories. Let us then consider the following, rather vague interaction Lagrangian, defined at the hadronic scale\footnote{See \eg Refs.~\cite{Hill:2011be, Haisch:2013uaa, Hill:2014yka, Kopp:2014tsa, Crivellin:2014qxa, Crivellin:2014gpa, Hill:2014yxa, DEramo:2014nmf, DEramo:2016gos, DEramo:2017zqw, Bishara:2018vix} for renormalization-group effects in theories defined at higher energies.} (about $1~\GeV$):
\beq\label{L quark}
\Lag = c_q \, \Op_\DM(\fvec{x}) \Op_q(\fvec{x}) \ ,
\eeq
with $\Op_\DM$ and $\Op_q$ operators built out of DM and quark fields, respectively, and $c_q$ a coupling constant. A simple, more concrete example, for a spin-$1/2$ DM field $\chi$, is $\Op_\DM = \bar{\chi} \gamma^\mu \chi$ and $\Op_q = \bar{q} \gamma_\mu q$, where $c_q$ has mass dimension $-2$. The quark field $q$ here is taken to be a Dirac fermion with both left and right components, as is appropriate in the Standard Model (SM) after integrating out electroweak-scale degrees of freedom. If a Lagrangian involves contraction of a DM spinor with a quark spinor, as in $\bar{\chi} \gamma^5 q \, \bar{q} \chi$, it should be possible to factor the $\chi$ and $q$ dependence into two separate operators by means of Fierz identities, so that we can recover the structure in \Eq{L quark}. Another example of an effective operator, this time involving gluons rather than quarks, can be
\beq\label{L gluon}
\Lag(\fvec{x}) = c_\text{g} \, \Op_\DM(\fvec{x}) \Op_\text{g}(\fvec{x}) \ ,
\eeq
with for example $\Op_\text{g}(\fvec{x}) = G^{a \mu\nu} G^a_{\mu\nu}$. Here the DM operator could be $\Op_\DM = \phi^\dagger \phi$ for a scalar DM field $\phi$, in which case $c_\text{g}$ is a parameter with mass dimension $-2$, or it could be $\Op_\DM = \bar{\chi} \chi$ for a spin-$1/2$ DM field $\chi$, in which case $c_\text{g}$ has mass dimension $-3$. A third example, this time with explicit couplings of DM and quarks to a scalar interaction mediator $S$, is
\beq\label{L mediator}
\Lag(\fvec{x}) = \Op_\DM(\fvec{x}) S(\fvec{x}) + \Op_q(\fvec{x}) S(\fvec{x}) \ ,
\eeq
again with $\Op_\DM$ and $\Op_q$ operators built out of DM and quark fields, respectively. Here, for instance, one could have $\Op_q = \bar{q} (a_q \unom_4 + i b_q \gamma^5) q$, with $a_q$, $b_q$ dimensionless coefficients. The DM operator could be, for scalar DM, $\Op_\DM = c \, \phi^\dagger \phi$, with $c$ a parameter with mass dimension $1$, or it could be $\Op_\DM = \bar{\chi} (c \unom_4 + i d \gamma^5) \chi$ for spin-$1/2$ DM, with $c$, $d$ dimensionless coefficients.

We define the DM-nucleon scattering amplitude $\Mel_N$ in relation to the DM-nucleon scattering matrix $S_N$ as
\beq\label{S & M matrix}
S_N = \braket{\DM', N'}{\DM, N} + i \, (2 \pi)^4 \delta^{(4)}(\fvec{p'} + \fvec{k'} - \fvec{p} - \fvec{k}) \, \Mel_N \ .
\eeq
In this Section we denote with $\fvec{k}$, $\fvec{k}'$ exclusively the initial and final nucleon (rather than nucleus) momenta, respectively. Notice that, due to momentum conservation, $\fvec{q}$ is both the momentum transferred by the DM to the nucleon and to the nucleus. We recall from \Sec{Notation} that $\ket{\DM^{(\prime)}}$ and $\ket{N^{(\prime)}}$ are a shorthand notation for $\ket{\DM^{(\prime)}(\bol{p}^{(\prime)}, s^{(\prime)})}$ and $\ket{N(\bol{k}^{(\prime)}, r^{(\prime)})}$, respectively. Similarly, we use $u^{(\prime)}_\chi$, $u^{(\prime)}_N$ as shorthand for $u^{(\prime)}_\chi(\bol{p}^{(\prime)}, s^{(\prime)})$ and $u_N(\bol{k}^{(\prime)}, r^{(\prime)})$, respectively (notice that $u^{(\prime)}_\chi$ refers to a DM particle with mass $\mDM^{(\prime)}$). Operator matrix elements are understood to be evaluated at the origin, unless the position is indicated explicitly. For instance, in our sloppy notation $\Nmatel{\Op_{q, \text{g}}}$ actually means $\matel{N(\bol{k}', r')}{\Op_{q, \text{g}}(\fvec{0})}{N(\bol{k}, r)}$. Notice that, in \Eq{S & M matrix}, the $\braket{\DM', N'}{\DM, N}$ term only contributes for $q = 0$, below the sensitivity limit of actual detectors, and therefore can be disregarded for practical purposes. The scattering matrix can be perturbatively expanded as
\begin{multline}
\label{S perturbative exp}
S_N = \sum_{n = 0}^\infty \frac{i^n}{n!} \int \ud^4 \fvec{x}_1 \cdots \ud^4 \fvec{x}_n \, \matel{\DM', N'}{\CMcal{T} (\Lag(\fvec{x}_1) \cdots \Lag(\fvec{x}_n))}{\DM, N} = \braket{\DM', N'}{\DM, N}
\\
+ i \int \ud^4 \fvec{x} \, \matel{\DM', N'}{\Lag(\fvec{x})}{\DM, N} - \frac{1}{2} \int \ud^4 \fvec{x} \, \ud^4 \fvec{y} \, \matel{\DM', N'}{\CMcal{T} (\Lag(\fvec{x}) \Lag(\fvec{y}))}{\DM, N} + \Ord(\Lag^3) \ ,
\end{multline}
where $\CMcal{T}$ denotes the time-ordered product. The first and second term in the last line are the first- and second-order contributions to the perturbative expansion, respectively, whereas $\Ord(\Lag^3)$ indicates third- and higher-order terms. For the two examples in Eqs.~\eqref{L quark},~\eqref{L gluon} we then have, at first order,
\beq
\Mel_N = \matel{\DM', N'}{\Lag}{\DM, N} = c_{q, \text{g}} \, \matel{\DM'}{\Op_\DM}{\DM} \, \Nmatel{\Op_{q, \text{g}}} \ ,
\eeq
with all operators evaluated at $\fvec{x} = \fvec{0}$. For the example in \Eq{L mediator}, the tree-level DM-nucleon scattering amplitude reads at second order in the perturbative expansion
\beq
\Mel_N = - \frac{1}{\fvec{q}^2 - m_S^2} \matel{\DM'}{\Op_\DM}{\DM} \, \Nmatel{\Op_q} \ ,
\eeq
with the operators again evaluated at $\fvec{x} = \fvec{0}$ (this will be understood from now on), and with $m_S$ the scalar mediator mass.

The quark/gluon matrix element can be parametrized in terms of nucleon-spinor bilinears,
\beq\label{< O_qg >}
\Nmatel{\Op_{q, \text{g}}} = \sum_{\tilde{\Gamma}} F_{\tilde{\Gamma}}^N(\fvec{q}^2) \, \bar{u}'_N \tilde{\Gamma}(\fvec{q}, \fvec{K}) u_N \ ,
\eeq
with the $\tilde{\Gamma}$'s matrices in spinor space depending on the two linearly-independent four-vectors $\fvec{q} = \fvec{k}' - \fvec{k}$ and $\fvec{K} \equiv \fvec{k} + \fvec{k}'$. Their explicit form matches the transformation properties of $\Op_q$ and $\Op_\text{g}$ under the Lorentz symmetry, parity and time reversal (to the extent that these are good symmetries), and possible internal symmetries, and is restricted by conservation laws and equations of motion. In particular, the equations of motion lead to identities such as
\begin{subequations}
\label{EOMs}
\begin{align}
i \fvec{q}_\mu \, \bar{u}'_N \gamma^\mu u_N &= 0 \ ,
&
i \fvec{q}_\mu \, \bar{u}'_N \gamma^\mu \gamma^5 u_N &= 2 \mN \, \bar{u}'_N \, i \gamma^5 u_N \ ,
\\
\fvec{K}_\mu \, \bar{u}'_N \gamma^\mu u_N &= 2 \mN \, \bar{u}'_N u_N \ ,
&
\fvec{K}_\mu \, \bar{u}'_N \gamma^\mu \gamma^5 u_N &= 0 \ ,
\\
i \fvec{q}_\mu \, \bar{u}'_N \sigma^{\mu\nu} u_N &= \bar{u}'_N [- 2 \mN \gamma^\nu + \fvec{K}^\nu] u_N \ ,
&
i \fvec{q}_\mu \, \bar{u}'_N \, i \sigma^{\mu\nu} \gamma^5 u_N &= \fvec{K}^\nu \, \bar{u}'_N \, i \gamma^5 u_N \ ,
\\
\fvec{K}_\mu \, \bar{u}'_N \sigma^{\mu\nu} u_N &= - i \fvec{q}^\nu \, \bar{u}'_N u_N \ ,
&
\fvec{K}_\mu \, \bar{u}'_N \, i \sigma^{\mu\nu} \gamma^5 u_N &= \bar{u}'_N [- 2 \mN \gamma^\nu + \fvec{q}^\nu] \gamma^5 u_N \ ,
\end{align}
\end{subequations}
see \eg Refs.~\cite{Lorce:2017isp, DelNobile:2018dfg}. For instance, if all we knew about $\Op_{q, \text{g}}$ is that it is a four-vector, such as $\bar{q} \gamma^\mu \gamma^5 q$ or $\partial_\nu (\bar{q} \sigma^{\mu\nu} q)$, examples of possible $\tilde{\Gamma}$ matrices entering \Eq{< O_qg >} would be $\gamma^\mu$, $\fvec{q}^\mu \gamma^5$, $\fvec{q}^\mu \unom_4$ with $\unom_4$ the unit matrix in spinor space, and $\fvec{K}_\nu \sigma^{\mu\nu}$, the latter being redundant due to the equations of motion. Knowledge of the $P$ and $T$ transformation properties of $\Op_{q, \text{g}}$ would further limit the form the $\tilde{\Gamma}$'s can take. The (operator-specific) hadronic form factors $F_{\tilde{\Gamma}}^N$ are functions of all independent Lorentz scalars one can build with $\fvec{q}$ and $\fvec{K}$, \ie they are just functions of $\fvec{q}^2$ since $\fvec{q} \cdot \fvec{K} = 0$ and $\fvec{K}^2 = 4 \mN^2 - \fvec{q}^2$.

To be concrete, we consider for $\Op_q$ the following set of color-neutral and electric charge-neutral, hermitian, and flavor-diagonal quark bilinears:
\begin{align}
\label{Gammas}
\bar{q} \Gamma q
&&
\text{with}
&&
\Gamma = \unom_4, i \gamma^5, \gamma^\mu, \gamma^\mu \gamma^5, \sigma^{\mu\nu},
\end{align}
where the sixteen $\Gamma$ matrices form a basis of hermitian $4 \times 4$ matrices. We employ the following definitions,
\begin{align}
\label{gamma5 + sigma}
\gamma^5 \equiv - \frac{i}{4!} \varepsilon_{\mu \nu \rho \sigma} \gamma^\mu \gamma^\nu \gamma^\rho \gamma^\sigma = i \, \gamma^0 \gamma^1 \gamma^2 \gamma^3 \ ,
&&&
\sigma^{\mu\nu} \equiv \frac{i}{2} [\gamma^\mu, \gamma^\nu] \ ,
\end{align}
which obey the relation
\beq\label{epsilon sigma}
i \sigma^{\mu\nu} \gamma^5 = - \frac{1}{2} \, \varepsilon^{\mu\nu\rho\tau} \sigma_{\rho\tau} \ ,
\eeq
with the completely anti-symmetric tensor $\varepsilon^{\mu \nu \rho \sigma}$ defined so that
\beq\label{LCepsilon}
\varepsilon^{0 1 2 3} = - \varepsilon_{0 1 2 3} = 1 \ .
\eeq
Moreover we take the gluon operator $\Op_\text{g}$ to be
\begin{align}
G^{a \mu\nu} G^a_{\mu\nu} \ ,
&&&
G^{a \mu\nu} \tilde{G}^a_{\mu\nu} \ ,
\end{align}
with the dual gluon field strength defined as
\beq\label{Gtilde}
\tilde{G}^a_{\mu\nu} \equiv \frac{1}{2} \varepsilon_{\mu\nu\rho\sigma} G^{a \rho\sigma} \ .
\eeq
The above operators transform under $P$ and $T$ as
\begin{align}
\label{P T}
P \Op(\fvec{x}) P^{-1} = \eta^P \Op(\mathscr{P} \fvec{x}) \ ,
&&&
T \Op(\fvec{x}) T^{-1} = \eta^T \Op(\mathscr{T} \fvec{x}) \ ,
\end{align}
with $\mathscr{P}^\mu_{\phantom{\mu} \nu} = - \mathscr{T}^\mu_{\phantom{\mu} \nu} = \text{diag}(+1, -1, -1, -1)$. The operator-specific coefficients $\eta^P$ and $\eta^T$ are provided in \Tab{tab: PT}, where $\psi$ is a generic spin-$1/2$ field and
\beq\label{(-1)^mu}
(-1)^\mu \equiv
\begin{cases}
+1 & \mu = 0,
\\
-1 & \mu = 1, 2, 3.
\end{cases}
\eeq

\begin{table}[t]
\begin{center}
\begin{tabular}{c |>{\pnt} c c c c c c c c}
& $\bar{\psi} \psi$ & $\bar{\psi} \, i \gamma^5 \psi$ & $\bar{\psi} \gamma^\mu \psi$ & $\bar{\psi} \gamma^\mu \gamma^5 \psi$ & $\bar{\psi} \, \sigma^{\mu\nu} \psi$ & $\bar{\psi} \, i \sigma^{\mu\nu} \gamma^5 \psi$ & $G^{a \mu\nu} G^a_{\mu\nu}$ & $G^{a \mu\nu} \tilde{G}^a_{\mu\nu}$
\\
\hline
$\eta^P$ & $+1$ & $-1$ & $(-1)^\mu$ & $-(-1)^\mu$ & $(-1)^\mu (-1)^\nu$ & $-(-1)^\mu (-1)^\nu$ & $+1$ & $-1$
\\
$\eta^T$ & $+1$ & $-1$ & $(-1)^\mu$ & $(-1)^\mu$ & $-(-1)^\mu (-1)^\nu$ & $(-1)^\mu (-1)^\nu$ & $+1$ & $-1$
\end{tabular}
\end{center}
\caption[$P$ and $T$ transformation coefficients of field bilinears]{\label{tab: PT}\emph{Transformation coefficients of spin-\figmath{1/2} bilinears \figmath{\bar{\psi} \Gamma \psi} and gluon field-strength bilinears under \figmath{P} and \figmath{T}, see \Eq{P T}. Here \figmath{(-1)^\mu \equiv 1} for \figmath{\mu = 0} while \figmath{(-1)^\mu \equiv -1} for \figmath{\mu = 1, 2, 3}, see \Eq{(-1)^mu}.}}
\end{table}

In the following we evaluate the above quark and gluon operators between nucleon states, determining the set of $\tilde{\Gamma}$ matrices featured in \Eq{< O_qg >} and the values of the hadronic form factors $F_{\tilde{\Gamma}}^N$ for different $\Op_q$'s and $\Op_\text{g}$'s. The main information about the form factors is their value at $\fvec{q}^2 = 0$, since their variation often occurs at hadronic-scale energies and therefore they can be approximated as constant at the low energies of interest to direct DM detection. We will therefore mostly focus on the $\fvec{q}^2 = 0$ value of the hadronic form factors, while also providing their $\fvec{q}^2$ dependence where known or relevant.

\subsection{Scalar couplings}
\label{Scalar couplings}
As can be seen in \Tab{tab: PT}, the scalar operators
\begin{align}
\label{scalar qg}
\Op_q = \bar{q} q \ ,
&&&
\Op_\text{g} = \frac{\aS}{12 \pi} \, G^{a \mu\nu} G^a_{\mu\nu} \ ,
\end{align}
have the same $P$ and $T$ quantum numbers, therefore we will deal with them together (the numerical factors in $\Op_\text{g}$ have been chosen for later convenience). Given the transformation properties of these operators under the Lorentz symmetry, spatial parity and time reversal, their nucleon matrix element can be parametrized in terms of a single operator-specific form factor:
\begin{align}
\label{scalar current parametrization}
\Nmatel{\bar{q} q} = F_\text{S}^{q, N}(\fvec{q}^2) \, \bar{u}'_N u_N \ ,
&&&
\frac{\aS}{12 \pi} \Nmatel{G^{a \mu\nu} G^a_{\mu\nu}} = F_\text{S}^{\text{g}, N}(\fvec{q}^2) \, \bar{u}'_N u_N \ .
\end{align}
$\Op_q$ and $\Op_\text{g}$ being hermitian implies that $F_\text{S}^{q, N}$, $F_\text{S}^{\text{g}, N}$ are real functions of $\fvec{q}^2$. Other Lorentz scalars that can be constructed with the available ingredients (\ie the nucleon spinors, the Dirac matrices and the nucleon momentum four-vectors) either have the wrong transformation properties under parity (\eg $\bar{u}'_N \, i \gamma^5 u_N$) or can be reduced to the above by means of the equations of motion (\eg $\fvec{K}_\mu \, \bar{u}'_N \gamma^\mu u_N$), see \Eq{EOMs}.

Here is an example to see how the parity transformation properties of $\Op_{q, \text{g}}$ can be used to constrain its matrix element (see \Sec{Vector couplings} for an example using time reversal). Using parity, one has
\beq\label{scalar P}
\Nmatel{\Op_{q, \text{g}}} = \Nmatel{P^{-1} P \Op_{q, \text{g}} P^{-1} P} = \matel{P N'}{\Op_{q, \text{g}}}{P N} \ ,
\eeq
where we used \Eq{P T} and we indicated with $\ket{P N^{(\prime)}} \equiv P \ket{N^{(\prime)}}$ the parity-transformed nucleon state. Let us now check whether (a term proportional to) $\bar{u}'_N \, i \gamma^5 u_N = \Nmatel{\bar{N} \, i \gamma^5 N}$ could appear on the right-hand side of the equal signs in \Eq{scalar current parametrization}. We can do so by noticing that
\beq
\matel{P N'}{\bar{N} \, i \gamma^5 N}{PN} = \Nmatel{P \bar{N} \, i \gamma^5 N P^{-1}} = - \Nmatel{\bar{N} \, i \gamma^5 N} \ ,
\eeq
where we used again \Eq{P T} and $P^{-1} = P$. This can only be compatible with \Eq{scalar P} if $\bar{u}'_N \, i \gamma^5 u_N$ appears in \Eq{scalar current parametrization} with null coefficient.

The nucleon matrix elements of $\Op_q$, $\Op_\text{g}$ can be computed at zero momentum transfer following Ref.~\cite{Shifman:1978zn} (see also Refs.~\cite{Cheng:1988im, Cheng:2012qr}). First we notice that, while the light quarks can provide sizeable contributions to the scalar nucleon current, heavy quarks contribute mostly by connecting to gluon lines (starting with a $1$-loop triangle diagram in a perturbative expansion). Therefore, couplings to the heavy quarks in \Eq{scalar qg} approximately probe the gluon content of the nucleon. Integrating out the heavy quarks $h = c, b, t$ via the heavy-quark expansion (see \eg Ref.~\cite{Zyla:2020zbs}) yields, at lowest order in $\aS$ and $\Lambda_\text{QCD} / m_h$, a result that is effectively reproduced by the substitution
\beq\label{heavyquarks scalar}
m_h \, \Nmatel{\bar{h} h} \rightarrow - \frac{\aS}{12 \pi} \Nmatel{G^{a \mu\nu} G^a_{\mu\nu}} \ .
\eeq
We will see below a more precise version of this result. While the $b$ and $t$ quarks are sufficiently heavy that this perturbative treatment is appropriate, this is not so clear for the $c$ quark, see \eg Ref.~\cite{Ellis:2018dmb}. However, recent lattice calculations also allow to take the charm-quark contribution explicitly into account, see \eg Ref.~\cite{Ellis:2018dmb}.

We can then relate the gluon matrix element to that of the light quarks in the following way. At zero momentum transfer, the nucleon mass can be written as
\beq\label{nucleon mass}
\mN \, \bar{u}_N(\bol{k}, r') u_N(\bol{k}, r) = \mN \, \matel{N(\bol{k}, r')}{\bar{N} N}{N(\bol{k}, r)} = \matel{N(\bol{k}, r')}{\Theta^\mu_{\phantom{\mu} \mu}}{N(\bol{k}, r)} \ ,
\eeq
where $\Theta^{\mu \nu}$ is the energy-momentum tensor. One way to see this is that, at zero momentum ($\bol{k} = \bol{0}$), the only non-zero component of $\Nmatel{\Theta^{\mu \nu}}$ is given by $\Theta^{0 0}$, which is the energy density of the system. Since the system is just a single nucleon at rest, its energy density is its mass times the particle number density, which with our state normalization~\eqref{statenormhere} is $\rho(k) = 2 E_k$. Therefore $\matel{N(\bol{0}, r)}{\Theta^\mu_{\phantom{\mu} \mu}}{N(\bol{0}, r)} = \matel{N(\bol{0}, r)}{\Theta^{0 0}}{N(\bol{0}, r)} = 2 \mN^2$, which matches the fact that $\bar{u}_N(\bol{k}, r) u_N(\bol{k}, r) = 2 \mN$. Another way to check this result is to take the zero-momentum limit of the relation $\matel{N(\bol{k}, r)}{\Theta^{\mu \nu}}{N(\bol{k}, r)} = 2 \fvec{k}^\mu \fvec{k}^\nu$~\cite{Jaffe:1989jz}. In Quantum Chromodynamics (QCD), upon application of the equations of motion, the trace of the energy-momentum tensor can be written as
\beq\label{thetamumu}
\Theta^\mu_{\phantom{\mu} \mu} = \sum_q m_q \, \bar{q} q + \frac{\beta(\aS)}{4 \aS} \, G^{a \mu\nu} G^a_{\mu\nu} \ ,
\eeq
where the gluon contribution is due to the trace anomaly. Comparison of \Eq{thetamumu} with \Eq{nucleon mass} clarifies why the matrix elements of the $\bar{q} q$ and $G^{a \mu\nu} G^a_{\mu\nu}$ operators are said to contribute to the nucleon mass. Truncating the beta function at lowest order in powers of $\aS$,
\beq
\beta(\aS) = - (11 - \tfrac{2}{3} N_\text{f}) \frac{\aS^2}{2 \pi} + \Ord(\aS^3) \ ,
\eeq
with $N_\text{f} = 6$ the number of quark flavors, and using \Eq{heavyquarks scalar}, we get
\beq
\Theta^\mu_{\phantom{\mu} \mu} = \sum_{q = u, d, s} m_q \, \bar{q} q - \frac{9 \aS}{8 \pi} \, G^{a \mu\nu} G^a_{\mu\nu} \ .
\eeq
In practice, the heavy-quark mass contribution to $\Theta^\mu_{\phantom{\mu} \mu}$ cancels exactly with the trace-anomaly contribution due to heavy quarks running in the loop, as the triangle diagram for the two processes is the same though with a relative minus sign~\cite{Shifman:1978zn}. The gluon contribution can then be expressed in terms of light quarks by means of \Eq{nucleon mass}. To do so we define
\begin{align}
f_{Tq}^{(N)} \equiv \frac{\matel{N(\bol{k}, r)}{m_q \, \bar{q} q}{N(\bol{k}, r)}}{2 \mN^2} = \frac{m_q}{\mN} F_\text{S}^{q, N}(0) \ ,
&&&
f_{TG}^{(N)} \equiv 1 - \sum_{q = u, d, s} f_{Tq}^{(N)} \ ,
\end{align}
where the $f_{Tq}^{(N)}$'s express the quark-mass contributions to the nucleon mass. \Eq{nucleon mass} then implies
\beq
f_{TG}^{(N)} = - \frac{27}{2} \, \frac{F_\text{S}^{\text{g}, N}(0)}{\mN} \ ,
\eeq
while \Eq{heavyquarks scalar} implies
\beq\label{f_Th}
f_{Th}^{(N)} \to \frac{2}{27} f_{TG}^{(N)}
\eeq
for the heavy quarks $h = c, b, t$ (see \Eq{f_Th Ellis} below for a more refined result). These formulas are usually found in the literature with a different state normalization than the one employed here, see \Eq{statenormhere}. Defining the kets $\ket{\tilde{N}(\bol{k}, r)} \equiv \frac{1}{\sqrt{2 E_k}} \ket{N(\bol{k}, r)}$, normalized so that
\beq\label{Ntilde statenorm}
\braket{\tilde{N}(\bol{k}', r')}{\tilde{N}(\bol{k}, r)} = \delta_{r r'} \, (2 \pi)^3 \delta^{(3)}(\bol{k} - \bol{k}') \ ,
\eeq
we have in the NR limit
\beq
f_{Tq}^{(N)} \NReq \frac{\matel{\tilde{N}(\bol{k}, r)}{m_q \, \bar{q} q}{\tilde{N}(\bol{k}, r)}}{\mN} \ .
\eeq

Two other, alternative parametrizations of the $\fvec{q}^2 = 0$ matrix element of the quark scalar currents are often encountered in the literature,
\begin{align}
B_q^N \equiv \frac{\matel{N(\bol{k}, r)}{\bar{q} q}{N(\bol{k}, r)}}{2 \mN} = \frac{\mN}{m_q} f_{Tq}^{(N)} \ ,
&&&
\sigma_q \equiv m_q B_q^p = \mN f_{Tq}^{(p)} \ ,
\end{align}
where by isospin symmetry
\begin{align}
B_u^p = B_d^n \ ,
&&
B_d^p = B_u^n \ ,
&&
B_s^p = B_s^n \ .
\end{align}
Combinations of these quantities that can be extracted from data are \eg
\begin{align}
\sigma_{\pi N} &\equiv \frac{1}{2} (m_u + m_d) (B_u^p + B_d^p) \ ,
\\
\sigma_0 &\equiv \frac{1}{2} (m_u + m_d) (B_u^p + B_d^p - 2 B_s^p) \ ,
\\
z &\equiv \frac{B_u^p - B_s^p}{B_d^p - B_s^p} \ ,
\end{align}
with $\sigma_{\pi N}$ the \emph{pion-nucleon $\sigma$ term}. Other combinations often used to characterize the strange-quark content of the proton are
\begin{align}
y \equiv \frac{2 B_s^p}{B_u^p + B_d^p} = 1 - \frac{\sigma_0}{\sigma_{\pi N}} \ ,
&&
\sigma_s = \frac{m_s}{m_u + m_d} (\sigma_{\pi N} - \sigma_0) \ .
\end{align}

In Ref.~\cite{Ellis:2018dmb}, the following ``simple but fair representation'' of current lattice estimates of the above matrix elements is proposed:
\begin{align}
\sigma_{\pi N} &= 46 \pm 11~\MeV \ ,
&
\sigma_s &= 35 \pm 16~\MeV \ ,
&
z &= 1.5 \pm 0.5 \ ,
\end{align}
yielding, for the fixed value $z = 1.49$,
\begin{subequations}
\label{f_Tq Ellis}
\begin{align}
f_{Tu}^{(p)} &= 0.018(5) \ ,
&
f_{Td}^{(p)} &= 0.027(7) \ ,
&
f_{Ts}^{(p)} &= 0.037(17) \ ,
&
f_{TG}^{(p)} &= 0.917(19) \ ,
\\
f_{Tu}^{(n)} &= 0.013(3) \ ,
&
f_{Td}^{(n)} &= 0.040(10) \ ,
&
f_{Ts}^{(n)} &= 0.037(17) \ ,
&
f_{TG}^{(n)} &= 0.910(20) \ .
\end{align}
\end{subequations}
Despite the fact that a perturbative computation of the $\bar{c} c$ hadronic matrix elements may not be applicable, as mentioned above, the authors of Ref.~\cite{Ellis:2018dmb} advocate using perturbative estimates for all three heavy quarks as a means to minimize the uncertainty on the final result. The perturbative result in Eqs.~\eqref{heavyquarks scalar},~\eqref{f_Th} can be improved at $\Ord(\aS^3)$, yielding (again at leading order in the heavy-quark expansion)~\cite{Vecchi:2013iza}
\begin{align}
\label{f_Th Ellis}
f_{Tc}^{(N)} &= \frac{2}{27} \left( - 0.3 + 1.48 f_{TG}^{(N)} \right) ,
&
f_{Tb}^{(N)} &= \frac{2}{27} \left( - 0.16 + 1.23 f_{TG}^{(N)} \right) ,
&
f_{Tt}^{(N)} &= \frac{2}{27} \left( - 0.05 + 1.07 f_{TG}^{(N)} \right) .
\end{align}
Using the values of $f_{TG}^{(N)}$ in \Eq{f_Tq Ellis}, this results in
\begin{subequations}
\label{f_Th Ellis 2}
\begin{align}
f_{Tc}^{(p)} &= 0.078(2) \ ,
&
f_{Tb}^{(p)} &= 0.072(2) \ ,
&
f_{Tt}^{(p)} &= 0.069(1) \ ,
\\
f_{Tc}^{(n)} &= 0.078(2) \ ,
&
f_{Tb}^{(n)} &= 0.071(2) \ ,
&
f_{Tt}^{(n)} &= 0.068(2) \ .
\end{align}
\end{subequations}
The $2 + 1 + 1$ flavors FLAG averages of lattice results are~\cite{Aoki:2019cca, Freeman:2012ry, Alexandrou:2014sha}
\begin{align}
\sigma_{\pi N} &= 64.9(1.5)(13.2)~\MeV \ ,
&
\sigma_s &= 41.0(8.8)~\MeV \ ,
\end{align}
which for $z = 1.49$ lead to
\begin{subequations}
\label{f_Tq FLAG}
\begin{align}
f_{Tu}^{(p)} &\approx 0.026 \ ,
&
f_{Td}^{(p)} &\approx 0.038 \ ,
&
f_{Ts}^{(p)} &\approx 0.044 \ ,
&
f_{TG}^{(p)} &\approx 0.89 \ ,
\\
f_{Tu}^{(n)} &\approx 0.018 \ ,
&
f_{Td}^{(n)} &\approx 0.056 \ ,
&
f_{Ts}^{(n)} &\approx 0.044 \ ,
&
f_{TG}^{(n)} &\approx 0.88 \ ,
\end{align}
\end{subequations}
where we used $m_u / m_d \approx 0.47$ and $m_s / m_d \approx 19.5$ from Ref.~\cite{Zyla:2020zbs}. For the heavy quarks we have, using \Eq{f_Th}, $f_{Th}^{(N)} \approx 0.066$, while the more precise \Eq{f_Th Ellis} yields
\begin{align}
\label{f_Th FLAG}
f_{Tc}^{(N)} &\approx 0.075 \ ,
&
f_{Tb}^{(N)} &\approx 0.069 \ ,
&
f_{Tt}^{(N)} &\approx 0.067 \ .
\end{align}
See \eg Ref~\cite{DelNobile:2013sia} for a compilation of numerical values of the $f_{Tq}^{(N)}$'s from older standard references of the direct DM detection literature, Refs.~\cite{Ellis:2000ds, Gondolo:2004sc, Ellis:2008hf, Belanger:2008sj, Cheng:2012qr, Belanger:2013oya}. An estimate of the leading NR corrections (of order $q^2$) to the $f_{Tq}^{(N)}$'s is provided \eg in Ref.~\cite{Hoferichter:2016nvd}.

\subsection{Pseudo-scalar couplings}
\label{Pseudo-scalar couplings}
\Tab{tab: PT} shows that the pseudo-scalar operators
\begin{align}
\Op_q = \bar{q} \, i \gamma^5 q \ ,
&&&
\Op_\text{g} = G^{a \mu\nu} \tilde{G}^a_{\mu\nu} \ ,
\end{align}
have the same $P$ and $T$ quantum numbers, therefore we will deal with them together (as for the scalar operator, the numerical factors in $\Op_\text{g}$ have been chosen for later convenience). Given the transformation properties of these operators under the Lorentz symmetry, spatial parity and time reversal, their nucleon matrix elements can be parametrized in terms of a single operator-specific form factor:
\begin{align}
\label{pseudo-scalar current parametrization}
\Nmatel{\bar{q} \, i \gamma^5 q} = F_\text{PS}^{q, N}(\fvec{q}^2) \, \bar{u}'_N \, i \gamma^5 u_N \ ,
&&&
\Nmatel{G^{a \mu\nu} \tilde{G}^a_{\mu\nu}} = F_\text{PS}^{\text{g}, N}(\fvec{q}^2) \, \bar{u}'_N \, i \gamma^5 u_N \ .
\end{align}
$\Op_q$ and $\Op_\text{g}$ being hermitian implies that $F_\text{PS}^{q, N}$, $F_\text{PS}^{\text{g}, N}$ are real functions of $\fvec{q}^2$. Other Lorentz scalars that can be constructed with the available ingredients (\ie the nucleon spinors, the Dirac matrices and the nucleon momentum four-vectors) either have the wrong transformation properties under parity (\eg $\bar{u}'_N u_N$) or can be reduced to the above by means of the equations of motion (\eg $i \fvec{q}_\mu \, \bar{u}'_N \gamma^\mu \gamma^5 u_N$), see \Eq{EOMs}. Notice that \Eq{pseudo-scalar current parametrization} does not formally define $F_\text{PS}^{q, N}(0)$ and $F_\text{PS}^{\text{g}, N}(0)$ because $\bar{u}'_N \, i \gamma^5 u_N$ vanishes at zero momentum transfer, as one can easily verify using the equivalent of \Eq{u spinor} for the nucleon four-spinor together with \Eq{Dirac matrices}.

To compute the matrix element of the pseudo-scalar operators $\Op_q$, $\Op_\text{g}$ we can proceed as done above for the scalar operators~\cite{Shifman:1978zn} (see also Refs.~\cite{Cheng:1988im, Cheng:2012qr, Dienes:2013xya}). Integrating out the heavy quarks $h = c, b, t$, a loop-induced gluon operator is generated whose contribution to the matrix element is reproduced by the following substitution valid at lowest order in $\aS$ and $\Lambda_\text{QCD} / m_h$,
\beq\label{heavyquarks pseudo-scalar}
m_h \, \Nmatel{\bar{h} \, i \gamma^5 h} \rightarrow \frac{\aS}{8 \pi} \Nmatel{G^{a \mu\nu} \tilde{G}^a_{\mu\nu}} \ .
\eeq
This is compatible with the chiral-anomaly relation
\beq\label{axial current divergence}
\Nmatel{\partial_\mu (\bar{q} \gamma^\mu \gamma^5 q)} = 2 m_q \Nmatel{\bar{q} \, i \gamma^5 q} - \frac{\aS}{4 \pi} \Nmatel{G^{a \mu\nu} \tilde{G}^a_{\mu\nu}} \ ,
\eeq
where the last term is due to the chiral anomaly.\footnote{\Eq{heavyquarks pseudo-scalar} and \Eq{axial current divergence} can be found in the literature with different signs with respect to those featured here. This depends on the different definitions of $\varepsilon^{\mu\nu\rho\sigma}$ and $\gamma^5$ employed. Our definitions are reported in \Eq{LCepsilon} and \Eq{gamma5 + sigma}. A factor of $2$ difference in \Eq{axial current divergence} could be explained by a different definition of $\tilde{G}^a_{\mu\nu}$, see \Eq{Gtilde}. Eqs.~\eqref{heavyquarks pseudo-scalar},~\eqref{axial current divergence}, together with other formulas derived in this subsection, correct the corresponding formulas derived in Ref~\cite{DelNobile:2013sia}.} Sufficiently heavy quarks have no appreciable dynamics in the nucleon and therefore the matrix element of the derivative term on the left-hand side vanishes~\cite{Dienes:2013xya}, yielding \Eq{heavyquarks pseudo-scalar}.

Evaluating the nucleon matrix element of the $G^{a \mu\nu} \tilde{G}^a_{\mu\nu}$ operator is a problematic task. We rely on the analysis performed in Refs.~\cite{Cheng:1988im, Cheng:2012qr}, based on the relation
\beq\label{Large Nc chiral limit}
\Nmatel{\bar{u} \, i \gamma^5 u + \bar{d} \, i \gamma^5 d + \bar{s} \, i \gamma^5 s} = 0 \ ,
\eeq
valid in the large-$N_\text{c}$ and chiral limits. We now use
\beq\label{psi(0) psi(x)}
\Op(\fvec{x}) = e^{+ i \fvec{P} \cdot \fvec{x}} \, \Op(\fvec{0}) \, e^{- i \fvec{P} \cdot \fvec{x}} \ ,
\eeq
with $\Op$ denoting a generic operator and $\fvec{P}$ denoting the four-momentum operator. Employing also \Eq{<qbar gamma^mu gamma^5 q>} below and \Eq{EOMs}, we have
\beq
\Nmatel{\partial_\mu (\bar{q} \gamma^\mu \gamma^5 q)} = i \fvec{q}_\mu \Nmatel{\bar{q} \gamma^\mu \gamma^5 q} \qeq 2 \mN G_q^N(\fvec{q}^2) \, \bar{u}'_N \, i \gamma^5 u_N \ ,
\eeq
with
\beq\label{G_q^N}
G_q^N(\fvec{q}^2) \equiv \Delta_q^{(N)} - \fvec{q}^2 \left( \frac{a_{q, \pi}^N}{\fvec{q}^2 - m_\pi^2} + \frac{a_{q, \eta}^N}{\fvec{q}^2 - m_\eta^2} \right) ,
\eeq
see \Sec{Axial-vector couplings} for further details. The $\qeq$ sign means the equality is only valid at some finite order in an expansion in powers of $q / \mN$, see below \Eq{<p'|A^a|p>} where it is introduced. Comparison with \Eq{axial current divergence} results in
\beq\label{gluons and lightq}
\frac{\aS}{4 \pi} \, \Nmatel{G^{a \mu\nu} \tilde{G}^a_{\mu\nu}} \qeq - 2 \mN G_q^N(\fvec{q}^2) \, \bar{u}'_N \, i \gamma^5 u_N + 2 m_q \, \Nmatel{\bar{q} \, i \gamma^5 q} \ ,
\eeq
so that dividing by $m_q$, summing over the light quarks and using \Eq{Large Nc chiral limit} yields
\beq
\frac{\aS}{8 \pi} \, \Nmatel{G^{a \mu\nu} \tilde{G}^a_{\mu\nu}} \qeq - \mN \bar{m} \left( \sum_{q = u, d, s} \frac{G_q^N(\fvec{q}^2)}{m_q} \right) \bar{u}'_N \, i \gamma^5 u_N \ ,
\eeq
with
\beq
\bar{m} \equiv \left( \frac{1}{m_u} + \frac{1}{m_d} + \frac{1}{m_s} \right)^{-1} \ .
\eeq
Using this result in \Eq{gluons and lightq} we then get for $q = u, d, s$
\beq
\Nmatel{\bar{q} \, i \gamma^5 q} \qeq \frac{\mN}{m_q} \left( G_q^N(\fvec{q}^2) - \bar{m} \sum_{q' = u, d, s} \frac{G_{q'}^N(\fvec{q}^2)}{m_{q'}} \right) \bar{u}'_N \, i \gamma^5 u_N \ .
\eeq

\subsection{Vector couplings}
\label{Vector couplings}
The nucleon matrix element of the quark vector currents,
\beq
\Op_q = V_q^\mu \equiv \bar{q} \gamma^\mu q \ ,
\eeq
can be parametrized in the following way by means of its transformation properties under the Lorentz symmetry, spatial parity and time reversal:
\beq\label{vector current parametrization}
\Nmatel{V_q^\mu} = \bar{u}'_N \left( F_1^{q, N}(\fvec{q}^2) \gamma^\mu + F_2^{q, N}(\fvec{q}^2) \frac{i \sigma^{\mu\nu} \fvec{q}_\nu}{2 \mN} \right) u_N \ .
\eeq
$\Op_q$ being hermitian implies that the \emph{Dirac form factor} $F_1^{q, N}$ and the \emph{Pauli form factor} $F_2^{q, N}$ are real. Other four-vectors that can be constructed with the available ingredients (\ie the nucleon spinors, the Dirac matrices and the nucleon momentum four-vectors) either have the wrong transformation properties under parity (\eg $\bar{u}'_N \gamma^\mu \gamma^5 u_N$) and/or under time reversal (\eg $i \fvec{q}^\mu \, \bar{u}'_N u_N$), or can be reduced to the above by means of the equations of motion, see \Eq{EOMs} (\eg $\fvec{K}^\mu \, \bar{u}'_N u_N$ can be rewritten using the Gordon identity).

Here is an example to see how the time-reversal transformation properties of $\Op_q$ can be used to constrain the right-hand side of \Eq{vector current parametrization} (see \Sec{Scalar couplings} for an example using parity instead). One has
\beq\label{vector T}
\Nmatel{\Op_q} = \Nmatel{T^{-1} T \Op_q T^{-1} T} = (-1)^\mu \matel{T N'}{\Op_q}{T N} \ ,
\eeq
where we used \Eq{P T} and we indicated with $\ket{T N^{(\prime)}} \equiv T \ket{N^{(\prime)}}$ the time-reversed nucleon state. Let us now check whether (a term proportional to) $i \fvec{q}^\mu \, \bar{u}'_N u_N = \Nmatel{\partial^\mu (\bar{N} N)}$ could appear on the right-hand side of \Eq{vector current parametrization}. We can do so by noticing that
\beq\label{T matel}
(-1)^\mu \matel{T N'}{\partial^\mu (\bar{N} N)}{T N} = (-1)^\mu \Nmatel{T \partial^\mu (\bar{N} N) T^{-1}} = - \Nmatel{\partial^\mu (\bar{N} N)} \ ,
\eeq
where we used again \Eq{P T}, the transformation properties of the derivative, and the fact that, when restricted to fermion states, $T^{-1} = - T$. This can only be compatible with \Eq{vector T} if $i \fvec{q}^\mu \, \bar{u}'_N u_N$ appears in \Eq{vector current parametrization} with null coefficient. Terms that, like this, have reversed $G$-parity assignment with respect to those on the right-hand side of \Eq{vector current parametrization}, are called (``for obscure historical reasons''~\cite{Georgi:1985kw}) \emph{second-class currents} in the context of processes involving charged currents such as $\bar{u}'_p \cdots u_n$~\cite{Weinberg:1958ut, Holstein:1976mw}.

The $F_1^{q, N}$, $F_2^{q, N}$ form factors entering \Eq{vector current parametrization} can be obtained as follows. Neglecting the heavy quarks $c$, $b$, $t$, QCD (and therefore nucleons) respect an approximate $U(3)$ flavor symmetry that is only (explicitly) broken by the $u$, $d$, $s$ quark mass differences. When including QED, this symmetry is also (explicitly) broken by the different up-type and down-type quark electric charges, but the approximate charge independence of the nucleon structure suggests we can ignore this effect for our practical purposes. Neglecting the breaking induces an error smaller than $25 \%$ on the predicted relations between couplings and matrix elements~\cite{Langacker:2010zza}. Assuming an unbroken $U(3)$ flavor symmetry, the nine flavor vector currents
\beq\label{vector currents}
\mathscr{V}^a_\mu \equiv \bar{f} \gamma_\mu T^a f \ ,
\eeq
with $f \equiv (u, d, s)^\tr$ and the $T^a$'s the nine $U(3) = U(1) \times SU(3)$ generators, are conserved (\emph{conserved vector current} or \emph{CVC}). Since we are only interested in the flavor-diagonal quark bilinears $V_q^\mu$, we only need the diagonal currents. We write as customary the $SU(3)$ generators in the fundamental representation as $T^a = \lambda^a / 2$, with $\lambda^a$ the Gell-Mann matrices; in this way the $T^a$'s are normalized so that $\Tr (T^a T^b) = \frac{1}{2} \delta_{a b}$. With $T^0$ normalized so that hadrons have baryon number (the conserved charge associated to $\mathscr{V}^0_\mu$) equal to $1$, the diagonal $U(3)$ generators are then determined by
\begin{align}
T^0 &\equiv \frac{1}{3}
\begin{pmatrix}
1 & 0 & 0
\\
0 & 1 & 0
\\
0 & 0 & 1
\end{pmatrix}
,
&
\lambda^3 &=
\begin{pmatrix}
1 & 0 & 0
\\
0 & -1 & 0
\\
0 & 0 & 0
\end{pmatrix}
,
&
\lambda^8 &= \frac{1}{\sqrt{3}}
\begin{pmatrix}
1 & 0 & 0
\\
0 & 1 & 0
\\
0 & 0 & -2
\end{pmatrix}
.
\end{align}
It will be convenient to use the flavor-diagonal operators
\begin{align}
V_3^\mu &\equiv 2 \, \mathscr{V}^{3 \mu} = V_u^\mu - V_d^\mu \ ,
\\
V_8^\mu &\equiv \frac{2}{\sqrt{3}} \, \mathscr{V}^{8 \mu} = \frac{1}{3} (V_u^\mu + V_d^\mu - 2 V_s^\mu) \ ,
\intertext{as well as the flavor-singlet operator}
\label{V_0 current}
V_0^\mu &\equiv \mathscr{V}^{0 \mu} = \frac{1}{3} (V_u^\mu + V_d^\mu + V_s^\mu) \ .
\end{align}
The electromagnetic current and the neutral vector current, defined by the interaction Lagrangian (part of the SM Lagrangian)
\beq\label{JA}
\Lag_{\gamma Z} = - e \, J_\text{EM}^\mu A_\mu - \frac{g}{2 c_\text{W}} \, J_\text{NC}^\mu Z_\mu \ ,
\eeq
can be written in terms of the flavor vector currents as
\begin{align}
\label{J_EM}
J_\text{EM}^\mu &= \frac{2}{3} V_u^\mu - \frac{1}{3} (V_d^\mu + V_s^\mu) = \frac{1}{2} (V_3^\mu + V_8^\mu) \ ,
\\
J_\text{NC}^\mu &= \left( \frac{1}{2} - \frac{4}{3} s_\text{W}^2 \right) V_u^\mu + \left( - \frac{1}{2} + \frac{2}{3} s_\text{W}^2 \right) (V_d^\mu + V_s^\mu) = (1 - 2 s_\text{W}^2) J_\text{EM}^\mu - \frac{1}{2} V_0^\mu \ .
\end{align}
Here $e$ is the electric-charge unit, $g$ is the $SU(2)_\text{L}$ gauge coupling, and $c_\text{W}$ and $s_\text{W}$ are the cosine and sine of the electroweak gauge bosons mixing angle, respectively. We can then apply \Eq{vector current parametrization} to define the $F_i^p, F_i^\text{NC}, F_i^0, F_i^3, F_i^8$ proton and $F_i^n$ neutron form factors, for $i = 1, 2$, as follows:
\begin{align}
\label{<J_EM>}
\Nmatel{J_\text{EM}^\mu} &= \bar{u}'_N \left( F_1^N(\fvec{q}^2) \gamma^\mu + F_2^N(\fvec{q}^2) \frac{i \sigma^{\mu\nu} \fvec{q}_\nu}{2 \mN} \right) u_N \ ,
\\
\matel{p'}{J_\text{NC}^\mu, V_0^\mu, V_3^\mu, V_8^\mu}{p} &= \bar{u}'_p \left( F_1^{\text{NC}, 0, 3, 8}(\fvec{q}^2) \gamma^\mu + F_2^{\text{NC}, 0, 3, 8}(\fvec{q}^2) \frac{i \sigma^{\mu\nu} \fvec{q}_\nu}{2 \mN} \right) u_p \ .
\end{align}
The other neutron form factors can be derived from the proton ones using isospin symmetry, which implies, since proton and neutron form a doublet under the $SU(2)$ isospin subgroup of $SU(3)$,
\begin{align}
\matel{p'}{V_3^\mu}{p} &= - \matel{n'}{V_3^\mu}{n} \ ,
&
\matel{p'}{V_8^\mu}{p} &= \matel{n'}{V_8^\mu}{n} \ ,
&
\matel{p'}{V_0^\mu}{p} &= \matel{n'}{V_0^\mu}{n} \ ,
\\
\label{V_q isospin}
\matel{p'}{V_u^\mu}{p} &= \matel{n'}{V_d^\mu}{n} \ ,
&
\matel{p'}{V_d^\mu}{p} &= \matel{n'}{V_u^\mu}{n} \ ,
&
\matel{p'}{V_s^\mu}{p} &= \matel{n'}{V_s^\mu}{n} \ .
\end{align}
The above relations among currents and among current matrix elements translate into
\begin{align}
F_i^p = \frac{1}{2} (F_i^8 + F_i^3) \ ,
&&
F_i^n = \frac{1}{2} (F_i^8 - F_i^3) \ ,
&&
F_i^\text{NC} = (1 - 2 s_\text{W}^2) F_i^p - \frac{1}{2} F_i^0 \ ,
\end{align}
which can be inverted to yield
\begin{align}
F_i^3 = F_i^p - F_i^n \ ,
&&
F_i^8 = F_i^p + F_i^n \ ,
&&
F_i^0 = 2 \left[ (1 - 2 s_\text{W}^2) F_i^p - F_i^\text{NC} \right] .
\end{align}
We also have
\begin{align}
F_i^{u, p} &= \frac{1}{2} (3 F_i^8 + F_i^3) + F_i^{s, p} = 2 F_i^p + F_i^n + F_i^{s, p} \ ,
\\
F_i^{d, p} &= \frac{1}{2} (3 F_i^8 - F_i^3) + F_i^{s, p} = F_i^p + 2 F_i^n + F_i^{s, p} \ ,
\\
F_i^{s, p} &= F_i^0 - F_i^8 = (1 - 4 s_\text{W}^2) F_i^p - F_i^n - 2 F_i^\text{NC} \ .
\end{align}

At zero momentum transfer, the $F_1$'s take the following values. For $F_1^N$ we have
\beq
F_1^N(0) = Q_N \ ,
\eeq
with
\begin{align}
\label{Q_N}
Q_p = 1 \ ,
&&&
Q_n = 0
\end{align}
the electric charge of proton and neutron in units of $e$, respectively, while
\beq
F_1^0(0) = 1
\eeq
is the baryon number of the proton. This is compatible with what we expect from the fact that, at zero momentum transfer, vector currents simply `count' the valence quarks in the nucleon, \ie
\begin{align}
\label{F_1^{q, p}}
F_1^{u, p}(0) = 2 \ ,
&&
F_1^{d, p}(0) = 1 \ ,
&&
F_1^{s, p}(0) = 0 \ .
\end{align}
This is because $\partial_\mu V_0^\mu = 0$ (up to a negligible anomaly) implies that $\int \ud^3 x \, V_0^0(\fvec{x})$ is the baryon-number conserved charge and therefore, using \Eq{psi(0) psi(x)},
\beq
\braket{p'}{p} = \int \ud^3 x \, \matel{p'}{V_0^0(\fvec{x})}{p} = \matel{p'}{V_0^0(\fvec{0})}{p} \int \ud^3 x \, e^{i \fvec{q} \cdot \fvec{x}} \ ,
\eeq
so that using
\beq\label{delta function integral rep}
\int \ud^n x \, e^{i x z} = (2 \pi)^n \delta^{(n)}(z)
\eeq
and Eqs.~\eqref{V_0 current},~\eqref{vector current parametrization},~\eqref{spinor normalization},~\eqref{statenormhere} yields
\beq
1 = \frac{1}{3} \sum_{q = u, d, s} F_1^{q, N}(0) \ .
\eeq
From the above we get the tree-level values
\begin{align}
F_1^3(0) = 1 \ ,
&&
F_1^8(0) = 1 \ ,
&&
F_1^\text{NC}(0) = \frac{1}{2} - 2 s_\text{W}^2 \approx 0.02 \ ,
\end{align}
where we used
\beq\label{s_W}
s_\text{W}^2 \approx 0.24
\eeq
at zero momentum transfer~\cite{Zyla:2020zbs}; including radiative corrections yields\footnote{$F_1^\text{NC}(0)$ equals half of the proton's weak charge $Q_\text{W}^p \approx 0.07$~\cite{Androic:2018kni, Zyla:2020zbs}.}
\beq\label{F_1^NC(0)}
F_1^\text{NC}(0) \approx 0.035 \ ,
\eeq
their sizeable impact being ascribable to the large cancellation in the tree-level computation of $F_1^\text{NC}(0)$, which occurs due to $s_\text{W}^2$ being close to $1/4$. The zero momentum transfer value of the $F_2$'s can be computed as follows. As explained later on in \Sec{Electromagnetic interactions},
\begin{align}
F_2^p(0) = \kappa_p \ ,
&&&
F_2^n(0) = \kappa_n
\end{align}
are the anomalous magnetic moments of respectively proton and neutron in units of the nuclear magneton $\hat{\mu}_\text{N}$, defined in \Eq{nuclear magneton}. This leads to
\begin{align}
F_2^3(0) = \kappa_p - \kappa_n \ ,
&&
F_2^8(0) = \kappa_p + \kappa_n \ ,
&&
F_2^0(0) = 2 [(1 - 2 s_\text{W}^2) \kappa_p - F_2^\text{NC}(0)] \ ,
\end{align}
or
\begin{subequations}
\label{F_2^{q, p}}
\begin{align}
F_2^{u, p}(0) &= 2 \kappa_p + \kappa_n + F_2^{s, p}(0) \ ,
\\
F_2^{d, p}(0) &= \kappa_p + 2 \kappa_n + F_2^{s, p}(0) \ ,
\\
F_2^{s, p}(0) &= (1 - 4 s_\text{W}^2) \kappa_p - \kappa_n - 2 F_2^\text{NC}(0) \ .
\end{align}
\end{subequations}
$F_2^{s, p}(0)$ can also be directly measured experimentally or computed on the lattice, the latter method yielding quite smaller values with considerably smaller error with respect to experiments; some results are:
\beq
F_2^{s, p}(0) =
\begin{cases}
- 0.19 \pm 0.14 & \text{Ref.~\cite{Androic:2018kni} (experiment)},
\\
- 0.26 \pm 0.26 & \text{Ref.~\cite{Gonzalez-Jimenez:2014bia} (global fit to experimental data)},
\\
- 0.022 \pm 0.008 & \text{Ref.~\cite{Green:2015wqa} (lattice)},
\\
- 0.064 \pm 0.017 & \text{Ref.~\cite{Sufian:2016pex} (lattice)},
\end{cases}
\eeq
where we added the different uncertainties on the result of Ref.~\cite{Green:2015wqa} in quadrature.

The approximate $\fvec{q}^2$ dependence of the form factors can be derived as follows. It is customary to employ the \emph{electric} and \emph{magnetic Sachs form factors} $G_\text{E}^N$, $G_\text{M}^N$ in place of $F_1^N$, $F_2^N$, defined as
\begin{align}
\label{Sachs FF}
G_\text{E}^N(\fvec{q}^2) \equiv F_1^N(\fvec{q}^2) + \frac{\fvec{q}^2}{4 \mN^2} F_2^N(\fvec{q}^2) \ ,
&&&
G_\text{M}^N(\fvec{q}^2) \equiv F_1^N(\fvec{q}^2) + F_2^N(\fvec{q}^2) \ ,
\end{align}
which inverted yield
\begin{align}
F_1^N(\fvec{q}^2) = \frac{G_\text{E}^N(\fvec{q}^2) - \frac{\fvec{q}^2}{4 \mN^2} G_\text{M}^N(\fvec{q}^2)}{1 - \fvec{q}^2 / 4 \mN^2} \ ,
&&&
F_2^N(\fvec{q}^2) = \frac{G_\text{M}^N(\fvec{q}^2) - G_\text{E}^N(\fvec{q}^2)}{1 - \fvec{q}^2 / 4 \mN^2} \ .
\end{align}
Their value at zero momentum transfer is
\begin{align}
\label{G_E,M(0)}
G_\text{E}^N(0) = Q_N \ ,
&&&
G_\text{M}^N(0) = \frac{g_N}{2} \ ,
\end{align}
the nucleon $g$-factors $g_N$ being given in \Eq{g_N} below. The first-order $\fvec{q}^2$ contribution to $G_\text{E}^N$ ($G_\text{M}^N$) in an expansion in powers of $\fvec{q}^2$ yields the nucleon \emph{charge} (\emph{magnetic}) \emph{radius},
\beq
\left. \frac{\ud G_{\text{E}, \text{M}}^N}{\ud \fvec{q}^2} \right|_{\fvec{q}^2 = 0} = \frac{1}{6} \langle r_N^2 \rangle_{\text{E}, \text{M}} \ .
\eeq
The $1/6$ factor can be seen originating from \Eq{Taylor expanded F} below (see \eg Ref.~\cite{Povh}), taking into account that $\fvec{q}^2 \NReq - q^2$, see \Eq{t = - q^2}. The $\fvec{q}^2$ dependence of $G_\text{E}^p$ and $G_\text{M}^N$ can be described to a good approximation by a so-called \emph{dipole fit},
\beq\label{dipole FF}
G_\text{D}(\fvec{q}^2, M) \equiv \frac{1}{(1 - \fvec{q}^2 / M^2)^2} \ ,
\eeq
which indicates an exponentially falling charge distribution (see \eg Ref.~\cite{Povh}). Taking into account the normalization in \Eq{G_E,M(0)}, with $Q_p = 1$, we thus have for the proton electric and magnetic form factors and for the neutron magnetic form factor~\cite{Walecka, Povh} (see also Ref.~\cite{Ye:2017gyb}),
\begin{align}
\label{G's}
G_\text{E}^p(\fvec{q}^2), \frac{G_\text{M}^N(\fvec{q}^2)}{g_N / 2} \approx G_\text{D}(\fvec{q}^2, M_\text{V})
&&
\text{with}
&&
M_\text{V}^2 \approx 0.71~\GeV^2 \ .
\end{align}
The respective charge and magnetic radii can be derived by noting that, in general, $\left. \ud G_\text{D} / \ud \fvec{q}^2 \right|_{\fvec{q}^2 = 0} = 2 / M^2$, which returns in this case a root mean square radius of $\sqrt{12 / M_\text{V}^2} \approx 0.81~\fm$; for reference, Ref.~\cite{Zyla:2020zbs} reports the following values:\footnote{The reported value of the proton charge radius, also recommended by Ref.~\cite{CODATA-2018}, is from muon-proton data. Refs.~\cite{Mohr:2015ccw, Mohr:2018hvt} recommend instead the value $\sqrt{\langle r_p^2 \rangle_\text{E}} \approx 0.88~\fm$, which is obtained from electron-proton data. The discrepancy between different ways of measuring $\langle r_p^2 \rangle_\text{E}$ is known as the \emph{proton radius puzzle}, see \eg Ref.~\cite{Zyla:2020zbs} for a concise summary or Refs.~\cite{Pohl:2013yb, Carlson:2015jba, Krauth:2017ijq} for more complete reviews. In any case, the discrepancy arises at a level of precision that is much higher than what is currently needed for DM detection.}
\begin{align}
\sqrt{\langle r_p^2 \rangle_\text{E}} \approx 0.84~\fm \ ,
&&
\sqrt{\langle r_p^2 \rangle_\text{M}} \approx 0.85~\fm \ ,
&&
\sqrt{\langle r_n^2 \rangle_\text{M}} \approx 0.86~\fm \ ,
\end{align}
where the small differences with the value derived from \Eq{G's} highlight slight deviations from the dipole parametrization. The neutron electric form factor, which equals $Q_n = 0$ at zero momentum transfer, can instead be approximately fitted by~\cite{Galster:1971kv} (see also Ref.~\cite{Kelly:2004hm})
\beq
G_\text{E}^n(\fvec{q}^2) \approx \frac{\kappa_n \, \fvec{q}^2 / 4 m_n^2}{1 - 5.6 \, \fvec{q}^2 / 4 m_n^2} G_\text{D}(\fvec{q}^2, M_\text{V}) \ ,
\eeq
with magnetic radius $3 \kappa_n / 2 \mN^2 \approx - 0.13~\fm^2$, where for reference Ref.~\cite{Zyla:2020zbs} reports
\beq
\langle r_n^2 \rangle_\text{E} \approx - 0.12~\fm^2 \ .
\eeq
The strange Sachs form factors $G_{\text{E}, \text{M}}^{s, p}$ can be defined analogously to $G_{\text{E}, \text{M}}^p$ in \Eq{Sachs FF}, with $F_{1, 2}^{s, p}$ in place of $F_{1, 2}^p$. $G_\text{E}^{s, p}$ can be approximated as (see \eg Ref.~\cite{Androic:2018kni} and references therein)
\beq
G_\text{E}^{s, p}(\fvec{q}^2) \approx \frac{1}{6} \langle r_{p, s}^2 \rangle_\text{E} \fvec{q}^2 \, G_\text{D}(\fvec{q}^2, 1~\GeV) \ ,
\eeq
where some determinations of the strange radius of the proton are
\beq
\langle r_{p, s}^2 \rangle_\text{E} =
\begin{cases}
- 0.013 \pm 0.007~\fm^2 & \text{Refs.~\cite{Androic:2018kni, Horowitz:2018yxh} (experiment)},
\\
- 0.061 \pm 0.038~\fm^2 & \text{Ref.~\cite{Gonzalez-Jimenez:2014bia} (global fit to experimental data)},
\\
- 0.0067 \pm 0.0025~\fm^2 & \text{Ref.~\cite{Green:2015wqa} (lattice)},
\\
-0.0043 \pm 0.0021~\fm^2 & \text{Ref.~\cite{Sufian:2016pex} (lattice)}.
\end{cases}
\eeq
Here we turned the values $\rho_s = 0.20 \pm 0.11$~\cite{Androic:2018kni} and $\rho_s = 0.92 \pm 0.58$~\cite{Gonzalez-Jimenez:2014bia} into values for the strange radius by using $G_\text{E}^{s, p} = - \rho_s \fvec{q}^2 G_\text{D} / 4 m_p^2$, yielding $\langle r_{p, s}^2 \rangle_\text{E} = - 6 \rho_s / 4 m_p^2$, and we added the different uncertainties on the result of Ref.~\cite{Green:2015wqa} in quadrature. $G_\text{M}^{s, p}$ can instead be approximated as~\cite{Androic:2018kni}
\beq
\frac{G_\text{M}^{s, p}(\fvec{q}^2)}{G_\text{M}^{s, p}(0)} \approx G_\text{D}(\fvec{q}^2, 1~\GeV) \ ,
\eeq
with $G_\text{M}^{s, p}(0) = F_2^{s, p}(0)$ (also denoted $\mu_s$ in the literature) in accordance with Eqs.~\eqref{F_1^{q, p}},~\eqref{Sachs FF}. For reference, Ref.~\cite{Green:2015wqa} finds
\beq
\langle r_{p, s}^2 \rangle_\text{M} = - 0.018 \pm 0.009~\fm^2 \ ,
\eeq
pointing to a mass parameter in the dipole parametrization of $26 \pm 13~\GeV^2$. A study of the $\fvec{q}^2$ dependence of the $F_{1, 2}^\text{NC}$ form factors for both proton and neutron can be found \eg in Ref.~\cite{Sufian:2016vso}.

\subsection{Axial-vector couplings}
\label{Axial-vector couplings}
The nucleon matrix element of the quark axial-vector currents,
\beq
\Op_q = A_q^\mu \equiv \bar{q} \gamma^\mu \gamma^5 q \ ,
\eeq
can be parametrized in the following way by means of its transformation properties under the Lorentz symmetry, spatial parity and time reversal:
\beq\label{axial current parametrization}
\Nmatel{A_q^\mu} = \bar{u}'_N \left( G_\text{A}^{q, N}(\fvec{q}^2) \gamma^\mu \gamma^5 + G_\text{P}^{q, N}(\fvec{q}^2) \frac{\fvec{q}^\mu}{2 \mN} \gamma^5 \right) u_N \ .
\eeq
$\Op_q$ being hermitian implies that the form factors $G_\text{A}^{q, N}$, $G_\text{P}^{q, N}$ are real. For the other four-vectors that can be constructed with the available ingredients (\ie the nucleon spinors, the Dirac matrices and the nucleon momentum four-vectors), those with wrong transformation properties under the space-time discrete symmetries can be discarded: for instance, $\bar{u}'_N \gamma^\mu u_N$ has the wrong parity while $i \fvec{q}_\mu \, \bar{u}'_N \, i \sigma^{\mu\nu} \gamma^5 u_N$ does not transform like an axial vector under time reversal, the latter also being a second-class current because of its reversed $G$-parity assignment~\cite{Weinberg:1958ut, Holstein:1976mw, Georgi:1985kw} (see the brief discussion after \Eq{T matel}). The remaining terms with the correct transformation properties can be reduced to the above by means of the equations of motion, see \Eq{EOMs} (\eg $\fvec{K}_\mu \, \bar{u}'_N \, i \sigma^{\mu\nu} \gamma^5 u_N$ can be rewritten using a Gordon-like identity).

At zero momentum transfer, the nucleon matrix element of the axial-vector quark bilinear is usually parametrized as
\beq\label{Deltaq}
\matel{N(\bol{k}, r')}{A_q^\mu}{N(\bol{k}, r)} = 2 \mN \Delta_q^{(N)} \, \fvec{s}^\mu \ ,
\eeq
where
\beq
\fvec{s}^\mu = \left( \frac{\bol{k} \cdot \hat{\bol{n}}}{\mN}, \hat{\bol{n}} + \bol{k} \frac{\bol{k} \cdot \hat{\bol{n}}}{\mN (E_k + \mN)} \right)^\tr = \frac{1}{2 \mN} \, \bar{u}_N(\bol{k}, r') \gamma^\mu \gamma^5 u_N(\bol{k}, r)
\eeq
is the nucleon polarization four-vector. The last equality can be proved by direct computation using the equivalent of \Eq{u spinor} for the nucleon four-spinor. The polarization unit three-vector $\hat{\bol{n}}$, normalized so that $\hat{\bol{n}} \cdot \hat{\bol{n}}^* = 1$, is identified with twice the nucleon spin, $\hat{\bol{n}} = 2 \bol{s}_N = {\xi^{r'}}^\dagger \bol{\sigma} \xi^r$ (see \Eq{I s} below). The polarization four-vector $\fvec{s}^\mu$ satisfies $\fvec{s} \cdot \fvec{k} = 0$ and $\fvec{s} \cdot \fvec{s}^* = - 1$ (the latter property can be easily checked in the nucleon rest frame, while $4 \mN \, \fvec{s} \cdot \fvec{k} = \bar{u}'_N \slashed{\fvec{k}} \gamma^5 u_N - \bar{u}'_N \gamma^5 \slashed{\fvec{k}} u_N = 0$ follows from the equations of motion). To make contact with the literature, we notice that with the state normalization~\eqref{Ntilde statenorm}, \Eq{Deltaq} reads in the NR limit
\beq
\matel{\tilde{N}(\bol{k}, r')}{A_q^\mu}{\tilde{N}(\bol{k}, r)} \NReq \Delta_q^{(N)} \, \fvec{s}^\mu \ .
\eeq

The coefficients $G_\text{A}^{q, N}(0) = \Delta_q^{(N)}$ parametrize the quark spin content of the nucleon. These coefficients are argued to be negligible for heavy quarks~\cite{Polyakov:1998rb}, while for light quarks they satisfy the following relations in the isospin-symmetric limit:
\begin{align}
\label{Delta isospin}
\Delta_u^{(p)} = \Delta_d^{(n)} \ ,
&&
\Delta_d^{(p)} = \Delta_u^{(n)} \ ,
&&
\Delta_s^{(p)} = \Delta_s^{(n)} \ .
\end{align}
The $2 + 1 + 1$ flavors FLAG averages of lattice results are~\cite{Aoki:2019cca, Lin:2018obj}
\begin{align}
\label{Delta_q}
\Delta_u^{(p)} &= 0.777(25)(30) \ ,
&
\Delta_d^{(p)} &= - 0.438(18)(30) \ ,
&
\Delta_s^{(p)} &= - 0.053(8) \ .
\end{align}
See \eg Ref~\cite{DelNobile:2013sia} for a compilation of numerical values of the $\Delta_q^{(p)}$'s from older standard references~\cite{Ellis:2000ds, Gondolo:2004sc, Ellis:2008hf, Belanger:2008sj, Cheng:2012qr, Belanger:2013oya} of the direct DM detection literature.

An alternative notation used in the literature involves the nine axial-vector flavor currents
\beq\label{axial currents}
\mathscr{A}^a_\mu \equiv \bar{f} \gamma_\mu \gamma^5 T^a f
\eeq
(see below \Eq{vector currents} for our notation). We will only need the flavor-diagonal operators
\begin{align}
A^{3 \mu} &\equiv 2 \, \mathscr{A}^{3 \mu} = A_u^\mu - A_d^\mu \ ,
\\
A^{8 \mu} &\equiv 2 \sqrt{3} \, \mathscr{A}^{8 \mu} = A_u^\mu + A_d^\mu - 2 A_s^\mu \ ,
\intertext{as well as the flavor-singlet operator}
A^{0 \mu} &\equiv 3 \, \mathscr{A}^{3 \mu} = A_u^\mu + A_d^\mu + A_s^\mu \ ,
\end{align}
with the inverse relations being
\begin{align}
\label{qbar gamma^mu gamma^5 q}
A_{u \mu} &= \frac{1}{6} (2 A^0_\mu + 3 A^3_\mu + A^8_\mu) \ ,
&
A_{d \mu} &= \frac{1}{6} (2 A^0_\mu - 3 A^3_\mu + A^8_\mu) \ ,
&
A_{s \mu} &= \frac{1}{3} (A^0_\mu - A^8_\mu) \ .
\end{align}
From \Eq{axial current parametrization} we can write for the proton matrix element of these axial currents
\beq\label{axial form factors}
\matel{p'}{A^a_\mu}{p} = \bar{u}'_p \left( G_\text{A}^a(\fvec{q}^2) \gamma_\mu \gamma^5 + G_\text{P}^a(\fvec{q}^2) \frac{\fvec{q}_\mu}{2 \mN} \gamma^5 \right) u_p \ ,
\eeq
where $G_\text{A}^a$, $G_\text{P}^a$ are the \emph{axial} and \emph{induced pseudo-scalar form factor}, respectively. The proton's axial-vector charges $g_\text{A}^a$ are defined as the zero-momentum value of the nucleon axial form factors $G_\text{A}^a$, namely $g_\text{A}^a \equiv G_\text{A}^a(0)$. We thus have, at zero momentum transfer,
\beq
\matel{p(\bol{k}, r')}{A^a_\mu}{p(\bol{k}, r)} = g_\text{A}^a \, \bar{u}_p(\bol{k}, r') \gamma_\mu \gamma^5 u_p(\bol{k}, r) \ ,
\eeq
so that comparison with \Eq{Deltaq} yields
\begin{subequations}
\begin{align}
g_\text{A}^3 &= \Delta_u^{(p)} - \Delta_d^{(p)} \ ,
\\
g_\text{A}^8 &= \Delta_u^{(p)} + \Delta_d^{(p)} - 2 \Delta_s^{(p)} \ ,
\\
g_\text{A}^0 &= \Delta_u^{(p)} + \Delta_d^{(p)} + \Delta_s^{(p)} \ .
\end{align}
\end{subequations}
$g_\text{A}^3$ is often denoted simply $g_\text{A}$ in the literature.

The $G_\text{P}^{q, N}$ form factors in \Eq{axial current parametrization} can be determined as follows. In the chiral limit (\ie when the quark masses are set to zero), the QCD Lagrangian enjoys a large $U(3)_\text{L} \times U(3)_\text{R}$ chiral symmetry. Away from the chiral limit, as long as the light quark masses are kept equal, this symmetry gets explicitly broken down to a vector $U(3)$, whose conserved currents are those appearing in \Eq{vector currents}. Eight of the nine remaining currents, which can be combined into the axial currents defined in \Eq{axial currents}, are only conserved in the chiral limit (the axial $U(1)$ is anomalous and therefore the divergence of the relative current, $\mathscr{A}^0_\mu$, receives non-zero quantum corrections, see \Eq{axial current divergence}). Chiral symmetry being spontaneously broken by the QCD vacuum, there exist eight pseudo-Goldstone bosons, the light pseudo-scalar mesons among which are the $\pi$'s and the $\eta$. These mesons, which would be exactly massless in the chiral limit, are light with respect to the rest of the hadronic spectrum due to the smallness of the light-quark masses. The above discussion, related to the issue of \emph{partially conserved axial current} or \emph{PCAC}, implies that
\beq\label{partial A^a}
\matel{p'}{\partial^\mu A^a_\mu}{p} = \left( 2 \mN G_\text{A}^a(\fvec{q}^2) + G_\text{P}^a(\fvec{q}^2) \frac{\fvec{q}^2}{2 \mN} \right) \bar{u}'_p \, i \gamma^5 u_p
\eeq
should vanish in the chiral limit for $a = 3, 8$. Since we already know that $G_\text{A}^a(0) = g_\text{A}^a \neq 0$, the only possibility for \Eq{partial A^a} to vanish is that the two terms on the right-hand side cancel each other at zero momentum transfer, but this can only happen if $G_\text{P}^a$ contains a pole. This suggests that the second term on the right-hand side of \Eq{partial A^a} is generated by the exchange of a light pseudo-scalar meson with the same flavor structure of $A^a_\mu$: the pion (part of an isospin triplet) for $A^3_\mu$, in which case $G_\text{P}^3(\fvec{q}^2) \propto (\fvec{q}^2 - m_\pi^2)^{-1}$, or the (isosinglet) $\eta$ for $A^8_\mu$, meaning $G^8_\text{P}(\fvec{q}^2) \propto (\fvec{q}^2 - m_\eta^2)^{-1}$. The neutral mesons masses are~\cite{Zyla:2020zbs}
\begin{align}
m_\pi \approx 135~\MeV \ ,
&&&
m_\eta \approx 548~\MeV \ ,
\end{align}
although some authors (see \eg Ref.~\cite{Klos:2013rwa}) adopt for the neutral pion the average mass $(m_\pi + m_{\pi^+} + m_{\pi^-}) / 3 \approx 138~\MeV$, with $m_{\pi^\pm} \approx 140~\MeV$~\cite{Zyla:2020zbs}. The $\pi - \eta$ mixing can be neglected, see \eg Ref.~\cite{Gross:1979ur}. Chiral-symmetry breaking terms spoiling these relations, if present, have no meson poles and can therefore be neglected at low momentum transfer, being suppressed by the $\fvec{q}^\mu / \mN$ factor in \Eq{axial form factors}. The vanishing of \Eq{partial A^a} in the chiral limit then completely fixes the induced pseudo-scalar form factors to
\begin{align}
G_\text{P}^3(\fvec{q}^2) = - \frac{4 \mN^2}{\fvec{q}^2 - m_\pi^2} G_\text{A}^3(\fvec{q}^2) \ ,
&&&
G_\text{P}^8(\fvec{q}^2) = - \frac{4 \mN^2}{\fvec{q}^2 - m_\eta^2} G_\text{A}^8(\fvec{q}^2) \ ,
\end{align}
up to unimportant chiral-symmetry breaking contributions. No such a relation is available for $G^0_\text{P}$ due to the anomaly (there is in fact no (pseudo-)Goldstone boson relative to the axial $U(1)$); for this reason, we do not expect $G^0_\text{P}$ to feature a meson pole and therefore we can neglect it at low momentum transfer in \Eq{axial form factors} due to its suppressed $\fvec{q}^\mu / \mN$ coefficient. Therefore we have, for the proton matrix elements of the axial currents,
\begin{subequations}
\label{<p'|A^a|p>}
\begin{align}
\matel{p'}{A^3_\mu}{p} &\qeq g_\text{A}^3 \, \bar{u}'_p \left( \gamma_\mu \gamma^5 - \frac{2 \mN \fvec{q}_\mu}{\fvec{q}^2 - m_\pi^2} \gamma^5 \right) u_p \ ,
\\
\matel{p'}{A^8_\mu}{p} &\qeq g_\text{A}^8 \, \bar{u}'_p \left( \gamma_\mu \gamma^5 - \frac{2 \mN \fvec{q}_\mu}{\fvec{q}^2 - m_\eta^2} \gamma^5 \right) u_p \ ,
\\
\matel{p'}{A^0_\mu}{p} &\qeq g_\text{A}^0 \, \bar{u}'_p \gamma_\mu \gamma^5 u_p \ ,
\end{align}
\end{subequations}
where the $\qeq$ sign means that the above equalities are only valid at some finite order in an expansion in powers of $q / \mN$ (in this specific case we ignored $\Ord(q^2 / \mN^2)$ corrections). Notice that $\fvec{q}^2 \leqslant 0$ implies that $|q^0| \leqslant q$ and thus that $q^0 / \mN$ is at least of $\Ord(q / \mN)$. Also, $\bar{u}_N \, i \gamma^5 u_N / \mN \sim \Ord(q / \mN)$: as can be easily verified using the equivalent of \Eq{u spinor} for the nucleon, $\bar{u}_N \, i \gamma^5 u_N$ vanishes at zero momentum transfer, and from \Eq{Nfermionbilinears} one can see that it is of order $q$. The meson-pole terms in \Eq{<p'|A^a|p>}, which are of order $\frac{q^2}{q^2 + m_{\pi, \eta}^2} = \frac{q^2 / m_{\pi, \eta}^2}{q^2 / m_{\pi, \eta}^2 + 1}$, vary significantly over momentum transfer scales $q \sim m_{\pi, \eta}$, in the reach of direct detection experiments, therefore we avoid truncating them in the $q / \mN$ power series. In fact, the pole terms are $q$-suppressed for $q \ll m_{\pi, \eta}$ but stop being suppressed for $q \gtrsim m_{\pi, \eta}$, a feature that is not captured by the first few terms of the expansion alone. On the contrary, the $G_\text{A}^a$'s are expected to vary smoothly for values of momentum transfer up to the hadronic scale and therefore we can expand them in powers of $q^2 / \mN^2$ and safely truncate them, as we did at lowest order. The above results can be combined to obtain, using \Eq{qbar gamma^mu gamma^5 q},
\beq\label{<qbar gamma^mu gamma^5 q>}
\Nmatel{A_q^\mu} \qeq \Delta_q^{(N)} \, \bar{u}'_N \gamma^\mu \gamma^5 u_N + 2 \mN \, i \fvec{q}^\mu \left( \frac{a_{q, \pi}^N}{\fvec{q}^2 - m_\pi^2} + \frac{a_{q, \eta}^N}{\fvec{q}^2 - m_\eta^2} \right) \bar{u}'_N \, i \gamma^5 u_N \ ,
\eeq
with
\begin{align}
a_{u, \pi}^p &= - a_{d, \pi}^p = \frac{1}{2} g_\text{A}^3 \ ,
&
a_{s, \pi}^N &= 0 \ ,
&
a_{u, \eta}^N &= a_{d, \eta}^N = - \frac{1}{2} a_{s, \eta}^N = \frac{1}{6} g_\text{A}^8 \ ,
\end{align}
where isospin symmetry implies $a_{u, \pi}^p = a_{d, \pi}^n$, $a_{d, \pi}^p = a_{u, \pi}^n$. The form factors entering \Eq{axial current parametrization} are then given by
\begin{align}
G_\text{A}^{q, N}(\fvec{q}^2) \qeq \sum_{q = u, d, s} c_q \, \Delta_q^{(N)} \ ,
&&&
G_\text{P}^{q, N}(\fvec{q}^2) \qeq - 4 \mN^2 \sum_{q = u, d, s} c_q \left( \frac{a_{q, \pi}^N}{\fvec{q}^2 - m_\pi^2} + \frac{a_{q, \eta}^N}{\fvec{q}^2 - m_\eta^2} \right) .
\end{align}
As for the form factors relative to the vector currents, the $\fvec{q}^2$ dependence of the isovector axial form factor can be approximated by a dipole fit (see \eg Refs.~\cite{Anikin:2016teg, Hill:2017wgb, Hoferichter:2015ipa, Hoferichter:2016nvd} and references therein),
\begin{align}
\frac{G_\text{A}^3(\fvec{q}^2)}{G_\text{A}^3(0)} \approx G_\text{D}(\fvec{q}^2, M_\text{A})
&&
\text{with}
&&
M_\text{A}^2 \approx 1~\GeV^2 \ ,
\end{align}
$G_\text{D}$ being defined in \Eq{dipole FF}.

\subsection{Tensor couplings}
Inspection of the symmetry and transformation properties of the quark tensor currents,
\beq
\Op_q = \bar{q} \sigma^{\mu\nu} q \ ,
\eeq
tells us that their nucleon matrix element can be parametrized in terms of three form factors (see \eg Ref~\cite{Bishara:2017pfq}):
\beq
\Nmatel{\bar{q} \sigma^{\mu\nu} q} = \bar{u}'_N \left( F_{\text{T}, 0}^{q, N}(\fvec{q}^2) \, \sigma^{\mu\nu} + F_{\text{T}, 1}^{q, N}(\fvec{q}^2) \, \frac{i \, \fvec{q}^{[ \mu} \gamma^{\nu ]}}{2 \mN} + F_{\text{T}, 2}^{q, N}(\fvec{q}^2) \, \frac{i \, \fvec{q}^{[ \mu} \fvec{K}^{\nu ]}}{\mN^2} \right) u_N \ .
\eeq
Notice that $\varepsilon_{\mu \nu \rho \sigma} \fvec{K}^\rho \, \bar{u}'_N \gamma^\sigma \gamma^5 u_N$ can be cast in terms of the above via the equations of motion, see \eg Ref.~\cite{Lorce:2017isp}. Using \Eq{epsilon sigma} one can also find the relative expression for the nucleon matrix element of the axial-tensor current involving the $\bar{q} \, i \sigma^{\mu\nu} \gamma^5 q$ quark bilinears.

Here we will only be concerned with the zero momentum transfer limit, for which all we need is
\beq
F_{\text{T}, 0}^{q, N}(0) = \delta_q^{(N)} \ ,
\eeq
see \eg Ref.~\cite{He:1994gz}. The tensor charges $\delta_q^{(N)}$ indicate the difference between the spin of quarks and anti-quarks in the nucleon. The $2 + 1 + 1$ flavors FLAG averages of lattice results are~\cite{Aoki:2019cca, Gupta:2018lvp}
\begin{align}
\delta_u^{(p)} &\approx 0.784(28)(10) \ ,
&
\delta_d^{(p)} &\approx - 0.204(11)(10) \ ,
&
\delta_s^{(p)} &\approx - 0.027(16) \ .
\end{align}
In the limit of isospin symmetry one has
\begin{align}
\delta_u^{(p)} = \delta_d^{(n)} \ ,
&&
\delta_d^{(p)} = \delta_u^{(n)} \ ,
&&
\delta_s^{(p)} = \delta_s^{(n)} \ .
\end{align}
Some other values of the $\delta_q^{(N)}$'s can be found \eg in Refs.~\cite{Anselmino:2008jk, Bacchetta:2012ty, Anselmino:2013vqa, Belanger:2008sj, Belanger:2013oya}, as reported in Ref~\cite{DelNobile:2013sia}.

\section{DM-nucleon interaction}
\label{DM-N}
Now that we know how to compute hadronic matrix elements of (some) quark and gluon operators, we remain with turning the DM-nucleon scattering amplitude $\Mel_N$ into a DM-nucleus cross section. To do so, we start by performing a NR expansion in powers of the DM-nucleon relative speed $v$, which allows to identify contributions from different types of NR interactions: some involving the nucleon spin, some involving the DM spin, some involving $q$, etc. In this Section we first introduce the NR expansion for spin-$0$ and spin-$1/2$ DM (see \eg Refs.~\cite{Gondolo:2020wge, Gondolo:2021fqo} and references therein for DM with higher spins), then we show how to expand $\Mel_N$ and express the result in terms of $16$ Galilean-invariant building blocks. We will then see in \Sec{Form factors} how different NR interactions involve different nuclear properties and correspond to different nuclear responses and form factors.

\subsection{Non-relativistic expansion}
\label{Non-relativistic expansion}
The NR expansion of $\Mel_N$ allows to make contact with the nuclear physics involved in the scattering process. Not only can different relativistic interactions involve different nuclear properties, but it can also happen that a single relativistic interaction involves multiple aspects of the nucleus, which can respond in very different ways to the scattering process. These contributions can be told apart with a NR expansion. Let us take, for example, the SI and SD interactions, discussed in \Sec{SI interaction} and \Sec{SD interaction}. As we will see, even taking the DM-nucleon scattering cross section (see Eqs.~\eqref{SI sigma_N},~\eqref{SD sigma_N}) to have the same size in the two cases, the DM-nucleus cross section can be orders of magnitude larger for the SI interaction. For this interaction, in fact, $\Mel_N$ does not depend on the nucleon spin at leading order in the NR expansion, while it does for the SD interaction (hence its name). The DM-nucleus cross section then depends on the nuclear matter density for the SI interaction, and on the spin density of bound nucleons for the SD interaction. Thus in one case one has a potentially large $A^2$ enhancement with respect to the DM-nucleon cross section, while in the other the cross section is hardly enhanced and actually vanishes for spinless nuclei. A less standard, yet educative example is a DM particle with a magnetic dipole moment. As we will see in more detail at the end of this Section and in \Sec{Magnetic-dipole DM}, the DM dipole interacts through photon exchange with both the electric charge (charge-dipole interaction) and the magnetic dipole moment (dipole-dipole interaction) of the nucleon. A NR expansion of $\Mel_N$ allows to identify different contributions to the DM-nucleus cross section, all potentially relevant, among which both a SI-like and a SD-like terms, together with a coupling of the DM spin to the nuclear orbital angular momentum.

Instead of the NR expansion of the scattering amplitude, one could perform a NR expansion of the Lagrangian. Using the \emph{heavy-particle effective theory}, the DM particle mass can be in a sense integrated out of the theory without completely integrating out the DM field~\cite{Hill:2011be, Hill:2013hoa, Hill:2014yka, Berlin:2015njh, Bishara:2016hek, Chen:2018uqz}. Comparing Refs.~\cite{DelNobile:2013sia} and~\cite{Bishara:2016hek}, one can see the two methods yield the same leading-order result for a spin-$1/2$ DM particle, singlet under the SM gauge group.

Our approach consists in Taylor-expanding the DM-nucleon scattering amplitude in powers of $v$, here indicating the DM-nucleon relative speed. Truncating the expansion at leading order constitutes a good approximation provided no speed scales other than the DM-nucleon relative speed and the speed of light appear in the model of DM-nucleon interactions, which is often but not necessarily the case (see \eg Ref.~\cite{Bai:2009cd} for an exception). $q$ being smaller than $\mN$ in most cases (see \eg \Fig{fig: TypicalER} for elastic scattering), one may simultaneously perform a Taylor-Laurent expansion in powers of $q / \mN$, as done \eg in \Sec{Axial-vector couplings} (however, terms varying significantly over momentum-transfer scales smaller than $\mN$ are not truncated, see discussion after \Eq{<p'|A^a|p>}). The `Laurent' part of this expansion is due to the fact that the propagators of massless particles can cause the appearance of negative powers of $q^2$ (see \eg \Sec{Electromagnetic interactions}). Analogously to what done in \Sec{scattering kinematics}, we treat $q / \mu_\text{N}, v^\perp_N \sim \Ord(v)$ and $\delta / \mu_\text{N} \sim \Ord(v^2)$, where $\bol{v}^\perp_N$ is the DM-nucleon transverse velocity. Four-momenta are expanded at first order in the particle speed, \ie expanding the Lorentz factor as $\gamma \NReq 1$ (see discussion related to \Eq{NR gamma}), which leaves the NR Galilean symmetry intact. We then have for $\fvec{P} = \fvec{p} + \fvec{p}'$, $\fvec{K} = \fvec{k} + \fvec{k}'$, $\fvec{q} = \fvec{p} - \fvec{p}'$ (see \Sec{scattering kinematics}),
\begin{align}
\fvec{P}^\mu &\NReq
\begin{pmatrix}
2 \mDM
\\
\bol{P}
\end{pmatrix}
,
&
\fvec{K}^\mu &\NReq
\begin{pmatrix}
2 \mN
\\
\bol{K}
\end{pmatrix}
,
&
\fvec{q}^\mu &\NReq
\begin{pmatrix}
q^0
\\
\bol{q}
\end{pmatrix}
,
\end{align}
where at the lowest non-zero order (see \Eq{q.P neq 0, q.K = 0})
\beq\label{q^0}
q^0 \NReq \frac{\bol{K} \cdot \bol{q}}{2 \mN} \NReq \frac{\bol{P} \cdot \bol{q}}{2 \mDM} - \delta \ .
\eeq
$\Mel_N$ shall feature elements like $\bar{u}'_\chi \Gamma_\chi u_\chi$ and $\bar{u}'_N \Gamma_N u_N$, with $\Gamma_\chi$ and $\Gamma_N$ gamma-matrix structures possibly involving momenta. Their NR expression can be obtained by expanding the Dirac spinors in powers of momenta. For the DM particle we have, in chiral representation,
\beq\label{u spinor}
u_\chi(\bol{p}, s) =
\begin{pmatrix}
\sqrt{\fvec{p} \cdot \sigma} \, \xi^s
\\
\sqrt{\fvec{p} \cdot \bar{\sigma}} \, \xi^s
\end{pmatrix}
=
\frac{1}{\sqrt{2 (E_p + \mDM)}}
\begin{pmatrix}
(\fvec{p} \cdot \sigma + \mDM \unom_2) \, \xi^s
\\
(\fvec{p} \cdot \bar{\sigma} + \mDM \unom_2) \, \xi^s
\end{pmatrix}
,
\eeq
and expanding $\fvec{p}^\mu \NReq (\mDM, \bol{p})^\tr$ we get
\beq
u_\chi(\bol{p}, s) \NReq
\frac{1}{\sqrt{4 \mDM}}
\begin{pmatrix}
(2 \mDM \unom_2 - \bol{p} \cdot \bol{\sigma}) \, \xi^s
\\
(2 \mDM \unom_2 + \bol{p} \cdot \bol{\sigma}) \, \xi^s
\end{pmatrix}
.
\eeq
Analogous expressions hold for the nucleon. Here $\xi^s$ is a two-spinor and
\begin{align}
\sigma^\mu \equiv (\unom_2, \bol{\sigma})^\tr \ ,
&&&
\bar{\sigma}^\mu \equiv (\unom_2, - \bol{\sigma})^\tr \ ,
\end{align}
with the Pauli matrices
\begin{align}
\label{Pauli matrices}
\sigma^1 =
\begin{pmatrix}
0 & 1
\\
1 & 0
\end{pmatrix} ,
&&
\sigma^2 =
\begin{pmatrix}
0 & -i
\\
i & 0
\end{pmatrix} ,
&&
\sigma^3 =
\begin{pmatrix}
1 & 0
\\
0 & -1
\end{pmatrix} .
\end{align}
In the chiral representation the Dirac matrices read
\begin{align}
\label{Dirac matrices}
\gamma^\mu =
\begin{pmatrix}
0 & \sigma^\mu
\\
\bar{\sigma}^\mu & 0
\end{pmatrix} ,
&&
\gamma^5 =
\begin{pmatrix}
- \unom_2 & 0
\\
0 & \unom_2
\end{pmatrix} ,
&&
\sigma^{\mu \nu} =
\begin{pmatrix}
\sigma_2^{\mu \nu} & 0
\\
0 & \bar{\sigma}_2^{\mu \nu}
\end{pmatrix} ,
\end{align}
with
\begin{align}
\sigma_2^{\mu \nu} \equiv \frac{i}{2} (\sigma^\mu \bar{\sigma}^\nu - \sigma^\nu \bar{\sigma}^\mu) \ ,
&&&
\bar{\sigma}_2^{\mu \nu} \equiv \frac{i}{2} (\bar{\sigma}^\mu \sigma^\nu - \bar{\sigma}^\nu \sigma^\mu) \ .
\end{align}
Products of Pauli matrices can be simplified by exploiting their algebraic properties,
\begin{align}
\left\{ \sigma^i, \sigma^j \right\} &= 2 \delta^{i j} \unom_2 \ ,
&
\left[ \sigma^i, \sigma^j \right] &= 2 i \, \varepsilon^{i j k} \sigma^k \ ,
\end{align}
resulting in
\beq\label{sigmaisigmaj}
\sigma^i \sigma^j = \delta^{i j} \unom_2 + i \, \varepsilon^{i j k} \sigma^k \ .
\eeq
For a DM fermion bilinear we then have, up to and including the first order in momenta,
\begin{subequations}
\label{fermionbilinears}
\begin{align}
\bar{u}'_\chi u_\chi &\NReq 2 \mDM \CMcal{I}_\chi \ ,
\\
\bar{u}'_\chi i \gamma^5 u_\chi &\NReq 2 i \bol{q} \cdot \bol{s}_\chi \ ,
\\
\bar{u}'_\chi \gamma^\mu u_\chi &\NReq
\begin{pmatrix}
2 \mDM \CMcal{I}_\chi
\\
\bol{P} \CMcal{I}_\chi + 2 i \bol{q} \times \bol{s}_\chi
\end{pmatrix} ,
\\
\bar{u}'_\chi \gamma^\mu \gamma^5 u_\chi &\NReq
\begin{pmatrix}
2 \bol{P} \cdot \bol{s}_\chi
\\
4 \mDM \bol{s}_\chi
\end{pmatrix} ,
\\
\bar{u}'_\chi \sigma^{\mu \nu} u_\chi &\NReq
\begin{pmatrix}
0 & i \bol{q} \CMcal{I}_\chi - 2 \bol{P} \times \bol{s}_\chi
\\
- i \bol{q} \CMcal{I}_\chi + 2 \bol{P} \times \bol{s}_\chi
&
4 \mDM \, \varepsilon_{i j k} s_\chi^k
\end{pmatrix} ,
\\
\bar{u}'_\chi i \sigma^{\mu \nu} \gamma^5 u_\chi &\NReq
\begin{pmatrix}
0
&
- 4 \mDM \bol{s}_\chi
\\
4 \mDM \bol{s}_\chi
&
i \, \varepsilon_{i j k} q^k \CMcal{I}_\chi - 2 P^i s_\chi^j + 2 P^j s_\chi^i
\end{pmatrix} ,
\end{align}
\end{subequations}
where we defined
\begin{align}
\label{I s}
\CMcal{I}_\chi \equiv {\xi^{s'}}^\dagger \xi^s = \delta_{s s'} \ ,
&&&
\bol{s}_\chi \equiv {\xi^{s'}}^\dagger \frac{\bol{\sigma}}{2} \xi^s \ .
\end{align}
Notice that the DM mass splitting $\delta$ does not appear in the above expressions at the considered expansion order. Similarly, we have for the nucleon
\begin{subequations}
\label{Nfermionbilinears}
\begin{align}
\bar{u}'_N u_N &\NReq 2 \mN \CMcal{I}_N \ ,
\\
\bar{u}'_N i \gamma^5 u_N &\NReq - 2 i \bol{q} \cdot \bol{s}_N \ ,
\\
\bar{u}'_N \gamma^\mu u_N &\NReq
\begin{pmatrix}
2 \mN \CMcal{I}_N
\\
\bol{K} \CMcal{I}_N - 2 i \bol{q} \times \bol{s}_N
\end{pmatrix} ,
\\
\bar{u}'_N \gamma^\mu \gamma^5 u_N &\NReq
\begin{pmatrix}
2 \bol{K} \cdot \bol{s}_N
\\
4 \mN \bol{s}_N
\end{pmatrix} ,
\\
\bar{u}'_N \sigma^{\mu \nu} u_N &\NReq
\begin{pmatrix}
0 & - i \bol{q} \CMcal{I}_N - 2 \bol{K} \times \bol{s}_N
\\
i \bol{q} \CMcal{I}_N + 2 \bol{K} \times \bol{s}_N
&
4 \mN \, \varepsilon_{i j k} s_N^k
\end{pmatrix} ,
\\
\bar{u}'_N i \sigma^{\mu \nu} \gamma^5 u_N &\NReq
\begin{pmatrix}
0
&
- 4 \mN \bol{s}_N
\\
4 \mN \bol{s}_N
&
- i \, \varepsilon_{i j k} q^k \CMcal{I}_N - 2 K^i s_N^j + 2 K^j s_N^i
\end{pmatrix} ,
\end{align}
\end{subequations}
with $\CMcal{I}_N$, $\bol{s}_N$ defined analogously to $\CMcal{I}_\chi$, $\bol{s}_\chi$ in \Eq{I s}.

One can notice that, in the bilinears in Eqs.~\eqref{fermionbilinears},~\eqref{Nfermionbilinears}, the momentum transfer $\bol{q}$ is always accompanied by the imaginary unit. This can be explained as follows. If the matrix element of an operator $\Op$ features a component of $\bol{q}$ in one of its pieces, $\matel{\bol{p}', \bol{k}'}{\Op}{\bol{p}, \bol{k}} \sim q^i$, we will also have $\matel{\bol{p}, \bol{k}}{\Op}{\bol{p}', \bol{k}'} \sim - q^i$ since $\bol{q} = \bol{p} - \bol{p}'$. Notice that the bilinears we are considering come from hermitian operators, $\Op = \bar{\chi} \Gamma \chi$ or $\Op = \bar{N} \Gamma N$ with $\Gamma$ given by \Eq{Gammas}, for which
\beq\label{Hermitian O}
\matel{\bol{p}', \bol{k}'}{\Op}{\bol{p}, \bol{k}}^* = \matel{\bol{p}, \bol{k}}{\Op^\dagger}{\bol{p}', \bol{k}'} = \matel{\bol{p}, \bol{k}}{\Op}{\bol{p}', \bol{k}'} \ .
\eeq
For this relation to be compatible with the above, $\bol{q}$ must always appear together with the imaginary unit. The same can be said of $\delta$, which can \eg appear in the scattering amplitude through the $\bol{q} \cdot \bol{v}^\perp_N$ scalar product (see below).

Bilinears for antiparticles, which feature the four-spinor
\beq
v_\chi(\bol{p}, s) =
\begin{pmatrix}
+ \sqrt{\fvec{p} \cdot \sigma} \, \eta^s
\\
- \sqrt{\fvec{p} \cdot \bar{\sigma}} \, \eta^s
\end{pmatrix}
=
\frac{1}{\sqrt{2 (E_p + \mDM)}}
\begin{pmatrix}
+ (\fvec{p} \cdot \sigma + \mDM \unom_2) \, \eta^s
\\
- (\fvec{p} \cdot \bar{\sigma} + \mDM \unom_2) \, \eta^s
\end{pmatrix} ,
\eeq
can be derived from the above by means of the relations
\begin{align}
\label{vu}
v_\chi(\bol{p}, s) &= \Cmat \bar{u}_\chi(\bol{p}, s)^\tr \ ,
&
u_\chi(\bol{p}, s) &= \Cmat \bar{v}_\chi(\bol{p}, s)^\tr \ .
\end{align}
The charge-conjugation matrix $\Cmat$ reads, in chiral representation,
\begin{align}
\Cmat =
\begin{pmatrix}
- \epsilon & 0
\\
0 & \epsilon
\end{pmatrix} ,
&&
\text{with}
&&
\epsilon = i \sigma^2 =
\begin{pmatrix}
0 & 1
\\
-1 & 0
\end{pmatrix} .
\end{align}
\Eq{vu} means that the two-spinors $\xi^s$ and $\eta^s$ are related by
\beq\label{eta xi relation}
\eta^s = - \epsilon \, \xi^{s *} \ ,
\eeq
as can be seen by using
\begin{align}
\sigma_\mu^* = - \epsilon \, \bar{\sigma}_\mu \, \epsilon \ ,
&&&
\bar{\sigma}_\mu^* = - \epsilon \, \sigma_\mu \, \epsilon \ .
\end{align}
From all of the above follows
\beq\label{charconj}
\bar{v}_\chi(\bol{p}, s) \Gamma v'_\chi(\bol{p}', s') = - \eta_\Gamma^C \, \bar{u}'_\chi(\bol{p}', s') \Gamma u_\chi(\bol{p}, s) \ ,
\eeq
with the $\eta_\Gamma^C$ coefficients provided in \Tab{tab: C}. Notice that this result allows to express $\bar{v} \Gamma v$ in terms of $\xi^s$ and $\xi^{s'}$. To obtain an expression in terms of $\eta^s$ and $\eta^{s'}$ one can use \Eq{eta xi relation} to get
\begin{align}
{\xi^{s'}}^\dagger \xi^s = {\eta^s}^\dagger \eta^{s'} \ ,
&&&
{\xi^{s'}}^\dagger \frac{\bol{\sigma}}{2} \xi^s = - {\eta^s}^\dagger \frac{\bol{\sigma}}{2} \eta^{s'} \ .
\end{align}

\begin{table}[t]
\begin{center}
\begin{tabular}{c |>{\pnt} c c c c c c}
& $\unom_4$ & $i \gamma^5$ & $\gamma^\mu$ & $\gamma^\mu \gamma^5$ & $\sigma^{\mu\nu}$ & $i \sigma^{\mu\nu} \gamma^5$
\\
\hline
$\eta_\Gamma^C$ & $+1$ & $+1$ & $-1$ & $+1$ & $-1$ & $-1$
\end{tabular}
\end{center}
\caption[Charge-conjugation coefficients for spin-$1/2$ DM bilinears]{\label{tab: C}\emph{Charge-conjugation coefficients for spin-\figmath{1/2} DM bilinears \figmath{\bar{u}'_\chi \Gamma u_\chi} for different \figmath{\Gamma} matrices, see \Eq{charconj}.}}
\end{table}

\subsection{Non-relativistic operators}
\label{NR operators}
We are now ready to compute the NR expression of a DM-nucleon matrix element, given the interaction Lagrangian. A NR operator with matching matrix element can then be identified, which allows to compute the full DM-nucleus scattering amplitude. However, before rushing to the computation for some specific cases, which is carried out in \Sec{Examples}, we start by building an understanding of the general features of the expansion, and anticipating how the result can look like.

In general we can expect the matrix element to be a function of the dynamical quantities such as particles spin and momenta. Momentum conservation implies that only three out of the four (two initial and two final) momenta are independent, which we can take to be $\bol{P}$, $\bol{K}$ and $\bol{q}$. As a leftover from the Lorentz symmetry of the full theory, the NR limit preserves rotational and Galilean invariance, the latter being the NR version of relativistic boost invariance. To make rotational invariance explicit we need to build the NR amplitude taking scalar products of our vectors, while for Galilean invariance to be manifest we should use Galilean-invariant quantities, such as spins and velocity differences.

The momentum transfer $\bol{q}$ is itself invariant under NR boosts (this is only strictly true for elastic scattering, however the non boost-invariant correction is subleading for $|\delta| \ll \mDM$~\cite{Barello:2014uda}). With the remaining momenta we can build the Galilean-invariant relative velocity
\beq
\bol{v} \equiv \bol{v}_\DM - \bol{v}_N = \frac{\bol{p}}{\mDM} - \frac{\bol{k}}{\mN} \ .
\eeq
This variable however is still not good enough. As we saw already concerning the $\bol{q}$ dependence of fermion bilinears, as long as our interaction operators are hermitian we should expect \Eq{Hermitian O} to hold. If the matrix element of an operator $\Op$ features a component of $\bol{v}$ in one of its pieces, however, we have $\matel{\bol{p}', \bol{k}'}{\Op}{\bol{p}, \bol{k}} \sim v^i$ and also $\matel{\bol{p}, \bol{k}}{\Op}{\bol{p}', \bol{k}'} \sim v'^i$, with $\bol{v}'$ the final DM-nucleon relative velocity (which is, too, a Galilean invariant quantity). Now for these relations to be compatible with \Eq{Hermitian O}, $\bol{v}$ and $\bol{v}'$ may only appear together in the form
\beq
\bol{v} + \bol{v}' = \frac{\bol{P}}{\mDM} - \frac{\bol{K}}{\mN} \ ,
\eeq
as well as in the alternative combination $i (\bol{v} - \bol{v}')$ which is however proportional to $i \bol{q}$ up to subleading inelastic contributions. We can then use $\bol{v} + \bol{v}'$ or, alternatively,
\beq\label{vperp_N}
\bol{v}^\perp_N \NReq \frac{1}{2} (\bol{v}_\DM + \bol{v}'_\DM - \bol{v}_N - \bol{v}'_N) \ ,
\eeq
where the equality can be proven by closely matching the discussion on $\bol{v}^\perp_T$ in \Sec{scattering kinematics}. $\bol{v}^\perp_N$ is the component of $\bol{v}$ orthogonal to $\bol{q}$ for elastic scattering, thus $\bol{v}^\perp_N \cdot \bol{q} = 0$ for $\delta = 0$. Following Ref.~\cite{DelNobile:2018dfg}, we employ here $\bol{v}^\perp_N$ also for inelastic scattering, instead of the nucleon equivalent of \Eq{vperpinel}, to make direct contact with the formalism of Refs.~\cite{Fitzpatrick:2012ix, Anand:2013yka} and the nuclear form factors provided therein. For inelastic scattering, using Eqs.~\eqref{vperp},~\eqref{vperpinel} we then have $\bol{v}^\perp_N \cdot \bol{q} \NReq \delta$.

Finally, among the ingredients that build up our DM-nucleon matrix element, we also have to take into account the spin vectors. We will focus on the case of a spin-$1/2$ DM, which encompasses that of a spin-$0$ DM for what concerns the present discussion (for spin-$0$ DM, one just needs to set $\bol{s}_\chi = \bol{0}$ in the following). From the above discussion, our DM-nucleon matrix element must be a rotationally-invariant function of $\bol{s}_\chi$, $\bol{s}_N$, $i \bol{q}$, and $\bol{v}^\perp_N$. Notice that, owing to \Eq{sigmaisigmaj}, each DM (nucleon) fermion bilinear in \Eq{fermionbilinears} (\Eq{Nfermionbilinears}) can be expressed as a sum of two terms, one independent of the DM (nucleon) spin vector and the other linear in $\bol{s}_\chi$ ($\bol{s}_N$). Therefore, when contracting a DM and a nucleon bilinear in the amplitude, the only possible spin structures are
\begin{align}
\CMcal{I}_\chi \CMcal{I}_N \ ,
&&
\CMcal{I}_\chi s_N^i \ ,
&&
\CMcal{I}_N s_\chi^i \ ,
&&
s_\chi^i s_N^j \ .
\end{align}
Any matrix element will therefore be a linear combination of these, each contracted in all possible ways with factors of $i \bol{q}$ and $\bol{v}^\perp_N$, and multiplied by a function of $q^2$ and ${v^\perp_N}^2$ (and of the non-dynamical constants $\mN$, $\mDM$, $\bol{v}^\perp_N \cdot \bol{q} \NReq \delta$). Notice that there is only a finite number of ways this contraction can be performed to produce new independent scalars: in fact, one can only contract through the $\delta_{ij}$ and $\varepsilon_{ijk}$ $SU(2)$-invariant tensors (\ie using scalar and vector products), however any product or contraction of two epsilon tensors returns a sum of products of Kronecker deltas, as in
\begin{align}
\label{contractions3}
\varepsilon_{ijk} \varepsilon_{ilm} = \delta_{jl} \delta_{km} - \delta_{jm} \delta_{kl} \ ,
& & &
\varepsilon_{ijk} \varepsilon_{ijl} = 2 \delta_{kl} \ ,
& & &
\varepsilon_{ijk} \varepsilon_{ijk} = 6 \ ,
\end{align}
and therefore only combinations featuring a single vector product can be independent. As a matter of fact, it was found in Ref.~\cite{Dobrescu:2006au} that one can only construct $16$ linearly-independent, rotationally-invariant combinations of our vectors, though each combination can be multiplied by an arbitrary function of $q^2$ and ${v^\perp_N}^2$. Our classification of possible combinations employing Galilean invariance, following Refs.~\cite{Fitzpatrick:2012ix, DelNobile:2018dfg}, generalizes that of Refs.~\cite{Dobrescu:2006au, Fan:2010gt} which is restricted to the CM frame. The scattering amplitude is then, in general, a linear combination of these rotationally-invariant terms with coefficients that depend on $q^2$ and ${v^\perp_N}^2$ (see Ref.~\cite{DelNobile:2018dfg} for a more in-depth discussion).

Each term can be unambiguously matched to a NR DM-$p$ or DM-$n$ interaction operator constructed as described in Ref.~\cite{Fitzpatrick:2012ix} (see \Sec{Scattering amplitude}), whose matrix element between DM-nucleus states yields the full DM-nucleus scattering amplitude. For a contained and effective discussion, such a NR operator can be basically obtained by promoting $\bol{q}$, $\bol{v}^\perp_N$, $\bol{s}_\chi$, $\bol{s}_N$ to operators, so that an amplitude term given \eg by $4 i \mN \, \bol{s}_\chi \cdot \bol{q}$ corresponds to the NR operator $4 i \mN \, \bol{s}_\chi \cdot \bol{q}$ (confusingly enough, we use the same symbol for the c-number quantities and their operator counterparts). From the above discussion, all possible operators can be described as a linear combination of $16$ independent basic operators (denoted \emph{building blocks} in Refs.~\cite{Fitzpatrick:2012ix, DelNobile:2018dfg}), each multiplied by $q^2$- and ${v^\perp_N}^2$-dependent operators. A possible criterion to sort out these building blocks can be to first write all rotationally-invariant combinations of $i \bol{q}$, $\bol{v}^\perp_N$, $\bol{s}_\chi$, $\bol{s}_N$ containing neither of the two spins, then those linear in one spin or the other, and finally write all terms linear in both spins. The latter category can be further divided into operators containing zero, one, two, and three instances of $\bol{v}^\perp_N$ and/or $\bol{q}$, using both scalar and vector products. These include all linearly-independent combinations one can build. Vector equalities such as those reported in Appendix A of Ref.~\cite{Dobrescu:2006au} can then be used to express some of these combinations in terms of the others. A set of building blocks then is, following the numbering introduced in Refs.~\cite{Fitzpatrick:2012ix, Anand:2013yka, DelNobile:2018dfg},
\beq\label{NR building blocks}
\begin{aligned}
\Op^N_1 &\equiv \unop \ ,
\hspace{35mm}
&
\\
\Op^N_3 &\equiv i \, \bol{s}_N \cdot (\bol{q} \times \bol{v}^\perp_N) \ ,
&
\Op^N_4 &\equiv \bol{s}_\chi \cdot \bol{s}_N \ ,
\\
\Op^N_5 &\equiv i \, \bol{s}_\chi \cdot (\bol{q} \times \bol{v}^\perp_N) \ ,
&
\Op^N_6 &\equiv (\bol{s}_\chi \cdot \bol{q}) (\bol{s}_N \cdot \bol{q}) \ ,
\\
\Op^N_7 &\equiv \bol{s}_N \cdot \bol{v}^\perp_N \ ,
&
\Op^N_8 &\equiv \bol{s}_\chi \cdot \bol{v}^\perp_N \ ,
\\
\Op^N_9 &\equiv i \, \bol{s}_\chi \cdot (\bol{s}_N \times \bol{q}) \ ,
&
\Op^N_{10} &\equiv i \, \bol{s}_N \cdot \bol{q} \ ,
\\
\Op^N_{11} &\equiv i \, \bol{s}_\chi \cdot \bol{q} \ ,
&
\Op^N_{12} &\equiv \bol{v}^\perp_N \cdot (\bol{s}_\chi \times \bol{s}_N) \ ,
\\
\Op^N_{13} &\equiv i (\bol{s}_\chi \cdot \bol{v}^\perp_N) (\bol{s}_N \cdot \bol{q}) \ ,
&
\Op^N_{14} &\equiv i (\bol{s}_\chi \cdot \bol{q}) (\bol{s}_N \cdot \bol{v}^\perp_N) \ ,
\\
\Op^N_{15} &\equiv (\bol{s}_\chi \cdot \bol{q}) [\bol{s}_N \cdot (\bol{q} \times \bol{v}^\perp_N)] \ ,
&
\Op^N_{16} &\equiv (\bol{s}_\chi \cdot \bol{v}^\perp_N) (\bol{s}_N \cdot \bol{v}^\perp_N) \ ,
\\
\Op^N_{17} &\equiv i [\bol{s}_\chi \cdot (\bol{q} \times \bol{v}^\perp_N)] (\bol{s}_N \cdot \bol{v}^\perp_N) \ .
&
\end{aligned}
\eeq
Notice that, as shown \eg in Ref.~\cite{DelNobile:2018dfg}, the two operators obtained by exchanging $\bol{s}_\chi$ and $\bol{s}_N$ in $\Op^N_{15}$ and $\Op^N_{17}$ can be expressed in terms of the above building blocks; and that $\Op^N_{16}$ and $\Op^N_{17}$ are usually neglected in the literature, some other operators (not independent of the other building blocks) being sometimes spuriously introduced in their place. The most general nucleon-specific interaction operator can then be written as
\beq\label{general NR Op}
\Op_\NR^N = \sum_i f_i^N(q^2, {v^\perp_N}^2) \, \Op^N_i \ ,
\eeq
with the $f_i^N$'s in principle arbitrary functions of $q^2$, ${v^\perp_N}^2$, and of the non-dynamical constants. Notice that $\Op^N_2 \equiv {v^\perp_N}^2 \Op^N_1$, first introduced in Ref.~\cite{Fitzpatrick:2012ix}, is not an actual building block since, as said above, we store all dependence on ${v^\perp_N}^2$ in the $f_i^N$'s (see Ref.~\cite{DelNobile:2018dfg} for a dedicated discussion on this operator). For a scalar DM $\phi$ we only have a subset of the above building blocks, namely those that do not depend on $\bol{s}_\chi$:
\begin{align*}
\Op^N_1, \Op^N_3, \Op^N_7, \Op^N_{10} &&& \text{spin-$0$ DM}.
\end{align*}
As we will see in \Sec{Examples}, the building blocks from $\Op^N_1$ to $\Op^N_{12}$ are enough to describe the NR limit of many of the quantum field theory operators often encountered in the literature. In fact, it was shown in Ref.~\cite{DelNobile:2018dfg} that the building blocks from $\Op^N_{13}$ to $\Op^N_{17}$, with the addition of $\Op^N_3$, do not arise below dimension $7$ in a DM-nucleon effective field theory (EFT) for a spin-$0$ or spin-$1/2$ DM singlet under the SM.

Each of the $\Op^N_i$'s contributes to a certain DM-nucleus coupling, realizing a specific interaction. As we will see in \Sec{sigma}, for instance, scattering through $\Op^N_1$ results in coupling the DM particle to the nucleon number density inside the nucleus, thus realizing the SI interaction, see \Sec{SI interaction}. $\Op^N_4$ instead couples the DM spin with the (nucleon-spin contribution to the) nuclear spin, realizing the SD interaction, see \Sec{SD interaction}. $\Op^N_1$ and $\Op^N_4$ are the only building blocks that are not NR suppressed by powers of $\bol{q}$ or $\bol{v}^\perp_N$, and are therefore the dominant terms in \Eq{general NR Op} unless their coefficients are themselves suppressed by small coupling constants or powers of $q^2$ and/or ${v^\perp_N}^2$, and/or other terms are enhanced by negative powers of $q^2$, as \eg in the example discussed in \Sec{Magnetic-dipole DM}. For this reason, $\Op^N_1$ and $\Op^N_4$ are the most important interactions which is why they alone have been named \emph{spin-independent} and \emph{spin-dependent} interactions, although clearly other interactions also exist that are independent or dependent on the nucleon or nuclear spin. $\Op^N_6$, for instance, which contributes to the SD interaction (see \Sec{SD interaction}), involves the nuclear spin, though only through the contribution of the nucleon spin-component longitudinal to $\bol{q}$ (see \Sec{Multipoles}). Factors of $\bol{v}^\perp_N$, too, can realize a coupling to the nuclear spin though through the orbital angular momentum of nucleons, while also contributing a coupling to the overall DM-nucleus relative motion through $\bol{v}^\perp_T$~\cite{Fitzpatrick:2012ix} (see \Sec{Form factors}).

If our interaction operator features more than one building block, we may get interference terms among some of them when we square the amplitude to obtain the cross section. To understand which of the $\Op^N_i$'s can interfere, it is instructive to study how they transform under parity and time-reversal transformations. Our fundamental ingredients transform as
\begin{align}
\label{P:qvs}
P:&
&
i \bol{q} &\to - i \bol{q},
&
\bol{v}^\perp_N &\to - \bol{v}^\perp_N,
&
\bol{s}_{\chi, N} &\to + \bol{s}_{\chi, N},
\\
T:&
&
i \bol{q} &\to + i \bol{q},
&
\bol{v}^\perp_N &\to - \bol{v}^\perp_N,
&
\bol{s}_{\chi, N} &\to - \bol{s}_{\chi, N},
\end{align}
(both $P$ and $T$ change the sign of $\bol{q}$ and $\bol{v}^\perp_N$, but $T$ being anti-linear also changes the sign of the imaginary unit). This implies the following transformation properties for the NR operators:
\begin{align*}
& \Op^N_1, \Op^N_3, \Op^N_4, \Op^N_5, \Op^N_6, \Op^N_{16} & \text{$P$-even and $T$-even},
\\
& \Op^N_7, \Op^N_8, \Op^N_9, \Op^N_{17} & \text{$P$-odd and $T$-even},
\\
& \Op^N_{13}, \Op^N_{14} & \text{$P$-even and $T$-odd},
\\
& \Op^N_{10}, \Op^N_{11}, \Op^N_{12}, \Op^N_{15} & \text{$P$-odd and $T$-odd}.
\end{align*}
Of course only operators with the same quantum numbers are able to interfere, provided the nuclear ground state is an eigenstate of $P$ and $T$ (as it is usually the case to a good approximation). Notice that the SI and SD interaction operators, $\Op^N_1$ and $\Op^N_4$ (and also $\Op^N_6$, which contributes to the SD interaction, see \Sec{SD interaction}), have the same quantum numbers and are therefore allowed to interfere, in principle.

Moreover, if DM particles are unpolarized, the interference term between an operator depending on $\bol{s}_\chi$ and one not depending on it vanishes. We can show this with an explicit example for the case of a spin-$1/2$ DM particle (analogous formulas can be derived for higher-spin DM with the help of \Eq{Tr[J_i J_j]} below). We will use the following equalities, which are useful when squaring the matrix element and summing over initial and final spins:
\begin{subequations}
\label{spinsum}
\begin{align}
\sum_s {\xi^s}^\dagger \xi^s &= 2 \ ,
&
\sum_s \xi^s {\xi^s}^\dagger &= \unom_2 \ ,
\\
\sum_s {\xi^s}^\dagger \sigma^i \xi^s &= \Tr \sigma^i = 0 \ ,
&
\sum_s {\xi^s}^\dagger \sigma^i \sigma^j \xi^s &= 2 \delta_{ij} \ ,
\end{align}
\end{subequations}
where for the last equality we used \Eq{sigmaisigmaj}. Now, if both operators either depend or do not depend on $\bol{s}_\chi$, the interference term in the averaged matrix element squared is proportional to, respectively,
\begin{align}
\label{spinsum 2}
\sum_{s, s'} {s_\chi^i}^* s_\chi^j &= \sum_{s, s'} {\xi^s}^\dagger \frac{\sigma^i}{2} \xi^{s'} {\xi^{s'}}^\dagger \frac{\sigma^j}{2} \xi^s = \frac{1}{2} \delta_{i j} \ ,
&
\sum_{s, s'} \CMcal{I}_\chi^* \CMcal{I}_\chi &= 2 \ ,
\end{align}
where we used the definition of $\CMcal{I}_\chi$ and $\bol{s}_\chi$ in \Eq{I s}. These quantities are non-zero, as we would expect. On the contrary, the interference term between an operator depending on $\bol{s}_\chi$ and one not depending on it is proportional to
\beq\label{spinsum 3}
\sum_{s, s'} \bol{s}_\chi^* \CMcal{I}_\chi = \sum_s {\xi^s}^\dagger \frac{\bol{\sigma}}{2} \xi^s = \bol{0} \ ,
\eeq
where again we used \Eq{spinsum}. Hence, the interference term vanishes. This would not happen for polarized DM particles, in which case the spin sum should include the density matrix describing the DM polarization. If, for instance, the DM particles were all polarized along the positive direction of the spin quantization axis, $s = + \frac{1}{2}$, the interference term would be proportional to
\beq
\sum_{s, s'} \delta_{s, + \frac{1}{2}} \bol{s}_\chi^* \CMcal{I}_\chi = {\xi^{+ \frac{1}{2}}}^\dagger \frac{\bol{\sigma}}{2} \xi^{+ \frac{1}{2}} = (0, 0, \tfrac{1}{2})^\tr \ ,
\eeq
which is non-zero. These considerations do not apply to the nucleus, for which $\bol{s}_N$-depending operators can interfere with $\bol{s}_N$-independent operators, even for unpolarized nuclei. This is due to the relevant nuclear quantity, \ie the nuclear spin, depending on both the nucleon spins and the nucleon orbital angular momenta, to which $\bol{s}_N$ and $\bol{v}^\perp_N$ respectively contribute. Therefore an operator featuring $\bol{s}_N$, as $\Op^N_9$, can interfere with a $\bol{s}_N$-independent operator featuring $\bol{v}^\perp_N$, as $\Op^N_8$, since both couple the DM to the nuclear spin. Notice moreover that, for unpolarized DM, \Eq{spinsum 2} implies that two operators where $\bol{s}_\chi$ couples to mutually orthogonal vectors do not interfere: taking for example $\Op^N_5$ and $\Op^N_6$, the interference term depends on the scalar product of the matrix elements of $\bol{q} \times \bol{v}^\perp_N$ with $\bol{q} (\bol{s}_N \cdot \bol{q})$, which clearly vanishes. All in all, the families of potentially interfering operators in the standard scenario of unpolarized DM are
\begin{align*}
& (\Op^N_1, \Op^N_3), (\Op^N_4, \Op^N_5), (\Op^N_4, \Op^N_6), (\Op^N_4, \Op^N_{16}), [\Op^N_6, \Op^N_{16}] & \text{$P$-even and $T$-even},
\\
& (\Op^N_8, \Op^N_9), (\Op^N_9, \Op^N_{17}) & \text{$P$-odd and $T$-even},
\\
& (\Op^N_{11}, \Op^N_{12}, \Op^N_{15}) & \text{$P$-odd and $T$-odd}.
\end{align*}
$(\Op^N_1, \Op^N_3)$ is the only family of interfering $\bol{s}_\chi$-independent operators. $[\Op^N_6, \Op^N_{16}]$, indicated with square brackets, can only interfere for inelastic scattering (we recall that $\bol{v}^\perp_N \cdot \bol{q} \NReq \delta$, see above). In the unpolarized cross section, the interference term between the SI and SD interactions vanishes.

\subsection{Examples}
\label{Examples}
Enough with the general considerations, we now take on some concrete quantum field theory interaction operators commonly considered in the direct detection literature and derive their NR decomposition in terms of the $\Op^N_i$'s. As summarized previously, this can be done by computing the DM-nucleon scattering amplitude and taking its NR limit using the techniques presented above. This can then be easily matched to a NR operator, which can be expressed in terms of the building blocks~\eqref{NR building blocks}. To ease this chain of computations, we report in \Tab{tab: DM-N bilinears} the matrix-element structures we will encounter in this subsection, together with the NR operators they match to and their $P$ and $T$ parity. While this list captures the matrix elements most commonly found in the literature for spin-$0$ and spin-$1/2$ DM, it by no means covers the whole range of matrix-element structures one may encounter in a generic theory; a comprehensive list of structures, together with the matching NR operators, can be found in Ref.~\cite{DelNobile:2018dfg}.

\begin{table}[t]
\begin{center}
\begin{tabular}{| c | c |>{\pnt} c | c c |}
\hline
& DM-$N$ bilinears & NR operator & $P$ & $T$
\\
\hline
\multirow{4}{*}{\rotatebox{90}{\hspace{-4mm}spin-$0$ DM}}
&
$\bar{u}_N u_N$
&
$2 \mN \, \Op^N_1$
&
$+$
&
$+$
\\
&
$\bar{u}'_N \, i \gamma^5 u_N$
&
$- 2 \, \Op^N_{10}$
&
$-$
&
$-$
\\
&
$\fvec{P}_\mu (\bar{u}_N \gamma^\mu u_N)$
&
$4 \mDM \mN \, \Op^N_1$
&
$+$
&
$+$
\\
&
$\fvec{P}_\mu (\bar{u}_N \gamma^\mu \gamma^5 u_N)$
&
$- 8 \mDM \mN \, \Op^N_7$
&
$-$
&
$+$
\\
\hline
\multirow{11}{*}{\rotatebox{90}{\hspace{-8mm}spin-$1/2$ DM}}
&
$(\bar{u}_\chi u_\chi) (\bar{u}_N u_N)$
&
$4 \mDM \mN \, \Op^N_1$
&
$+$
&
$+$
\\
&
$(\bar{u}_\chi \, i \gamma^5 u_\chi) (\bar{u}_N u_N)$
&
$4 \mN \, \Op^N_{11}$
&
$-$
&
$-$
\\
&
$(\bar{u}_\chi u_\chi) (\bar{u}_N \, i \gamma^5 u_N)$
&
$- 4 \mDM \, \Op^N_{10}$
&
$-$
&
$-$
\\
&
$(\bar{u}_\chi \, i \gamma^5 u_\chi) (\bar{u}_N \, i \gamma^5 u_N)$
&
$4 \, \Op^N_6$
&
$+$
&
$+$
\\
&
$(\bar{u}_\chi \gamma^\mu u_\chi) (\bar{u}_N \gamma_\mu u_N)$
&
$4 \mDM \mN \, \Op^N_1$
&
$+$
&
$+$
\\
&
$(\bar{u}_\chi \gamma^\mu \gamma^5 u_\chi) (\bar{u}_N \gamma_\mu u_N)$
&
$8 \mDM (\mN \, \Op^N_8 - \Op^N_9)$
&
$-$
&
$+$
\\
&
$(\bar{u}_\chi \gamma^\mu u_\chi) (\bar{u}_N \gamma_\mu \gamma^5 u_N)$
&
$- 8 \mN (\mDM \, \Op^N_7 + \Op^N_9)$
&
$-$
&
$+$
\\
&
$(\bar{u}_\chi \gamma^\mu \gamma^5 u_\chi) (\bar{u}_N \gamma_\mu \gamma^5 u_N)$
&
$- 16 \mDM \mN \, \Op^N_4$
&
$+$
&
$+$
\\
&
$(\bar{u}_\chi \gamma^\mu \gamma^5 u_\chi) (\bar{u}_N \sigma_{\mu\nu} u_N) i \fvec{q}^\nu$
&
$- 16 \mDM \mN \, \Op^N_9$
&
$-$
&
$+$
\\
&
$(\bar{u}_\chi \sigma^{\mu\nu} u_\chi) (\bar{u}_N \sigma_{\mu\nu} u_N)$
&
$32 \mDM \mN \, \Op^N_4$
&
$+$
&
$+$
\\
&
$(\bar{u}_\chi \, i \sigma^{\mu\nu} \gamma^5 u_\chi) (\bar{u}_N \sigma_{\mu\nu} u_N)$
&
$8 (\mN \, \Op^N_{10} - \mDM \, \Op^N_{11} - 4 \mDM \mN \, \Op^N_{12})$
&
$-$
&
$-$
\\
\hline
\end{tabular}
\end{center}
\caption[DM-nucleon matrix-element structures and corresponding NR operators]{\label{tab: DM-N bilinears}\emph{A collection of DM-nucleon matrix-element structures and the NR operators they map to (see \Eq{NR building blocks}), together with their \figmath{P} and \figmath{T} parity, for spin-\figmath{0} and spin-\figmath{1/2} DM.}}
\end{table}

As concrete examples we will first consider some effective DM-quark and DM-gluon operators for spin-$0$ and spin-$1/2$ DM, and then move to the realm of DM-photon interactions. Few other examples not covered here can be found \eg in Ref.~\cite{DelNobile:2018dfg}.

\subsubsection{DM-quark and DM-gluon effective operators}
\label{EFT operators}
We list in \Tab{tab: spin-0 DM EFT} a number of dimension-$5$ and -$6$ effective interaction operators of a scalar DM field $\phi$ with quarks and gluons. The $G_q^N$ form factor, defined in \Eq{G_q^N}, reads (see \Eq{t = - q^2})
\beq\label{G_q^N NR}
G_q^N(- q^2) = \Delta_q^{(N)} - q^2 \left( \frac{a_{q, \pi}^N}{q^2 + m_\pi^2} + \frac{a_{q, \eta}^N}{q^2 + m_\eta^2} \right) .
\eeq
The matrix element of derivative operators can be easily computed by means of relations such as
\begin{align}
\label{d phi matrix element}
i \matel{\phi'}{\phi^\dagger \overleftrightarrow{\partial^\mu} \phi}{\phi} &= \fvec{P}^\mu \ ,
&
\matel{\phi'}{\partial^\mu (\phi^\dagger \phi)}{\phi} &= - i \fvec{q}^\mu \ ,
\end{align}
from which the $\partial_\mu (\phi^\dagger \phi) \, \bar{q} \gamma^\mu q$ operator is seen to vanish upon application of the equations of motion, see \eg \Eq{EOMs}.

\begin{table}[t]
\begin{center}
\begin{tabular}{| c |>{\pnt} c | c c |}
\hline
EFT operator & NR operator & $P$ & $T$
\\
\hline
$\phi^\dagger \phi \, \bar{q} q$
&
$\displaystyle{2 \frac{\mN^2}{m_q} f_{Tq}^{(N)} \, \Op^N_1}$ \rule{0mm}{7mm}
&
$+$
&
$+$
\\
$\phi^\dagger \phi \, \bar{q} \, i \gamma^5 q$
&
$\displaystyle{- 2 \frac{\mN}{m_q} \left( G_q^N(- q^2) - \bar{m} \sum_{q' = u, d, s} \frac{G_{q'}^N(- q^2)}{m_{q'}} \right) \Op^N_{10}}$
&
$-$
&
$-$
\\
$i (\phi^\dagger \overleftrightarrow{\partial_\mu} \phi) \, \bar{q} \gamma^\mu q$
&
$4 \mDM \mN F_1^{q, N}(0) \, \Op^N_1$
&
$+$
&
$+$
\\
$i (\phi^\dagger \overleftrightarrow{\partial_\mu} \phi) \, \bar{q} \gamma^\mu \gamma^5 q$
&
$- 8 \mDM \mN \Delta_q^{(N)} \, \Op^N_7$
&
$-$
&
$+$
\\
$\partial_\mu (\phi^\dagger \phi) \, \bar{q} \gamma^\mu \gamma^5 q$
&
$4 \mN G_q^N(- q^2) \, \Op^N_{10}$
&
$-$
&
$-$
\\
$\displaystyle{\frac{\aS}{12 \pi} \, \phi^\dagger \phi \, G^{a \mu\nu} G^a_{\mu\nu}}$
&
$\displaystyle{- \frac{4}{27} \mN^2 f_{TG}^{(N)} \, \Op^N_1}$
&
$+$
&
$+$
\\
$\displaystyle{\frac{\aS}{8 \pi} \, \phi^\dagger \phi \, G^{a \mu\nu} \tilde{G}^a_{\mu\nu}}$
&
$\displaystyle{2 \mN \bar{m} \left( \sum_{q = u, d, s} \frac{G_q^N(- q^2)}{m_q} \right) \Op^N_{10}}$ \rule[-7mm]{0mm}{15mm}
&
$-$
&
$-$
\\
\hline
\end{tabular}
\end{center}
\caption[DM-quark/gluon operators and corresponding NR operators for spin-$0$ DM]{\label{tab: spin-0 DM EFT}\emph{A collection of DM-quark and DM-gluon effective operators and the NR operators they map to (see \Eq{NR building blocks}), together with their \figmath{P} and \figmath{T} parity, for a scalar DM field \figmath{\phi}. For a real scalar, a factor of \figmath{\frac{1}{2}} should be applied to the EFT operators in the first column, keeping also in mind that not all operators exist in this case \eg due to \Eq{Real scalar}. Information on the hadron-physics coefficients and form factors is provided in \Sec{qg to N}, with \figmath{G_q^N(- q^2)} given in \Eq{G_q^N NR}.}}
\end{table}

For a self-conjugated DM field, \ie a real $\phi = \phi^\dagger$, a factor of $\frac{1}{2}$ should be applied to all EFT operators in the first column of \Tab{tab: spin-0 DM EFT}. Not all operators exist in this case, \eg due to
\beq\label{Real scalar}
\phi \overleftrightarrow{\partial_\mu} \phi = 0 \ .
\eeq

In \Tab{tab: spin-1/2 DM EFT} we list a number of dimension-$6$ effective interaction operators of a spin-$1/2$ DM field $\chi$ with quarks, and with gluons at dimension $7$. Owing to \Eq{epsilon sigma}, we have
\begin{align}
\bar{\chi} \, \sigma^{\mu\nu} \chi \, \bar{q} \, i \sigma_{\mu\nu} \gamma^5 q &= + \bar{\chi} \, i \sigma^{\mu\nu} \gamma^5 \chi \, \bar{q} \, \sigma_{\mu\nu} q \ ,
\\
\bar{\chi} \, i \sigma^{\mu\nu} \gamma^5 \chi \, \bar{q} \, i \sigma_{\mu\nu} \gamma^5 q &= - \bar{\chi} \sigma^{\mu\nu} \chi \, \bar{q} \, \sigma_{\mu\nu} q \ .
\end{align}
The matrix element of derivative operators (which we do not consider explicitly here) can be easily computed by means of relations such as
\begin{align}
\label{d chi matrix element}
i \matel{\chi'}{\bar{\chi} \overleftrightarrow{\partial_\mu} \Gamma \chi}{\phi} &= \fvec{P}_\mu \, \bar{u}_\chi \Gamma u_\chi \ ,
&
\matel{\chi'}{\partial^\mu (\bar{\chi} \Gamma \chi)}{\chi} &= - i \fvec{q}^\mu \, \bar{u}_\chi \Gamma u_\chi \ .
\end{align}

\begin{table}[t!]
\begin{center}
\begin{tabular}{| c |>{\pnt} c | c c |}
\hline
EFT operator & NR operator & $P$ & $T$
\\
\hline
$\bar{\chi} \chi \, \bar{q} q$
&
$\displaystyle{4 \frac{\mDM \mN^2}{m_q} f_{Tq}^{(N)} \, \Op^N_1}$ \rule{0mm}{7mm}
&
$+$
&
$+$
\\
$\bar{\chi} \, i \gamma^5 \chi \, \bar{q} q$
&
$\displaystyle{4 \frac{\mN^2}{m_q} f_{Tq}^{(N)} \, \Op^N_{11}}$ \rule{0mm}{7mm}
&
$-$
&
$-$
\\
$\bar{\chi} \chi \, \bar{q} \, i \gamma^5 q$
&
$\displaystyle{- 4 \frac{\mDM \mN}{m_q} \left( G_q^N(- q^2) - \bar{m} \sum_{q' = u, d, s} \frac{G_{q'}^N(- q^2)}{m_{q'}} \right) \Op^N_{10}}$
&
$-$
&
$-$
\\
$\bar{\chi} \, i \gamma^5 \chi \, \bar{q} \, i \gamma^5 q$
&
$\displaystyle{4 \frac{\mN}{m_q} \left( G_q^N(- q^2) - \bar{m} \sum_{q' = u, d, s} \frac{G_{q'}^N(- q^2)}{m_{q'}} \right) \Op^N_6}$
&
$+$
&
$+$
\\
$\bar{\chi} \gamma^\mu \chi \, \bar{q} \gamma_\mu q$
&
$4 \mDM \mN F_1^{q, N}(0) \, \Op^N_1$
&
$+$
&
$+$
\\
$\bar{\chi} \gamma^\mu \gamma^5 \chi \, \bar{q} \gamma_\mu q$
&
$8 \mDM \mN F_1^{q, N}(0) \, \Op^N_8 - 8 \mDM (F_1^{q, N}(0) + F_2^{q, N}(0)) \Op^N_9$
&
$-$
&
$+$
\\
$\bar{\chi} \gamma^\mu \chi \, \bar{q} \gamma_\mu \gamma^5 q$
&
$- 8 \mN \Delta_q^{(N)} (\mDM \, \Op^N_7 + \Op^N_9)$
&
$-$
&
$+$
\\
$\bar{\chi} \gamma^\mu \gamma^5 \chi \, \bar{q} \gamma_\mu \gamma^5 q$
&
$\displaystyle{- 16 \mDM \mN \left[ \Delta_q^{(N)} \, \Op^N_4 + \left( \frac{a_{q, \pi}^N}{\fvec{q}^2 - m_\pi^2} + \frac{a_{q, \eta}^N}{\fvec{q}^2 - m_\eta^2} \right) \Op^N_6 \right]}$
&
$+$
&
$+$
\\
$\bar{\chi} \, \sigma^{\mu\nu} \chi \, \bar{q} \, \sigma_{\mu\nu} q$
&
$32 \mDM \mN \delta_q^{(N)} \, \Op^N_4$
&
$+$
&
$+$
\\
$\bar{\chi} \, i \sigma^{\mu\nu} \gamma^5 \chi \, \bar{q} \, \sigma_{\mu\nu} q$
&
$8 \delta_q^{(N)} (\mN \, \Op^N_{10} - \mDM \, \Op^N_{11} - 4 \mDM \mN \, \Op^N_{12})$
&
$-$
&
$-$
\\
$\displaystyle{\frac{\aS}{12 \pi} \, \bar{\chi} \chi \, G^{a \mu\nu} G^a_{\mu\nu}}$
&
$\displaystyle{- \frac{8}{27} \mDM \mN^2 f_{TG}^{(N)} \, \Op^N_1}$ \rule{0mm}{7mm}
&
$+$
&
$+$
\\
$\displaystyle{\frac{\aS}{12 \pi} \, \bar{\chi} \, i \gamma^5 \chi \, G^{a \mu\nu} G^a_{\mu\nu}}$
&
$\displaystyle{- \frac{8}{27} \mN^2 f_{TG}^{(N)} \, \Op^N_{11}}$ \rule{0mm}{7mm}
&
$-$
&
$-$
\\
$\displaystyle{\frac{\aS}{8 \pi} \, \bar{\chi} \chi \, G^{a \mu\nu} \tilde{G}^a_{\mu\nu}}$
&
$\displaystyle{4 \mDM \mN \bar{m} \left( \sum_{q = u, d, s} \frac{G_q^N(- q^2)}{m_q} \right) \Op^N_{10}}$ \rule{0mm}{9mm}
&
$-$
&
$-$
\\
$\displaystyle{\frac{\aS}{8 \pi} \, \bar{\chi} \, i \gamma^5 \chi \, G^{a \mu\nu} \tilde{G}^a_{\mu\nu}}$
&
$\displaystyle{- 4 \mN \bar{m} \left( \sum_{q = u, d, s} \frac{G_q^N(- q^2)}{m_q} \right) \Op^N_6}$ \rule[-7mm]{0mm}{15mm}
&
$+$
&
$+$
\\
\hline
\end{tabular}
\end{center}
\caption[DM-quark/gluon operators and corresponding NR operators for spin-$1/2$ DM]{\label{tab: spin-1/2 DM EFT}\emph{A collection of DM-quark and DM-gluon effective operators and the NR operators they map to (see \Eq{NR building blocks}), together with their \figmath{P} and \figmath{T} parity, for a spin-\figmath{1/2} DM field \figmath{\chi}. For a Majorana fermion, a factor of \figmath{\frac{1}{2}} should be applied to the EFT operators in the first column, keeping also in mind that not all operators exist in this case \eg due to \Eq{Majorana}. Information on the hadron-physics coefficients and form factors is provided in \Sec{qg to N}, with \figmath{G_q^N(- q^2)} given in \Eq{G_q^N NR}.}}
\end{table}

For a self-conjugated DM field, \ie a Majorana $\chi = \chi^\text{c}$, a factor of $1/2$ should be applied to all EFT operators in the first column of \Tab{tab: spin-1/2 DM EFT}. Not all operators exist in this case, \eg due to
\beq\label{Majorana}
\bar{\chi} \gamma^\mu \chi = \bar{\chi} \, \sigma^{\mu\nu} \chi = \bar{\chi} \, i \sigma^{\mu\nu} \gamma^5 \chi = 0 \ ,
\eeq
thus only the bilinears $\bar{\chi} \chi$, $\bar{\chi} \, i \gamma^5 \chi$ and $\bar{\chi} \gamma^\mu \gamma^5 \chi$ are non-zero among those considered here.

If the DM does not couple to the $u$ and $d$ quarks, the leading NR DM-nucleon operator corresponding to the $i (\phi^\dagger \overleftrightarrow{\partial_\mu} \phi) \, \bar{q} \gamma^\mu q$ and $\bar{\chi} \gamma^\mu \chi \, \bar{q} \gamma_\mu q$ operators vanishes (see \Tab{tab: spin-0 DM EFT} and \Tab{tab: spin-1/2 DM EFT}, respectively, and \Eq{F_1^{q, p}}). One should then consider higher-order corrections, which include a Darwin-Foldy and a spin-orbit term (see discussion below \Eq{<J_EM> NR}) and can be computed from Eqs.~\eqref{fermionbilinears},~\eqref{Nfermionbilinears}, see \eg Refs.~\cite{Hoferichter:2015ipa, Hoferichter:2016nvd}. If instead the leading NR operator vanishes only for one nucleon species (either protons or neutrons), these corrections to the DM-nucleus scattering are outweighed by the contribution of the other species and can therefore be neglected. In any case, one should be aware of what contributions may become important when a leading-order result, whether the outcome of a NR truncation or other approximations, is suppressed or vanishes (see \eg discussions related to QCD and $2$-body corrections in Secs.~\ref{SI interaction},~\ref{SD interaction}).

\subsubsection{Electromagnetic interactions}
\label{Electromagnetic interactions}
The interaction operators considered above describe DM couplings to quarks and gluons through contact interactions: the massive mediators have been integrated out and the interaction region is point-like. A different set of interesting interactions can be determined by studying the possible electromagnetic properties of the DM (see \eg Refs.~\cite{Bagnasco:1993st, Pospelov:2000bq, Sigurdson:2004zp, Barger:2010gv, Chang:2010en, DelNobile:2011je, Fornengo:2011sz, DelNobile:2012tx, Ho:2012bg, Fitzpatrick:2012ib, Ibarra:2015fqa, DelNobile:2015bqo} and references therein). Of course the DM being `dark' means that it cannot have sizeable couplings with photons, but nothing prevents it to have some amount of interaction (provided it is weak enough to have escaped detection so far). The simplest option may be to endow the DM particles with a tiny electric charge, a model frequently called \emph{millicharged DM} regardless of the fact that DM particles with electric charge above $10^{-3} e$ are experimentally excluded for a wide range of DM masses by present-day constraints, and that current direct detection experiments have the sensitivity to constrain charges down to $10^{-9} e$ (see \eg Ref.~\cite{DelNobile:2015bqo}). Such a tiny electric charge may derive from a theory of dark Electromagnetism, where the dark photon mixes with the SM photon so that the DM particle acquires a minimal coupling to the standard photon proportional to the small mixing parameter~\cite{Holdom:1985ag} (see also \eg Appendix B of Ref.~\cite{Feldman:2007wj}); or it may just so happen that the DM has such a peculiar charge assignment (see \eg Ref.~\cite{DelNobile:2015bqo} for a discussion). Another way for DM to interact with photons, even if electrically neutral, is through a charge radius, indicating an extended charge distribution. Other possibilities for DM particles with non-zero spin are to have an anomalous magnetic dipole moment, an electric dipole moment or an anapole moment. Such electromagnetic properties may stem for instance from the DM particle being a bound state of electrically charged particles, as the neutron, or from the DM particle coupling with heavy charged states which then generate the DM-photon coupling via loop processes, as it happens for neutrinos.

From a phenomenological standpoint these interactions are interesting in that they produce DM-nucleus scattering cross sections (and therefore recoil-energy spectra) that are quite different from those of models with heavy mediators. In these processes the DM exchanges photons with the nucleus, so that the interaction can be long range as opposed to the contact interactions in \Eq{L quark}. The scattering amplitude will then feature propagators of a massless particle (the photon), which go like $1 / q^2$ and thus enhance the low-energy part of the spectrum, corresponding to an enhancement of the scattering in the forward direction. So, while contact interactions feature matrix elements with non-negative powers of $q$, as we saw above, photon-mediated processes can yield negative powers of the momentum transfer thus producing a different recoil spectrum, as we will see more in detail in \Sec{Pheno}. Finally, some DM candidates with electromagnetic properties have a scattering cross section featuring a non-trivial interplay between a SI-like and a SD-like interaction, as we will see in the following.

Once the DM is coupled to the photon, it interacts with nucleons (and thus with nuclei) through the nucleon-photon coupling in \Eq{JA}, whereas the matrix element of the electromagnetic nucleon current is given in \Eq{<J_EM>} as a function of the \emph{Dirac form factor} $F_1^N$ and the \emph{Pauli form factor} $F_2^N$. As seen already in \Sec{Vector couplings}, $F_1^N$ is normalized to the nucleon electric charge $Q_N$ in units of $e$, see \Eq{Q_N}, while $F_2^N$ is normalized to the anomalous magnetic moment in units of the nuclear magneton $\hat{\mu}_\text{N}$~\eqref{nuclear magneton}:
\begin{align}
F_1^N(0) = Q_N \ ,
&&&
F_2^N(0) = \kappa_N \ .
\end{align}
Notice that $\kappa_N \hat{\mu}_\text{N}$ is not necessarily the whole nucleon magnetic moment. Using the Gordon identity (already quoted in \Eq{EOMs}),
\beq
\bar{u}'_N \gamma^\mu u_N = \bar{u}'_N \left( \frac{\fvec{K}^\mu}{2 \mN} + \frac{i \sigma^{\mu\nu} \fvec{q}_\nu}{2 \mN} \right) u_N \ ,
\eeq
\Eq{<J_EM>} can be written as
\beq
\Nmatel{J_\text{EM}^\mu} = \bar{u}'_N \left( F_1^N(\fvec{q}^2) \frac{\fvec{K}^\mu}{2 \mN} + \left( F_1^N(\fvec{q}^2) + F_2^N(\fvec{q}^2) \right) \frac{i \sigma^{\mu\nu} \fvec{q}_\nu}{2 \mN} \right) u_N \ .
\eeq
From this, the full magnetic dipole moment of the nucleon can be read off being $\lambda_N \equiv (Q_N + \kappa_N) \hat{\mu}_\text{N}$, yielding~\cite{Zyla:2020zbs}
\begin{align}
\lambda_p \approx + 2.79 \, \hat{\mu}_\text{N} \ ,
&&&
\lambda_n \approx - 1.91 \, \hat{\mu}_\text{N} \ ,
\end{align}
\ie
\begin{align}
\kappa_p \approx + 1.79 \ ,
&&&
\kappa_n \approx - 1.91 \ .
\end{align}
In general, the gyromagnetic ratio of a particle is defined as the ratio of its magnetic dipole moment $\bol{\mu}$ and spin $\bol{s}$. The Land\'e $g$-factor is the gyromagnetic ratio in units of $e / 2 \genp{m}$ with $\genp{m}$ the particle mass, \ie $\bol{\mu} \equiv g \frac{e}{2 \genp{m}} \bol{s}$. For the nucleon we then have $\lambda_N = g_N \hat{\mu}_\text{N} s_N$ with $s_N = 1/2$, implying
\beq\label{g_N}
g_N = 2 (Q_N + \kappa_N) \ ,
\eeq
thus
\begin{align}
\label{g_N values}
g_p \approx + 5.59 \ ,
&&&
g_n \approx - 3.83 \ .
\end{align}
The first part, $2 Q_N$, is the $g$-factor obtained at tree-level by using the Dirac equation. Therefore, the electric charge contributes to the particle magnetic moment. The rest, also known as `$g - 2$' (at least for unit-charged particles such as the electron, the muon and the proton), gives rise to the anomalous magnetic moment $\kappa_N \hat{\mu}_\text{N}$, which is due to loop corrections and to the composite nature of nucleons. The NR form of \Eq{<J_EM>} can be computed using \Eq{Nfermionbilinears},
\beq\label{<J_EM> NR}
\Nmatel{J_\text{EM}^\mu} \NReq
\begin{pmatrix}
2 \mN \CMcal{I}_N Q_N
\\
\bol{K} \CMcal{I}_N Q_N + i g_N \bol{s}_N \times \bol{q}
\end{pmatrix}
.
\eeq
Notice that this entails $\matel{n'}{J_\text{EM}^0}{n} \NReq 0$ for neutrons: we neglect higher-order corrections (the Darwin-Foldy and spin-orbit terms, see \eg Refs.~\cite{DeForest:1966ycn, Bertozzi:1972jff, Donnelly:1975ze, Donnelly:1984rg} and Refs.~\cite{Hoferichter:2015ipa, Hoferichter:2016nvd}) as these are outweighed by the leading DM-proton interactions.

For concreteness, we consider here the DM-photon couplings listed in \Tab{tab: EM interactions}. $Q_\DM$ is the DM electric charge in units of $e$, $\langle r_\DM^2 \rangle_\text{E}$ is the mean-square DM charge radius, whereas $\mu_\chi$, $d_\chi$, and $a_\chi$ are the magnetic dipole, electric dipole, and anapole moments of a spin-$1/2$ DM field $\chi$, respectively. Among these interactions, the millicharge operator is the only renormalizable as it has mass dimension $4$; the magnetic and electric-dipole operators have dimension $5$, while the charge-radius and the anapole-moment operators have dimension $6$. The couplings $\mu_\chi$, $d_\chi$, have dimensions of an inverse mass and are usually expressed in units of $e~\cm$, while $a_\chi$ and $\langle r_\DM^2 \rangle_\text{E}$ have dimensions of a squared inverse mass. Of the considered operators, only the anapole does not vanish for a self-conjugated DM field, for which \Eq{Majorana} holds (for Majorana DM, the operator should appear in the effective Lagrangian multiplied by a factor of $1/2$ to account for the field being self-conjugated). The spatial and time-reversal parities of the operators, displayed in the last two columns of \Tab{tab: EM interactions}, can be derived from \Eq{P T}, \Tab{tab: PT} and from the fact that
\begin{align}
P A^\mu(\fvec{x}) P^{-1} &= + \mathscr{P}^\mu_{\phantom{\mu} \nu} A^\nu(\mathscr{P} \fvec{x}) \ ,
&
T A^\mu(\fvec{x}) T^{-1} &= - \mathscr{T}^\mu_{\phantom{\mu} \nu} A^\nu(\mathscr{T} \fvec{x}) \ .
\end{align}
Other possible DM-photon interaction operators, such as $\phi^\dagger \phi \, F^{\mu\nu} F_{\mu\nu}$ and $\phi^\dagger \phi \, F^{\mu\nu} \tilde{F}_{\mu\nu}$ ($\bar{\chi} \chi \, F^{\mu\nu} F_{\mu\nu}$ and $\bar{\chi} \chi \, F^{\mu\nu} \tilde{F}_{\mu\nu}$) for spin-$0$ (spin-$1/2$) DM, arising at mass dimension $6$ ($7$) in an EFT, were studied \eg in Refs.~\cite{Weiner:2012cb, Frandsen:2012db, Crivellin:2014gpa, Ovanesyan:2014fha}.

\begin{table}[t]
\begin{center}
\begin{tabular}{| c |>{\pnt} c | c | c c |}
\hline
& Field operator & NR operator & $P$ & $T$
\\
\hline
\raisebox{-3.5mm}{\multirow{2}{*}{Millicharge}}
&
$- Q_\DM e \, i (\phi^\dagger \overleftrightarrow{\partial^\mu} \phi) A_\mu$ \rule[-5mm]{0mm}{12mm}
&
\raisebox{-3.5mm}{\multirow{2}{*}{$\displaystyle{- 4 \frac{\mDM \mN}{q^2} e^2 Q_\DM Q_N \, \Op^N_1}$}}
&
\raisebox{-3.5mm}{\multirow{2}{*}{$+$}}
&
\raisebox{-3.5mm}{\multirow{2}{*}{$+$}}
\\
\cline{2-2}
&
$- Q_\DM e \, \bar{\chi} \gamma^\mu \chi \, A_\mu$ \rule[-5mm]{0mm}{12mm}
&
&
&
\\
\hline
\raisebox{-3.5mm}{\multirow{2}{*}{Charge radius}}
&
$\displaystyle{- \frac{1}{6} \langle r_\DM^2 \rangle_\text{E} e \, i (\phi^\dagger \overleftrightarrow{\partial^\mu} \phi) \partial^\nu F_{\mu\nu}}$ \rule[-5mm]{0mm}{12mm}
&
\raisebox{-3.5mm}{\multirow{2}{*}{$\displaystyle{\frac{4}{6} \mDM \mN e^2 \langle r_\DM^2 \rangle_\text{E} Q_N \, \Op^N_1}$}} \rule[-5mm]{0mm}{12mm}
&
\raisebox{-3.5mm}{\multirow{2}{*}{$+$}}
&
\raisebox{-3.5mm}{\multirow{2}{*}{$+$}}
\\
\cline{2-2}
&
$\displaystyle{- \frac{1}{6} \langle r_\DM^2 \rangle_\text{E} e \, \bar{\chi} \gamma^\mu \chi \, \partial^\nu F_{\mu\nu}}$ \rule[-5mm]{0mm}{12mm}
&
&
&
\\
\hline
\makecell{Magnetic \\ dipole moment}
&
$\displaystyle{- \frac{\mu_\chi}{2} \, \bar{\chi} \, \sigma^{\mu\nu} \chi \, F_{\mu\nu}}$
&
\makecell{$\displaystyle{2 e \mu_\chi \left[ \mN Q_N \, \Op^N_1 + 4 \frac{\mDM \mN}{q^2} Q_N \, \Op^N_5 \right.}$ \\ $\displaystyle{\left. + 2 \mDM g_N \left( \Op^N_4 - \frac{\Op^N_6}{q^2} \right) \right]}$} \rule[-11mm]{0mm}{24mm}
&
$+$
&
$+$
\\
\hline
\makecell{Electric \\ dipole moment}
&
$\displaystyle{- \frac{d_\chi}{2} \, \bar{\chi} \, i \sigma^{\mu\nu} \gamma^5 \chi \, F_{\mu\nu}}$
&
$\displaystyle{8 \frac{\mDM \mN}{q^2} e d_\chi Q_N \, \Op^N_{11}}$ \rule[-5mm]{0mm}{12mm}
&
$-$
&
$-$
\\
\hline
Anapole moment
&
$a_\chi \, \bar{\chi} \gamma^\mu \gamma^5 \chi \, \partial^\nu F_{\mu\nu}$
&
$4 \mDM e a_\chi \left[ g_N \, \Op^N_9 - 2 \mN Q_N \, \Op^N_8 \right]$ \rule[-3mm]{0mm}{9mm}
&
$-$
&
$+$
\\
\hline
\end{tabular}
\end{center}
\caption[DM-photon operators and corresponding NR operators]{\label{tab: EM interactions}\emph{A collection of DM-photon interaction operators and the NR operators they map to, together with their \figmath{P} and \figmath{T} parity, for a spin-\figmath{0} DM field \figmath{\phi} and a spin-\figmath{1/2} DM field \figmath{\chi}. Only the anapole operator does not vanish for a self-conjugated DM field.}}
\end{table}

The NR electromagnetic potentials induced by the interactions in \Tab{tab: EM interactions} can be derived by computing the transition matrix $T$ for DM scattering off an external field $A^\mu$. The $T$ matrix is related to the scattering matrix $S$ by
\beq
S = \braket{\text{f}}{\text{i}} - i \, 2 \pi \, \delta(E_\text{f} - E_\text{i}) \, T \ ,
\eeq
where $\ket{\text{i}}$, $\ket{\text{f}}$ are the initial and final state, respectively, and $E_\text{i}$, $E_\text{f}$ their energies. In the following we will consider $\ket{\text{i}} = \ket{\bol{p}}$, $\ket{\text{f}} = \ket{\bol{p}'}$ mutually different one-particle DM states. The first-order $S$ matrix expansion (Born approximation),
\beq
S \simeq i \int \ud^4 \fvec{x} \, \matel{\bol{p}'}{\Lag(\fvec{x})}{\bol{p}} = i \, 2 \pi \, \delta(E_\text{f} - E_\text{i}) \int \ud^3 x \, \matel{\bol{p}'}{\Lag(\bol{x})}{\bol{p}} \ ,
\eeq
then yields
\beq\label{T & Lag}
T \simeq - \int \ud^3 x \, \matel{\bol{p}'}{\Lag(\bol{x})}{\bol{p}} = \int \ud^3 x \, \matel{\bol{p}'}{\mathscr{V}(\bol{x})}{\bol{p}} = \sqrt{\rho(p') \rho(p)} \int \ud^3 x \, V(\bol{x}) \, e^{i \bol{q} \cdot \bol{x}} \ ,
\eeq
with $\mathscr{V} = - \Lag$ the interaction Hamiltonian. In the last equality we inserted twice the identity operator in \Eq{unit op}, used \Eq{NR free wave function}, and denoted (for a local potential $\mathscr{V}$)
\beq
\matel{\bol{y}'}{\mathscr{V}(\bol{x})}{\bol{y}} = V(\bol{x}) \, \delta^{(3)}(\bol{x} - \bol{y}') \, \delta^{(3)}(\bol{x} - \bol{y}) \ .
\eeq
Finally we obtain, inverting \Eq{T & Lag},
\beq
V(\bol{x}) \simeq \int \frac{\ud^3 q}{(2 \pi)^3} \, \frac{T}{\sqrt{\rho(p') \rho(p)}} \, e^{- i \bol{q} \cdot \bol{x}} \ ,
\eeq
that is, $V$ is the Fourier transform of $T$, divided by the square root of the initial and final-state normalization factors appearing in \Eq{statenorm}. Now we notice that the interactions listed in \Tab{tab: EM interactions} can be cast as a sum of terms of the form $J^\mu A_\mu$ by performing a number of integration by parts. For each term we then have, using the first identity in \Eq{T & Lag},
\beq
T \simeq - \int \ud^3 x \, \matel{\bol{p}'}{J^\mu(\bol{x})}{\bol{p}} A_\mu(\bol{x}) = - \matel{\bol{p}'}{J^\mu(\bol{0})}{\bol{p}} \tilde{A}_\mu(\bol{q}) \ ,
\eeq
with
\beq
\tilde{A}^\mu(\bol{q}) \equiv \int \ud^3 x \, e^{i \bol{q} \cdot \bol{x}} A^\mu(\bol{x}) \ ,
\eeq
which leads to
\beq
V(\bol{x}) \simeq - \int \frac{\ud^3 q}{(2 \pi)^3} \, \frac{\matel{\bol{p}'}{J^\mu(\bol{0})}{\bol{p}}}{\sqrt{\rho(p') \rho(p)}} \tilde{A}_\mu(\bol{q}) \, e^{- i \bol{q} \cdot \bol{x}} \ .
\eeq
Neglecting the terms whose contribution to the DM-nucleon scattering amplitude vanishes by virtue of the conservation of the nucleon electric current (which amounts to discarding the $\partial^\mu A_\mu$ terms), the `currents' for the millicharge (indicated below by the label C for Coulomb), charge radius (CR), magnetic dipole (MDM), electric dipole (EDM), and anapole moment (AM) interactions are
\begin{subequations}
\label{EM currents}
\begin{align}
J_\text{C}^\mu &\equiv - Q_\DM e \, i (\phi^\dagger \overleftrightarrow{\partial^\mu} \phi) \ ,
\\
J_\text{CR}^\mu &\equiv + \frac{1}{6} \langle r_\DM^2 \rangle_\text{E} e \, \square \, i (\phi^\dagger \overleftrightarrow{\partial^\mu} \phi) \ ,
\\
\intertext{for spin-$0$ DM, and}
J_\text{C}^\mu &\equiv - Q_\DM e \, \bar{\chi} \gamma^\mu \chi \ ,
\\
J_\text{CR}^\mu &\equiv + \frac{1}{6} \langle r_\DM^2 \rangle_\text{E} e \, \square (\bar{\chi} \gamma^\mu \chi) \ ,
\\
J_\text{MDM}^\mu &\equiv - \mu_\chi \, \partial_\nu (\bar{\chi} \, \sigma^{\mu\nu} \chi) \ ,
\\
J_\text{EDM}^\mu &\equiv - d_\chi \, \partial_\nu (\bar{\chi} \, i \sigma^{\mu\nu} \gamma^5 \chi) \ ,
\\
J_\text{AM}^\mu &\equiv - a_\chi \, \square (\bar{\chi} \gamma^\mu \gamma^5 \chi) \ ,
\end{align}
\end{subequations}
for spin-$1/2$ DM.
Using Eqs.~\eqref{d phi matrix element},~\eqref{d chi matrix element},~\eqref{fermionbilinears} one can compute their matrix elements as well as their leading-order contribution in the NR expansion:
\begin{subequations}
\label{<EM currents>}
\begin{align}
\matel{\phi'}{J_\text{C}^\mu}{\phi} &= - Q_\DM e \, \fvec{P}^\mu \NReq - Q_\DM e
\begin{pmatrix}
2 \mDM
\\
\bol{0}
\end{pmatrix}
,
\\
\matel{\phi'}{J_\text{CR}^\mu}{\phi} &= - \frac{1}{6} \langle r_\DM^2 \rangle_\text{E} e \, \fvec{q}^2 \fvec{P}^\mu \NReq \frac{1}{6} \langle r_\DM^2 \rangle_\text{E} e \, q^2
\begin{pmatrix}
2 \mDM
\\
\bol{0}
\end{pmatrix}
,
\\
\intertext{for spin-$0$ DM, and}
\matel{\chi'}{J_\text{C}^\mu}{\chi} &= - Q_\DM e \, \bar{u}'_\chi \gamma^\mu u_\chi \NReq - Q_\DM e
\begin{pmatrix}
2 \mDM \CMcal{I}_\chi
\\
\bol{0}
\end{pmatrix}
,
\\
\matel{\chi'}{J_\text{CR}^\mu}{\chi} &= - \frac{1}{6} \langle r_\DM^2 \rangle_\text{E} e \, \fvec{q}^2 \, \bar{u}'_\chi \gamma^\mu u_\chi \NReq \frac{1}{6} \langle r_\DM^2 \rangle_\text{E} e \, q^2
\begin{pmatrix}
2 \mDM \CMcal{I}_\chi
\\
\bol{0}
\end{pmatrix}
,
\\
\matel{\chi'}{J_\text{MDM}^\mu}{\chi} &= + \mu_\chi \, \bar{u}'_\chi \sigma^{\mu\nu} u_\chi \, i \fvec{q}_\nu \NReq \mu_\chi
\begin{pmatrix}
q^2 \CMcal{I}_\chi + 2 i \bol{s}_\chi \cdot (\bol{q} \times \bol{P})
\\
4 i \mDM \bol{s}_\chi \times \bol{q}
\end{pmatrix}
,
\\
\matel{\chi'}{J_\text{EDM}^\mu}{\chi} &= + d_\chi \, \bar{u}'_\chi \, i \sigma^{\mu\nu} \gamma^5 u_\chi \, i \fvec{q}_\nu \NReq d_\chi
\begin{pmatrix}
4 i \mDM \bol{s}_\chi \cdot \bol{q}
\\
\bol{0}
\end{pmatrix}
,
\\
\matel{\chi'}{J_\text{AM}^\mu}{\chi} &= + a_\chi \, \fvec{q}^2 \, \bar{u}'_\chi \gamma^\mu \gamma^5 u_\chi \NReq - a_\chi \, q^2
\begin{pmatrix}
2 \bol{P} \cdot \bol{s}_\chi
\\
4 \mDM \bol{s}_\chi
\end{pmatrix}
,
\end{align}
\end{subequations}
for spin-$1/2$ DM. Some next-to-leading terms in the NR expansion have been kept as they contribute to the DM-nucleon scattering amplitude at leading order (see below). Finally, employing \Eq{rhohere} and the solutions to Maxwell's equations,
\begin{align}
\bol{E} = - \dot{\bol{A}} - \bol{\nabla} A^0 \ ,
&&&
\bol{B} = \bol{\nabla} \times \bol{A} \ ,
\end{align}
we obtain the NR static potentials
\begin{align}
V_\text{C} = Q_\DM e \, A^0 \ ,
&&
V_\text{CR} = - \frac{1}{6} \langle r_\DM^2 \rangle_\text{E} e \, \bol{\nabla} \cdot \bol{E} \ ,
\end{align}
for spin-$0$ DM, and
\begin{align}
V_\text{C} = Q_\DM e \CMcal{I}_\chi \, A^0 \ ,
&&
V_\text{CR} = - \frac{1}{6} \langle r_\DM^2 \rangle_\text{E} e \CMcal{I}_\chi \, \bol{\nabla} \cdot \bol{E} \ ,
\end{align}
\begin{align}
V_\text{MDM} = - \mu_\chi \, \frac{\bol{s}_\chi}{s_\chi} \cdot \bol{B} \ ,
&&
V_\text{EDM} = - d_\chi \, \frac{\bol{s}_\chi}{s_\chi} \cdot \bol{E} \ ,
&&
V_\text{AM} = - a_\chi \, \frac{\bol{s}_\chi}{s_\chi} \cdot \bol{\nabla} \times \bol{B} \ ,
\end{align}
for spin-$1/2$ DM, with
\beq
s_\chi = \frac{1}{2} \ .
\eeq
A probably quicker way to guess the structure of the potentials, without performing all the aforementioned steps, is to use \Eq{fermionbilinears} to compute the NR limit of the DM matrix element of the operators in \Tab{tab: EM interactions}, substituting
\begin{align}
F_{0 0} = F_{i i} = 0 \ ,
&&
F_{0 i} = - F_{i 0} = E^i \ ,
&&
F_{i j} = - \varepsilon_{ijk} B^k
\end{align}
together with $\partial^\mu = (\ud / \ud t, - \bol{\nabla})^\tr$. Upon changing the sign to account for the sign difference between interaction Lagrangian and Hamiltonian, this trick allows to determine the structure of the DM coupling to an external electromagnetic field sourced by the nucleus. Notice that, while $\bol{E}$ is generated by the electric charges of the nucleons, $\bol{B}$ is generated by their slow movement within the nucleus, alongside with their spins, and should therefore be taken of the same order as $\Ord(v) \bol{E}$ in the NR expansion. For this reason, we have retained the next-to-leading order terms of the right-hand sides in \Eq{<EM currents>} that couple to $\bol{E}$, when the leading NR order couples to $\bol{B}$. One can anticipate that, for DM particles with a magnetic dipole or anapole moment, both the coupling with the nuclear charge (which generates $\bol{E}$) and with the nuclear magnetic moment (which generates $\bol{B}$) contribute to the scattering process, as we will see below.

The NR operators describing electromagnetic DM-nucleon interactions in \Tab{tab: EM interactions} can be computed as done for the EFT operators in \Sec{EFT operators}, \ie computing the NR expansion of the DM-nucleon scattering amplitude using Eqs.~\eqref{fermionbilinears},~\eqref{Nfermionbilinears} and matching it to a quantum mechanical operator. The only difference is that now the $S$ matrix needs to be perturbatively expanded to second order (for a tree-level computation, so-called \emph{one-photon exchange approximation}) rather than first order, with the photon propagator being accounted for explicitly. The computation is straightforward for the millicharge and electric-dipole interactions. For the other interactions, as discussed above, one can \eg suitably integrate the interaction operator by parts until a sum of terms of the form $J^\mu A_\mu$ is obtained (see \Eq{EM currents}, where non-contributing terms have been neglected), and then use \Eq{<EM currents>}. For the magnetic-dipole interaction, as an example, we can use the relevant formula in \Eq{<EM currents>} to obtain, at second order in the $S$-matrix perturbative expansion,
\begin{multline}
\label{magnetic dipole Mel_N}
\Mel_N = - i e \matel{\chi'}{J_\text{MDM}^\mu}{\chi} \left( - \frac{i}{\fvec{q}^2} g^{\mu \nu} \right) \Nmatel{J_\text{EM}^\nu}
\\
\NReq 2 e \mu_\chi \frac{1}{q^2} \left[ \mN Q_N \CMcal{I}_\chi \CMcal{I}_N q^2 + 4 i \mDM \mN Q_N \CMcal{I}_N \bol{s}_\chi \cdot (\bol{q} \times \bol{v}^\perp_N) + 2 \mDM g_N (\bol{s}_\chi \times \bol{q}) \cdot (\bol{s}_N \times \bol{q}) \right] .
\end{multline}
Interestingly, the expression in square brackets is second order (rather than zeroth or first order) in the NR expansion. Using \Eq{contractions3} we get
\beq
(\bol{s}_\chi \times \bol{q}) \cdot (\bol{s}_N \times \bol{q}) = q^2 \bol{s}_\chi \cdot \bol{s}_N - (\bol{s}_\chi \cdot \bol{q}) (\bol{s}_N \cdot \bol{q}) \ ,
\eeq
from which one derives the associated NR operator
\begin{multline}
\label{magnetic dipole ONR}
\Op_\NR^N = 2 e \mu_\chi \left[ \mN Q_N \unop + 4 i \frac{\mDM \mN}{q^2} Q_N \bol{s}_\chi \cdot (\bol{q} \times \bol{v}^\perp_N) + 2 \mDM g_N \left( \bol{s}_\chi \cdot \bol{s}_N - \frac{(\bol{s}_\chi \cdot \bol{q}) (\bol{s}_N \cdot \bol{q})}{q^2} \right) \right]
\\
= 2 e \mu_\chi \left[ \mN Q_N \, \Op^N_1 + 4 \frac{\mDM \mN}{q^2} Q_N \, \Op^N_5 + 2 \mDM g_N \left( \Op^N_4 - \frac{\Op^N_6}{q^2} \right) \right] ,
\end{multline}
with the NR building blocks $\Op^N_i$'s defined in \Eq{NR building blocks}. We will use this result in \Sec{Magnetic-dipole DM}. For the charge-radius and anapole-moment interactions, alternatively, one can also substitute in the Lagrangian the (lowest order in the EFT) equations of motion for $A^\mu$,
\beq
\partial_\nu F^{\mu\nu} = - e J_\text{EM}^\mu \ ,
\eeq
and then proceed with a first-order $S$-matrix perturbative expansion as done above for the EFT operators (see \Eq{S perturbative exp}), employing \Eq{<J_EM> NR} for taking the NR limit. For instance, the DM-nucleon scattering amplitude induced by the anapole operator can be computed from the interaction Lagrangian
\beq\label{anapole Lag}
\Lag = - e a_\chi \, \bar{\chi} \gamma_\mu \gamma^5 \chi \, J_\text{EM}^\mu \ ,
\eeq
from which, using Eqs.~\eqref{fermionbilinears},~\eqref{<J_EM> NR}, and~\eqref{vperp pk}, one gets the DM-nucleon scattering amplitude (see \Eq{S & M matrix})
\beq
\Mel_N = - e a_\chi \, \matel{\chi'}{\bar{\chi} \gamma_\mu \gamma^5 \chi}{\chi} \, \Nmatel{J_\text{EM}^\mu} \NReq 4 \mDM e a_\chi \left[ i g_N \bol{s}_\chi \cdot (\bol{s}_N \times \bol{q}) - 2 \mN \CMcal{I}_N Q_N \bol{s}_\chi \cdot \bol{v}^\perp_N \right] .
\eeq
This first-order result in the NR expansion finally maps onto the NR operator
\beq
\Op_\NR^N = 4 \mDM e a_\chi \left[ i g_N \bol{s}_\chi \cdot (\bol{s}_N \times \bol{q}) - 2 \mN Q_N \bol{s}_\chi \cdot \bol{v}^\perp_N \right] = 4 \mDM e a_\chi \left( g_N \, \Op^N_9 - 2 \mN Q_N \, \Op^N_8 \right) .
\eeq
The same result can be also obtained from \Eq{anapole Lag} by using \Eq{J_EM} for $J_\text{EM}^\mu$, then taking the NR operators relative to the $\bar{\chi} \gamma^\mu \gamma^5 \chi \, \bar{q} \gamma_\mu q$ DM-quark operators from \Tab{tab: spin-1/2 DM EFT}, and finally using Eqs.~\eqref{F_1^{q, p}},~\eqref{F_2^{q, p}}, and~\eqref{g_N}.

\section{From nucleons to nuclei}
\label{Form factors}
We have seen in \Sec{DM-N} how to compute the DM scattering amplitude, and associated NR operator, for a single target nucleon. We can now take a look at how nuclear physics gets involved in the computation of the scattering amplitude with the full nucleus. Its contribution can be usually factored within nuclear form factors that can be conveniently incorporated in the DM-nucleus differential cross section even with little knowledge of what's behind. A systematic approach was taken \eg in Refs.~\cite{Fitzpatrick:2012ix, Anand:2013yka}, where a recipe was provided to promptly obtain the DM-nucleus differential cross section based on the decomposition in \Eq{general NR Op} of any NR interaction into the building blocks in \Eq{NR building blocks}. Here we follow this approach. Our discussion is aimed at illustrating qualitatively how multi-body nuclear matrix elements of the single-nucleon NR operators in \Eq{general NR Op} can be computed, following Refs.~\cite{Donnelly:1978tz, Donnelly:1984rg, Fitzpatrick:2012ix, Anand:2013yka, Walecka}. Other references are \eg Refs.~\cite{DeForest:1966ycn, Hughes:1975eg, Donnelly:1975ze, Donnelly:1979ezn}.

\subsection{Nuclear and single-nucleon matrix elements}
\label{single-nucleon matel}
In this Section we indicate with $\Op$ a Lorentz-scalar operator acting on both the DM and nuclear degrees of freedom. In particular, we have in mind a position-dependent operator related to the $S$ matrix as in \Eq{S = O exp} below, although for now it can be a generic operator. Since we are mainly concerned here with the physics of the nucleus, it is useful to evaluate $\Op$ over all degrees of freedom not pertaining to the internal nuclear state, that is, the DM degrees of freedom as well as those relative to the overall nuclear motion. In this way one obtains an operator $\bar{\Op}$ acting exclusively over the internal nuclear degrees of freedom, with all other variables encoded in c-numbers (see below).

We take internal nuclear states to be eigenstates of total angular momentum, normalized as in \Eq{amom norm}, and denote them simply $\ket{\amom{J}^{(\prime)}, \amom{M}^{(\prime)}}$, with a $'$ indicating as usual final-state quantities. It is also useful to introduce single-nucleon states $\ket{\alpha} \equiv \ket{n, \ell, j, m}$, with $n$ a node number, $\ell$ the orbital angular momentum, $j = \ell \pm \frac{1}{2}$ the total single-particle angular momentum, and $m$ its projection along the quantization axis. These states describe a single nucleon inside the nucleus. We further define state vectors without the magnetic quantum number $m$, \ie $\ket{a} \equiv \ket{n, \ell, j}$. These definitions can be generalized to include isospin, but for our purposes it is enough to simply keep in mind whether each single-nucleon state describes a proton or a neutron.

A second-quantized operator $\bar{\Op}$ can be expanded as
\beq
\bar{\Op} = \bar{\Op}^{(0)} + \bar{\Op}^{(1)} + \bar{\Op}^{(2)} + \dots \ ,
\eeq
with
\begin{align}
\bar{\Op}^{(0)} &\propto \unop \ ,
\\
\label{O^(1)}
\bar{\Op}^{(1)} &= \sum_{\alpha, \alpha'} \matel{\alpha'}{\bar{\Op}^{(1)}}{\alpha} \, c_{\alpha'}^\dagger c_\alpha \ ,
\\
\bar{\Op}^{(2)} &= \sum_{\alpha, \alpha', \beta, \beta'} \matel{\alpha', \beta'}{\bar{\Op}^{(2)}}{\alpha, \beta} \, c_{\alpha'}^\dagger c_{\beta'}^\dagger c_\beta c_\alpha \ ,
\end{align}
and so on, where $c_\alpha^\dagger$ and $c_\alpha$ are the creation and annihilation operators for the state $\ket{\alpha}$, respectively. The $\matel{\alpha'}{\bar{\Op}^{(1)}}{\alpha}$ factors are single-nucleon matrix elements, the $\matel{\alpha', \beta'}{\bar{\Op}^{(2)}}{\alpha, \beta}$'s are two-nucleon matrix elements, etc. These matrix elements can be computed using the first-quantized version of the operator, as we will see more explicitly in \Sec{Scattering amplitude}. The zero-body operator, $\bar{\Op}^{(0)}$, if non-zero, only contributes to forward ($q = 0$) scattering, which is not of interest to us. The one-body operator $\bar{\Op}^{(1)}$ often provides the largest contribution, though sometimes the two-body operator $\bar{\Op}^{(2)}$ needs to be taken into account as well. Here we focus on $\bar{\Op}^{(1)}$, making the implicit assumption that it provides the main contribution to $\bar{\Op}$; two-body contributions can be computed \eg within Chiral EFT, see \eg Refs.~\cite{Prezeau:2003sv, Cirigliano:2012pq, Menendez:2012tm, Cannoni:2012jq, Divari:2013fx, Klos:2013rwa, Beane:2013kca, Cirigliano:2013zta, Hoferichter:2015ipa, Hoferichter:2016nvd, Hoferichter:2018acd} and the relevant discussions in Secs.~\ref{SI interaction},~\ref{SD interaction}. From now on we deliberately confuse operators acting over nuclear degrees of freedom with their one-body component, unless otherwise stated.

Any one-body operator is related, in the NR limit, to a specific DM-nucleon NR operator. We have seen in \Sec{DM-N} that such NR operators, whose general form is given in \Eq{general NR Op}, are built out of $\bol{s}_\chi$, $\bol{s}_N$, $i \bol{q}$, and $\bol{v}^\perp_N$. Let us now look at what, among these ingredients, depends non-trivially on the internal nuclear degrees of freedom. Easy to imagine, the DM spin $\bol{s}_\chi$ has nothing to do with the nucleus, therefore any $\bol{s}_\chi$ factor can be promptly evaluated into a c-number quantity when computing the DM-nucleus matrix element. The nucleon spin $\bol{s}_N$, on the contrary, is a purely internal nuclear degree of freedom, thus it remains an operator when computing $\bar{\Op}$ from $\Op$. $\bol{k}' - \bol{k}$, the momentum transferred to the nucleon in the scattering, coincides, by momentum conservation, with the momentum transferred to the whole nucleus; this combination then only depends on the overall nuclear motion and not on the details of its internal state. Moreover, again due to momentum conservation, $\bol{k}' - \bol{k}$ can be confused with $\bol{q} = \bol{p} - \bol{p}'$, as we have been doing so far. The other combination of nucleon momenta, $\bol{k}' + \bol{k}$, always appears encoded in $\bol{v}^\perp_N$ (see \Eq{vperp pk}), as long as the relativistic interactions are hermitian and the NR expansion is kept at an order where the Galilean symmetry is respected (see discussion in \Sec{NR operators}). $\bol{v}^\perp_N$ has a component related to the overall nuclear motion as well as an intrinsic component due to the motion of the nucleon relative to the nuclear CM, and we will see in \Eq{v^perp_T} below how to take care of their separation. With this in mind, as an example, if $\Op$ involves in the NR limit the NR operator $\Op_\NR^N = \Op^N_{10} = i \, \bol{s}_N \cdot \bol{q}$ (see \Eq{NR building blocks}), with $\bol{s}_N$ and $\bol{q}$ operators, $\bar{\Op}$ involves $\bar{\Op}_\NR^N = i \, \CMcal{I}_\chi \, \bol{s}_N \cdot \bol{q}$, where $\bol{s}_N$ is still an operator while $\bol{q}$ is a c-number (vector) and $\CMcal{I}_\chi$ is the (c-number) unit matrix on DM-spin space, see \Eq{I s}. Further examples can be found in \Sec{Scattering amplitude}. Notice that, as in \Sec{DM-N}, we use the same symbol for operator and c-number quantities, so for instance $\bol{s}_\chi$ can be either the DM spin operator or the matrix defined in \Eq{I s}.

As suggested by \Eq{general NR Op}, a scalar operator $\Op$ can be decomposed in the NR limit as a sum of nucleon-specific ($N = p, n$) terms of the form $\CMcal{K}^N \CMcal{O}^N$, with $\CMcal{O}^N$ an operator depending on the internal nuclear degrees of freedom and $\CMcal{K}^N$ an operator depending on all other variables. In each term, $\CMcal{K}^N$ and $\CMcal{O}^N$ can be scalars under rotations, in which case $\CMcal{K}^N \CMcal{O}^N$ is just their product, or can be three-vectors $\bol{\CMcal{K}}^N$ and $\bol{\CMcal{O}}^N$, for which $\CMcal{K}^N \CMcal{O}^N$ stands for a scalar product, or can be irreducible higher-rank tensors combined into a scalar. Correspondingly, evaluating all degrees of freedom but those related to the internal nuclear state, we can write in the NR limit $\bar{\Op}$ as a sum of terms of the form $K^N \CMcal{O}^N$, with $K^N$ a c-number quantity. More in detail, $K^N$ is in general a (c-number) matrix acting over the DM-spin space, \eg it is a simple c-number for spin-$0$ DM and a linear combination of $\CMcal{I}_\chi$ and $\bol{s}_\chi$ for spin-$1/2$ DM. $K^N$ also incorporates the NR dependence of $\bar{\Op}$ on $\bol{q}$ and on the CM component of $\bol{v}^\perp_N$, so that $\CMcal{O}^N$ only depends on $\bol{s}_N$ as well as on the intrinsic component of $\bol{v}^\perp_N$. We also include in $K^N$ any nucleon-specific numerical coefficient, such as coupling constants and hadron-physics coefficients, see \Sec{qg to N}. A more concrete discussion is carried out in \Sec{Multipoles}.

Since we are dealing with nuclear angular-momentum eigenstates, it is convenient to work with operators that behave as objects of definite angular momentum. These are the \emph{spherical tensor operators}, denoted $\CMcal{T}_{JM}$, with the rank $J$ playing the role of the angular momentum quantum number, while $M$, which indexes the tensor components, plays the role of the angular momentum projection along the quantization axis (see \eg Ref.~\cite{Merzbacher}). $J$ and $M$ can be `summed' to other angular momenta quantum numbers in the usual way, thus making computations of matrix elements involving angular momentum eigenstates easier and the use of selection rules more transparent. Spherical tensor operators can be obtained as a linear combination of the corresponding tensor's Cartesian components, as we will see shortly for some specific cases.

To work with spherical tensor operators, instead of using Cartesian components, it is useful to project $K^N$ onto a spherical basis. We can then write
\beq\label{KO}
K^N \CMcal{O}^N = \sum_{M = -J}^J K^N_{JM} \CMcal{O}^N_{JM} \ ,
\eeq
with $J$ fixed by the rank of the irreducible tensors involved: $J = 0$ for scalars, $J = 1$ for vectors, and so on. Determining the $K^N_{JM}$ and $\CMcal{O}^N_{JM}$ spherical components is trivial for scalars, for which we have $K^N = K^N_{0 0}$ and $\CMcal{O}^N = \CMcal{O}^N_{0 0}$. For vectors $\bol{K}^N$ and $\bol{\CMcal{O}}^N$, given a Cartesian coordinate basis of unit vectors $\hat{\bol{e}}_1$, $\hat{\bol{e}}_2$, $\hat{\bol{e}}_3$, we introduce the spherical vector basis
\begin{align}
\label{spherical vector basis}
\hat{\bol{e}}_{\pm 1} = \mp \frac{1}{\sqrt 2} (\hat{\bol{e}}_1 \pm i \hat{\bol{e}}_2) \ ,
&&&
\hat{\bol{e}}_0 = \hat{\bol{e}}_3 \ .
\end{align}
These basis vectors satisfy
\begin{align}
\hat{\bol{e}}_\lambda^* = (-1)^\lambda \, \hat{\bol{e}}_{- \lambda} \ ,
&&&
\hat{\bol{e}}_\lambda^* \cdot \hat{\bol{e}}_{\lambda'} = \delta_{\lambda \lambda'} \ ,
\end{align}
and allow to express a generic vector $\bol{V}$ as~\cite{Hughes:1975eg, Edmonds, Walecka}
\begin{align}
\label{V spherical compo}
\bol{V} = \sum_{\lambda = \pm 1, 0} V_\lambda \hat{\bol{e}}_\lambda^* \ ,
&&
\text{with}
&&
V_\lambda \equiv \bol{V} \cdot \hat{\bol{e}}_\lambda \ .
\end{align}
For later convenience, we also notice that the scalar product of two vectors $\bol{V}$, $\bol{W}$ reads in spherical components
\begin{align}
\label{scalarproduct}
\bol{V} \cdot \bol{W} = \sum_{\lambda = \pm 1, 0} (-1)^\lambda \, V_\lambda W_{- \lambda} \ ,
&&&
\bol{V}^* \cdot \bol{W} = \sum_{\lambda = \pm 1, 0} V_\lambda^* W_\lambda \ ,
\end{align}
and that
\beq\label{vectorproduct}
i (\bol{V}^* \times \bol{W}) \cdot \hat{\bol{e}}_0 = \sum_{\lambda = \pm 1} \lambda \, V_\lambda^* W_\lambda \ .
\eeq
Expressing $\bol{K}^N$ in terms of spherical components we can then write
\beq
\bol{K}^N \cdot \bol{\CMcal{O}}^N = \sum_{\lambda = \pm 1, 0} K^N_{1 \lambda} \CMcal{O}^N_{1 \lambda} \ ,
\eeq
where we defined
\begin{align}
K^N_{1 \lambda} \equiv K^N_\lambda = \bol{K}^N \cdot \hat{\bol{e}}_\lambda \ ,
&&&
\CMcal{O}^N_{1 \lambda} \equiv \bol{\CMcal{O}}^N \cdot \hat{\bol{e}}_\lambda^* = (-1)^\lambda \, \CMcal{O}^N_{- \lambda} \ .
\end{align}
Higher-rank tensors can be projected onto spherical components in a similar way, but we will be mostly interested in scalars and vectors, which are the main cases of interest in the NR limit: in fact, as can be seen in \Eq{O_NR scalar-vector dec} below, they are the only possible cases if we limit ourselves to NR operators at most linear in $\bol{v}^\perp_N$.

$\CMcal{O}^N_{JM}$ is a spherical tensor operator if $\CMcal{O}^N$ transforms as a standard tensor under rotations. However, this does not apply if $\CMcal{O}^N$ depends on position (thus transforming as a tensor field rather than a tensor), as it will be the case for us. We will then see in Secs.~\ref{nuc form factors},~\ref{Multipoles}, and~\ref{Nuclear matrix element} how to express the DM-nucleus scattering amplitude in terms of spherical tensors $\CMcal{T}_{JM}$, or alternatively how to derive $\CMcal{T}_{JM}$ from $\CMcal{O}^N$. For instance, as we will see in \Sec{nuc form factors}, the scattering amplitude in the limit of a point-like nucleus depends on the matrix element of the spatial integral of $\CMcal{O}^N_{JM}$, which is a spherical tensor. We will generalize this analysis to a spatially-extended nucleus in \Sec{Multipoles} and present the relative scattering amplitude in \Sec{Nuclear matrix element}.

A powerful tool in the computation of matrix elements $\matel{\amom{J}', \amom{M}'}{\CMcal{T}_{JM}}{\amom{J}, \amom{M}}$ of spherical tensor operators is the Wigner-Eckart theorem, which states that the $\amom{M}$ and $\amom{M}'$ dependence of the matrix element is entirely encoded in the Clebsch-Gordan coefficient $\CG{\amom{J}, \amom{M}}{J, M}{\amom{J}', \amom{M}'}$, independent of the actual operator. The remaining part of the matrix element, which does not depend on $\amom{M}$ and $\amom{M}'$, is called \emph{reduced matrix element} and is denoted with double vertical bars:
\beq\label{Wigner-Eckart}
\matel{\amom{J}', \amom{M}'}{\CMcal{T}_{JM}}{\amom{J}, \amom{M}} = \CG{\amom{J}, \amom{M}}{J, M}{\amom{J}', \amom{M}'} \, \redmatel{\amom{J}'}{\CMcal{T}_J}{\amom{J}} \ .
\eeq
This is essentially the definition of $\redmatel{\amom{J}'}{\CMcal{T}_J}{\amom{J}}$ (we follow the notation of Ref.~\cite{Merzbacher}, other slightly different definitions may be found in the literature).

Applying now \Eq{O^(1)} to a spherical tensor operator $\CMcal{T}_{JM}$, we can use the Wigner-Eckart theorem~\eqref{Wigner-Eckart} to obtain
\beq
\CMcal{T}_{JM}^{(1)} = \sum_{a, a'} \redmatel{a'}{\CMcal{T}_J^{(1)}}{a} \, \psi_{JM}^\dagger(a', a) \ ,
\eeq
where the one-body operator $\psi_{JM}^\dagger(a', a)$ is an appropriate combination of the $c_{\alpha'}^\dagger$'s and $c_\alpha$'s with definite angular momentum. Notice that the magnetic quantum numbers of $\ket{\alpha}$ and $\ket{\alpha'}$ are summed over within $\psi_{JM}^\dagger(a', a)$. $\CMcal{T}_{JM}^{(2)}$ and the other multi-body operators can be treated analogously. Assuming the interaction to be dominated by one-nucleon contributions, we are only interested in the reduced matrix element of $\CMcal{T}_{JM}^{(1)}$,
\beq
\redmatel{\amom{J}'}{\CMcal{T}_J^{(1)}}{\amom{J}} = \sum_{a, a'} \redmatel{a'}{\CMcal{T}_J^{(1)}}{a} \, \redmatel{\amom{J}'}{\psi_J^\dagger(a', a)}{\amom{J}} \ .
\eeq
Thus all of the single-nucleon dependence has been factored in the single-nucleon reduced matrix elements, $\redmatel{a'}{\CMcal{T}_J^{(1)}}{a}$, while the complexities of the nuclear many-body problem have been isolated in the so-called \emph{one-nucleon density-matrix elements} $\redmatel{\amom{J}'}{\psi_J^\dagger(a', a)}{\amom{J}}$.

\subsection{Scattering amplitude}
\label{Scattering amplitude}
The $S$ matrix for a fundamental DM particle scattering off a target nucleus $T$ can be written as
\beq\label{S = O exp}
S = \braket{\DM', T'}{\DM, T} + \int \ud^4 \fvec{x} \, \matel{\DM', T'}{\Op(\fvec{x})}{\DM, T} = \braket{\DM', T'}{\DM, T} + \int \ud^4 \fvec{x} \, \matel{T'}{\bar{\Op}(\fvec{x})}{T} \, e^{- i \fvec{q} \cdot \fvec{x}} \ ,
\eeq
where $\ket{T}$, $\ket{T'}$ are the initial and final nuclear states, respectively, and we recall that $\fvec{q} \equiv \fvec{p} - \fvec{p}'$. For our purposes, these expressions can be basically considered as defining $\Op(\fvec{x})$ and $\bar{\Op}(\fvec{x})$. The whole $S$ matrix perturbative expansion can be cast in this form, with $\Op(\fvec{x}) = i \, \Lag_T(\fvec{x}) + \dots$, see \eg the first equality in \Eq{S perturbative exp}; $\Op(\fvec{x})$ may be a simple-looking operator if we truncate the expansion at tree level, otherwise it may have a more involved expression. The exponential in the last equality comes from the initial and final DM particle free wave functions, which originate from evaluating $\Op(\fvec{x})$ over the DM degrees of freedom. Evaluating $\Op(\fvec{x})$ also over the degrees of freedom pertaining to the overall nuclear motion then yields the $\bar{\Op}(\fvec{x})$ operator, acting exclusively over internal nuclear degrees of freedom. Since the nucleus initial and final states are energy eigenstates, the time dependence of the operator $\bar{\Op}(\fvec{x})$ can be factored out by means of \Eq{psi(0) psi(x)}~\cite{Walecka}, yielding
\beq\label{S 1delta}
S = \braket{\DM', T'}{\DM, T} + \int \ud x^0 \, e^{- i (q^0 + k^0 - {k'}^0) x^0} \int \ud^3 x \, \matel{T'}{\bar{\Op}(\bol{x})}{T} \, e^{i \bol{q} \cdot \bol{x}} \ .
\eeq
The integral over $x^0$ yields an energy-conservation delta function, see \Eq{delta function integral rep}.

As an example of \Eq{S = O exp}, we can consider the DM-nucleon contact interaction
\beq
\Lag_N(\fvec{x}) = c_N \, \Op_\DM(\fvec{x}) \Op_N(\fvec{x}) \ ,
\eeq
so that the full DM-nucleus Lagrangian is
\beq\label{DM-nucleus Lag}
\Lag_T(\fvec{x}) = \sum_{N = p, n} \sum_i \Lag_{N_i}(\fvec{x}) \ ,
\eeq
with $N_i$ the $i^\text{th}$ proton ($N = p$) or neutron ($N = n$) field. At first order in a perturbative expansion, see \eg \Eq{S perturbative exp}, we get
\begin{multline}
S = \braket{\DM', T'}{\DM, T} + i \int \ud^4 \fvec{x} \, \matel{\DM', T'}{\Lag_T(\fvec{x})}{\DM, T}
\\
= \braket{\DM', T'}{\DM, T} + i \int \ud^4 \fvec{x} \, \matel{\DM'}{\Op_\DM(\fvec{x})}{\DM} \, \matel{T'}{\sum_{N = p, n} \sum_i c_N \Op_{N_i}(\fvec{x})}{T} \ ,
\end{multline}
so that using \Eq{psi(0) psi(x)} we have
\beq
S = \braket{\DM', T'}{\DM, T} + i \, \matel{\DM'}{\Op_\DM(\fvec{0})}{\DM} \int \ud^4 \fvec{x} \, \matel{T'}{\sum_{N = p, n} \sum_i c_N \Op_{N_i}(\fvec{x})}{T} \, e^{- i \fvec{q} \cdot \fvec{x}} \ .
\eeq
$\Op(\fvec{x})$ in \Eq{S = O exp} may then be taken to be approximately $\Op(\fvec{x}) = i \, \Lag_T(\fvec{x})$, while $\bar{\Op}(\fvec{x})$ is approximated by $i \, \matel{\DM'}{\Op_\DM(\fvec{0})}{\DM} \sum_{N, i} c_N \Op_{N_i}(\fvec{x})$ evaluated over the degrees of freedom related to the overall nuclear motion. To be more explicit we may take for instance $\Lag_N = c_N \, \phi^\dagger \phi \, \bar{N} N$ for spin-$0$ DM or $\Lag_N = c_N \, \bar{\chi} \chi \, \bar{N} N$ for spin-$1/2$ DM, two effective Lagrangians inducing the SI interaction (see \Sec{SI interaction}). $\bar{\Op}(\fvec{x})$ can then be respectively identified with $i \sum_{N, i} c_N \, \bar{N}_i(\fvec{x}) N_i(\fvec{x})$ or $i \, \bar{u}'_\chi(\bol{p}', s') u_\chi(\bol{p}, s) \sum_{N, i} c_N \, \bar{N}_i(\fvec{x}) N_i(\fvec{x})$ evaluated over the overall nuclear motion. This identification is perhaps more transparent in the NR limit, where the separation of CM and intrinsic nuclear degrees of freedom can be easily taken care of as discussed in \Sec{single-nucleon matel}. For a single nucleon of type $N$, following \Sec{DM-N}, we obtain from $\Op(\fvec{x})$ the NR operator $\Op_\NR^N = 2 \mN c_N \, \Op^N_1$ for spin-$0$ DM and $\Op_\NR^N = 4 \mDM \mN c_N \, \Op^N_1$ for spin-$1/2$ DM, both multiplied by $i$. The NR limit of the Dirac spinor bilinears, which can be computed using Eqs.~\eqref{fermionbilinears},~\eqref{Nfermionbilinears}, gives rise to the $2 \mDM$ and $2 \mN$ factors that reflect our spinor normalization in \Eq{spinor normalization}. In general, the NR limit of $\Op(\fvec{x})$ may then be schematically written as the sum over nucleons of $\Op_\NR^N$ times the DM and nucleon number-density operators:
\beq\label{NR EFT Op O}
\Op(\fvec{x}) \NReq
\begin{cases}
\displaystyle{\frac{i}{2 \mN} \sum_{N, i} \Op_\NR^{N_i} \, \phi^\dagger \phi \, \bar{N}_i \gamma^0 N_i} & \text{spin-$0$ DM},
\\[5mm]
\displaystyle{\frac{i}{4 \mDM \mN} \sum_{N, i} \Op_\NR^{N_i} \, \bar{\chi} \gamma^0 \chi \, \bar{N}_i \gamma^0 N_i} & \text{spin-$1/2$ DM},
\end{cases}
\eeq
with all field operators taken at position $\fvec{x}$. Evaluating over the DM degrees of freedom we then have for the $\Op^N_1$ example above $\bar{\Op}(\fvec{x}) \NReq \frac{i}{2 \mN} \sum_{N, i} c_N \, \bar{N}_i(\fvec{x}) \gamma^0 N_i(\fvec{x})$ for spin-$0$ DM and $\bar{\Op}(\fvec{x}) \NReq \frac{i}{2 \mN} \CMcal{I}_\chi \sum_{N, i} c_N \, \bar{N}_i(\fvec{x}) \gamma^0 N_i(\fvec{x})$ for spin-$1/2$ DM.

The construction in \Eq{NR EFT Op O} matches that of Ref.~\cite{Fitzpatrick:2012ix}, which in the following we employ as a case example. There, an effective NR field-theoretic Lagrangian is schematically built by multiplying the NR operator describing the system's dynamics, \Eq{general NR Op}, by the DM and nucleon number-density operators,
\beq\label{NR EFT Op}
\Lag_N(\fvec{x}) =
\begin{cases}
\displaystyle{\frac{1}{2 \mN} \, \Op_\NR^N \, \phi^\dagger \phi \, \bar{N} \gamma^0 N} & \text{spin-$0$ DM},
\\[5mm]
\displaystyle{\frac{1}{4 \mDM \mN} \, \Op_\NR^N \, \bar{\chi} \gamma^0 \chi \, \bar{N} \gamma^0 N} & \text{spin-$1/2$ DM},
\end{cases}
\eeq
with all field operators evaluated in $\fvec{x}$. The full DM-nucleus Lagrangian is then again given by \Eq{DM-nucleus Lag}. This schematic expression represents an operator whose matrix element between DM and nucleon momentum eigenstates yields in the NR limit
\beq
\matel{\DM', N'}{\Lag_N(\fvec{0})}{\DM, N} \NReq \langle \Op_\NR^N \rangle \ ,
\eeq
with $\langle \Op_\NR^N \rangle$ the NR matrix element of $\Op_\NR^N$ between momentum and spin eigenstates. For instance, $\langle \Op^N_8 \rangle = \CMcal{I}_N \bol{s}_\chi \cdot \bol{v}^\perp_N$ and $\langle \Op^N_{10} \rangle = i \, \CMcal{I}_\chi \, \bol{s}_N \cdot \bol{q}$ (see \Eq{NR building blocks}), with $\bol{s}_\chi$, $\bol{s}_N$, $\bol{q}$, and $\bol{v}^\perp_N$ all c-number quantities. The $S$ matrix can be computed at first order as above, resulting in \Eq{NR EFT Op O} and
\beq\label{Obar(x)}
\bar{\Op}(\fvec{x}) \NReq \frac{i}{2 \mN} \sum_{N = p, n} \sum_i \bar{\Op}_\NR^{N_i} \, \bar{N}_i(\fvec{x}) \gamma^0 N_i(\fvec{x}) \ ,
\eeq
with $\bar{\Op}_\NR^N$ being $\Op_\NR^N$ evaluated over the DM degrees of freedom and those pertaining to the overall nuclear motion. As an example, we can choose in \Eq{NR EFT Op} $\Op_\NR^N = c_N \, \Op^N_{13} = i c_N (\bol{s}_\chi \cdot \bol{v}^\perp_N) (\bol{s}_N \cdot \bol{q})$ for a spin-$1/2$ DM particle, with $c_N$ a numerical coefficient and $\bol{s}_\chi$, $\bol{s}_N$, $\bol{q}$, $\bol{v}^\perp_N$ being operators. We then have $\bar{\Op}_\NR^N = i c_N (\bol{s}_\chi \cdot \bol{v}^\perp_N) (\bol{s}_N \cdot \bol{q})$, with $\bol{s}_\chi$, $\bol{q}$, and the CM component of $\bol{v}^\perp_N$ c-number vectors, while $\bol{s}_N$ and the intrinsic component of $\bol{v}^\perp_N$ remain operators.

The separation of the CM and intrinsic $\bol{v}^\perp_N$ components can be taken care of as follows. Let us denote with
\beq
\bol{v}^\perp_i \NReq \frac{1}{2} (\bol{v}_\DM + \bol{v}'_\DM - \bol{v}_i - \bol{v}'_i)
\eeq
the transverse velocity relative to the $i^\text{th}$ nucleon, where $\bol{v}_\DM$ and $\bol{v}_i$ ($\bol{v}'_\DM$ and $\bol{v}'_i$) are the initial (final) DM and nucleon velocities (see \Eq{vperp_N}). We then have
\beq
\bol{v}^\perp_T \NReq \frac{1}{2} (\bol{v}_\DM + \bol{v}'_\DM - \bol{v}_T - \bol{v}'_T) = \frac{1}{A} \sum_{i = 1}^A \bol{v}^\perp_i
\eeq
for the DM-nucleus transverse velocity, which depends on the nuclear-CM velocity
\beq
\bol{v}^{(\prime)}_T = \frac{1}{A} \sum_{i = 1}^A \bol{v}^{(\prime)}_i \ .
\eeq
A NR operator of the form $\bol{v}^\perp_N \cdot \bol{O}$, with $\bol{O}$ any vector operator (dependent on isospin for simplicity), enters the Lagrangian schematically as $\sum_i \bol{v}^\perp_i \cdot \bol{O}_i$, where $\bol{O}_i$ is $\bol{O}$ applied to the $i^\text{th}$ nucleon. For instance, in the example above we have $\bol{O}_i = i c_i (\bol{s}_\chi \cdot \bol{v}^\perp_i) (\bol{s}_i \cdot \bol{q})$ with $\bol{s}_i$ the spin operator of the $i^\text{th}$ nucleon and $c_i$ an isospin-dependent coefficient. We can then write
\beq\label{v^perp_T}
\sum_{i = 1}^A \bol{v}^\perp_i \cdot \bol{O}_i = \underbrace{\bol{v}^\perp_T \vphantom{\sum_{j > i = 1}^A}}_\text{CM} \cdot \sum_{i = 1}^A \bol{O}_i + \underbrace{\frac{1}{A} \sum_{j > i = 1}^A (\bol{v}^\perp_i - \bol{v}^\perp_j) \cdot (\bol{O}_i - \bol{O}_j)}_\text{intrinsic} \ ,
\eeq
where $\bol{v}^\perp_T$ depends solely on the motion of the nuclear CM (much like the momentum transferred to the nucleon in the scattering, see \Sec{single-nucleon matel}), while the last term, dubbed \emph{intrinsic}, acts exclusively over the internal nuclear degrees of freedom. The above formula can be proved by counting the $\bol{v}^\perp_i \cdot \bol{O}_j$ terms: the left-hand side counts every diagonal ($i = j$) term once, the first term on the right-hand side counts $1 / A$ times every diagonal and non-diagonal term, while the second term adds $(A - 1) / A$ times every diagonal term while subtracting $1 / A$ times every non-diagonal term.

For a NR system as the nucleus, the overall motion can be factored out from the state vectors. For this discussion we follow Appendix A of Ref.~\cite{Foldy:1969rk} and Appendix B of Ref.~\cite{Walecka} (see also Appendix B of Ref.~\cite{Fitzpatrick:2012ix}). Modelling the nucleus as a system of point-like nucleons, we denote with $\bol{x}_i$ the coordinates of each nucleon, with $\bol{r}$ the nuclear CM coordinates,
\beq
\bol{r} = \frac{1}{A} \sum_{i = 1}^A \bol{x}_i \ ,
\eeq
and with $\bar{\bol{x}}_i$ the coordinates relative to the CM,
\begin{align}
\bar{\bol{x}}_i \equiv \bol{x}_i - \bol{r} \ ,
&&
\text{satisfying}
&&
\sum_{i = 1}^A \bar{\bol{x}}_i = \bol{0} \ .
\end{align}
Position eigenstates, which we denote $\ket{\bol{x}_1, \dots, \bol{x}_A} = \ket{\bar{\bol{x}}_1, \dots, (\bar{\bol{x}}_A)} \otimes \ket{\bol{r}}$ with $\bar{\bol{x}}_A \equiv - \sum_{i = 1}^{A - 1} \bar{\bol{x}}_i$ indicated in parentheses not being an independent variable, can be defined to satisfy
\begin{align}
\label{r norm}
\braket{\bol{r}'}{\bol{r}} = \frac{1}{A^3} \delta^{(3)}(\bol{r} - \bol{r}') \ ,
&&&
\braket{\bar{\bol{x}}'_1, \dots, (\bar{\bol{x}}'_A)}{\bar{\bol{x}}_1, \dots, (\bar{\bol{x}}_A)} = \prod_{i = 1}^{A - 1} \delta^{(3)}(\bar{\bol{x}}_i - \bar{\bol{x}}'_i) \ .
\end{align}
This can be derived from
\beq
\braket{\bol{x}'_1, \dots, \bol{x}'_A}{\bol{x}_1, \dots, \bol{x}_A} = \prod_{i = 1}^A \delta^{(3)}(\bol{x}_i - \bol{x}'_i)
\eeq
by noting that
\beq\label{deltas}
\prod_{i = 1}^A \delta^{(3)}(\bol{x}_i - \bol{x}'_i) = \prod_{i = 1}^A \delta^{(3)}(\bar{\bol{x}}_i + \bol{r} - \bar{\bol{x}}'_i - \bol{r}') = \frac{1}{A^3} \delta^{(3)}(\bol{r} - \bol{r}') \prod_{i = 1}^{A - 1} \delta^{(3)}(\bar{\bol{x}}_i - \bar{\bol{x}}'_i) \ ,
\eeq
where we used $\delta(A z) = \delta(z) / A$. The CM and internal unit operators are
\begin{align}
\unop_\text{CM} &= A^3 \int \ud^3 r \, \ket{\bol{r}} \bra{\bol{r}} \ ,
\\
\label{uno_int}
\unop_\text{int} &= \int \ud^3 \bar{x}_1 \, \cdots \, \ud^3 \bar{x}_{A - 1} \, \ket{\bar{\bol{x}}_1, \dots, (\bar{\bol{x}}_A)} \bra{\bar{\bol{x}}_1, \dots, (\bar{\bol{x}}_A)} \ ,
\end{align}
as can be derived from
\beq\label{uno_target}
\unop_\text{nucleus} = \int \ud^3 x_1 \, \cdots \, \ud^3 x_A \, \ket{\bol{x}_1, \dots, \bol{x}_A} \bra{\bol{x}_1, \dots, \bol{x}_A}
\eeq
by noting that
\beq
\ud^3 x_1 \, \cdots \, \ud^3 x_A = A^3 \, \ud^3 \bar{x}_1 \, \cdots \, \ud^3 \bar{x}_{A - 1} \, \ud^3 r \ .
\eeq
The latter equality results from $A^3$ being the Jacobian of the basis change, or alternatively from
\begin{multline}
\ud^3 x_1 \, \cdots \, \ud^3 x_A = \ud^3 x_1 \, \cdots \, \ud^3 x_A \, \delta^{(3)} \! \left( \bol{r} - \frac{1}{A} \sum_{i = 1}^A \bol{x}_i \right) \ud^3 r = \ud^3 \bar{x}_1 \, \cdots \, \ud^3 \bar{x}_A \, \delta^{(3)} \! \left( \frac{1}{A} \sum_{i = 1}^A \bar{\bol{x}}_i \right) \ud^3 r
\\
= A^3 \, \ud^3 \bar{x}_1 \, \cdots \, \ud^3 \bar{x}_A \, \delta^{(3)} \! \left( \sum_{i = 1}^A \bar{\bol{x}}_i \right) \ud^3 r = A^3 \, \ud^3 \bar{x}_1 \, \cdots \, \ud^3 \bar{x}_{A - 1} \, \ud^3 r \ .
\end{multline}
The overall nuclear motion and the internal nuclear state can be factored as $\ket{T^{(\prime)}} = \ket{\bol{k}^{(\prime)}} \otimes \ket{T_\text{int}^{(\prime)}}$, where the nuclear CM wave function is $\braket{\bol{r}}{\bol{k}^{(\prime)}} = \sqrt{\rho(k^{(\prime)}) / A^3} \, e^{i \bol{k}^{(\prime)} \cdot \bol{r}}$, see Eqs.~\eqref{NR free wave function},~\eqref{r norm}. $\ket{T^{(\prime)}_\text{int}}$ is a many-body state describing solely the internal nuclear state, which we will later take to have definite total angular momentum $\amom{J}^{(\prime)}$ and projection along the quantization axis $\amom{M}^{(\prime)}$. We can now insert $\unop_\text{CM}$ twice in $\matel{T'}{\bar{\Op}(\bol{x})}{T}$, which, if $\bar{\Op}(\bol{x})$ does not depend on $\bol{r}$, as we will assume, returns
\beq
\matel{T'}{\bar{\Op}(\bol{x})}{T} = \sqrt{\rho(k) \rho(k')} \int \ud^3 r \, e^{i (\bol{k} - \bol{k}') \cdot \bol{r}} \, \matel{T'_\text{int}}{\bar{\Op}(\bol{x})}{T_\text{int}} \ .
\eeq
We can then write \Eq{S 1delta} as
\begin{multline}
S = \braket{\DM', T'}{\DM, T} + (2 \pi) \delta(q^0 + k^0 - {k'}^0) \int \ud^3 r \, e^{i (\bol{q} + \bol{k} - \bol{k}') \cdot \bol{r}} \, \sqrt{\rho(k) \rho(k')} \int \ud^3 \bar{x} \, \matel{T'_\text{int}}{\widetilde{\Op}(\bar{\bol{x}})}{T_\text{int}} \, e^{i \bol{q} \cdot \bar{\bol{x}}}
\\
= \braket{\DM', T'}{\DM, T} + (2 \pi)^4 \delta^{(4)}(\fvec{q} + \fvec{k} - \fvec{k}') \, \sqrt{\rho(k) \rho(k')} \int \ud^3 \bar{x} \, \matel{T'_\text{int}}{\widetilde{\Op}(\bar{\bol{x}})}{T_\text{int}} \, e^{i \bol{q} \cdot \bar{\bol{x}}} \ ,
\end{multline}
where we denoted with
\beq
\widetilde{\Op}(\bar{\bol{x}}) = \bar{\Op}(\bol{x})
\eeq
the operator $\bar{\Op}(\bol{x})$ as a function of $\bar{\bol{x}} \equiv \bol{x} - \bol{r}$, and we used \Eq{delta function integral rep}. The DM-nucleus scattering amplitude $\Mel$, defined by
\beq
S = \braket{\DM', T'}{\DM, T} + i \, (2 \pi)^4 \delta^{(4)}(\fvec{p'} + \fvec{k'} - \fvec{p} - \fvec{k}) \, \Mel \ ,
\eeq
reads then for mutually different initial and final states
\beq\label{Mel = Otilde exp}
\Mel = - i \sqrt{\rho(k) \rho(k')} \int \ud^3 \bar{x} \, \matel{T'_\text{int}}{\widetilde{\Op}(\bar{\bol{x}})}{T_\text{int}} \, e^{i \bol{q} \cdot \bar{\bol{x}}} \ .
\eeq
Since $\matel{T'_\text{int}}{\widetilde{\Op}(\bar{\bol{x}})}{T_\text{int}}$ has support only within the nucleus, the Riemann-Lebesgue lemma tells us that $\Mel$ vanishes for $q$ much larger than the inverse nuclear radius.

It can be convenient to express $\bar{\Op}(\bol{x})$ in terms of first-quantized position operators. Exploiting the fact that the operators we are interested in satisfy
\beq\label{O(x)}
\matel{\bol{x}'_1, \dots, \bol{x}'_A}{\bar{\Op}(\bol{x})}{\bol{x}_1, \dots, \bol{x}_A} \propto \prod_{i = 1}^A \delta^{(3)}(\bol{x}_i - \bol{x}'_i) \ ,
\eeq
we can write
\begin{align}
\label{O(x) def}
\matel{T'}{\bar{\Op}(\bol{x})}{T} &\equiv \int \ud^3 x_1 \, \cdots \, \ud^3 x_A \, \braket{T'}{\bol{x}_1, \dots, \bol{x}_A} \, O(\bol{x}) \, \braket{\bol{x}_1, \dots, \bol{x}_A}{T} \ ,
\\
\matel{T'_\text{int}}{\bar{\Op}(\bol{x})}{T_\text{int}} &= \int \ud^3 \bar{x}_1 \, \cdots \, \ud^3 \bar{x}_{A - 1} \, \braket{T'_\text{int}}{\bar{\bol{x}}_1, \dots, (\bar{\bol{x}}_A)} \, O(\bol{x}) \, \braket{\bar{\bol{x}}_1, \dots, (\bar{\bol{x}}_A)}{T_\text{int}} \ ,
\end{align}
where $O(\bol{x})$ is essentially the `proportionality factor' in \Eq{O(x)}. To obtain the first result we inserted $\unop_\text{nucleus}$ twice in $\matel{T'}{\bar{\Op}(\bol{x})}{T}$, see \Eq{uno_target}, and integrated away the $\delta^{(3)}(\bol{x}_i - \bol{x}'_i)$'s in \Eq{O(x)}. For the second result we inserted $\unop_\text{int}$ twice in $\matel{T'_\text{int}}{\bar{\Op}(\bol{x})}{T_\text{int}}$, see \Eq{uno_int}, and then used \Eq{r norm} to write
\beq
\matel{\bol{x}'_1, \dots, \bol{x}'_A}{\bar{\Op}(\bol{x})}{\bol{x}_1, \dots, \bol{x}_A} = \matel{\bar{\bol{x}}'_1, \dots, (\bar{\bol{x}}'_A)}{\bar{\Op}(\bol{x})}{\bar{\bol{x}}_1, \dots, (\bar{\bol{x}}_A)} \, \frac{1}{A^3} \delta^{(3)}(\bol{r} - \bol{r}') \ ,
\eeq
which by Eqs.~\eqref{deltas},~\eqref{O(x)} implies
\beq
\matel{\bar{\bol{x}}'_1, \dots, (\bar{\bol{x}}'_A)}{\bar{\Op}(\bol{x})}{\bar{\bol{x}}_1, \dots, (\bar{\bol{x}}_A)} \propto \prod_{i = 1}^{A - 1} \delta^{(3)}(\bar{\bol{x}}_i - \bar{\bol{x}}'_i) \ ,
\eeq
again with $O(\bol{x})$ the proportionality factor. As an example, $\bar{\Op}(\bol{x})$ could correspond to the nucleon number-density operator for point-like nucleons $O(\bol{x}) = \sum_i \delta^{(3)}(\bol{x} - \bol{x}_i) = \sum_i \delta^{(3)}(\bar{\bol{x}} - \bar{\bol{x}}_i)$. We will see below that the operators we are interested in all involve a similar sum of delta functions.

We can now express the internal nuclear matrix element in terms of single-nucleon matrix elements, see \Sec{single-nucleon matel}. Taking our case example, the interaction described in \Eq{NR EFT Op}, we have the single-nucleon matrix element
\begin{multline}
\label{<y'|O|y>}
\matel{\bol{x}'_i}{\bar{\Op}(\bol{x})}{\bol{x}_i} = \int \frac{\ud^3 k_i}{(2 \pi)^3} \, \frac{\ud^3 k'_i}{(2 \pi)^3} \, e^{i \, \bol{k}'_i \cdot (\bol{x}'_i - \bol{x})} \frac{\matel{\bol{k}'_i}{\bar{\Op}(\bol{0})}{\bol{k}_i}}{\sqrt{\rho(k'_i) \rho(k_i)}} e^{- i \, \bol{k}_i \cdot (\bol{x}_i - \bol{x})}
\\
\NReq \frac{i}{2 \mN} \int \frac{\ud^3 k_i}{(2 \pi)^3} \, \frac{\ud^3 k'_i}{(2 \pi)^3} \, e^{i \, \bol{k}'_i \cdot (\bol{x}'_i - \bol{x})} \, e^{- i \, \bol{k}_i \cdot (\bol{x}_i - \bol{x})} \, \langle \Op_\NR^{N_i} \rangle \ ,
\end{multline}
where in the first equality we inserted twice the momentum identity operator~\eqref{unit op} and used Eqs.~\eqref{NR free wave function},~\eqref{psi(0) psi(x)}, and in the last equality we assumed the $S$ matrix to be truncated at first order in the perturbative expansion. We omitted spin degrees of freedom for simplicity. To be concrete, let us take as a specific example $\Op_\NR^N$ to be linear in $\bol{v}^\perp_N$, \eg $\Op_\NR^N = \Op^N_8 = \bol{s}_\chi \cdot \bol{v}^\perp_N$ for spin-$1/2$ DM. Given \Eq{vperp_N}, we can separate $\bar{\Op}(\bol{x})$ into two terms, one depending on $\bol{v}_\DM + \bol{v}'_\DM$ and the other depending on $\bol{v}_N + \bol{v}'_N$. For the first term, the NR matrix element in \Eq{<y'|O|y>} does not depend on the nucleon momenta and the $\ud^3 k_i$, $\ud^3 k'_i$ integrals return position delta functions: in fact, it suffices to consider for a nucleon of type $N$ at position $\bol{x}_i$
\beq
\matel{\bol{x}'_i}{\bar{N}_i(\bol{x}) \gamma^0 N_i(\bol{x})}{\bol{x}_i} \NReq \int \frac{\ud^3 k_i}{(2 \pi)^3} \, \frac{\ud^3 k'_i}{(2 \pi)^3} \, e^{i \, \bol{k}'_i \cdot (\bol{x}'_i - \bol{x})} \, e^{- i \, \bol{k}_i \cdot (\bol{x}_i - \bol{x})} = \delta^{(3)}(\bol{x} - \bol{x}'_i) \, \delta^{(3)}(\bol{x} - \bol{x}_i) \ ,
\eeq
where in the last equality we employed \Eq{delta function integral rep}. For the second term we have
\begin{multline}
\matel{\bol{x}'_i}{(\bol{v}_{N_i} + \bol{v}'_{N_i}) \, \bar{N}_i(\bol{x}) \gamma^0 N_i(\bol{x})}{\bol{x}_i} \NReq
\\
\int \frac{\ud^3 k_i}{(2 \pi)^3} \, \frac{\ud^3 k'_i}{(2 \pi)^3} \, e^{i \, \bol{k}'_i \cdot (\bol{x}'_i - \bol{x})} \left( - i \frac{\overleftarrow{\bol{\nabla}}_{\bol{x}'_i} - \overrightarrow{\bol{\nabla}}_{\bol{x}_i}}{2 \mN} \right) e^{- i \, \bol{k}_i \cdot (\bol{x}_i - \bol{x})}
= \delta^{(3)}(\bol{x} - \bol{x}'_i) \left( - i \frac{\overleftarrow{\bol{\nabla}}_{\bol{x}'_i} - \overrightarrow{\bol{\nabla}}_{\bol{x}_i}}{2 \mN} \right) \delta^{(3)}(\bol{x} - \bol{x}_i) \ ,
\end{multline}
so that putting all together we get
\beq\label{O(x) for O^NR_8}
\matel{\bol{x}'_i}{\bar{\Op}(\bol{x})}{\bol{x}_i} \NReq \frac{i}{4 \mN} \, \delta^{(3)}(\bol{x} - \bol{x}'_i) \, \bol{s}_\chi \cdot \left( \bol{v}_\DM + \bol{v}'_\DM + i \frac{\overleftarrow{\bol{\nabla}}_{\bol{x}'_i} - \overrightarrow{\bol{\nabla}}_{\bol{x}_i}}{2 \mN} \right) \delta^{(3)}(\bol{x} - \bol{x}_i) \ .
\eeq
Exploiting the fact that the matrix element, upon integration by parts if necessary, is in general proportional to $\delta^{(3)}(\bol{x} - \bol{x}'_i) \, \delta^{(3)}(\bol{x} - \bol{x}_i) = \delta^{(3)}(\bol{x}_i - \bol{x}'_i) \, \delta^{(3)}(\bol{x} - \bol{x}_i)$, we can write a generic single-nucleon matrix element for the $i^\text{th}$ nucleon as
\beq
\matel{\alpha'_i}{\bar{\Op}(\bol{x})}{\alpha_i} = \int \ud^3 x_i \, \ud^3 x'_i \, \braket{\alpha'_i}{\bol{x}'_i} \matel{\bol{x}'_i}{\bar{\Op}(\bol{x})}{\bol{x}_i} \braket{\bol{x}_i}{\alpha_i} \equiv \int \ud^3 x_i \, \braket{\alpha'_i}{\bol{x}_i} \, O_i(\bol{x}) \, \braket{\bol{x}_i}{\alpha_i} \ ,
\eeq
where in the first equality we inserted twice the position identity operator~\eqref{unit op}. $O_i(\bol{x})$ is essentially $\matel{\bol{x}'_i}{\bar{\Op}(\bol{x})}{\bol{x}_i}$ upon integrating away the $\delta^{(3)}(\bol{x}_i - \bol{x}'_i)$, or, in other words, it is the same as $O(\bol{x})$ defined in \Eq{O(x) def} but for the single, $i^\text{th}$ nucleon: for instance, if $O(\bol{x}) = \sum_i \delta^{(3)}(\bol{x} - \bol{x}_i)$, then $O_i(\bol{x}) = \delta^{(3)}(\bol{x} - \bol{x}_i)$. \Eq{O^(1)} then allows to write for one-body operators
\beq
\bar{\Op}(\bol{x}) = \sum_i \sum_{\alpha_i, \alpha'_i} c_{\alpha'_i}^\dagger c_{\alpha_i} \int \ud^3 x_i \, \braket{\alpha'_i}{\bol{x}_i} \, O_i(\bol{x}) \, \braket{\bol{x}_i}{\alpha_i} \ .
\eeq
In the above example with $\Op_\NR^N = \Op^N_8$ we obtain, integrating \Eq{O(x) for O^NR_8} by parts where necessary,
\beq\label{O_i example}
O_i(\bol{x}) \NReq \frac{i}{4 \mN} \, \bol{s}_\chi \cdot \left[ (\bol{v}_\DM + \bol{v}'_\DM) \delta^{(3)}(\bol{x} - \bol{x}_i) - \frac{i}{2 \mN} \left( \overleftarrow{\bol{\nabla}}_{\bol{x}_i} \delta^{(3)}(\bol{x} - \bol{x}_i) - \delta^{(3)}(\bol{x} - \bol{x}_i) \overrightarrow{\bol{\nabla}}_{\bol{x}_i} \right) \right] .
\eeq
Extending our example to
\beq
\Op_\NR^N = c_1^N \Op^N_1 + c_4^N \Op^N_4 + c_7^N \Op^N_7 + c_8^N \Op^N_8 + c_{10}^N \Op^N_{10} + c_{11}^N \Op^N_{11} \ ,
\eeq
with the $c_i^N$'s numerical coefficients, we get
\begin{multline}
O(\bol{x}) \NReq \frac{i}{2 \mN} \sum_{N = p, n} \left[ \left( c_1^N + i c_{11}^N \bol{s}_\chi \cdot \bol{q} \right) \sum_i \delta^{(3)}(\bol{x} - \bol{x}_i) + \left( c_4^N \bol{s}_\chi + i c_{10}^N \bol{q} \right) \cdot \sum_i \frac{\bol{\sigma}_i}{2} \delta^{(3)}(\bol{x} - \bol{x}_i) \vphantom{\frac{c_7^N}{2} \frac{c_8^N}{2} \left[ \frac{i}{2 \mN} \right]} \right.
\\
+ \frac{c_7^N}{2} \sum_i \left[ (\bol{v}_\DM + \bol{v}'_\DM) \cdot \frac{\bol{\sigma}_i}{2} \delta^{(3)}(\bol{x} - \bol{x}_i) - \frac{i}{2 \mN} \left( \overleftarrow{\bol{\nabla}}_{\bol{x}_i} \cdot \frac{\bol{\sigma}_i}{2} \delta^{(3)}(\bol{x} - \bol{x}_i) - \delta^{(3)}(\bol{x} - \bol{x}_i) \frac{\bol{\sigma}_i}{2} \cdot \overrightarrow{\bol{\nabla}}_{\bol{x}_i} \right) \right]
\\
\left. + \frac{c_8^N}{2} \bol{s}_\chi \cdot \sum_i \left[ (\bol{v}_\DM + \bol{v}'_\DM) \delta^{(3)}(\bol{x} - \bol{x}_i) - \frac{i}{2 \mN} \left( \overleftarrow{\bol{\nabla}}_{\bol{x}_i} \delta^{(3)}(\bol{x} - \bol{x}_i) - \delta^{(3)}(\bol{x} - \bol{x}_i) \overrightarrow{\bol{\nabla}}_{\bol{x}_i} \right) \right] \vphantom{\frac{c_7^N}{2} \left[ \frac{i}{2 \mN} \right]} \right] ,
\end{multline}
with the index $i$ running over all nucleons of type $N$, and $\bol{\sigma}_i$ the Pauli matrices acting on the spin of the $i^\text{th}$ nucleon.

As another example we can adapt this formalism to considering the electromagnetic current-density operator $J_\text{EM}^\mu(\fvec{x})$, through which nucleons couple to photons according to the first term in the Lagrangian in \Eq{JA}. The matrix element of $J_\text{EM}^\mu(\fvec{0})$ between single-nucleon states of definite linear momentum is given in \Eq{<J_EM>}, whose NR expression is provided in \Eq{<J_EM> NR}. We can then identify, in the language of \Eq{NR EFT Op}, the operators
\begin{align}
\label{JEM NR 1}
J_\text{EM}^0(\bol{x}) \NReq \rho(\bol{x}) \ ,
&&&
\bol{J}_\text{EM}(\bol{x}) \NReq \bol{J}_\text{c}(\bol{x}) + \bol{\nabla} \times \bol{\mu}(\bol{x}) \ ,
\end{align}
with
\begin{subequations}
\label{JEM NR 2}
\begin{align}
\rho(\bol{x}) &= \sum_{N = p, n} Q_N \sum_i \bar{N}_i(\bol{x}) \gamma^0 N_i(\bol{x}) \ ,
\\
\bol{J}_\text{c}(\bol{x}) &= \frac{1}{2 \mN} \sum_{N = p, n} Q_N \sum_i \bol{K}_i \, \bar{N}_i(\bol{x}) \gamma^0 N_i(\bol{x}) \ ,
\\
\bol{\mu}(\bol{x}) &= \frac{1}{2 \mN} \sum_{N = p, n} g_N \sum_i \bol{s}_{N_i} \, \bar{N}_i(\bol{x}) \gamma^0 N_i(\bol{x}) \ ,
\end{align}
\end{subequations}
where again the index $i$ runs over all nucleons of type $N$. $\rho$ is the electric-charge density operator, $\bol{J}_\text{c}$ is the convection-current density operator due to the motion of charged nucleons inside the nucleus, and $\bol{\mu}$ is the intrinsic-magnetization density operator due to the nucleon magnetic moments, all in units of the electric-charge unit $e$. By taking position-eigenstate matrix elements we can then identify the `equivalent' of $O(\bol{x})$ for the different components in the NR limit,
\begin{align}
\rho(\bol{x}) &\longrightarrow \sum_{N = p, n} Q_N \sum_i \delta^{(3)}(\bol{x} - \bol{x}_i) \ ,
\\
\bol{J}_\text{c}(\bol{x}) &\longrightarrow \frac{i}{2 \mN} \sum_{N = p, n} Q_N \sum_i \left( \overleftarrow{\bol{\nabla}}_{\bol{x}_i} \delta^{(3)}(\bol{x} - \bol{x}_i) - \delta^{(3)}(\bol{x} - \bol{x}_i) \overrightarrow{\bol{\nabla}}_{\bol{x}_i} \right) ,
\\
\bol{\mu}(\bol{x}) &\longrightarrow \frac{1}{2 \mN} \sum_{N = p, n} g_N \sum_i \frac{\bol{\sigma}_i}{2} \delta^{(3)}(\bol{x} - \bol{x}_i) \ .
\end{align}
Finally, we can derive the following known results for a single nucleon of type $N$ with wave function $\varphi(\bol{x}_i)$:
\beq
\matel{\varphi}{\rho(\bol{x})}{\varphi} \NReq Q_N \int \ud^3 x_i \, \varphi(\bol{x}_i)^* \, \delta^{(3)}(\bol{x} - \bol{x}_i) \, \varphi(\bol{x}_i) = Q_N \, |\varphi(\bol{x})|^2 \ ,
\eeq
\begin{multline}
\matel{\varphi}{\bol{J}_\text{c}(\bol{x})}{\varphi} \NReq \frac{i}{2 \mN} Q_N \int \ud^3 x_i \, \varphi(\bol{x}_i)^* \left( \overleftarrow{\bol{\nabla}}_{\bol{x}_i} \delta^{(3)}(\bol{x} - \bol{x}_i) - \delta^{(3)}(\bol{x} - \bol{x}_i) \overrightarrow{\bol{\nabla}}_{\bol{x}_i} \right) \varphi(\bol{x}_i)
\\
= - \frac{i}{2 \mN} Q_N \, \varphi(\bol{x})^* \overleftrightarrow{\bol{\nabla}} \varphi(\bol{x}) \ .
\end{multline}

\subsection{Nuclear form factors}
\label{nuc form factors}
We now take the nucleus to be in an eigenstate of total angular momentum, $\ket{T^{(\prime)}_\text{int}} = \ket{\amom{J}^{(\prime)}, \amom{M}^{(\prime)}}$. We also recall from \Sec{single-nucleon matel} that an operator $\widetilde{\Op}(\bar{\bol{x}})$ can be expressed in the NR limit as a sum of nucleon-specific terms of the form $K^N \CMcal{O}^N(\bar{\bol{x}})$, where, schematically, $\CMcal{O}^N(\bar{\bol{x}})$ indicates a scalar or vector operator depending on the internal nuclear degrees of freedom, and $K^N$ indicates a c-number quantity, related to a scalar or vector operator $\CMcal{K}^N$, depending on all other variables (see \Sec{Multipoles} for a more concrete discussion). As anticipated in \Sec{single-nucleon matel}, we will see in \Eq{O_NR scalar-vector dec} below that scalar and vector operators are enough for our purposes, when restricting our attention to NR operators at most linear in $\bol{v}^\perp_N$. By decomposing $K^N$ into spherical components we can then write in the NR limit $\widetilde{\Op}(\bar{\bol{x}})$ as a sum of terms of the form $\sum_M K^N_{JM} \CMcal{O}^N_{JM}(\bar{\bol{x}})$, each with $J = 0, 1$ fixed (see \Eq{KO}). Considering only one of such generic terms for definiteness, we can write the NR limit of \Eq{Mel = Otilde exp} as
\begin{multline}
\label{Mel pre nuc FF}
\Mel \NReq - 2 m_T i \sum_{N = p, n} K^N \int \ud^3 \bar{x} \, \matel{\amom{J}', \amom{M}'}{\CMcal{O}^N(\bar{\bol{x}})}{\amom{J}, \amom{M}} \, e^{i \bol{q} \cdot \bar{\bol{x}}}
\\
= - 2 m_T i \sum_{N = p, n} \sum_M K^N_{JM} \int \ud^3 \bar{x} \, \matel{\amom{J}', \amom{M}'}{\CMcal{O}^N_{JM}(\bar{\bol{x}})}{\amom{J}, \amom{M}} \, e^{i \bol{q} \cdot \bar{\bol{x}}} \ .
\end{multline}

It is instructive to consider the limit in which the nucleus behaves as a point-like object, with no spatial extension nor internal structure, to understand how it, as a whole, responds to the interaction. In this limit, the matrix element in \Eq{Mel = Otilde exp} is proportional to $\delta^{(3)}(\bar{\bol{x}})$, which entitles us to neglect $e^{i \bol{q} \cdot \bar{\bol{x}}}$, \ie to set $q = 0$ in the exponential, as already discussed in \Sec{Rate preliminaries}. This is different from setting $q = 0$ everywhere, as $\widetilde{\Op}(\bar{\bol{x}})$ may depend itself on $q$ because of the properties of the DM-nucleon interaction (which does not depend on whether the nucleus is point-like or extended). If, for instance, $\bar{\Op}_\NR^N = (\bol{s}_\chi \cdot \bol{q}) (\bol{s}_N \cdot \bol{q})$ in \Eq{Obar(x)}, setting $q = 0$ would imply no interaction, rather than interaction with a point-like nucleon. The separation between the $q$ dependence inherent to the DM-nucleon interaction and that due to the spatial extension of the nucleus becomes perhaps clearer by looking at \Eq{Mel pre nuc FF}, where the former is encoded in the $K^N$ and $K^N_{JM}$ coefficients while the latter is due to $e^{i \bol{q} \cdot \bar{\bol{x}}}$. We can then write for a point-like nucleus (PLN)
\begin{multline}
\Mel_\text{PLN} \NReq - 2 m_T i \sum_{N = p, n} K^N \int \ud^3 \bar{x} \, \matel{\amom{J}', \amom{M}'}{\CMcal{O}^N(\bar{\bol{x}})}{\amom{J}, \amom{M}}
\\
= - 2 m_T i \sum_{N = p, n} \sum_M K^N_{JM} \int \ud^3 \bar{x} \, \matel{\amom{J}', \amom{M}'}{\CMcal{O}^N_{JM}(\bar{\bol{x}})}{\amom{J}, \amom{M}} \ .
\end{multline}
Since $\CMcal{O}^N(\bar{\bol{x}})$ does not transform as a simple scalar or vector under rotations, but as a scalar or vector field due to its position dependence, $\CMcal{O}^N_{JM}(\bar{\bol{x}})$ is not a spherical tensor operator such as those discussed in \Sec{single-nucleon matel}. However, its spatial integral is. We can therefore apply \Eq{Wigner-Eckart} to the matrix element of
\beq\label{T^N_JM}
\CMcal{T}^N_{JM} \equiv \frac{1}{\sqrt{4 \pi (2 J + 1)}} \int \ud^3 \bar{x} \, \CMcal{O}^N_{JM}(\bar{\bol{x}}) \ ,
\eeq
where the $1 / \sqrt{4 \pi (2 J + 1)}$ factor is introduced for later convenience, obtaining
\beq\label{M_PLN}
\Mel_\text{PLN} \NReq - 2 m_T i \sqrt{4 \pi (2 J + 1)} \sum_{N = p, n} \left( \sum_M \CG{\amom{J}, \amom{M}}{J, M}{\amom{J}', \amom{M}'} \, K^N_{JM} \right) \redmatel{\amom{J}'}{\CMcal{T}^N_J}{\amom{J}} \ .
\eeq
In the following, as anticipated in \Sec{single-nucleon matel}, we will have the chance to appreciate the convenience of expressing the scattering amplitude in terms of (reduced) matrix elements of spherical tensors. We will explain in Secs.~\ref{Multipoles} and~\ref{Nuclear matrix element} how to extend this analysis to finite $q$ values, \ie taking into account the finite size of the nucleus.

The squared amplitude averaged over initial spins and summed over final spins (\Eq{Mel^2} below), needed to obtain the unpolarized scattering cross section (see \Sec{sigma}), can be computed in the point-like nucleus limit with the help of the Clebsch-Gordan symmetry properties (see \eg Sec.~3.5 of Ref.~\cite{Edmonds}),
\begin{align}
\label{Clebsch-Gordan symm 1}
\CG{J_1, M_1}{J_2, M_2}{J_3, M_3} &= (-1)^{J_1 + J_2 - J_3} \CG{J_2, M_2}{J_1, M_1}{J_3, M_3} \ ,
\\
\label{Clebsch-Gordan symm 2}
\CG{J_1, M_1}{J_2, M_2}{J_3, M_3} &= (-1)^{J_1 - M_1} \sqrt{\frac{2 J_3 + 1}{2 J_2 + 1}} \CG{J_3, M_3}{J_1, - M_1}{J_2, M_2} \ ,
\end{align}
and of their orthogonality relation,
\beq\label{Clebsch-Gordan orthogonality}
\sum_{\amom{M}, \amom{M}'} \braket{J', M'}{\amom{J}', \amom{M}'; \amom{J}, \amom{M}} \CG{\amom{J}', \amom{M}'}{\amom{J}, \amom{M}}{J, M} = \delta_{J J'} \delta_{M M'} \delta(\amom{J}, \amom{J}', J) \ ,
\eeq
with
\beq
\delta(\amom{J}, \amom{J}', J) \equiv
\begin{cases}
1 & |\amom{J} - \amom{J}'| \leqslant J \leqslant \amom{J} + \amom{J}',
\\
0 & \text{otherwise}.
\end{cases}
\eeq
We then have
\begin{multline}
\overline{| \Mel_\text{PLN} |^2} \NReq \frac{1}{2 s_\DM + 1} \frac{1}{2 \amom{J} + 1} 16 \pi m_T^2 (2 J + 1)
\\
\times \sum_{s, s'} \sum_{\amom{M}, \amom{M}'} \left| \sum_M \CG{\amom{J}, \amom{M}}{J, M}{\amom{J}', \amom{M}'} \sum_N K^N_{JM} \redmatel{\amom{J}'}{\CMcal{T}^N_J}{\amom{J}} \right|^2
\\
= \frac{1}{2 s_\DM + 1} \frac{2 \amom{J}' + 1}{2 \amom{J} + 1} \, \delta(\amom{J}, \amom{J}', J) \, 16 \pi m_T^2 \sum_{N, N'} \left( \sum_{s, s'} \sum_M {K^N_{JM}}^* K^{N'}_{JM} \right) \redmatel{\amom{J}'}{\CMcal{T}^N_J}{\amom{J}}^* \redmatel{\amom{J}'}{\CMcal{T}^{N'}_J}{\amom{J}} \ ,
\end{multline}
with $s_\DM$ the DM spin and $s$, $s'$ the initial and final DM spin indices. For a scalar operator ($J = 0$) we have $\sum_M {K^N_{JM}}^* K^{N'}_{JM} = {K^N_{00}}^* K^{N'}_{00} = {K^N}^* K^{N'}$, while for a vector operator ($J = 1$) we have $\sum_M {K^N_{JM}}^* K^{N'}_{JM} = {\bol{K}^N}^* \cdot \bol{K}^{N'}$, see \Eq{scalarproduct}.

The full momentum-transfer dependence of the amplitude is often parametrized within nuclide- and operator-dependent nuclear form factors. Deferring a more general treatment to Secs.~\ref{Multipoles} and~\ref{Nuclear matrix element}, we can illustrate the nature of the nuclear form factors considering a simple example where the DM scatters elastically off a spin-$0$ nucleus ($\amom{J} = \amom{J}' = 0$). We also assume that $\CMcal{O}^N(\bar{\bol{x}})$ is a scalar operator, which is the case for instance of the SI interaction (see \Sec{SI interaction}). The nuclear form factor is sometimes introduced in the nuclear physics literature in the context of electron-nucleus Rutherford scattering (see \eg Refs.~\cite{Krane:1987ky, Povh}), which also occurs via a scalar operator ($\rho$ in \Eq{JEM NR 2}) in DM-nucleus scattering if the DM has a tiny electric charge (see \Sec{Electromagnetic interactions}). There is only one relevant nuclear matrix element, which can be parametrized as
\beq
\matel{0, 0}{\CMcal{O}^N(\bar{\bol{x}})}{0, 0} = \mathscr{N}_\CMcal{O}^N \varrho_\CMcal{O}^N(\bar{\bol{x}}) \ ,
\eeq
where $\varrho_\CMcal{O}^N$ may be intended as a sort of nuclear density and $\mathscr{N}_\CMcal{O}^N$ is a normalization constant defined below. We may then define the nuclear form factor $F_\CMcal{O}^N(\bol{q})$ as the Fourier transform of $\varrho_\CMcal{O}^N$,
\beq
F_\CMcal{O}^N(\bol{q}) \equiv \int \ud^3 \bar{x} \, \varrho_\CMcal{O}^N(\bar{\bol{x}}) \, e^{i \bol{q} \cdot \bar{\bol{x}}} \ ,
\eeq
so that \Eq{Mel pre nuc FF} reads
\beq
\Mel \NReq - 2 m_T i \sum_{N = p, n} K^N \mathscr{N}_\CMcal{O}^N F_\CMcal{O}^N(\bol{q}) \ .
\eeq
As already explained after \Eq{Mel = Otilde exp}, the rapid oscillations of the exponential suppress the form factor (and thus $\Mel$) for $q$ much larger than the inverse nuclear radius. Beside depending on the specific nuclide, $F_\CMcal{O}^N$ depends on the properties of either protons (for $N = p$) or neutrons (for $N = n$). For instance, in the example of Rutherford scattering, $\varrho_\CMcal{O}^N(\bar{\bol{x}})$ represents the electric-charge density of protons or neutrons within the nucleus (the latter essentially vanishing), while for the SI interaction it represents their number density (see \Sec{SI interaction}). We fix the normalization constant $\mathscr{N}_\CMcal{O}^N$ by setting
\beq\label{F normalization}
F_\CMcal{O}^N(\bol{0}) = \int \ud^3 \bar{x} \, \varrho_\CMcal{O}^N(\bar{\bol{x}}) = 1 \ ,
\eeq
so that $\mathscr{N}_\CMcal{O}^N = \int \ud^3 \bar{x} \, \matel{0, 0}{\CMcal{O}^N(\bar{\bol{x}})}{0, 0}$, unless
\beq\label{0 PLN}
\int \ud^3 \bar{x} \, \matel{0, 0}{\CMcal{O}^N(\bar{\bol{x}})}{0, 0} = 0 \ ,
\eeq
in which case we set $\mathscr{N}_\CMcal{O}^N = 1$ (implying $F_\CMcal{O}^N(\bol{0}) = 0$). Notice however that form factors are sometimes defined as to incorporate a multiplicative normalization factor and can be therefore normalized differently from here. For instance, in the example of Rutherford scattering, the form factor relative to interactions with nuclear protons may be found normalized to their total electric charge, $Z$ (in units of $e$), which here is factored within $\mathscr{N}_\CMcal{O}^N$. When \Eq{0 PLN} holds, unless $\matel{0, 0}{\CMcal{O}^N(\bar{\bol{x}})}{0, 0} = 0$, the leading $q^2$ dependence at small $q^2$ may be factored out of the form factor via Taylor expansion, so to make this dependence explicit while obtaining a finite value for $F_\CMcal{O}^N(\bol{0})$. As a consequence of our normalization choice we can write
\beq
\Mel_\text{PLN} \NReq - 2 m_T i \sum_{N = p, n} K^N \mathscr{N}_\CMcal{O}^N \ ,
\eeq
with the understanding that, if \Eq{0 PLN} holds, $\Mel_\text{PLN}$ does not represent the actual point-like nucleus amplitude (which vanishes), since its vanishing part in \Eq{0 PLN} has been factored within $F_\CMcal{O}^N$. It is apparent that the $\bol{q}$ dependence of the form factors provides information on the internal structure of the nucleus, while taking the form factors at $q = 0$ provides information on the properties of the nucleus as a whole.

For isotropic densities, $\varrho_\CMcal{O}^N(\bar{\bol{x}}) = \varrho_\CMcal{O}^N(\bar{x})$, the form factor can only depend on $q^2$ and we get
\beq
F_\CMcal{O}^N(q^2) = \int_0^\infty \ud \bar{x} \, 4 \pi \bar{x}^2 j_0(q \bar{x}) \varrho_\CMcal{O}^N(\bar{x}) \ ,
\eeq
where
\beq
j_0(q \bar{x}) = \frac{\sin(q \bar{x})}{q \bar{x}}
\eeq
is the order-$0$ spherical Bessel function of the first kind, and we used
\begin{multline}
\int \ud^3 \bar{x} \, \varrho_\CMcal{O}^N(\bar{x}) \, e^{\pm i \bol{q} \cdot \bar{\bol{x}}}
= \int_0^\infty \ud \bar{x} \, \bar{x}^2 \varrho_\CMcal{O}^N(\bar{x}) \int_0^{2 \pi} \ud\phi \int_{-1}^{+1} \ud\cos\theta \, e^{i q \bar{x} \cos\theta}
\\
= \frac{2 \pi}{i q} \int_0^\infty \ud \bar{x} \, \bar{x} \left( e^{i q \bar{x}} - e^{- i q \bar{x}} \right) \varrho_\CMcal{O}^N(\bar{x})
= \frac{4 \pi}{q} \int_0^\infty \ud \bar{x} \, \bar{x} \sin(q \bar{x}) \varrho_\CMcal{O}^N(\bar{x}) \ .
\end{multline}
If $\CMcal{O}^N$ is hermitian, $\varrho_\CMcal{O}^N(\bar{x})$ can be defined to be real, so that $F_\CMcal{O}^N(q^2)$ is also real. Taylor-expanding $\sin \alpha = \alpha - \alpha^3 / 3! + \Ord(\alpha^5)$, and defining the (mean square) \emph{radius} relative to the density $\varrho_\CMcal{O}^N$ as the average $\bar{x}^2$,
\beq
\langle \bar{x}_N^2 \rangle_\CMcal{O} \equiv \int \ud^3 \bar{x} \, \bar{x}^2 \varrho_\CMcal{O}^N(\bar{x}) = 4 \pi \int_0^\infty \ud \bar{x} \, \bar{x}^4 \varrho_\CMcal{O}^N(\bar{x}) \ ,
\eeq
we can finally write
\beq\label{Taylor expanded F}
F_\CMcal{O}^N(q^2) = F_\CMcal{O}^N(0) - \frac{1}{6} q^2 \langle \bar{x}_N^2 \rangle_\CMcal{O} + \Ord(q^4) \ .
\eeq

\subsection{Multipole expansion and nuclear responses}
\label{Multipoles}
In \Sec{nuc form factors} we exploited the fact that $\widetilde{\Op}(\bar{\bol{x}})$ in \Eq{Mel = Otilde exp} can be written in the NR limit as a sum of terms of the form $K^N \CMcal{O}^N(\bar{\bol{x}})$, each of which can be written as $\sum_M K^N_{JM} \CMcal{O}^N_{JM}(\bar{\bol{x}})$ (with $J$ fixed) by projecting the tensor $K^N$ onto a spherical-tensor basis. The advantage of this approach is that the integral of $\CMcal{O}^N_{JM}(\bar{\bol{x}})$ is a spherical tensor operator, an object which behaves as having definite angular momentum (see \Sec{single-nucleon matel}), which greatly simplifies the computation of the scattering amplitude in the limit of point-like nucleus. Such limit corresponds to setting $q = 0$ in the $e^{i \bol{q} \cdot \bar{\bol{x}}}$ exponential in \Eq{Mel pre nuc FF}, or in other words to disregarding the exponential altogether. Here we generalize this analysis to take into account the full $\bol{q}$ dependence of the amplitude, thus considering the effect of the finite size of the nucleus. This can be done by performing a multipole expansion, obtained expanding the exponential into a series of terms that give rise to spherical tensors, which again simplifies the treatment of the scattering amplitude. Before delving into the specifics of this expansion, however, we need to take a more concrete look at the nature of the $K^N \CMcal{O}^N(\bar{\bol{x}})$ terms.

For the (somewhat schematic) contact Lagrangian in \Eq{NR EFT Op}, $\widetilde{\Op}(\bar{\bol{x}})$ depends in the NR limit on $\bar{\Op}_\NR^N$, that is, $\Op_\NR^N$ evaluated over all degrees of freedom but those pertaining to the internal nuclear state (see \Eq{Obar(x)}). In other words, $\bar{\Op}_\NR^N$ has the same form of $\Op_\NR^N$ in \Eq{general NR Op}, but with $\bol{q}$ a c-number quantity, the dependence on the DM spin described by the matrices defined in \Eq{I s} rather than by operators, and the DM-nucleon transverse velocity $\bol{v}^\perp_N$ separated into its CM component, a c-number quantity, and its intrinsic component, an operator (see \Eq{v^perp_T}). To ease dealing with the latter separation, in the following we restrict our attention to the $\Op_\NR^N$ terms at most linear in $\bol{v}^\perp_N$ in \Eq{general NR Op}, that is, we take
\beq\label{f(q^2) ONR^N}
\Op_\NR^N = \sum_i f_i^N(q^2) \, \Op^N_i \ ,
\eeq
limiting the sum to the NR building blocks up to and including $\Op^N_{15}$. This can be then parametrized as
\beq\label{O_NR scalar-vector dec}
\Op_\NR^N = \CMcal{S}^N_1(\bol{q}, \bol{s}_\chi) \, \unop_N + \CMcal{S}^N_2(\bol{q}, \bol{s}_\chi) \, \bol{v}^\perp_N \cdot \bol{s}_N + \bol{\CMcal{V}}^N_1(\bol{q}, \bol{s}_\chi) \cdot \bol{v}^\perp_N + \bol{\CMcal{V}}^N_2(\bol{q}, \bol{s}_\chi) \cdot \bol{s}_N + \bol{\CMcal{V}}^N_3(\bol{q}, \bol{s}_\chi) \cdot (\bol{v}^\perp_N \times \bol{s}_N) \ ,
\eeq
with the $\CMcal{S}^N_i$'s and $\bol{\CMcal{V}}^N_i$'s scalar and vector operators, respectively, and $\unop_N$ the identity operator over nucleons of type $N$; notice that $\Op^N_{13}$ can be cast in terms of a $\bol{\CMcal{V}}^N_3$-type term (plus a $\bol{\CMcal{V}}^N_2$-type term for inelastic scattering) by using \Eq{contractions3}:
\beq\label{ONR_13 recast}
\Op^N_{13} \NReq i (\bol{s}_\chi \times \bol{q}) \cdot (\bol{v}^\perp_N \times \bol{s}_N) + i \delta \, \Op^N_4 \ .
\eeq
Being at most linear in $\bol{s}_\chi$ (see \Sec{NR operators}), the $\CMcal{S}^N_i$ scalars are linear combinations of $\unop_\DM$ (the identity operator over DM states) and $\bol{q} \cdot \bol{s}_\chi$, while the $\bol{\CMcal{V}}^N_i$ vectors are linear combinations of $\bol{q}$, $\bol{s}_\chi$, $\bol{q} \times \bol{s}_\chi$, and $(\bol{q} \cdot \bol{s}_\chi) \, \bol{q}$, in both cases the coefficients being in principle $q^2$-dependent. Comparison with \Eq{f(q^2) ONR^N} returns (see the list of NR building blocks in \Eq{NR building blocks})
\begin{subequations}
\label{S's and V's}
\begin{align}
\CMcal{S}^N_1 &= f_1^N \, \unop_\DM + i f_{11}^N \, (\bol{s}_\chi \cdot \bol{q}) \ ,
\\
\CMcal{S}^N_2 &= f_7^N \, \unop_\DM + i f_{14}^N \, (\bol{s}_\chi \cdot \bol{q}) \ ,
\\
\bol{\CMcal{V}}^N_1 &= i f_5^N \, (\bol{s}_\chi \times \bol{q}) + f_8^N \, \bol{s}_\chi \ ,
\\
\bol{\CMcal{V}}^N_2 &= \left( f_4^N + i \delta f_{13}^N \right) \bol{s}_\chi + f_6^N \, (\bol{s}_\chi \cdot \bol{q}) \, \bol{q} + i f_9^N \, (\bol{q} \times \bol{s}_\chi) + i f_{10}^N \, \bol{q} \ ,
\\
\bol{\CMcal{V}}^N_3 &= i f_3^N \, \bol{q} - f_{12}^N \, \bol{s}_\chi + i f_{13}^N \, (\bol{s}_\chi \times \bol{q}) + f_{15}^N \, (\bol{s}_\chi \cdot \bol{q}) \, \bol{q} \ .
\end{align}
\end{subequations}
For the $\bol{s}_N$ and $\bol{v}^\perp_N \times \bol{s}_N$ terms, a further subdivision can be operated between the components of $\bol{O} = \bol{s}_N, \bol{v}^\perp_N \times \bol{s}_N$ along $\hat{\bol{q}}$, namely the longitudinal components $\bol{O}^\parallel = (\bol{O} \cdot \hat{\bol{q}}) \, \hat{\bol{q}}$, and the transverse components, $\bol{O}^\perp = \bol{O} - \bol{O}^\parallel$. By evaluating \Eq{O_NR scalar-vector dec} over the DM degrees of freedom and those pertaining to the overall nuclear motion we obtain
\beq\label{O_NR scalar-vector dec 2}
\bar{\Op}_\NR^N = S^N_1(\bol{q}, \bol{s}_\chi) \, \unop_N + S^N_2(\bol{q}, \bol{s}_\chi) \, \bol{v}^\perp_N \cdot \bol{s}_N + \bol{V}^N_1(\bol{q}, \bol{s}_\chi) \cdot \bol{v}^\perp_N + \bol{V}^N_2(\bol{q}, \bol{s}_\chi) \cdot \bol{s}_N + \bol{V}^N_3(\bol{q}, \bol{s}_\chi) \cdot (\bol{v}^\perp_N \times \bol{s}_N) \ ,
\eeq
with $S^N_i = \langle \CMcal{S}^N_i \rangle$ and $\bol{V}^N_i = \langle \bol{\CMcal{V}}^N_i \rangle$ matrices over DM-spin space, and the $\bol{v}^\perp_N$ factors implicitly separated into their operator, intrinsic component and their c-number, CM component. Summing over all nucleons and using Eqs.~\eqref{Obar(x)},~\eqref{v^perp_T} we see that $\widetilde{\Op}(\bar{\bol{x}})$ depends on $\bol{v}^\perp_T$ through the NR combination
\beq
\sum_{N = p, n} \tilde{S}^N_1(\bol{q}, \bol{s}_\chi, \bol{v}^\perp_T) \sum_i \bar{N}_i \gamma^0 N_i + \sum_{N = p, n} \tilde{\bol{V}}^N_2(\bol{q}, \bol{s}_\chi, \bol{v}^\perp_T) \cdot \sum_i \bol{s}_{N_i} \, \bar{N}_i \gamma^0 N_i \ ,
\eeq
with
\begin{align}
\tilde{S}^N_1 = \frac{i}{2 \mN} \left( S^N_1 + \bol{V}^N_1 \cdot \bol{v}^\perp_T \right) ,
&&&
\tilde{\bol{V}}^N_2 = \frac{i}{2 \mN} \left( \bol{V}^N_2 + S^N_2 \, \bol{v}^\perp_T + \bol{V}^N_3 \times \bol{v}^\perp_T \right) .
\end{align}
More completely, defining
\begin{align}
\tilde{S}^N_2 \equiv \frac{i}{2 \mN} S^N_2 \ ,
&&
\tilde{\bol{V}}^N_1 \equiv \frac{i}{2 \mN} \bol{V}^N_1 \ ,
&&
\tilde{\bol{V}}^N_3 \equiv \frac{i}{2 \mN} \bol{V}^N_3 \ ,
\end{align}
we can write
\begin{multline}
\label{Obar scalar-vector dec}
\bar{\Op}(\bol{x}) \NReq \sum_{N, i} \tilde{S}^N_1 \, \bar{N}_i \gamma^0 N_i + \left[ \sum_{N, i} \tilde{S}^N_2 \left( \bol{v}^\perp_{N_i} \cdot \bol{s}_{N_i} \right) \bar{N}_i \gamma^0 N_i \right]_\text{intr} + \left[ \sum_{N, i} \tilde{\bol{V}}^N_1 \cdot \bol{v}^\perp_{N_i} \, \bar{N}_i \gamma^0 N_i \right]_\text{intr}
\\
+ \sum_{N, i} \tilde{\bol{V}}^N_2 \cdot \bol{s}_{N_i} \, \bar{N}_i \gamma^0 N_i + \left[ \sum_{N, i} \tilde{\bol{V}}^N_3 \cdot \left( \bol{v}^\perp_{N_i} \times \bol{s}_{N_i} \right) \bar{N}_i \gamma^0 N_i \right]_\text{intr} \ ,
\end{multline}
where we indicated within square brackets the intrinsic components given by the last term in \Eq{v^perp_T}. Therefore, every $\bol{V}^N_1$ term in \Eq{O_NR scalar-vector dec 2} (\eg $\bol{s}_\chi \cdot \bol{v}^\perp_N$) induces in \Eq{Obar scalar-vector dec} a $\tilde{\bol{V}}^N_1$ term with the intrinsic component of $\bol{v}^\perp_N$ as well as a $\tilde{S}^N_1$ term with its CM component. Analogously, every $S^N_2$ (\eg $q^2 \, \bol{v}^\perp_N \cdot \bol{s}_N$) and $\bol{V}^N_3$ (\eg $i \, \bol{q} \cdot (\bol{v}^\perp_N \times \bol{s}_N)$) terms induce respectively a $\tilde{S}^N_2$ and $\tilde{\bol{V}}^N_3$ term with the $\bol{v}^\perp_N$ intrinsic component, and a $\tilde{\bol{V}}^N_2$ term with the $\bol{v}^\perp_T$ component.

\Eq{Obar scalar-vector dec} shows that, when expressing $\widetilde{\Op}(\bar{\bol{x}})$ as a sum of terms of the form $K^N \CMcal{O}^N(\bar{\bol{x}})$ in the NR limit, the $K^N$ factors can be identified with $\tilde{S}^N_1$, $\tilde{S}^N_2$, $\tilde{\bol{V}}^N_1$, $\tilde{\bol{V}}^N_2$, and $\tilde{\bol{V}}^N_3$, so that the $\CMcal{O}^N(\bar{\bol{x}})$ operators are essentially related to $\unop_N$, $\bol{v}^\perp_N \cdot \bol{s}_N$, $\bol{v}^\perp_N$, $\bol{s}_N$, and $\bol{v}^\perp_N \times \bol{s}_N$, respectively. For the purposes of studying the multipole decomposition for scalar and vector operators, in the following we take for simplicity the representative interaction
\beq\label{representative Op}
\widetilde{\Op}(\bar{\bol{x}}) \NReq \tilde{S} \CMcal{O}_\text{s}(\bar{\bol{x}}) + \tilde{\bol{V}} \cdot \bol{\CMcal{O}}_\text{v}(\bar{\bol{x}}) \ ,
\eeq
which is understood to hold for one scalar operator $\CMcal{O}_\text{s}(\bar{\bol{x}})$ plus one vector operator $\bol{\CMcal{O}}_\text{v}(\bar{\bol{x}})$, and for a single nucleon type (either proton or neutron); generalization to more operators and to both nucleon types is straightforward (see \Sec{Nuclear matrix element}). We also restrict ourselves to elastic scattering (thus setting $\delta = 0$), and to the standard assumption the nucleus does not get excited in the scattering: while many of our formulas will be derived for a generic $\amom{J}'$, we will only be interested in the consequences for $\amom{J}' = \amom{J}$. To make contact with Refs.~\cite{DeForest:1966ycn, Hughes:1975eg, Donnelly:1975ze, Donnelly:1978tz, Walecka}, where $\fvec{q}$ is defined as $\fvec{p}' - \fvec{p}$ rather than $\fvec{p} - \fvec{p}'$ as here, we shall denote in this Section $- \fvec{q}$ with $\fvec{b}$ (which conveniently resembles a rotated $\fvec{q}$):
\beq\label{b = - q}
(b^0, \bol{b})^\tr = \fvec{b}^\mu \equiv - \fvec{q}^\mu \ .
\eeq
As we will see later on, the definition of $\fvec{q}$ affects the very definition of the nuclear responses, and using $\fvec{b}$ will allow for a simpler comparison of our formulas with the rest of the literature. The NR scattering amplitude in \Eq{Mel pre nuc FF} can then, using \Eq{V spherical compo}, be seen to depend on the nuclear matrix element of
\beq\label{S Os + V Ov}
\int \ud^3 x \left( \tilde{S} \CMcal{O}_\text{s}(\bol{x}) + \bol{\CMcal{O}}_\text{v}(\bol{x}) \cdot \hat{\bol{e}}_0^* \tilde{V}_0 + \bol{\CMcal{O}}_\text{v}(\bol{x}) \cdot \sum_{\lambda = \pm 1} \hat{\bol{e}}_\lambda^* \tilde{V}_\lambda \right) e^{- i \bol{b} \cdot \bol{x}} \ ,
\eeq
where here and in the following we denote for simplicity $\bar{\bol{x}}$ with $\bol{x}$. For the time being we define our spherical-vector basis in \Eq{spherical vector basis} so that $\hat{\bol{e}}_0 = \hat{\bol{b}}$, and take this to be the angular-momentum quantization axis.

The multipole decomposition of the operator~\eqref{S Os + V Ov} can be obtained by using
\begin{align}
\label{exponential}
e^{i \bol{b} \cdot \bol{x}} &= \sum_{\ell = 0}^\infty i^\ell \sqrt{4 \pi (2 \ell + 1)} \, j_\ell(bx) Y_{\ell 0}(\Omega_{\bol{x}}) \ ,
\\
\label{exponentiale}
\hat{\bol{e}}_\lambda e^{i \bol{b} \cdot \bol{x}} &= \sum_{\ell = 0}^\infty \sum_{J = 0}^\infty i^\ell \sqrt{4 \pi (2 \ell + 1)} \, j_\ell(bx) \CG{\ell, 0}{1, \lambda}{J, \lambda} \bol{Y}_{J \ell 1}^\lambda(\Omega_{\bol{x}}) \ ,
\end{align}
where $j_\ell(bx)$ are the spherical Bessel functions of the first kind, $Y_{\ell m}(\Omega_{\bol{x}})$ are the spherical harmonics, $\Omega_{\bol{x}}$ is a shorthand notation for the angles describing the orientation of $\bol{x}$ relative to $\bol{b}$, and
\beq\label{bol Y}
\bol{Y}_{J \ell 1}^M(\Omega_{\bol{x}}) \equiv \sum_{m, \lambda} \CG{\ell, m}{1, \lambda}{J, M} Y_{\ell m}(\Omega_{\bol{x}}) \, \hat{\bol{e}}_\lambda
\eeq
are the vector spherical harmonics. The latter definition simply couples the spherical harmonics, with angular momentum $\ell$ and projection along the quantization axis $m$, to a spherical vector which has angular momentum $1$ and projection $\lambda$, to form an object with definite angular momentum $J$ and projection $M$. In complex-conjugating the above expressions it will be useful to know that
\begin{align}
\label{spherical harmonics*}
j_\ell(z)^* = j_\ell(z^*) \ ,
&&
Y_{\ell m}(\Omega_{\bol{x}})^* = (-1)^m \, Y_{\ell, -m}(\Omega_{\bol{x}}) \ ,
&&
\bol{Y}_{J J 1}^M(\Omega_{\bol{x}})^* = (-1)^{M+1} \, \bol{Y}_{J J 1}^{-M}(\Omega_{\bol{x}}) \ .
\end{align}
To prove \Eq{exponentiale}, we can simply invert \Eq{bol Y} by exploiting the orthogonality properties of the Clebsch-Gordan coefficients,
\beq
Y_{\ell m}(\Omega_{\bol{x}}) \, \hat{\bol{e}}_\lambda = \sum_{J, M} \CG{\ell, m}{1, \lambda}{J, M} \bol{Y}_{J \ell 1}^M(\Omega_{\bol{x}}) \ ,
\eeq
and then use \Eq{exponential}. The latter can be verified starting from the following integral representation of the spherical Bessel function~\cite{NIST DLMF},
\beq
j_\ell(z) = \frac{1}{2 i^\ell} \int_{-1}^{+1} e^{i z y} P_\ell(y) \, \ud y \ ,
\eeq
with $P_\ell(y)$ the Legendre polynomials. Multiplying by $P_\ell(y')$ and using the following series representation of the Dirac delta~\cite{NIST DLMF},
\beq
\delta(y - y') = \sum_{\ell = 0}^\infty (\ell + \tfrac{1}{2}) \, P_\ell(y) P_\ell(y') \ ,
\eeq
we get
\beq\label{exp P_ell}
e^{i z y'} = \sum_{\ell = 0}^\infty i^\ell (2 \ell + 1) \, j_\ell(z) P_\ell(y') \ .
\eeq
We can then use the addition theorem of the spherical harmonics,
\beq
\sum_{m = - \ell}^{+ \ell} Y^*_{\ell m}(\Omega_{\bol{b}}) Y_{\ell m}(\Omega_{\bol{x}}) = \frac{2 \ell + 1}{4 \pi} P_\ell(\hat{\bol{b}} \cdot \hat{\bol{x}}) \ ,
\eeq
and the fact that $\hat{\bol{b}}$ having being chosen as the angular-momentum quantization axis implies
\beq
Y_{\ell m}(\Omega_{\bol{b}}) = \sqrt{\frac{2 \ell + 1}{4 \pi}} \delta_{m 0} \ ,
\eeq
to recover \Eq{exponential} upon identification of $z = b x$ and $y' = \hat{\bol{b}} \cdot \hat{\bol{x}}$ in \Eq{exp P_ell}.

As we will see in the following, the multipole expansion for the different terms in \Eq{S Os + V Ov} can be expressed in terms of a number of \emph{nuclear responses}, defined in terms of
\begin{align}
\label{M'n'M}
M_{JM}(b \bol{x}) \equiv j_J(b x) Y_{JM}(\Omega_{\bol{x}}) \ ,
&&&
\bol{M}_{JL}^M(b \bol{x}) \equiv j_L(b x) \bol{Y}_{JL1}^M(\Omega_{\bol{x}}) \ .
\end{align}
Notice that, while only being functions of the product $b \bol{x}$, $M_{JM}$ and $\bol{M}_{JL}^M$ intrinsically also depend on $\hat{\bol{b}}$ through the (vector) spherical harmonics, whose very definition hinges on the direction of $\bol{b}$; in other words, $\hat{\bol{b}}$ determines what functions (of $b \bol{x}$) $M_{JM}$ and $\bol{M}_{JL}^M$ are. The nuclear responses of interest to direct DM detection, assuming the nucleus remains in the ground state during the scattering, are:
\begin{align}
M_{JM}(b \bol{x}) & \ ,
\\
\Delta_{JM}(b \bol{x}) &\equiv \bol{M}_{JJ}^M(b \bol{x}) \cdot \frac{1}{b} \bol{\nabla} \ ,
\\
\Sigma'_{JM}(b \bol{x}) &\equiv - i \left( \frac{1}{b} \bol{\nabla} \times \bol{M}_{JJ}^M(b \bol{x}) \right) \cdot \bol{\sigma} \ ,
\\
\Sigma''_{JM}(b \bol{x}) &\equiv \left( \frac{1}{b} \bol{\nabla} M_{JM}(b \bol{x}) \right) \cdot \bol{\sigma} \ ,
\\
\tilde{\Phi}'_{JM}(b \bol{x}) &\equiv \left( \frac{1}{b} \bol{\nabla} \times \bol{M}_{JJ}^M(b \bol{x}) \right) \cdot \left( \bol{\sigma} \times \frac{1}{b} \bol{\nabla} \right) + \frac{1}{2} \bol{M}_{JJ}^M(b \bol{x}) \cdot \bol{\sigma} \ ,
\\
\Phi''_{JM}(b \bol{x}) &\equiv i \left( \frac{1}{b} \bol{\nabla} M_{JM}(b \bol{x}) \right) \cdot \left( \bol{\sigma} \times \frac{1}{b} \bol{\nabla} \right) .
\intertext{\Tab{tab: nuclear responses} collects the nuclear responses induced for elastic scattering by each of the terms of the most general NR operator at most linear in $\bol{v}^\perp_N$, see \Eq{f(q^2) ONR^N}. For completeness we also list here the nuclear responses that only connect nuclear states with mutually different spatial or time-reversal parity, and thus do not contribute unless the nucleus gets excited (see \eg Refs.~\cite{Fitzpatrick:2012ix, Anand:2013yka}):}
\Delta'_{JM}(b \bol{x}) &\equiv - i \left( \frac{1}{b} \bol{\nabla} \times \bol{M}_{JJ}^M(b \bol{x}) \right) \cdot \frac{1}{b} \bol{\nabla} \ ,
\\
\Sigma_{JM}(b \bol{x}) &\equiv \bol{M}_{JJ}^M(b \bol{x}) \cdot \bol{\sigma} \ ,
\\
\tilde{\Omega}_{JM}(b \bol{x}) &\equiv \Omega_{JM}(b \bol{x}) + \frac{1}{2} \Sigma_{JM}''(b \bol{x}) = M_{JM}(b \bol{x}) \bol{\sigma} \cdot \frac{1}{b} \bol{\nabla} + \frac{1}{2} \left( \frac{1}{b} \bol{\nabla} M_{JM}(b \bol{x}) \right) \cdot \bol{\sigma} \ ,
\\
\tilde{\Phi}_{JM}(b \bol{x}) &\equiv \Phi_{JM}(b \bol{x}) - \frac{1}{2} \Sigma'_{JM}(b \bol{x}) = i \bol{M}_{JJ}^M(b \bol{x}) \cdot \left( \bol{\sigma} \times \frac{1}{b} \bol{\nabla} \right) + \frac{i}{2} \left( \frac{1}{b} \bol{\nabla} \times \bol{M}_{JJ}^M(b \bol{x}) \right) \cdot \bol{\sigma} \ ,
\\
\tilde{\Delta}''_{JM}(b \bol{x}) &\equiv \Delta''_{JM}(b \bol{x}) - \frac{1}{2} M_{JM}(b \bol{x}) = \left( \frac{1}{b} \bol{\nabla} M_{JM}(b \bol{x}) \right) \cdot \frac{1}{b} \bol{\nabla} - \frac{1}{2} M_{JM}(b \bol{x}) \ .
\end{align}

\begin{table}[t!]
\footnotesize
\begin{center}
\begin{tabular}{|>{\rule[-3mm]{0mm}{7.6mm}} l | c c c c | c c c c |}
\hline
\multicolumn{1}{| c |}{\raisebox{-2mm}{\multirow{2}{*}{NR building block}}} & \multicolumn{4}{c |}{CM \rule[-3mm]{0mm}{7.6mm}} & \multicolumn{4}{c |}{intrinsic}
\\
\cline{2-9}
& $\CMcal{O}$ & multipoles & responses & $J$ & $\CMcal{O}$ & multipoles & responses & $J$
\\
\hline
$\Op^N_1 = \unop$ & $\unop_N$ & $\mulop{M}$ & $M$ & $+$ & & & &
\\
$\Op^N_3 = i \, \bol{s}_N \cdot (\bol{q} \times \bol{v}^\perp_N)$ & $\bol{s}_N^\perp$ & $\mulop{T}^\text{el}$ & $\Sigma'$ & $-$ & $(\bol{v}^\perp_N \times \bol{s}_N)^\parallel$ & $\mulop{L}$ & $\Phi''$ & $+$
\\
$\Op^N_4 = \bol{s}_\chi \cdot \bol{s}_N$ & $\bol{s}_N$ & $\mulop{T}^\text{el}$, $\mulop{L}$ & $\Sigma'$, $\Sigma''$ & $-$ & & & &
\\
$\Op^N_5 = i \, \bol{s}_\chi \cdot (\bol{q} \times \bol{v}^\perp_N)$ & $\unop_N$ & $\mulop{M}$ & $M$ & $+$ & $\bol{v}^\perp_N$ & $\mulop{T}^\text{mag}$ & $\Delta$ & $-$
\\
$\Op^N_6 = (\bol{s}_\chi \cdot \bol{q}) (\bol{s}_N \cdot \bol{q})$ & $\bol{s}_N^\parallel$ & $\mulop{L}$ & $\Sigma''$ & $-$ & & & &
\\
$\Op^N_7 = \bol{s}_N \cdot \bol{v}^\perp_N$ & $\bol{s}_N^\perp$ & $\mulop{T}^\text{el}$ & $\Sigma'$ & $-$ & & & &
\\
$\Op^N_8 = \bol{s}_\chi \cdot \bol{v}^\perp_N$ & $\unop_N$ & $\mulop{M}$ & $M$ & $+$ & $\bol{v}^\perp_N$ & $\mulop{T}^\text{mag}$ & $\Delta$ & $-$
\\
$\Op^N_9 = i \, \bol{s}_\chi \cdot (\bol{s}_N \times \bol{q})$ & $\bol{s}_N^\perp$ & $\mulop{T}^\text{el}$ & $\Sigma'$ & $-$ & & & &
\\
$\Op^N_{10} = i \, \bol{s}_N \cdot \bol{q}$ & $\bol{s}_N^\parallel$ & $\mulop{L}$ & $\Sigma''$ & $-$ & & & &
\\
$\Op^N_{11} = i \, \bol{s}_\chi \cdot \bol{q}$ & $\unop_N$ & $\mulop{M}$ & $M$ & $+$ & & & &
\\
$\Op^N_{12} = \bol{v}^\perp_N \cdot (\bol{s}_\chi \times \bol{s}_N)$ & $\bol{s}_N$ & $\mulop{T}^\text{el}$, $\mulop{L}$ & $\Sigma'$, $\Sigma''$ & $-$ & $\bol{v}^\perp_N \times \bol{s}_N$ & $\mulop{T}^\text{el}$, $\mulop{L}$ & $\tilde{\Phi}'$, $\Phi''$ & $+$
\\
$\Op^N_{13} = i (\bol{s}_\chi \cdot \bol{v}^\perp_N) (\bol{s}_N \cdot \bol{q})$ & $\bol{s}_N^\parallel$ & $\mulop{L}$ & $\Sigma''$ & $-$ & $(\bol{v}^\perp_N \times \bol{s}_N)^\perp$ & $\mulop{T}^\text{el}$ & $\tilde{\Phi}'$ & $+$
\\
$\Op^N_{14} = i (\bol{s}_\chi \cdot \bol{q}) (\bol{s}_N \cdot \bol{v}^\perp_N)$ & $\bol{s}_N^\perp$ & $\mulop{T}^\text{el}$ & $\Sigma'$ & $-$ & & & &
\\
$\Op^N_{15} = (\bol{s}_\chi \cdot \bol{q}) [\bol{s}_N \cdot (\bol{q} \times \bol{v}^\perp_N)]$ & $\bol{s}_N^\perp$ & $\mulop{T}^\text{el}$ & $\Sigma'$ & $-$ & $(\bol{v}^\perp_N \times \bol{s}_N)^\parallel$ & $\mulop{L}$ & $\Phi''$ & $+$
\\
\hline
\end{tabular}
\caption[Multipoles and nuclear responses]{\label{tab: nuclear responses}\emph{Multipoles and nuclear responses generated by the NR building blocks up to \figmath{\Op^N_{15}}, for elastic scattering and in the assumption that the nucleus remains in the ground state. These building blocks are those entering the most general NR operator at most linear in \figmath{\bol{v}^\perp_N}, see \Eq{f(q^2) ONR^N}. For each building block, its part depending on the internal nuclear degrees of freedom is shown in the column marked with \figmath{\CMcal{O}}, where the \figmath{\parallel} and \figmath{\perp} symbols mean that only the component longitudinal or orthogonal to \figmath{\bol{q}} contributes, respectively. For operators depending on \figmath{\bol{v}^\perp_N}, both the CM component (depending on \figmath{\bol{v}^\perp_T}) and the intrinsic component are taken into account. Due to parity and time reversal constraints, the intrinsic component of \figmath{\bol{v}^\perp_N \cdot \bol{s}_N} does not contribute to the scattering unless the nucleus gets excited, see \Sec{PT selection rules}. Parity constraints select either \figmath{J} even (\figmath{+}) or odd (\figmath{-}), as indicated in the column marked with \figmath{J}. The intrinsic component of \figmath{\Op^N_{13}} can be easily determined by using \Eq{ONR_13 recast}.}}
\end{center}
\end{table}

We now proceed with performing the multipole expansion for the terms in \Eq{S Os + V Ov}. The final result, which will be further explored in \Sec{Nuclear matrix element}, is that the $\amom{J}' = \amom{J}$ scattering amplitude can be expressed as a combination of the above nuclear responses,
\begin{multline}
\Mel \NReq \sum_{N = p, n} \sum_{\substack{X = M, \Delta, \\ \Sigma', \Sigma'', \tilde{\Phi}', \Phi''}} c_X^N \, \matel{\amom{J}, \amom{M}'}{\int \ud^3 x \sum_i \delta^{(3)}(\bol{x} - \bol{x}_i) \, X_{JM}^N(b \bol{x})}{\amom{J}, \amom{M}}
\\
= \sum_{N = p, n} \sum_{\substack{X = M, \Delta, \\ \Sigma', \Sigma'', \tilde{\Phi}', \Phi''}} c_X^N \, \matel{\amom{J}, \amom{M}'}{\sum_i X_{JM}^N(b \bol{x}_i)}{\amom{J}, \amom{M}} \ ,
\end{multline}
where $i$ runs over all nucleons of type $N = p, n$ and we made the nuclear response dependence on the nucleon type explicit. The quantity
\beq
\sum_{\substack{X = M, \Delta, \\ \Sigma', \Sigma'', \tilde{\Phi}', \Phi''}} c_X^N \, \delta^{(3)}(\bol{x} - \bol{x}_i) \, X_{JM}^N(b \bol{x})
\eeq
plays here the role in the NR limit of $O_i(\bol{x})$ in \Sec{Scattering amplitude}, see example in \Eq{O_i example}.

\subsubsection{Multipole decomposition for a scalar operator}
Using the complex conjugate of \Eq{exponential} we get
\beq\label{Coulomb multipole expansion}
\int \ud^3 x \, \CMcal{O}_\text{s}(\bol{x}) \, e^{- i \bol{b} \cdot \bol{x}} = \sum_{J = 0}^\infty (-i)^J \sqrt{4 \pi (2 J + 1)} \, \mulop{M}_{J 0}(b) \ ,
\eeq
where we defined the \emph{charge density} or \emph{Coulomb multipole} operators
\beq
\mulop{M}_{JM}(b) \equiv \int \ud^3 x \, M_{JM}(b \bol{x}) \, \CMcal{O}_\text{s}(\bol{x}) \ ,
\eeq
also denoted $\mulop{C}_{JM}(b)$ in the literature (see \eg Ref.~\cite{Engel:1992bf}). Comparison with \Eq{Obar scalar-vector dec} shows that the integrand can take the form of the $M$ nuclear response if $\CMcal{O}_\text{s}(\bol{x})$ is an $\tilde{S}^N_1$ term. This happens for the $\Op^N_1$, $\Op^N_5$, $\Op^N_8$, and $\Op^N_{11}$ terms in \Eq{f(q^2) ONR^N}, see \Tab{tab: nuclear responses} (for $\Op^N_5$ and $\Op^N_8$, which feature $\bol{v}^\perp_N$, only the CM component contributes to the $M$ response). A common example of this nuclear response is provided, in the context of electromagnetic electron-nucleus interactions, by the nuclear electric-charge density $\CMcal{O}_\text{s}(\bol{x}) = \rho(\bol{x})$ in \Eq{JEM NR 2}, mediating the common electrostatic interaction described by the Coulomb potential which lends its name to the multipoles (see \Sec{examples and applications}). The intrinsic component of $\tilde{S}^N_2$ terms, namely the $\Op^N_7$ and $\Op^N_{14}$ terms in \Eq{f(q^2) ONR^N}, gives rise to the $\tilde{\Omega}$ response, which however does not contribute to the scattering if the nucleus remains in the ground state, due to parity and time-reversal constraints (see \Sec{PT selection rules}).

\subsubsection{Multipole decomposition for a vector operator: $\lambda = 0$}
To compute \Eq{exponentiale} with $\lambda = 0$ we need the following Clebsch-Gordan coefficients:
\beq
\sqrt{2 \ell + 1} \, \CG{\ell, 0}{1, 0}{J, 0} =
\begin{cases}
- \sqrt{J+1} & \ell = J+1,
\\
+ \sqrt{J} & \ell = J-1,
\\
0 & \text{otherwise},
\end{cases}
\eeq
so that we get
\beq
\hat{\bol{e}}_0 e^{i \bol{b} \cdot \bol{x}} = - i \sqrt{4 \pi} \sum_{J = 0}^\infty i^J \left( \sqrt{J + 1} \, j_{J + 1}(bx) \bol{Y}_{J, J + 1, 1}^0(\Omega_{\bol{x}}) + \sqrt{J} \, j_{J - 1}(bx) \bol{Y}_{J, J - 1, 1}^0(\Omega_{\bol{x}}) \right) .
\eeq
We now use the following properties of the spherical Bessel functions~\cite{DeForest:1966ycn, Hughes:1975eg, NIST DLMF},
\begin{align}
\label{Besselprop}
\left[ \frac{J}{z} - \frac{\ud}{\ud z} \right] j_J(z) = j_{J+1}(z) \ ,
&&&
\left[ \frac{J+1}{z} + \frac{\ud}{\ud z} \right] j_J(z) = j_{J-1}(z) \ ,
\end{align}
and of the vector spherical harmonics~\cite{DeForest:1966ycn, Hughes:1975eg, Edmonds},
\begin{multline}
\bol{\nabla} (\phi(x) Y_{JM}(\Omega_{\bol{x}})) =
\\
\sqrt{\frac{J+1}{2J+1}} \left[ \frac{J}{x} - \frac{\ud}{\ud x} \right] \phi(x) \bol{Y}_{J, J+1, 1}^M(\Omega_{\bol{x}}) + \sqrt{\frac{J}{2J+1}} \left[ \frac{J+1}{x} + \frac{\ud}{\ud x} \right] \phi(x) \bol{Y}_{J, J-1, 1}^M(\Omega_{\bol{x}}) \ ,
\end{multline}
with $\phi(x)$ an arbitrary function. We then have
\beq
\hat{\bol{e}}_0 e^{i \bol{b} \cdot \bol{x}} = - \frac{i}{b} \sum_{J = 0}^\infty i^J \sqrt{4 \pi (2J+1)} \, \bol{\nabla} ( j_J(bx) Y_{J0}(\Omega_{\bol{x}})) \ .
\eeq
Finally, complex-conjugating this expression we get
\beq\label{Longitudinal multipole expansion}
\int \ud^3 x \, \bol{\CMcal{O}}_\text{v}(\bol{x}) \cdot \hat{\bol{e}}_0^* \, e^{- i \bol{b} \cdot \bol{x}} = \sum_{J = 0}^\infty (-i)^J \sqrt{4 \pi (2 J + 1)} \, \mulop{L}_{J0}(b) \ ,
\eeq
where we defined the \emph{longitudinal multipoles}
\beq
\mulop{L}_{JM}(b) \equiv \frac{i}{b} \int \ud^3 x \, (\bol{\nabla} M_{JM}(b \bol{x})) \cdot \bol{\CMcal{O}}_\text{v}(\bol{x}) \ .
\eeq
The integrand of the longitudinal multipoles can take the form of the $\Sigma''$ and $\Phi''$ responses, respectively related to the longitudinal components of $\bol{s}_N$ and $\bol{v}^\perp_N \times \bol{s}_N$. $\Sigma''$ ($\Phi''$) can arise from $\tilde{\bol{V}}^N_2$ ($\tilde{\bol{V}}^N_3$) terms in \Eq{Obar scalar-vector dec}, whenever the longitudinal component of $\bol{s}_N$ (of $\bol{v}^\perp_N \times \bol{s}_N$) is selected. Only the $\Op^N_4$, $\Op^N_6$, $\Op^N_{10}$, $\Op^N_{12}$, and $\Op^N_{13}$ ($\Op^N_3$, $\Op^N_{12}$, and $\Op^N_{15}$) terms in \Eq{f(q^2) ONR^N} induce the $\Sigma''$ ($\Phi''$) response, see \Tab{tab: nuclear responses}.

\subsubsection{Multipole decomposition for a vector operator: $\lambda = \pm 1$}
To compute \Eq{exponentiale} with $\lambda = \pm 1$ we need the following Clebsch-Gordan coefficients:
\beq
\CG{\ell, 0}{1, \lambda}{J, \lambda} =
\begin{cases}
- \frac{\lambda}{\sqrt{2}} & \ell = J,
\\
\frac{1}{\sqrt{2}} \sqrt{\frac{J}{2J+3}} & \ell = J+1,
\\
\frac{1}{\sqrt{2}} \sqrt{\frac{J+1}{2J-1}} & \ell = J-1,
\\
0 & \text{otherwise},
\end{cases}
\eeq
so that
\begin{multline}
\hat{\bol{e}}_\lambda e^{i \bol{b} \cdot \bol{x}} =
\sqrt{2 \pi} \sum_{J = 1}^\infty i^J
\Big(
- \lambda \sqrt{2 J + 1} \, j_J(bx) \bol{Y}_{J J 1}^\lambda(\Omega_{\bol{x}})
\\
+ i \sqrt{J} \, j_{J+1}(bx) \bol{Y}_{J, J+1, 1}^\lambda(\Omega_{\bol{x}})
- i \sqrt{J+1} \, j_{J-1}(bx) \bol{Y}_{J, J-1, 1}^\lambda(\Omega_{\bol{x}})
\Big) \, .
\end{multline}
Notice that here $J$ must be at least $1$ for $\lambda$ to take values $\pm 1$. Using \Eq{Besselprop} as well as~\cite{DeForest:1966ycn, Hughes:1975eg, Edmonds, Walecka}
\begin{multline}
\bol{\nabla} \times (\phi(x) \bol{Y}_{JJ1}^M(\Omega_{\bol{x}})) =
\\
- i \sqrt{\frac{J}{2J+1}} \left[ \frac{J}{x} - \frac{\ud}{\ud x} \right] \phi(x) \, \bol{Y}_{J, {J+1}, 1}^M(\Omega_{\bol{x}}) + i \sqrt{\frac{J+1}{2J+1}} \left[ \frac{J+1}{x} + \frac{\ud}{\ud x} \right] \phi(x) \, \bol{Y}_{J, {J-1}, 1}^M(\Omega_{\bol{x}}) \ ,
\end{multline}
we get
\beq
\hat{\bol{e}}_\lambda e^{i \bol{b} \cdot \bol{x}} = - \sum_{J = 1}^\infty i^J \sqrt{2 \pi (2J+1)} \left( \lambda \, j_J(bx) \bol{Y}_{J J 1}^\lambda(\Omega_{\bol{x}}) + \frac{1}{b} \bol{\nabla} \times (j_J(bx) \bol{Y}_{JJ1}^\lambda(\Omega_{\bol{x}})) \right) .
\eeq
Complex-conjugating we then obtain
\beq
\int \ud^3 x \, \bol{\CMcal{O}}_\text{v}(\bol{x}) \cdot \hat{\bol{e}}_\lambda^* \, e^{- i \bol{b} \cdot \bol{x}} = - \sum_{J = 1}^\infty (-i)^J \sqrt{2 \pi (2J+1)} \left( \mulop{T}^\text{el}_{J, - \lambda}(b) + \lambda \, \mulop{T}^\text{mag}_{J, - \lambda}(b) \right) ,
\eeq
where we defined the \emph{transverse electric} and \emph{transverse magnetic multipoles}
\begin{subequations}
\label{Tel + Tmag}
\begin{align}
\mulop{T}^\text{el}_{JM}(b) & \equiv \frac{1}{b} \int \ud^3 x \, (\bol{\nabla} \times \bol{M}_{JJ}^M(b \bol{x})) \cdot \bol{\CMcal{O}}_\text{v}(\bol{x}) \ ,
\\
\mulop{T}^\text{mag}_{JM}(b) & \equiv \int \ud^3 x \, \bol{M}_{JJ}^M(b \bol{x}) \cdot \bol{\CMcal{O}}_\text{v}(\bol{x}) \ .
\end{align}
\end{subequations}
The integrand of $\mulop{T}^\text{el}_{JM}$ ($\mulop{T}^\text{mag}_{JM}$) can take the form of the $\Sigma'$ ($\Sigma$) response, related to the transverse
part of $\bol{s}_N$, or the $\Delta'$ and $\tilde{\Phi}'$ ($\Delta$ and $\tilde{\Phi}$) responses, related to the transverse part of the intrinsic component of $\bol{v}^\perp_N$ and $\bol{v}^\perp_N \times \bol{s}_N$, respectively. The $\Sigma$, $\Delta'$, and $\tilde{\Phi}$ responses, however, do not contribute to the scattering if the nucleus remains in the ground state, due to parity and time-reversal constraints (see \Sec{PT selection rules}). The $\Sigma'$ response can arise as $\mulop{T}^\text{el}_{JM}$ from $\tilde{\bol{V}}^N_2$ terms in \Eq{Obar scalar-vector dec}; $\Delta$ can arise as $\mulop{T}^\text{mag}_{JM}$ from the intrinsic component of $\tilde{\bol{V}}^N_1$ terms; and $\tilde{\Phi}'$ can arise as $\mulop{T}^\text{el}_{JM}$ from the intrinsic component of $\tilde{\bol{V}}^N_3$ terms. The $\Op^N_3$, $\Op^N_4$, $\Op^N_7$, $\Op^N_9$, $\Op^N_{12}$, $\Op^N_{14}$, and $\Op^N_{15}$ terms in \Eq{f(q^2) ONR^N} induce $\Sigma'$, the $\Op^N_5$ and $\Op^N_8$ terms induce $\Delta$, and the $\Op^N_{12}$ and $\Op^N_{13}$ terms induce $\tilde{\Phi}'$, see \Tab{tab: nuclear responses}.

\subsubsection{Parity and time-reversal selection rules}
\label{PT selection rules}
Parity and time-reversal, while not necessarily good symmetries of the interaction, can provide useful constraints. Since $\bol{v}^\perp_N$ is a polar vector while $\bol{s}_N$ is an axial vector, see \Eq{P:qvs}, we can formally define a parity coefficient $\eta_\CMcal{O}^P$ as being $+1$ for $\unop_N$ and $\bol{s}_N$, and $-1$ for $\bol{v}^\perp_N$, $\bol{v}^\perp_N \cdot \bol{s}_N$, and $\bol{v}^\perp_N \times \bol{s}_N$. From \Eq{P T} and \Tab{tab: PT} we see that $P \bar{N}(\bol{x}) \gamma^0 N(\bol{x}) P^{-1} = \bar{N}(- \bol{x}) \gamma^0 N(- \bol{x})$, which then implies
\beq\label{P transf O}
P \CMcal{O}^N(\bol{x}) P^{-1} = \eta_\CMcal{O}^P \, \CMcal{O}^N(- \bol{x}) \ .
\eeq
We will also need to know that
\begin{align}
\bol{\nabla}_{- \bol{x}} = - \bol{\nabla}_{\bol{x}} \ ,
&&
Y_{\ell m}(\Omega_{- \bol{x}}) = (-1)^\ell \, Y_{\ell m}(\Omega_{\bol{x}}) \ ,
&&
\bol{Y}_{J \ell 1}^M(\Omega_{- \bol{x}}) = (-1)^\ell \, \bol{Y}_{J \ell 1}^M(\Omega_{\bol{x}}) \ ,
\end{align}
which implies
\begin{align}
M_{JM}(- b \bol{x}) = (-1)^J \, M_{JM}(b \bol{x}) \ ,
&&&
\bol{M}_{JL}^M(- b \bol{x}) = (-1)^J \, \bol{M}_{JL}^M(b \bol{x}) \ .
\end{align}
If the nucleus remains in the ground state after the scattering, which we take having definite parity, we have for the matrix element of the $\mulop{M}_{JM}$ multipoles
\begin{multline}
\matel{\amom{J}, \amom{M}'}{\int \ud^3 x \, M_{JM}(b \bol{x}) \, \CMcal{O}_\text{s}(\bol{x})}{\amom{J}, \amom{M}} = \matel{\amom{J}, \amom{M}'}{P^{-1} P \int \ud^3 x \, M_{JM}(b \bol{x}) \, \CMcal{O}_\text{s}(\bol{x}) \, P^{-1} P}{\amom{J}, \amom{M}} =
\\
\eta_\CMcal{O}^P \, \matel{\amom{J}, \amom{M}'}{\int \ud^3 x \, M_{JM}(b \bol{x}) \, \CMcal{O}_\text{s}(- \bol{x})}{\amom{J}, \amom{M}} = (-1)^J \eta_\CMcal{O}^P \, \matel{\amom{J}, \amom{M}'}{\int \ud^3 x \, M_{JM}(b \bol{x}) \, \CMcal{O}_\text{s}(\bol{x})}{\amom{J}, \amom{M}} \ ,
\end{multline}
where in the last equality we changed integration variable. Therefore, the $\mulop{M}_{JM}$ multipoles contribute to the scattering only with the $J$ values for which $(-1)^J \eta_\CMcal{O}^P = +1$, that is, $J$ even if $\eta_\CMcal{O}^P = +1$ and $J$ odd if $\eta_\CMcal{O}^P = -1$. Analogous computations can be performed for the other multipoles, concluding that only $J$ odd (even) contributes for the $\mulop{L}_{JM}$ and $\mulop{T}^\text{el}_{JM}$ multipoles with $\eta_\CMcal{O}^P = +1$ ($\eta_\CMcal{O}^P = -1$), and only $J$ even (odd) contributes for the $\mulop{T}^\text{mag}_{JM}$ multipoles with $\eta_\CMcal{O}^P = +1$ ($\eta_\CMcal{O}^P = -1$). The two transverse multipoles can be promptly confirmed to have opposite parity by analysing their very definition in \Eq{Tel + Tmag}. \Tab{tab: nuclear responses} summarizes the parity constraints on the multipoles and nuclear responses relative to the relevant NR building blocks. The multipoles involving vector operators with abnormal parity (\ie axial vectors like $\bol{s}_N$) can be found denoted $\mulop{L}^\text{5}_{JM}$, $\mulop{T}^\text{el5}_{JM}$, $\mulop{T}^\text{mag5}_{JM}$ in the literature, to distinguish them from the multipoles involving polar vectors (as $\bol{v}^\perp_N$ and $\bol{v}^\perp_N \times \bol{s}_N$).

The constraints from time reversal can be derived as follows, where we extend the proof of Appendix E of Ref.~\cite{Walecka} (see otherwise Appendix B of Ref.~\cite{DeForest:1966ycn}) to general operators transforming under time reversal as
\beq\label{T transf O}
T \CMcal{O}^N(\bol{x}) T^{-1} = \eta_\CMcal{O}^T \, \CMcal{O}^N(\bol{x}) \ .
\eeq
For the operators of interest to us we have $\eta_\CMcal{O}^T = +1$ for $\unop_N$, $\bol{v}^\perp_N \cdot \bol{s}_N$, and $\bol{v}^\perp_N \times \bol{s}_N$, and $\eta_\CMcal{O}^T = -1$ for $\bol{v}^\perp_N$ and $\bol{s}_N$. The action of $T$ on a generic multipole operator $\CMcal{T}_{JM} = \mulop{M}_{JM}, \mulop{L}_{JM}, \mulop{T}^\text{el}_{JM}, \mulop{T}^\text{mag}_{JM}$ relative to an hermitian operator $\CMcal{O}^N$ transforming as in \Eq{T transf O} can be derived by noticing that, due to its anti-unitary properties (most notably $T i T^{-1} = - i$), $T$ complex-conjugates every c-number quantity. Using \Eq{spherical harmonics*} we thus obtain
\beq
T \CMcal{T}^\dagger_{JM}(b) T^{-1} = \eta_\CMcal{O}^T \, \CMcal{T}_{JM}(b) \ .
\eeq
We now exploit again the anti-unitary properties of $T$ (most notably $\braket{\text{f}}{\text{i}} = \matel{\text{f}}{T^{-1} T}{\text{i}} = \braket{T \text{i}}{T \text{f}}$), which yield for a generic matrix element of an operator $O$
\beq
\matel{\text{f}}{O}{\text{i}} = \braket{O^\dagger \text{f}}{\text{i}} = \braket{T \text{i}}{T O^\dagger \text{f}} = \matel{T \text{i}}{T O^\dagger T^{-1}}{T \text{f}} \ ,
\eeq
where in the last equality we inserted again the unit operator $T^{-1} T$. Our phase convention is fixed by
\beq
T \ket{J, M} = (-1)^{J + M} \ket{J, - M} \ .
\eeq
This corresponds, given \Eq{spherical harmonics*} and assuming $T \ket{\bol{x}} = \ket{\bol{x}}$, to $\braket{\bol{x}}{\ell, m} = i^\ell Y_{\ell m}(\Omega_{\bol{x}})$ for integer $\ell$~\cite{Edmonds}: in fact, $\braket{\bol{x}}{\ell, m} = \braket{T \bol{x}}{T (\ell, m)}^* = (-1)^{\ell + m} \braket{\bol{x}}{\ell, - m}^*$, which is only compatible with \Eq{spherical harmonics*} for the above phase choice. Using the above and the Wigner-Eckart theorem in \Eq{Wigner-Eckart} we get
\begin{multline}
\CG{\amom{J}, \amom{M}}{J, M}{\amom{J}', \amom{M}'} \redmatel{\amom{J}'}{\CMcal{T}_J}{\amom{J}} = \matel{\amom{J}', \amom{M}'}{\CMcal{T}_{JM}}{\amom{J}, \amom{M}}
\\
= (-1)^{\amom{J} + \amom{M}} (-1)^{\amom{J}' + \amom{M}'} \matel{\amom{J}, - \amom{M}}{T \CMcal{T}^\dagger_{JM} T^{-1}}{\amom{J}', - \amom{M}'}
\\
= (-1)^{\amom{J} + \amom{M}} (-1)^{\amom{J}' + \amom{M}'} \eta_\CMcal{O}^T \, \matel{\amom{J}, - \amom{M}}{\CMcal{T}_{JM}}{\amom{J}', - \amom{M}'}
\\
= (-1)^{\amom{J} + \amom{M}} (-1)^{\amom{J}' + \amom{M}'} \eta_\CMcal{O}^T \, \CG{\amom{J}', - \amom{M}'}{J, M}{\amom{J}, - \amom{M}} \redmatel{\amom{J}}{\CMcal{T}_J}{\amom{J}'} \ .
\end{multline}
We can relate the Clebsch-Gordan coefficients appearing in the first and last line by using \Eq{Clebsch-Gordan symm 2} twice and then \Eq{Clebsch-Gordan symm 1}, obtaining
\beq
\CG{\amom{J}, \amom{M}}{J, M}{\amom{J}', \amom{M}'} = (-1)^{J + 2 \amom{J}' - \amom{M} - \amom{M}'} \sqrt{\frac{2 \amom{J}' + 1}{2 \amom{J} + 1}} \CG{\amom{J}', - \amom{M}'}{J, M}{\amom{J}, - \amom{M}} \ .
\eeq
Since $\amom{M} + \amom{M}'$ and $J$ are both integers, we can finally write
\beq
\redmatel{\amom{J}'}{\CMcal{T}_J}{\amom{J}} = (-1)^{J + \amom{J} - \amom{J}'} \eta_\CMcal{O}^T \sqrt{\frac{2 \amom{J} + 1}{2 \amom{J}' + 1}} \, \redmatel{\amom{J}}{\CMcal{T}_J}{\amom{J}'} \ .
\eeq
For $\amom{J}' = \amom{J}$ the reduced matrix element can then only be non-zero if $(-1)^J = \eta_\CMcal{O}^T$. Joined with the parity selection rules derived above, this constraint eliminates the $\mulop{T}^\text{el}_{JM}$ multipoles of the intrinsic component of $\bol{v}^\perp_N$, the $\mulop{T}^\text{mag}_{JM}$ multipoles of $\bol{s}_N$ and $\bol{v}^\perp_N \times \bol{s}_N$ (again the intrinsic component), and eliminates the intrinsic component of the $\bol{v}^\perp_N \cdot \bol{s}_N$ interaction altogether, as reflected in \Tab{tab: nuclear responses}.

\subsubsection{Examples and applications}
\label{examples and applications}
The multipole expansion can be found extensively discussed in the context of electron-nucleus interactions. In particular, Refs.~\cite{DeForest:1966ycn, Donnelly:1975ze, Donnelly:1984rg, Walecka} focus on the electromagnetic interactions, which can also occur with DM particles through the electromagnetic nuclear current in Eqs.~\eqref{JEM NR 1},~\eqref{JEM NR 2} (see \Sec{Electromagnetic interactions}). The conservation of the four-vector current can be used to express the matrix element of longitudinal multipoles in terms of matrix elements of Coulomb multipoles. In fact, the Fourier transform of $\partial_\mu \matel{T'}{J_\text{EM}^\mu(\fvec{x})}{T} = 0$ is $- i \fvec{b}_\mu \int \ud^4 \fvec{x} \, \matel{T'}{J_\text{EM}^\mu(\fvec{x})}{T} \, e^{i \fvec{b} \cdot \fvec{x}} = 0$, where we integrated by parts; we then have, repeating the procedure leading to \Eq{S 1delta},
\beq
\int \ud^3 x \, \matel{T'}{\bol{J}_\text{EM}(\bol{x}) \cdot \hat{\bol{b}}}{T} \, e^{- i \bol{b} \cdot \bol{x}} = \frac{b^0}{b} \int \ud^3 x \, \matel{T'}{\rho(\bol{x})}{T} \, e^{- i \bol{b} \cdot \bol{x}} \ ,
\eeq
which relates the matrix elements of \Eq{Coulomb multipole expansion} and \Eq{Longitudinal multipole expansion}. We then just need to consider the $\mulop{M}_{JM}$, $\mulop{T}^\text{el}_{JM}$, and $\mulop{T}^\text{mag}_{JM}$ multipoles. As seen in \Sec{PT selection rules}, parity and time-reversal constraints cause the odd-$J$ multipoles of $\mulop{M}_{JM}$ to vanish, together with the even-$J$ multipoles of $\mulop{T}^\text{mag}_{JM}$ and all $\mulop{T}^\text{el}_{JM}$ multipoles. One then remains with the even $\mulop{M}_{JM}$ and odd $\mulop{T}^\text{mag}_{JM}$ multipoles.

Another example, more common in the direct DM detection literature, is that of the SI and SD interactions. As we will see in \Sec{SI interaction}, the SI interaction involves exclusively the $\Op^N_1$ NR building block, thus inducing the $\mulop{M}_{JM}$ multipoles in the form of the $M$ response. The SD interaction involves instead the $\Op^N_4$ building block, with also a contribution from $\Op^N_6$ due to meson exchange (see \Sec{SD interaction}); in this case, as already commented above, the $\mulop{T}^\text{mag}_{JM}$ multipoles vanish due to parity and time-reversal constraints, and one remains with the $\mulop{L}_{JM}$, $\mulop{T}^\text{el}_{JM}$ multipoles, also denoted $\mulop{L}^\text{5}_{JM}$, $\mulop{T}^\text{el5}_{JM}$ due to the abnormal parity of the $\bol{s}_N$ operator. The relevant nuclear responses for the SD interaction are $\Sigma'$ and $\Sigma''$.

Some operators, such as the intrinsic components of $\bol{v}^\perp_N$ and $\bol{v}^\perp_N \times \bol{s}_N$, can only contribute to the scattering if the nucleus is spatially extended. This can be intuitively explained by noticing that, while the CM component of $\bol{v}^\perp_N$ involves the overall nuclear motion, its intrinsic component involves exclusively relative nucleon velocities (see \Eq{v^perp_T}), which vanish in the limit of point-like nucleus. More formally, we saw in \Sec{nuc form factors} that this limit is achieved by neglecting the exponential in \Eq{S Os + V Ov}. This implies that, for a generic operator $\CMcal{O}^N(\bol{x})$ transforming under parity as in \Eq{P transf O}, the scattering amplitude depends in this limit on
\begin{multline}
\matel{\amom{J}, \amom{M}'}{\int \ud^3 x \, \CMcal{O}^N(\bol{x})}{\amom{J}, \amom{M}} = \matel{\amom{J}, \amom{M}'}{P^{-1} P \int \ud^3 x \, \CMcal{O}^N(\bol{x}) P^{-1} P}{\amom{J}, \amom{M}} =
\\
\eta_\CMcal{O}^P \, \matel{\amom{J}, \amom{M}'}{\int \ud^3 x \, \CMcal{O}^N(- \bol{x})}{\amom{J}, \amom{M}} = \eta_\CMcal{O}^P \, \matel{\amom{J}, \amom{M}'}{\int \ud^3 x \, \CMcal{O}^N(\bol{x})}{\amom{J}, \amom{M}} \ ,
\end{multline}
where in the last equality we changed integration variable. Therefore, operators with $\eta_\CMcal{O}^P = -1$ as $\bol{v}^\perp_N$ and $\bol{v}^\perp_N \times \bol{s}_N$ do not contribute in the limit of point-like nucleus, as expected. For this reason, we can expect that the terms of the scattering amplitude involving the nuclear responses related to these operators, namely $\Delta$, $\tilde{\Phi}'$, and $\Phi''$, vanish at zero momentum transfer. This effect is taken into account by the fact these responses always appear multiplied by $b / \mN$ in the scattering amplitude: in this way, even if they are defined to be finite at $b = 0$, the terms of the scattering amplitude where they are featured vanish in that limit. The limit of point-like nucleus can be further explored by noticing that $j_\ell(z) \simeq z^\ell / (2 \ell + 1)!!$ for $z \to 0$~\cite{NIST DLMF} implies that all but the $\mulop{M}_{00}$, $\mulop{L}_{10}$, and $\mulop{T}^\text{el}_{1, \pm 1}$ multipoles vanish at $b = 0$. Incidentally, this also provides another reason why amplitude terms featuring the $\Delta$ response, which is related to the $\mulop{T}^\text{mag}_{JM}$ multipoles, should vanish in this limit.

Finally, one immediate application of the multipole expansion concerns spin-$0$ nuclei. In this case, due to angular-momentum conservation, only the $J = 0$ multipoles can contribute to the scattering: in fact, for $\amom{J} = \amom{J}' = 0$, the Clebsch-Gordan coefficients in \Eq{Multipole expansion amplitude} below vanish unless $J = 0$. Therefore, only the $\mulop{M}_{JM}$ and $\mulop{L}_{JM}$ multipoles (more precisely, only $\mulop{M}_{00}$ and $\mulop{L}_{00}$) can contribute, as $\mulop{T}^\text{el}_{JM}$ and $\mulop{T}^\text{mag}_{JM}$ start from $J = 1$. Of the nuclear responses related to the $\mulop{L}_{JM}$ multipoles, however, $\Sigma''_{JM}$ is only non-zero for $J$ odd (due to the abnormal parity of $\bol{s}_N$) and thus does not contribute, see \Tab{tab: nuclear responses}. The SD interaction, which features the $\mulop{T}^\text{el}_{JM}$ and $\mulop{L}_{JM}$ multipoles in the form of $\Sigma'$ and $\Sigma''$ responses respectively (see \Sec{examples and applications}), is therefore entirely forbidden for spin-$0$ nuclei. The only responses contributing to DM scattering off spinless nuclei are then $M_{00}$ and $\Phi''_{00}$. Moreover, regarding the latter, we saw above that the terms of the scattering amplitude featuring $\Phi''_{00}$ should vanish in the limit of point-like nucleus. One more reason for that to be the case for spin-$0$ nuclei is that, in this limit, a vector operator as $\bol{v}^\perp_N \times \bol{s}_N$ cannot mediate a $0 \to 0$ transition by angular momentum conservation (in other words, the $\mulop{L}_{00}$ multipole vanishes at zero momentum transfer). With similar arguments we can also conclude that the $\mulop{T}^\text{el}_{JM}$ multipoles in the form of the $\tilde{\Phi}'$ response, which start from $J = 2$ due to their quantum numbers (see \Tab{tab: nuclear responses}), only contribute to the scattering for $\amom{J} = \amom{J}' \geqslant 1$ in the limit of point-like nucleus.

\subsection{Scattering amplitude in the multipole expansion}
\label{Nuclear matrix element}
Collecting the above results, we can write the operator~\eqref{S Os + V Ov} in terms of the spherical tensor operators $\mulop{M}_{JM}$, $\mulop{L}_{JM}$, $\mulop{T}^\text{el}_{JM}$, $\mulop{T}^\text{mag}_{JM}$, so that the scattering amplitude (see \eg \Eq{Mel pre nuc FF}) reads
\begin{multline}
\label{Multipole expansion amplitude}
\Mel \NReq - 2 m_T i \, \matel{\amom{J}', \amom{M}'}{\int \ud^3 x \, (\tilde{S} \CMcal{O}_\text{s}(\bol{x}) + \tilde{\bol{V}} \cdot \bol{\CMcal{O}}_\text{v}(\bol{x})) \, e^{- i \bol{b} \cdot \bol{x}}}{\amom{J}, \amom{M}}
\\
= - 2 m_T i \left[ \sum_{J = 0}^\infty (-i)^J \sqrt{4 \pi (2 J + 1)} \, \CG{\amom{J}, \amom{M}}{J, 0}{\amom{J}', \amom{M}'} \, \redmatel{\amom{J}'}{\mulop{M}_J(b) \, \tilde{S} + \mulop{L}_J(b) \, \tilde{V}_0}{\amom{J}}
\vphantom{\sum_{\lambda = \pm 1} \sum_{J = 1}^\infty} \right.
\\
\left. \vphantom{\sum_{J = 0}^\infty}
- \sum_{\lambda = \pm 1} \sum_{J = 1}^\infty (-i)^J \sqrt{2 \pi (2J+1)} \, \CG{\amom{J}, \amom{M}}{J, - \lambda}{\amom{J}', \amom{M}'} \, \redmatel{\amom{J}'}{\mulop{T}^\text{el}_J(b) + \lambda \, \mulop{T}^\text{mag}_J(b)}{\amom{J}} \, \tilde{V}_\lambda \right] ,
\end{multline}
where we used \Eq{Wigner-Eckart}. While not manifest here, the scattering amplitude depends on $\hat{\bol{b}}$ through the $\tilde{S}$, $\tilde{V}_0$, and $\tilde{V}_\lambda$ coefficients as well as through the very definition of the multipole operators (see discussion after \Eq{M'n'M}). The limit of point-like nucleus is obtained by taking all multipole operators at $b = 0$, meaning that only $\mulop{M}_{00}$, $\mulop{L}_{10}$, and $\mulop{T}^\text{el}_{1, \pm 1}$ survive as discussed in \Sec{examples and applications}. The above expression matches in this limit \Eq{M_PLN}, which is valid for only one (scalar or vector) operator but encompasses DM interactions with both protons and neutrons: for a scalar operator, for instance, we can identify $\CMcal{O}^N = \CMcal{O}^N_{00} = \CMcal{O}_\text{s}$ and $K^N = K^N_{00} = \tilde{S}$, so that we see from Eqs.~\eqref{T^N_JM},~\eqref{M'n'M} that $\mulop{M}_{00}(0) = \CMcal{T}^N_{00}$ since $j_0(0) = 1$ and $Y_{00}(\Omega_{\bol{x}}) = 1 / \sqrt{4 \pi}$.

The squared amplitude averaged over initial states and summed over final states (\Eq{Mel^2} below), needed to obtain the unpolarized scattering cross section (see \Sec{sigma}), can be computed with the help of Eqs.~\eqref{Clebsch-Gordan symm 2},~\eqref{Clebsch-Gordan orthogonality}, yielding
\begin{multline}
\label{Multipole expansion amplitude 2}
\overline{| \Mel |^2} \NReq 4 m_T^2 \frac{1}{2 s_\DM + 1} \frac{1}{2 \amom{J} + 1} \sum_{s, s'} \sum_{\amom{M}, \amom{M}'} \left| \matel{\amom{J}', \amom{M}'}{\int \ud^3 x \, (\tilde{S} \CMcal{O}_\text{s}(\bol{x}) + \tilde{\bol{V}} \cdot \bol{\CMcal{O}}_\text{v}(\bol{x})) \, e^{- i \bol{b} \cdot \bol{x}}}{\amom{J}, \amom{M}} \right|^2
\\
= 16 \pi m_T^2 \frac{1}{2 s_\DM + 1} \frac{2 \amom{J}' + 1}{2 \amom{J} + 1} \sum_{s, s'} \left( \sum_{J = 0}^\infty \left| \redmatel{\amom{J}'}{\mulop{M}_J(b) \, \tilde{S} + \mulop{L}_J(b) \, \tilde{V}_0}{\amom{J}} \right|^2 \delta(\amom{J}, \amom{J}', J) \right.
\\
\left. + \frac{1}{2} \sum_{\lambda = \pm 1} \sum_{J = 1}^\infty \left| \redmatel{\amom{J}'}{\mulop{T}^\text{el}_J(b) + \lambda \, \mulop{T}^\text{mag}_J(b)}{\amom{J}} \, \tilde{V}_\lambda \right|^2 \delta(\amom{J}, \amom{J}', J) \right) .
\end{multline}
For each given $\bol{\CMcal{O}}_\text{v}(\bol{x})$ operator (relative to a single NR building block in \Eq{NR building blocks}), $\mulop{T}^\text{el}_{JM}(b)$ and $\mulop{T}^\text{mag}_{JM}(b)$ have opposite parity, as discussed in \Sec{PT selection rules} and made clear by their very definition in \Eq{Tel + Tmag}, and therefore do not interfere. However, parts of these responses can interfere if relative to different $\bol{\CMcal{O}}_\text{v}(\bol{x})$ operators with opposite parity: for instance, the transverse magnetic multipoles relative to the $\Delta$ response can interfere with the transverse electric multipoles of the $\Sigma'$ response (see below). The interference term between transverse electric and magnetic responses can be computed via \Eq{vectorproduct}. In the absence of interference between the two transverse responses, the last term in parentheses in \Eq{Multipole expansion amplitude 2} becomes
\beq
\frac{1}{2} \left| \tilde{\bol{V}}^\perp \right|^2 \sum_{J = 1}^\infty \left( \left| \redmatel{\amom{J}'}{\mulop{T}^\text{el}_J(b)}{\amom{J}} \right|^2 + \left| \redmatel{\amom{J}'}{\mulop{T}^\text{mag}_J(b)}{\amom{J}} \right|^2 \right) \delta(\amom{J}, \amom{J}', J) \ ,
\eeq
where we defined (see \Eq{V spherical compo})
\beq
\tilde{\bol{V}}^\perp \equiv \tilde{\bol{V}} - (\tilde{\bol{V}} \cdot \hat{\bol{b}}) \, \hat{\bol{b}} = \tilde{\bol{V}} - \tilde{V}_0 \, \hat{\bol{b}} \ ,
\eeq
so that (see \Eq{scalarproduct})
\beq
\left| \tilde{\bol{V}}^\perp \right|^2 = |\tilde{\bol{V}}|^2 - |\tilde{V}_0|^2 = \sum_{\lambda = \pm 1} | \tilde{V}_\lambda |^2 \ .
\eeq

In the standard assumption the nucleus does not get excited in the scattering, which we will enforce from now on, $\amom{J}' = \amom{J}$ and the $\delta(\amom{J}, \amom{J}', J)$ factor implies that only values of $J \leqslant 2 \amom{J}$ contribute to the sums, so that we have
\begin{multline}
\label{M^2 multipole}
\overline{| \Mel |^2} \NReq 16 \pi m_T^2 \frac{1}{2 s_\DM + 1} \sum_{s, s'} \left( \sum_{J = 0}^{2 \amom{J}} \left| \redmatel{\amom{J}}{\mulop{M}_J(b) \, \tilde{S} + \mulop{L}_J(b) \, \tilde{V}_0}{\amom{J}} \right|^2 \right.
\\
\left. + \frac{1}{2} \sum_{\lambda = \pm 1} \sum_{J = 1}^{2 \amom{J}} \left| \redmatel{\amom{J}}{\mulop{T}^\text{el}_J(b) + \lambda \, \mulop{T}^\text{mag}_J(b)}{\amom{J}} \, \tilde{V}_\lambda \right|^2 \right) .
\end{multline}
Being rotationally invariant, $\overline{| \Mel |^2}$ does not depend on $\hat{\bol{b}}$ if the nuclear ground state is spherically symmetric, as we will assume. It is apparent that, due to angular-momentum conservation, interference can only occur between the Coulomb and longitudinal multipoles and between the two transverse multipoles, and within these cases, only among multiples with the same $J$. With reference to the families of potentially interfering NR operators identified at the end of \Sec{NR operators}, we can then see from \Tab{tab: nuclear responses} that the $\Op^N_1$--$\Op^N_3$ interference can only occur between the $\mulop{L}_{JM}$ multipoles relative to the intrinsic component of $\Op^N_3$ and the $\Op^N_1$ $\mulop{M}_{JM}$ multipoles; $\Op^N_4$ cannot interfere with $\Op^N_5$ due to the different parity of their same-$J$ $\mulop{M}_{JM}$ and $\mulop{L}_{JM}$ multipoles, while it can interfere with $\Op^N_6$ through the $\mulop{L}_{JM}$ multipoles alone; the $\Op^N_8$--$\Op^N_9$ interference can only occur between the $\mulop{T}^\text{mag}_{JM}$ multipoles relative to the intrinsic component of $\Op^N_8$ and the $\Op^N_9$ $\mulop{T}^\text{el}_{JM}$ multipoles; the NR operator $\Op^N_{11}$, which gives rise to $\mulop{M}_{JM}$ multipoles, can only interfere with the $\mulop{L}_{JM}$ multipoles relative to the intrinsic components of $\Op^N_{12}$ and $\Op^N_{15}$, the interference with those relative to the $\Op^N_{12}$ CM component being forbidden by parity constraints; also due to parity constraints, the $\Op^N_{12}$--$\Op^N_{15}$ interference only occurs among their CM and intrinsic components separately. We thus conclude that only the $M$--$\Phi''$ and $\Sigma'$--$\Delta$ interferences play a role in elastic DM-nucleus scattering with $\amom{J}' = \amom{J}$.

\Eq{M^2 multipole} has been derived for the representative operator in \Eq{representative Op}, featuring one scalar operator (either the $\tilde{S}^N_1$ or $\tilde{S}^N_2$ term in \Eq{Obar scalar-vector dec}) plus one vector operator (either the $\tilde{\bol{V}}^N_1$, $\tilde{\bol{V}}^N_2$ or $\tilde{\bol{V}}^N_3$ term), and for either protons or neutrons interacting with the DM. In the general case with all terms in \Eq{Obar scalar-vector dec} and both nucleon types involved in the interaction, we can obtain a relatively compact form for $\Mel$ by substituting the multipole (spherical tensor) operators appearing in \Eq{Multipole expansion amplitude} with
\beq
\CMcal{T}^N_{JM}[X] \equiv \int \ud^3 x \sum_i \delta^{(3)}(\bol{x} - \bol{x}_i) \, X_{JM}^N(b \bol{x}) = \sum_i X_{JM}^N(b \bol{x}_i)
\eeq
for $X = M, \Delta, \Sigma', \Sigma'', \tilde{\Phi}', \Phi''$, where $i$ runs over all nucleons of type $N = p, n$ and we made the nuclear-response dependence on $N$ explicit. More in detail, for $\amom{J}' = \amom{J}$, we substitute in \Eq{Multipole expansion amplitude}~\cite{Anand:2013yka}
\begin{subequations}
\label{multipoles to responses}
\begin{align}
\mulop{M}_J(b) \, \tilde{S} &\longrightarrow \frac{i}{2 \mN} \sum_{N = p, n} \CMcal{T}^N_J[M] \, S^N \ ,
\\
\mulop{L}_J(b) \, \tilde{V}_0 &\longrightarrow \frac{i}{2 \mN} \sum_{N = p, n} \left( \frac{i}{2} \, \CMcal{T}^N_J[\Sigma''] \, U_0^N - \frac{1}{2} \frac{i b}{\mN} \, \CMcal{T}^N_J[\Phi''] \, V_0^N \right) ,
\\
\mulop{T}^\text{el}_J(b) \, \tilde{V}_\lambda &\longrightarrow \frac{i}{2 \mN} \sum_{N = p, n} \left( \frac{i}{2} \, \CMcal{T}^N_J[\Sigma'] \, U_\lambda^N - \frac{1}{2} \frac{i b}{\mN} \, \CMcal{T}^N_J[\tilde{\Phi}'] \, V_\lambda^N \right) ,
\\
\mulop{T}^\text{mag}_J(b) \, \tilde{V}_\lambda &\longrightarrow \frac{i}{2 \mN} \sum_{N = p, n} \frac{i b}{\mN} \, \CMcal{T}^N_J[\Delta] \, W_\lambda^N \ ,
\end{align}
\end{subequations}
where the $1/2$ factors are due to the fact that the nuclear responses are defined in terms of Pauli matrices rather than $\bol{s}_N$, and the $i$ and $\pm i b / \mN$ factors are also due to the nuclear responses's very definition. Regarding the latter, the $b / \mN$ factors are responsible for the amplitude terms featuring the $\Delta$, $\tilde{\Phi}'$, $\Phi''$ responses to vanish at zero momentum transfer, as explained in \Sec{examples and applications}. The coefficients can be checked from \Eq{S's and V's} and subsequent discussion to be (the components of)
\begin{subequations}
\label{S and U, V, W}
\begin{align}
S^N &= \frac{2 \mN}{i} \, \tilde{S}^N_1 = f_1^N \, \CMcal{I}_\chi + i f_5^N \, (\bol{s}_\chi \times \bol{q}) \cdot \bol{v}^\perp_T + f_8^N \, \bol{s}_\chi \cdot \bol{v}^\perp_T + i f_{11}^N \, (\bol{s}_\chi \cdot \bol{q}) \ ,
\\
\begin{split}
\bol{U}^N &= \frac{2 \mN}{i} \, \tilde{\bol{V}}^N_2 = i f_3^N \, \CMcal{I}_\chi \, (\bol{q} \times \bol{v}^\perp_T) + f_4^N \, \bol{s}_\chi + f_6^N \, (\bol{s}_\chi \cdot \bol{q}) \, \bol{q} + f_7^N \, \CMcal{I}_\chi \, \bol{v}^\perp_T
\\
&\phantom{= \frac{2 \mN}{i} \, \tilde{\bol{V}}^N_2 =} + i f_9^N \, (\bol{q} \times \bol{s}_\chi) + i f_{10}^N \, \CMcal{I}_\chi \, \bol{q} - f_{12}^N \, (\bol{s}_\chi \times \bol{v}^\perp_T) + i f_{13}^N \, (\bol{s}_\chi \cdot \bol{v}^\perp_T) \, \bol{q}
\\
&\phantom{= \frac{2 \mN}{i} \, \tilde{\bol{V}}^N_2 =} + i f_{14}^N \, (\bol{s}_\chi \cdot \bol{q}) \, \bol{v}^\perp_T + f_{15}^N \, (\bol{s}_\chi \cdot \bol{q}) (\bol{q} \times \bol{v}^\perp_T) \ ,
\end{split}
\\
\bol{V}^N &= \frac{2 \mN}{i} \, \tilde{\bol{V}}^N_3 = i f_3^N \, \CMcal{I}_\chi \, \bol{q} - f_{12}^N \, \bol{s}_\chi + i f_{13}^N \, (\bol{s}_\chi \times \bol{q}) + f_{15}^N \, (\bol{s}_\chi \cdot \bol{q}) \, \bol{q} \ ,
\\
\bol{W}^N &= \frac{2 \mN}{i} \, \tilde{\bol{V}}^N_1 = i f_5^N \, (\bol{s}_\chi \times \bol{q}) + f_8^N \, \bol{s}_\chi \ ,
\end{align}
\end{subequations}
where we used \Eq{contractions3} to write $(\bol{s}_\chi \times \bol{q}) \times \bol{v}^\perp_T = (\bol{s}_\chi \cdot \bol{v}^\perp_T) \, \bol{q}$. As discussed in \Sec{PT selection rules}, the $\tilde{S}^N_2$ term in \Eq{Obar scalar-vector dec} does not contribute to the scattering unless the nucleus gets excited, due to parity and time-reversal selection rules. Notice that we regard the functions $f_i^N(q^2)$ as hermitian operators if their argument is an operator, as in Eqs.~\eqref{f(q^2) ONR^N},~\eqref{S's and V's}, but otherwise, as here, they are understood to be real c-numbers.

To write the unpolarized squared scattering amplitude using this formalism we define, following Ref.~\cite{Fitzpatrick:2012ix}, the `squared form factors'
\beq
F_{X, Y}^{(N, N')}(q^2) \equiv 4 \pi \sum_J^{2 \amom{J}} \redmatel{\amom{J}}{\CMcal{T}^N_J[X]}{\amom{J}} \redmatel{\amom{J}}{\CMcal{T}^{N'}_J[Y]}{\amom{J}} \ .
\eeq
The sum starts from $J = 0$ for the Coulomb and longitudinal responses ($M, \Sigma'', \Phi''$), and from $J = 1$ for the transverse responses ($\Delta, \Sigma', \tilde{\Phi}'$), although in practice selection rules imply that the sum starts from $J = 1$ for $\Sigma''$ and from $J = 2$ for $\tilde{\Phi}'$ because of their quantum numbers (see \Tab{tab: nuclear responses}). The only non-vanishing non-diagonal squared form factors are $F_{M, \Phi''}^{(N, N')}$ and $F_{\Sigma', \Delta}^{(N, N')}$, as explained above. Compared with the form factors introduced in the general discussion in \Sec{nuc form factors}, the $F_{X, Y}^{(N, N')}$'s are in a sense products of two form factors that do not follow our normalization prescription~\eqref{F normalization}. They obey
\begin{align}
\label{F_XY}
F_{X, Y}^{(N, N')} = F_{Y, X}^{(N', N)} \ ,
&&&
F_X^{(N, N')} = F_X^{(N', N)} \ ,
\end{align}
where we defined $F_X^{(N, N')} \equiv F_{X, X}^{(N, N')}$ to make contact with the notation of Ref.~\cite{Fitzpatrick:2012ix}. The $\redmatel{\amom{J}}{\CMcal{T}^N_J[X]}{\amom{J}}$ reduced matrix elements are real, as can be seen by noticing from \Eq{spherical harmonics*} that the multipole operators $\CMcal{T}_{JM} = \mulop{M}_{JM}, \mulop{L}_{JM}, \mulop{T}^\text{el}_{JM}, \mulop{T}^\text{mag}_{JM}$ satisfy
\beq
\CMcal{T}_{JM}^\dagger = (-1)^{M + \eta} \, \CMcal{T}_{J, -M} \ ,
\eeq
with $\eta = 0$ for scalar operators $\CMcal{O}_\text{s}$ (thus for $\mulop{M}_{JM}$) and $\eta = 1$ for vector operators $\bol{\CMcal{O}}_\text{v}$ (thus for $\mulop{L}_{JM}$, $\mulop{T}^\text{el}_{JM}$, $\mulop{T}^\text{mag}_{JM}$). We then have, using \Eq{Wigner-Eckart},
\begin{multline}
\label{reduced reality}
\CG{\amom{J}, \amom{M}}{J, M}{\amom{J}', \amom{M}'} \, \redmatel{\amom{J}'}{\CMcal{T}_J}{\amom{J}}^* = \matel{\amom{J}', \amom{M}'}{\CMcal{T}_{JM}}{\amom{J}, \amom{M}}^* = \matel{\amom{J}, \amom{M}}{\CMcal{T}_{JM}^\dagger}{\amom{J}', \amom{M}'}
\\
= (-1)^{M + \eta} \matel{\amom{J}, \amom{M}}{\CMcal{T}_{J, -M}}{\amom{J}', \amom{M}'} = (-1)^{M + \eta} \CG{\amom{J}', \amom{M}'}{J, - M}{\amom{J}, \amom{M}} \, \redmatel{\amom{J}}{\CMcal{T}_J}{\amom{J}'} \ ,
\end{multline}
where we exploited the Clebsch-Gordan coefficients being real (as customarily defined \eg via the Condon-Shortley phase convention, see \eg Ref.~\cite{Edmonds}). Exploiting now their symmetry properties in Eqs.~\eqref{Clebsch-Gordan symm 1},~\eqref{Clebsch-Gordan symm 2}, we have
\beq
(-1)^M \CG{\amom{J}', \amom{M}'}{J, - M}{\amom{J}, \amom{M}} = (-1)^{2 J + 2 M + \amom{J}' - \amom{J}} \sqrt{\frac{2 \amom{J} + 1}{2 \amom{J}' + 1}} \CG{\amom{J}, \amom{M}}{J, M}{\amom{J}', \amom{M}'} \ ,
\eeq
which plugged into \Eq{reduced reality} shows that
\beq
\redmatel{\amom{J}}{\CMcal{T}_J}{\amom{J}}^* = (-1)^\eta \, \redmatel{\amom{J}}{\CMcal{T}_J}{\amom{J}} \ .
\eeq
Therefore, with the phases in \Eq{multipoles to responses} (and keeping in mind that the $i / 2 \mN$ factors come from a redefinition of the coefficients, see \Eq{S and U, V, W}), one obtains that $\redmatel{\amom{J}}{\CMcal{T}^N_J[X]}{\amom{J}}$ is real for $X = M, \Delta, \Sigma', \Sigma'', \tilde{\Phi}', \Phi''$.

Equipped with this formalism we can write \Eq{M^2 multipole} as
\beq\label{M^2 form factors RXY}
\overline{| \Mel |^2} \NReq \frac{m_T^2}{\mN^2} \sum_{N, N' = p, n} \sum_{\substack{X, Y = M, \Delta, \\ \Sigma', \Sigma'', \tilde{\Phi}', \Phi''}} R_{X Y}^{N N'}(q^2, {v^\perp_T}^2) \, F_{X, Y}^{(N, N')}(q^2) \ ,
\eeq
where, defining $R_X^{N N'} \equiv R_{X X}^{N N'}$,
\begin{align}
R_{M}^{N N'} &= \frac{1}{2 s_\DM + 1} \sum_{s, s'} {S^N}^* S^{N'} = f_1^N f_1^{N'} + \frac{q^2 {v^\perp_T}^2}{4} f_5^N f_5^{N'} + \frac{{v^\perp_T}^2}{4} f_8^N f_8^{N'} + \frac{q^2}{4} f_{11}^N f_{11}^{N'} \ ,
\\
\begin{split}
R_{\Sigma'}^{N N'} &= \frac{1}{8} \frac{1}{2 s_\DM + 1} \sum_{s, s'} \sum_{\lambda = \pm 1} {U_\lambda^N}^* U_\lambda^{N'} = \frac{1}{8} \left[ q^2 {v^\perp_T}^2 f_3^N f_3^{N'} + \frac{1}{2} f_4^N f_4^{N'} + {v^\perp_T}^2 f_7^N f_7^{N'} + \frac{q^2}{2} f_9^N f_9^{N'}
\vphantom{\frac{{v^\perp_T}^2}{4}} \right.
\\
&\phantom{= \frac{1}{8} \frac{1}{2 s_\DM + 1} \sum_{s, s'} \sum_{\lambda = \pm 1} {U_\lambda^N}^* U_\lambda^{N'} =} \left. + \frac{{v^\perp_T}^2}{4} (f_{12}^N - q^2 f_{15}^N) (f_{12}^{N'} - q^2 f_{15}^{N'}) + \frac{q^2 {v^\perp_T}^2}{4} f_{14}^N f_{14}^{N'} \right] ,
\end{split}
\\
\begin{split}
R_{\Sigma''}^{N N'} &= \frac{1}{4} \frac{1}{2 s_\DM + 1} \sum_{s, s'} {U_0^N}^* U_0^{N'} = \frac{1}{4} \left[ \frac{1}{4} (f_4^N + q^2 f_6^N) (f_4^{N'} + q^2 f_6^{N'})
\vphantom{\frac{{v^\perp_T}^2}{4}} \right.
\\
&\phantom{= \frac{1}{4} \frac{1}{2 s_\DM + 1} \sum_{s, s'} {U_0^N}^* U_0^{N'} =} \left. + q^2 f_{10}^N f_{10}^{N'} + \frac{{v^\perp_T}^2}{4} f_{12}^N f_{12}^{N'} + \frac{q^2 {v^\perp_T}^2}{4} f_{13}^N f_{13}^{N'} \right] ,
\end{split}
\\
R_{\tilde{\Phi}'}^{N N'} &= \frac{q^2}{8 \mN^2} \frac{1}{2 s_\DM + 1} \sum_{s, s'} \sum_{\lambda = \pm 1} {V_\lambda^N}^* V_\lambda^{N'} = \frac{q^2}{16 \mN^2} \left[ f_{12}^N f_{12}^{N'} + q^2 f_{13}^N f_{13}^{N'} \right] ,
\\
R_{\Phi''}^{N N'} &= \frac{q^2}{4 \mN^2} \frac{1}{2 s_\DM + 1} \sum_{s, s'} {V_ 0^N}^* V_ 0^{N'} = \frac{q^2}{4 \mN^2} \left[ q^2 f_3^N f_3^{N'} + \frac{1}{4} (f_{12}^N - q^2 f_{15}^N) (f_{12}^{N'} - q^2 f_{15}^{N'}) \right] ,
\\
R_{\Delta}^{N N'} &= \frac{q^2}{2 \mN^2} \frac{1}{2 s_\DM + 1} \sum_{s, s'} \sum_{\lambda = \pm 1} {W_\lambda^N}^* W_\lambda^{N'} = \frac{q^2}{4 \mN^2} \left[ q^2 f_5^N f_5^{N'} + f_8^N f_8^{N'} \right] ,
\\
R_{M \Phi''}^{N N'} &= - \frac{i q}{2 \mN} \frac{1}{2 s_\DM + 1} \sum_{s, s'} {S^N}^* V_ 0^{N'} = - \frac{q^2}{2 \mN} \left[ f_1^N f_3^{N'} + \frac{1}{4} f_{11}^N (f_{12}^{N'} - q^2 f_{15}^{N'}) \right] ,
\\
R_{\Sigma' \Delta}^{N N'} &= \frac{q}{4 \mN} \frac{1}{2 s_\DM + 1} \sum_{s, s'} \sum_{\lambda = \pm 1} \lambda \, {U_\lambda^N}^* W_\lambda^{N'} = \frac{q^2}{8 \mN} \left[ - f_4^N f_5^{N'} + f_9^N f_8^{N'} \right] ,
\end{align}
with ${v^\perp_T}^2$ given in \Eq{vperp^2}. Here we used Eqs.~\eqref{scalarproduct},~\eqref{vectorproduct}, as well as
\begin{align}
\frac{1}{2 s_\DM + 1} \sum_{s, s'} | \CMcal{I}_\chi |^2 &= 1 \ ,
&
\frac{1}{2 s_\DM + 1} \sum_{s, s'} \bol{s}_\chi^* \CMcal{I}_\chi &= \bol{0} \ ,
\\
\label{spinsum s_chi}
\frac{1}{2 s_\DM + 1} \sum_{s, s'} | \bol{s}_\chi \cdot \hat{\bol{n}} |^2 &= \frac{1}{4} \ ,
&
\frac{1}{2 s_\DM + 1} \sum_{s, s'} | \bol{s}_\chi |^2 &= \frac{3}{4} \ ,
\end{align}
and other similar algebraic relations derived from Eqs.~\eqref{spinsum 2},~\eqref{spinsum 3},~\eqref{contractions3}. The $f_i^N$'s being real implies
\begin{align}
R_{X Y}^{N N'} = R_{Y X}^{N' N} \ ,
&&&
R_X^{N N'} = R_X^{N' N} \ .
\end{align}
The $m_T^2 / \mN^2$ factor in \Eq{M^2 form factors RXY} expresses the change in state normalization in going from nucleon to nuclear states, see \Eq{statenormhere NR}. Eventually we can also write
\beq\label{M^2 form factors fifj}
\overline{| \Mel |^2} \NReq \frac{m_T^2}{\mN^2} \sum_{i, j} \sum_{N, N' = p, n} f_i^N(q^2) f_j^{N'}(q^2) \, F_{i, j}^{(N, N')}(q^2, {v^\perp_T}^2) \ ,
\eeq
with
\begin{subequations}
\label{F_ij}
\begin{align}
F_{1, 1}^{(N, N')} &= F_M^{(N, N')} \ ,
\\
F_{3, 3}^{(N, N')} &= \frac{q^2}{4} \left( \frac{{v^\perp_T}^2}{2} F_{\Sigma'}^{(N, N')} + \frac{q^2}{\mN^2} F_{\Phi''}^{(N, N')} \right) ,
\\
F_{4, 4}^{(N, N')} &= \frac{1}{16} \left( F_{\Sigma'}^{(N, N')} + F_{\Sigma''}^{(N, N')} \right) ,
\\
F_{5, 5}^{(N, N')} &= \frac{q^2}{4} \left( {v^\perp_T}^2 F_M^{(N, N')} + \frac{q^2}{\mN^2} F_{\Delta}^{(N, N')} \right) ,
\\
F_{6, 6}^{(N, N')} &= \frac{q^4}{16} F_{\Sigma''}^{(N, N')} \ ,
\\
F_{7, 7}^{(N, N')} &= \frac{{v^\perp_T}^2}{8} F_{\Sigma'}^{(N, N')} \ ,
\\
F_{8, 8}^{(N, N')} &= \frac{1}{4} \left( {v^\perp_T}^2 F_M^{(N, N')} + \frac{q^2}{\mN^2} F_{\Delta}^{(N, N')} \right) ,
\\
F_{9, 9}^{(N, N')} &= \frac{q^2}{16} F_{\Sigma'}^{(N, N')} \ ,
\\
F_{10, 10}^{(N, N')} &= \frac{q^2}{4} F_{\Sigma''}^{(N, N')} \ ,
\\
F_{11, 11}^{(N, N')} &= \frac{q^2}{4} F_M^{(N, N')} \ ,
\\
F_{12, 12}^{(N, N')} &= \frac{{v^\perp_T}^2}{16} \left( \frac{1}{2} F_{\Sigma'}^{(N, N')} + F_{\Sigma''}^{(N, N')} \right) + \frac{q^2}{16 \mN^2} \left( F_{\tilde{\Phi}'}^{(N, N')} + F_{\Phi''}^{(N, N')} \right) ,
\\
F_{13, 13}^{(N, N')} &= \frac{q^2}{16} \left( {v^\perp_T}^2 F_{\Sigma''}^{(N, N')} + \frac{q^2}{\mN^2} F_{\tilde{\Phi}'}^{(N, N')} \right) ,
\\
F_{14, 14}^{(N, N')} &= \frac{q^2 {v^\perp_T}^2}{32} F_{\Sigma'}^{(N, N')} \ ,
\\
F_{15, 15}^{(N, N')} &= \frac{q^4}{16} \left( \frac{{v^\perp_T}^2}{2} F_{\Sigma'}^{(N, N')} + \frac{q^2}{\mN^2} F_{\Phi''}^{(N, N')} \right) ,
\\
F_{1, 3}^{(N, N')} &= - \frac{q^2}{2 \mN} F_{M, \Phi''}^{(N, N')} \ ,
\\
F_{4, 5}^{(N, N')} &= - \frac{q^2}{8 \mN} F_{\Sigma', \Delta}^{(N, N')} \ ,
\\
F_{4, 6}^{(N, N')} &= \frac{q^2}{16} F_{\Sigma''}^{(N, N')} \ ,
\\
F_{9, 8}^{(N, N')} &= \frac{q^2}{8 \mN} F_{\Sigma', \Delta}^{(N, N')} \ ,
\\
F_{11, 12}^{(N, N')} &= - \frac{q^2}{8 \mN} F_{M, \Phi''}^{(N, N')} \ ,
\\
F_{11, 15}^{(N, N')} &= \frac{q^4}{8 \mN} F_{M, \Phi''}^{(N, N')} \ ,
\\
F_{12, 15}^{(N, N')} &= - \frac{q^2}{16} \left( \frac{{v^\perp_T}^2}{2} F_{\Sigma'}^{(N, N')} + \frac{q^2}{\mN^2} F_{\Phi''}^{(N, N')} \right) .
\end{align}
\end{subequations}
Apparent sign differences with the results of Ref.~\cite{Anand:2013yka} can be traced back to the different definition of $\bol{q}$: for instance, here $V_0^N = - \bol{V}^N \cdot \hat{\bol{q}}$ (see discussion related to \Eq{b = - q}). \Eq{M^2 form factors fifj} (or alternatively \Eq{M^2 form factors RXY}) is the formula on which Refs.~\cite{Fitzpatrick:2012ix, DelNobile:2013sia, Anand:2013yka} base their results and the simplicity and efficacy of their approach, which rests on the separation between the particle physics (model dependent, parametrized by the $f_i^N$'s) and the nuclear physics (completely factored within the $F_{i, j}^{(N, N')}$'s, computed once and for all) of DM-nucleus scattering.

\section{Scattering cross section}
\label{sigma}
In \Sec{Form factors} we saw how to derive the DM-nucleus squared scattering amplitude from the DM-nucleon interaction. Equipped with this knowledge, we are now ready to compute the DM-nucleus scattering cross section. In this Section we start by recalling the definition of cross section, and then work out in detail some specific examples: the SI and SD interactions, general interactions mediated by spin-$0$ and spin-$1$ heavy and light bosons (\eg the SM Higgs or the $Z$ boson), and electromagnetic interactions via the DM magnetic dipole moment. Our introduction in \Sec{Differential cross section} encompasses both elastic and inelastic scattering, while we restrict the discussion to elastic scattering when dealing with the specific examples in the rest of this Section.

\subsection{Differential cross section}
\label{Differential cross section}
In terms of the scattering amplitude $\Mel$ between states normalized as in \Eq{statenorm}, the DM-nucleus differential scattering cross section can be written as
\beq\label{cross section}
\ud \sigma_T = \frac{| \Mel |^2}{\rho(p) \rho(k) \, v_\text{M}} \ud \Phi^{(2)} \ ,
\eeq
with
\beq\label{phase space}
\ud \Phi^{(2)} \equiv (2 \pi)^4 \delta^{(4)}(\fvec{p}' + \fvec{k}' - \fvec{p} - \fvec{k}) \frac{\ud^3 p'}{(2 \pi)^3 \rho(p')} \frac{\ud^3 k'}{(2 \pi)^3 \rho(k')}
\eeq
the $2$-body phase space. The M{\o}ller velocity factor
\beq
v_\text{M} \equiv \sqrt{(\bol{v}_\DM - \bol{v}_T)^2 - (\bol{v}_\DM \times \bol{v}_T)^2}
\eeq
reduces to the relative speed in any reference frame where the DM-particle and nuclear motions are collinear, including the CM and lab frames. If the experiment does not control the initial polarizations of the colliding particles, we can average the cross section over the initial spins (which are usually assumed to be randomly distributed). If the experiment does not measure the final polarizations, we can sum the cross sections corresponding to the different DM and nuclear spins. Since DM and nuclear spins are not known in direct detection experiments, we then will, for practical applications, substitute to $| \Mel |^2$ in \Eq{cross section} the unpolarized squared matrix element
\beq\label{Mel^2}
\overline{| \Mel |^2} \equiv \frac{1}{2 s_\DM + 1} \frac{1}{2 \amom{J} + 1} \sum_{\substack{\text{initial \&} \\ \text{final spins}}} | \Mel |^2 \ ,
\eeq
with $s_\DM$ and $\amom{J}$ the DM and nuclear spin, respectively.

With our choice~\eqref{rhohere} for the state normalization, both $\braket{\bol{\genp{p}}}{\bol{\genp{p}}'}$, the integral measure $\ud^3 \genp{p} / (2 \pi)^3 \rho(\genp{p})$, the unpolarized squared matrix element $\overline{| \Mel |^2}$, and the phase space $\ud \Phi^{(2)}$, are Lorentz invariants. This implies that the unpolarized cross section is also Lorentz invariant, in fact the denominator in \Eq{cross section} is proportional to the Lorentz scalar
\beq
E_p E_k \, v_\text{M} = \sqrt{(\fvec{p} \cdot \fvec{k})^2 - \mDM^2 m_T^2} = \tilde{p} \sqrt{s} \ ,
\eeq
where $\tilde{p}$ is the momentum of the two initial particles in the CM frame, see \Eq{tildep^2}. Notice however that the normalization factor $\rho$ is unphysical, and physical observables as the cross section in \Eq{cross section} do not depend on it. All normalization factors appearing in the denominator of \Eq{cross section} and in the phase space in \Eq{phase space} cancel in fact with those implicit in the squared matrix element.

With the adopted normalization~\eqref{rhohere}, and denoting $\ud^3 \tilde{p}' = {\tilde{p}}^{\prime 2} \, \ud \tilde{p}' \, \ud \Omega$, with $\ud \Omega \equiv \ud \cos\theta \, \ud \phi$ the differential solid scattering angle in the CM frame, the energy-momentum conservation delta functions in \Eq{phase space} can be integrated to obtain
\beq
\ud \Phi^{(2)} = \frac{1}{4 \pi} \frac{\tilde{p}'}{\sqrt{s}} \frac{\ud \Omega}{4 \pi} = \frac{\ud \Omega}{32 \pi^2} \sqrt{1 + \frac{({\mDM'}^2 - m_T^2)^2}{s^2} - 2 \frac{{\mDM'}^2 + m_T^2}{s}} \ ,
\eeq
with $\tilde{p}'$ given by \Eq{tildep^2} provided one substitutes $\mDM$ with the final DM particle mass $\mDM'$. The differential cross section in \Eq{cross section} is therefore
\beq
\ud \sigma_T = \frac{| \Mel |^2}{4 E_p E_k \, v_\text{M}} \ud \Phi^{(2)} = \frac{\tilde{p}'}{\tilde{p}} \frac{| \Mel |^2}{16 \pi s} \frac{\ud \Omega}{4 \pi} \ ,
\eeq
with $\tilde{p}' = \tilde{p}$ for elastic scattering. At leading order in the NR limit we get for the phase space
\beq
\ud \Phi^{(2)} \NReq \frac{\ud \Omega}{16 \pi^2} \frac{\mu_T}{\mDM + m_T} v \sqrt{1 - \frac{2 \delta}{\mu_T v^2}} \ ,
\eeq
where as specified in \Sec{Inelastic scattering} we treat the mass splitting $\delta = \mDM' - \mDM$ as a parameter of order $\Op(\mu_T v^2)$. The NR cross section is then, in the lab frame,
\beq
\ud \sigma_T \NReq \frac{\ud \Omega}{64 \pi^2} \frac{| \Mel |^2}{(\mDM + m_T)^2} \sqrt{1 - \frac{2 \delta}{\mu_T v^2}} \ .
\eeq
Since the unpolarized squared matrix element can only depend, in terms of CM-frame variables, on $\tilde{\bol{p}}$ and $\tilde{\bol{p}}'$, and thus, through $\cos\theta \propto \tilde{\bol{p}} \cdot \tilde{\bol{p}}'$, on the $\theta$ angle alone, we can promptly integrate over $\ud \phi$ in $[0, 2 \pi]$ obtaining
\beq
\frac{\ud \sigma_T}{\ud \cos\theta} \NReq \frac{1}{32 \pi} \frac{\mu_T^2}{\mDM^2 m_T^2} \overline{| \Mel |^2} \sqrt{1 - \frac{2 \delta}{\mu_T v^2}} \ .
\eeq
The total cross section is obtained by integrating over $\ud \cos\theta$ in $[-1, 1]$.

To obtain the differential cross section in nuclear recoil energy, we can use \Eq{ERinel} to write
\beq
\ud \ER \NReq - \frac{\mu_T^2 v^2}{m_T} \sqrt{1 - \frac{2 \delta}{\mu_T v^2}} \, \ud \cos\theta \ .
\eeq
The overall minus sign on the right-hand side is in agreement with the fact that $\ER$ decreases as $\cos\theta$ increases, and in fact the lower (upper) integration extremum $\ER^-$ ($\ER^+$) for $\ER$ at fixed $v$ corresponds to $\cos\theta = +1$ ($-1$), see \Eq{ER^pm}.\footnote{We are using here the notation of \Sec{Inelastic scattering}, for generic mass splitting $\delta$. For elastic scattering ($\delta = 0$, see \Sec{Elastic scattering}), we recall our notation $\ER^+(v) = \ER^\text{max}(v)$ and $\ER^-(v) = 0$, see \Eq{ERinterval}.} Therefore, when integrating, one has
\beq
\int_{\ER^-}^{\ER^+} \ud \ER \overset{\text{\tiny{NR}}}{\propto} - \int_{+1}^{-1} \ud \cos\theta = + \int_{-1}^{+1} \ud \cos\theta \ ,
\eeq
so that the minus sign is taken care of by arranging the integration extrema in the natural way. In the following, as customary, we neglect the minus sign with the understanding that the integration extrema must be naturally arranged. Finally, the differential cross section in $\ER$ reads
\beq\label{dsigmadER}
\frac{\ud \sigma_T}{\ud \ER} \NReq \frac{1}{32 \pi} \frac{1}{\mDM^2 m_T} \frac{1}{v^2} \overline{| \Mel |^2} \ .
\eeq

If the scattering amplitude does not depend on $\bol{v}$, the velocity dependence of the differential cross section is $\ud \sigma_T / \ud \ER \propto 1 / v^2$. The differential scattering rate in Eqs.~\eqref{diffrate},~\eqref{Taylor rate} then features uniquely the $\eta_0$ velocity integral defined in \Eq{eta},
\beq\label{eta_0}
\eta_0(\vmin, t) = \int_{v \geqslant \vmin} \ud^3 v \, \frac{\fE(\bol{v}, t)}{v} \ .
\eeq
This is the most common situation since both the SI and SD interactions are of this type, as one can see from the related NR operators~\eqref{ONR SI},~\eqref{ONR SD} not depending on $\bol{v}$. Different velocity dependences imply different velocity integrals, which, as can be seen in Eqs.~\eqref{diffrate},~\eqref{Taylor rate}, affect both the $\ER$ dependence of the differential rate (through the $\vmin(\ER)$ function) and its time dependence (see \Sec{Modulation}). The velocity integral is, in fact, the sole source of time dependence of the rate, so that, given $\fE(\bol{v}, t)$ (that is, given the local DM velocity distribution and knowing Earth's motion in the galactic frame with accuracy), the time dependence of a putative DM signal may be traced back to the velocity dependence of the differential scattering cross section.

If $\overline{| \Mel |^2}$ does not depend on $\cos \theta$ (\ie on $\ER$), $\ud \sigma_T / \ud \cos\theta$ does not depend on $\cos\theta$ either (equivalently, $\ud \sigma_T / \ud \ER$ does not depend on $\ER$) and the scattering is isotropic. This happens when the DM-nucleon scattering amplitude $\Mel_N$, or equivalently the NR interaction operator $\Op_\NR^N$~\eqref{general NR Op}, does not depend on $q$. For the DM-nucleus system, however, isotropic scattering can only be an approximation since the scattering probability gets suppressed at large values of the momentum transfer (thus at large $\ER$ values) due to coherence loss: since a larger $q$ probes a smaller area, at large momentum transfer the scattering occurs practically with single nucleons rather than coherently with the whole nucleus (see discussion in \Sec{Form factors}). Furthermore, even when interacting effectively with single nucleons, the scattering probability decreases with energy as the single-nucleon wave function is spread out over much of the nucleus~\cite{Engel:1991wq}. Exactly isotropic scattering can thus only occur in the limit of point-like nucleus. If, in this limit, $\ud \sigma_T / \ud \ER$ does not indeed depend on $\ER$, it is customary to define the \emph{zero-momentum transfer cross section}
\beq\label{sigma_0}
\sigma_0 \equiv \int_{\ER^-(v)}^{\ER^+(v)} \left. \frac{\ud \sigma_T}{\ud \ER} \right|_{\ER = 0} \, \ud \ER = \left. \frac{\ud \sigma_T}{\ud \ER} \right|_{\ER = 0} (\ER^+(v) - \ER^-(v)) \ .
\eeq
$\sigma_0$ is not the DM-nucleus total cross section (as also stressed in Ref.~\cite{Jungman:1995df}), as it does not take into account the size of the nucleus and the coherence loss at large momentum transfer. This effect is often parametrized by a single nuclear form factor $F_T(\ER)$, which is generally normalized as $F_T(0) = 1$ so that setting the form factor to $1$ in the differential cross section corresponds to taking the limit of point-like nucleus. If the scattering is isotropic in this limit we can then factorize $\ud \sigma_T / \ud \ER$ into $F_T^2(\ER)$ times an energy-independent factor that can be read off directly from \Eq{sigma_0}:
\beq\label{sigma_0 F_T}
\frac{\ud \sigma_T}{\ud \ER} \NReq \sigma_0 \frac{m_T}{2 \mu_T^2 v^2} \left( 1 - \frac{2 \delta}{\mu_T v^2} \right)^{- 1/2} F_T^2(\ER) \ .
\eeq
Since the loss of coherence causes the cross section to decrease at large energies, $F_T^2(\ER) \leqslant 1$ and thus $\sigma_0$ is larger than the DM-nucleus total cross section. The usefulness of introducing $\sigma_0$ stands in that it conveniently parametrizes the overall size of the differential cross section, as clear in \Eq{sigma_0 F_T}. This only happens if, in the limit of point-like nucleus, $\overline{| \Mel |^2}$ does not depend on $\ER$ at leading order in the NR expansion; in other words, the convenience of the linear parametrization in \Eq{sigma_0 F_T} relies on the scattering being (approximately) isotropic in the limit of point-like nucleus. \Eq{sigma_0} does not provide a useful parametrization if, for instance, the leading NR operator is proportional to a positive or negative power of $q$ (the latter possibility stemming \eg from a massless $t$-channel mediator as in \Eq{Millicharged sigma} below), or has a more involved $q$ dependence (see \eg Secs.~\ref{Vector-mediated interaction},~\ref{Magnetic-dipole DM}). A concrete example is provided by the SD interaction, which only features isotropic scattering in the limit where the induced pseudo-scalar contribution (the $\Op^N_6$ term in \Eq{ONR SD}) is negligible, see \Sec{SD interaction}: in fact, all the terms of the differential cross section that vanish at zero momentum transfer do not enter \Eq{sigma_0}, implying that $\sigma_0$ is only representative of the size of the differential cross section when those terms can be neglected. Other limitations to the usefulness of the parametrization in \Eq{sigma_0 F_T} concern $F_T(\ER)$, as we will now discuss more in general.

Given the restrictions on the applicability or the usefulness of \Eq{sigma_0 F_T}, one may avoid employing $\sigma_0$ while still parametrizing the $\ER$ dependence of the differential cross section away from the limit of point-like nucleus through a single nuclear form factor,
\beq\label{F_T}
\frac{\ud \sigma_T}{\ud \ER} = \left. \frac{\ud \sigma_T}{\ud \ER} \right|_\text{PLN} F_T^2(\ER) \ ,
\eeq
with $F_T(\ER)$ normalized as above. This returns \Eq{sigma_0 F_T} when the DM-nucleon interaction is inherently independent of $q$, as in that case the point-like nucleus limit coincides with the limit of zero momentum transfer. Two instances of the parametrization in \Eq{F_T} are \Eq{diffsigmaSI 1} and \Eq{diffsigmaSD-PS}, respectively for the SI and SD interaction, where, as mentioned above, the latter only features isotropic scattering when the contribution of the induced pseudo-scalar interaction (the $F_\text{PS}^{(N, N')}$ terms in \Eq{SDMel^2}) can be neglected. Notice that $F_T(\ER)$ in \Eq{F_T} is different from the form factors introduced in \Sec{Form factors}, which are specific to the proton or neutron content of a given nuclide, and also to the form of the interaction, since different interactions probe different properties of the nucleus, such as its electric charge, its mass number, its spin, its magnetic moment, and so on. Instead, $F_T^2(\ER)$ here merely parametrizes the $\ER$ dependence of the differential cross section away from the limit of point-like nucleus. Paradoxically, were the cross section to result from two distinct interactions of the DM with nuclei (\eg the SI and SD interactions together, or two interactions involving different mediators), $F_T^2(\ER)$ would be a non-trivial combination of the different DM and nucleon coupling constants as well as of the nucleon-specific and interaction-specific form factors of \Sec{Form factors}. Even with a single interaction, in general $F_T^2(\ER)$ depends non-trivially on the DM-proton and DM-neutron couplings (see \eg \Eq{diffsigmaSD-PS} and subsequent discussion for the SD interaction). However, this does not happen if the $\ER$ dependence of the DM-nucleus interaction is the same for DM interacting with protons and DM interacting with neutrons, as usually assumed for the SI interaction (see discussion above \Eq{SIMel^2}). Also, \Eq{F_T} may provide a convenient parametrization when the ratio of the DM-proton and DM-neutron couplings is fixed, in which case $F_T^2(\ER)$ does not depend on the one independent coupling (while depending on the coupling ratio): for instance when the interaction does not distinguish between protons and neutrons (so that the only difference stems from their distribution within the nucleus), or when the DM couples to only one type of nucleons. In general, in any event, one should bear in mind that, despite being usually called `nuclear form factor', $F_T(\ER)$ in the parametrization~\eqref{F_T} does not depend solely on the nuclear properties.

We presented here some general considerations about the DM-nucleus differential scattering cross section. In the following we consider for illustration some specific interactions and compute the relative cross section for elastic scattering.

\subsection{Spin-independent interaction}
\label{SI interaction}
The traditional DM-quark effective Lagrangian giving rise to the SI interaction is
\beq\label{SILag}
\Lag = \bar{\chi} \chi \sum_q c_q \, \bar{q} q \ ,
\eeq
with $\chi$ a spin-$1/2$ DM particle. This effective interaction may be obtained for instance by exchange of a scalar mediator between DM and quarks. As can be seen in \Tab{tab: spin-1/2 DM EFT}, \Sec{Scalar couplings}, and \Eq{NR building blocks}, the interaction in \Eq{SILag} gives rise to the NR operator
\beq\label{ONR SI}
\Op_\NR^N = 4 f_N \mDM \mN \, \Op^N_1
\eeq
with $\Op^N_1 = \unop$ and
\beq
f_N = \sum_q c_q \frac{\mN}{m_q} f_{Tq}^{(N)} \ .
\eeq

This same result can be also obtained with other interaction Lagrangians. For instance,
\beq\label{SILag GG}
\Lag = \frac{c_g \aS}{12 \pi} \, \bar{\chi} \chi \, G^{a \mu\nu} G^a_{\mu\nu}
\eeq
yields \Eq{ONR SI} with
\beq
f_N = - \frac{2}{27} c_g \mN f_{TG}^{(N)} \ .
\eeq
Also the Lagrangian
\beq
\Lag = \bar{\chi} \gamma^\mu \chi \sum_q c_q \, \bar{q} \gamma_\mu q \ ,
\eeq
which can be generated with quarks and Dirac DM coupling to a heavy vector boson (see \eg \Sec{Vector-mediated interaction}), generates \Eq{ONR SI} with
\beq
f_N = \sum_q c_q F_1^{q, N}(0) \ ,
\eeq
see \Tab{tab: spin-1/2 DM EFT} and \Sec{Vector couplings}. Notice that, in this case, the time component of $\bar{u}' \gamma^\mu u$ contributes at leading order in the NR expansion while the space components are NR suppressed, as can be seen in Eqs.~\eqref{fermionbilinears},~\eqref{Nfermionbilinears}. Finally, for a scalar DM particle $\phi$, the operator
\beq\label{SILag phi}
\Lag = \phi^\dagger \phi \sum_q c_q \, \bar{q} q + \frac{c_g \aS}{12 \pi} \, \phi^\dagger \phi \, G^{a \mu\nu} G^a_{\mu\nu}
\eeq
generates again \Eq{ONR SI} with
\beq
f_N = \frac{1}{2 \mDM} \sum_q \frac{\mN}{m_q} c_q f_{Tq}^{(N)} - \frac{1}{27} \frac{\mN}{\mDM} f_{TG}^{(N)} \ ,
\eeq
as can be seen from \Tab{tab: spin-0 DM EFT}. Notice that extending Eqs.~\eqref{SILag},~\eqref{SILag GG} as to include other (pseudo-)scalar operators,
\begin{multline}
\Lag = \bar{\chi} \chi \left( \sum_q c_q \, \bar{q} q + \sum_q \tilde{c}_q \, \bar{q} \, i \gamma^5 q + \frac{c_g \aS}{12 \pi} \, G^{a \mu\nu} G^a_{\mu\nu} + \frac{\tilde{c}_g \aS}{8 \pi} \, G^{a \mu\nu} \tilde{G}^a_{\mu\nu} \right)
\\
+ \bar{\chi} \, i \gamma^5 \chi \left( \sum_q c'_q \, \bar{q} q + \sum_q \tilde{c}'_q \, \bar{q} \, i \gamma^5 q + \frac{c'_g \aS}{12 \pi} \, G^{a \mu\nu} G^a_{\mu\nu} + \frac{\tilde{c}'_g \aS}{8 \pi} \, G^{a \mu\nu} \tilde{G}^a_{\mu\nu} \right) ,
\end{multline}
or \Eq{SILag phi} for a scalar DM field,
\beq
\Lag = \phi^\dagger \phi \left( \sum_q c_q \, \bar{q} q + \sum_q \tilde{c}_q \, \bar{q} \, i \gamma^5 q + \frac{c_g \aS}{12 \pi} \, \phi^\dagger \phi \, G^{a \mu\nu} G^a_{\mu\nu} + \frac{\tilde{c}_g \aS}{8 \pi} \, G^{a \mu\nu} \tilde{G}^a_{\mu\nu} \right) ,
\eeq
still yields \Eq{ONR SI} at leading order in the NR expansion. This is because the contributions from all involved NR operators other than $\Op^N_1$ are subdominant (unless of course $f_N = 0$), being suppressed by powers of the momentum transfer (see \eg \Sec{Scalar-mediated interaction}). The above scalar couplings to gluons and heavy quarks are source of potentially relevant $2$-body contributions to the DM-nucleus scattering amplitude, see \eg Refs.~\cite{Hoferichter:2016nvd, Hoferichter:2018acd}, while the pseudo-scalar gluon couplings induce a large isospin violation, see Ref.~\cite{Gross:1979ur}.

This interaction is historically called the \emph{spin-independent} (SI) interaction, as opposed to the \emph{spin-dependent} (SD) interaction discussed in \Sec{SD interaction} whose corresponding NR operator depends on the nucleon spin. It is easy to see that also the $\Op_\NR^N = q^2 \Op^N_1$ and the $\Op_\NR^N = \Op^N_{11} = i \, \bol{s}_\chi \cdot \bol{q}$ interactions, for instance, do not depend on the nucleon spin; similarly, the $\Op_\NR^N = q^2 \Op^N_4 = q^2 \, \bol{s}_\chi \cdot \bol{s}_N$ or the $\Op_\NR^N = \Op^N_{10} = i \, \bol{s}_N \cdot \bol{q}$ interactions, too, do depend on the nucleon spin. However, historically the \emph{spin-independent} and \emph{spin-dependent} terminology describes exclusively the $\Op_\NR^N = c_1^N \Op^N_1$ and $\Op_\NR^N = c_4^N \Op^N_4$ interactions respectively, with coefficients (approximately) independent of $q^2$ and ${v^\perp}^2$ as in \Eq{ONR SI} (the $\Op^N_6$ building block also contributes to the SD interaction, as we will see in \Sec{SD interaction}). As we will see below, the most archetypical SI interaction also features equal DM couplings to proton and neutron, $c_1^p = c_1^n$.

Given that the $\Op^N_1 = \unop$ operator acts trivially on nucleon states, its nuclear matrix element in the limit of point-like nucleus is given simply by a sum of diagonal amplitudes describing DM scattering with one nucleon at a time. Therefore, in this limit, which coincides with the limit of zero momentum transfer since the DM-nucleon interaction does not depend on $\bol{q}$, the $\Op^N_1$ interaction basically `counts' the number of nucleons in the nucleus. Another way to see this is by noticing that this same NR interaction is induced at zero momentum transfer by DM couplings to the nucleon number-density operator $\bar{N} \gamma^0 N$, \eg in the effective DM-nucleon Lagrangian operator $\bar{\chi} \gamma_\mu \chi \, \bar{N} \gamma^\mu N$ whose matrix element is dominated by the time component of the scalar product since the spatial part is NR suppressed and thus negligible (see \Sec{DM-N}). Therefore we expect the scattering amplitude to be proportional to the number of nucleons taking part in the interaction, and the cross section being proportional to its square, which means nuclei with large mass numbers have much larger interaction probability than nuclei with few nucleons. This large enhancement is due to the fact that DM interactions with any one nucleon interfere (constructively) with the interactions with all other nucleons. Interactions yielding constructive interference among nucleons (see \eg Refs.~\cite{Fitzpatrick:2012ix, Vietze:2014vsa, Hoferichter:2016nvd, Hoferichter:2018acd}) are denoted, in jargon, \emph{coherent}, despite the fact that interactions where the interference is destructive (\eg the SD interaction) are also coherent in a quantum-mechanical sense.

The DM-nucleus scattering amplitude can then be obtained in the limit of point-like nucleus (see \Sec{nuc form factors}) by naively summing over all nucleons the (trivial) DM-nucleon matrix element of $\Op_\NR^N$ in \Eq{ONR SI}, multiplied by a factor $2 m_T / 2 \mN$ due to the change in state normalization in going from nucleon to nuclear states (see \Eq{statenormhere NR}). One then obtains for a point-like nucleus
\beq\label{SIMel}
\Mel_\text{PLN} \NReq 4 \mDM m_T (Z f_p + (A - Z) f_n) \, \delta_{s s'} \delta_{\amom{M} \amom{M}'} \ ,
\eeq
where $Z$ and $A$ are the atomic and mass number of the target nucleus, respectively, and $s$, $\amom{M}$ ($s'$, $\amom{M}'$) are the initial (final) DM and nuclear spin projections along the quantization axis, respectively.

As we saw in \Sec{nuc form factors}, the effect of the nucleus being spatially extended can be taken into account with an appropriate form factor, which in this case corresponds to the Fourier transform of the nucleon number density inside the nucleus. While this density may in principle be different for protons and neutrons, it is usually assumed that the neutron and proton number densities are the same, which implies the same form factor for both. We denote this form factor as $F_\text{SI}(\ER)$, normalized so that $F_\text{SI}(0) = 1$. The unpolarized squared matrix element in \Eq{Mel^2} reads therefore
\beq\label{SIMel^2}
\overline{| \Mel |^2} \NReq 16 \mDM^2 m_T^2 (Z f_p + (A - Z) f_n)^2 F_\text{SI}^2(\ER) \ ,
\eeq
where both $f_N$ and $F_\text{SI}$ are real. We saw in Secs.~\ref{Multipoles},~\ref{Nuclear matrix element} that only the $M$ nuclear response contributes to the scattering in this case (see \Tab{tab: nuclear responses} and Refs.~\cite{Fitzpatrick:2012ix, Anand:2013yka}), and in fact a quick comparison with Eqs.~\eqref{M^2 form factors fifj},~\eqref{F_ij} shows that
\beq\label{F_11}
\CMcal{N}_N \CMcal{N}_{N'} F_\text{SI}^2(\ER) = F_{1, 1}^{(N, N')}(q^2, {v^\perp_T}^2) = F_M^{(N, N')}(q^2) \ ,
\eeq
with $\CMcal{N}_p \equiv Z$ and $\CMcal{N}_n \equiv A - Z$ the number of protons and neutrons, respectively.

The DM-nucleus differential scattering cross section averaged over initial spins and summed over final spins, \Eq{dsigmadER}, is
\beq\label{diffsigmaSI 1}
\frac{\ud \sigma_T}{\ud \ER} \NReq \frac{m_T}{2 \pi v^2} (Z f_p + (A - Z) f_n)^2 F_\text{SI}^2(\ER) \ .
\eeq
For a single, point-like nucleon we can easily integrate the differential scattering cross section (using \Eq{ERinterval}) to obtain the total cross section
\beq\label{SI sigma_N}
\sigma_N \NReq \frac{\mu_\text{N}^2}{\pi} f_N^2 \ .
\eeq
We can use this to parametrize the DM-nucleus differential cross section as
\beq\label{diffsigmaSI 2}
\frac{\ud \sigma_T}{\ud \ER} \NReq \frac{m_T}{2 \mu_\text{N}^2 v^2} \sigma_p \left( Z + (A - Z) \frac{f_n}{f_p} \right)^2 F_\text{SI}^2(\ER) \ .
\eeq
The zero-momentum transfer cross section in \Eq{sigma_0} reads
\beq
\sigma_0 = \frac{\mu_T^2}{\pi} (Z f_p + (A - Z) f_n)^2 \NReq \sigma_p \frac{\mu_T^2}{\mu_\text{N}^2} \left( Z + (A - Z) \frac{f_n}{f_p} \right)^2 \ .
\eeq
The parametrization~\eqref{sigma_0 F_T} of the differential cross section is justified in this case by the DM-nucleon interaction (\ie $\Op_\NR^N$ in \Eq{ONR SI}) being inherently independent of $q$, as well as by the assumption that the nuclear form factor is the same for DM interactions with protons and with neutrons, as discussed above.

The most studied choice of couplings is the isosinglet condition $f_p = f_n$, which is the model the experimental collaborations use to set bounds on the SI interaction. In this case we get
\begin{align}
\label{isosinglet sigma}
\frac{\ud \sigma_T}{\ud \ER} \NReq \frac{m_T}{2 \mu_\text{N}^2 v^2} \sigma_p A^2 F_\text{SI}^2(\ER) \ ,
&&&
\sigma_0 = \frac{\mu_T^2}{\pi} A^2 f_p^2 \NReq \sigma_p \frac{\mu_T^2}{\mu_\text{N}^2} A^2 \ .
\end{align}
The $A^2$ enhancement of the DM-nucleus cross section with respect to the DM-nucleon cross section can be quite sizeable, and is due to the fact that all nucleons contribute coherently to the scattering as mentioned above. Therefore, for fixed DM-nucleon cross section, nuclei with large $A$ such as Xe and I (see \Tab{tab: nuclides}) are clearly favored in searching for DM with SI interactions. As we will see in \Sec{Rate spectrum}, however, upon (somewhat artificially) fixing the minimum $\ER$ value that can be detected by experiments, the sensitivity of detectors featuring lighter targets extends to smaller DM masses, see Figs.~\ref{fig: minDMmass} and~\ref{fig: bounds}.

The model in \Eq{isosinglet sigma} has only two free parameters, $\mDM$ and $\sigma_p$ (or alternatively $\mDM$ and $f_p^2$). Experimental bounds on the SI interaction are thus usually expressed as an upper constraint on $\sigma_p$ at any given DM mass, for isosinglet couplings. This constraint, call it $\bar{\sigma}^\text{SI}_p(\mDM)$, can be recast in some circumstances to the case of isospin-violating couplings $f_n \neq f_p$ (note that for isospin invariance to be preserved with $f_n \neq f_p$, the DM should be charged under isospin). This can be done exactly if the target material only features one target nuclide, \ie one isotope of a single target element. This is the case for instance if the detector is entirely made of Ar, see \Tab{tab: nuclides}; or if the detector features a compound material but the main contribution to the signal can be ascribed for some reason to DM scattering off a single nuclide, \eg if the detector is made of NaI but the DM is too light to produce a signal above threshold for scattering off iodine, so that the entirety of the signal can be attributed to scattering off sodium. In this case a comparison of the detection rate~\eqref{dRdEd} for the interaction in \Eq{isosinglet sigma} with that for \Eq{diffsigmaSI 2} amounts to a direct comparison of the two differential cross sections, which returns
\beq\label{barsigma}
\bar{\sigma}^\text{SI}_p(\mDM) > \sigma_p \frac{(Z + (A - Z) f_n / f_p)^2}{A^2} = \frac{\mu_\text{N}^2}{\pi} \frac{(Z f_p + (A - Z) f_n)^2}{A^2} \ .
\eeq
The $\bar{\sigma}^\text{SI}_p(\mDM)$ constraint on the DM-proton cross section in the model of \Eq{isosinglet sigma} is then recast into a bound on a certain combination of $f_p$ and $f_n$, or alternatively $\sigma_p$ and $f_n / f_p$ (or even $\sigma_n$ and $f_p / f_n$), for the interaction in Eqs.~\eqref{diffsigmaSI 1},~\eqref{diffsigmaSI 2}. If more isotopes of a single element are present in the detector, meaning different nuclides with same $Z$ and similar values of $A$ as for xenon detectors, this procedure may be generalized to obtain an approximate bound, as illustrated \eg in Ref.~\cite{Feng:2011vu}. In fact, assuming the $\bar{\sigma}^\text{SI}_p(\mDM)$ constraint has been derived from the annual-average term alone in Eqs.~\eqref{eta bar tilde},~\eqref{Taylor R}, as is usually the case, and upon checking that the function $\overline{\eta}_0(\vmin(\ER)) \, F_\text{SI}^2(\ER) \, m_T / \mu_T^2$ (see \Eq{Rate SI} below) is approximately the same in the $\ER$ range of interest for all isotopes, one can approximately write
$\bar{\sigma}^\text{SI}_p(\mDM) > D \sigma_p$ with
\beq\label{D factor}
D \equiv \frac{\sum_{T} \zeta_T (Z + (A - Z) f_n / f_p)^2 \mu_T^2 / m_T}{\sum_{T} \zeta_T A^2 \mu_T^2 / m_T} = \frac{\sum_{T} \xi_T (Z + (A - Z) f_n / f_p)^2 \mu_T^2}{\sum_{T} \xi_T A^2 \mu_T^2}
\eeq
(see \Fig{fig: Dfactor+Rdelta} below), where in the last equality we used \Eq{zeta_T}. However, if two or more elements with very different mass number take part in the scattering, this simple recast of the isosinglet constraint cannot be straightforwardly performed.

Some phenomenological consequences of isospin-violating couplings $f_n \neq f_p$ were studied \eg in Refs.~\cite{Kurylov:2003ra, Giuliani:2005my, Feng:2011vu}, where it was pointed out that specific values of the $f_n / f_p$ ratio decrease the sensitivity of experiments employing a certain target (see the left panel of \Fig{fig: Dfactor+Rdelta} below). In fact, it can be seen from \Eq{diffsigmaSI 2} that the scattering cross section vanishes for a specific nuclide if one chooses $f_n / f_p = - Z / (A - Z)$. However, if the detector material contains more than one element or more than one isotope, the rate gets suppressed with respect to the isospin-conserving case but it does not vanish. Notice also that the actual value of $f_n / f_p$ for which the cross section is suppressed (the dips in the left panel of \Fig{fig: Dfactor+Rdelta}) receives potentially large long-distance QCD corrections that can be computed in a chiral expansion~\cite{Cirigliano:2012pq, Cirigliano:2013zta}.

\begin{figure}[t]
\begin{center}
\includegraphics[width=.49\textwidth]{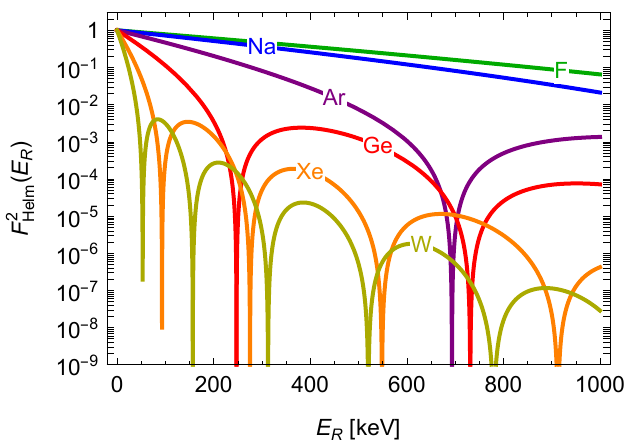}
\includegraphics[width=.49\textwidth]{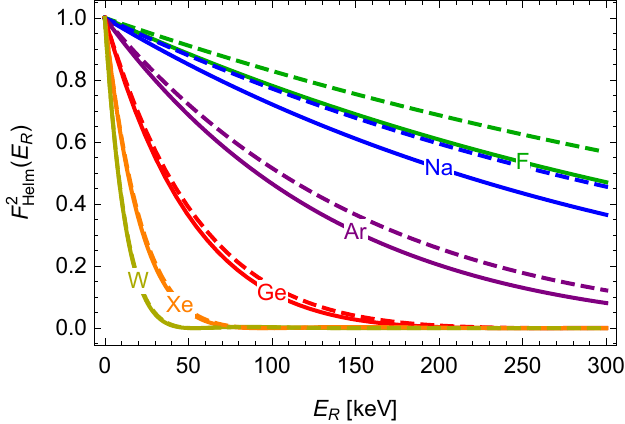}
\caption[Helm form factor (squared)]{\label{fig: HelmFF}\emph{Square of the Helm form factor in \Eq{HelmFF} with the choice of parameters in \Eq{Helm parameters 2}. For each element, the most abundant isotope (same as in the right panel of \Fig{fig: reduced mass + qR=1}) has been chosen as representative. \textbf{Left:} logarithmic scale. \textbf{Right:} linear scale. The dashed lines correspond to the choice of parameters in \Eq{Helm parameters 1}.}}
\figcode
\end{center}
\end{figure}

As mentioned above, the nuclear form factor $F_\text{SI}$ can be thought of as the Fourier transform of the nucleon number density~\cite{Jungman:1995df}, see \Sec{nuc form factors}. Since the nuclear number density is poorly known, it is usually approximated with the nuclear charge density (in the assumption the two approximately coincide, see however Refs.~\cite{Duda:2006uk, Orrigo:2016wgu}), which is determined through elastic electron scattering. The nuclear charge density for an isotropic nuclear ground state can be parametrized with a uniform density $\rho_\text{unif}$ (as if the nucleus were a hard sphere) convoluted with a Gaussian surface-smearing density $\rho_\text{Gauss}$~\cite{Helm:1956zz},
\begin{align}
\rho_\text{unif}(\bol{x}) \equiv \frac{1}{\frac{4}{3} \pi R^3} \, \Theta(R - x) \ ,
&&&
\rho_\text{Gauss}(\bol{x}) \equiv \frac{e^{- x^2 / 2 s^2}}{(2 \pi s^2)^{3/2}} \ ,
\end{align}
with $R$ a measure of the nuclear radius and $s$ a measure of the nuclear skin thickness. The respective Fourier transforms are
\begin{align}
F_\text{unif}(q^2) = \frac{3 j_1(q R)}{q R} \ ,
&&&
F_\text{Gauss}(q^2) \equiv e^{- (q s)^2 / 2} \ ,
\end{align}
with
\beq
j_1(x) \equiv \frac{\sin(x) - x \cos(x)}{x^2}
\eeq
the order-$1$ spherical Bessel function of the first kind. We have then for the nuclear charge density
\beq
\rho_\text{Helm}(\bol{x}) \equiv \int \ud^3 x' \, \rho_\text{unif}(\bol{x}') \, \rho_\text{Gauss}(\bol{x} - \bol{x}') \ ,
\eeq
whose form factor is
\beq\label{HelmFF}
F_\text{Helm}(q^2) = F_\text{unif}(q^2) F_\text{Gauss}(q^2) = \frac{3 j_1(q R)}{q R} \, e^{- (q s)^2 / 2} \ .
\eeq
The Helm form factor~\cite{Helm:1956zz} is the most used form factor for the SI interaction. The following parameter values were suggested in Ref.~\cite{Engel:1991wq} for the nuclei of interest in direct DM detection,
\begin{align}
\label{Helm parameters 1}
R &= \sqrt{\tilde{R}^2 - 5 s^2} \ ,
&
\tilde{R} &= 1.2 A^{1/3}~\fm \ ,
&
s &= 1~\fm \ ,
\end{align}
a choice also endorsed in Refs.~\cite{Jungman:1995df, Engel:1992bf}. Ref.~\cite{Lewin:1995rx} found instead the following values to provide a good fit to data:
\begin{subequations}
\label{Helm parameters 2}
\begin{align}
R &= \sqrt{c^2 + \tfrac{7}{3} \pi^2 a^2 - 5 s^2} \ ,
&
s &= 0.9~\fm \ ,
\\
a &= 0.52~\fm \ ,
&
c &= (1.23 A^{1/3} - 0.60)~\fm \ .
\end{align}
\end{subequations}
This choice, although with $s = 1~\fm$, agrees with the SI form factors computed in Ref.~\cite{Vietze:2014vsa}, which also show agreement with those of Ref.~\cite{Fitzpatrick:2012ix}. \Fig{fig: HelmFF} displays the squared Helm form factor with the choice of parameters in \Eq{Helm parameters 2} (logarithmic scale in the left panel, linear scale in the right panel). For each target element, the most abundant isotope (same as in the right panel of \Fig{fig: reduced mass + qR=1}) has been chosen as representative. The dashed lines in the right panel correspond to the choice of parameters in \Eq{Helm parameters 1}. Another model for the nuclear charge density is the two-parameter Fermi or Woods-Saxon\footnote{The name is probably due to the fact that this distribution has the same form of the Woods-Saxon potential for the nucleons inside the nucleus.} distribution, given by
\beq
\rho_\text{Woods-Saxon}(\bol{x}) \equiv \frac{\rho_0}{e^{(x - R) / a} + 1} \ ,
\eeq
with normalization $\rho_0 = - \left( 8 \pi a^3 \text{Li}_3 \! \left(- e^{R / a} \right) \right)^{-1}$ with $\text{Li}_3$ the polylogarithm of order $3$. $R$ is here the nuclear radius at half the central density $\rho_0$, while $a$ is the diffuseness of the nuclear surface. From nuclear data we have~\cite{Krane:1987ky}
\begin{align}
R = 1.2 A^{1/3}~\fm \ ,
&&&
t \approx 2.3~\fm \ ,
\end{align}
with $t$ the skin or surface thickness parameter defined as the distance over which the distribution falls from $90 \%$ to $10 \%$ of its value at $r = 0$; in the $R \gg a$ limit we get $t = 4 a \ln 3$, leading to $a \approx 0.52~\fm$. The form factor of the Woods-Saxon distribution, however, must be integrated numerically, which is why the Helm parametrization is often preferred (see however Refs.~\cite{Sprung, Maximon}). Other less model-dependent and more accurate form factors can be found \eg in Refs.~\cite{Duda:2006uk, Orrigo:2016wgu}, together with more information on the Helm and Woods-Saxon form factors. More advanced computations of the SI form factor using Chiral EFT, which account for the difference in proton and neutron distribution and also include $2$-body and other corrections, can be found \eg in Refs.~\cite{Hoferichter:2016nvd, Hoferichter:2018acd}. Further computations can be found \eg in Refs.~\cite{Chen:2011xp, Co:2012adt, Fitzpatrick:2012ix, Vietze:2014vsa}.

\subsection{Spin-dependent interaction}
\label{SD interaction}
The traditional effective DM-quark interaction Lagrangian giving rise to the SD interaction is (see \eg Refs.~\cite{Jungman:1995df, Engel:1992bf, Bednyakov:2004xq})
\beq\label{SDLag}
\Lag = \bar{\chi} \gamma^\mu \gamma^5 \chi \sum_q a_q \, \bar{q} \gamma_\mu \gamma^5 q \ ,
\eeq
with $\chi$ a spin-$1/2$ DM particle. This interaction may be obtained by integrating out a heavy vector mediator with axial couplings to both the DM and the quarks. As can be seen in Eqs.~\eqref{fermionbilinears},~\eqref{Nfermionbilinears}, the time component of $\bar{u}' \gamma^\mu \gamma^5 u$ is NR suppressed with respect to the space components, and therefore only the latter contribute to the nucleon matrix element at leading order in the NR expansion. Since the space components depend on the spin, we expect the NR operator induced by \Eq{SDLag} to depend on both the DM and the nucleon spin. By looking at \Tab{tab: spin-1/2 DM EFT}, \Sec{Axial-vector couplings}, and \Eq{NR building blocks}, in fact, one can see that \Eq{SDLag} gives rise to the NR operator
\beq\label{ONR SD}
\Op_\NR^N = - 16 \mDM \mN \left( a_N \, \Op^N_4 + \tilde{a}_N(q^2) \, \Op^N_6 \right) ,
\eeq
with
\begin{align}
a_N = \sum_q a_q \Delta_q^{(N)} \ ,
&&&
\tilde{a}_N(q^2) = - \sum_q a_q \left( \frac{a_{q, \pi}^N}{q^2 + m_\pi^2} + \frac{a_{q, \eta}^N}{q^2 + m_\eta^2} \right) .
\end{align}
Given its dependence on $\bol{s}_N$ which translates into a dependence on the nuclear spin, the $\Op^N_4$ interaction was dubbed \emph{spin-dependent} (SD), to distinguish it from the SI $\Op^N_1$ interaction discussed in \Sec{SI interaction}. As discussed already in \Sec{NR operators}, these are the only interactions which contribute at zeroth order in the NR expansion, and thus are traditionally the most considered interactions. Also traditionally the $\Op^N_6$ part of the interaction is neglected because vanishing at $q^2 = 0$, contrarily to $\Op^N_4$. However we retain it here since its contribution can become sizeable as $q$ grows to values of order $m_\pi$ and larger, as shown \eg in Ref.~\cite{Bishara:2017pfq}. As can be seen from \Tab{tab: spin-1/2 DM EFT}, another interaction Lagrangian yielding a NR operator proportional to $\Op^N_4$ at zero momentum transfer is
\beq
\Lag = \bar{\chi} \, \sigma^{\mu\nu} \chi \sum_q b_q \, \bar{q} \, \sigma_{\mu\nu} q \ ,
\eeq
while the effective Lagrangian
\beq
\Lag = \bar{\chi} \, i \gamma^5 \chi \sum_q c_q \, \bar{q} \, i \gamma^5 q
\eeq
generates at leading order a NR operator proportional to $\Op^N_6$. Integrating out a heavy vector mediator with general couplings to both the DM and the quarks,
\beq
\Lag = \bar{\chi} \gamma^\mu (v'_\chi + a'_\chi \gamma^5) \chi \sum_q \bar{q} \gamma_\mu (v'_q + a'_q \gamma^5) q \ ,
\eeq
\Eq{SDLag} is singled out if the DM is Majorana: in fact, the two terms proportional to $v'_\chi$ vanish due to \Eq{Majorana}, while the contribution to $\Op_\NR^N$ of the term proportional to $a'_\chi v'_q$ is NR suppressed with respect to that of the $a'_\chi a'_q$ term, see \Tab{tab: spin-1/2 DM EFT}.

It is sometimes more convenient to work in the isospin basis rather than in the $p$, $n$ basis. For instance, since the induced pseudo-scalar interaction and the $2$-body currents (to be discussed below) have mainly an isotriplet structure, they have simpler expressions in this basis. To this end one introduces the isoscalar and isovector coefficients
\begin{subequations}
\label{a0 a1}
\begin{align}
a_0 &\equiv a_p + a_n = (a_u + a_d) (\Delta u^{(p)} + \Delta d^{(p)}) + 2 a_s \Delta s^{(p)} \ ,
\\
a_1 &\equiv a_p - a_n = (a_u - a_d) (\Delta u^{(p)} - \Delta d^{(p)}) \ ,
\end{align}
\end{subequations}
where we used \Eq{Delta isospin} valid in the isospin-symmetric limit. In the form of matrices acting on the $SU(2)$ isospin doublet $(p, n)^\tr$, these coefficients can be written as
\beq
\begin{pmatrix}
a_p & 0
\\
0 & a_n
\end{pmatrix}
=
\frac{a_0 \unom_2 + a_1 \tau^3}{2} = \frac{1}{2}
\begin{pmatrix}
a_0 + a_1 & 0
\\
0 & a_0 - a_1
\end{pmatrix}
,
\eeq
with $\tau^3$ the third Pauli matrix, numerically equal to $\sigma^3$ in \Eq{Pauli matrices}. We also define for later convenience
\begin{subequations}
\label{tilde a0 a1}
\begin{align}
\tilde{a}_0(q^2) &\equiv \tilde{a}_p(q^2) + \tilde{a}_n(q^2) = - \frac{a_0 - \frac{2}{3} (a_u + a_d + a_s) (\Delta u^{(p)} + \Delta d^{(p)} + \Delta s^{(p)})}{q^2 + m_\eta^2} \ ,
\\
\tilde{a}_1(q^2) &\equiv \tilde{a}_p(q^2) - \tilde{a}_n(q^2) = - \frac{a_1}{q^2 + m_\pi^2} \ .
\end{align}
\end{subequations}
The contribution of the $\eta$ meson is often neglected, so that $\tilde{a}_0(q^2) \simeq 0$ and
\beq\label{tilde a0 a1 - eta}
\tilde{a}_p(q^2) \simeq - \tilde{a}_n(q^2) \simeq - \frac{1}{2} \frac{a_p - a_n}{q^2 + m_\pi^2} \ .
\eeq

The DM-nucleus scattering amplitude can be obtained in the limit of point-like nucleus (see \Sec{nuc form factors}) by naively summing the nucleon matrix element of $\Op_\NR^N$ over all nucleons, as already done for the SI interaction in \Sec{SI interaction}, and multiplying by $2 m_T / 2 \mN$ to take care of the change in state normalization in going from nucleon to nuclear states (see \Eq{statenormhere NR}). We then get
\beq\label{SDMel}
\Mel_\text{PLN} \NReq - 16 \mDM m_T \sum_{N = p, n} \left( a_N \, \bol{s}_\chi \cdot \matel{\amom{J}, \amom{M}'}{\bol{S}_N}{\amom{J}, \amom{M}} + \tilde{a}_N(q^2) \, \bol{s}_\chi \cdot \bol{q} \, \matel{\amom{J}, \amom{M}'}{\bol{S}_N}{\amom{J}, \amom{M}} \cdot \bol{q} \right) ,
\eeq
where $\bol{s}_\chi$ is given by \Eq{I s} and
\begin{align}
\label{S_p S_n}
\bol{S}_p \equiv \sum_{i = \text{protons}} \bol{s}_{p_i} \ ,
&&&
\bol{S}_n \equiv \sum_{i = \text{neutrons}} \bol{s}_{n_i} \ ,
\end{align}
are the total proton and neutron spin operators. The nuclear matrix element in \Eq{SDMel} quantifies the fraction of nuclear spin due to the spin of the nucleons. For nuclei with spin $\amom{J} = 0$, this contribution clearly vanishes and therefore $\Mel_\text{PLN} \NReq 0$. Indeed, we have seen in \Sec{examples and applications} that spin-$0$ nuclei have no SD interaction at any value of momentum transfer, due to angular momentum conservation and parity constraints; in the following we therefore assume $\amom{J} \neq 0$.

Notice that, owing to the dependence of $\tilde{a}_p$ and $\tilde{a}_n$ on both $a_p$ and $a_n$, one cannot isolate a `proton-only' or `neutron-only' contribution to the scattering amplitude. In fact, the DM would only interact with protons (neutrons) alone if both $a_n$ and $\tilde{a}_n$ ($a_p$ and $\tilde{a}_p$) vanished, but one can see from \Eq{tilde a0 a1} that setting $a_n = 0$ ($a_p = 0$) is only compatible with $\tilde{a}_n = 0$ ($\tilde{a}_p = 0$) if also $a_p$ ($a_n$) vanishes. In other words, setting $a_n$ or $\tilde{a}_n$ ($a_p$ or $\tilde{a}_p$) to zero does not prevent the DM from interacting with neutrons (protons), while setting both of them to zero erases the whole interaction. This mixing of DM-proton and DM-neutron couplings is more easily seen in \Eq{tilde a0 a1 - eta}, where the contribution of the $\eta$ meson is neglected as customary. As we will see below, $2$-body contributions also produce an analogous mixing, even at zero momentum transfer, whereas the pseudo-scalar interaction only affects the scattering at $q > 0$. Both effects can have important phenomenological consequences, as discussed further below. No parameter choice can yield DM interactions with only one nucleon species, unless both the induced pseudo-scalar interaction and $2$-body corrections can be neglected.

We now make use of the Wigner-Eckart theorem~\eqref{Wigner-Eckart} to parametrize $\matel{\amom{J}, \amom{M}'}{\bol{S}_N}{\amom{J}, \amom{M}}$ in terms of matrix elements of the total angular momentum operator $\bol{J}$. To do so, we exploit the fact that certain linear combinations of the Cartesian components of a generic vector operator, such as $\bol{S}_N$ or $\bol{J}$, form the components of a spherical tensor of rank $1$ (see \Eq{V spherical compo}). Applying \Eq{Wigner-Eckart} to the matrix elements of both $\bol{S}_N$ and $\bol{J}$ one gets two expressions featuring the same Clebsch-Gordan coefficient $\CG{\amom{J} \amom{M}}{1 M}{\amom{J} \amom{M}'}$, which can be then replaced giving rise to
\beq\label{Wigner-Eckart 0}
\matel{\amom{J}, \amom{M}'}{\bol{S}_N}{\amom{J}, \amom{M}} = \frac{\matel{\amom{J}, \amom{M}'}{\bol{J}}{\amom{J}, \amom{M}}}{\redmatel{\amom{J}}{\bol{J}}{\amom{J}}} \redmatel{\amom{J}}{\bol{S}_N}{\amom{J}} \ ,
\eeq
where we are adopting the standard notation. Notice that the reduced matrix elements $\redmatel{\amom{J}}{\bol{J}}{\amom{J}}$ and $\redmatel{\amom{J}}{\bol{S}_N}{\amom{J}}$ are scalars, contrary to the ordinary matrix elements, that are vectors. The $\redmatel{\amom{J}}{\bol{S}_N}{\amom{J}} / \redmatel{\amom{J}}{\bol{J}}{\amom{J}}$ ratio is usually re-formulated by projecting \Eq{Wigner-Eckart 0} onto an arbitrary quantization axis, customarily denoted the $z$ direction, and taking states of maximal angular momentum, $\amom{M} = \amom{M}' = \amom{J}$: we then obtain
\beq\label{SS_N}
\mathbb{S}_N \equiv \matel{\amom{J}, \amom{J}}{S_N^z}{\amom{J}, \amom{J}} = \amom{J} \, \frac{\redmatel{\amom{J}}{\bol{S}_N}{\amom{J}}}{\redmatel{\amom{J}}{\bol{J}}{\amom{J}}} \ ,
\eeq
so that \Eq{Wigner-Eckart 0} can be written as
\beq\label{Wigner-Eckart spin}
\matel{\amom{J}, \amom{M}'}{\bol{S}_N}{\amom{J}, \amom{M}} = \mathbb{S}_N \frac{\matel{\amom{J}, \amom{M}'}{\bol{J}}{\amom{J}, \amom{M}}}{\amom{J}} \ .
\eeq
We denote here $\matel{\amom{J}, \amom{J}}{S_N^z}{\amom{J}, \amom{J}}$ with $\mathbb{S}_N$ instead of $\mathbf{S}_N$, the customary notation, to avoid confusion with the total-spin operator $\bol{S}_N$. Determining $\mathbb{S}_N$ requires detailed calculations within realistic nuclear models, and as a consequence the values found in the literature are often model dependent and sometimes differ one from another for a given nuclide. Some indicative values can be found in \Tab{tab: Sp Sn}, obtained as averages of values from Refs.~\cite{Ellis:1987sh, Engel:1989ix, Iachello:1990ut, Ellis:1991ef, Ellis:1992vh, Nikolaev:1993dd, Ressell:1993qm, Dimitrov:1994gc, Engel:1995gw, Ressell:1997kx, Divari:2000dc, Bednyakov:2004xq, Kortelainen:2006rd, Toivanen:2009zza, Fitzpatrick:2012ix, Klos:2013rwa}, which were collected in Refs.~\cite{Bednyakov:2004xq, Toivanen:2009zza, Klos:2013rwa}. Also reported are the minimum and maximum values from the same collections, to give a flavor of the uncertainty attached to the average values.

\begin{table}[t]
\begin{center}
\begin{tabular}{|>{\rule[-2mm]{0mm}{6.6mm}} c | c c | c c |}
\hline
\raisebox{-0.4mm}{\multirow{2}{*}{Nuclide}} & \multicolumn{2}{c |}{$\mathbb{S}_p$} & \multicolumn{2}{c |}{$\mathbb{S}_n$}
\\
\cline{2-5}
& Average & [Min, Max] & Average & [Min, Max]
\\
\hline
$^{13}$C & $-0.009$ & $[-0.026, 0]$ & $-0.226$ & $[-0.327, -0.167]$
\\
\hline
$^{19}$F & $0.476$ & $[0.475, 0.478]$ & $-0.007$ & $[-0.009, -0.002]$
\\
\hline
$^{23}$Na & $0.243$ & $[0.224, 0.248]$ & $0.021$ & $[0.02, 0.024]$
\\
\hline
$^{27}$Al & $0.334$ & $[0.326, 0.343]$ & $0.034$ & $[0.03, 0.038]$
\\
\hline
$^{29}$Si & $0.004$ & $[-0.002, 0.016]$ & $0.140$ & $[0.13, 0.156]$
\\
\hline
$^{73}$Ge & $0.015$ & $[0.005, 0.031]$ & $0.436$ & $[0.378, 0.475]$
\\
\hline
$^{127}$I & $0.336$ & $[0.264, 0.418]$ & $0.051$ & $[0.03, 0.075]$
\\
\hline
$^{129}$Xe & $0.011$ & $[-0.002, 0.028]$ & $0.302$ & $[0.248, 0.359]$
\\
\hline
$^{131}$Xe & $-0.007$ & $[-0.012, -0.0007]$ & $-0.208$ & $[-0.272, -0.125]$
\\
\hline
$^{133}$Cs & $-0.289$ & $[-0.389, -0.2]$ & $0.005$ & $[0, 0.021]$
\\
\hline
$^{183}$W & $0$ & $[0, 0]$ & $-0.1$ & $[-0.17, -0.03]$
\\
\hline
\end{tabular}
\caption[$\mathbb{S}_p$ and $\mathbb{S}_n$ for the SD interaction]{\label{tab: Sp Sn}\emph{Indicative values of \figmath{\mathbb{S}_p} and \figmath{\mathbb{S}_n}, defined in \Eq{SS_N}. ``Average'' indicates the rounded average of the values collected in Refs.~\cite{Bednyakov:2004xq, Toivanen:2009zza} (\figmath{^{13}\text{C}}, \figmath{^{133}\text{Cs}}, \figmath{^{183}\text{W}}) and Ref.~\cite{Klos:2013rwa} (\figmath{^{19}\text{F}}, \figmath{^{23}\text{Na}}, \figmath{^{27}\text{Al}}, \figmath{^{29}\text{Si}}, \figmath{^{73}\text{Ge}}, \figmath{^{127}\text{I}}, \figmath{^{129}\text{Xe}}, \figmath{^{131}\text{Xe}}). The values within brackets are instead the minimum and maximum values from the same references, and are reported here to give a flavor of the uncertainty attached to the averages. The values with largest uncertainties are \figmath{\mathbb{S}_n} for \figmath{^{133}\text{Cs}} and \figmath{\mathbb{S}_p} for \figmath{^{29}\text{Si}} and \figmath{^{183}\text{W}}. The above references report values from Refs.~\cite{Ellis:1987sh, Engel:1989ix, Iachello:1990ut, Ellis:1991ef, Ellis:1992vh, Nikolaev:1993dd, Ressell:1993qm, Dimitrov:1994gc, Engel:1995gw, Ressell:1997kx, Divari:2000dc, Kortelainen:2006rd, Toivanen:2009zza, Fitzpatrick:2012ix}. See \Tab{tab: nuclides} for a list of other nuclear properties of the considered nuclides.}}
\end{center}
\end{table}

Since same-type nucleons (protons or neutrons) are packed in the nucleus so that they have pairwise opposite spins, the matrix elements of $\bol{S}_p$ and $\bol{S}_n$ receive their main contribution respectively from the single unpaired proton and neutron, if any. For this reason, nuclei with an even number of protons (neutrons) tend to have small values of $\mathbb{S}_p$ ($\mathbb{S}_n$). Nuclei with an odd number of protons, such as $^{19}$F, $^{23}$Na, and $^{127}$I, and/or of neutrons, such as $^{129}$Xe and $^{131}$Xe (see \Tab{tab: nuclides}), are therefore favored in searching for DM with SD interactions. Yet, the values of the $\mathbb{S}_N$'s are quite small compared to the potentially large $A$ enhancement of the DM-nucleus scattering amplitude with respect to the DM-nucleon matrix element for the SI interaction, see \Eq{SIMel}. This smaller enhancement is the reason of the widespread belief that the SD interaction is subdominant to the SI interaction, if both are present. This however does not need to be the case, as the relative size between the two interactions also depends on the ratio between the $f_N$ and $a_N$ couplings, see discussion below \Eq{diffsigmaSD}. Example models where the SI and SD interactions have the same size can be found \eg in Ref.~\cite{Marcos:2015dza}.

To compute the squared matrix element averaged over the initial nuclear spins and summed over the final nuclear spins, see \Eq{Mel^2}, we need
\beq\label{Tr[J_i J_j]}
\sum_{\amom{M}, \amom{M}'} \matel{\amom{J}, \amom{M}}{J^i}{\amom{J}, \amom{M}'} \matel{\amom{J}, \amom{M}'}{J^j}{\amom{J}, \amom{M}} = \underbrace{\sum_\amom{M} \matel{\amom{J}, \amom{M}}{J^i J^j}{\amom{J}, \amom{M}}}_{\displaystyle = \Tr[J^i J^j]_\amom{J}} = \frac{\amom{J} (\amom{J} + 1) (2 \amom{J} + 1)}{3} \delta_{ij} \ ,
\eeq
where in the first equality we used the fact that $\sum_{\amom{M}'} \ket{\amom{J}, \amom{M}'} \bra{\amom{J}, \amom{M}'}$ is the projector onto the subspace of total angular momentum $\amom{J}$. This result can be seen as the generalization to any spin of the first relation in \Eq{spinsum 2}. To prove the second identity we can notice that $\Tr[J^i J^j]_\amom{J}$ is rotationally invariant due to $R_{i j} J^j = U^\dagger J^i U$ and the trace being cyclic, with $R$ a special orthogonal $3 \times 3$ matrix, \ie in the adjoint representation of $SU(2)$, and $U$ a unitary $SU(2)$ operator. Therefore, $\Tr[J^i J^j]_\amom{J}$ can be built out of the $SU(2)$ invariant tensors $\delta_{ij}$ and $\varepsilon_{ijk}$. Being symmetric in $i \leftrightarrow j$, it can only be $\Tr[J^i J^j]_\amom{J} \propto \delta_{ij}$, with the proportionality factor determined by contracting with $\delta_{ij}$:
\beq
\Tr[\bol{J}^2]_\amom{J} = \sum_\amom{M} \amom{J} (\amom{J} + 1) = \amom{J} (\amom{J} + 1) (2 \amom{J} + 1) \ .
\eeq
We then have from \Eq{SDMel}, for a point-like nucleus,
\beq
\overline{| \Mel_\text{PLN} |^2} \NReq 64 \mDM^2 m_T^2 \frac{\amom{J} + 1}{\amom{J}} \sum_{N, N'} \mathbb{S}_N \mathbb{S}_{N'} \left( a_N a_{N'} + \frac{2}{3} q^2 a_N \tilde{a}_{N'}(q^2) + \frac{q^4}{3} \tilde{a}_N(q^2) \tilde{a}_{N'}(q^2) \right) ,
\eeq
where we used Eqs.~\eqref{Wigner-Eckart spin},~\eqref{spinsum s_chi} and the fact that both $a_N$, $\tilde{a}_N$, and $\mathbb{S}_N$ are real.

We now move away from the picture of a point-like nucleus and encode the effect of the finite nuclear size on the scattering cross section within the `squared form factors' $F_\text{pSD}^{(N, N')}(\ER)$ and $F_\text{PS}^{(N, N')}(\ER)$, normalized so that
\beq
F_\text{pSD}^{(N, N')}(0) = F_\text{PS}^{(N, N')}(0) = 1 \ .
\eeq
The PS label indicates here the induced pseudo-scalar interaction, see \Sec{Axial-vector couplings}, while pSD indicates the `pure SD' contribution of $\Op^N_4$ to \Eq{ONR SD}. Finally we get
\begin{multline}
\label{SDMel^2}
\overline{| \Mel |^2} \NReq 64 \mDM^2 m_T^2 \amom{J} (\amom{J} + 1) \sum_{N, N'} \left( \Lambda_N \Lambda_{N'} F_\text{pSD}^{(N, N')}(\ER) \vphantom{\frac{2}{3} \frac{q^4}{3}} \right.
\\
\left. + \frac{2}{3} q^2 \Lambda_N \tilde{\Lambda}_{N'}(q^2) F_\text{PS}^{(N, N')}(\ER) + \frac{q^4}{3} \tilde{\Lambda}_N(q^2) \tilde{\Lambda}_{N'}(q^2) F_\text{PS}^{(N, N')}(\ER) \right) ,
\end{multline}
where we defined
\begin{align}
\Lambda_N \equiv \frac{a_N \mathbb{S}_N}{\amom{J}} \ ,
&&&
\tilde{\Lambda}_N(q^2) \equiv \frac{\tilde{a}_N(q^2) \mathbb{S}_N}{\amom{J}} \ .
\end{align}
The $\Op^N_4$--$\Op^N_6$ interference term has the same squared form factor as the squared induced pseudo-scalar term, as explained in the following. The $\Op^N_6$ building block in \Eq{ONR SD} features the longitudinal components of $\bol{s}_\chi$ and $\bol{s}_N$, \ie their projections along $\hat{\bol{q}}$, which are also featured by $\Op^N_4$ together with their transverse components. Therefore, $\Op^N_4$ and $\Op^N_6$ interfere in the squared matrix element (see \Sec{NR operators}). As we saw in \Sec{Multipoles}, the longitudinal component of $\bol{s}_N$ gives rise to the $\Sigma''$ nuclear response in the form of $\mulop{L}_{JM}$ multipoles, whereas its transverse components give rise to the $\Sigma'$ response in the form of $\mulop{T}^\text{el}_{JM}$ multipoles, see \Tab{tab: nuclear responses} (the $\mulop{T}^\text{mag}_{JM}$ multipoles vanish as explained in \Sec{PT selection rules}). Therefore we can expect the squared form factor of the pure $\Op^N_4$ term in \Eq{SDMel^2} to depend on both responses, and that of the pure $\Op^N_6$ term to only involve $\Sigma''$ thus being proportional to $F_{\Sigma''}^{(N, N')}$ appearing in \Eq{M^2 form factors fifj}. As the $\Sigma'$ and $\Sigma''$ responses do not interfere, we expect the squared form factor of the $\Op^N_4$--$\Op^N_6$ interference term, too, to be proportional to $F_{\Sigma''}^{(N, N')}$. This can be verified by noticing that the squared form factors appearing in \Eq{SDMel^2} are in a one-to-one correspondence with $F_{4, 4}^{(N, N')}$, $F_{6, 6}^{(N, N')}$, and $F_{4, 6}^{(N, N')}$ in Eqs.~\eqref{M^2 form factors fifj},~\eqref{F_ij}. Therefore, with our normalization, the squared induced pseudo-scalar term and the interference term in \Eq{SDMel^2} have the same squared form factor. Comparison with \Eq{M^2 form factors fifj} yields
\begin{align}
\label{F_44}
F_{4, 4}^{(N, N')} &= \frac{1}{4} \frac{\amom{J} + 1}{\amom{J}} \mathbb{S}_N \mathbb{S}_{N'} F_\text{pSD}^{(N, N')} ,
\\
F_{6, 6}^{(N, N')} &= \frac{q^4}{12} \frac{\amom{J} + 1}{\amom{J}} \mathbb{S}_N \mathbb{S}_{N'} F_\text{PS}^{(N, N')} \ ,
\\
F_{4, 6}^{(N, N')} &= \frac{q^2}{12} \frac{\amom{J} + 1}{\amom{J}} \mathbb{S}_N \mathbb{S}_{N'} F_\text{PS}^{(N, N')} \ .
\end{align}
From \Eq{F_ij} we further find at zero momentum transfer
\beq\label{F_Sigma}
\frac{1}{2} F_{\Sigma'}^{(N, N')}(0) = F_{\Sigma''}^{(N, N')}(0) = \frac{4}{3} \frac{\amom{J} + 1}{\amom{J}} \mathbb{S}_N \mathbb{S}_{N'} \ ,
\eeq
which can be used to determine the $\mathbb{S}_p$ and $\mathbb{S}_n$ values implicit in the squared form factors of Refs.~\cite{Fitzpatrick:2012ix, Anand:2013yka}, up to a relative sign (see \eg Ref.~\cite{Klos:2013rwa}).

The differential cross section averaged over initial spins and summed over final spins, \Eq{dsigmadER}, is
\begin{multline}
\label{diffsigmaSD}
\frac{\ud \sigma_T}{\ud \ER} \NReq \frac{2 m_T}{\pi v^2} \amom{J} (\amom{J} + 1) \sum_{N, N'} \left( \Lambda_N \Lambda_{N'} F_\text{pSD}^{(N, N')}(\ER) \vphantom{\frac{2}{3} \frac{q^4}{3}} \right.
\\
\left. + \frac{2}{3} q^2 \Lambda_N \tilde{\Lambda}_{N'}(q^2) F_\text{PS}^{(N, N')}(\ER) + \frac{q^4}{3} \tilde{\Lambda}_N(q^2) \tilde{\Lambda}_{N'}(q^2) F_\text{PS}^{(N, N')}(\ER) \right) .
\end{multline}
Due to the dependence $(\amom{J} + 1) / \amom{J}$ on the nuclear spin, $\ud \sigma_T / \ud \ER$ is maximal for $\amom{J} = 1/2$ at fixed $a_N$'s and $\mathbb{S}_N$'s, and decreases slightly for larger values of $\amom{J}$. When compared among different isotopes (with $\amom{J} \neq 0$) of the same element, its main source of variation lies therefore in its dependence on $\mathbb{S}_N$, see \Tab{tab: Sp Sn}, whereas its dependence on $m_T$ also becomes important if one compares nuclear elements with very different masses. Comparing with \Eq{diffsigmaSI 1}, we see that the SD interaction can have the same size as the SI interaction if $4 (a_p \mathbb{S}_p + a_n \mathbb{S}_n)^2 (\amom{J} + 1) / \amom{J}$ is comparable with $(Z f_p + (A - Z) f_n)^2$ for at least one relevant target nuclide in the detector. This is easier to happen for a light target (so that $A$ and $Z$ are small), with small but non-zero spin (so that $(\amom{J} + 1) / \amom{J}$ is maximized), and with large $\mathbb{S}_p$ and/or $\mathbb{S}_n$. As can be seen from \Tab{tab: nuclides} and \Tab{tab: Sp Sn}, an ideal example is fluorine for which the SI and SD interactions are certainly comparable if $|a_p|$ is ten times larger than $|f_p|$ and $|f_n|$, or even five times larger if the contribution of either protons or neutrons to the SI interaction can be neglected. For heavier nuclei, this SD-SI ratio of coupling constants must be larger for the two interactions to be comparable, assuming destructive interference does not play a considerable role for the SI interaction (see discussion after \Eq{D factor}): for instance, with SI and SD couplings of the same size, the SI differential cross section can easily be $10^4$ times larger than that for the SD interaction for DM scattering off xenon nuclei.

As commented above, the induced pseudo-scalar interaction mixes the contributions of the DM-proton and DM-nucleon couplings in the scattering, as one can see by noting that \Eq{tilde a0 a1} implies that $a_n = 0$ ($a_p = 0$) is only compatible with $\tilde{a}_n = 0$ ($\tilde{a}_p = 0$) if $a_p = 0$ ($a_n = 0$). This mixing can give sizeable contributions to the cross section at finite momentum transfer, especially for heavy, odd-mass nuclei (\ie nuclei with $A$ odd), as explicitly shown \eg in Ref.~\cite{Bishara:2017pfq}. For example, if at $q = 0$ the DM has only pSD couplings with protons ($a_n = 0$), and interacts with a nucleus with an unpaired neutron but no unpaired protons ($Z$ even, $A$ odd), all $F_\text{pSD}^{(N, N')}$ terms above vanish apart from the $F_\text{pSD}^{(p, p)}$ term, which is suppressed by the small value of $\mathbb{S}_p$. On the other hand, the $q^2 F_\text{PS}^{(p, n)}$ and $q^4 F_\text{PS}^{(n, n)}$ terms can give a significant contribution, at sufficiently large values of $q$ (hence the need for both $m_T$ and $\mDM$ to be somewhat large), owing to the fact that $\tilde{\Lambda}_n$ has a dependence on $a_p \mathbb{S}_n \gg a_p \mathbb{S}_p$.

The energy dependence of the differential cross section is often parametrized, in the isospin basis of \Eq{a0 a1}, through the nuclear spin structure function
\beq\label{SD structure function}
S(q^2) = a_0^2 S_{00}(q^2) + a_1^2 S_{11}(q^2) + a_0 a_1 S_{01}(q^2) \ ,
\eeq
which features a pure isoscalar term $S_{00}(q^2)$, a pure isovector term $S_{11}(q^2)$, and the interference term $S_{01}(q^2)$. One can also define the following structure functions in the $p$, $n$ basis, used \eg in Ref.~\cite{Klos:2013rwa},
\begin{align}
\label{Spn}
S_{pp} &\equiv S_{00} + S_{11} + S_{01} \ ,
&
S_{nn} &\equiv S_{00} + S_{11} - S_{01} \ ,
&
S_{pn} = S_{np} &\equiv S_{00} - S_{11} \ ,
\end{align}
with which
\beq\label{S_pn}
S(q^2) = a_p^2 S_{pp}(q^2) + a_n^2 S_{nn}(q^2) + 2 a_p a_n S_{pn}(q^2) = \sum_{N, N' = p, n} a_N a_{N'} S_{N N'}(q^2) \ .
\eeq
This parametrization allows to distinguish the contributions of DM-proton and DM-neutron interactions to the cross section, while mixing the contributions of the pSD and PS interactions (as opposed to the form factors employed in \Eq{SDMel^2}, see discussion above). $S(q^2)$ is defined so that, at zero-momentum transfer,
\beq
S(0) = \frac{2 \amom{J} + 1}{\pi} \amom{J} (\amom{J} + 1) \Lambda^2 = \frac{1}{\pi} \frac{(2 \amom{J} + 1) (\amom{J} + 1)}{\amom{J}} (a_p \mathbb{S}_p + a_n \mathbb{S}_n)^2 \ ,
\eeq
where we defined
\beq
\Lambda \equiv \Lambda_p + \Lambda_n = \frac{a_p \mathbb{S}_p + a_n \mathbb{S}_n}{\amom{J}} \ .
\eeq
This implies
\begin{subequations}
\label{Sij(0)}
\begin{align}
S_{00}(0) &= 4 C(\amom{J}) \mathbb{S}_0^2 = C(\amom{J}) (\mathbb{S}_p + \mathbb{S}_n)^2 \ ,
&
S_{pp}(0) &= 4 C(\amom{J}) \mathbb{S}_p^2 \ ,
\\
S_{11}(0) &= 4 C(\amom{J}) \mathbb{S}_1^2 = C(\amom{J}) (\mathbb{S}_p - \mathbb{S}_n)^2 \ ,
&
S_{nn}(0) &= 4 C(\amom{J}) \mathbb{S}_n^2 \ ,
\\
S_{01}(0) &= 8 C(\amom{J}) \mathbb{S}_0 \mathbb{S}_1 = 2 C(\amom{J}) (\mathbb{S}_p^2 - \mathbb{S}_n^2) \ ,
&
S_{pn}(0)
&= 4 C(\amom{J}) \mathbb{S}_p \mathbb{S}_n \ ,
\end{align}
\end{subequations}
with
\begin{align}
\mathbb{S}_0 \equiv \frac{\mathbb{S}_p + \mathbb{S}_n}{2} \ ,
&&&
\mathbb{S}_1 \equiv \frac{\mathbb{S}_p - \mathbb{S}_n}{2} \ ,
\end{align}
and
\beq
C(\amom{J}) \equiv \frac{(2 \amom{J} + 1) (\amom{J} + 1)}{4 \pi \amom{J}} \ .
\eeq
The differential cross section reads
\beq\label{diffsigmaSD-PS}
\frac{\ud \sigma_T}{\ud \ER} \NReq \frac{2 m_T}{v^2} \frac{S(q^2)}{2 \amom{J} + 1} = \frac{2 m_T}{\pi v^2} \amom{J} (\amom{J} + 1) \Lambda^2 \frac{S(q^2)}{S(0)} \ .
\eeq
Notice that $S(q^2) / S(0)$ depends non-trivially on $a_p$ and $a_n$, unless their ratio is fixed in which case only one of them is an independent parameter ($S(q^2) / S(0)$ would then only depend on the coupling ratio). Unlike for the SI interaction, it is not possible here to factor the $a_p$ and $a_n$ dependence from the $q$ dependence in the differential cross section (compare \eg with \Eq{diffsigmaSI 1}). This is due to the assumption that the nuclear matrix elements of $\Op_\NR^p$ and $\Op_\NR^n$ have the same $q$ dependence (regardless of their normalization) being justified for the SI interaction in \Eq{ONR SI} but not for the SD interaction in \Eq{ONR SD}, not even for its pure $\Op^N_4$ part. One can check that the $S_{ij}$'s are related to the form factors appearing in \Eq{SDMel^2} by
\begin{align}
S_{00} &\simeq 4 C(\amom{J}) F_\text{pSD}^{(0, 0)} \ ,
\\
S_{11} &\simeq 4 C(\amom{J}) \left( F_\text{pSD}^{(1, 1)} - \frac{2}{3} \frac{q^2}{q^2 + m_\pi^2} F_\text{PS}^{(1, 1)} + \frac{1}{3} \frac{q^4}{(q^2 + m_\pi^2)^2} F_\text{PS}^{(1, 1)} \right) ,
\\
S_{01} &\simeq 4 C(\amom{J}) \left( F_\text{pSD}^{(0, 1)} + F_\text{pSD}^{(1, 0)} - \frac{2}{3} \frac{q^2}{q^2 + m_\pi^2} F_\text{PS}^{(0, 1)} \right) ,
\end{align}
where we neglected, as customary, the contribution of the $\eta$ meson, and we defined for $\text{X} = \text{pSD}, \text{PS}$
\begin{align}
F_\text{X}^{(0, 0)} &\equiv \frac{1}{4} \left( \mathbb{S}_p^2 F_\text{X}^{(p, p)} + \mathbb{S}_p \mathbb{S}_n F_\text{X}^{(p, n)} + \mathbb{S}_p \mathbb{S}_n F_\text{X}^{(n, p)} + \mathbb{S}_n^2 F_\text{X}^{(n, n)} \right) ,
\\
F_\text{X}^{(0, 1)} &\equiv \frac{1}{4} \left( \mathbb{S}_p^2 F_\text{X}^{(p, p)} - \mathbb{S}_p \mathbb{S}_n F_\text{X}^{(p, n)} + \mathbb{S}_p \mathbb{S}_n F_\text{X}^{(n, p)} - \mathbb{S}_n^2 F_\text{X}^{(n, n)} \right) ,
\\
F_\text{X}^{(1, 0)} &\equiv \frac{1}{4} \left( \mathbb{S}_p^2 F_\text{X}^{(p, p)} + \mathbb{S}_p \mathbb{S}_n F_\text{X}^{(p, n)} - \mathbb{S}_p \mathbb{S}_n F_\text{X}^{(n, p)} - \mathbb{S}_n^2 F_\text{X}^{(n, n)} \right) ,
\\
F_\text{X}^{(1, 1)} &\equiv \frac{1}{4} \left( \mathbb{S}_p^2 F_\text{X}^{(p, p)} - \mathbb{S}_p \mathbb{S}_n F_\text{X}^{(p, n)} - \mathbb{S}_p \mathbb{S}_n F_\text{X}^{(n, p)} + \mathbb{S}_n^2 F_\text{X}^{(n, n)} \right) ,
\end{align}
which satisfy $F_\text{X}^{(i, j)}(0) = \mathbb{S}_i \mathbb{S}_j$.
One can see that the $\Op^N_4$ and $\Op^N_6$ contributions are mixed within the $S_{ij}$'s. Some authors (see \eg Refs.~\cite{Engel:1991wq, Engel:1992bf, Menendez:2012tm, Klos:2013rwa}) provide form factors that are inclusive of the contribution of the induced pseudo-scalar interaction, while the $\Op^N_4$ and $\Op^N_6$ contributions are kept separated in Refs.~\cite{Fitzpatrick:2012ix, Anand:2013yka} as related to different NR building blocks. One can also check that, as already mentioned above, the contributions of protons and neutrons are mixed by the presence of the induced pseudo-scalar interaction: for instance, in the $a_p^2$ term in \Eq{S_pn}, the $S_{pp}$ structure function also contains finite-$q$ contributions from neutrons (\eg it depends on $\mathbb{S}_n^2 F_\text{PS}^{(n, n)}$).

To make contact with the portion of the literature which neglects the $\Op^N_6$ contribution to \Eq{ONR SD}, we forget temporarily about the induced pseudo-scalar interaction and focus on $\Op^N_4$ alone. The DM-nucleon scattering cross section can be obtained by setting $\mathbb{S}_N = \amom{J} = 1/2$ in \Eq{diffsigmaSD-PS}, since for a single nucleon $\bol{J} = \bol{S}_N$, and integrating over $\ER$. For a point-like nucleon this yields
\beq\label{SD sigma_N}
\sigma_N \NReq \frac{3}{\pi} \mu_\text{N}^2 a_N^2 \ .
\eeq
While the DM-nucleon cross section has this form only in the absence of the induced pseudo-scalar interaction, $\sigma_p$ or $\sigma_n$ as defined in \Eq{SD sigma_N} can still be used as a parameter expressing the overall size of the DM-nucleus cross section for the full interaction (at least if the $\eta$ meson is neglected), in the assumption the ratio between the pSD couplings $a_p$ and $a_n$ is fixed. The absence of the $\Op^N_6$ term in \Eq{ONR SD} makes the DM-nucleon interaction inherently independent of $q$, so that the DM-nucleus scattering is isotropic in the point-like nucleus limit and it makes sense to define the zero-momentum transfer cross section in \Eq{sigma_0},
\beq
\sigma_0 \NReq \frac{4 \mu_T^2}{\pi} \amom{J} (\amom{J} + 1) \Lambda^2 \ .
\eeq
One can than write the differential cross section as in \Eq{sigma_0 F_T}, where however the form factor depends in general on $a_p$ and $a_n$ as discussed below \Eq{diffsigmaSD-PS} (see also discussion below \Eq{F_T}). One can check that the $S_{N N'}$'s in \Eq{Spn} are related to the $F_\text{pSD}^{(N, N')}$'s by
\begin{align}
F_\text{pSD}^{(p, p)}(\ER) &= \frac{S_{pp}(q^2)}{S_{pp}(0)} \ ,
&
F_\text{pSD}^{(n, n)}(\ER) &= \frac{S_{nn}(q^2)}{S_{nn}(0)} \ ,
&
F_\text{pSD}^{(p, n)}(\ER) = F_\text{pSD}^{(n, p)}(\ER) &= \frac{S_{pn}(q^2)}{S_{pn}(0)}
\ .
\end{align}

As per the discussion in \Sec{nuc form factors}, $F_\text{pSD}^{(N, N)}$ may be thought of as the square of the Fourier transform of the spin density of the nucleon $N$. In the assumption the spin density of paired nucleons is negligible, since paired nucleons have opposite spins, one may consider a thin-shell density to approximate a single unpaired outer-shell nucleon in a spherically symmetric state~\cite{Lewin:1995rx},
\beq
\rho_\text{thin shell}(\bol{x}) \equiv \frac{1}{4 \pi R^2} \, \delta(x - R) \ ,
\eeq
leading to
\beq\label{thin-shell FF}
F_\text{thin shell}(q^2) = j_0(q R) = \frac{\sin(q R)}{q R} \ .
\eeq
More detailed calculations reveal that the early zeros of $j_0$ are at least partially filled, therefore a better choice for the squared form factor for $q R \leqslant 6$ is to replace the first dip with its value at the second maximum~\cite{Lewin:1995rx}:
\beq\label{thin-shell filled FF}
F_\text{filled}^2(q^2) =
\begin{cases}
j_0^2(q R) & q R < 2.55~\text{and}~q R > 4.49,
\\
0.047 & 2.55 \leqslant q R \leqslant 4.49,
\end{cases}
\eeq
with $R = 1.0 A^{1/3}~\fm$. However, this is not necessarily a good model and one should use the more advanced computations of the pSD form factor where available, although these results usually do not come in such a simple analytical form.

Some of the most recent computations can be found in Refs.~\cite{Menendez:2012tm, Klos:2013rwa}, which also include $2$-body corrections to the scattering cross section. These corrections effectively act as a $q$-dependent renormalization of the isovector couplings (see also Refs.~\cite{Cannoni:2012jq, Divari:2013fx}),
\begin{align}
a_0 &\to a_0 \ ,
&
a_1 &\to a_1 (1 + \delta a_1(q^2)) \ ,
\\
\tilde{a}_0(q^2) &\to \tilde{a}_0(q^2) \ ,
&
\tilde{a}_1(q^2) &\to \tilde{a}_1(q^2) (1 + \delta \tilde{a}_1(q^2)) \ .
\end{align}
Using \Eq{tilde a0 a1} and neglecting, as customary, the contribution of the $\eta$ meson, we then have for the $2$-body corrected induced pseudo-scalar couplings
\begin{align}
\tilde{a}_0(q^2) &\simeq 0 \ ,
&
\tilde{a}_1(q^2) &= - \frac{a_1}{q^2 + m_\pi^2} (1 + \delta \tilde{a}_1(q^2)) \ .
\end{align}
For simplicity, the extra factors in $a_1$ and $\tilde{a}_1$ can be absorbed into $S_{11}$ and $S_{10}$, so to obtain again \Eq{SD structure function} with couplings as defined by \Eq{a0 a1} but modified structure functions. At zero momentum transfer, $\delta a_1(0)$ can be estimated according to the results in Ref.~\cite{Klos:2013rwa} to lie in the range $[-0.32, -0.14]$ (depending on the nuclear model), while $\delta \tilde{a}_1$ does not contribute at $q = 0$ due to the induced pseudo-scalar interaction vanishing in this limit. These corrections can be sizeable, especially for odd-mass nuclei (\ie nuclei with $A$ odd), as they effectively induce DM interactions with both protons and neutrons at zero momentum transfer even if the DM has pSD couplings with only one species (\ie either $a_p$ or $a_n$ vanishes). This effect is analogous to what we discussed above concerning the induced pseudo-scalar interaction, although $2$-body corrections affect the cross section already at zero momentum transfer while the induced pseudo-scalar interaction only contributes at sufficiently large $q$. A negative $\delta a_1(0)$ in the aforementioned range implies that $|S_{01}(0)|$ and $S_{11}(0)$ decrease respectively by $23 \% \pm 9 \%$ and its square. One can then see from \Eq{Spn} that $S_{pp}(0)$ ($S_{nn}(0)$) decreases in those species where proton (neutron) spins provide the dominant contribution to the nuclear spin, and thus $S_{01}$ is positive (negative), see \Eq{Sij(0)}. On the other hand, again when proton (neutron) spins are dominant, $S_{nn}(0)$ ($S_{pp}(0)$) increases by a significant amount: in fact, $a_0$ not being affected by $2$-body effects implies that $a_p$ and $a_n$ receive equal-size (though opposite) additive corrections, which translate into relative modifications of the structure functions that are much larger for the subdominant species, given that $S_{pp}(0)$ and $S_{nn}(0)$ differ by orders of magnitude. Recalling the example already discussed above, let us take a DM particle with only pSD couplings with protons, scattering off a nucleus with an unpaired neutron but no unpaired protons ($Z$ even, $A$ odd). Here the $1$-body currents are suppressed at zero momentum transfer by the small value of $\mathbb{S}_p$, while $2$-body currents induce a contribution depending on $\mathbb{S}_n \gg \mathbb{S}_p$. To see this concretely, at zero momentum transfer $\overline{| \Mel |^2}$ in \Eq{SDMel^2} involves for the pSD interaction the factor
\begin{multline}
\frac{1}{2} \left[ (a_0 + a_1) \mathbb{S}_p + (a_0 - a_1) \mathbb{S}_n \right] \to \frac{1}{2} \left[ (a_0 + a_1 + a_1 \delta a_1(0)) \mathbb{S}_p + (a_0 - a_1 - a_1 \delta a_1(0)) \mathbb{S}_n \right]
\\
= a_p \left[ \mathbb{S}_p + \frac{\delta a_1(0)}{2} (\mathbb{S}_p - \mathbb{S}_n) \right] + a_n \left[ \mathbb{S}_n - \frac{\delta a_1(0)}{2} (\mathbb{S}_p - \mathbb{S}_n) \right] ,
\end{multline}
which, if $a_0 = a_1$ (\ie $a_n = 0$), only depends on $\mathbb{S}_n$ if $2$-body contributions are included.

\begin{figure}[t!]
\begin{center}
\includegraphics[width=.49\textwidth]{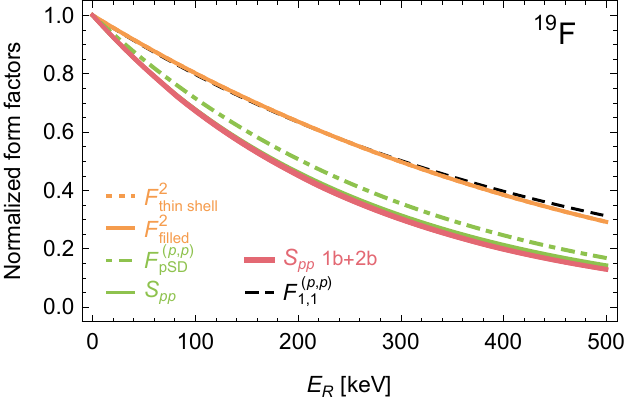}
\includegraphics[width=.49\textwidth]{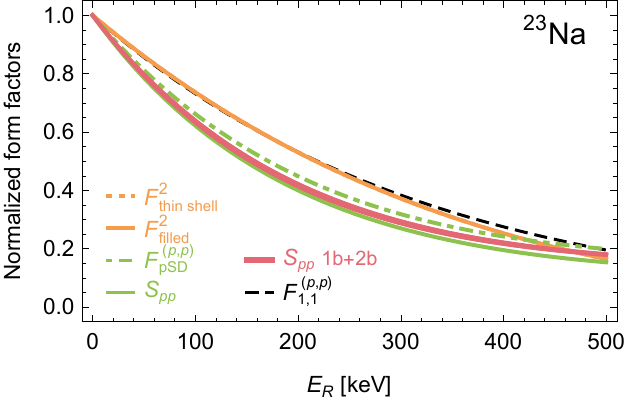}
\\[5mm]
\includegraphics[width=.49\textwidth]{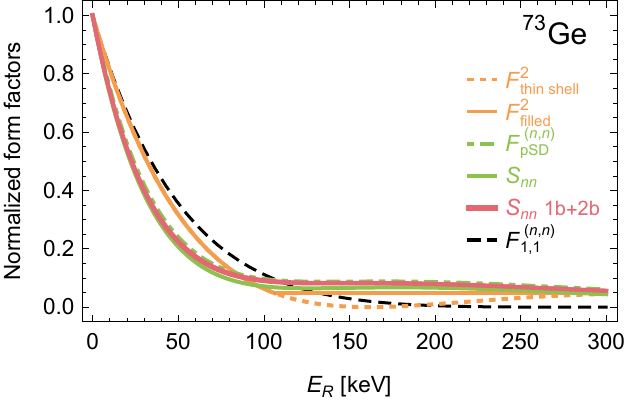}
\includegraphics[width=.49\textwidth]{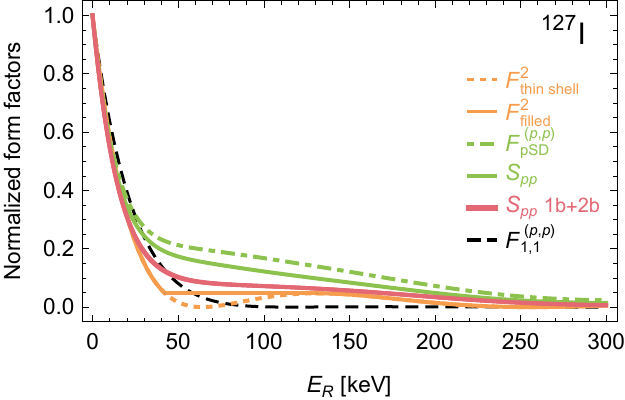}
\\[5mm]
\includegraphics[width=.49\textwidth]{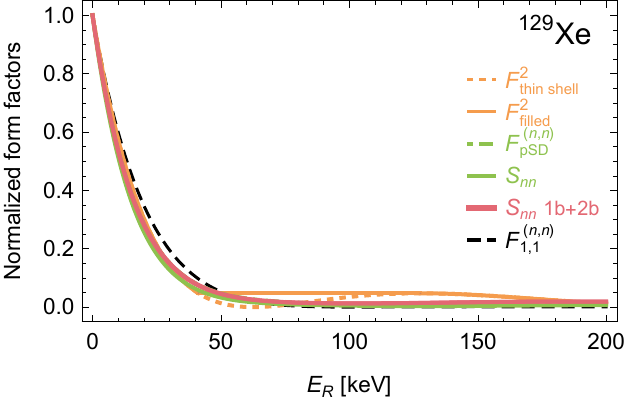}
\includegraphics[width=.49\textwidth]{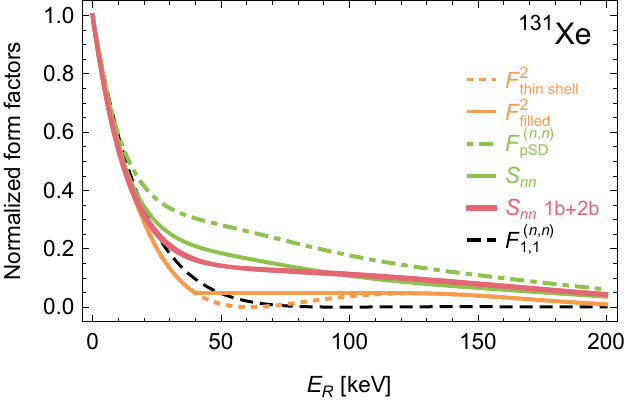}
\caption[SD form factors and structure functions]{\label{fig: SDFF}\emph{\figmath{\ER} dependence of SD form factors and structure functions for a representative set of nuclides (all curves are normalized to \figmath{1} at \figmath{\ER = 0}). The dotted (solid) orange curves represent the (filled) squared thin-shell form factor in \Eq{thin-shell FF} (\Eq{thin-shell filled FF}), while the solid (dot-dashed) green curves represent the \figmath{S_{NN}} (\figmath{F_\text{pSD}^{(N, N)}}) form factors from Ref.~\cite{Fitzpatrick:2012ix}, which does (not) include the induced pseudo-scalar contribution (the \figmath{F_\text{PS}^{(N, N')}} terms in \Eq{SDMel^2}). The red curves represent the form factors from Ref.~\cite{Klos:2013rwa}, which include the induced pseudo-scalar contribution as well as \figmath{2}-body corrections, whose uncertainty dictate the curve thickness. For a comparison, the \figmath{F_{1, 1}^{(N, N)}} SI form factors from Ref.~\cite{Fitzpatrick:2012ix} have been included in the plots as dashed black curves, see \Eq{F_11}.}}
\end{center}
\end{figure}

\Fig{fig: SDFF} displays SD form factors and structure functions from Refs.~\cite{Fitzpatrick:2012ix, Klos:2013rwa}, together with the thin-shell form factors in Eqs.~\eqref{thin-shell FF} (dotted orange lines) and~\eqref{thin-shell filled FF} (solid orange lines). For a comparison, the SI form factors from Ref.~\cite{Fitzpatrick:2012ix} are also shown as dashed black lines, see \Eq{F_11}. All curves are normalized to $1$ at $\ER = 0$. The structure functions from Ref.~\cite{Klos:2013rwa}, in red, include the contribution of the induced pseudo-scalar interaction (the $F_\text{PS}^{(N, N')}$ terms in \Eq{SDMel^2}), while those from Ref.~\cite{Fitzpatrick:2012ix} are shown both with (solid green lines) and without (dot-dashed green lines) this contribution. The form factors from Ref.~\cite{Klos:2013rwa} also include $2$-body corrections, whose uncertainty dictates the width of the curves (notice that it is not possible to see the $q = 0$ effect of these corrections due to the common normalization). It can be appreciated that the tail of the SD form factors and structure functions of the heavy nuclei is longer than that of the respective SI form factor, a rather general feature that in certain conditions could make the SD scattering more efficient than the SI scattering for detection~\cite{Engel:1991wq}. This enhanced sensitivity however can only be possibly achieved if the DM particles are very heavy, as the scattering rate for light DM is dampened at large recoil energies by the exponential fall-off of the velocity distribution, see \Sec{velocity}.

\subsection{Vector-mediated interaction}
\label{Vector-mediated interaction}
We now consider a DM particle scattering elastically off a nucleus through exchange of a vector mediator $V$. We restrict ourselves to spin-$1/2$ DM as it has a richer phenomenology than the spin-$0$ case, as one can see by \eg comparing \Tab{tab: spin-0 DM EFT} and \Tab{tab: spin-1/2 DM EFT}. We will first consider a generic vector mediator and then, in an almost real-world example, check the viability of a model where spin-$1/2$ DM couples at tree level with the $Z$ boson.

The interaction Lagrangian can be written as
\beq\label{DM V}
\Lag = \bar{\chi} (c_\chi \gamma^\mu + a_\chi \gamma^\mu \gamma^5) \chi \, V_\mu + \sum_q \bar{q} (c_q \gamma^\mu + a_q \gamma^\mu \gamma^5) q \, V_\mu \ ,
\eeq
where the $c_\chi$ term vanishes for Majorana DM, see \Eq{Majorana}. As can be seen by using \Eq{P T} and \Tab{tab: PT}, $a_\chi = a_q = 0$ ($c_\chi = c_q = 0$) is a necessary condition for the interaction to preserve parity, in which case $V$ can be assigned a definite parity:
\beq
P V^\mu(x) P^{-1} = - \eta \mathscr{P}^\mu_{\phantom{\mu} \nu} V^\nu(\mathscr{P} x) = - (-1)^\mu \eta V^\mu(\mathscr{P} x)
\eeq
with $\eta = -1$ ($\eta = 1$). With the results of \Sec{Examples} we can compute the DM-nucleon scattering amplitude at second order in the $S$-matrix expansion and at leading order in the NR expansion. Disregarding for simplicity the induced pseudo-scalar interaction (the $\Op^N_6$ term in \Eq{ONR SD}), we have
\beq
\Mel_N \NReq - \frac{4 \mDM \mN}{q^2 + m_V^2} (c_\chi c_N \, \CMcal{I}_\chi \CMcal{I}_N - 4 a_\chi a_N \, \bol{s}_\chi \cdot \bol{s}_N) \ ,
\eeq
where $m_V$ is the vector mediator mass, $\CMcal{I}_{\chi, N}$ and $\bol{s}_{\chi, N}$ are defined as in \Eq{I s}, and
\begin{align}
c_N \equiv \sum_q c_q F_1^{q, N}(0) \ ,
&&&
a_N \equiv \sum_q a_q \Delta_q^{(N)} \ ,
\end{align}
see Secs.~\ref{Vector couplings} and~\ref{Axial-vector couplings}. From this we get for the vector couplings, using Eqs.~\eqref{F_1^{q, p}},~\eqref{V_q isospin},
\begin{align}
c_p = 2 c_u + c_d \ ,
&&&
c_n = c_u + 2 c_d \ .
\end{align}
The above result yields the leading-order NR interaction operator
\beq
\Op_\NR^N = - \frac{4 \mDM \mN}{q^2 + m_V^2} \left( c_\chi c_N \, \Op^N_1 - 4 a_\chi a_N \, \Op^N_4 \right) ,
\eeq
which involves both a SI part (the $\Op^N_1$ term) and a SD part (the $\Op^N_4$ term), although with $q^2$-dependent coefficients. Assuming as usual the initial-state DM particle to be unpolarized, so that the interference term between the SI and SD interactions vanishes (see \Sec{NR operators}), and using \Eq{spinsum 2}, we obtain
\beq
\overline{| \Mel_N |^2} \NReq \frac{16 \mDM^2 \mN^2}{(q^2 + m_V^2)^2} (c_\chi^2 c_N^2 + 3 a_\chi^2 a_N^2) \ .
\eeq
The DM-nucleon total cross section is therefore
\beq\label{sigma_N with propagator}
\sigma_N \NReq \frac{\mu_\text{N}^2}{\pi} \frac{1}{m_V^2 (q_\text{max}^2 + m_V^2)} (c_\chi^2 c_N^2 + 3 a_\chi^2 a_N^2) \ ,
\eeq
where $q_\text{max}^2 \equiv 4 \mu_\text{N}^2 v^2$ is the maximum attainable $q^2$ in DM-nucleon scattering (see \Eq{ERinterval}). The factor of $3$ multiplying the SD part is a consequence of the sum over spins and can be seen appearing by comparing Eqs.~\eqref{SI sigma_N} with~\eqref{SD sigma_N}. For $m_V^2 \gg q_\text{max}^2$ we get
\beq\label{V sigma_N}
\sigma_N \approx (10^8~\pb = 10^{-28}~\cm^2) \times \left( \frac{\mu_\text{N}}{\GeV} \right)^2 \left( \frac{\GeV}{m_V} \right)^4 (c_\chi^2 c_N^2 + 3 a_\chi^2 a_N^2) \ .
\eeq

Using the results of Secs.~\ref{SI interaction} and~\ref{SD interaction} we obtain for the differential cross section
\beq\label{diffsigma V}
\frac{\ud \sigma_T}{\ud \ER} \NReq \frac{m_T}{2 \pi v^2} \frac{1}{(q^2 + m_V^2)^2} \left[ c_\chi^2 (Z c_p + (A - Z) c_n)^2 F_\text{SI}^2(\ER) + 4 a_\chi^2 \frac{\amom{J} + 1}{\amom{J}} \sum_{N, N'} a_N a_{N'} \mathbb{S}_N \mathbb{S}_{N'} F_\text{pSD}^{(N, N')}(\ER) \right] .
\eeq
We see that $\sigma_N$, as opposed to the SI and SD interactions (although with the limitations discussed in \Sec{SD interaction}), is not a convenient quantity to parametrize the size of the differential cross section. One reason is that the $\sigma_N$ dependence on the SI and SD coupling constants is quite different from that of the differential cross section. This could be dealt with by writing $\sigma_N$ as a sum of its SI and SD parts, which can then be used to parametrize the size of the SI and SD parts of the differential cross section, respectively, although it is probably simpler to just express $\ud \sigma_T / \ud \ER$ in terms of the coupling constants as in \Eq{diffsigma V}. The other reason is its non-trivial $v$ dependence, owing to the mediator of the $V$ boson, which is absent in \Eq{SI sigma_N} and \Eq{SD sigma_N} due to the contact nature of the SI and SD interactions. Such dependence indeed can be neglected in the limit of heavy mediator, \ie when $m_V^2$ is much larger than $q^2$ in the whole kinematical range of interest for the experiment. The maximum momentum transfer, which depends on the DM and nuclear masses, can be inferred from the left panel of \Fig{fig: TypicalER}, where we also see that $V$ can be always considered a heavy mediator (for the purposes of direct detection) if heavier than few $\GeV$.

We can see from \Eq{diffsigma V} that, for the purposes of direct detection, this model has $\mDM$, $m_V$, $c_\chi c_p$, $c_\chi c_n$, $a_\chi a_p$, and $a_\chi a_n$ as free parameters. Experimental constraints are most usually set in the literature on contact interactions, so that one can only hope to be able to recast such results for this model in the heavy-mediator limit, \ie when we can approximate $q^2 + m_V^2 \simeq m_V^2$. In this limit \Eq{diffsigma V} is basically the sum of the differential cross sections for the SI and SD interactions. This means, for instance, that experimental constraints on the SI interaction with isosinglet couplings can be applied to the relevant slice of parameter space of this model, where the SD interaction can be neglected (see discussion after \Eq{diffsigmaSD}) and $c_p = c_n$ (see \Eq{isosinglet sigma} and related discussion); in such a case the combination of parameters that can be constrained can be seen in \Eq{diffsigma V} to be $c_\chi^2 c_p^2 / m_V^4$ as a function of $\mDM$, see discussion after \Eq{Rate light med} below. It may not be possible to recast existing constraints to different choices of parameter values, in particular for a small $m_V$. Example bounds on a simple model with a light $t$-channel mediator are computed in \Sec{Constraints}.

We now identify $V$ with the $Z$ boson. The DM can couple at tree level with the $Z$ boson if for instance it is part of an electroweak multiplet, although suitable assignments of the multiplet quantum numbers can prevent this interaction as \eg in the model described in Refs.~\cite{Cirelli:2005uq, Cirelli:2009uv}. This coupling is indeed ruled out by direct detection constraints, as we check in the following. The interaction is described by \Eq{DM V} with
\begin{align}
c_{\chi, q} = - \frac{g}{2 c_\text{W}} \tilde{c}_{\chi, q} \ ,
&&&
a_{\chi, q} = - \frac{g}{2 c_\text{W}} \tilde{a}_{\chi, q} \ ,
\end{align}
where $g$ is the $SU(2)_\text{L}$ gauge coupling, $c_\text{W}$ is the cosine of the electroweak gauge bosons mixing angle, and
\begin{align}
\tilde{c}_{u, c, t} &= + \frac{1}{2} - \frac{4}{3} s_\text{W}^2 \ ,
&
\tilde{c}_{d, s, b} &= - \frac{1}{2} + \frac{2}{3} s_\text{W}^2 \ ,
&
\tilde{a}_{u, c, t} &= - \tilde{a}_{d, s, b} = - \frac{1}{2} \ ,
\end{align}
compare with \Eq{JA}. The $g / 2 c_\text{W}$ factor may be computed by means of the tree-level relation~\cite{Zyla:2020zbs}
\beq
\frac{g}{2 c_\text{W}} = \sqrt{\sqrt{2} G_\text{F} m_Z^2} = \frac{m_Z}{v_\text{EW}} \approx 0.37 \ ,
\eeq
where we used $m_Z \approx 91~\GeV$ and $v_\text{EW} \approx 246~\GeV$ (or alternatively $G_\text{F} \approx 1.17 \times 10^{-5}~\GeV^{-2}$). Using then \Eq{s_W} and, for concreteness, the numerical values in \Eq{Delta_q}, we get for the nucleon couplings
\begin{align}
c_p = - \frac{g}{2 c_\text{W}} \left( \frac{1}{2} - 2 s_\text{W}^2 \right) &\approx - 0.01 \ ,
&
c_n = \frac{1}{2} \frac{g}{2 c_\text{W}} &\approx 0.2 \ ,
\\
a_p &\approx 0.2 \ ,
&
a_n &\approx - 0.2 \ .
\end{align}
As already noted in \Sec{Vector couplings}, the value of $F_1^\text{NC}(0) = 1/2 - 2 s_\text{W}^2$ is sizeably affected by radiative corrections due to a large cancellation in the tree-level computation, with the above value of $c_p$ still providing a suitable approximation after applying the correction in \Eq{F_1^NC(0)}. Notice that the $Z$-boson vector coupling to the neutron is more than one order of magnitude larger than the vector coupling to the proton. Since $m_Z^2 \gg q_\text{max}^2$, we get from \Eq{V sigma_N} the rough estimate
\beq\label{Z sigma_N}
\sigma_N \approx 2~\pb \times \left( \frac{\mu_\text{N}}{\GeV} \right)^2 (c_\chi^2 c_N^2 + 3 a_\chi^2 a_N^2) \approx 10^{-2}~\pb \times \left( \frac{\mu_\text{N}}{\GeV} \right)^2 \left( \left(\frac{c_N}{c_n}\right)^2 \tilde{c}_\chi^2 + 3 \tilde{a}_\chi^2 \right) ,
\eeq
see \Eq{units} for the conversion from $\pb$ to $\cm^2$. The fact the $Z$ boson has SI interactions mainly with neutrons implies that the experimental constraints, which are usually set on the model of \Eq{isosinglet sigma} with equal DM-proton and DM-neutron couplings, need to be recast as explained in the discussion related to Eqs.~\eqref{barsigma},~\eqref{D factor} (or otherwise recomputed afresh). Simplifying, assuming that all nuclear species taking active part in the scattering within a given detector can be approximated as a single nuclide, we obtain the bound
\beq
\bar{\sigma}^\text{SI}_p(\mDM) > \frac{\mu_\text{N}^2}{\pi} \frac{c_\chi^2}{m_Z^4} \frac{(Z c_p + (A - Z) c_n)^2}{A^2} \approx 2 \times 10^{-3}~\pb \times \left( \frac{\mu_\text{N}}{\GeV} \right)^2 \tilde{c}_\chi^2 \ ,
\eeq
where $\bar{\sigma}^\text{SI}_p(\mDM)$ is the experimental constraint on the SI DM-proton interaction cross section of the model in \Eq{isosinglet sigma}, and in the last equality we made the rough approximation $A \approx 2 Z$. With $| \tilde{c}_\chi |$ a $\Ord(1)$ number, as one would expect were it non vanishing, a DM-nucleon cross section of this size is completely excluded over a wide range of DM masses by direct detection constraints on the SI interaction (see \Sec{Constraints} and in particular \Fig{fig: bounds} for the computation of an example but realistic xenon bound). The $\bar{\sigma}^\text{SI}_p(\mDM)$ constraint applies strictly if $a_\chi = 0$, and is otherwise conservative in that it allows for a portion of parameter space which a computation taking into proper account a non-zero $a_\chi$ would exclude. The cross section in \Eq{Z sigma_N} is excluded over a wide range of DM masses even if the $c_\chi^2$ term is absent, as for Majorana DM. In this case the interaction is purely SD, to which the experiments, while not as sensitive as for the SI interaction, are sensitive enough to set tight constraints. We recall from \Sec{SD interaction} that the induced pseudo-scalar interaction, which we neglected here for simplicity, may be needed taking into proper account in this case.

\subsection{Scalar-mediated interaction}
\label{Scalar-mediated interaction}
We now consider DM scattering elastically off a nucleus through exchange of a scalar boson $S$, referring the reader to \Sec{Vector-mediated interaction} for some detailed discussions that are not repeated here. We focus on spin-$1/2$ DM interacting with a generic scalar boson, also in the case of pseudo-scalar interaction, and we check the viability of a Higgs-boson mediated interaction for both spin-$0$ and spin-$1/2$ DM.

The interaction Lagrangian can be written as
\beq\label{Lag S}
\Lag = \bar{\chi} (c_\chi + a_\chi \, i \gamma^5) \chi \, S + \sum_q \bar{q} (c_q + a_q \, i \gamma^5) q \, S \ .
\eeq
$a_\chi = a_q = 0$ ($c_\chi = c_q = 0$) is a necessary condition for the scalar interactions to preserve parity, in which case $S$ can be assigned a definite parity,
\beq\label{S parity}
P S(x) P^{-1} = \eta S(\mathscr{P} x)
\eeq
with $\eta = 1$ ($\eta = -1$), and is called a scalar (pseudo-scalar) boson. Quark couplings to scalar bosons are often set to be proportional to the quark mass,
\begin{align}
\label{MFV couplings}
c_q = \frac{m_q}{\Lambda} \ ,
&&&
a_q = \frac{m_q}{\Lambda} \ ,
\end{align}
as it happens if the DM-quark coupling is mediated by the Higgs boson where the mass scale $\Lambda$ is the Higgs vacuum expectation value $v_\text{EW}$. This coupling structure also satisfies the Minimal Flavor Violating paradigm, see \eg Ref.~\cite{DAmbrosio:2002vsn}, while other choices of couplings may incur in strong constraints from flavor physics, see \eg Ref.~\cite{Dolan:2014ska}.

If $c_\chi$ and at least one among the $c_q$'s are non-zero, and unless the other couplings are significantly larger so to compensate for the NR suppression of neglected terms, the leading-order DM-nucleon scattering amplitude is
\beq
\Mel_N \NReq \frac{4 \mDM \mN}{q^2 + m_S^2} c_\chi c_N \, \CMcal{I}_\chi \CMcal{I}_N \ ,
\eeq
where $m_S$ is the scalar mediator mass and
\beq
c_N \equiv \sum_q \frac{\mN}{m_q} c_q f_{Tq}^{(N)} \xrightarrow{\text{\Eq{MFV couplings}}} \frac{\mN}{\Lambda} \sum_q f_{Tq}^{(N)} \ ,
\eeq
see \Sec{Examples} and \Sec{Scalar couplings}. The relevant NR operator is then
\beq\label{ONR S}
\Op_\NR^N = \frac{4 \mDM \mN}{q^2 + m_S^2} \, c_\chi c_N \, \Op^N_1 \ ,
\eeq
which is a SI interaction as explained in \Sec{SI interaction}. The DM-nucleus differential scattering cross section is
\beq
\frac{\ud \sigma_T}{\ud \ER} \NReq \frac{m_T}{2 \pi v^2} \frac{c_\chi^2}{(q^2 + m_S^2)^2} (Z c_p + (A - Z) c_n)^2 F_\text{SI}^2(\ER) \ .
\eeq
The DM-nucleon squared scattering amplitude averaged over initial spins and summed over final spins is
\beq
\overline{| \Mel_N |^2} \NReq \frac{16 \mDM^2 \mN^2}{(q^2 + m_S^2)^2} c_\chi^2 c_N^2 \ ,
\eeq
so that the DM-nucleon cross section reads
\beq
\sigma_N \NReq \frac{\mu_\text{N}^2}{\pi} \frac{1}{m_S^2 (q_\text{max}^2 + m_S^2)} c_\chi^2 c_N^2 \ ,
\eeq
see discussion related to \Eq{sigma_N with propagator}.

We now take $S$ to be the physical Higgs boson $h$, with $m_h \approx 125~\GeV$~\cite{Zyla:2020zbs} and $\Lambda = v_\text{EW}$. Using the values in Eqs.~\eqref{f_Tq Ellis},~\eqref{f_Th Ellis 2} or those in Eqs.~\eqref{f_Tq FLAG},~\eqref{f_Th FLAG} we obtain $\sum_q f_{Tq}^{(N)} \approx 0.3$, a result also confirmed by the more precise computation of Ref.~\cite{Hoferichter:2017olk}. The DM-nucleon scattering cross section then takes, for both proton and neutron, the approximate value
\beq
\sigma_N \simeq \frac{\mu_\text{N}^2}{\pi} \frac{1}{m_h^4} \frac{\mN^2}{v_\text{EW}^2} c_\chi^2 \left( \sum_q f_{Tq}^{(N)} \right)^2 \approx 10^{-6}~\pb \times \left( \frac{\mu_\text{N}}{\GeV} \right)^2 c_\chi^2 \ ,
\eeq
which is excluded over a wide range of DM masses by direct detection constraints on the SI interaction, unless $| c_\chi |$ is considerably smaller than $1$ (see \Sec{Constraints} and in particular \Fig{fig: bounds} for the computation of an example but realistic xenon bound, and see \Eq{units} for the conversion from $\pb$ to $\cm^2$). For scalar DM coupling to the Higgs boson we may write the interaction Lagrangian
\beq
\Lag = c_\phi v_\text{EW} \, \phi^\dagger \phi \, h + \sum_q \frac{m_q}{v_\text{EW}} \, \bar{q} q \, h \ ,
\eeq
so that
\beq
\Mel_N \NReq \frac{2 \mN^2}{q^2 + m_h^2} c_\phi \, \CMcal{I}_N \sum_q f_{Tq}^{(N)} \ .
\eeq
The DM-nucleon cross section is then, for both proton and neutron,
\beq
\sigma_N \simeq \frac{\mu_\text{N}^2}{\pi} \frac{1}{m_h^4} \frac{\mN^2}{4 \mDM^2} c_\phi^2 \left( \sum_q f_{Tq}^{(N)} \right)^2 \approx 10^{-2}~\pb \times \frac{\mu_\text{N}^2}{\mDM^2} c_\phi^2 \ ,
\eeq
which has a different dependence on the DM mass with respect to the spin-$1/2$ case. This cross section is also completely excluded by direct detection constraints unless $c_\phi$ is considerably smaller than $1$ and/or $\mDM$ is large enough (although light DM may also evade detection if outside of the sensitivity window of the experiment).

Going back to the case of spin-$1/2$ DM interacting with a general scalar field $S$, the leading SI interaction in \Eq{ONR S} is absent if $c_\chi = c_q = 0$ in \Eq{Lag S}. In this case, the dominant NR interaction can be seen from \Tab{tab: spin-1/2 DM EFT} to be
\beq\label{ONR PS}
\Op_\NR^N = \frac{4}{q^2 + m_S^2} \, a_\chi a_N(q^2) \, \Op^N_6 \ ,
\eeq
with
\beq
a_N(q^2) \equiv \sum_q a_q \frac{\mN}{m_q} \left( G_q^N(- q^2) - \bar{m} \sum_{q' = u, d, s} \frac{G_{q'}^N(- q^2)}{m_{q'}} \right) ,
\eeq
see \Sec{Pseudo-scalar couplings}. Parity is preserved by \Eq{Lag S} if $S$ is a pseudo-scalar, \ie if it is assigned $\eta = -1$ in \Eq{S parity}. Loop corrections to this or similar models are considered \eg in Refs.~\cite{Sanderson:2018lmj, Abe:2018emu, Azevedo:2018exj, Ishiwata:2018sdi, Ghorbani:2018pjh}. The DM-nucleus squared amplitude for unpolarized scattering reads
\beq
\overline{| \Mel |^2} \NReq \frac{4}{3} \frac{m_T^2}{\mN^2} \frac{q^4}{(q^2 + m_S^2)^2} \frac{\amom{J} + 1}{\amom{J}} a_\chi^2 \sum_{N, N'} a_N(q^2) a_{N'}(q^2) \mathbb{S}_N \mathbb{S}_{N'} F_\text{PS}^{(N, N')}(\ER) \ ,
\eeq
see \Sec{SD interaction}, and the differential scattering cross section is
\beq\label{diffsigmaPS}
\frac{\ud \sigma_T}{\ud \ER} \NReq \frac{1}{24 \pi v^2} \frac{m_T}{\mDM^2 \mN^2} \frac{q^4}{(q^2 + m_S^2)^2} \frac{\amom{J} + 1}{\amom{J}} a_\chi^2 \sum_{N, N'} a_N(q^2) a_{N'}(q^2) \mathbb{S}_N \mathbb{S}_{N'} F_\text{PS}^{(N, N')}(\ER) \ .
\eeq

\subsection{Magnetic-dipole DM}
\label{Magnetic-dipole DM}
An electrically neutral Dirac DM fermion $\chi$ can interact with photons through an anomalous magnetic moment $\mu_\chi$ (\ie a magnetic dipole moment not directly due to its electric charge, see \Sec{Electromagnetic interactions}). The effective Lagrangian describing this interaction is given in \Tab{tab: EM interactions}, see also Eqs.~\eqref{Q_N},~\eqref{g_N values}. We recall that, owing to \Eq{Majorana}, Majorana particles cannot have a (diagonal) magnetic moment, although they can have a \emph{transition magnetic moment} which couple two different fermion species to a photon. Nucleons interact with photons through the first term in \Eq{JA}, where the nucleon matrix element of the electromagnetic current is given in \Eq{<J_EM>} and its NR expansion in \Eq{<J_EM> NR}. The DM-nucleon NR operator describing the interaction, computed in \Eq{magnetic dipole ONR} and already presented in \Tab{tab: EM interactions}, is
\beq
\Op_\NR^N = 2 e \mu_\chi \left[ \mN Q_N \, \Op^N_1 + 4 \frac{\mDM \mN}{q^2} Q_N \, \Op^N_5 + 2 \mDM g_N \left( \Op^N_4 - \frac{\Op^N_6}{q^2} \right) \right] ,
\eeq
see \Eq{NR building blocks}. This operator presents several interesting aspects. The first part, featuring the $\Op^N_1$ and $\Op^N_5$ building blocks, is due to the \emph{charge-dipole interaction} of the DM magnetic moment with the nuclear electric charge, while the second part, featuring the $\Op^N_4$ and $\Op^N_6$ building blocks, is due to the \emph{dipole-dipole interaction} between the magnetic moments of DM and nucleus. Notice that the first part is independent of the nucleon spin $\bol{s}_N$ while the second part depends on it, and as such this operator can be seen as a combination of the SI and SD interactions discussed in Secs.~\ref{SI interaction},~\ref{SD interaction} (which feature the $\Op^N_1$, $\Op^N_4$, and $\Op^N_6$ building blocks); however, it goes beyond that with its unusual $q^2$ dependence and the presence of $\Op^N_5$, a building block describing the coupling of the DM spin to the nuclear orbital angular momentum, which also introduces an unconventional velocity dependence. Interestingly, the SD-like part of the interaction only involves the $\bol{s}_N$ component orthogonal to $\bol{q}$, as one can see by noticing that $\Op^N_4 - \Op^N_6 / q^2$ can be written as $\bol{s}_\chi \cdot [\bol{s}_N - (\bol{s}_N \cdot \hat{\bol{q}}) \, \hat{\bol{q}}]$; from the discussion in \Sec{Multipoles} we will then expect the scattering amplitude to only depend on the transverse $\Sigma'$ response, as opposed to the SD interaction which also involves the longitudinal $\Sigma''$ response. Finally, $\Op_\NR^N$ features a combination of building blocks that have different degrees of NR suppression: $\Op^N_1$ and $\Op^N_4$ are not suppressed, while $\Op^N_5$ and $\Op^N_6$ are suppressed by (positive) powers of $q$ and $v^\perp_T$. Yet all the terms appear at the same order of the NR expansion, owing to the $1 / q^2$ factors balancing the NR suppression of the $\Op^N_5$ and $\Op^N_6$ terms. Such a variety of structures has to do with the fact that, as already noted after \Eq{magnetic dipole Mel_N}, the DM-nucleon scattering amplitude is second order in the NR expansion, though multiplied by $1 / q^2$; in contrast, the SI and SD interactions all arise as zeroth-order terms.

To have a first idea of the DM-nucleus scattering amplitude, we can take a look at the simpler DM-nucleon scattering: the scattering amplitude is given in \Eq{magnetic dipole Mel_N}, and its square averaged over initial spins and summed over final spins is
\beq
\overline{| \Mel_N |^2} \NReq 4 \mN^2 e^2 \mu_\chi^2 \left[ \left( 1 + 4 \frac{\mDM^2 {v^\perp_N}^2}{q^2} \right) Q_N + \frac{1}{2} \frac{\mDM^2}{\mN^2} g_N^2 \right] ,
\eeq
where we used Eqs.~\eqref{spinsum 2},~\eqref{spinsum 3} and their nucleon equivalent, as well as \Eq{contractions3}. A result closer to the full DM-nucleus scattering amplitude can be obtained with the approximation of point-like nucleus, already adopted \eg in \Sec{SD interaction}. Summing over all nucleons and using \Eq{S_p S_n} we have
\begin{multline}
\Mel_\text{PLN} \NReq 2 m_T e \mu_\chi \left[ \left( \CMcal{I}_\chi + 4 i \frac{\mDM}{q^2} \, \bol{s}_\chi \cdot (\bol{q} \times \bol{v}^\perp_T) \right) Z \, \delta_{\amom{M} \amom{M}'} \right.
\\
\left. + 2 \frac{\mDM}{\mN} \sum_{N = p, n} g_N \left( \bol{s}_\chi \cdot \matel{\amom{J}, \amom{M}'}{\bol{S}_N}{\amom{J}, \amom{M}} - \frac{1}{q^2} \, \bol{s}_\chi \cdot \bol{q} \, \matel{\amom{J}, \amom{M}'}{\bol{S}_N}{\amom{J}, \amom{M}} \cdot \bol{q} \right) \right] ,
\end{multline}
where we know from the discussion in \Sec{examples and applications} (see also \Tab{tab: nuclear responses}) that only the CM (or $\bol{v}^\perp_T$) component of $\Op^N_5$ contributes for a point-like nucleus, see \Eq{v^perp_T}. The unpolarized squared amplitude can be computed in this limit following again \Sec{SD interaction}, obtaining
\beq
\overline{| \Mel_\text{PLN} |^2} \NReq 4 m_T^2 e^2 \mu_\chi^2 \left[ \left( 1 + 4 \frac{\mDM^2 {v^\perp_T}^2}{q^2} \right) Z^2 + \frac{2}{3} \frac{\mDM^2}{\mN^2} \frac{\amom{J} + 1}{\amom{J}} (g_p \mathbb{S}_p + g_n \mathbb{S}_n)^2 \right] .
\eeq
The full DM-nucleus unpolarized squared amplitude can be computed using the results of \Sec{Nuclear matrix element}, which yield
\begin{multline}
\overline{| \Mel |^2} \NReq 4 \frac{\mDM^2 m_T^2}{\mN^2} e^2 \mu_\chi^2 \left[ \left( \frac{1}{\mDM^2} - \frac{1}{\mu_T^2} + 4 \frac{v^2}{q^2} \right) \mN^2 F_M^{(p, p)}(q^2) + 4 F_{\Delta}^{(p, p)}(q^2) \right.
\\
\left. \vphantom{\left( 1 + 4 \frac{\mDM^2 {v^\perp_T}^2}{q^2} \right)}
- 2 \sum_N g_N F_{\Sigma' \Delta}^{(N, p)}(q^2) + \frac{1}{4} \sum_{N, N'} g_N g_{N'} F_{\Sigma'}^{(N, N')}(q^2) \right] ,
\end{multline}
where we used \Eq{vperp^2} to write the transverse speed $v^\perp_T$ in terms of the relative DM-nucleus speed $v$. The differential scattering cross section~\eqref{dsigmadER} is therefore
\begin{multline}
\label{Magnetic dipole sigma}
\frac{\ud \sigma_T}{\ud \ER} \NReq \frac{1}{8 \pi} \frac{m_T}{\mN^2} \frac{1}{v^2} e^2 \mu_\chi^2 \left[ \left( \frac{v^2}{\ER} - \frac{\mDM + 2 m_T}{2 \mDM m_T} \right) 2 \frac{\mN^2}{m_T} F_M^{(p, p)}(q^2) + 4 F_{\Delta}^{(p, p)}(q^2) \right.
\\
\left. \vphantom{\left( 1 + 4 \frac{\mDM^2 {v^\perp_T}^2}{q^2} \right)}
- 2 \sum_N g_N F_{\Sigma' \Delta}^{(N, p)}(q^2) + \frac{1}{4} \sum_{N, N'} g_N g_{N'} F_{\Sigma'}^{(N, N')}(q^2) \right] .
\end{multline}
Beside the $F_{\Delta}^{(p, p)}$ and $F_{\Sigma' \Delta}^{(N, p)}$ squared form factors, which do not appear for the other interactions discussed in this Section, the $v^2 / q^2$ (or $v^2 / \ER$) term introduces an unusual dependence on both $q^2$ and $v^2$. The $1 / q^2$ factor modifies the recoil-energy spectrum with respect to that of the SI and SD interactions, especially at low energy where the nuclear form factors do not significantly contribute to the shape of the spectrum. The $v^2$ factor introduces instead some unexpected features in the annual modulation of the signal (see \Sec{Modulation} and the discussion after \Eq{Taylor R}), examined \eg in Refs.~\cite{DelNobile:2015tza, DelNobile:2015rmp, DelNobile:2017fzy}. As argued also in Ref.~\cite{DelNobile:2012tx}, the $v^2 / q^2$ term can be sizeable and actually dominates at low momentum transfer (\ie at low recoil energy); similarly, the dipole-dipole terms can be non-negligible for targets with a large magnetic dipole moment. All these aspects make the magnetic-moment DM model an extremely instructive example, with several features not found in the most standard cases: the NR operator describing the interaction arises at second order in the NR expansion rather than zeroth or first order; it features building blocks with different degrees of NR suppression (see \Eq{NR building blocks}) contributing at the same order of the NR expansion; it features aspects of both the SI and SD interactions with the latter being not always negligible; the differential cross section features a non-standard $q^2$ and $v^2$ dependence and non-standard nuclear form factors.

\section{DM velocity distribution and velocity integral}
\label{velocity}
We have focused so far on the interaction of DM particles with nuclei. It is now time to take a deeper look into the DM velocity distribution at Earth's location and how it affects the scattering rate, which has been introduced in \Sec{Rate}. We will then have all needed ingredients for a qualitative and quantitative understanding of the phenomenology of direct DM detection, which will be discussed in the next Section.

\subsection{DM velocity distribution in Earth's frame}
\label{DM velocity distribution}
In \Sec{Scattering rate} we introduced the DM velocity distribution at Earth's location in the detector's rest frame, $\fE(\bol{v}, t)$. Neglecting the small effect of Earth's rotation around its own axis (see below), the detector's rest frame coincide with that of Earth. The DM distribution is not expected to change significantly over the timescale of an experiment (years), so that the time dependence of $\fE$ is mainly due to Earth's revolution around the Sun. $\fE$ can be obtained from the local DM velocity distribution in the rest frame of the galaxy, $\fG(\bol{w})$, where we denote with $\bol{w}$ the DM velocity in that reference frame. The velocity distribution is normalized as
\beq\label{f norm}
\int \ud^3 w \, \fG(\bol{w}) = 1 \ ,
\eeq
a relation that can be trivially boosted to any reference frame (see \eg \Eq{f norm E}). $\bol{v} = \bol{w} - \bol{v}_\text{E}(t)$ implies $\fE(\bol{v}, t) = \fG(\bol{v} + \bol{v}_\text{E}(t))$, where $\bol{v}_\text{E}(t)$ (sometimes denoted $\bol{v}_\text{obs}(t)$) is Earth's velocity with respect to the galactic rest frame. $\bol{v}_\text{E}(t)$ can be decomposed as
\beq\label{vec vE}
\bol{v}_\text{E}(t) = \bol{v}_\text{S} + \bol{v}_\oplus(t) \ ,
\eeq
with $\bol{v}_\text{S}$ the Sun's velocity in the galactic rest frame, and $\bol{v}_\oplus(t)$ the velocity of Earth with respect to the Sun. The time dependence is due to Earth's revolution around the Sun. In turn, $\bol{v}_\text{S}$ can be written as
\beq
\bol{v}_\text{S} = \bol{v}_\text{LSR} + \bol{v}_\odot \ ,
\eeq
with $\bol{v}_\text{LSR}$ the rotational velocity of the local standard of rest (LSR) and $\bol{v}_\odot$ the velocity of the Sun with respect to the local standard of rest (Sun's peculiar velocity). We can express these velocities in galactic coordinates, where $\hat{\bol{x}}$ points towards the galactic center, $\hat{\bol{y}}$ in the direction of galactic rotation and $\hat{\bol{z}}$ normal to the galactic plane in the direction of the galactic North pole. Then $\bol{v}_\text{LSR} \approx (0, v_\text{c} , 0)^\tr$, where the conventionally assumed value $v_\text{c} = 220~\km / \sd$ has an $\Ord(10 \%)$ uncertainty~\cite{Bovy:2009dr, Reid:2009nj, McMillan:2009yr, Bovy:2012ba} (see also Ref.~\cite{Green:2017odb}), and $\bol{v}_\odot \approx (11, 12, 7)^\tr~\km / \sd$ with less than $20 \%$ uncertainty~\cite{Schoenrich:2009bx} (summing the statistical and systematic uncertainties in quadrature). The Sun's speed in the rest frame of the galaxy is then $v_\text{S} \approx 232~\km / \sd$. The local circular speed $v_\text{c}$ is also denoted $v_\text{rot}$.

Neglecting the small eccentricity of Earth's orbit, the rotational velocity of Earth can be written as
\beq
\bol{v}_\oplus(t) = v_\oplus \left( \cos[\omega (t - t_\text{eq})] \, \hat{\bol{\epsilon}}_1 + \sin[\omega (t - t_\text{eq})] \, \hat{\bol{\epsilon}}_2 \right) ,
\eeq
with $v_\oplus \approx 30~\km / \sd$, $\omega \equiv 2 \pi / \yr$ and $t_\text{eq}$ the time of the March equinox, about March $21^\text{st}$ (vernal or spring equinox in the Northern hemisphere; the exact date depends on the year). The orthogonal unit vectors $\hat{\bol{\epsilon}}_1$ and $\hat{\bol{\epsilon}}_2$, spanning the ecliptic plane, have $\hat{\bol{x}}$, $\hat{\bol{y}}$, $\hat{\bol{z}}$ coordinates
\begin{align}
\hat{\bol{\epsilon}}_1 \approx (0.99, 0.11, 0.00)^\tr \ ,
&&&
\hat{\bol{\epsilon}}_2 \approx (- 0.05, 0.49, - 0.87)^\tr \ .
\end{align}
$\hat{\bol{\epsilon}}_1$ is anti-aligned with the projection of Earth's rotational axis onto the ecliptic plane, so that it points from the Sun to Earth during the June solstice, around June $21^\text{st}$, when Earth's rotational axis is maximally tilted toward the Sun in the Northern hemisphere (summer solstice). $\hat{\bol{\epsilon}}_2$ points instead from Earth to the Sun at $t_\text{eq}$. Because the equinoxes precess, the galactic coordinates of these vectors change in time, with the above approximate values being valid for at least few decades after $2020$~(see \eg Ref.~\cite{McCabe:2013kea}). We now denote with $t_0$ the time of maximal alignment between $\bol{v}_\text{S}$ and $\bol{v}_\oplus$, meaning the time when $\bol{v}_\oplus$ is parallel to the projection of $\bol{v}_\text{S}$ onto the ecliptic plane. $t_0$ is therefore the time when $\vE(t)$ is maximal, \ie Earth moves fastest in the rest frame of the galaxy and, seen from Earth, DM particles can reach their highest speeds. We then have
\beq
\hat{\bol{v}}_\text{S} \cdot \hat{\bol{v}}_\oplus(t) = b \cos[\omega (t - t_0)] = \hat{\bol{v}}_\text{S} \cdot \hat{\bol{\epsilon}}_1 \, \cos[\omega (t - t_\text{eq})] + \hat{\bol{v}}_\text{S} \cdot \hat{\bol{\epsilon}}_2 \, \sin[\omega (t - t_\text{eq})] \ ,
\eeq
with
\beq
b \equiv \sqrt{(\hat{\bol{v}}_\text{S} \cdot \hat{\bol{\epsilon}}_1)^2 + (\hat{\bol{v}}_\text{S} \cdot \hat{\bol{\epsilon}}_2)^2} \approx 0.5
\eeq
the cosine of the angle between $\bol{v}_\text{S}$ and the ecliptic plane (about $60^\circ$). Expressing this in terms of sines and cosines of $\omega t$ we find
\beq
t_0 = \frac{1}{\omega} \arctan \left( \frac{\hat{\bol{v}}_\text{S} \cdot \hat{\bol{\epsilon}}_1 \, \sin(\omega t_\text{eq}) + \hat{\bol{v}}_\text{S} \cdot \hat{\bol{\epsilon}}_2 \, \cos(\omega t_\text{eq})}{\hat{\bol{v}}_\text{S} \cdot \hat{\bol{\epsilon}}_1 \, \cos(\omega t_\text{eq}) - \hat{\bol{v}}_\text{S} \cdot \hat{\bol{\epsilon}}_2 \, \sin(\omega t_\text{eq})} \right) = t_\text{eq} + \frac{1}{\omega} \arctan \left( \frac{\hat{\bol{v}}_\text{S} \cdot \hat{\bol{\epsilon}}_2}{\hat{\bol{v}}_\text{S} \cdot \hat{\bol{\epsilon}}_1} \right) ,
\eeq
resulting around June $1^\text{st}$~(see \eg Ref.~\cite{McCabe:2013kea}). In the literature $t_0$ is most often found stated as June $2^\text{nd}$, which is within the inaccuracy of this discussion. From \Eq{vec vE} we can write Earth's speed in the rest frame of the galaxy as
\beq\label{vE}
\vE(t) = \sqrt{v_\text{S}^2 + v_\oplus^2 + 2 b \, v_\text{S} v_\oplus \cos[\omega (t - t_0)]} \ .
\eeq
More detailed information on $\bol{v}_\oplus(t)$ and $\bol{v}_\text{E}(t)$ can be found \eg in Refs.~\cite{Fornengo:2003fm, Green:2003yh, Lee:2013xxa, McCabe:2013kea}.

The speed of DM particles that are gravitationally bound to our galaxy is limited by the galactic escape speed, which depends on the distance from the galactic center. For the local escape speed, $\vesc$, the RAVE survey found $\vesc = 533^{+54}_{-41}~\km / \sd$ (at $90 \%$ confidence, with an additional $4 \%$ systematic uncertainty)~\cite{Piffl:2013mla}, while $\vesc = 580 \pm 63~\km / \sd$ was inferred from data from the Gaia survey~\cite{Brown:2018dum} in Ref.~\cite{Monari:2018}. Notice that the local astrophysical parameters $\vesc$, the circular speed of the Sun $v_\text{c}$, and the local DM density $\rho$, are all correlated, as stressed \eg in Refs.~\cite{Belli:2002yt, McCabe:2010zh, Lavalle:2014rsa}.

We adopt the values
\begin{align}
\label{speeds}
v_\text{c} = 220~\km / \sd \ ,
&&
v_\text{S} = 232~\km / \sd \ ,
&&
v_\oplus = 30~\km / \sd \ ,
&&
\vesc = 533~\km / \sd \ ,
&&
b = 0.5 \ ,
\end{align}
for our numerical results and plots, unless otherwise noted. The impact of astrophysical uncertainties on direct DM searches is studied \eg in Refs.~\cite{McCabe:2010zh, Green:2010gw, Wu:2019nhd}. The local DM density $\rho$ can be seen from \Eq{diffrate} to be a multiplicative parameter of the rate (thus completely degenerate with the overall size of the cross section, as already noted in \Sec{Scattering rate}). A variation in the value of $\rho$, whose uncertainty was discussed in \Sec{Scattering rate}, changes then the overall size of the scattering rate, which affects equally all the different direct DM searches, so for instance an increase in $\rho$ would proportionally enhance the sensitivity of all experiments while leaving their relative sensitivities unaltered. On the contrary, a change in $\vesc$ can affect the sensitivity of some experiments more than others. The detection of light DM, in fact, relies on the DM particles having large speeds for the nuclear recoil energy to be large enough to lie within the energy range the experiment is sensitive to (see discussion in \Sec{Rate spectrum}). Varying the maximum possible speed DM particles can have thus changes the minimum mass with which an experiment can detect them. Even in a hypothetical situation where different experiments are sensitive to the same recoil energies, as explained in Secs.~\ref{Elastic scattering},~\ref{Inelastic scattering}, the maximum $\ER$ at given DM mass depends on the target mass: for light DM, it is smaller for heavier nuclei provided the mass splitting $\delta$, if positive, is sufficiently small. Therefore, a smaller escape speed is likely to cause a larger reduction in the sensitivity to light DM particles for experiments employing heavier targets, unless $\delta$ is positive and sizeable. Some aspects of this behavior can be appreciated in Figs.~\ref{fig: ElasticKinematics},~\ref{fig: InelasticKinematics}.

$\fE(\bol{v}, t)$ enters the scattering rate through a velocity integral weighting the DM-nucleus differential scattering cross section $\ud \sigma_T / \ud \ER$ (times a flux factor) with the probability of the DM particle having a certain velocity in Earth's frame, see \Eq{diffrate}. In analogy with \Eq{eta}, we define here a `generic' velocity integral
\begin{align}
\label{generic eta}
\eta(\vmin, t) = \int_{v \geqslant \vmin} \ud^3 v \, H(\bol{v}, t)
&&
\text{with}
&&
H(\bol{v}, t) \equiv \fE(\bol{v}, t) \, v \, h(\bol{v}) \ ,
\end{align}
where
\beq
\ud^3 v = \ud v \, (v \, \ud \vartheta) (v \sin \vartheta \, \ud \phi) = \ud v \, v^2 \, \ud \cos \vartheta \, \ud \phi \ .
\eeq
Although the nature of $\vmin$ is of little relevance here, we recall from \Sec{Scattering rate} that this is the minimum speed a DM particle must have in order to be able to transfer a certain energy $\ER$ to the nucleus, see \Eq{vmin} for elastic scattering and \Eq{vmin_inelastic} for inelastic scattering. The non-negative function $h(\bol{v})$ represents the velocity dependence of $\ud \sigma_T / \ud \ER$. The $\eta_n$'s defined in \Eq{eta} (see also \Eq{Taylor rate}) are all specific instances of this generic velocity integral, with
\beq\label{eta_n h(v)}
h(\bol{v}) = v^{2 (n - 1)} \ .
\eeq
By definition, the velocity integral is non-negative (because the integrand is non-negative), is a non-increasing function of $\vmin$ (because raising $\vmin$ reduces the domain of the integral), and vanishes for $\vmin > \vesc + \vE(t)$ (because $\fE(\bol{v}, t)$ vanishes for $v > \vesc + \vE(t)$, \ie for galactic-frame speeds larger than $\vesc$). Neglecting Earth's rotation around its own axis, which accounts for a maximum speed at Earth's surface of roughly $0.5~\km / \sec$ in Earth's rest frame, the time dependence of the velocity integral is entirely due to the $\fE(\bol{v}, t)$ dependence on $\bol{v}_\text{E}(t)$.

In the standard assumption the differential cross section only depends on $v$ and not on $\hat{\bol{v}}$, which is certainly the case if both the DM particle and the target nucleus are unpolarized, we denote $h(\bol{v})$ in \Eq{generic eta} with $h(v)$. In this situation the angular integrals in \Eq{generic eta} only involve $\fE(\bol{v}, t)$, so that it makes sense to define the one-dimensional speed distribution in Earth's frame,
\beq\label{1D speed distro}
F_\text{E}(v, t) \equiv v^2 \int_{-1}^{+1} \ud \cos \vartheta \int_0^{2 \pi} \ud \phi \, \fE(\bol{v}, t) \ ,
\eeq
which is normalized as
\beq
\int_0^\infty \ud v \, F_\text{E}(v, t) = \int \ud^3 v \, \fE(\bol{v}, t) = 1 \ ,
\eeq
see \Eq{f norm E}. The velocity integral in \Eq{generic eta} can then be written as
\beq\label{generic eta h(v)}
\eta(\vmin, t) = \int_{\vmin}^\infty \ud v \, F_\text{E}(v, t) \, v \, h(v) \ .
\eeq

The above discussion can help understanding how the $\eta_n$ integrals defined in \Eq{eta} are mutually related, as anticipated in \Sec{Scattering rate} (see \eg Ref.~\cite{DelNobile:2015rmp}). Use of \Eq{eta_n h(v)} in \Eq{generic eta h(v)} and differentiation with respect to $\vmin$ leads to
\beq
F_\text{E}(v, t) = - \frac{1}{v^{2 n - 1}} \frac{\ud \eta_n(v, t)}{\ud v} \ ,
\eeq
independently of $n$, which can be plugged back into \Eq{generic eta h(v)} to obtain for $m \geqslant 0$
\beq
\eta_m(\vmin, t) = - \int_{\vmin}^\infty \ud v \, v^{2 (m - n)} \frac{\ud \eta_n(v, t)}{\ud v} \ ,
\eeq
yielding upon integration by parts
\beq\label{eta_n relation}
\eta_m(\vmin, t) = \vmin^{2 (m - n)} \, \eta_n(\vmin, t) + 2 (m - n) \int_{\vmin}^\infty \ud v \, v^{2 (m - n) - 1} \, \eta_n(v, t) \ .
\eeq
This provides a relation between any two $\eta_n$'s, which, involving only a one-dimensional integral, may prove convenient when computing the $\eta_n$'s for a velocity distribution where the three-dimensional $\ud^3 v$ integral in \Eq{eta} is particularly demanding. \Eq{eta_n relation} can also be cast into a formula for the indefinite integral of $v^{2 (m - n) - 1} \, \eta_n(v, t)$: in fact, evaluating at two different $\vmin$ values and subtracting the two expressions yields
\beq\label{eta integral}
2 (m - n) \int \ud \vmin \, \vmin^{2 (m - n) - 1} \, \eta_n(\vmin, t) = \vmin^{2 (m - n)} \, \eta_n(\vmin, t) - \eta_m(\vmin, t)
\eeq
up to a constant. The same result can also be obtained starting from the definite integral of $v^{2 (m - n) - 1} \, \eta_n(v, t)$ with generic extrema $v_1$ and $v_2$, substituting Eqs.~\eqref{generic eta h(v)},~\eqref{eta_n h(v)}, and then exchanging the $\vmin$ and $v$ integrals through
\beq\label{integral exchange}
\int_{v_1}^{v_2} \ud \vmin \int_{\vmin}^\infty \ud v = \int_{v_1}^{v_2} \ud v \int_{v_1}^v \ud \vmin + \int_{v_2}^\infty \ud v \int_{v_1}^{v_2} \ud \vmin \ .
\eeq
An immediate application of \Eq{eta_n relation} is, for $r > 0$,
\beq\label{eta_n+r(0)}
\eta_{n + r}(0, t) = 2 r \int_0^\infty \ud v \, v^{2 r - 1} \, \eta_n(v, t) \ ,
\eeq
which allows to quantify the characteristic size of $\eta_{n + r}$ while also providing a chain of integral equalities for different values of $n, r$ at fixed $n + r$. One example, extending the definition of the $\eta_n$ integrals in \Eq{eta} to half-integer indices, is
\beq
\eta_{n + 1/2}(0, t) = \int_0^\infty \ud v \, \eta_n(v, t) \ ,
\eeq
which immediately yields with \Eq{f norm E} a normalization condition for $\eta_0$,
\beq\label{int eta0}
\int_0^\infty \ud \vmin \, \eta_0(\vmin, t) = 1 \ ,
\eeq
also obtainable by applying Eqs.~\eqref{generic eta h(v)},~\eqref{eta_n h(v)} and then \Eq{integral exchange}.

\subsection{Annual modulation}
\label{Modulation}
The velocity integral, and thus the scattering rate, depends on time through the variation of the DM flux on Earth due to Earth's motion around the Sun. DM signals are therefore expected to be modulated on a (mostly) annual basis, see \eg Ref.~\cite{Freese:2012xd} for a review. This time dependence has distinctive (though model-dependent) features that can help telling a putative DM signal from mismodeled or unaccounted for backgrounds, a likely crucial test in establishing the actual DM origin of any signal. The time dependence of known backgrounds, in fact, is different from what is expected from a DM signal, see \eg Refs.~\cite{Chang:2011eb, FernandezMartinez:2012wd, Pradler:2012kv, Bernabei:2014tqa, Davis:2014ama, Klinger:2015vga}. The time dependence can also help discriminating between different models of DM interactions and of DM halos, see \eg Refs.~\cite{DelNobile:2015nua, DelNobile:2015uua, Witte:2016ydc}.

Given the annual periodicity of $\fE(\bol{v}, t)$ (to a very good approximation), the time dependence of a generic velocity integral $\eta(\vmin, t)$ can be meaningfully parametrized in terms of a Fourier series,
\beq\label{Fourier eta}
\begin{split}
\eta(\vmin, t) &= a_0(\vmin) + \sum_{n = 1}^\infty \big( a_n(\vmin) \cos[n \omega (t - \tau)] + b_n(\vmin) \sin[n \omega (t - \tau)] \big)
\\
&= A_0(\vmin) + \sum_{n = 1}^\infty A_n(\vmin) \cos[n \omega (t - t_n(\vmin))] \ ,
\end{split}
\eeq
with $\tau$ an arbitrary phase parameter. The $t_n$'s can be chosen so that $A_{n \neq 0} \geqslant 0$, while $A_0 \geqslant 0$ is ensured by $\eta(\vmin, t)$ being by definition non-negative (it is the integral of a non-negative function). Expressing the above in terms of sines and cosines of $n \omega (t - t_n)$ and comparing the two parametrizations one has
\begin{align}
A_0 = a_0 \ ,
&&
A_{n \neq 0} = \sqrt{a_n^2 + b_n^2} \ ,
&&
t_n = \tau + \frac{1}{n \omega} \arctan \left( \frac{b_n}{a_n} \right) .
\end{align}
For $\eta_0$ defined in \Eq{eta} (see also \Eq{eta_0}), \Eq{int eta0} implies $\int_0^\infty \ud \vmin \, a_0(\vmin) = 1$, while the sum of all higher modes integrates to zero.

For a locally isotropic DM velocity distribution in the galactic rest frame, the velocity integral depends on $\vE(t)$ but not on $\hat{\bol{v}}_\text{E}(t)$: in fact, $\eta(\vmin, t)$ is invariant under rotations but there are no available three-vectors to form a rotational invariant with $\hat{\bol{v}}_\text{E}(t)$ (apart of course from $\hat{\bol{v}}_\text{E}(t)$ itself). This implies that $\vE(t)$ is the only source of time dependence of $\eta(\vmin, t)$. Assuming Earth's orbit to be perfectly circular, $\vE(t)$ is symmetric about $t = t_0$, or in other words $\vE(t)$ is an even function of $\Delta t \equiv t - t_0$ (see \Eq{vE}): Earth's speed in the galactic rest frame is the same a time $\Delta t$ before and after $t_0$. It follows that, under these assumptions, also $\eta(\vmin, t)$ is an even function of $t - t_0$, which means that choosing $\tau = t_0$ automatically sets all $b_n = 0$. All the $t_n(\vmin)$'s would then equal $t_0$ ($t_0 + \yr / 2$) for those $\vmin$ values where $a_n(\vmin)$ is positive (negative). Anisotropies in the DM velocity distribution modify this picture, most strikingly by endowing the $t_n(\vmin)$'s with a marked $\vmin$ dependence. If not by possible DM velocity substructures (\eg DM streams), a local anisotropy in $\fG(\bol{w})$ is certainly induced by the Sun acting as a gravitational lens, which focusses the DM particles depending on their velocity (see \eg Refs.~\cite{Griest:1987vc, Sikivie:2002bj, Alenazi:2006wu, Patla:2013vza, Lee:2013wza, Bozorgnia:2014dqa, DelNobile:2015nua}). While the effect on the rate annual average (see below) can be negligible, the annual modulation and higher modulation harmonics can be sizeably influenced, see \eg Ref.~\cite{DelNobile:2015nua}. The eccentricity of Earth's orbit, too, affects the time dependence of the velocity integral, although to a lesser extent (see \eg Refs.~\cite{Bozorgnia:2014dqa, DelNobile:2015nua}).

If $v_\oplus$ were zero, Earth would be constantly experiencing the same DM flux, and the rate would simply be constant in time. Given that $v_\oplus \ll v_\text{S}$, we can expect the annual average
\beq
A_0(\vmin) = \frac{1}{1~\yr} \int_{1~\yr} \eta(\vmin, t) \, \ud t
\eeq
to be the main mode, followed by the annual modulation amplitude $A_1$. The higher modes can be relevant in the presence of large anisotropies in the DM velocity distribution (see \eg Ref.~\cite{DelNobile:2015nua}). Assuming that these higher modes can be neglected, a Taylor expansion of the velocity integral in powers of $v_\oplus / v_\text{S} \simeq 0.1$ returns
\beq\label{Taylor eta}
\eta(\vmin, t) \simeq \overline{\eta}(\vmin) + \widetilde{\eta}(\vmin) \cos[\omega (t - t_0)] \ ,
\eeq
with
\begin{align}
\label{etabartilde def}
\overline{\eta}(\vmin) \equiv \left. \eta(\vmin, t) \right|_{\vE(t) = v_\text{S}} \ ,
&&&
\widetilde{\eta}(\vmin) \equiv \left. \frac{\ud \eta(\vmin, t)}{\ud \vE} \right|_{\vE(t) = v_\text{S}} b \, v_\oplus \ .
\end{align}
Here we exploited the fact that $\eta$ depends on $v_\oplus(t)$ only through $\vE(t)$, given in \Eq{vE}, which can be approximated with less than $1 \%$ error as
\beq\label{Taylor vE}
\vE(t) \simeq v_\text{S} + b \, v_\oplus \cos[\omega (t - t_0)] \ .
\eeq
Setting $\tau = t_0$ in \Eq{Fourier eta}, we can then match
\begin{align}
A_0(\vmin) = a_0(\vmin) \simeq \overline{\eta}(\vmin) \ ,
&&&
a_1(\vmin) \simeq \widetilde{\eta}(\vmin) \ .
\end{align}
$\overline{\eta}$ and $\widetilde{\eta}$ constitute convenient approximations to the annual average and annual modulation amplitude, respectively, whose computation would otherwise entail performing integrals. In the regime of validity of \Eq{Taylor eta} one also has the useful approximations
\begin{align}
\label{etabartilde approx}
\overline{\eta}(\vmin) \simeq \frac{\eta(\vmin, t_0) + \eta(\vmin, t_0 + \yr / 2)}{2} \ ,
&&&
\widetilde{\eta}(\vmin) \simeq \frac{\eta(\vmin, t_0) - \eta(\vmin, t_0 + \yr / 2)}{2} \ .
\end{align}
It will be useful for the rest of this discussion to define
\begin{subequations}
\label{vmax^+-}
\begin{align}
\vmax^+ &\equiv \vesc + \max_t \vE(t) = \vesc + \vE(t_0) \simeq \vesc + v_\text{S} + b \, v_\oplus \ ,
\\
\vmax^- &\equiv \vesc + \min_t \vE(t) = \vesc + \vE(t_0 + \yr / 2) \simeq \vesc + v_\text{S} - b \, v_\oplus \ ,
\end{align}
\end{subequations}
respectively the maximum and minimum value of the maximum DM speed on Earth during the year. All speeds $\vmax^- < v < \vmax^+$ only contribute to the velocity integral for part of the year (see \Eq{computing velocity integral} below and subsequent discussion). For $\vmax^- < \vmin < \vmax^+$, the velocity integral $\eta(\vmin, t)$ is then only non-zero for a time interval around $t_0$ (see the bottom-left panel of \Fig{fig: eta} below for an illustration). This is a feature \Eq{Taylor eta} cannot reproduce without the higher Fourier modes being included, a sign that the adopted approximation breaks down for $\vmin > \vmax^-$. Such large $\vmin$ values can be notably relevant for light DM particles, especially when scattering off heavy targets (or off light targets for a positive and sufficiently large mass splitting $\delta$, see \Sec{Inelastic scattering}): in this case, in fact, only DM particles with very large speeds can kick the target nucleus hard enough for its recoil energy to be in the sensitivity range of the experiments (see dedicated discussion in \Sec{Rate spectrum}).

One can see from Eqs.~\eqref{Taylor eta},~\eqref{etabartilde def} that, unless there are reasons to expect $\left| \ud \eta / \ud \vE \right|_{\vE = v_\text{S}}$ to be significantly larger than $\overline{\eta} / v_\text{S}$, the annual modulation amplitude $|\widetilde{\eta}|$ is suppressed with respect to the annual average $\overline{\eta}$ by a factor $b \, v_\oplus / v_\text{S} \approx 0.06$. This provides an order of magnitude expectation for the fractional annual modulation $| \widetilde{\eta} / \overline{\eta}| $. However, large departures from this value can occur in at least two cases, other than the presence of large anisotropies in the DM velocity distribution. First, $| \widetilde{\eta} / \overline{\eta} |$ reaches $100 \%$ at $\vmin = \vmax^-$ due to the fact that, as explained above, the constant term in \Eq{Fourier eta} stops being dominant against the other Fourier modes. Another indication that $| \widetilde{\eta} / \overline{\eta} |$ approaches unity in this regime can be derived from \Eq{etabartilde approx} by setting $\eta(\vmin, t_0 + \yr / 2) = 0$, a consequence of $\eta(\vmin > \vmax^-, t)$ only being non-zero within a time interval (symmetric) around $t_0$ (see discussion above). Second, $\widetilde{\eta}$ may vanish at a given $\vmin$ value. This can happen if the one-dimensional speed distribution $F_\text{E}$~\eqref{1D speed distro} at a given time increases with $t$ at some $v$ value while decreasing at some other $v$ value, a feature that the velocity integral may inherit depending on $h(v)$. If this is the case, $\widetilde{\eta}$ has opposite signs at two such $\vmin$ values, and therefore it must vanish somewhere in between. This occurs for instance with the $\eta_0$ velocity integral in the Standard Halo Model, see Figs.~\ref{fig: eta},~\ref{fig: etabartilde} below. In general, one can expect higher-order corrections and otherwise small effects (such as higher Fourier modes, or the gravitational focussing effect of the Sun) to become relevant, or even more so, when $\widetilde{\eta}$ vanishes within the adopted approximation.

\subsection{Computing the velocity integral}
\label{Computing the velocity integral}
The boost from the galactic frame to Earth's frame causes $\hat{\bol{v}}_\text{E}(t)$ to be a preferential direction in the DM velocity distribution $\fE(\bol{v}, t)$, even in the assumption $\fG(\bol{w})$ is completely isotropic. It is then natural to take this as a reference direction to measure angles. In the following we take $\vartheta$ to be the angle between $\bol{v}$ and $\bol{v}_\text{E}(t)$ (zenith angle), and $\phi$ to be the angle of rotation about the direction of $\hat{\bol{v}}_\text{E}(t)$ (azimuth).

\begin{figure}[t!]
\begin{center}
\includegraphics[width=.49\textwidth]{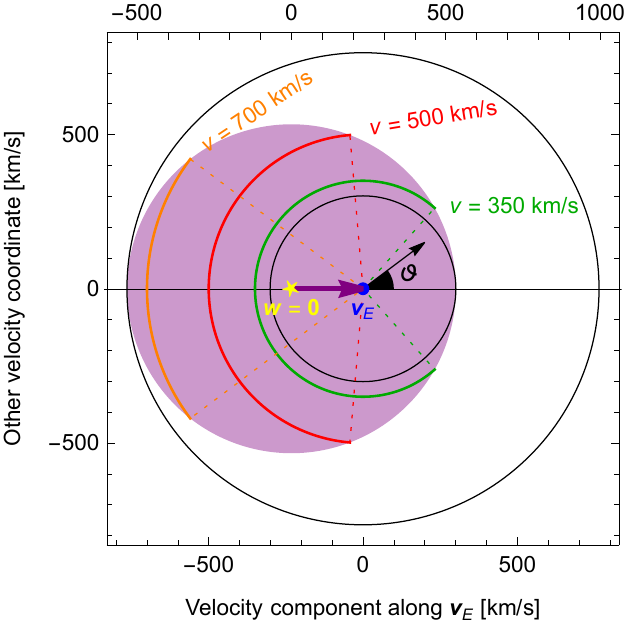}
\includegraphics[width=.49\textwidth]{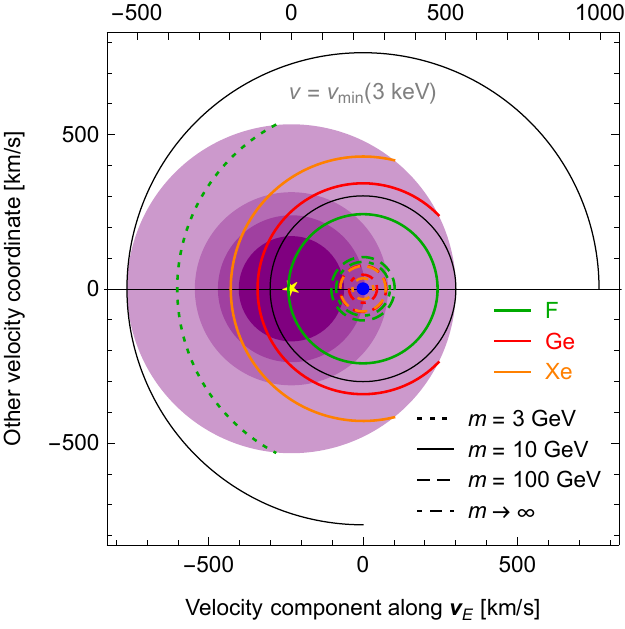}
\caption[Domain of the velocity integral]{\label{fig: f_E volume}\emph{Two-dimensional slice of the three-dimensional domain of the velocity integral (the plots have cylindrical symmetry about the horizontal axis). We set \figmath{\vE(t) = v_\text{S}} for definiteness. Coordinates on the bottom refer to Earth's frame, those on the top to the galactic frame. The yellow star indicates the \figmath{\bol{w} = \bol{0}} point, and the lightest purple disk indicates the \figmath{w \leqslant \vesc} region (where \figmath{\fE} can be non-zero). The blue point indicates \figmath{\bol{v}_\text{E}(t)}. In the left panel, the purple arrow indicates the direction of \figmath{\bol{v}_\text{E}(t)} in the galactic frame, and the black arrow indicates a generic velocity in Earth's frame, with the corresponding \figmath{\vartheta} angle also indicated. The two thin black circles in both panels indicate \figmath{v = \vesc - \vE} (inner circle) and \figmath{v = \vesc + \vE} (outer circle). The inner circle is entirely contained in the purple disk, so for \figmath{v \leqslant \vesc - \vE} the \figmath{\cos\vartheta} integral covers the whole round angle. All points outside the outer circle correspond to speeds larger than \figmath{\vesc} in the galactic frame, where \figmath{\fE = 0}. Between the two circles, both speeds with \figmath{w < \vesc} (purple disc) and \figmath{w > \vesc} (white space) are present, and circles of fixed \figmath{v} meet both. The \figmath{\cos\vartheta} integral at fixed \figmath{v} is thus only performed over angles where such circles overlap with the purple disk. \textbf{Left:} the green, red and orange circular arcs, corresponding to \figmath{v = 350}, \figmath{500}, and \figmath{700~\km / \sd}, respectively, are limited to \figmath{\cos\vartheta} values between \figmath{-1} and \figmath{c_\text{max}(v)}, see \Eq{c_max} (the top and bottom parts of each arc correspond to \figmath{\phi} values mutually differing by \figmath{\pi}). \textbf{Right:} the concentric purple discs are slices of three-dimensional spheres in velocity space containing \figmath{1/4}, \figmath{2/4}, \figmath{3/4}, and \figmath{4/4} of galactic DM particles in the Standard Halo Model (see \Sec{SHM}, in particular \Eq{SHM f} with \figmath{\beta = 0}). The green, red, orange circles and arcs indicate \figmath{v = \vmin(3~\keV)} for elastic scattering for the lightest stable fluorine, germanium, and xenon targets in \Tab{tab: nuclides}, respectively. Dotted, solid, dashed, and dot-dashed lines correspond to \figmath{\mDM = 3}, \figmath{10}, \figmath{100~\GeV} and \figmath{m \to \infty}, respectively (notice that \figmath{\vmin(3~\keV) > \vesc + \vE} for \figmath{\text{Xe}} and \figmath{\text{Ge}} targets with \figmath{\mDM = 3~\GeV}, and that the dashed \figmath{\text{Xe}} and \figmath{\text{Ge}} circles overlap accidentally). Only the velocities outside these circles and arcs contribute to elastic scattering yielding a nuclear recoil energy of \figmath{3~\keV}. Higher speeds are needed in Earth's frame for sufficiently light (heavy) DM particles scattering elastically off heavier (lighter) targets.}}
\figcode
\end{center}
\end{figure}

The integration domain in \Eq{generic eta} encompasses the whole range of speeds larger than $\vmin$, as well as the whole solid angle. $\fE(\bol{v}, t)$, however, is only non-zero within some finite region of velocity space. The boundaries of this region can be simple in the galactic frame (\eg $w \leqslant \vesc$ within the whole solid angle), but have a more complicated description in Earth's frame, see \Fig{fig: f_E volume}. They are derived in the following, starting, as a warm-up exercise, from the simplest case where $\vesc$ can be ignored. This is the case for instance if $\fE(\bol{v}, t)$ drops off so rapidly at large speeds that the contribution of these speeds to the velocity integral is negligible. Then $\vesc$ can be approximated as infinite, and $\fE(\bol{v}, t)$ can be considered non-zero for $\cos\vartheta \in [-1, 1]$ and $\phi \in [0, 2 \pi]$, with no restrictions on $v$. While all this information is naturally enclosed in $\fE(\bol{v}, t)$, it can be made more directly manifest by specifying the extrema of the velocity integral (which is otherwise unbounded, apart from the requirement $v \geqslant \vmin$). The velocity integral then reads, in the $\vesc \to \infty$ approximation,
\beq\label{infinite vesc}
\int_{\vmin}^\infty \ud v \, v^2 \int_{-1}^{+1} \ud \cos \vartheta \int_0^{2 \pi} \ud \phi \, H \ .
\eeq

Including the effect of a finite escape speed in the analysis introduces some complications, as we see in the following starting from the case $\vmin = 0$ and then considering the general case. The maximum allowed DM speed in Earth's frame is $\vesc + \vE(t)$; particles with this speed are those that reach the escape speed in the galactic frame and arrive on Earth only from the opposite direction with respect to Earth's motion, $\cos\vartheta = -1$. Particles with speed in Earth's frame between $\vesc + \vE(t)$ and $\vesc - \vE(t)$ arrive from within a certain angle about that direction: in fact, $w^2 = v^2 + \vE^2(t) + 2 v \vE(t) \cos\vartheta$ implies that, for fixed $v$, $\cos\vartheta$ grows with $w$ until $w = \vesc$, corresponding to the maximum value for $\cos\vartheta$
\beq\label{c_max}
c_\text{max}(v) \equiv \frac{\vesc^2 - v^2 - \vE^2(t)}{2 v \vE(t)} \ .
\eeq
$c_\text{max}(v) = 1$, corresponding to $v = \vesc - \vE(t)$, means that particles arrive on Earth from the whole solid angle. This also happens for slower DM particles, since particles with $v < \vesc - \vE(t)$ do not reach the escape speed in the galactic frame. Therefore, omitting for simplicity the integral over $\phi$ which plays no role in this discussion, the velocity integral at $\vmin = 0$ can be written as
\beq\label{thetaintegrals}
\int_0^{\vesc - \vE} \ud v \, v^2 \int_{-1}^{+1} \ud\cos\vartheta \, H
+
\int_{\vesc - \vE}^{\vesc + \vE} \ud v \, v^2 \int_{-1}^{c_\text{max}(v)} \ud\cos\vartheta \, H \ .
\eeq
At finite $\vmin$, we can notice that: for $\vmin \leqslant \vesc - \vE$, the first speed integral becomes $\int_{\vmin}^{\vesc - \vE} \ud v$; for $\vesc - \vE \leqslant \vmin \leqslant \vesc + \vE$, the first speed integral vanishes while the second becomes $\int_{\vmin}^{\vesc + \vE} \ud v$; for $\vmin \geqslant \vesc + \vE$, both integrals vanish. Therefore, defining
\beq
v_\pm \equiv \min(\vmin, \vesc \pm \vE) \ ,
\eeq
we can write the velocity integral as
\begin{multline}
\label{computing velocity integral}
\int_{v_-}^{\vesc - \vE} \ud v \, v^2 \int_{-1}^{+1} \ud\cos\vartheta \, H + \left[ \int_{\vesc - \vE}^{v_-} \ud v \, v^2 + \int_{v_+}^{\vesc + \vE} \ud v \, v^2 \right] \int_{-1}^{c_\text{max}(v)} \ud\cos\vartheta \, H
\\
= \int_{v_-}^{\vesc - \vE} \ud v \, v^2 \int_{c_\text{max}(v)}^{+1} \ud\cos\vartheta \, H + \int_{v_+}^{\vesc + \vE} \ud v \, v^2 \int_{-1}^{c_\text{max}(v)} \ud\cos\vartheta \, H
\\
= \sum_\text{signs} \pm \int_{v_\pm}^{\vesc \pm \vE} \ud v \, v^2 \int_{\mp 1}^{c_\text{max}(v)} \ud\cos\vartheta \, H \ ,
\end{multline}
where again we omitted for brevity the integral over $\phi$, and the sum in the last line involves the two expressions with the upper and lower sign. It is easily checked that this result reduces to \Eq{thetaintegrals} for $\vmin = 0$, and to \Eq{infinite vesc} for $\vesc \to \infty$. Notice that, for $\vmax^- < \vmin < \vmax^-$ (see \Eq{vmax^+-}), the speed integral corresponding to the lower sign in the last line vanishes, while the other is only non-zero for part of the year, namely when $\vesc + \vE(t) > \vmin$.

In the standard assumption the differential cross section only depends on $v$ and not on $\hat{\bol{v}}$, $h(\bol{v}) = h(v)$ in \Eq{generic eta} and one can use the one-dimensional speed distribution in Earth's frame defined in \Eq{1D speed distro}. Using the above results, this can be written as
\beq\label{1D F}
F_\text{E}(v, t) = \sum_\text{signs} \pm \Theta(\vesc \pm \vE - v) \, v^2 \int_{\mp 1}^{c_\text{max}(v)} \ud\cos\vartheta \int_0^{2 \pi} \ud \phi \, \fE(\bol{v}, t) \ ,
\eeq
where the theta functions specify explicitly the boundaries of the integration region (already encoded in $\fE(\bol{v}, t)$). The velocity integral in \Eq{generic eta h(v)} can then be written as
\beq\label{eta with 1D F}
\eta(\vmin, t) = \sum_\text{signs} \pm \int_{v_\pm}^{\vesc \pm \vE} \ud v \, F_\text{E}(v, t) \, v \, h(v) \ .
\eeq

For DM velocity distributions that are locally isotropic in the galactic rest frame, $\fE(\bol{v}, t)$ has cylindrical symmetry about $\hat{\bol{v}}_\text{E}(t)$. If $h(\bol{v}) = h(v)$, $H(\bol{v}, t)$ is also symmetric about $\hat{\bol{v}}_\text{E}(t)$. Therefore, the velocity integral can be performed trivially in $\phi$:
\beq\label{phi integral}
\int_0^{2 \pi} \ud \phi \, H = 2 \pi \, H \ .
\eeq
However, the assumption of locally isotropic velocity distribution is spoiled at the very least by the gravitational focussing effect of the Sun, as already commented above.

\subsection{Standard Halo Model}
\label{SHM}
The Standard Halo Model (SHM) is a non-rotating, isotropic DM distribution, falling with the distance from the galactic center $r$ as $r^{-2}$ in the vicinities of the Sun. A self-gravitating gas of (effectively) collisionless particles such as the galactic DM may reach thermal equilibrium through the \emph{violent relaxation} mechanism, see \eg Ref.~\cite{Binney:2008}. The corresponding local velocity distribution is an isotropic, isothermal (Maxwell-Boltzmann or Maxwellian) velocity distribution:
\beq
f_\text{SHM}(\bol{w}) \propto e^{- w^2 / v_0^2} \ .
\eeq
This model features a flat rotation curve at large $r$, with the root-mean-square speed (also often called velocity dispersion) $\sqrt{3/2} \, v_0$ related to the asymptotic value of the circular speed. It is usually assumed that the rotation curve has already reached its asymptotic value at the Sun's location, so that one has $v_0 = v_\text{c}$ (see \eg Ref.~\cite{Binney:2008}). For this reason, $v_0$ is often used in place of $v_\text{c}$. Both the density and velocity distributions formally extend to infinite values, which makes it necessary to cut off speeds above the local escape speed $\vesc$ as faster particles are not gravitationally bound to the finite Milky Way halo. For this reason, the SHM is also described as an isothermal sphere. The two most common implementations of this cutoff can be parametrized as
\begin{align}
\label{SHM f}
f_\text{SHM}(\bol{w}) &\equiv \frac{e^{- w^2 / v_0^2} - \beta \, e^{- \vesc^2 / v_0^2}}{(v_0 \sqrt{\pi})^3 N_\text{esc}} \, \Theta(\vesc - w) \ ,
\\
N_\text{esc} &\equiv \erf(\vesc / v_0) - \frac{2}{3 \sqrt{\pi}} \frac{\vesc}{v_0} \left( 3 + 2 \beta \, \vesc^2 / v_0^2 \right) e^{- \vesc^2 / v_0^2} \ ,
\end{align}
for $\beta = 0$ and $\beta = 1$, with the error function defined as
\beq
\erf(x) \equiv \frac{2}{\sqrt{\pi}} \int_0^x \ud z \, e^{- z^2} \ .
\eeq
The normalization factor $N_\text{esc}$ can be computed from \Eq{f norm} by noting that the angular integrals are trivial and that\footnote{See \eg Ref.~\cite{Lewin:1995rx} for a more pedagogical derivation of this and other results of this subsection.}
\begin{multline}
\label{D erf}
\int_a^b \ud z \, z^2 \, e^{- z^2} = \left. - \frac{\ud}{\ud \alpha} \int_a^b \ud z \, e^{- \alpha z^2} \right|_{\alpha = 1} = \left. - \frac{\ud}{\ud \alpha} \left[ \frac{1}{\sqrt{\alpha}} \int_{\sqrt{\alpha} a}^{\sqrt{\alpha} b} \ud t \, e^{- t^2} \right] \right|_{\alpha = 1}
\\
= \left. \frac{\alpha^{-3/2}}{2} \int_{\sqrt{\alpha} a}^{\sqrt{\alpha} b} \ud t \, e^{- t^2} - \frac{b \, e^{- \alpha b^2} - a \, e^{- \alpha a^2}}{2 \alpha} \right|_{\alpha = 1} = \frac{\sqrt{\pi}}{4} (\erf(b) - \erf(a)) - \frac{b \, e^{- b^2} - a \, e^{- a^2}}{2} \ ,
\end{multline}
thus
\beq
\frac{4 \pi}{(v_0 \sqrt{\pi})^3} \int_0^{\vesc} \ud w \, w^2 \, e^{- w^2 / v_0^2} = \frac{4}{\sqrt{\pi}} \int_0^{\vesc / v_0} \ud z \, z^2 \, e^{- z^2} = \erf(\vesc / v_0) - \frac{2}{\sqrt{\pi}} \frac{\vesc}{v_0} \, e^{- \vesc^2 / v_0^2} \ ,
\eeq
with $\erf(0) = 0$ by the definition of error function. $N_\text{esc}$ is defined so that $N_\text{esc} \xrightarrow{\vesc \to \infty} 1$, as can be proven by noting that $\erf(x) \xrightarrow{x \to \infty} 1$. $\beta = 0$ entails unphysically truncating the velocity distribution at $w = \vesc$. The distribution obtained with the alternative, but still ad hoc, truncation prescription $\beta = 1$ has a form also found in the King models (see \eg Ref.~\cite{Binney:2008}).

While the SHM is not entirely theoretically consistent, it nevertheless constitutes a useful approximation where the velocity integral takes a conveniently analytical form (see below). Other halo models, including theoretically consistent ones, and their impact on direct DM detection have been explored \eg in Refs.~\cite{Ullio:2000bf, Evans:2000gr, Vergados:2002hc, Green:2002ht, Fornengo:2003fm, Freese:2003na, Savage:2006qr, Vogelsberger:2008qb, Kuhlen:2009vh, McCabe:2010zh, Chaudhury:2010hj, Green:2010gw, Lisanti:2010qx, Catena:2011kv, Purcell:2012sh, Fairbairn:2012zs, Bhattacharjee:2012xm, Mao:2012hf, Bozorgnia:2013pua, Fornasa:2013iaa, Lavalle:2014rsa, Bozorgnia:2016ogo, Bozorgnia:2017brl, Lacroix:2018qqh, OHare:2019qxc}. Halo-independent analyses designed to obtain information about the DM velocity distribution from direct detection data, rather than assuming a given $\fE$ to infer the particle physics properties of the DM (\eg mass and couplings), have been developed \eg in Refs.~\cite{Fox:2010bz, Frandsen:2011gi, HerreroGarcia:2011aa, Gondolo:2012rs, HerreroGarcia:2012fu, DelNobile:2013cta, Bozorgnia:2013hsa, DelNobile:2013cva, Kavanagh:2013eya, DelNobile:2014eta, Feldstein:2014gza, Fox:2014kua, Scopel:2014kba, Cherry:2014wia, Feldstein:2014ufa, Anderson:2015xaa, Blennow:2015gta, Ferrer:2015bta, Herrero-Garcia:2015kga, Gelmini:2015voa, Gelmini:2016pei, Kahlhoefer:2016eds, Ibarra:2017mzt, Gelmini:2017aqe, Catena:2018ywo, Kahlhoefer:2018knc, Ibarra:2018yxq}. It should be noted that, while DM-only cosmological simulations (\ie with no baryons) return anisotropic velocity distributions which deviate significantly from the SHM, recent simulations including baryons produce halos with more isotropic DM velocity distributions, thus making the SHM a better fit (see \eg Ref.~\cite{Bozorgnia:2017brl}).

The SHM DM velocity distribution in Earth's frame is
\beq
f_{\text{SHM}, \text{E}}(\bol{v}, t) = \frac{e^{- (\bol{v} + \bol{v}_\text{E})^2 / v_0^2} - \beta \, e^{- \vesc^2 / v_0^2}}{(v_0 \sqrt{\pi})^3 N_\text{esc}} \, \Theta(\vesc - |\bol{v} + \bol{v}_\text{E}|) \ .
\eeq
The effect of the theta function on the domain of the velocity integral has already been discussed above. We assume $h(\bol{v}) = h(v)$ in \Eq{generic eta}, in which case the $\phi$ integral is trivial as in \Eq{phi integral}, and so is the $\cos\vartheta$ integral of the $\beta$ term. The indefinite angular integral for the other term yields
\beq
\int \ud \cos\vartheta \, e^{- (\bol{v} + \bol{v}_\text{E})^2 / v_0^2}
=
e^{- (v^2 + \vE^2) / v_0^2} \int \ud \cos\vartheta \, e^{- 2 v \vE \, \cos\vartheta / v_0^2}
=
- \frac{e^{- (v^2 + \vE^2 + 2 v \vE \, \cos\vartheta) / v_0^2}}{2 v \vE / v_0^2} \ .
\eeq
The one-dimensional speed distribution in Earth's rest frame is then (see \Eq{1D F})
\begin{multline}
F_{\text{SHM}, \text{E}}(v, t) =
\\
\frac{1}{\sqrt{\pi} N_\text{esc} v_0 \vE} \sum_\text{signs} \pm \Theta(\vesc \pm \vE - v) \, v \left[ e^{- (v \mp \vE)^2 / v_0^2} - e^{- \vesc^2 / v_0^2} \left( 1 + \beta \frac{\vesc^2 - (v \mp \vE)^2}{v_0^2} \right) \right]
\\
=
\frac{1}{\sqrt{\pi} N_\text{esc} v_0 \zE} \sum_\text{signs} \pm \Theta(\zesc \pm \zE - z) \, z \left[ e^{- (z \mp \zE)^2} - e^{- \zesc^2} \left( 1 + \beta \left( \zesc^2 - (z \mp \zE)^2 \right) \right) \right] ,
\end{multline}
where we defined in the last line the following convenient dimensionless variables:
\begin{align}
z \equiv \frac{v}{v_0} \ ,
&&
\zE \equiv \frac{\vE}{v_0} \ ,
&&
\zesc \equiv \frac{\vesc}{v_0} \ .
\end{align}
Noticing that $F_{\text{SHM}, \text{E}}$ has the structure
\beq
\sum_\text{signs} \pm \Theta(\zesc - (z \mp \zE)) \, z [f(z \mp \zE) - f(\zesc)] = \sum_\text{signs} \pm z \, f(\min(z \mp \zE, \zesc)) \ ,
\eeq
with $f(x) = e^{- x^2} + \beta \, e^{- \zesc^2} \, x^2$, we can also obtain (extending to the case $\beta = 1$ the formula in Ref.~\cite{Barger:2010gv}) the particularly compact expression
\beq
F_{\text{SHM}, \text{E}}(v, t) = \frac{z}{\sqrt{\pi} N_\text{esc} v_0 \zE} \left[ e^{- x_-^2} - e^{- x_+^2} + \beta \, e^{- \zesc^2} \left( x_-^2 - x_+^2 \right) \right] ,
\eeq
where we defined
\beq
x_\pm \equiv \min(z \pm \zE, \zesc) \ .
\eeq
In the three separate speed regimes of interest here we get
\begin{multline}
F_{\text{SHM}, \text{E}}(v, t) =
\\
\frac{z}{\sqrt{\pi} N_\text{esc} v_0 \zE} \times
\begin{cases}
e^{- (z - \zE)^2} - e^{- (z + \zE)^2} - 4 \beta z \zE \, e^{- \zesc^2} & z \leqslant \zesc - \zE,
\\
e^{- (z - \zE)^2} - e^{- \zesc^2} \left[ 1 + \beta \left( \zesc^2 - (z - \zE)^2 \right) \right] & \zesc - \zE \leqslant z \leqslant \zesc + \zE,
\\
0 & \zesc + \zE \leqslant z.
\end{cases}
\end{multline}
In the approximation of infinite escape speed we have
\beq\label{FSHME infinite vesc}
F_{\text{SHM}, \text{E}}^\infty(v, t) \equiv \lim_{\vesc \to \infty} F_{\text{SHM}, \text{E}}(v, t) = \frac{z}{\sqrt{\pi} v_0 \zE} \left( e^{- (z - \zE)^2} - e^{- (z + \zE)^2} \right) .
\eeq
As for the velocity integral, the time dependence of $F_{\text{SHM}, \text{E}}(v, t)$ is entirely encoded in $\vE(t)$ (see discussion above).

\begin{figure}[t]
\begin{center}
\includegraphics[width=.49\textwidth]{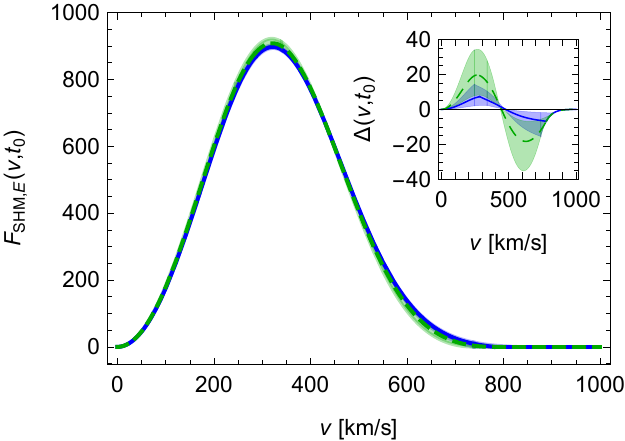}
\includegraphics[width=.49\textwidth]{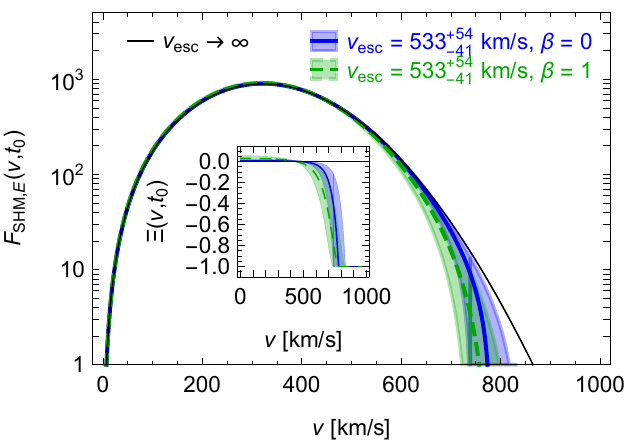}
\caption[SHM one-dimensional speed distribution in Earth's rest frame]{\label{fig: FSHME}\emph{The SHM one-dimensional DM speed distribution in Earth's rest frame, \figmath{F_{\text{SHM}, \text{E}}}, for \figmath{\vesc = 533^{+54}_{-41}~\km / \sd}~\cite{Piffl:2013mla} (colored lines, with shaded regions representing the uncertainty) and \figmath{\vesc \to \infty} (thin black line). The solid blue line is for \figmath{\beta = 0}, whereas the dashed green line is for \figmath{\beta = 1}. \figmath{F_{\text{SHM}, \text{E}}} is made dimensionless in the plots by multiplication with the speed of light. For definiteness, the time is set to \figmath{t = t_0}, when Earth reaches its maximum speed in the galactic rest frame, \figmath{\vE(t_0) \simeq v_\text{S} + b v_\oplus} (see \Eq{Taylor vE}). \textbf{Left:} linear scale. The inset shows the difference between \figmath{F_{\text{SHM}, \text{E}}} at finite and infinite \figmath{\vesc}: \figmath{\Delta \equiv F_{\text{SHM}, \text{E}} - F_{\text{SHM}, \text{E}}^\infty}, see \Eq{FSHME infinite vesc}. \textbf{Right:} logarithmic scale, to better appreciate the effect that varying \figmath{\vesc} has on the high-speed tail of the distribution. The inset shows the relative size of the difference between \figmath{F_{\text{SHM}, \text{E}}} at finite and infinite \figmath{\vesc}: \figmath{\Xi \equiv \Delta / F_{\text{SHM}, \text{E}}^\infty}. The velocity integral in \Eq{generic eta h(v)} samples only the tail of \figmath{F_{\text{SHM}, \text{E}}} for large \figmath{\vmin} values (\ie relatively large nuclear recoil energies, and also small energies for inelastic scattering), in which case the precise shape of the high-speed portion of the distribution becomes crucial. This is relevant for sufficiently light DM particles, especially with heavy targets (or light targets for \figmath{\delta} positive and sizeable, see \Sec{Inelastic scattering}), if the recoil energy has to lie above a certain value for the experiments to be able to detect the scattering.}}
\figcode
\end{center}
\end{figure}

$F_{\text{SHM}, \text{E}}(v, t)$ is shown as a dimensionless function of $v$ in \Fig{fig: FSHME} (all functions plotted in this subsection are made dimensionless by multiplication with the appropriate power of the speed of light). The time is set for definiteness to $t = t_0$, when Earth reaches its maximum speed in the galactic rest frame, $\vE(t_0) \simeq v_\text{S} + b v_\oplus$ (see \Eq{Taylor vE}). The escape speed is taken to be $\vesc = 533^{+54}_{-41}~\km / \sd$~\cite{Piffl:2013mla} (colored lines, with shaded regions representing the uncertainty) and $\vesc \to \infty$ (thin black line). The solid blue line is for $\beta = 0$, whereas the dashed green line is for $\beta = 1$. The difference between these cases can be better appreciated in the inset in the left panel, showing $\Delta \equiv F_{\text{SHM}, \text{E}} - F_{\text{SHM}, \text{E}}^\infty$. The velocity integral in \Eq{generic eta h(v)} samples only the tail of $F_{\text{SHM}, \text{E}}$ for large $\vmin$ values (\ie relatively large nuclear recoil energies, and also small energies for inelastic scattering), in which case the precise shape of the high-speed portion of the distribution becomes crucial. This is relevant for sufficiently light DM particles, especially with heavy targets (or light targets for $\delta$ positive and sizeable, see \Sec{Inelastic scattering}), if the recoil energy has to lie above a certain value for the experiments to be able to detect the scattering. The tails of the speed distributions for the different cases mentioned above can be better discerned in the right panel of \Fig{fig: FSHME}, in logarithmic scale, and in the inset, showing their relative size $\Xi \equiv \Delta / F_{\text{SHM}, \text{E}}^\infty$.

The velocity integral can now be obtained \eg from \Eq{eta with 1D F} and reads, using the different sets of variables introduced above,
\begin{multline}
\eta(\vmin, t) =
\\
\frac{1}{\sqrt{\pi} N_\text{esc} v_0 \vE} \sum_\text{signs} \pm \int_{v_\pm}^{\vesc \pm \vE} \ud v \, v^2 h(v) \left[ e^{- (v \mp \vE)^2 / v_0^2} - e^{- \vesc^2 / v_0^2} \left( 1 + \beta \frac{\vesc^2 - (v \mp \vE)^2}{v_0^2} \right) \right]
\\
= \frac{v_0}{\sqrt{\pi} N_\text{esc} \zE} \sum_\text{signs} \pm \int_{z_\pm}^{\zesc \pm \zE} \ud z \, z^2 h(v_0 z) \left[ e^{- (z \mp \zE)^2} - e^{- \zesc^2} \left( 1 + \beta \left( \zesc^2 - (z \mp \zE)^2 \right) \right) \right]
\\
= \frac{v_0}{\sqrt{\pi} N_\text{esc} \zE} \sum_\text{signs} \pm \int_{\tilde{z}_\mp}^{\zesc} \ud \tilde{z} \, (\tilde{z} \pm \zE)^2 h(v_0 (\tilde{z} \pm \zE)) \left[ e^{- \tilde{z}^2} - e^{- \zesc^2} \left( 1 + \beta \left( \zesc^2 - \tilde{z}^2 \right) \right) \right]
\\
= \frac{v_0}{\sqrt{\pi} N_\text{esc} \zE} \int_{\zmin}^\infty \ud z \, z^2 h(v_0 z) \left[ e^{- x_-^2} - e^{- x_+^2} + \beta \, e^{- \zesc^2} \left( x_-^2 - x_+^2 \right) \right] ,
\end{multline}
where we defined
\begin{align}
\zmin \equiv \frac{\vmin}{v_0} \ ,
&&&
z_\pm \equiv \frac{v_\pm}{v_0} = \min(\zmin, \zesc \pm \zE) \ ,
\end{align}
and, separately for each sign in the sum in the second to last line,
\begin{align}
\tilde{z} \equiv z \mp \zE \ ,
&&&
\tilde{z}_\pm \equiv z_\mp \pm \zE = \min(\zmin \pm \zE, \zesc)
\end{align}
(notice that, with these definitions, $\tilde{z}_+ \geqslant \tilde{z}_-$). The velocity dependence of $\ud \sigma_T / \ud \ER$ that is most often encountered is $h(v) = 1 / v^2$, occurring when the scattering amplitude does not depend on velocity as for both the SI and SD interactions, see Secs.~\ref{SI interaction},~\ref{SD interaction} and the related discussion in \Sec{Scattering rate}. Therefore, the velocity integral most often featured in the scattering rate is $\eta_0$, defined in \Eq{eta} (see also \Eq{eta_0}). In the SHM this is (see \eg Refs.~\cite{Lewin:1995rx, Savage:2006qr, McCabe:2010zh, Barger:2010gv, Freese:2012xd})
\begin{multline}
\label{eta_0 v1}
\eta_0(\vmin, t)
=
\frac{1}{\sqrt{\pi} N_\text{esc} v_0 \zE} \sum_\text{signs} \pm \int_{\tilde{z}_\mp}^{\zesc} \ud \tilde{z} \left[ e^{- \tilde{z}^2} - e^{- \zesc^2} \left( 1 + \beta \left( \zesc^2 - \tilde{z}^2 \right) \right) \right]
\\
= \frac{1}{2 N_\text{esc} v_0 \zE} \left[ \erf(\tilde{z}_+) - \erf(\tilde{z}_-) - \frac{2}{\sqrt{\pi}} e^{- \zesc^2} (\tilde{z}_+ - \tilde{z}_-) \left( 1 - \frac{\beta}{3} \left( \tilde{z}_+^2 + \tilde{z}_+ \tilde{z}_- + \tilde{z}_-^2 - 3 \zesc^2 \right) \right) \right] ,
\end{multline}
which also reads, separately in the three speed regimes of interest,
\begin{multline}
\label{eta_0 v2}
\eta_0(\vmin, t) = \frac{1}{2 N_\text{esc} v_0 \zE}
\\
\times
\begin{cases}
\begin{aligned}
& \erf(\zmin + \zE) - \erf(\zmin - \zE)
\\
& \hspace{10mm} - \frac{4}{\sqrt{\pi}} \zE \, e^{- \zesc^2} \left[ 1 - \beta \left( \zmin^2 + \frac{\zE^2}{3} - \zesc^2 \right) \right] \rule{0pt}{20pt}
\end{aligned}
& \zmin \leqslant \zesc - \zE,
\\
\begin{aligned}
& \erf(\zesc) - \erf(\zmin - \zE) + \frac{2}{\sqrt{\pi}} e^{- \zesc^2} (\zmin - \zE - \zesc) \rule{0pt}{30pt}
\\
& \hspace{10mm} \times \left[ 1 - \frac{\beta}{3} (\zmin - \zE - \zesc) (\zmin - \zE + 2 \zesc) \right] \rule{0pt}{20pt}
\end{aligned}
& \zesc - \zE \leqslant \zmin \leqslant \zesc + \zE,
\\
0 \rule{0pt}{30pt} & \zesc + \zE \leqslant \zmin.
\end{cases}
\end{multline}
In the limit of infinite escape speed we have
\beq
\eta_0(\vmin, t) \xrightarrow{\vesc \to \infty} \frac{\erf(\zmin + \zE) - \erf(\zmin - \zE)}{2 v_0 \zE} \ .
\eeq
Another velocity integral that is sometimes encountered is $\eta_1$ (again defined in \Eq{eta}), arising when $h(v) = 1$ (see \Eq{eta_n h(v)}). This happens for instance for DM particles with a magnetic dipole moment and/or an anapole moment which interact with nuclei through photon exchange, see Secs.~\ref{Electromagnetic interactions},~\ref{Magnetic-dipole DM} and related discussion in \Sec{Scattering rate}. In the SHM $\eta_1$ reads
\begin{multline}
\label{eta_1}
\eta_1(\vmin, t)
=
\frac{v_0}{\sqrt{\pi} N_\text{esc} \zE} \left[ \frac{\sqrt{\pi}}{4} \left( 1 + 2 \zE^2 \right) (\erf(\tilde{z}_+) - \erf(\tilde{z}_-))
\vphantom{\frac{(\tilde{z}_+ - 2 \zE) e^{- \tilde{z}_+^2} - (\tilde{z}_- + 2 \zE) e^{- \tilde{z}_-^2}}{2}} \right.
\\
- \frac{(\tilde{z}_+ - 2 \zE) e^{- \tilde{z}_+^2} - (\tilde{z}_- + 2 \zE) e^{- \tilde{z}_-^2}}{2} - e^{- \zesc^2} \left[ \frac{z_-^3 - z_+^3}{3} + 2 \zE \left( 1 + \frac{\zE^2}{3} + \zesc^2 \right)
\vphantom{\left[ \left( \frac{z_+^5 - z_-^5}{5} \right) \right]} \right.
\\
\left. \vphantom{\frac{(\tilde{z}_+ - 2 \zE) e^{- \tilde{z}_+^2} - (\tilde{z}_- + 2 \zE) e^{- \tilde{z}_-^2}}{2}} \left. \vphantom{\left( \frac{\zE^2}{3} \right)}
+ \beta \left( \frac{z_+^5 - z_-^5}{5} + \frac{\zE}{2} \left( 2 \zesc^4 - z_+^4 - z_-^4 \right) + \left( \zesc^2 - \zE^2 \right) \frac{z_-^3 - z_+^3}{3} + \frac{\zE^3}{15} \left( 10 \zesc^2 - \zE^2 \right) \right) \right] \right] ,
\end{multline}
where we used \Eq{D erf}, and we mixed the $z$ and $\tilde{z}$ notations for brevity.
In the limit of infinite escape speed we have
\beq
\eta_1(\vmin, t) \xrightarrow{\vesc \to \infty} \frac{v_0}{2 \sqrt{\pi} \zE} \sum_\text{signs} \pm \left[ \frac{\sqrt{\pi}}{2} \left( 1 + 2 \zE^2 \right) \erf(\zmin \pm \zE) \pm (\zmin \pm \zE) \, e^{- (\zmin \mp \zE)^2} \right] .
\eeq
$\eta_2$ (see \Eq{eta}), corresponding to $h(v) = v^2$, is needed to study NR operators with two powers of $v$ such as the $\Op^N_{16}$ and $\Op^N_{17}$ building blocks in \Eq{NR building blocks}. While $\eta_2$ can be computed analytically (see \eg the \website~\cite{Appe-website}), we only report here its relatively compact expression in the limit of infinite $\vesc$:
\begin{multline}
\label{eta_2 inf vesc}
\eta_2(\vmin, t) \xrightarrow{\vesc \to \infty} \frac{v_0^3}{8 \sqrt{\pi} \zE} \sum_\text{signs} \pm \left[ e^{- (\zmin \mp \zE)^2} \left[ 4 \left( \zmin^2 + \zE^2 \right) (\zmin \pm \zE) + 2 (3 \zmin \pm 5 \zE) \right]
\vphantom{+ \sqrt{\pi} \left( 3 + 12 \zE^2 + 4 \zE^4 \right) \erf(\zmin \pm \zE)}
\right.
\\
\left. + \sqrt{\pi} \left( 3 + 12 \zE^2 + 4 \zE^4 \right) \erf(\zmin \pm \zE)
\vphantom{e^{- (\zmin \mp \zE)^2} \left[ 4 \left( \zmin^2 + \zE^2 \right) (\zmin \pm \zE) + 2 (3 \zmin \pm 5 \zE) \right]}
\right] .
\end{multline}

\begin{figure}[t!]
\begin{center}
\includegraphics[width=.32\textwidth]{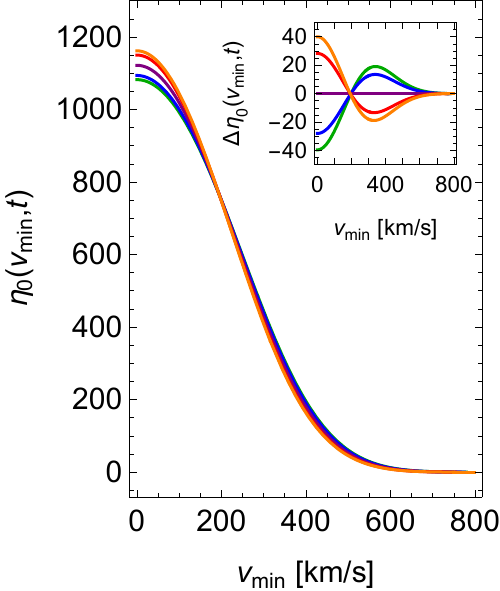}
\includegraphics[width=.32\textwidth]{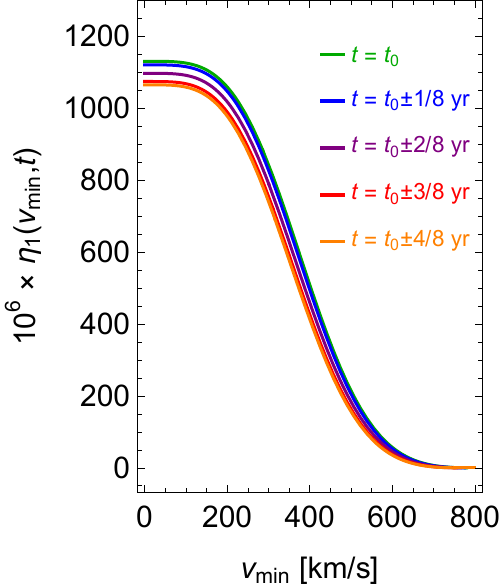}
\includegraphics[width=.32\textwidth]{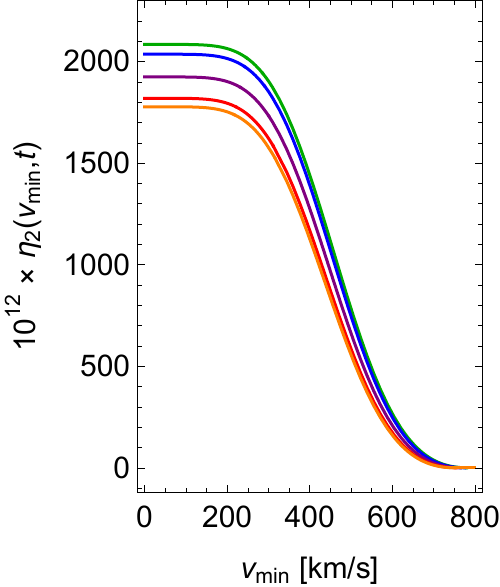}
\\[5mm]
\includegraphics[width=.49\textwidth]{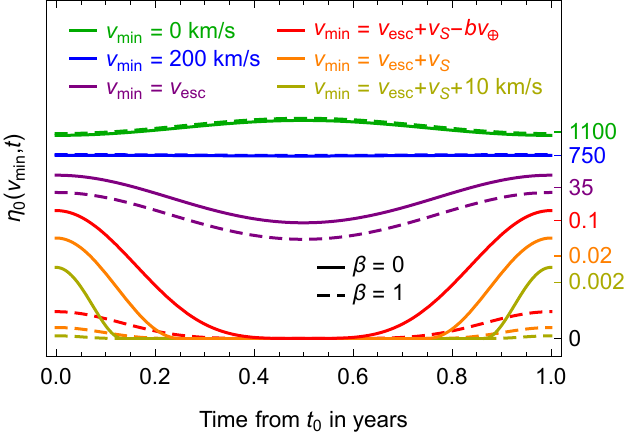}
\includegraphics[width=.49\textwidth]{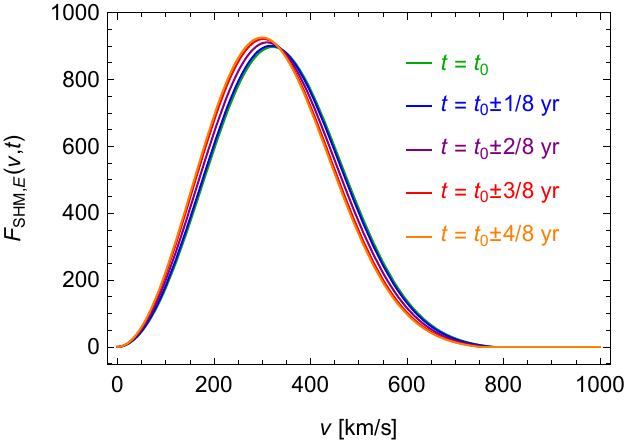}
\caption[Temporal variation of SHM speed distribution and velocity integrals]{\label{fig: eta}\emph{Temporal variation of velocity integrals and one-dimensional speed distribution in the SHM. All functions are made dimensionless by multiplication with appropriate powers of the speed of light. \textbf{Top:} the SHM velocity integrals \figmath{\eta_0(\vmin, t)} (\textbf{left}), \figmath{\left( 10^6 \times \right) \eta_1(\vmin, t)} (\textbf{center}), and \figmath{\left( 10^{12} \times \right) \eta_2(\vmin, t)} (\textbf{right}), as functions of \figmath{\vmin}, at eight equispaced times of the year, including \figmath{t = t_0} (\figmath{t_0 \pm \yr / 2} is considered as a single time of the year). Their time dependence implies that the velocity integral is the same at equal \figmath{|t - t_0| \mod \yr}. \figmath{\beta = 0} is taken for definiteness. The (purple) lines at \figmath{t = t_0 \pm \yr / 4}, when \figmath{\vE(t) \simeq v_\text{S}} (see \Eq{Taylor vE}), approximately correspond to the annual averages \figmath{\overline{\eta}_{0, 1, 2}}, defined in \Eq{etabartilde def} and depicted in the top panels of \Fig{fig: etabartilde}. The inset for \figmath{\eta_0} shows the difference \figmath{\Delta \eta_0(\vmin, t) \equiv \eta_0(\vmin, t) - \eta_0(\vmin, t_0 \pm \yr / 4)}. \textbf{Bottom left:} \figmath{\eta_0} plotted as a function of time for different values of \figmath{\vmin}, with solid lines for \figmath{\beta = 0} and dashed lines for \figmath{\beta = 1}. The plot has a different vertical scale for each \figmath{\vmin} value, indicated by the tick and number with corresponding color to the right (the \figmath{0} is in common). Notice the opposite oscillations for small and large \figmath{\vmin} values, and the extended periods of time where \figmath{\eta_0} vanishes for \figmath{\vmin > \vmax^- \simeq \vesc + v_\text{S} - b v_\oplus} (see discussion in \Sec{Modulation}). \textbf{Bottom right:} \figmath{F_{\text{SHM}, \text{E}}(v, t)} as a function of \figmath{v}, at the same times of the year as the velocity integrals in the top panels.}}
\figcode
\end{center}
\end{figure}

\begin{figure}[t!]
\begin{center}
\includegraphics[width=.32\textwidth]{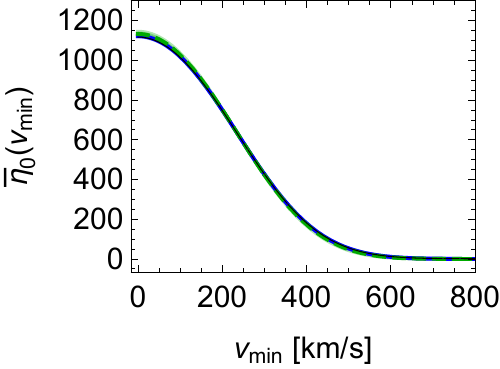}
\includegraphics[width=.32\textwidth]{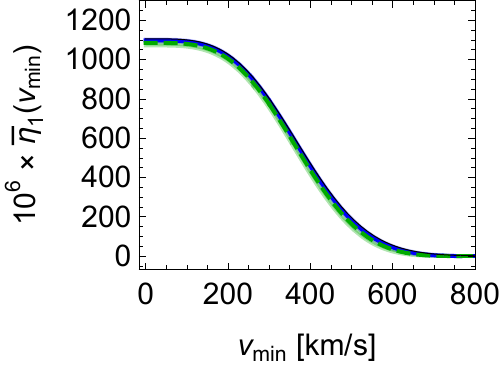}
\includegraphics[width=.32\textwidth]{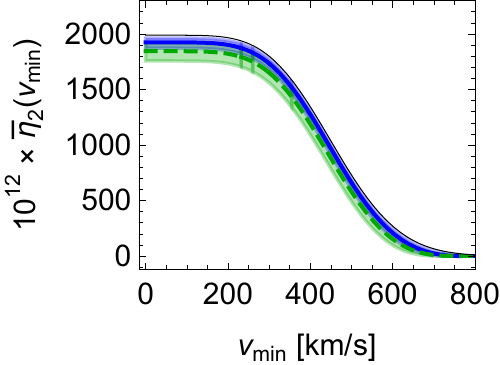}
\\[4mm]
\includegraphics[width=.32\textwidth]{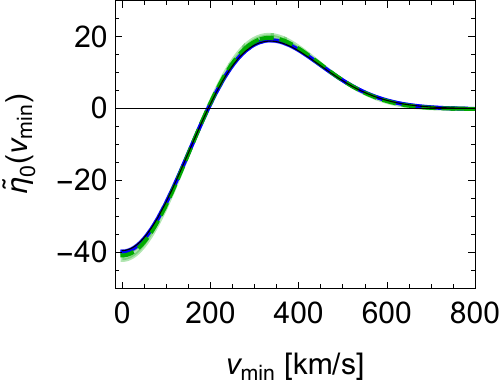}
\includegraphics[width=.32\textwidth]{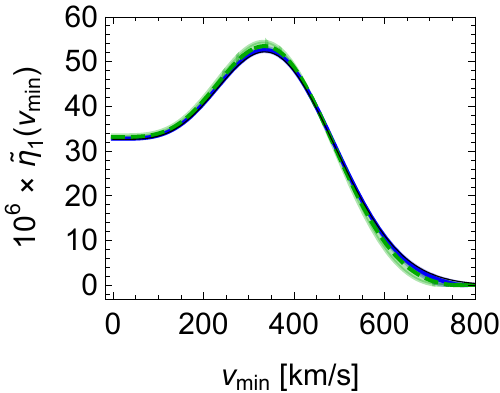}
\includegraphics[width=.32\textwidth]{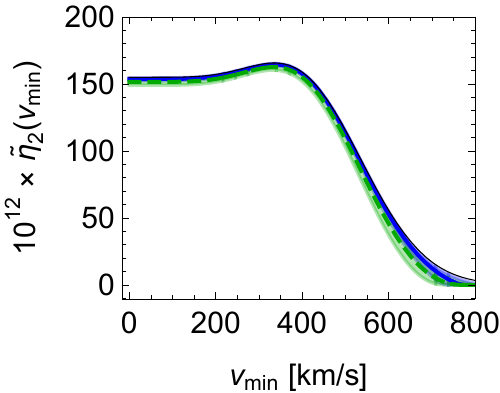}
\\[4mm]
\includegraphics[width=.32\textwidth]{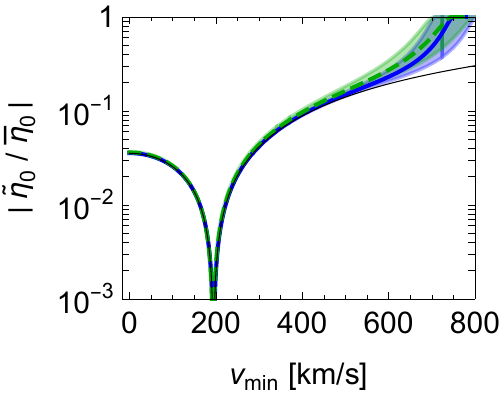}
\includegraphics[width=.32\textwidth]{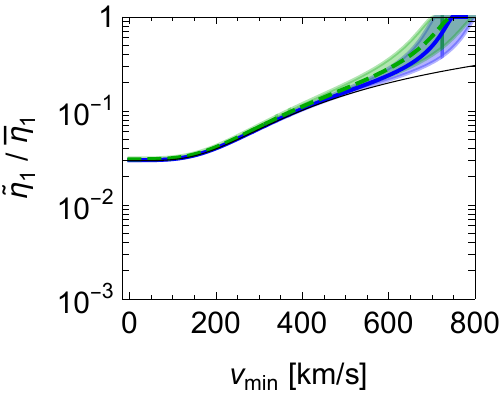}
\includegraphics[width=.32\textwidth]{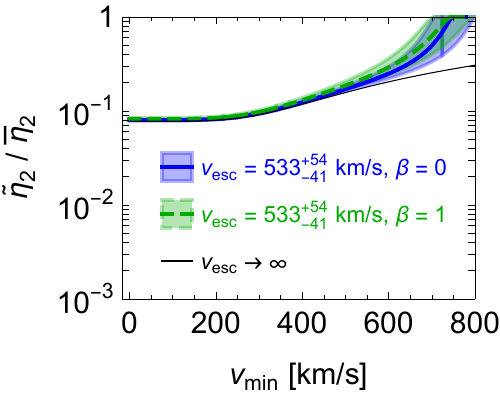}
\caption[Properties of SHM velocity integrals]{\label{fig: etabartilde}\emph{Properties of the SHM velocity integrals \figmath{\eta_0} (\textbf{left}), \figmath{\left( 10^6 \times \right) \eta_1} (\textbf{center}), and \figmath{\left( 10^{12} \times \right) \eta_2} (\textbf{right}), made dimensionless by multiplication with appropriate powers of the speed of light. Colored lines are for \figmath{\vesc = 533^{+54}_{-41}~\km / \sd}~\cite{Piffl:2013mla}, with the shaded regions representing the uncertainty, while the thin black line is for \figmath{\vesc \to \infty}; the solid blue line is for \figmath{\beta = 0}, whereas the dashed green line is for \figmath{\beta = 1}, as in \Fig{fig: FSHME}. \textbf{Top:} the annual averages \figmath{\overline{\eta}_{0, 1, 2}(\vmin)} defined in \Eq{etabartilde def}. As commented in the text, the higher power of \figmath{v} in the velocity integral makes \figmath{\eta_n} with larger \figmath{n} more sensitive to \figmath{\vesc} and to the truncation prescription (\figmath{\beta = 0, 1}). \textbf{Middle:} the modulation amplitudes \figmath{\widetilde{\eta}_{0, 1, 2}}, defined in \Eq{etabartilde def} and computed with \Eq{etabartilde approx}. Due to \Eq{Taylor eta}, the zero of \figmath{\widetilde{\eta}_0} implies that \figmath{\eta_0} does not (approximately) depend on time in that point and therefore all lines in the top-left panel of \Fig{fig: eta} and its inset cross at (approximately) the same point. However, as commented in the text, anisotropies in the local DM velocity distribution such as the Sun's gravitational focussing become more relevant in that point, spoiling this feature. \textbf{Bottom:} the fractional amplitudes \figmath{\widetilde{\eta}_{0, 1, 2} / \overline{\eta}_{0, 1, 2}} computed with \Eq{etabartilde approx}. Since \figmath{\widetilde{\eta}_0 / \overline{\eta}_0} is negative at low \figmath{\vmin}, we plot its absolute value so that the entire function can be appreciated in logarithmic scale. As commented in \Sec{Modulation}, a natural order of magnitude of the fractional modulation is set by \figmath{b \, v_\oplus / v_\text{S} \simeq 0.06}, however \figmath{| \widetilde{\eta} / \overline{\eta} |} reaches \figmath{100 \%} at \figmath{\vmin} values close to \figmath{\vmax^- \simeq \vesc + v_\text{S} - b v_\oplus} due to \figmath{\overline{\eta}} no longer being dominant against the other Fourier modes.}}
\figcode
\end{center}
\end{figure}

Figs.~\ref{fig: eta},~\ref{fig: etabartilde} show different, complementing aspects of the SHM $\eta_0$, $\eta_1$, $\eta_2$ velocity integrals. All functions are made dimensionless by multiplication with appropriate powers of the speed of light; $\eta_1$ and $\eta_2$ are also multiplied by a factor $10^6$ and $10^{12}$, respectively, so that all functions have readily comparable sizes. In the top panels of \Fig{fig: eta} the velocity integrals are shown, as functions of $\vmin$ (with $\beta = 0$ for definiteness), at eight equispaced times of the year (considering $t_0 \pm \yr / 2$ as a single time of the year), including $t = t_0$. For a perfectly isotropic local DM velocity distribution as the SHM, the velocity integral depends on time only through $\vE(t)$, given in the approximation of circular Earth's orbit in \Eq{Taylor vE}; the time dependence of $\vE(t)$ then implies that the velocity integral is the same at equal $|t - t_0| \mod \yr$. The annual averages $\overline{\eta}_{0, 1, 2}$, defined in \Eq{etabartilde def}, are plotted in the top panels of \Fig{fig: etabartilde} for $\vesc = 533^{+54}_{-41}~\km / \sd$~\cite{Piffl:2013mla} (colored lines, with shaded regions representing the uncertainty) and $\vesc \to \infty$ (thin black line); the solid blue line is for $\beta = 0$, whereas the dashed green line is for $\beta = 1$, as in \Fig{fig: FSHME}. Given that $\vE(t_0 \pm \yr / 4) \simeq v_\text{S}$ (see \Eq{Taylor vE}), the solid blue lines in the top panels of \Fig{fig: etabartilde} approximately correspond to the purple lines in the top panels of \Fig{fig: eta}.

One can see in \Fig{fig: eta} that, at low $\vmin$ values, $\eta_0$ is minimum at $t = t_0$, growing with time up to its maximum at $t = t_0 + \yr / 2$, and decreasing again in the next half year, while at large $\vmin$ values the opposite is true. Correspondingly, the annual modulation amplitude $\widetilde{\eta}_0$, plotted in the middle-left panel of \Fig{fig: etabartilde} as approximated in \Eq{etabartilde approx}, is negative at low $\vmin$ and positive at large $\vmin$ (before vanishing completely), which implies it has a zero in between. There, due to \Eq{Taylor eta}, $\eta_0(\vmin, t)$ does not (approximately) depend on time, which explains why the different lines in the top-left panel of \Fig{fig: eta} all cross at (approximately) the same point. This feature, which is spoiled by anisotropies in the local DM velocity distribution such as the Sun's gravitational focussing (see \eg Ref.~\cite{DelNobile:2015rmp}), is highlighted in the inset, showing the difference $\Delta \eta_0(\vmin, t) \equiv \eta_0(\vmin, t) - \eta_0(\vmin, t_0 \pm \yr / 4)$. An alternative way of representing the opposite temporal oscillations of $\eta_0(\vmin, t)$ at low and high $\vmin$ consists in having opposite modulation phases in the two regimes (while we fixed the phase to $t_0$ in \Eq{Taylor eta}), thus making the modulation amplitude non-negative. The latter approach is a consequence of describing the annual modulation with $A_1 \simeq |\widetilde{\eta}_0|$ and $t_1 = t_0$ ($t_1 = t_0 \pm \yr / 2$) at large (small) $\vmin$ values in \Eq{Fourier eta}, rather than with $a_1 \simeq \widetilde{\eta}_0$ and a fixed phase as in \Eq{Taylor eta}. The bottom-left panel of \Fig{fig: eta}, where $\eta_0$ is plotted as a function of time for different values of $\vmin$ (individually normalized for each $\vmin$ value), provides a further illustration of the different temporal behavior of $\eta_0(\vmin, t)$ at low and high $\vmin$. This behavior is inherited from the time dependence of $F_{\text{SHM}, \text{E}}$, depicted for reference in the bottom-right panel of \Fig{fig: eta}. Such a feature is absent for $\eta_1$ and $\eta_2$, which are maximum at $t = t_0$ (the time of maximum Earth's speed with respect to the DM) regardless of $\vmin$. Correspondingly, $\widetilde{\eta}_{n > 0}$ is non-negative at all $\vmin$ values. This is because, for $\eta_{n > 0}$, the integrand favors larger speeds with respect to the $\eta_0$ integrand $F_{\text{SHM}, \text{E}} / v$, thus setting the integral more in phase with the time dependence of $F_{\text{SHM}, \text{E}}$ at large speeds. Larger powers of $v$ in the integrand also imply a larger relative spread in the $\eta_n$ curves at different times, as can be seen comparing the different top panels of \Fig{fig: eta}, thus increasing the fractional annual modulation $| \widetilde{\eta}_n / \overline{\eta}_n |$ (see bottom panels of \Fig{fig: etabartilde}). Finally, higher powers of $v$ in the integrand also make the velocity integral more sensitive to the shape of the high-speed tail of the distribution, thus on the value of $\vesc$ and its uncertainty, as well as to the truncation prescription ($\beta = 0, 1$). The precise shape of the high-$\vmin$ tail of the velocity integral can be important for sufficiently light DM particles, especially with heavy targets (or with light targets if the mass splitting $\delta$ is positive and sizeable, see \Sec{Inelastic scattering}); in this case, in fact, nuclear recoil energies can only be large enough to enter the sensitivity range of the experiment for sufficiently large $\vmin$ values.

\section{Phenomenology of direct DM detection}
\label{Pheno}
We now resume our discussion on the rate from where we left off in \Sec{Rate}. Having delved into scattering kinematics (\Sec{scattering kinematics}), the NR expansion (\Sec{DM-N}), the scattering amplitude (\Sec{Form factors}) and cross section (\Sec{sigma}), hadronic (\Sec{qg to N}) and nuclear (\Sec{Form factors}) responses to the scattering and related form factors, and the DM velocity distribution and velocity integral (\Sec{velocity}), we have all the needed ingredients to explore the features of the differential rate and the constraints direct detection experiment can set on the DM properties. We will do so by studying a number of example cases with qualitatively different recoil-energy spectra: a very much standard SI contact interaction, see \Sec{SI interaction}, for both elastic and inelastic scattering; and, less standard, a SI interaction with a light mediator, see Secs.~\ref{Vector-mediated interaction} and~\ref{Scalar-mediated interaction}; the millicharged DM model, see \Sec{Electromagnetic interactions}; and the case of DM with anomalous magnetic dipole moment, see Secs.~\ref{Electromagnetic interactions} and~\ref{Magnetic-dipole DM}.

\subsection{Setup and example models}
\label{Example models and setup}
We consider for simplicity an ideal detector with $\epsilon(\Ed) = 1$ (perfect efficiency) and $\mathcal{K}_T(\ER, \Ed) = \delta(\ER - \Ed)$ (perfect energy resolution) in \Eq{dRdEd}, so that the detection rate $\ud R / \ud \Ed$ is the same as the scattering rate $\ud R_T / \ud \ER$ summed over all target nuclides:
\begin{align}
\frac{\ud R}{\ud \Ed}(\Ed, t) = \CMcal{R}(\Ed, t)
&&
\text{with}
&&
\CMcal{R}(\ER, t) \equiv \sum_{T} \frac{\ud R_T}{\ud \ER}(\ER, t) \ .
\end{align}
For definiteness we assume that a given detector features all stable isotopes of a single nuclear element, see \Tab{tab: nuclides}, so that $\xi_T = \tilde{\xi}_T$ (see discussion after \Eq{zeta_T}). We recall from \Sec{Rate} that, under certain quite standard circumstances (see \Sec{velocity}), the time dependence of the velocity integrals can be approximated as in \Eq{eta bar tilde}. As a consequence, the scattering rate enjoys an analogous approximation, see \Eq{Taylor R}, from which we have
\beq
\CMcal{R}(\ER, t) \simeq \overline{\CMcal{R}}(\ER) + \widetilde{\CMcal{R}}(\ER) \cos \left[ 2 \pi \frac{t - t_0}{\yr} \right] ,
\eeq
where $t_0$ is the time of maximum Earth's speed in the galactic frame and we distinguished the annual-average and annual-modulation components
\begin{align}
\overline{\CMcal{R}}(\ER) \equiv \sum_T \frac{\ud \overline{R}_T}{\ud \ER}(\ER) \ ,
&&&
\widetilde{\CMcal{R}}(\ER) \equiv \sum_T \frac{\ud \widetilde{R}_T}{\ud \ER}(\ER) \ .
\end{align}
Following \Eq{etabartilde def}, $\overline{\CMcal{R}}(\ER)$ can be obtained by evaluating (the velocity integrals featured in) $\CMcal{R}(\ER, t)$ at $\vE(t) = v_\text{S}$, while $\widetilde{\CMcal{R}}(\ER)$ may be conveniently obtained through the approximation in Eqs.~\eqref{etabartilde approx},~\eqref{Taylor vE}, as we do here (we recall from \Sec{Modulation} that for a locally isotropic DM velocity distribution in the galactic rest frame the velocity integral only depends on time through $\vE(t)$).

For concreteness we assume in the following the SHM velocity distribution of DM particles, see \Sec{SHM}. $\eta_0(\vmin, t)$ is then given by \Eq{eta_0 v1} or~\eqref{eta_0 v2}, while $\eta_1(\vmin, t)$ is given in \Eq{eta_1} (see Figs.~\ref{fig: eta},~\ref{fig: etabartilde}); we set $\beta = 0$ for definiteness, see \Eq{SHM f} and subsequent discussion. For our numerical results and plots we adopt the values in \Eq{speeds}, together with $v_0 = v_\text{c}$. We also assume a single DM species with local density $\rho = 0.3~\GeV / \cm^3$, see \Eq{rho}. For the nuclear form factor $F_\text{SI}(\ER)$ we adopt the Helm formula in \Eq{HelmFF} with the parameters specified in \Eq{Helm parameters 2} (solid lines in \Fig{fig: HelmFF}), while for the other form factors we adopt the results of Ref.~\cite{Fitzpatrick:2012ix}.

As a first example model we consider the SI interaction discussed in detail in \Sec{SI interaction}. From the differential cross section in \Eq{diffsigmaSI 2} we can write for the rate
\beq\label{Rate SI}
\CMcal{R}(\ER, t) \NReq \sigma_p \frac{\rho}{\mDM} \frac{1}{2 \mu_\text{N}^2} \sum_T \zeta_T (Z + (A - Z) f_n / f_p)^2 F_\text{SI}^2(\ER) \, \eta_0(\vmin(\ER), t) \ ,
\eeq
see Eqs.~\eqref{Taylor rate},~\eqref{eta}. $\vmin(\ER)$ is given for elastic scattering in \Eq{vmin}. The DM-proton total cross section, $\sigma_p$, is specified in \Eq{SI sigma_N}. As already discussed in more general terms in \Sec{Scattering rate}, it is apparent in \Eq{Rate SI} that the local DM density $\rho$ is completely degenerate with $\sigma_p$, so that direct detection experiments are only sensitive to their product. Analogous considerations hold for other interactions as well, where $\sigma_p$ is to be substituted by whatever quantity parametrizes the overall size of the scattering cross section.

\begin{figure}[t!]
\begin{center}
\includegraphics[width=.49\textwidth]{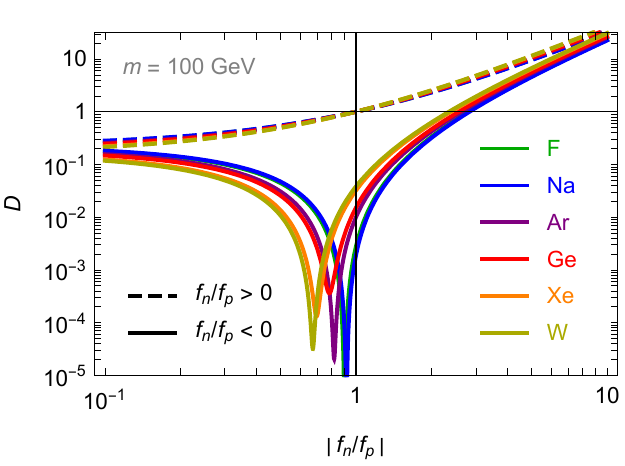}
\includegraphics[width=.49\textwidth]{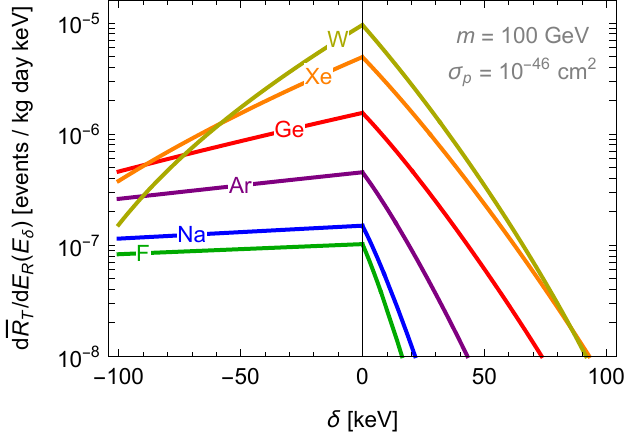}
\caption[The $D$ factor parametrizing the dependence of the SI rate on $f_n / f_p$]{\label{fig: Dfactor+Rdelta}\emph{\textbf{Left:} The \figmath{D} factor, defined in \Eq{D factor}, expressing the dependence of the SI rate in \Eq{Rate SI} on \figmath{f_n / f_p}, the ratio of the DM-neutron and DM-proton couplings. Lines of different colors correspond to different target elements. The horizontal axis, in log scale, reports the absolute value \figmath{| f_n / f_p |}; dashed (solid) lines correspond to positive (negative) \figmath{f_n / f_p} values. Owing to stable nuclei having roughly the same amount of protons and neutrons, \figmath{D} becomes independent of \figmath{f_n / f_p} for \figmath{| f_n / f_p | \ll 1}, where the contribution of neutrons become negligible, while it grows linearly with \figmath{(f_n / f_p)^2} for \figmath{| f_n / f_p | \gg 1}, where the neutron contribution is dominant. In between, one can notice the effect of the constructive (destructive) interference between protons and neutrons for \figmath{f_n / f_p > 0} (\figmath{f_n / f_p < 0}). The DM mass is set to \figmath{100~\GeV} for concreteness, although \figmath{D} barely depends on \figmath{\mDM}. Notice that \figmath{D} can receive long-distance QCD corrections that are especially relevant for the location of the dip that occurs for \figmath{f_n / f_p < 0}, see Refs.~\cite{Cirigliano:2012pq, Cirigliano:2013zta}: in this respect, the lowest-order result in \Eq{D factor} (so as the position of the dips in the plot) should only be thought of as indicative. \textbf{Right:} \figmath{\ud \overline{R}_T / \ud \ER (E_\delta)}, for the SI interaction, as a function of the DM mass splitting \figmath{\delta}. \figmath{\ud \overline{R}_T / \ud \ER (\ER)} is the annual average of the nuclide-specific differential rate \figmath{\ud R_T / \ud \ER (\ER, t)}, which for the SI interaction can be read off from the right-hand side of \Eq{Rate SI} by removing the sum over targets. \figmath{E_\delta} is the \figmath{\ER} value for which \figmath{\vmin} reaches its minimum value and the velocity integral is therefore maximum (see discussion related to \Eq{v_delta}). \figmath{\ud \overline{R}_T / \ud \ER (E_\delta)} may thus be thought of as representative of the dependence of the differential rate on \figmath{\delta}. Lines of different colors correspond to different target elements, where the most abundant isotope (same as in the right panel of \Fig{fig: reduced mass + qR=1}) has been chosen as representative for each element and its numerical abundance set to \figmath{\xi_T = 100 \%}, meaning \figmath{\zeta_T = 1} (see \Eq{zeta_T}). With this modification, the size of the nuclide-specific rate \figmath{\ud R_T / \ud \ER (\ER, t)} is comparable to that of \figmath{\CMcal{R}(\ER, t)}. For concreteness we set \figmath{\mDM = 100~\GeV}, \figmath{f_p = f_n}, and \figmath{\sigma_p = 10^{-46}~\cm^2}.}}
\figcode
\end{center}
\end{figure}

As commented upon in \Sec{SI interaction}, the standard assumption for the DM-proton and DM-neutron couplings is the isosinglet condition $f_p = f_n$, for which the $(Z + (A - Z) f_n / f_p)^2$ factor in \Eq{Rate SI} reduces to $A^2$. More in general, $|f_p|$ can be used to parametrize the overall size of the differential scattering cross section, \eg through the $\sigma_p$ parameter, with the $f_n / f_p$ ratio parametrizing the interplay of protons and neutrons: as we saw in \Sec{SI interaction}, for instance, protons and neutrons interfere destructively for $f_n / f_p < 0$, and the rate of DM scattering off a certain nuclear target can be severely reduced for a specific value of this ratio. The dependence of the rate on $f_n / f_p$ can be expressed \eg through the $D$ factor defined in \Eq{D factor}, plotted here in the left panel of \Fig{fig: Dfactor+Rdelta}. Notice that this factor, which can be checked to barely depend on the DM mass, can receive long-distance QCD corrections that are especially relevant for the location of the dip that occurs for $f_n / f_p < 0$, see Refs.~\cite{Cirigliano:2012pq, Cirigliano:2013zta}. From now on we adopt the standard assumption $f_p = f_n$, so that we can more easily focus our attention on the dependence of the rate on $\ER$.

\Eq{Rate SI} also holds for inelastic scattering, in which case $\vmin(\ER)$ depends on the DM mass splitting $\delta \neq 0$, see \Eq{vmin_inelastic}. While for elastic scattering ($\delta = 0$) the rate for the SI interaction is maximum at $\ER = 0$, as both $\eta_0$ and $F_\text{SI}^2$ are maximum there, for $\delta \neq 0$ the rate vanishes at zero recoil energy due to the fact that $\ER = 0$ actually corresponds to infinite $\vmin$ (see \Fig{fig: InelasticKinematics}), where $\eta_0$ is zero (see \eg the right panel of \Fig{fig: eta(ER)} below). A qualitative understanding of the rate dependence on $\delta$ may then be gained by rather looking at $\ER = E_\delta$, where $\vmin$ reaches its minimum value $v_\delta$ (see Eqs.~\eqref{E_delta},~\eqref{v_delta}). The velocity integral is maximum here, since the domain of the $\ud^3 v$ integral in \Eq{diffrate} is largest and the integrand is non-negative; the rate would then also be maximum, were it not for the nuclear form factor. $\ud \overline{R}_T / \ud \ER (E_\delta)$ is shown in the right panel of \Fig{fig: Dfactor+Rdelta} for a representative set of nuclides (same as in the right panel of \Fig{fig: reduced mass + qR=1}), with $\mDM = 100~\GeV$ and $\sigma_p = 10^{-46}~\cm^2$ for definiteness. We artificially set the numerical abundance of each employed nuclide to $\xi_T = 100 \%$ (which implies $\zeta_T = 1$, see \Eq{zeta_T}), so to make the size of $\ud \overline{R}_T / \ud \ER$ representative of that of $\overline{\CMcal{R}}$. The annual-average rate can be seen to decrease with increasing $|\delta|$ within the plotted range for both positive and negative values of $\delta$. One of the causes is the fact that $E_\delta$ increases with $|\delta|$ and therefore $\ud R_T / \ud \ER (E_\delta, t)$ experiences a reduction due to the decrease of $F_\text{SI}^2(E_\delta)$ occurring for the moderate $\delta$ values of interest here. Moreover, for $\delta > 0$, $v_\delta$ grows with $\delta$ thus reducing the domain of the $\ud^3 v$ integral (see the top-right panel of \Fig{fig: InelasticKinematics}); in fact, as already noted in \Sec{Inelastic scattering}, endothermic scattering is less kinematically favored for larger values of $\delta$. For $\delta < 0$, instead, $v_\delta = 0$ and the (domain of the) velocity integral does not change with $\delta$.

Another model we take into account here is a SI interaction (with isosinglet couplings) mediated by a light particle in a tree-level $t$-channel diagram. Our reference differential cross section is \Eq{diffsigma V} with $c_p = c_n$ and $a_\chi = 0$ (see the Lagrangian in \Eq{DM V}). The $a_\chi = 0$ condition eliminates the SD interaction, including the induced pseudo-scalar contribution which for simplicity was omitted in \Eq{diffsigma V}, as explained in \Sec{Vector-mediated interaction}. The differential rate is then
\beq\label{Rate light med}
\CMcal{R}(\ER, t) \NReq \frac{\rho}{\mDM} \frac{1}{2 \pi} \frac{\lambda^2}{(q^2 + m_V^2)^2} \sum_T \zeta_T A^2 F_\text{SI}^2(\ER) \, \eta_0(\vmin(\ER), t) \ ,
\eeq
with $\lambda \equiv c_\chi c_p$ and $m_V$ the mediator mass. This reduces to \Eq{Rate SI} (with isosinglet couplings $f_p = f_n$) in the $m_V^2 \gg q^2$ limit, with $\sigma_p$ identified with $\lambda^2 \mu_\text{N}^2 / \pi m_V^4$. As argued in \Sec{Vector-mediated interaction}, inspection of \Fig{fig: TypicalER} reveals that the $t$-channel mediator in a tree-level exchange can always be considered heavy for the purposes of direct detection (meaning that $m_V^2 \gg q^2$ is satisfied for all kinematically allowed values of momentum transfer and for all targets commonly employed in direct searches) if heavier than few $\GeV$. Due to the momentum-transfer dependence of the mediator propagator, we can expect the $\ER$ dependence of the rate in \Eq{Rate light med} to be steeper than that in \Eq{Rate SI}, at least for a sufficiently light mediator (see \Fig{fig: spectra} below). This feature may be used to distinguish the hypothesis of light mediator from that of contact interaction in case of a putative DM signal. However, as pointed out \eg in Ref.~\cite{DelNobile:2015uua}, a steepening of the spectrum for the SI interaction in \Eq{Rate SI} may also be obtained for a smaller value of the DM mass, so that the pronounced slope of the high-speed tail of the velocity integral is moved to lower energies (see \Fig{fig: SI spectra} below). A possible degeneracy may be solved by means of a second experiment employing a different target material or by measuring the annual-modulation spectrum $\widetilde{\CMcal{R}}(\ER)$~\cite{DelNobile:2015uua}.

As explained above, the contact SI interaction in \Eq{Rate SI} can be considered as the heavy-mediator limit of \Eq{Rate light med}, where the DM-nucleon scattering amplitude does not depend on energy. In the opposite limit, the characteristic $q$ dependence of the amplitude due to the light-mediator propagator in the $t$ channel can turn into an even steeper dependence when the mediator is massless, as the photon. Arguably the simplest model with this feature is millicharged DM, already discussed in \Sec{Electromagnetic interactions}. The Lagrangian and NR operator are provided in \Tab{tab: EM interactions}, from which it is easy to derive the differential scattering cross section
\beq\label{Millicharged sigma}
\frac{\ud \sigma_T}{\ud \ER} \NReq \frac{8 \pi m_T}{v^2} \frac{1}{q^4} \alpha_\text{EM}^2 Q_\DM^2 Z^2 F_\text{SI}^2(\ER) \ ,
\eeq
and subsequently the rate
\beq\label{Rate millicharged}
\CMcal{R}(\ER, t) \NReq \frac{\rho}{\mDM} 8 \pi \frac{\alpha_\text{EM}^2 Q_\DM^2}{q^4} \sum_T \zeta_T Z^2 F_\text{SI}^2(\ER) \, \eta_0(\vmin(\ER), t) \ ,
\eeq
with $\alpha_\text{EM} \equiv e^2 / 4 \pi$ the fine-structure constant. Numerically, $\alpha_\text{EM} \approx 1 / 137$ at the low energy scale of interest to direct DM detection, corresponding to $e \approx 3$. The nuclear form factor for Coulomb interactions with protons is the same as that of the SI interaction (actually, the other way around), as explained in \Sec{SI interaction}.

Finally, we consider an electrically neutral Dirac DM field interacting with photons through an anomalous magnetic moment, see \Sec{Magnetic-dipole DM}. Nuclei interact with photons through their electric charge and magnetic moment, both being relevant here to the DM-nucleus scattering (featuring both charge-dipole and dipole-dipole interactions), as opposed to the millicharged DM model where the Coulomb (\ie charge-charge) interaction dominates. The differential scattering cross section is given at tree level in \Eq{Magnetic dipole sigma}, and the differential rate is
\begin{multline}
\label{Rate magnetic}
\CMcal{R}(\ER, t) \NReq \frac{\rho}{\mDM} \frac{\alpha_\text{EM} \mu_\chi^2}{2 \mN^2} \sum_T \zeta_T \left[ \left( \frac{1}{\ER} \eta_1(\vmin(\ER), t) - \frac{\mDM + 2 m_T}{2 \mDM m_T} \eta_0(\vmin(\ER), t) \right) 2 \frac{\mN^2}{m_T} F_M^{(p, p)}(q^2) \right.
\\
\left. \vphantom{\left( 1 + 4 \frac{\mDM^2 {v^\perp_T}^2}{q^2} \right)}
+ \left( 4 F_{\Delta}^{(p, p)}(q^2) - 2 \sum_N g_N F_{\Sigma' \Delta}^{(N, p)}(q^2) + \frac{1}{4} \sum_{N, N'} g_N g_{N'} F_{\Sigma'}^{(N, N')}(q^2) \right) \eta_0(\vmin(\ER), t) \right] ,
\end{multline}
where $\mu_\chi$ is the DM magnetic dipole moment while $g_N$ is the nucleon Land\'e $g$-factor, see \Eq{g_N values}. Here the steep $q$ dependence of the photon propagator is balanced by $q$- and $v$-dependent factors arising at second order in the NR expansion (by contrast, the millicharged-DM and SI scattering amplitudes arise at zeroth order). These factors are also responsible for the appearance in \Eq{Rate magnetic} of the $\eta_1$ velocity integral, defined in \Eq{eta}, as opposed to the previous models that only involve $\eta_0$. We refer the reader to \Sec{Magnetic-dipole DM} for a discussion on these and other characteristic (and instructive) features of this interaction.

\subsection{Rate spectrum}
\label{Rate spectrum}
To get a qualitative understanding of the rate and its dependence on the model parameters, it should be kept in mind that the energy dependence of the differential scattering rate $\ud R_T / \ud \ER$ is due to the interplay of three key ingredients:
\begin{itemize}
\item The possible $q$ dependence of the DM-nucleon scattering amplitude (and thus of the NR interaction operator $\Op_\NR^N$~\eqref{general NR Op}), as when the interaction is NR suppressed or is mediated by a light or massless particle in a tree-level $t$-channel exchange.
\item The nuclear form factor(s), which encodes the effect of nuclear compositeness.
\item The velocity integral(s), which as a function of $\vmin$ only depends on the astrophysical properties of the local DM distribution, but acquires dependence on the scattering kinematics (thus on $m_T$, $\mDM$, and $\delta$) when expressed in terms of $\ER$ through the $\vmin(\ER)$ function.
\end{itemize}
We discuss these three ingredients in the following.

The first ingredient, the $q$ dependence of the DM-nucleon scattering amplitude, is not there for the SI interaction, as can be seen by noticing that \Eq{ONR SI} does not depend on momentum transfer, and the $\ER$ dependence of \Eq{Rate SI} is entirely due to the nuclear form factor and the velocity integral. For the other examples we considered above, a $q$ dependence arises already at the level of the DM-nucleon scattering amplitude due to the $t$-channel propagator of a light or massless mediator; in addition, the magnetic-dipole DM model features NR suppression factors that reduce the leading $q$ dependence at low momentum transfer from $q^{-4}$ as for the millicharged-DM model, see \Eq{Rate millicharged}, to $q^{-2}$, see \Eq{Rate magnetic}. All these factors change the spectral shape of the rate, which can in principle be observed. We will compare below the different recoil spectra arising from these interactions.

As explained in \Sec{Form factors}, nuclear compositeness also induces a $q$ dependence in the scattering amplitude, that is usually conveniently encoded within nuclear form factors in the differential cross section. Form factors are most often normalized to a finite value at zero momentum transfer and, especially when expressed in terms of $\ER$, decrease exponentially at a faster rate for larger nuclei, see \eg Figs.~\ref{fig: HelmFF},~\ref{fig: SDFF}. Therefore, form factors suppress the rate for DM scattering off heavier nuclei more than for lighter nuclei, see \eg \Fig{fig: SI spectra} below. Despite the exponential suppression and possibly some occasional zero, form factors can be essentially thought of as non-zero functions, as opposed to the velocity integrals which actually vanish for sufficiently large values of $\vmin$ (thus of $\ER$).

The velocity integral is the integral over $\bol{v}$ of a function proportional to the local DM velocity distribution in Earth's rest frame, see \Eq{diffrate}. The domain of the integral is determined by a single variable, $\vmin$, which depends on $\ER$ and on the DM and nuclear masses. As such, if the integrand does not depend on these parameters, as for the $\eta_n$ functions defined in \Eq{eta}, the velocity integral is a function of $\vmin$ that is uniquely determined by the local DM velocity distribution (see \Sec{SHM} for the SHM). The dependence of the velocity integral on $m_T$, $\mDM$, and $\delta$ is then ascribed exclusively to the $\vmin(\ER)$ function (see \Sec{scattering kinematics}): in this sense it is useful to think of the velocity integral as a single function of $\vmin$ that is stretched onto the $\ER$ axis in a $m_T$- and $\mDM$- (and $\ER$-) dependent way. Since DM particles that are gravitationally bound to the halo of our galaxy have speeds below the local escape speed in the galactic rest frame, the DM velocity distribution drops to zero for large speeds and so does the velocity integral. This implies that, when expressed in terms of $\ER$, the velocity integral vanishes for large enough recoil energies as the DM speeds that would be needed to impart such energies to a target nucleus exceed the maximum allowed speed. In other words, a maximum speed (and thus a maximum $\vmin$) corresponds to a maximum kinematically attainable nuclear recoil energy, whose actual value depends on the DM and target masses. Such maximum $\ER$ value, where the velocity integral becomes zero, is smaller for lighter DM, see discussion on the $\ER$--$\vmin$ mapping in Secs.~\ref{Elastic scattering},~\ref{Inelastic scattering}. Therefore direct DM search experiments, which due to their threshold may be thought of as being effectively sensitive only to nuclear recoil energies above a certain minimum value, are precluded from detecting light enough DM particles. The largest possible DM speed in Earth's frame then determines (in a way that depends on the scattering kinematics) the lower end of an experiment's reach in DM mass. We will discuss this point more quantitatively below.

\begin{figure}[t!]
\begin{center}
\includegraphics[width=.49\textwidth]{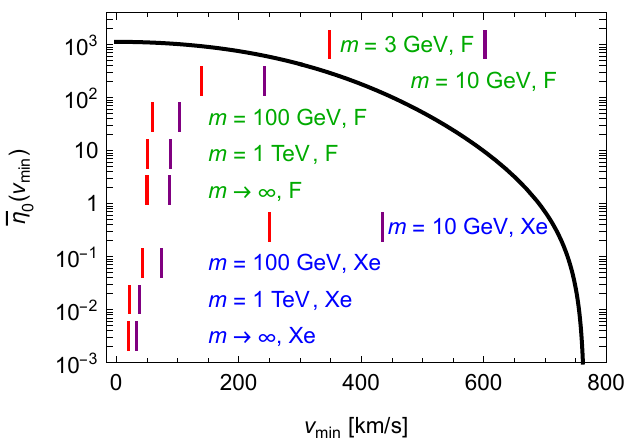}
\includegraphics[width=.49\textwidth]{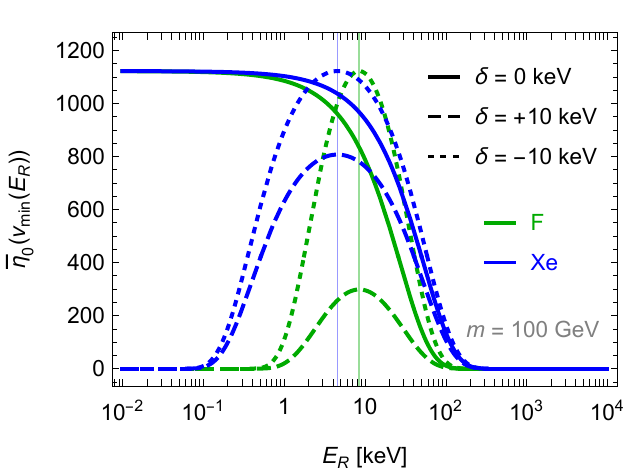}
\caption[The SHM $\overline{\eta}_0$ and the $\ER$--$\vmin$ mapping for elastic and inelastic scattering]{\label{fig: eta(ER)}\emph{The annual-average velocity integral \figmath{\overline{\eta}_0}, see \Eq{etabartilde def}, in the SHM, see \Eq{eta_0 v1} or~\eqref{eta_0 v2} and the top-left panel of \Fig{fig: etabartilde}. We set \figmath{\beta = 0} in \Eq{SHM f} for concreteness. \textbf{Left:} as a function of \figmath{\vmin} (black line). The vertical bars indicate the \figmath{\vmin} values of \figmath{\ER = 1~\keV} (red bars) and \figmath{\ER = 3~\keV} (purple bars) for two nuclear targets (fluorine, with green labels, and xenon, with blue labels) and different values of DM mass, for elastic scattering (see also \Fig{fig: ElasticKinematics}). This can give an idea of the relevant portion of the velocity integral contributing to the rate in a specific energy interval. For fluorine we employ the only stable isotope, \figmath{^{19}\text{F}}, while for xenon we employ the lightest stable isotope in \Tab{tab: nuclides}, \figmath{^{128}\text{Xe}}, which yields the smallest \figmath{\vmin} value at given \figmath{\ER} for elastic scattering of light DM. The DM mass is taken to be \figmath{3~\GeV}, \figmath{10~\GeV}, \figmath{100~\GeV}, and \figmath{1~\TeV}, together with an infinite value which is representative of very large masses, \figmath{\mDM \gg m_T}. For a \figmath{3~\GeV} DM particle scattering off xenon, recoil energies of \figmath{1~\keV} or larger correspond to \figmath{\vmin} values above \figmath{\vmax} in \Eq{vmax}, where \figmath{\overline{\eta}_0} vanishes, and thus they are not displayed; this also holds for a DM particle with mass \figmath{1~\GeV}, for both \figmath{\text{F}} and \figmath{\text{Xe}}. \textbf{Right:} as a function of \figmath{\ER}, for \figmath{\mDM = 100~\GeV} and three values of the DM mass splitting: \figmath{\delta = 0} (elastic scattering, solid lines), \figmath{\delta = 10~\keV} (endothermic scattering, dashed lines), and \figmath{\delta = - 10~\keV} (exothermic scattering, dotted lines). Green lines are for a \figmath{^{19}\text{F}} target, blue lines for \figmath{^{128}\text{Xe}}; the thin vertical lines indicate the value of \figmath{E_\delta} for \figmath{|\delta| = 10~\keV} for the two nuclides, \figmath{E_\delta \approx 8.5~\keV} for \figmath{^{19}\text{F}} and \figmath{E_\delta \approx 4.6~\keV} for \figmath{^{128}\text{Xe}}. For \figmath{\delta \neq 0}, \Eq{vmin E_delta} implies that \figmath{\vmin(\ER)} appears symmetric about \figmath{E_\delta} when plotted on a logarithmic \figmath{\ER} scale, and the same of course holds for functions of \figmath{\vmin(\ER)} such as the velocity integral. \figmath{\overline{\eta}_0(\vmin(\ER))} vanishing for large enough \figmath{\ER} values then implies that it also vanishes for small enough \figmath{\ER} values, yielding the bell shape of the dashed and dotted curves.}}
\figcode
\end{center}
\end{figure}

An illustration of the $\ER$--$\vmin$ relation is presented in the left panel of \Fig{fig: eta(ER)}, where $\ER = 1~\keV$ (red vertical bars) and $\ER = 3~\keV$ (purple vertical bars) are mapped for elastic scattering onto $\vmin$ for different values of the DM mass, including $\mDM \to \infty$ representative of the $\mDM \gg m_T$ limit, and for two target nuclides: $^{19}$F (green labels), the only stable isotope of fluorine (see \Tab{tab: nuclides}), and $^{128}$Xe (blue labels), the lightest stable isotope of xenon in \Tab{tab: nuclides}, thus the one allowing for the smallest $\vmin$ value at given $\ER$ for elastic scattering of light DM (see \Sec{Elastic scattering}). For reference, the SHM $\overline{\eta}_0(\vmin)$ velocity integral is also shown, so that one can see what part of it contributes to the differential rate at the aforementioned values of recoil energy. Other pedagogical depictions of the effect of the $\ER$--$\vmin$ mapping can be found \eg in Ref.~\cite{DelNobile:2015uua}. Notice that $\overline{\eta}_n(\vmin)$ vanishes for $\vmin$ greater than
\beq\label{vmax}
\vmax \equiv \vesc + v_\text{S} \approx 765~\km / \sd \ ,
\eeq
see \Eq{speeds}. As already discussed in \Sec{Elastic scattering}, for elastic scattering, heavier DM is kinematically favored over light DM, and lighter (heavier) targets are favored for sufficiently light (heavy) DM (see \Fig{fig: ElasticKinematics}). For inelastic scattering, as discussed in \Sec{Inelastic scattering}, each $\vmin > v_\delta$ corresponds to two values of $\ER$, one larger and one smaller than $E_\delta$, related as $\ER^+$ and $\ER^-$ in \Eq{E_delta} (see also \Eq{vmin E_delta}). $\vmin(\ER < E_\delta)$ may then be seen as a mirrored and (unevenly) stretched version of $\vmin(\ER > E_\delta)$, and the same holds for $\eta_n(\vmin(\ER), t)$. As a consequence, the velocity integral vanishing for sufficiently large $\ER$ values implies that it also vanishes for sufficiently small $\ER$ values. This behavior is depicted in the right panel of \Fig{fig: eta(ER)}, showing again $\overline{\eta}_0$ this time as a function of $\ER$, for $\mDM = 100~\GeV$ and three values of the DM mass splitting: $\delta = 0$ (elastic scattering, solid lines), $\delta = 10~\keV$ (endothermic scattering, dashed lines), and $\delta = - 10~\keV$ (exothermic scattering, dotted lines). Green lines are for a $^{19}$F target, blue lines for $^{128}$Xe; the thin vertical lines indicate the value of $E_\delta$ for $|\delta| = 10~\keV$ for the two nuclides, $E_\delta \approx 8.5~\keV$ for $^{19}$F and $E_\delta \approx 4.6~\keV$ for $^{128}$Xe. For $\delta \neq 0$, the discussion related to \Eq{vmin E_delta} implies that functions of $\vmin(\ER)$ appear symmetric about $E_\delta$ when plotted on a logarithmic $\ER$ scale (dashed and dotted lines). As already discussed in \Sec{Inelastic scattering}, elastic scattering is always kinematically favored over endothermic scattering, implying that the velocity integral is larger for $\delta = 0$ (solid line) than for $\delta > 0$ (dashed line of same color). In fact, the larger $\delta \geqslant 0$, the smaller the domain of the velocity integral. For the same reason we see, as already commented in \Sec{Inelastic scattering}, that exothermic scattering is always kinematically favored (disfavored) over elastic scattering for $\ER$ values larger (smaller) than $E_\delta / 2$, the point where the dotted line meets the solid line of the same color. Exothermic scattering being always more kinematically favored than endothermic scattering at given $|\delta|$, the dotted line ($\delta = - 10~\keV$) lies entirely above the dashed one ($\delta = + 10~\keV$) of the same color.

\begin{figure}[t]
\begin{center}
\includegraphics[width=.32\textwidth]{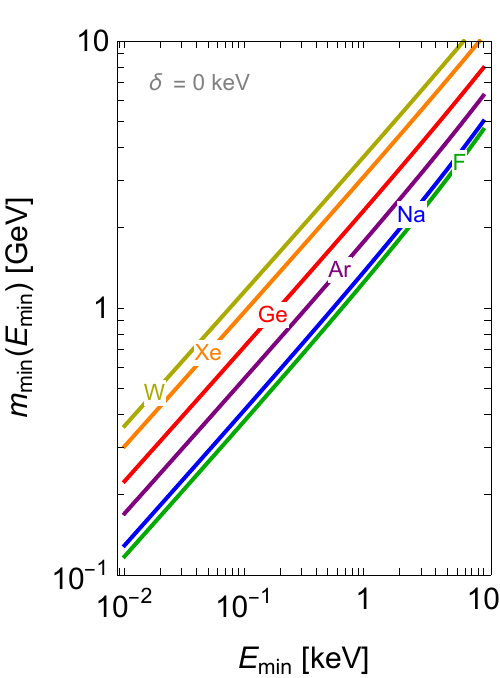}
\includegraphics[width=.32\textwidth]{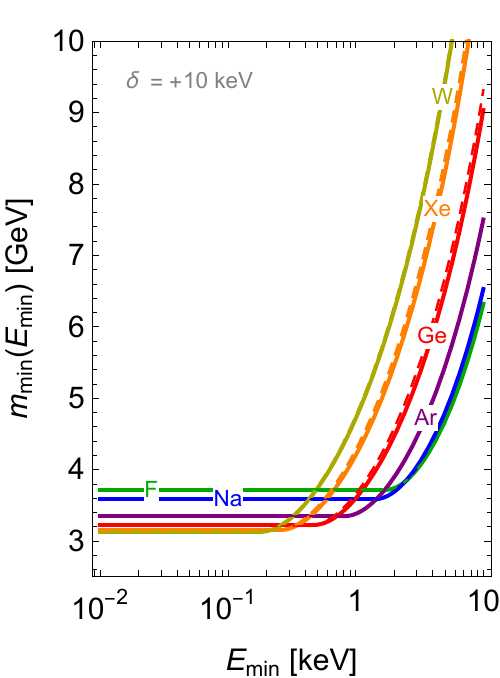}
\includegraphics[width=.32\textwidth]{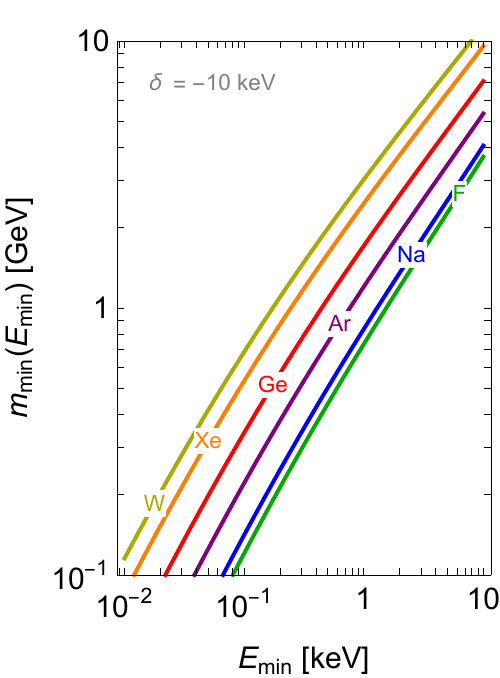}
\caption[Minimum DM mass that can be probed for a given minimum recoil energy]{\label{fig: minDMmass}\emph{Minimum DM mass that can be probed with an experiment effectively sensitive only to nuclear recoil energies above a certain \figmath{\Emin}, see \Eq{minDMmass}. Notice that the finite energy resolution of actual experiments makes this quantity purely indicative. Lines of different colors correspond to different target elements (\figmath{\text{F}}, \figmath{\text{Na}}, \figmath{\text{Ar}}, \figmath{\text{Ge}}, \figmath{\text{Xe}}, \figmath{\text{W}}), see \Tab{tab: nuclides} for details. The lightest isotope in \Tab{tab: nuclides} is employed for each element, as for the small DM masses of interest here \figmath{\mDM_\text{min}(\Emin)} increases with increasing \figmath{m_T} for \figmath{\delta \leqslant 0}, and for \figmath{\delta > 0} at large enough \figmath{\Emin} values (decreasing instead with \figmath{m_T} for \figmath{\delta > 0} and sufficiently small \figmath{\Emin}). While for elastic and exothermic scattering, in principle, the sensitivity of an experiment can always be extended to lighter DM by lowering its threshold (and thus \figmath{\Emin}), this is not possible for endothermic scattering. \textbf{Left:} for elastic scattering, \figmath{\delta = 0}. \textbf{Center:} for endothermic scattering, \figmath{\delta = 10~\keV}. The thin, dashed lines are for the heaviest isotope of each element (note that \figmath{\text{F}} and \figmath{\text{Na}} only have one stable isotope), which at sufficiently low \figmath{\Emin} yields a smaller \figmath{\mDM_\text{min}(\Emin)} value with respect to the other isotopes. \textbf{Right:} for exothermic scattering, \figmath{\delta = - 10~\keV}.}}
\figcode
\end{center}
\end{figure}

As discussed above, experiments can only detect sufficiently heavy DM particles, due to their finite threshold and to the speed of halo DM particles being limited from above. To have an idea of the minimum mass DM particles must have in order to be detected by a certain experiment, we denote with $\Emin$ the minimum nuclear recoil energy the experiment is effectively sensitive to. Notice that $\Emin$ is distinct from the experimental threshold, which is the minimum value of $\Ed$ rather than $\ER$ (see \Sec{Rate}), and that the finite energy resolution of actual experiments makes this discussion purely indicative. In fact, $\ER$ is only statistically related to the detected signal $\Ed$, and thus $\Emin$ should only be considered as a convenient theoretical device. With this in mind, since the maximum attainable recoil energy increases with the mass and speed of the DM particle, the minimum DM mass a given experiment is sensitive to can be determined by equating the detector $\Emin$ value with the maximum recoil energy ($\ER^+$ in \Eq{ER^pm}, corresponding to $\ER^\text{max}$ in \Eq{ERinterval} for elastic scattering) attained at the maximum possible DM speed. The maximum speed halo DM particles can have in Earth's rest frame is $\vmax^+$, given in \Eq{vmax^+-}, however speeds between $\vmax^-$ and $\vmax^+$ can only be achieved during part of the year due to the change in Earth's velocity as it rotates around the Sun. Moreover, standard analyses of direct DM detection data approximate the differential rate $\CMcal{R}(\ER, t)$ with its annual average $\overline{\CMcal{R}}(\ER)$, thus employing the annual-average velocity integrals $\overline{\eta}_n(\vmin)$ which vanish for $\vmin \geqslant \vmax$. In this discussion we will therefore adopt $\vmax$ as a (typical, at least) maximum value for the DM speed. With this prescription, the minimum DM mass that can be probed with an experiment only sensitive to recoil energies above $\Emin$, call it $\mDM_\text{min}(\Emin)$, can be determined by solving $\ER^+(\vmax) = \Emin$ for $\mDM$. However, it is probably simpler to solve the equivalent equation
\beq\label{minDMmass condition}
\vmin(\Emin) = \vmax \ ,
\eeq
see \Eq{vmin_inelastic}, with the supplementary condition $\Emin > E_\delta$, see \Eq{E_delta}; this condition ensures that we are considering the maximum kinematically allowed recoil energies, \ie the $\ER^+$ branch in \Eq{ER^pm}. Before solving \Eq{minDMmass condition} we need to pay special care to its domain, which is non-trivial for $\delta > 0$ (see the top panels of \Fig{fig: InelasticKinematics}): imposing that the minimum possible $\vmin$ value, \ie $v_\delta$ in \Eq{v_delta}, is not larger than $\vmax$, implies for $\delta > 0$ the requirement $\mu_T > 2 \delta / \vmax^2$, which combined with the above supplementary condition yields $\Emin > 2 \delta^2 / m_T \vmax^2$ as a validity condition on \Eq{minDMmass condition}. For DM masses such that $\mu_T$ is below the above minimum value, the annual-average rate is zero as the domain of the velocity integral lies entirely within the region where the integrand vanishes. This means that, while for elastic and exothermic scattering, in principle, the sensitivity of an experiment can always be extended to lighter DM by lowering its threshold (and thus $\Emin$), this is not possible for endothermic scattering. Solving \Eq{minDMmass condition} for $\mu_T$ and using \Eq{inverse mu_T} then results in the following expression for the minimum DM mass that can be probed with an experiment effectively sensitive only to recoil energies above $\Emin$:
\beq\label{minDMmass}
\mDM_\text{min}(\Emin) =
\begin{cases}
\dfrac{2 m_T \delta}{m_T \vmax^2 - 2 \delta} & \delta > 0~\text{and}~\Emin < \dfrac{2 \delta^2}{m_T \vmax^2},
\\
\dfrac{m_T \Emin}{\vmax \sqrt{2 m_T \Emin} - \Emin - \delta} & \text{otherwise}. \rule{0pt}{20pt}
\end{cases}
\eeq
Notice that the above requirement on $\mu_T$ and the fact that $m_T > \mu_T$ (see \Sec{kinematics notation}) implies that $m_T \vmax^2 > 2 \delta$, no scattering being kinematically possible for values of $\delta$ that violate this inequality. As already mentioned in \Sec{Inelastic scattering}, the minimum DM mass current experiments are sensitive to is of order of few $\GeV$ or lower. For these DM masses, the scattering at $\ER > E_\delta$ is more kinematically favored for lighter targets, assuming $\delta \leqslant 0$, while for $\delta > 0$ lighter targets are more kinematically efficient only for sufficiently large recoil energies, heavier targets becoming more efficient at smaller energies (see \Sec{Inelastic scattering}). Therefore, in these conditions, in \Eq{minDMmass} one should adopt for elastic and exothermic scattering the lightest of the detector nuclides taking part in the interaction, for which the resulting $\mDM_\text{min}(\Emin)$ is minimum. For endothermic scattering, instead, the smallest $\mDM_\text{min}(\Emin)$ value is obtained with the lightest (heaviest) interacting target at sufficiently large (small) $\Emin$. $\mDM_\text{min}(\Emin)$ is pictured in \Fig{fig: minDMmass} for different target elements and for $\delta = 0$ (elastic scattering, left panel), $\delta = 10~\keV$ (endothermic scattering, center panel), and $\delta = - 10~\keV$ (exothermic scattering, right panel). The lightest isotope in \Tab{tab: nuclides} is employed for each element, although for endothermic scattering the heaviest isotope is also employed yielding the thin, dashed lines (note that F and Na only have one stable isotope, see \Tab{tab: nuclides}). As a consequence of the above discussion, lighter targets yield a smaller $\mDM_\text{min}(\Emin)$ for $\delta \leqslant 0$ and, for sufficiently large $\Emin$ values, also for $\delta > 0$, while heavier targets are more efficient for $\delta > 0$ and sufficiently small $\Emin$.

We now analyse the interplay of the three ingredients discussed above (inherent momentum dependence of the DM-nucleon interaction, nuclear form factors, and velocity integrals) and the effects it has on the rate. We start by focusing, for elastic scattering alone, on the SI interaction, which lacks the first ingredient and only features a single form factor and velocity integral; such a model is thus ideal to inspect the interplay of form factor and velocity integral on the spectral shape of the rate, namely its $\ER$ dependence, as well as its dependence on the nuclear target and the model parameters such as the DM mass. We then draw a comparison among the example interactions introduced in \Sec{Example models and setup} to better understand how the rate spectrum is modified as a consequence of their inherent dependence on momentum transfer and of a non-zero DM mass splitting.

\Fig{fig: SI spectra} displays the annual-average differential rate $\overline{\CMcal{R}}(\ER)$ for the SI interaction with elastic scattering, see \Eq{Rate SI}, for F, Na, Ar, Ge, Xe, W nuclear targets and four values of the DM mass, $\mDM = 3~\GeV$, $10~\GeV$, $100~\GeV$, and $1~\TeV$. The $f_n / f_p$ coupling ratio is set to $1$ while the $\sigma_p$ parameter (the DM-proton cross section) is fixed to $\sigma_p = 10^{-46}~\cm^2$, a choice of model parameters that will be later labelled M$1$. The dashed lines are computed with no form factor, \ie setting $F_\text{SI}(\ER) = 1$. The noticeable dips for Xe and W are due to the Helm form factor vanishing at specific, nuclide-dependent values of momentum transfer, as one can see by looking at \Fig{fig: HelmFF}. The dips, which are present also for the other targets though outside of the plotted range, are partially filled by the sum over different isotopes when more than one isotope is present, see \eg \Tab{tab: nuclides}. Comparing the different panels of \Fig{fig: SI spectra}, corresponding to different DM masses, one can see that the nuclear form factor has a negligible or small effect on the rate for light DM and/or light target nuclei, but has a large impact for heavy DM scattering off heavy nuclei. For light nuclei, in fact, it is clear from \Fig{fig: HelmFF} that the SI form factor has only a mild $\ER$ dependence in the energy range of interest to direct DM detection experiments, which can be understood as a consequence of the small nuclear size (see discussion in \Sec{Rate preliminaries}). The negligible impact of the form factor on the rate for light DM, even with heavy target nuclei, can instead be explained with the fact that, for small $\mDM$, the rate spectrum is dominated by the pronounced $\ER$ dependence of the velocity integral. In fact, for elastic scattering $\vmin$ gets mapped to smaller $\ER$ values for lighter DM (see the right panel of \Fig{fig: ElasticKinematics} and the left panel of \Fig{fig: eta(ER)}). The sharp drop-off of the velocity integral at $\vmin$ values close to $\vmax$ turns therefore into a sharp decrease of the rate spectrum occurring at lower energies for lighter DM particles. For sufficiently light DM, the rate is then only non-zero across recoil-energy scales where the nuclear form factor does not vary much with respect to the velocity integral. For heavy DM, instead, the rate at $\ER$ values of interest to the experiments only probes the low-speed portion of the velocity integral (as opposed to its high-speed tail), which has a smoother dependence on $\vmin$. For $\mDM \gg m_T$, small and intermediate values of $\ER$ map onto small $\vmin$ values (see the right panel of \Fig{fig: ElasticKinematics} and the left panel of \Fig{fig: eta(ER)}), where the SHM velocity integral is fairly constant. The spectral shape of the differential scattering rate (regardless of its normalization) is then entirely determined by the nuclear form factor. While the spectral shape has a sizeable dependence on $\mDM$ for light DM, it becomes fairly independent of $\mDM$ for large values thereof, as can be seen by comparing the different panels of \Fig{fig: SI spectra}. In this case, the main appreciable dependence of the rate on $\mDM$ is through the $\rho / \mDM$ factor in \Eq{Rate SI}, which implies that for $\mDM \gg m_T$ the scattering rate is inversely proportional to $\mDM$. This explains why the experimental bounds on $\sigma_p$ as a function of $\mDM$ takes the form of a straight line for large DM masses in the usual log-log plot, see \eg \Fig{fig: bounds} and the discussion related to \Eq{large mDM rate integral}).

\begin{figure}[t!]
\begin{center}
\includegraphics[width=.49\textwidth]{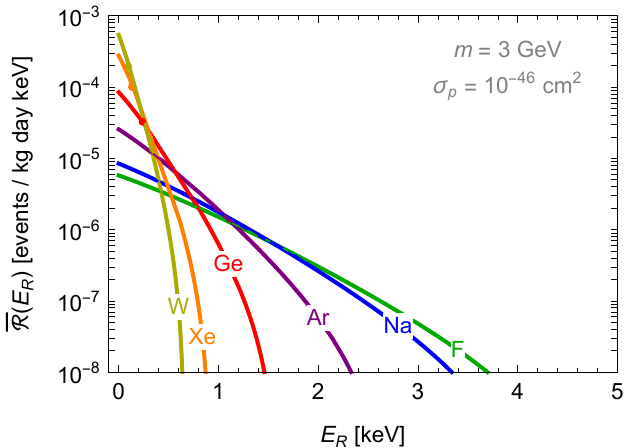}
\includegraphics[width=.49\textwidth]{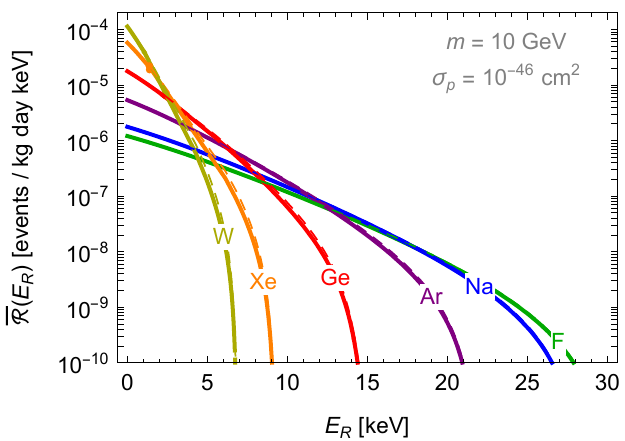}
\\[5mm]
\includegraphics[width=.49\textwidth]{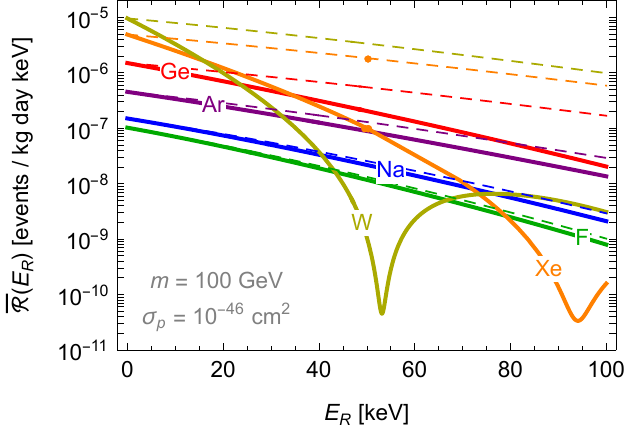}
\includegraphics[width=.49\textwidth]{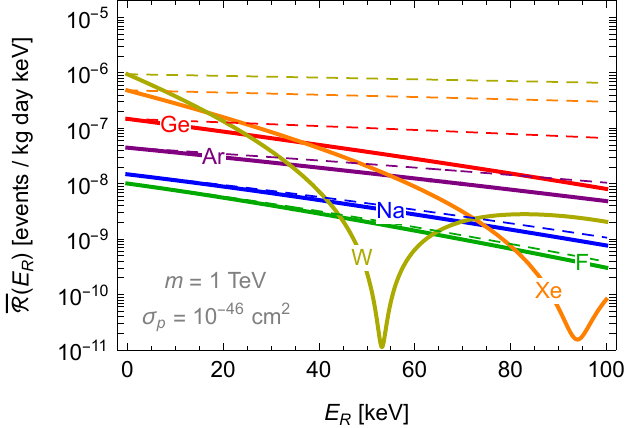}
\caption[Annual-average rate spectra for the SI interaction]{\label{fig: SI spectra}\emph{Annual-average rate spectra \figmath{\overline{\CMcal{R}}(\ER)} for the SI interaction discussed in detail in \Sec{SI interaction} (solid lines), see \Eq{Rate SI}. We assume elastic scattering and isosinglet couplings \figmath{f_p = f_n}, and fix the DM-proton cross section at \figmath{\sigma_p = 10^{-46}~\cm^2}; this choice of model and parameters is dubbed \figmath{\text{M$1$}} in \Fig{fig: spectra}. Lines of different colors correspond to different target elements (\figmath{\text{F}}, \figmath{\text{Na}}, \figmath{\text{Ar}}, \figmath{\text{Ge}}, \figmath{\text{Xe}}, \figmath{\text{W}}), see \Tab{tab: nuclides} for details. The DM mass has values \figmath{\mDM = 3~\GeV} (\textbf{top left}), \figmath{10~\GeV} (\textbf{top right}), \figmath{100~\GeV} (\textbf{bottom left}), and \figmath{1~\TeV} (\textbf{bottom right}). The noticeable dips for \figmath{\text{Xe}} and \figmath{\text{W}} are due to the Helm form factor vanishing at specific, nuclide-dependent values of momentum transfer, as one can see by looking at \Fig{fig: HelmFF}. The dips, which are present also for the other targets though outside of the plotted range, are partially filled by the sum over different isotopes when more than one isotope is present (see \eg \Tab{tab: nuclides}). The dashed lines are computed with no form factor, \ie setting \figmath{F_\text{SI}(\ER) = 1} in \Eq{Rate SI}. As detailed in the text, the spectral shape of the rate is determined by the steep \figmath{\ER} dependence of the velocity integral for light DM and/or light targets, while it is dominated by the nuclear form factor for heavy DM particles scattering off heavy targets. For \figmath{\mDM \gg m_T}, the main appreciable dependence of the rate on the DM mass is through the \figmath{\rho / \mDM} flux factor in \Eq{Rate SI}, which causes the scattering rate to be inversely proportional to \figmath{\mDM}.}}
\figcode
\end{center}
\end{figure}

\begin{figure}[t]
\begin{center}
\includegraphics[width=.49\textwidth]{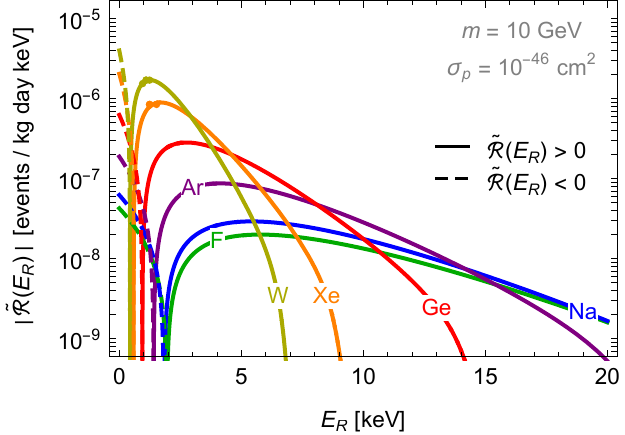}
\includegraphics[width=.49\textwidth]{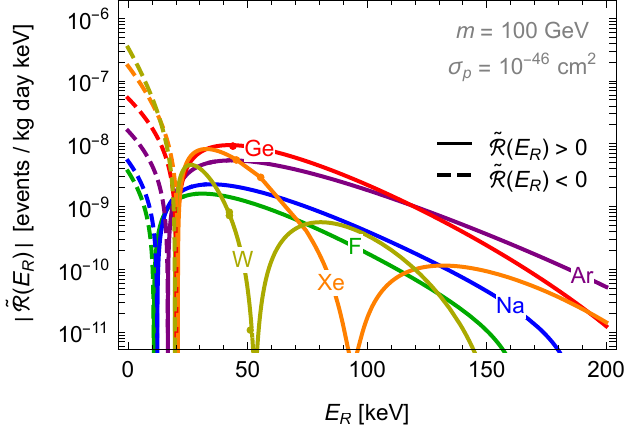}
\caption[Annual-modulation rate spectra for the SI interaction]{\label{fig: SI modulated spectra}\emph{Absolute value of the annual-modulation rate spectra \figmath{\widetilde{\CMcal{R}}(\ER)} for the SI interaction discussed in detail in \Sec{SI interaction}, see \Eq{Rate SI}. Solid lines are for \figmath{\widetilde{\CMcal{R}}(\ER) > 0}, dashed lines are for \figmath{\widetilde{\CMcal{R}}(\ER) < 0}. As in \Fig{fig: SI spectra}, we assume elastic scattering and isosinglet couplings \figmath{f_p = f_n}, and fix the DM-proton cross section at \figmath{\sigma_p = 10^{-46}~\cm^2}. Lines of different colors correspond to different target elements (\figmath{\text{F}}, \figmath{\text{Na}}, \figmath{\text{Ar}}, \figmath{\text{Ge}}, \figmath{\text{Xe}}, \figmath{\text{W}}), see \Tab{tab: nuclides} for details. The DM mass has values \figmath{10~\GeV} (\textbf{left}) and \figmath{100~\GeV} (\textbf{right}). In the region where \figmath{| \widetilde{\CMcal{R}} |} has a zero as a consequence of its change in sign, otherwise small effects that we neglect here, such as the Sun's gravitational focussing or other anisotropies in the local DM velocity distribution, become relevant.}}
\figcode
\end{center}
\end{figure}

Analogous considerations can be made for the annual-modulation rate $\widetilde{\CMcal{R}}(\ER)$, depicted in \Fig{fig: SI modulated spectra} for the SI interaction with the same parameters as above and for the two DM mass values $\mDM = 10~\GeV$ (left panel) and $\mDM = 100~\GeV$ (right panel). The two plots, in logarithmic scale, actually depict the absolute value $| \widetilde{\CMcal{R}}(\ER) |$: solid lines indicate the positive portion of $\widetilde{\CMcal{R}}(\ER)$, while dashed lines indicate the absolute value of its negative part. The change of sign of $\widetilde{\CMcal{R}}(\ER)$ indicates that, at a given time, the rate $\CMcal{R}(\ER, t)$ increases with $t$ at some $\ER$ value while decreasing at some other $\ER$ value. This is determined by an analogous behavior of the velocity integral, which is the source of time dependence of the rate. This behavior, as noted in \Sec{Modulation}, depends on both the DM velocity distribution and the rest of the integrand: for instance, for the SHM, of the $\eta_n$ velocity integrals defined in \Eq{eta} only $\eta_0$ presents this uneven time dependence (see Figs.~\ref{fig: eta},~\ref{fig: etabartilde}). Also to be noted, as commented in \Sec{SHM}, otherwise small effects that we neglect here, such as the Sun's gravitational focussing or other anisotropies in the local DM velocity distribution, become relevant in the region where $\widetilde{\eta}_0$ has a zero due to its sign change.

\begin{figure}[t!]
\begin{center}
\includegraphics[width=.49\textwidth]{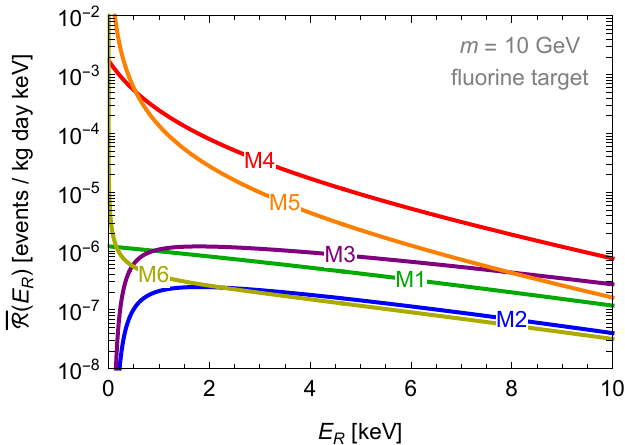}
\includegraphics[width=.49\textwidth]{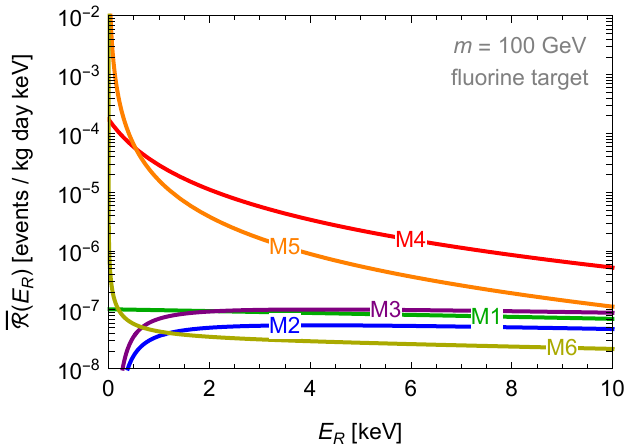}
\\[4mm]
\includegraphics[width=.49\textwidth]{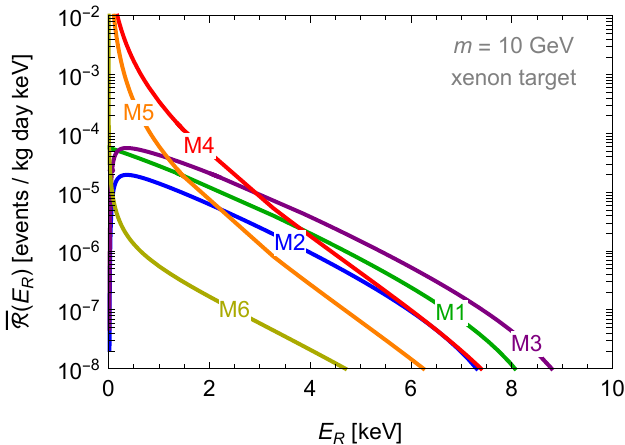}
\includegraphics[width=.49\textwidth]{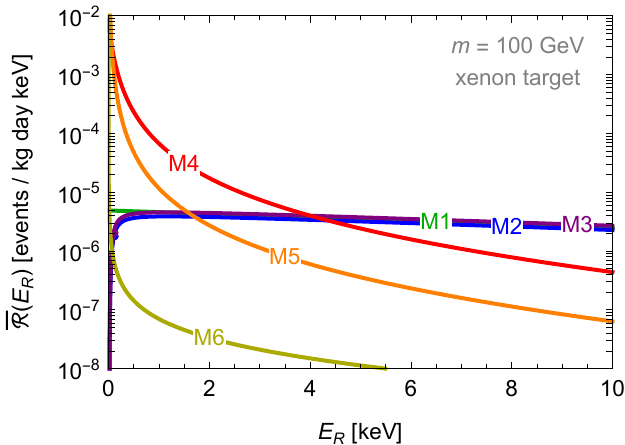}
\caption[Annual-average rate spectra for different models]{\label{fig: spectra}\emph{Annual-average rate spectra \figmath{\overline{\CMcal{R}}(\ER)} for the models summarized in \Tab{tab: Models}, labelled \figmath{\text{M$1$}} through \figmath{\text{M$6$}}. We employ two values of DM mass, \figmath{\mDM = 10~\GeV} (\textbf{left}) and \figmath{100~\GeV} (\textbf{right}), and two nuclear targets, fluorine (\textbf{top}) and xenon (\textbf{bottom}). \figmath{\text{M$1$}} incarnates the standard SI interaction with elastic scattering, whereas \figmath{\text{M$2$}} features endothermic scattering (\figmath{\delta > 0}) and \figmath{\text{M$3$}} exothermic scattering (\figmath{\delta < 0}). The rate for \figmath{\text{M$2$}} is thus suppressed at all energies with respect to that for \figmath{\text{M$1$}} (and also to that for \figmath{\text{M$3$}}, since \figmath{|\delta|} is the same), and both the \figmath{\text{M$2$}} and \figmath{\text{M$3$}} rates vanish at sufficiently low energies as a consequence of the inelastic-scattering kinematics. The millicharged DM model \figmath{\text{M$5$}} features a massless mediator (the photon), whose \figmath{t}-channel propagator makes the rate steeper than that for \figmath{\text{M$1$}} or in other words enhances it at low energies. \figmath{\text{M$4$}} features instead a light mediator, whose mass is comparable to kinematically accessible values of momentum transfer; the spectral shape of its rate bridges between that of \figmath{\text{M$1$}} at very low energies (not visible on the scales of the plots) and that of \figmath{\text{M$5$}} at large energies. Finally, \figmath{\text{M$6$}}, a model of DM with magnetic-dipole moment, also features the same propagator as \figmath{\text{M$5$}}, but NR-suppression factors reduce the rate steepness with respect to that of \figmath{\text{M$5$}}, thus also reducing the low-energy enhancement of the rate with respect to the \figmath{\text{M$1$}} rate.}}
\end{center}
\end{figure}

We devote the last part of this discussion on the rate spectrum to interactions with qualitatively different $\ER$ dependences, see \Sec{Example models and setup}. The above analysis of the spectrum produced by elastic DM-nucleus scattering through the SI interaction has clarified the interplay of nuclear form factor(s) and velocity integral(s), two of the three ingredients contributing to the rate $\ER$ dependence. Considering different interactions allows now to illustrate the effect on the rate of the last ingredient, the inherent momentum dependence of the DM-nucleon interaction, while taking into account the possibility of inelastic scattering also contributes to the discussion of the different spectral shapes the rate can take. We consider the models listed in \Tab{tab: Models}, labeled M$1$ through M$6$: a standard SI interaction with elastic (M$1$), endothermic (M$2$), or exothermic (M$3$) scattering, for which the rate is specified in \Eq{Rate SI}; a SI interaction mediated by a light particle (M$4$), see \Eq{Rate light med}; a millicharged DM model (M$5$), see \Eq{Rate millicharged}; and a model of DM with magnetic dipole moment (M$6$), see \Eq{Rate magnetic}. The annual-average rate $\overline{\CMcal{R}}(\ER)$ for these models is depicted in \Fig{fig: spectra} for two values of DM mass, $\mDM = 10~\GeV$ (left panels) and $100~\GeV$ (right panels), and two nuclear targets, fluorine (top panels) and xenon (bottom panels). For reference, M$1$ is the same model discussed above, whose rate is illustrated in Figs.~\ref{fig: SI spectra},~\ref{fig: SI modulated spectra}. M$2$ and M$3$ represent models that are related to M$1$ through extension of the SI interaction with elastic-scattering kinematics to non-zero values of the DM mass splitting $\delta$: in other words, M$1$, M$2$, and M$3$ are all instances of a continuum of models obtained by varying the value of $\delta$ from $0$ to positive and negative values. Likewise, M$1$ can be thought of as the heavy-mediator limit of a continuum of models obtained by varying the mass of a $t$-channel mediator exchanged at tree level, whereas M$5$ incarnates the opposite limit of massless mediator while M$4$ works as an intermediate example with a light mediator. Finally, M$5$ can also be seen as the leading-order term in the NR expansion of a tree-level interaction cross section with a $t$-channel photon propagator, whereas M$6$ is a second-order term (see discussion in \Sec{Magnetic-dipole DM}) which becomes important in the absence or suppression of the otherwise leading terms.

\begin{table}[t]
\begin{center}
\begin{tabular}{|>{\rule[-2mm]{0mm}{6.6mm}} c | c | c | c | c | c |}
\hline
Label & Model & Kinematics & $\ud \sigma_T / \ud \ER$ & $\CMcal{R}(\ER, t)$ & Parameters
\\
\hline
\hline
M$1$ & SI interaction & \makecell{Elastic \\ $\delta = 0~\keV$} & & &
\\
\cline{1-3}
M$2$ & SI interaction & \makecell{Endothermic \\ $\delta = + 5~\keV$} & \Eq{diffsigmaSI 2} & \Eq{Rate SI} & \makecell{$f_p = f_n$ \\ $\sigma_p = 10^{-46}~\cm^2$}
\\
\cline{1-3}
M$3$ & SI interaction & \makecell{Exothermic \\ $\delta = - 5~\keV$} & & &
\\
\hline
M$4$ & \makecell{SI interaction \\ with light mediator} & Elastic & \Eq{diffsigma V} & \Eq{Rate light med} & \makecell{$a_\chi = 0$, $c_p = c_n$ \\ $\lambda \equiv c_\chi c_p = 10^{-12}$ \\ $m_V = 5~\MeV$}
\\
\hline
M$5$ & Millicharged DM & Elastic & \Eq{Millicharged sigma} & \Eq{Rate millicharged} & $Q_\DM = 10^{-11}$
\\
\hline
M$6$ & \makecell{Magnetic dipole \\ moment DM} & Elastic & \Eq{Magnetic dipole sigma} & \Eq{Rate magnetic} & $\mu_\chi = 10^{-21} e~\cm$
\\
\hline
\end{tabular}
\caption[Some models with mutually different rate spectra]{\label{tab: Models}\emph{Models considered in the text and in \Fig{fig: spectra} for an analysis and comparison of their rate spectra. The first column indicates the model label. The second column provides a short description of the model. The third column indicates whether the scattering is elastic or inelastic and, in the latter case, the value of the non-zero DM mass splitting \figmath{\delta}. The fourth and fifth column indicates the reference formula for the differential scattering cross section \figmath{\ud \sigma_T / \ud \ER} and for the rate \figmath{\CMcal{R}(\ER, t)}, respectively. Finally, the last column indicates the parameters values chosen for representation in \Fig{fig: spectra}.}}
\end{center}
\end{table}

As one can see in \Fig{fig: spectra}, compared to M$1$'s elastic scattering ($\delta = 0$), the endothermic scattering ($\delta > 0$) in M$2$ causes a decrease in the rate, due to the reduction of the domain of the velocity integral already commented above and illustrated in \Fig{fig: InelasticKinematics} (compare with \Fig{fig: ElasticKinematics}) and the right panel of \Fig{fig: eta(ER)}. The exothermic scattering ($\delta < 0$) in M$3$, instead, increases the velocity-integral domain (and thus the rate) at large enough $\ER$ values while reducing it at lower energies, with respect to elastic scattering. Comparing endothermic and exothermic scattering, the velocity-integral domain (and thus the rate) for M$3$ is always larger than that for M$2$ given that $|\delta|$ is the same in the two cases (see explanation in \Sec{Inelastic scattering} and example illustrated in the right panel of \Fig{fig: eta(ER)}). For $\delta \neq 0$, as explained above, the velocity integral somewhat mirrors at low energies its behavior at large energies, where it goes to zero due to the vanishing of the DM velocity distribution at large speeds; the rate for inelastic scattering (M$2$ and M$3$) thus also vanishes at low $\ER$ values, its spectral shape being given by a bell-shaped curve (as that shown in the right panel of \Fig{fig: eta(ER)}) multiplied by the SI nuclear form factor.

Comparison of M$4$ and M$1$ in \Fig{fig: spectra} illustrates the effect on the elastic-scattering rate of a light mediator in a tree-level $t$-channel exchange. Here light means the mediator mass, $m_V$ in \Eq{Rate light med}, being comparable to (or smaller than) the momentum transfer $q$: in fact, were $m_V^2 \gg q^2$ for all kinematically allowed values of momentum transfer (see \Eq{ERinterval} and \Fig{fig: TypicalER}), the rate for M$4$ in \Eq{Rate light med} would basically have the same $\ER$ dependence as that for M$1$ in \Eq{Rate SI}. While not immediately visible on the scales of the plots, for $q^2 \ll m_V^2$ the energy dependence of the mediator propagator does not significantly contribute to the M$4$ rate, which thus matches the M$1$ rate apart from the different normalization (the former being $(\lambda^2 / \pi m_V^4) / (\sigma_p / \mu_\text{N}^2) \approx 2 \times 10^3$ times larger than the latter at $\ER = 0$). The propagator modifies the spectral shape of the rate at higher energies, when $q^2$ becomes comparable to $m_V^2$, by causing it to become much steeper than that of M$1$. For $q^2 \gg m_V^2$, assuming such large values of momentum transfer are kinematically accessible, the spectral shape of the M$4$ rate matches that of the M$5$ rate, where the mediator is massless and the squared propagator drops off as $1 / q^4$. In other words, the $t$-channel mediator propagator enhances the M$4$ and M$5$ rates at low energies with respect to the M$1$ rate; the largest enhancement is there for a massless mediator, as in M$5$, while M$4$ bridges between M$1$ at low energies and M$5$ at large energies. Despite the photon propagator being there also for M$6$, the scattering amplitude arising at second order in the NR expansion causes the presence of NR-suppression factors that partially compensate the $1 / q^4$ low-energy enhancement of the cross section, which indeed increases only as $1 / q^2$ at low energies. Finally, the rates for all models drop to zero at sufficiently large energies as a consequence of the velocity integral(s) vanishing, which happens at lower energies for lighter DM as explained above.

With such a diverse set of possibilities, of which we presented here only some examples, it may be possible for two or more models, in the limited range of energies probed by an experiment, to give equally good fits to a putative signal. For instance, the drop of the M$1$ rate for light DM (due to the sharp decrease of the velocity integral) may resemble that caused by a light mediator in M$4$. To avert such degeneracies, as argued \eg in Refs.~\cite{DelNobile:2015uua, Witte:2016ydc}, one may use the results of a second experiment employing a different target, or, with a single experiment, the time-information on the rate, most notably the annual-modulation rate spectrum (see \Fig{fig: SI modulated spectra} for M$1$).

\subsection{Constraining DM properties}
\label{Constraints}
Constraints on the DM properties can be set by measuring the rate of scattering of DM particles with nuclei in a detector, and comparing the outcome with the prediction of a specific model. A model should specify the nature and interactions of the DM particles as well as their local density and velocity distribution. For concreteness, let us assume that a given statistical data analysis disfavors (in a quantifiable way) the possibility that the dataset features $1$ or more events of DM origin in the considered energy window (we recall from \Sec{Example models and setup} that our idealized detector directly measures recoil energies with infinite precision). We will then be interested in determining what DM models predict $1$ or more events, so that we can deduce what choices of parameter values are at odds with the experiment. In this simplified setup, a model predicting more (less) than $1$ DM event would perform equally as bad (good) as a model with a different spectrum that predicts the same amount of DM events. Our analysis is therefore insensitive to the spectral shape of the rate, and only depends on its overall size. While a certain sensitivity to the shape of the spectrum may be obtained \eg by sorting the events into more than one bin, we assume all events to be grouped together for the sake of simplicity. We take the energy window employed in the data analysis to be $[\Emin, \infty]$: since the rate is quickly suppressed at large enough energies by the nuclear form factor(s) and the velocity integral(s), an infinite-energy upper limit simply represents a sufficiently large value for the experiment not to miss a significant fraction of scattering events (see \eg Ref.~\cite{Bozorgnia:2018jep}). Looking at \Eq{Nevents}, the only models that are not disfavored by the statistical analysis in our simplified setup are then those satisfying
\beq\label{N < 1}
1 > N_{[\Emin, \infty]} \simeq w \int_{\Emin}^\infty \ud \ER \, \overline{\CMcal{R}}(\ER) \ ,
\eeq
where $w$ is the experimental exposure and in the last equality we neglected the second term in \Eq{Taylor R(Ed)}.

\begin{figure}[t]
\begin{center}
\includegraphics[width=.49\textwidth]{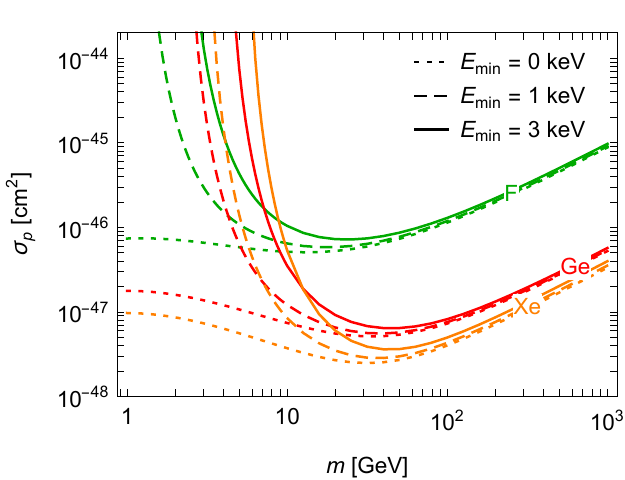}
\includegraphics[width=.49\textwidth]{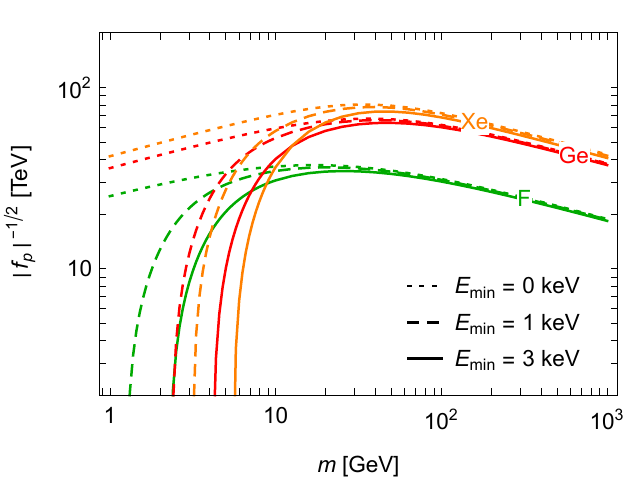}
\caption[Constraint curves for the SI interaction with isosinglet couplings]{\label{fig: bounds}\emph{\figmath{N_{[\Emin, \infty]} = 1} curves for the SI interaction with isosinglet couplings and elastic scattering, see \Eq{Rate SI} with \figmath{f_p = f_n}. \figmath{N_{[\Emin, \infty]} = 1} means that the models corresponding to the parameter-space points on the curves predict \figmath{1}~DM event in the \figmath{[\Emin, \infty]} energy window. We adopt three target materials (\figmath{\text{F}}, \figmath{\text{Ge}}, and \figmath{\text{Xe}}) and the three minimum-energy values \figmath{\Emin = 0}, \figmath{1}, and \figmath{3~\keV} (dotted, dashed, and solid lines, respectively), and an exposure \figmath{w = 1~\ton~\yr}. \textbf{Left:} as upper bounds on \figmath{\sigma_p} for each value of \figmath{\mDM}, where only choices of parameter values below the curves are acceptable. The conversion from square centimeters to picobarn can be obtained through \Eq{units}, \figmath{10^{-36}~\cm^2 = 1~\pb}. \textbf{Right:} as lower bounds on \figmath{|f_p|^{- 1/2}}, related to \figmath{\sigma_p} through \Eq{SI sigma_N}. In an EFT approach, \figmath{f_p} may be schematically thought of as the product of some fundamental coupling constants, times an hadronic matrix element (see \Sec{qg to N}), divided by an energy scale squared: \figmath{|f_p|^{- 1/2}} is then proportional to this energy scale.}}
\figcode
\end{center}
\end{figure}

\Fig{fig: bounds} shows $N_{[\Emin, \infty]} = 1$ curves for the SI interaction with isosinglet couplings and elastic scattering, see \Eq{Rate SI} with $f_p = f_n$. Constraint curves are shown for three target materials, F, Ge, and Xe, and the three minimum-energy values $\Emin = 0$, $1$, and $3~\keV$ (dotted, dashed, and solid lines, respectively). The exposure is set to $w = 1~\ton~\yr$; easy to imagine, a variation in the exposure would have the effect of proportionally shifting the constraint curves upwards or downwards, so as a variation in the value of the local DM density $\rho$. A less trivial effect of a variation in $w$ would be a change in the maximum DM mass the constraints apply to: assuming a data-taking time of $1$ to $10~\yr$ and taking the density of the detector material of order $1~\ton / \mt^3 = 1~\gr / \cm^3$ for definiteness, the argument carried out in \Sec{Scattering rate} implies that any constraint would only possibly apply up to $\mDM$ of order $10^{18}~\GeV$. As already commented upon in \Sec{SI interaction}, the particle-physics model taken into account here has only two free parameters, the DM mass $\mDM$ and the DM-proton scattering cross section $\sigma_p$, the latter controlling the overall size of the rate. The $N_{[\Emin, \infty]} < 1$ constraint can then be translated into an upper bound on $\sigma_p$ for each value of $\mDM$ (the $\bar{\sigma}^\text{SI}_p(\mDM)$ function in \Sec{SI interaction}, see \eg \Eq{barsigma}), so that results can be plotted on a $\mDM$--$\sigma_p$ plane where only choices of parameter values below the curves are acceptable (left panel of \Fig{fig: bounds}). To do so, the integral in \Eq{N < 1} may be computed for each value of DM mass at a fixed, arbitrary value of $\sigma_p$, say $10^{-46}~\cm^2$, and then used as a scaling factor for that $\sigma_p$ value so that the model reproduces the maximum number of DM events allowed by the statistical analysis of the data: in formulas, within our example where only $1$ DM event is allowed, the bound would then read $\bar{\sigma}^\text{SI}_p(\mDM) = 10^{-46}~\cm^2 \times 1 / N_{[\Emin, \infty]}$. This constraint on $\sigma_p$ may also be translated via \Eq{SI sigma_N} onto variables that can more immediately and transparently yield information on the underlying DM model, see \eg examples in Secs.~\ref{Vector-mediated interaction},~\ref{Scalar-mediated interaction}. For instance, in an EFT approach, the DM-proton coupling $f_p$ in \Eq{SI sigma_N} may be schematically thought of as the product of some fundamental coupling constants, times an hadronic matrix element (see \Sec{qg to N}), divided by an energy scale squared. This combination is constrained in the right panel of \Fig{fig: bounds}, where the $\sigma_p$ bound shown in the left panel is translated into a bound on $|f_p|^{- 1/2}$, which is proportional to the above-mentioned energy scale. The only acceptable choices of parameter values are those above the curves.

For very heavy DM, $\mDM \gg m_T$, one can see in \Fig{fig: bounds} that the constraint curves scale as $\sigma_p / \mDM =$~constant, as already mentioned in \Sec{Rate spectrum}. To understand this, one can notice that the $\ER$--$\vmin$ relation only depends on the DM mass through the DM-nucleus reduced mass $\mu_T$, which becomes approximately independent of $\mDM$ for $\mDM \gg m_T$ (see \Sec{kinematics notation}). Therefore the velocity integral stops varying with $\mDM$ in this regime, and we obtain, for any $E_1 < E_2$,
\beq\label{large mDM rate integral}
\int_{E_1}^{E_2} \ud \ER \, \overline{\CMcal{R}}(\ER) \xrightarrow{\mDM \gg m_T} \sigma_p \frac{\rho}{\mDM} \frac{1}{2 \mN^2} \sum_T \zeta_T A^2 \int_{E_1}^{E_2} \ud \ER \, F_\text{SI}^2(\ER) \, \overline{\eta}_0(\vmin(\ER)) \ ,
\eeq
where we used the fact that $\mu_N$ approximates to $\mN$ for $\mDM \gg \mN$. For very heavy DM, therefore, the rate only depends on the DM mass through the $\rho / \mDM$ flux factor, thus explaining why the constraint curves scale as $\sigma_p / \mDM =$~constant. The sensitivity difference of the constraints for different target elements is driven, in this regime, by the $A^2$ factor favoring the heavier targets, while being also affected by the velocity integral, which also favors heavier targets for $\mDM \gg m_T$ (see \Fig{fig: ElasticKinematics} and the left panel of \Fig{fig: eta(ER)}), and by $F_\text{SI}^2$, which instead favors lighter targets (see \Fig{fig: HelmFF}). As a consequence, the constraint curves scale for different targets with a factor that is not exactly the naive ratio of the respective (isotope-averaged) $A^2$ factors.

In the opposite regime of very light DM, we observe in the left panel of \Fig{fig: bounds} an apparent plateau of the dotted ($\Emin = 0$) lines at $\mDM \sim 1~\GeV$. This is not indicative of the behavior of the limits at smaller DM masses, which in fact go again as $\sigma_p / \mDM =$~constant for small enough $\mDM$ values. In fact, as explained in \Sec{Rate spectrum}, for small DM masses the rate spectral shape is dominated by the velocity integral, which is only non-zero below a certain $\ER$ value that decreases for decreasing $\mDM$. For light enough DM the nuclear form factor can be approximated as constant at the small energies where the velocity integral is non-zero, so that the integral in \Eq{N < 1} is just the $\ER$ integral of $\overline{\eta}_0(\vmin(\ER))$ times $\ER$-independent factors. Noticing that $\mu_T$ and $\mu_\text{N}$ approximate to $\mDM$ for $\mDM \ll \mN$, see \Sec{kinematics notation}, we then obtain
\beq\label{small mDM rate integral}
\int_0^{E_2} \ud \ER \, \overline{\CMcal{R}}(\ER) \xrightarrow{\text{sufficiently small}~\mDM} 2 \sigma_p \frac{\rho}{\mDM} \sum_T \frac{\zeta_T}{m_T} A^2 F_\text{SI}^2(0) \int_0^{\vmin(E_2)} \ud \vmin \, \vmin \, \overline{\eta}_0(\vmin) \ ,
\eeq
where we changed integration variable from $\ER$ to $\vmin$ using \Eq{vmin}. For generality, we kept a generic upper limit $E_2$ for the energy integral rather than immediately setting it to $\infty$. Now, given that the $\overline{\eta}_0(\vmin)$ function is entirely determined by the astrophysical properties of the local DM halo, the only quantity that depends on $\mDM$ in the $\vmin$ integral is $\vmin(E_2)$. Since $\overline{\eta}_0$ vanishes for $\vmin \geqslant \vmax$, the $\vmin$ integral does not change with $\mDM$ as long as $\vmin(E_2) \geqslant \vmax$, which is always true for small enough $\mDM$ values.\footnote{Quantitatively, given the incidental correspondence with \Eq{minDMmass condition}, we can say that $\vmin(E_2) \geqslant \vmax$ is satisfied for $\mDM \leqslant \mDM_\text{min}(E_2)$, see \Eq{minDMmass} with $\delta = 0$ and the left panel of \Fig{fig: minDMmass}. Also incidentally, the $\vmin$ integral in \Eq{small mDM rate integral} can be expressed through \Eq{eta integral} as a combination of $\overline{\eta}_0$ and $\overline{\eta}_1$ that does not involve any integrals: for instance, for $E_2 \to \infty$ it equals $\overline{\eta}_1(0) / 2$, see \Eq{eta_n+r(0)}.} This amounts to saying that for light enough DM the $\ER$ integral of $\overline{\eta}_0$ in the $[0, E_2]$ interval is quadratic in $\mDM$, so that its dependence on the DM mass approximately cancels out with that of the $1 / \mu_\text{N}^2$ factor in \Eq{Rate SI} provided $\mDM \ll \mN$. Therefore, for sufficiently light DM the integrated rate in \Eq{small mDM rate integral} (and thus $N_{[\Emin, \infty]}$ for $\Emin = 0$) scales as $\sigma_p / \mDM$.

The effect of realistically imposing a finite minimum energy can be observed in \Fig{fig: bounds}, for $\Emin = 1~\keV$ (dashed lines) and $\Emin = 3~\keV$ (solid lines). The constraint curves are seen to quickly lose sensitivity to small DM masses. This limitation, as already explained in \Sec{Rate spectrum}, is due to the fact that lighter DM needs larger speeds to impart target nuclei with a recoil energy above a certain, finite $\Emin$ value, but the speed of halo DM particles is limited from above as too fast DM particles are not gravitationally bound to our galaxy. $\overline{\eta}_0(\vmin)$ vanishing abruptly for $\vmin$ values close to $\vmax$ (see the left panel of \Fig{fig: eta(ER)}) then causes experimental bounds to quickly lose sensitivity as $\vmin(\Emin)$ grows towards $\vmax$ (and eventually larger than that) for lighter and lighter DM, see discussion related to \Fig{fig: minDMmass}.

\begin{figure}[t!]
\begin{center}
\includegraphics[width=.49\textwidth]{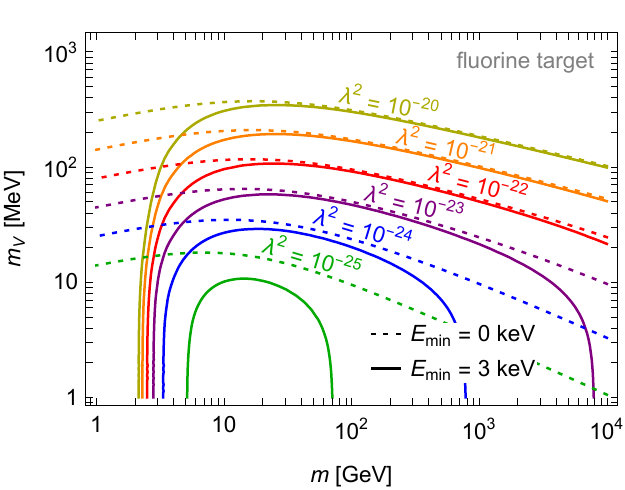}
\includegraphics[width=.49\textwidth]{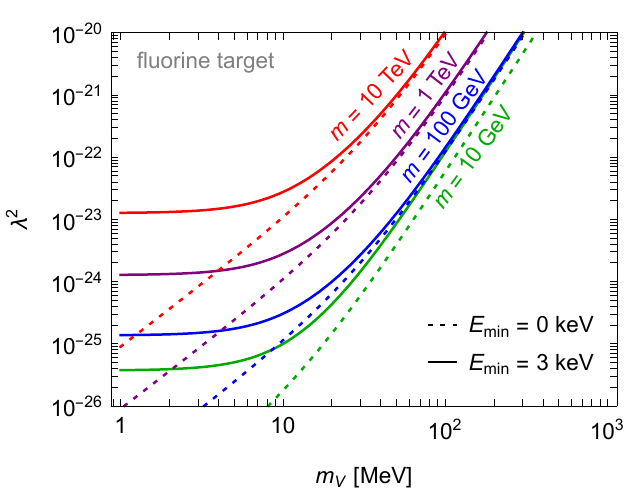}
\\[2mm]
\includegraphics[width=.49\textwidth]{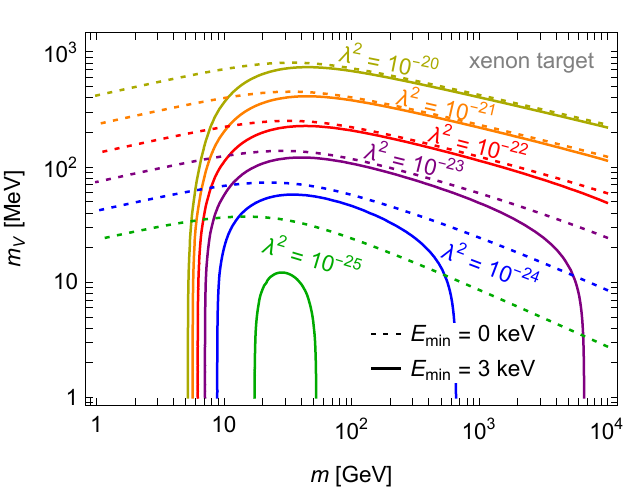}
\includegraphics[width=.49\textwidth]{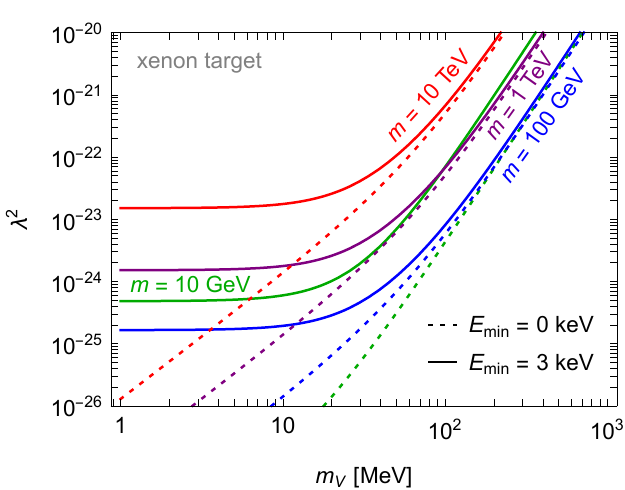}
\caption[Constraint curves for a model with a $t$-channel mediator]{\label{fig: boundslightmed}\emph{\figmath{N_{[\Emin, \infty]} = 1} curves for a model with a \figmath{t}-channel mediator of mass \figmath{m_V} with isosinglet couplings to nucleons, see \Eq{Rate light med}. \figmath{N_{[\Emin, \infty]} = 1} means that the models corresponding to the parameter-space points on the curves predict \figmath{1}~DM event in the \figmath{[\Emin, \infty]} energy window. We adopt the two minimum-energy values \figmath{\Emin = 0} and \figmath{3~\keV} (dotted and solid lines, respectively), and an exposure \figmath{w = 1~\ton~\yr}. \textbf{Top:} for a \figmath{\text{F}} target. \textbf{Bottom:} for a \figmath{\text{Xe}} target. \textbf{Left:} on the \figmath{\mDM}--\figmath{m_V} plane for fixed values of \figmath{\lambda^2}, where only choices of parameter values above the curves are acceptable. \textbf{Right:} on the \figmath{m_V}--\figmath{\lambda^2} plane for fixed values of \figmath{\mDM}, where only parameter-space points below the curves are acceptable.}}
\figcode
\end{center}
\end{figure}

We end this Section with the analysis of the constraint curves in a model with a light $t$-channel mediator in a tree level scattering, see \Eq{Rate light med}. These constraints are computationally more expensive than those for the SI interaction, as we have one more parameter ($m_V$) affecting non-trivially the spectral shape of the rate. This implies that the integral in \Eq{N < 1} needs to be computed not only for each value of $\mDM$, as for the SI interaction, but also for each value of $m_V$. Luckily, we have already seen above that no new integral needs to be performed for the parameters that have no influence on the spectral shape of the rate but only on its overall size, as $\sigma_p$ in \Eq{Rate SI} and $\lambda^2$ in \Eq{Rate light med}, as these can be fixed to some arbitrary value that can then be scaled according to the resulting number of DM events predicted by the model. Moreover, beside being computationally more demanding, the light-mediator model has a $3$-dimensional parameter space, as opposed to the $2$-dimensional SI parameter space, so that the results of the analysis are usually displayed in two-dimensional slices of the full parameter space. Two such slices are shown in \Fig{fig: boundslightmed}, where $N_{[\Emin, \infty]} = 1$ curves for an exposure $w = 1~\ton~\yr$ are displayed for a fluorine (top panels) and xenon (bottom panels) target at fixed values of $\lambda^2$ in the $\mDM$--$m_V$ plane (left panels) and at fixed values of $\mDM$ in the $m_V$--$\lambda^2$ plane (right panels). In the $\mDM$--$m_V$ ($m_V$--$\lambda^2$) plane, for a given $\lambda^2$ value ($\mDM$ value), models corresponding to parameter-space points below (above) the respective curve predict more than $1$ event in the $[\Emin, \infty]$ energy window. $w$ and the local DM density $\rho$ are degenerate with $\lambda^2$ in \Eq{N < 1}, so that a variation in the exposure and/or in the value of $\rho$ may be compensated by an appropriate scaling of $\lambda^2$. As per the above discussion on the effects of varying $w$ on the SI-interaction constraints, and under the same assumptions, constraints can only possibly apply up to DM masses of order $10^{18}~\GeV$.

For $\Emin = 0$ (dotted curves in \Fig{fig: boundslightmed}), the constraints scale as $\lambda^2 / \mDM =$~constant at fixed $m_V$ for sufficiently heavy DM: in fact, we have seen in the discussion related to \Eq{large mDM rate integral} that $\overline{\eta}_0(\vmin(\ER))$ does not (approximately) depend on the DM mass for $\mDM \gg m_T$, so that the $\ER$ integral of the rate~\eqref{Rate light med} only depends on $\mDM$ through the $\rho / \mDM$ factor. Furthermore, for $m_V$ large enough so that $1 / (q^2 + m_V^2)^2$ can be approximated as $1 / m_V^4$ in \Eq{N < 1}, the constraints scale as $\lambda^2 / \mDM m_V^4=$~constant. For sufficiently light DM, instead, only very small energies contribute to the integral (due to $\overline{\eta}_0(\vmin(\ER))$ quickly falling to zero), so that the $q$ dependence of the mediator propagator can be neglected in the integrated rate, the nuclear form factor can be approximated with its $\ER = 0$ value as in \Eq{small mDM rate integral}, and, again as in \Eq{small mDM rate integral}, the $\ER$ integral of $\overline{\eta}_0$ can be seen to be quadratic in $\mDM$. This implies that the $\Emin = 0$ constraint curves scale approximately as $\lambda^2 \mDM / m_V^4 =$~constant for sufficiently light DM.

To understand the shape of the constraints for $\Emin > 0$ (the solid lines in \Fig{fig: boundslightmed} are for $\Emin = 3~\keV$), we can proceed as follows~\cite{DelNobile:2015uua}. At fixed $\lambda^2$ and $m_V$, the $\ER$ integral of the rate~\eqref{Rate light med} vanishes for DM so light that $\vmin(\Emin) \geqslant \vmax$. It also approaches zero for $\mDM \to \infty$: to see this, one may repeat the reasoning related to \Eq{large mDM rate integral} or otherwise notice that the integrated rate for finite $\Emin$ is smaller than for $\Emin = 0$, which as seen above scales as $\lambda^2 / \mDM$ at fixed $m_V$ for $\mDM \gg m_T$ and thus vanishes in the limit of infinte DM mass. For intermediate values of $\mDM$, the integrated rate is non-zero and has therefore a global maximum. Since the rate in \Eq{Rate light med} is larger for smaller $m_V$ values, such global maximum increases for lighter mediators and reaches its peak in the limit of massless mediator, which is effectively attained for $m_V^2 \ll 2 m_T \Emin$. This is the maximum value the integrated rate can reach by varying both $\mDM$ and $m_V$ at fixed $\lambda^2$. The presence of this maximum implies the existence of a minimum value of $\lambda^2$ below which no bounds can be set: such minimum value may be determined, in our setting, by imposing $\max_{\mDM, m_V} N_{[\Emin, \infty]} = 1$, as for smaller $\lambda^2$ values \Eq{N < 1} is always satisfied.

Above this minimum $\lambda^2$ value, the shape of the constraints can be figured as follows. Starting from a parameter-space point where $\mDM$ is very large and $m_V = 0$ (bottom-right portion of the left panels of \Fig{fig: boundslightmed}), we can decrease $\mDM$ at fixed $\lambda^2$, thus increasing the number of predicted events, until we reach $N_{[\Emin, \infty]} = 1$. The $\lambda^2 =$~constant constraint curve in parameter space can be charted from this starting point by increasing $m_V$: the curve raises vertically in the $\mDM$--$m_V$ plane until $m_V^2$ is of the same order of magnitude as $2 m_T \Emin$ and the integrated-rate dependence on the mediator mass becomes non negligible. At this point, an increase in $m_V$ can be compensated by a decrease in $\mDM$, so that the $N_{[\Emin, \infty]} = 1$ condition remains satisfied. We may then enter the contact-interaction regime, where $m_V^2 \gg 2 m_T \ER$ for all energies relevant to the $\ER$ integral; here, if the DM mass is still much larger than the target mass, so that $\overline{\eta}_0(\vmin(\ER))$ is approximately independent of $\mDM$, the integrated rate scales as $\lambda^2 / \mDM m_V^4$, similarly to what discussed above for $\Emin = 0$. Eventually, If we keep lowering the DM mass, the dependence of the $\vmin(\ER)$ function on $\mDM$ becomes non-negligible and $\vmin(\ER)$ starts increasing significantly, so that the energy value above which $\overline{\eta}_0(\vmin(\ER))$ vanishes gets smaller and smaller and the integrated rate decreases quickly. If we are to keep $\lambda^2$ constant, this reduction in the integrated rate can only be compensated by a decrease in $m_V$, until we enter again the long-range regime where the rate is no longer sensitive to $m_V$ and the constraint curve drops vertically in the $\mDM$--$m_V$ plane. For large values of $\lambda^2$, away from the minimum value that can be constrained, this drop in experimental sensitivity occurs for a value of DM mass close to that saturating \Eq{minDMmass condition}, as for the SI interaction discussed above; for smaller values of $\lambda^2$, this drop occurs at larger DM masses as a consequence of the rate being smaller.

The above analysis of parameter-space constraints relies on a detailed knowledge of the key aspects of the different ingredients entering the rate (differential cross section, velocity integral, scattering kinematics). The general arguments we used, having applied to such qualitatively different example cases as a contact interaction and an interaction mediated by a light particle, may as well be employed for other interactions that have not been discussed here. Our analysis highlights the importance of a comprehensive understanding of all aspects of direct DM detection: from the modelling of the particle-physics properties and interactions of the DM (see \Sec{qg to N}) to computing the DM-nucleus differential scattering cross section (see \Sec{sigma}); from the NR physics of DM-nucleon interactions (see \Sec{DM-N}) to the nuclear form factors (see \Sec{Form factors}); from the astrophysical properties of the DM halo, such as the DM velocity distribution (see \Sec{velocity}), to the scattering kinematics and the recoil-energy dependence of the velocity integral (see \Sec{scattering kinematics}); from the construction of the differential rate (see \Sec{Rate}) to understanding the interplay of its factors (discussed in this Section, \Sec{Pheno}). The reader interested in a detailed discussion of any of these topics may jump to the relevant Section, while the one looking for a brief overview of the content of these notes may benefit from the two-page summary and the handy Q\&A subsection featured in \Sec{Summary}, which may also work as entry points to the various sections.

\section{Summary}
\label{Summary}

\subsection{A kind of afterword}
In these notes we tried to present in some detail the theory of direct DM detection. One aim was to provide a comprehensive explanation for all those known arguments and formulas that, while widely used, are hardly ever fully proved or explained in the recent and less recent literature. Another was to gather in a single reference a somewhat complete set of information and tools that would otherwise be scattered across a number of reviews and papers, and to attempt a systematic exposition of the subject. Ideally, such a presentation would equip readers with the conceptual as well as concrete machinery to readily start their direct DM detection analysis. In this direction, we also made a code to generate most of the figures of these notes publicly available on the \website~\cite{Appe-website}, which already contains some of the tools needed for a direct DM detection analysis and can be used as a playground or as a starting point for an actual study. The theory of direct DM detection lies at the interface of multiple disciplines, from particle physics to nuclear physics to astrophysics, mastering all of which requires quite some time and dedication; these notes should help in making that easier.

In spirit, the notes are close to Ref.~\cite{Lewin:1995rx}, although unfortunately way less concise. On the plus side, however, they touch upon a number of subjects of interest in the more recent literature, are possibly more updated, and offer a wider, more general, and in-depth treatment of DM-nucleon and DM-nucleus interactions. The idea was to provide a pedagogical guide to who approaches the subject for the first time, that was both general, as much as possible complete, and possibly also easy to use, so to also serve as a reference for the well-versed practitioner.

In our attempt to be as general as possible, we tried to separate the abstract concepts from the concrete, particular cases, and to state systematically all the assumptions and their consequences. A good practice on its own, this aims also at avoiding giving wrong impressions, \eg that the DM-nucleus interactions can only possibly be of SI or SD type, or that the velocity distribution coincides with the SHM, or that $\eta_0$ is the only possible velocity integral (or that WIMPs are the only DM candidate, although for the purposes of these notes they actually are). Even for those who are only interested in the more standard scenarios, our approach has the added value of enriching one's understanding of the conventional assumptions, beside of course the greater generality and its applicability to more and more diverse cases.

The pedagogical character of the notes is highlighted, for instance, by the computations being worked out in all their crucial steps, and by a number of examples being presented throughout to complement and illustrate the theoretical arguments. Moreover, a code to generate most of the figures is publicly available on the \website~\cite{Appe-website}, which allows for an applied and hands-on approach. An attempt has been made to keep the discussion as simple as possible, although a number of advanced subjects are also discussed in some detail.

Regarding ease of use, an effort was made to present the material of these notes in a form compatible with the different notations adopted in the literature, so to be readily comparable with results found elsewhere. The single sections are conceived as self-contained and as much as possible independent of one another, with \Sec{Rate} and \Sec{Pheno} working as a frame to the various parts. This should possibly allow to more easily navigate through the different topics and to find the information one is looking for. To the advantage of the reader seeking quick responses, a two-page summary and a Q\&A section are provided below which may also work as entry points to the other sections of these notes.

The word `Appendiciario' used in the title has no meaning, to the best of our knowledge. In Italian it may sound like a long collection of appendices, and in this sense it was ironically coined by P.~Panci as a nickname for Ref.~\cite{DelNobile:2011uf}. Regardless, it seems like a perfect word to describe these notes: appendices are sometimes employed in research papers as brief but pedagogical introductions to some basic aspects of their topic, and these notes may well be regarded as an ideal collection of such appendices. Besides, they genuinely are an appendix to the title page.

\subsection{Two-pages summary}
Our notation is described in \Sec{Notation}. Direct DM (Dark Matter) searches aim at detecting nuclear recoils due to DM particles impinging on detector nuclei (see \Fig{fig: basics}). They do so by attempting to measure the recoil energy $\ER = q^2 / 2 m_T$ of target nuclei $T$, which depends on the DM mass $\mDM$ and the DM-nucleus relative speed, $v \sim 10^{-3}$ for halo DM (see \Fig{fig: TypicalER}). The main quantity of interest is then arguably the differential scattering rate per unit detector mass,
\begin{equation*}
\frac{\ud R_T}{\ud \ER}(\ER, t) = \frac{\rho}{\mDM} \frac{\zeta_T}{m_T} \int_{v \geqslant \vmin(\ER)} \ud^3 v \, \fE(\bol{v}, t) \, v \, \frac{\ud \sigma_T}{\ud \ER}(\ER, \bol{v}) \ ,
\tag{\ref{diffrate}}
\end{equation*}
which also serves as a list of the ingredients needed for a phenomenological analysis:
\begin{itemize}
\item $m_T$, the mass of the target nucleus (see \Tab{tab: nuclides}),
\item $\zeta_T$, the target mass fraction (see \Eq{zeta_T} and \Tab{tab: nuclides}),
\item $\rho$, the local DM mass density (see \Sec{Scattering rate}),
\item $\fE(\bol{v}, t)$, the DM velocity distribution in Earth's frame (see \Sec{velocity}); it depends on time because of Earth's motion around the Sun.
\end{itemize}
The other ingredients are discussed more in detail in the following.

The DM-nucleus differential cross section,
\begin{equation*}
\frac{\ud \sigma_T}{\ud \ER} \NReq \frac{1}{32 \pi} \frac{1}{\mDM^2 m_T} \frac{1}{v^2} \overline{| \Mel |^2} \ ,
\tag{\ref{dsigmadER}}
\end{equation*}
can be computed following the several examples discussed in detail in \Sec{sigma}: SI (spin-independent) interaction in \Sec{SI interaction}, SD (spin-dependent) interaction in \Sec{SD interaction}, generic tree-level interactions mediated by a vector (\eg the $Z$ boson) or a scalar (\eg the Higgs boson) in \Sec{Vector-mediated interaction} and~\ref{Scalar-mediated interaction}, respectively, and DM with magnetic dipole moment in \Sec{Magnetic-dipole DM}. More in general, one may start from a Lagrangian describing the interactions of a DM candidate with SM (Standard Model) particles, \eg quarks and gluons: the first step would then be to compute the DM-nucleon scattering amplitude $\Mel_N$ and related hadronic matrix element(s), see \Sec{qg to N} (chances are that the computation has already been performed, see below). One may then take the NR (non-relativistic) limit as detailed in \Sec{DM-N}, most notably through Eqs.~\eqref{fermionbilinears},~\eqref{Nfermionbilinears}, and express the result as an operator
\begin{equation*}
\Op_\NR^N = \sum_i f_i^N(q^2) \, \Op^N_i
\tag{\ref{f(q^2) ONR^N}}
\end{equation*}
(or more in general as in \Eq{general NR Op}) in terms of the NR building blocks $\Op^N_i$~\eqref{NR building blocks} for spin-$0$ and spin-$1/2$ DM. The result of this operation can be promptly found in Tabs.~\ref{tab: DM-N bilinears},~\ref{tab: spin-0 DM EFT},~\ref{tab: spin-1/2 DM EFT},~\ref{tab: EM interactions} for an extensive catalog of conventional cases, including electromagnetic interactions induced by DM-photon couplings (see \Sec{Electromagnetic interactions}). One may then add up the DM interactions with all nucleons within the target nucleus, an operation that is carried out in \Sec{Form factors}: the effect of nuclear compositeness may be parametrized within nuclide- and operator-dependent nuclear form factors, whose size gets reduced for $q > 0$ as a consequence of the diminished scattering coherence (see \Sec{Rate preliminaries}). Eventually, the DM-nucleus unpolarized squared scattering amplitude can be conveniently written as
\begin{equation*}
\overline{| \Mel |^2} \NReq \frac{m_T^2}{\mN^2} \sum_{i, j} \sum_{N, N' = p, n} f_i^N(q^2) f_j^{N'}(q^2) \, F_{i, j}^{(N, N')}(q^2, {v^\perp_T}^2) \ ,
\tag{\ref{M^2 form factors fifj}}
\end{equation*}
with ${v^\perp_T}^2$ given in \Eq{vperp^2} and the $F_{i, j}^{(N, N')}$'s given in \Eq{F_ij} for elastic scattering, see Refs.~\cite{Fitzpatrick:2012ix, Anand:2013yka} (we assume that the nucleus remains in the ground state during the scattering).

Assuming that the differential cross section can be Taylor-expanded in powers of $v$ and safely truncated at leading order, the rate can be expressed in terms of the $\eta_n(\vmin)$ velocity integrals defined in \Eq{eta} (see Eqs.~\eqref{Taylor cross section},~\eqref{Taylor rate}), which only depend on the astrophysical properties of the local DM distribution. The most prominent example (though not the only case of interest) occurs when the scattering amplitude does not depend on $v$ at leading order, so that to a good approximation $\ud \sigma_T / \ud \ER \propto 1 / v^2$ (see \Eq{dsigmadER}) and the rate is proportional to
\begin{equation*}
\eta_0(\vmin, t) = \int_{v \geqslant \vmin} \ud^3 v \, \frac{\fE(\bol{v}, t)}{v} \ .
\tag{\ref{eta_0}}
\end{equation*}
The velocity integral may be computed following \Sec{Computing the velocity integral}, with its time dependence being analysed in \Sec{Modulation}. While not entirely theoretically consistent, the SHM (Standard Halo Model) conveniently allows for the $\eta_n$'s to be computed analytically, see \Sec{SHM} where $\eta_0$, $\eta_1$, and $\eta_2$ are given in Eqs.~\eqref{eta_0 v1},~\eqref{eta_0 v2}~\eqref{eta_1},~\eqref{eta_2 inf vesc} (see also the \website~\cite{Appe-website}) and illustrated in Figs.~\ref{fig: eta},~\ref{fig: etabartilde}. A crucial property of any halo-DM velocity distribution in Earth's frame is that the related velocity integrals (their annual average) vanish for speeds larger than $\vmax^+$ in \Eq{vmax^+-} ($\vmax$ in \Eq{vmax}), due to particles faster than $\vmax^+$ in Earth's frame not being gravitationally bound to our galaxy. This causes the rate to also vanish for $\ER$ values such that $\vmin(\ER)$ exceeds this maximum speed.

$\vmin(\ER)$ is the minimum speed a DM particle must have in order to be able to transfer an energy $\ER$ to the nucleus,
\begin{equation*}
\vmin(\ER) \equiv \left| \frac{q}{2 \mu_T} + \frac{\delta}{q} \right| = \frac{1}{\sqrt{2 m_T \ER}} \left| \frac{m_T \ER}{\mu_T} + \delta \right| ,
\tag{\ref{vmin},~\ref{vmin_inelastic}}
\end{equation*}
with $\mu_T$ the DM-nucleus reduced mass and $\delta$ the DM mass splitting ($\delta = 0$ implies elastic scattering). $\vmin$ controls the domain of the rate velocity integral, so that models with smaller $\vmin$ values at given $\ER$ have a larger rate (since the integrand is non-negative) and thus are more kinematically favored. The dependence of the $\vmin(\ER)$ function on the model parameters (such as the DM mass), discussed in Secs.~\ref{Elastic scattering},~\ref{Inelastic scattering} and illustrated in Figs.~\ref{fig: ElasticKinematics},~\ref{fig: InelasticKinematics},~\ref{fig: eta(ER)}, is instrumental in understanding the properties of the rate. For instance, DM speeds being limited from above implies that the largest $\ER$ value kinematically allowed is finite, and Eqs.~\eqref{vmin},~\eqref{vmin_inelastic} entail that this value is smaller for lighter DM; experiments being only effectively sensitive to recoil energies above a minimum value $\Emin$ then imply that they are unable to detect sufficiently light DM (see \Fig{fig: minDMmass} and related discussions in \Sec{Rate spectrum}).

Finally, the interplay of the different ingredients entering the differential rate (inherent momentum dependence of $\Mel_N$, nuclear form factors, and velocity integrals with their dependence on the scattering kinematics) is explored in \Sec{Pheno}. The spectral shape of the rate is discussed on general grounds in \Sec{Rate spectrum}, and is depicted in Figs.~\ref{fig: SI spectra},~\ref{fig: SI modulated spectra} for the SI interaction and in \Fig{fig: spectra} for the qualitatively different models listed in \Tab{tab: Models}. Moreover, an example phenomenological analysis of a (pretend) experimental result is carried out in \Sec{Constraints}, where parameter-space constraints are derived for the two qualitatively different example cases of contact interaction (\Fig{fig: bounds}) and interaction mediated by a light particle (\Fig{fig: boundslightmed}). The shape of the constraints is discussed on the basis of general properties of the rate and the scattering kinematics, highlighting the importance of a comprehensive understanding of all aspects of direct DM detection.

\subsection{Q\&A}
\etocsetnexttocdepth{5}
\etocsettocstyle{}{}
\cftsubsubsecindent 0pt
\localtableofcontents

\subsubsection{Why are nuclei effective targets?}
As discussed in Secs.~\ref{Rate preliminaries},~\ref{Scattering amplitude}, the amplitude for DM scattering off a spatially extended target drops off quickly for values of momentum transfer much larger than the inverse target radius: in other words, the interaction is coherent across distances of order $1 / q$ or smaller, and is thus suppressed when $1 / q$ is smaller than the target size. Quantitatively, $q \sim 200~\MeV$ roughly entails coherence across distances of order $1~\fm$, see \Eq{units}, guaranteeing at least some degree of coherence in the scattering of halo DM particles with Earth-borne nuclei (see the right panel of \Fig{fig: reduced mass + qR=1} and the left panel of \Fig{fig: TypicalER} for elastic scattering). Notice that, while it may be tempting to refer to $\sim 1 / q$ as a wavelength, this can only be meaningfully done in the context of a one-particle exchange approximation, as no one intermediate particle is required to have momentum $q$ in a loop diagram. It may also be tempting to approximate $q$ with the initial DM momentum, so that $1 / q$ corresponds to the de Broglie wavelength of the incoming DM particle (divided by $2 \pi$), but this can be a very poor approximation, see \Sec{Rate preliminaries}.

The typical nuclear recoil energies induced by elastic scattering of halo DM particles with detector nuclei are displayed in the right panel of \Fig{fig: TypicalER} (note however that these may not be representative of the values relevant to specific models, as discussed in \Sec{Rate preliminaries}). Crucially, experiments could be developed that are at least partially sensitive to these recoil energies. Nuclei are thus effective targets as their scattering with galactic DM particles yields recoil energies that are both large enough for detection and small enough for the scattering to be at least partially coherent, so that the signal is not overly suppressed.

\subsubsection{What nuclear properties are relevant to direct DM detection?}
\label{Q&A nuc prop}
From an experimental point of view, only nuclear elements or compounds satisfying certain technical requirements related to the experimental design can be employed in direct DM searches. Therefore, not all nuclei constitute good targets. A selection of nuclides of interest for direct DM detection experiments is reported in \Tab{tab: nuclides}, which also details some of their properties.

One of the key properties of nuclei is their mass, which determines the kinematics of the DM-nucleus scattering together with the DM mass and the mass splitting $\delta$, see \Sec{scattering kinematics} (we assume that the nucleus remains in the ground state during the scattering). The other nuclear properties relevant to direct detection depend on the interactions the DM may have with nuclei. In fact, different DM-nucleon interactions probe different properties of the nucleus: the SI interaction probes the nucleon distribution inside the nucleus (see \Sec{SI interaction}), the SD interaction probes the nucleon spin distribution (see \Sec{SD interaction}), electromagnetic interactions probe the electric-charge and magnetic dipole moment distributions of nucleons (see \Sec{Electromagnetic interactions}), and so on. Before looking more systematically at the other nuclear properties of interest to direct DM detection, we can be a little more quantitative through an in-depth example.

As mentioned above, the SI interaction probes the distribution of nucleons inside the nucleus, or more precisely their number density (see \Sec{SI interaction}); since the DM couplings to protons and neutrons may be different, the proton number density and the neutron number density are both separately relevant nuclear properties. Each of them has two key aspects: its overall size or normalization, here set by the number of nucleons of a given species, and its spatial features (see \Sec{nuc form factors}). In the limit of point-like nucleus, which for the SI interaction coincides with the limit of zero momentum transfer (see \eg \QnA{Q&A nuc PLN limit}), the DM interacts coherently with all $Z$ protons and all $A - Z$ neutrons within the nucleus, so that the atomic and mass numbers are relevant nuclear properties for this interaction. Away from the point-like nucleus limit, when the full size of the nucleus (and consequent loss of scattering coherence) is taken into account, the spatial distribution of protons and neutrons also becomes important, as scattering at different values of $q$ probes the nuclear interior at different length scales. This effect is usually represented in the differential cross section by `squared form factors', in principle one relative to protons, one to neutrons, and one to the proton-neutron interference, although for the SI interaction it is usually assumed that they are the same function of $q$, see \Sec{SI interaction} (see also \QnA{Q&A form factors}). Indicatively, in the point-like nucleus limit, these three terms have size respectively $Z^2$, $(A - Z)^2$, and $2 Z (A - Z)$ times the relevant DM-nucleon couplings; their size is reduced for $q > 0$ as a consequence of the diminished scattering coherence.

Since different interactions probe in general different nuclear features, a systematic exploration of the set of relevant nuclear properties may start from looking at the possible DM-nucleon interactions. An analysis of the NR DM-nucleon interaction dynamics allows to parametrize the most general NR interaction operator as in \Eq{general NR Op}. This is written in terms of the DM and nucleon spin vectors, $\bol{s}_\chi$ and $\bol{s}_N$, and of the two kinematical variables $\bol{q}$ and $\bol{v}^\perp_N$, the momentum transfer vector and the DM-nucleon transverse velocity; this choice of variables reflects a certain easiness in constructing rotationally- and Galilean-invariant, hermitian NR operators, as explained in \Sec{NR operators}. In \Sec{single-nucleon matel} it was then discussed what are the degrees of freedom pertaining to the internal nuclear state, as their different combinations constitute the different ways the nucleus can respond to the scattering. Such combinations were determined in \Sec{Multipoles} to be, restricting the discussion to NR operators at most linear in $\bol{v}^\perp_N$ for simplicity, the components of the $\unop_N$, $\bol{v}^\perp_N \cdot \bol{s}_N$, $\bol{v}^\perp_N$, $\bol{s}_N$, and $\bol{v}^\perp_N \times \bol{s}_N$ operators, where $\bol{v}^\perp_N$ represents here only the intrinsic component of the DM-nucleon transverse velocity (see \Eq{f(q^2) ONR^N} and subsequent discussion). Notice that these are only scalar and vector operators, so that the formulas to project onto spherical coordinates presented in \Sec{single-nucleon matel} are enough for conveniently writing the scattering amplitude in terms of spherical tensor operators. Parity and time-reversal selection rules, as described in \Sec{PT selection rules}, reduce the number of components contributing to DM-nucleus scattering, for instance excluding the $\bol{v}^\perp_N \cdot \bol{s}_N$ operator altogether.

The surviving components define six nuclear responses, denoted $M, \Delta, \Sigma', \Sigma'', \tilde{\Phi}', \Phi''$ (see \Tab{tab: nuclear responses}), which determine the (nucleon-specific) nuclear properties relevant to direct DM detection. The $M$ response, featured \eg in the SI interaction (see \Sec{SI interaction}), is related to the nucleon number densities (separately of protons and neutrons). The $\Delta$ response, featured \eg in the interaction described in \Sec{Magnetic-dipole DM}, is related to the distribution of nucleon orbital angular momentum. The $\Sigma'$ and $\Sigma''$ responses, featured \eg in the SD interaction (see \Sec{SD interaction}), are related to the nucleon spin densities, in particular to the nucleon-spin component transverse and longitudinal to $\bol{q}$, respectively. The $\tilde{\Phi}'$, $\Phi''$ responses are related to the transverse and longitudinal components of the $\bol{v}^\perp_N \times \bol{s}_N$ nucleon operator, respectively, with $\Phi''$ realizing a nuclear spin-orbit coupling (see \eg Ref.~\cite{Fitzpatrick:2012ix}). As mentioned above, each of these responses enters the DM-nucleus differential scattering cross section with three squared form factors (four for the interference among different responses), see \Sec{Nuclear matrix element}, whose prominent features are the overall size, or normalization at $q = 0$, and the $q$ dependence, the latter representing the effect of the reduced scattering coherence at $q > 0$.
Regarding the overall size of the respective terms in the differential cross section, the $\Delta$, $\tilde{\Phi}'$, and $\Phi''$ responses produce terms that vanish in the point-like nucleus limit, as explained in \Sec{examples and applications}; the $M$ response, as per the above example on the SI interaction, leads to squared form factors whose overall size at $q = 0$ is basically the squared number of nucleons of a given species (thus $Z^2$ for protons, $(A - Z)^2$ for neutrons, and $2 Z (A - Z)$ for their interference, see \Eq{F_11}); and the $\Sigma'$ and $\Sigma''$ responses lead to squared form factors whose overall size is determined by the (square of the) values in \Tab{tab: Sp Sn}, see \Eq{F_Sigma}.

\subsubsection{Can the DM have electromagnetic interactions?}
Despite being characterized as `dark', nothing in principle prevents the DM from having electromagnetic interactions, as long as these are sufficiently weak to avoid all present constraints. Such interactions may stem for instance from the DM particle being a bound state of electrically charged particles, as the neutron, or from the DM coupling with heavy charged states which then generate the DM-photon coupling via loop processes, as it happens for neutrinos. An interesting aspect of DM with electromagnetic interactions is that its direct detection phenomenology is often different from that of models with heavy mediators, as can be seen \eg in \Fig{fig: spectra}; a number of candidates are explored in \Sec{Electromagnetic interactions}, see also \Sec{Magnetic-dipole DM}.

\subsubsection{What is the energy scale of the DM-nucleus interaction?}
The energy scale of the DM-nucleus interaction in direct detection is arguably set by the momentum transfer $q$. Renormalization-group effects should therefore be accounted for when considering theories defined at higher energy scales and/or when comparing results of direct DM detection searches with those of experiments operating at different energy scales, as high-energy particle colliders.

\subsubsection{Can the momentum transfer $q$ be approximated with the DM momentum?}
It can be a very poor approximation, even as an order of magnitude estimate. A better approach could be approximating $q$ with about its maximum value, $q \sim \mu_T v$ for elastic scattering, however this may also turn out to be a poor approximation, depending on the DM model. See discussion with examples in \Sec{Rate preliminaries}.

\subsubsection{Does setting $q = 0$ in the DM-nucleus differential cross section coincide with taking its point-like nucleus limit?}
\label{Q&A nuc PLN limit}
The $q = 0$ limit of the DM-nucleus differential cross section entails that the nucleus can effectively be thought of as point-like, but the opposite is not true: in fact, the point-like nucleus approximation is enforced by neglecting the exponential in \Eq{Mel = Otilde exp} (or in the more schematic \Eq{support}), \ie by only setting $q = 0$ within that exponential. Thus setting $q = 0$ in the DM-nucleus differential cross section can only possibly coincide with taking its point-like nucleus limit if the only $q$ (or $\ER$) dependence of the scattering amplitude is that of the aforementioned exponential. This is the case for instance of the differential cross section for the SI interaction, see \Eq{diffsigmaSI 1} where the recoil-energy dependence of the exponential is encoded by $F_\text{SI}^2$. An example where instead $q = 0$ does not coincide with the point-like nucleus limit is the SD-interaction differential cross section when the induced pseudo-scalar contribution to the DM-nucleon scattering is taken into account, see \eg \Eq{diffsigmaSD} where the point-like nucleus limit is attained by setting $F_\text{pSD}^{(N, N')} = F_\text{PS}^{(N, N')} = 1$. Another example in this sense is any interaction where the DM-nucleon scattering amplitude vanishes at zero momentum transfer, \eg in a model of spin-$1/2$ DM interacting with nucleons through tree-level exchange of a pseudo-scalar mediator (see \Eq{diffsigmaPS}).

\subsubsection{Can the DM-nucleus differential cross section be approximated with its $q = 0$ value?}
The momentum-transfer dependence of the DM-nucleus differential cross section has two sources: the inherent $q$ dependence of the DM-nucleon interaction and the $q$ dependence induced by nuclear compositeness, the latter being usually encoded within nuclear form factors. In the following we discuss whether $q = 0$ can be a good approximation for each of these ingredients.

The DM-nucleon interaction depending on momentum transfer means that the DM-nucleon scattering amplitude $\Mel_N$ depends on $q$ at leading order in the NR expansion, and so does the NR operator $\Op_\NR^N$~\eqref{general NR Op} describing the interaction. The DM-nucleon interaction does not necessarily depend on momentum transfer, an example being the SI interaction discussed in \Sec{SI interaction} (see \Eq{ONR SI}). However, other interactions do depend on $q$, see \eg \Eq{ONR SD} for the SD interaction (which includes the induced pseudo-scalar interaction), \Eq{ONR PS} for a model of spin-$1/2$ DM interacting with nucleons through tree-level exchange of a pseudo-scalar mediator, and \Tab{tab: EM interactions} for a host of DM-nucleon electromagnetic interactions most of which depend on $q$. When this happens, whether $\Mel_N$ (and thus $\Op_\NR^N$) can be approximated with its $q = 0$ value must be checked on a case by case basis. For instance, the induced pseudo-scalar contribution to the SD interaction may be neglected in certain circumstances, as explained in \Sec{SD interaction} (see also \QnA{Q&A induced pseudo-scalar}), effectively setting $q = 0$ in \Eq{ONR SD}. Also the $q$ dependence of a $t$-channel propagator in a tree-level exchange may be neglected in certain conditions, as explained in \Sec{Vector-mediated interaction} (see also \QnA{Q&A max q}). Examples where setting $q = 0$ does not provide a good approximation are instead the interaction described by \Eq{ONR PS}, as it vanishes entirely at zero momentum transfer, and some of the models in \Tab{tab: EM interactions}, where $\Op_\NR^N$ is undefined in this regime.

The other source of momentum-transfer dependence of the DM-nucleus differential cross section, namely the nuclear form factors, is discussed in \Sec{Rate spectrum} in relation to \Fig{fig: SI spectra}, where it is concluded that nuclear form factors have a negligible or small effect on the rate for light DM and/or light target nuclei, but have a large impact for heavy DM scattering off heavy nuclei.

\subsubsection{When can the $q$ dependence of a $t$-channel mediator propagator in a tree-level diagram be neglected?}
\label{Q&A max q}
The momentum-transfer dependence of a $t$-channel propagator in a tree-level exchange may be neglected if $- t \NReq q^2$ (see \Eq{t = - q^2}) is much smaller than the squared mediator mass for all kinematically allowed values of momentum transfer, see \eg \Eq{diffsigma V} and the discussion related to \Fig{fig: TypicalER}. The maximum value of momentum transfer in an elastic scattering can be easily derived from the left panel of \Fig{fig: TypicalER}, displaying the typical momentum transfer as a function of the DM mass for different target elements used in direct detection experiments. As mentioned in \Sec{Rate preliminaries}, all curves have to be scaled up by $\sqrt{2}$ to obtain the maximum kinematically allowed value of $q$ at fixed $v$ in an elastic scattering. Also, keeping into account that the the maximum value of $q$ at fixed $v$ is proportional to $v$ (see \Eq{ERinterval}), and taking $\vmax \approx 765~\km / \sd$ as a typical value for the maximum DM speed in Earth's frame (see \Eq{vmax} and the discussion above \Eq{minDMmass condition}), all curves have to be further scaled up by a factor about $3.3$. We then conclude that a $t$-channel mediator in an elastic-scattering tree-level exchange can always be considered heavy for the purposes of direct detection (meaning for all nuclear targets commonly employed in direct searches) if heavier than few $\GeV$, although it can also be lighter than that if one is only concerned with a specific nuclear target (see \Fig{fig: TypicalER}).

\subsubsection{When can the DM-nucleus differential cross section be parametrized in terms of the DM-nucleon cross section $\sigma_{p, n}$?}
\label{Q&A sigma_p,n}
Experimental collaborations normally express the sensitivity of their setups and their data as a constraint on the DM-proton cross section $\sigma_p$ or the DM-neutron cross section $\sigma_n$. They do so in the context of two models (see \QnA{Q&A experiments}): the SI interaction with isosinglet couplings $f_p = f_n$ (see \Eq{isosinglet sigma} and related discussion), meaning that the DM does not distinguish between protons and neutrons, and the SD interaction with either $a_p = 0$ or $a_n = 0$, meaning that the DM interacts with only one nucleon species when neglecting both the induced pseudo-scalar interaction and $2$-body corrections (see \Sec{SD interaction}). In both cases, the ratio of the DM-proton and DM-neutron couplings is fixed and the model has only two free parameters, the DM mass and one of the couplings. For the SI interaction, the simple form of the DM-nucleon total cross section~\eqref{SI sigma_N}, which is in a one-to-one correspondence with the DM-nucleon coupling squared, allows to easily parametrize the DM-nucleus differential cross section in terms of, say, $\sigma_p$ instead of $f_p^2$, as in \Eq{isosinglet sigma}; results of an analysis may then be equally presented in terms of $\sigma_p$ or $f_p^2$, or any other function of $f_p^2$, see \eg the two panels of \Fig{fig: bounds}. The same holds for the SD interaction, where however the DM-nucleon cross section has only the simple form in \Eq{SD sigma_N} when the induced pseudo-scalar contribution to the interaction is neglected; nevertheless, one may still employ, say, $\sigma_p$ as defined in \Eq{SD sigma_N} as a parameter substituting $a_p$ in the complete DM-nucleus differential cross section (in the assumption $a_n$ vanishes or that its ratio with $a_p$ is fixed), with the understanding that $\sigma_p$ is not the DM-nucleus total cross section.

From this discussion it is clear that, in the considered models, the DM-nucleon cross section is just a suitable parameter that can be used to present the results of an analysis, in alternative to the DM-nucleon coupling or any function thereof. However, it is not always as convenient as in the above examples. Models where the DM-nucleus differential cross section cannot be easily expressed in terms of the DM-nucleon cross section are featured \eg in Secs.~\ref{Vector-mediated interaction},~\ref{Scalar-mediated interaction},~\ref{Magnetic-dipole DM}. In those examples, as briefly discussed below \Eq{diffsigma V}, the presence of multiple interactions can cause the coupling-constant dependence of the DM-nucleon total cross section to be quite different from that of the differential cross section, so that parametrizing the DM-nucleus differential cross section in terms of $\sigma_p$ and/or $\sigma_n$ may not be easy or convenient. Likewise, a non-trivial $q$ and/or $v$ dependence of the differential cross section may cause the DM-nucleon total cross section to depend on $v$, making its use quite inconvenient.

\subsubsection{When is it useful to define the zero-momentum transfer cross section $\sigma_0$?}
The zero-momentum transfer cross section $\sigma_0$, defined in \Eq{sigma_0}, is different from the DM-nucleus total cross section as it does not take into account the size of the nucleus and the coherence loss at large momentum transfer. Since this effect causes the differential cross section to decrease at large energies, $\sigma_0$ is actually larger than the DM-nucleus total cross section.

The usefulness of introducing $\sigma_0$ stands in that it can conveniently parametrize the overall size of the DM-nucleus differential cross section, as clear from \Eq{sigma_0 F_T}. This however only happens if the DM-nucleon interaction (\ie the NR interaction operator~\eqref{general NR Op}) does not inherently depend on $q$, in which case the DM-nucleus differential scattering cross section also does not depend on $q$ in the limit of point-like nucleus. In other cases it may not even be possible to define $\sigma_0$, since it may diverge or vanish as in the examples in Eqs.~\eqref{Millicharged sigma},~\eqref{diffsigmaPS}, or it may be possible but not useful because it fails to represent in a convenient form the overall size of the differential cross section, as would be the case for \eg \Eq{diffsigma V}.

\subsubsection{When can the $\ER$ dependence of the DM-nucleus differential cross section be parametrized with a single nuclear form factor?}
\label{Q&A form factors}
Nuclear form factors are functions of momentum transfer describing the effect on the differential cross section of the loss of scattering coherence due to the finite nuclear size, or in other words they characterize the momentum-transfer dependence of the differential cross section away from the limit of point-like nucleus. Ideally, therefore, nuclear form factors only depend on nuclear properties. Also, they are specific to a given nuclide, as different nuclides have different nuclear properties; to a given interaction, as different interactions probe different properties of the nucleus (\eg its electric charge, its mass number, its spin, its magnetic moment, and so on), see \QnA{Q&A nuc prop}; and to a given nucleon type, as different nuclear properties pertain in different ways to the proton and neutron distributions (\eg protons are electrically charged while neutrons are not, etc.). So, for instance, in the simple example in \Sec{nuc form factors}, $F_\CMcal{O}^N(\bol{q})$ depends on the interaction operator $\CMcal{O}$, on the nucleon type $N$, and of course on the specific nuclide. More in general, given that different operators and even different nucleon species can interfere in the squared scattering amplitude, we introduced in \Sec{Nuclear matrix element} the `squared form factors' $F_{X, Y}^{(N, N')}(q^2)$, depending on two nuclear responses $X$, $Y$ (related to a NR interaction operator each) and on two nucleon types $N$, $N'$. Notice that these form factors are summed over initial and final spins, in a sense, while $F_\CMcal{O}^N(\bol{q})$ does not depend on spin because it was only defined for a spin-$0$ nucleus.

From this discussion we see that there exist several nuclear form factors, and even if only one interaction were present there should be three (possibly independent) squared form factors: $F_X^{(p, p)}$, $F_X^{(n, n)}$, and $F_X^{(p, n)} = F_X^{(n, p)}$ (see \Eq{F_XY}). As mentioned in \Sec{SI interaction}, these can be argued for the SI interaction to be the same function of momentum transfer (aside from their normalization, see \Eq{F_11}), assuming the number density of neutrons in the nucleus to be equal to that of protons; in this way the loss of scattering coherence can be effectively characterized by a single function of recoil energy, which we denoted $F_\text{SI}(\ER)$ (see \eg \Eq{diffsigmaSI 1}). This is usually not the case for other interactions, certainly not so for the SD interaction, even when the induced pseudo-scalar contribution is neglected (see discussion after \Eq{diffsigmaSD-PS}). Nevertheless, one may choose to parametrize the $q$ dependence of the differential scattering cross section within a single function similarly to what done with $F_T(\ER)$ in \Eq{F_T}, or analogously with $S(q^2)$ in \Eq{diffsigmaSD-PS} for the SD interaction. Such a function, however, while often called `nuclear form factor' (or `nuclear structure function' for $S(q^2)$), in general does not depend solely on the nuclear properties, contrary to the $F_{X, Y}^{(N, N')}$'s. For the SD interaction, this is obvious from the definition of $S(q^2)$ in \Eq{SD structure function} (see also \Eq{S_pn}), whose dependence on the DM-nucleon couplings does not even cancel in general when divided by $S(0)$, see \Eq{diffsigmaSD-PS} and subsequent discussion.

All in all, it may always be possible to parametrize the $\ER$ dependence of the DM-nucleus differential cross section, or even just its $\ER$ dependence away from the point-like nucleus limit, within a single function $F_T^2(\ER)$, although this function may be an inconveniently complicated combination of the DM-nucleon coupling constants and the interaction- and nucleon-specific nuclear form factors $F_{X, Y}^{(N, N')}(q^2)$. As explained after \Eq{F_T}, and also touched upon above regarding the SI interaction, one instance in which the dependence on the coupling constants cancels is when the $F_{X, Y}^{(N, N')}$'s all have the same $q$ dependence, regardless of their normalization. Another instance is when the ratio of the DM-proton and DM-neutron couplings is fixed, in which case $F_T^2(\ER)$ does not depend on the one independent coupling (while depending on the coupling ratio). For the SD interaction, for instance, we can easily see from \Eq{S_pn} that $S(q^2) / S(0)$ does not depend \eg on $a_p$ when we fix $a_n$ to be a certain fraction of $a_p$ (or the other way around). Aside from these special cases, however, parametrizing the $\ER$ dependence of the DM-nucleus differential cross section within a single function may not be convenient.

\subsubsection{Does neglecting the nuclear form factors lead to large errors?}
Neglecting nuclear form factors normalized so that $F(0) = 1$ (see \Eq{F normalization}) may provide an order of magnitude estimate of the DM-nucleus differential scattering cross section for sufficiently small $q$ values, but it may grossly overestimate the cross section otherwise. The effect of neglecting nuclear form factors (at least for the SI interaction) can be observed in \Fig{fig: SI spectra}, where the dashed lines are obtained with no form factor. One can see that, as explained in the text, nuclear form factors have a negligible or small effect on the rate for light DM and/or light target nuclei, but have a large impact for heavy DM scattering off heavy nuclei.

\subsubsection{Can the SD interaction be disregarded when the SI interaction is also present?}
If the DM-nucleon coupling constants relative to the SI and SD interactions in the model have similar size, or if the SI coupling constants are larger than the SD ones, the SD interaction may be safely neglected for the nuclear targets currently used in direct DM searches. Here and in the following we are actually comparing the coupling-constant absolute values, and are assuming that destructive interference does not play a considerable role for the SI interaction (see discussion after \Eq{D factor}). In the opposite situation, it should be checked whether the SD couplings are sufficiently small for the SD interaction to be neglected. This means that, barring destructive interference for the SI interaction, they should be sufficiently larger than the SI couplings to compensate for the coherent enhancement of the SI DM-nucleus cross section with respect to the SD cross section. For instance, as discussed after \Eq{diffsigmaSD}, for a fluorine target the SI and SD interactions are certainly comparable if the size of the SD coupling constants is a factor ten larger than that of the SI coupling constants. This SD-SI coupling-constant ratio must be higher for heavier nuclei, for the two interactions to be comparable: for instance, with SI and SD couplings of the same size, the SI differential cross section can easily be $10^4$ times larger than that for the SD interaction for DM scattering off xenon nuclei.

\subsubsection{Does neglecting the induced pseudo-scalar contribution to the SD interaction lead to large errors?}
\label{Q&A induced pseudo-scalar}
The induced pseudo-scalar contribution to the SD interaction (see \Sec{SD interaction}) arises from the DM-nucleon exchange of light pseudo-scalar mesons such as the neutral pion and the $\eta$ meson. In the NR limit it is represented by the $\Op^N_6$ term in \Eq{ONR SD}, which only contributes at non-zero values of momentum transfer (see \Eq{NR building blocks}) and for this reason it is often neglected. It can be seen arising together with the $\Op^N_4$ term by computing the NR expression of the DM-nucleon scattering amplitude from the effective Lagrangian for the SD interaction in \Eq{SDLag}, see result in \Tab{tab: spin-1/2 DM EFT}. As explained in more detail after \Eq{<p'|A^a|p>}, its contribution varies significantly over momentum transfer scales $q \sim m_{\pi, \eta}$, in the reach of direct detection experiments, and therefore it should not be naively truncated in the $q / \mN$ power-series expansion. In short, it can become sizeable for $q^2$ of order $m_\pi^2$ and larger, values that can be attained for heavy DM especially when scattering off heavy target nuclei, as can be seen for elastic scattering in \Fig{fig: TypicalER} (see discussion on the maximum attainable momentum transfer in \QnA{Q&A max q}).

\subsubsection{Can the parameters of a model be tuned so to cancel or severely suppress the cross section?}
\label{Q&A tuning}
The differential scattering cross section is often computed through a number of approximations, \eg by neglecting next-to-leading order terms in the $S$ matrix perturbative expansion and in the NR expansion, subleading interactions and nuclear-physics contributions, the breaking of certain approximate symmetries, and so on. When an approximated formula for the DM-nucleus differential cross section vanishes or receives a large suppression as a result of a cancellation among different contributions, it is thus in order to determine what otherwise subleading contributions may become relevant. Here are some examples.
\begin{itemize}
\item For the SI interaction with generic couplings $f_p$ and $f_n$ (see relevant discussion in \Sec{SI interaction}), DM interactions with protons and neutrons in the nucleus interfere constructively for positive values of $f_n / f_p$ and destructively for negative values. For destructive interference, the leading contribution to the DM-nucleus differential cross section~\eqref{diffsigmaSI 1} vanishes for the nuclide-dependent choice $f_n / f_p = - Z / (A - Z)$ (see the dips in the left panel of \Fig{fig: Dfactor+Rdelta}). In this regime, long-distance QCD corrections that can otherwise be considered subleading and thus neglected become important, shifting the actual value of $f_n / f_p$ for which the differential cross section is suppressed: in this respect, the lowest-order result in \Eq{diffsigmaSI 1}, so as the position of the dips in the left panel of \Fig{fig: Dfactor+Rdelta}, should only be thought of as indicative.
\item For the SD interaction, the DM is often assumed to have `pure SD' couplings to only one nucleon species, \ie either $a_p$ or $a_n$ in \Eq{ONR SD} is set to zero. In the absence of the induced pseudo-scalar interaction and of $2$-body corrections, see \Sec{SD interaction}, this would imply that the DM interacts with either protons or neutrons in the nucleus, which may cause the DM-nucleus interaction to be particularly suppressed. Recalling the example presented in \Sec{SD interaction}, for instance, we may consider a target nucleus whose spin is mainly due to the contribution of neutron spins, \eg because it has $Z$ even but $A$ odd and therefore it only features an unpaired neutron (all other nucleons having pairwise opposite spins that thus contribute little to the overall nuclear spin). The DM-nucleus scattering would then mainly occur through DM-neutron interactions, and would therefore be suppressed were $a_n$ to vanish. In this context, otherwise subleading contributions that effectively couple the DM to neutrons may become important. As discussed in \Sec{SD interaction}, the induced pseudo-scalar interaction and $2$-body corrections mix the contributions of DM-proton and DM-neutron interactions to the DM-nucleus scattering, so that $a_n = 0$ in the above example does not bar DM interactions with neutrons in the nucleus. Here these corrections have an especially large effect thanks to the considerable sensitivity of our target to DM-nucleon SD interactions.
\item Another example where otherwise subleading corrections to a certain quantity, although not a cross section, become important when the leading-order contribution vanishes, is provided by the annual-modulation velocity integral $\widetilde{\eta}_0$ in the SHM, see \Eq{eta_0}, Eqs.~\eqref{etabartilde def},~\eqref{etabartilde approx}, and \Eq{eta_0 v1} or~\eqref{eta_0 v2}. As can be seen in the middle-left panel of \Fig{fig: etabartilde}, $\widetilde{\eta}_0$ in the SHM vanishes at a certain value of $\vmin$ (see explanation in \Sec{SHM}). However, the presence of anisotropies in the local DM velocity distribution as that caused by the Sun's gravitational focussing, which may otherwise be neglected, become relevant here and in fact spoil this feature.
\end{itemize}

\subsubsection{What range in DM mass can be covered with direct DM detection techniques?}
As explained in \Sec{Rate spectrum}, the reach in DM mass of a given experiment is limited from below by its finite threshold. In fact, halo-DM speeds being limited from above entails a maximum possible amount of energy in the DM-nucleus system and thus a maximum possible $\ER$ value, which is lower the lighter the DM. This implies that an experiment effectively sensitive only to recoil energies above a minimum value $\Emin$ cannot detect sufficiently light DM, as scattering of such particles with nuclei would only induce nuclear recoil energies below $\Emin$. Notice that, as a recoil energy, $\Emin$ is distinct from the experimental threshold, and that, since $\ER$ is only statistically related to the signal that is actually recorded by the detector, $\Emin$ should only be considered as a convenient theoretical device. This being said, the lowest possible DM mass a given experiment can be sensitive to is quantified as a function of $\Emin$ in \Eq{minDMmass} and depicted in \Fig{fig: minDMmass}. While for elastic and exothermic scattering, in principle, the sensitivity of an experiment can always be extended to lighter DM by lowering its threshold (and thus $\Emin$), this is not possible for endothermic scattering.

The upper reach in DM mass of an experiment is limited by the fact that, given the value of the local DM mass density $\rho$ inferred by observations (see discussion related to \Eq{rho}), heavier DM has a lower number density at the Sun's location. For instance, as discussed in \Sec{Scattering rate}, for a hypothetical detector with linear size of order $10~\cm$ and a data-taking period of $10~\yr$ we can expect (on average) less than $1$ DM particle crossing the detector during the time of its operations for DM heavier than roughly $10^{17}~\GeV$.

\subsubsection{What are the assumptions behind standard experimental constraints?}
\label{Q&A experiments}
In the absence of a signal that can be interpreted as of DM origin, experimental collaborations usually express the sensitivity of their setups and their data as a constraint on a parameter expressing the overall size of the DM-nucleon interaction at given DM mass. This entails making assumptions on the following matters.
\begin{itemize}
\item The nature of (particle) DM, \eg how many different types of particles contribute to the DM and what properties (such as mass and spin) they have. It is usually assumed that one single particle type constitutes the whole of the DM.
\item A model for the DM interactions, most often the SI interaction (see \Sec{SI interaction}) or sometimes the SD interaction (see \Sec{SD interaction}). These interactions are effectively of contact type, \ie short range. For the SI interaction, the DM-proton and DM-neutron couplings are assumed to be equal, thus enforcing $f_p = f_n$ in \Eq{diffsigmaSI 1} to obtain \Eq{isosinglet sigma}. In this way the DM-proton and DM-neutron total scattering cross sections in \Eq{SI sigma_N} are equal, $\sigma_p = \sigma_n$, and $\sigma_p$ can be taken to parametrize the overall size of the DM-nucleus differential cross section (see \QnA{Q&A sigma_p,n}); $\sigma_p$ is then used as the parameter to be constrained for each value of $\mDM$ (see \eg \Fig{fig: bounds}). For the SD interaction, where the induced pseudoscalar interaction is often neglected thus effectively setting $F_\text{PS}^{(N, N')} = 0$ in \Eq{diffsigmaSD} (see also discussion below \Eq{SD sigma_N}), constraints are produced for the $a_p = 0$ or $a_n = 0$ assumption (sometimes both cases are separately presented), so that $\sigma_n$ or $\sigma_p$~\eqref{SD sigma_N} is used as parameter to be constrained, respectively. Here, as discussed in \Sec{SD interaction} (see also \QnA{Q&A sigma_p,n} and \QnA{Q&A tuning}), the induced pseudo-scalar interaction and $2$-body corrections mix the proton and neutron contributions to the DM-nucleus scattering (only at finite momentum transfer for the former), so that DM-proton couplings also contribute to DM-neutron scattering and vice-versa.
\item Nuclear form factors. The Helm form factor, whose functional form depends on two parameters, is most usually adopted for the SI interaction, see \Eq{HelmFF} and \Fig{fig: HelmFF}. No realistic form factors for other interactions enjoy such a conveniently analytic form. Some form factors for the SD interaction from the DM literature are shown in \Fig{fig: SDFF}. Form factors for other interactions can be found \eg in Refs.~\cite{Fitzpatrick:2012ix, Anand:2013yka}, see \Sec{Nuclear matrix element}.
\item A value for the local DM mass density $\rho$, with $\rho = 0.3~\GeV / \cm^3$ the value most usually adopted in the direct detection literature (see \Eq{rho} and related discussion). This parameter is completely degenerate with the overall size of the DM-nucleus scattering cross section, thus with either among $\sigma_p$ and $\sigma_n$ is chosen as independent parameter for the SI and SD interactions (see above).
\item A model for the DM velocity distribution, usually a truncated Maxwell-Boltzmann as in the SHM (see \Sec{SHM}). The parameters of this model are $\vesc$ (the local escape speed in our galaxy) and $v_0$, with a truncation prescription (parametrized by $\beta = 0, 1$ in \Eq{SHM f}) also being required. $v_0$ is related to the root-mean-square speed at $\vesc \to \infty$ and thus to the asymptotic value of the circular speed; as such, it is usually equated to the local circular speed $v_\text{c}$ in the assumption the rotation curve has already reached its asymptotic value at the Sun's location. The numerical values adopted in these notes are reported in \Eq{speeds}, above which it is also noted that, independently of the particular model for the DM distribution, the values of the $\vesc$, $v_\text{c}$, and $\rho$ parameters are all correlated. The effects on the scattering rate of a variation in their values are discussed below \Eq{speeds}.
\end{itemize}

\subsubsection{Is there a way to recast the experimental constraints on the SI or SD interaction for other models?}
\label{Q&A recast}
In general, it is possible to recast an experimental constraint from a model to another if the differential detection rate $\ud R / \ud \Ed$ in \Eq{dRdEd} has the same $\Ed$ dependence (and, if relevant, time dependence) in the two models. This means that the two rates only differ by an overall multiplicative factor $\alpha$, in which case recasting the constraint consists in a trivial rescaling. A sufficient condition is that the nuclide-specific differential scattering rate $\ud R_T / \ud \ER$ in \Eq{diffrate} is the same in the two models, up to $\alpha$, for each detector nuclide taking part in the interaction. That is certainly the case, $\rho \fE$ being equal, if $\vmin(\ER)$ is the same in the two models and $\mDM^{-1} \, \ud \sigma_T / \ud \ER$ is the same up to $\alpha$. The nuclides taking part in the interaction should also be the same.

Under these constraints, it is easy to imagine that recasting the SI or SD experimental constraints is only possible for a limited number of models. Some example cases of practical relevance are discussed in the following.
\begin{itemize}
\item If the model is the same as the SI interaction but has isospin-violating couplings $f_p \neq f_n$ in \Eq{diffsigmaSI 1}, there is a chance the experimental constraints on the SI interaction with isosinglet couplings $f_p = f_n$ (see \Eq{isosinglet sigma}) may be recast approximately or even exactly, depending on the detector material. This possibility is examined in \Sec{SI interaction}, see analysis related to Eqs.~\eqref{barsigma},~\eqref{D factor} and the subsequent discussion on the phenomenology of models with $f_p \neq f_n$.
\item Performing a similar recasting for the SD interaction, even assuming that the pseudo-scalar interaction and other corrections (see \Sec{SD interaction}) can be neglected, is complicated by the fact that the spin distribution of protons and neutrons in the nucleus are in general qualitatively different; this leads to the presence of three distinct `squared form factors' for the `pure SD' contribution, $F_\text{pSD}^{(N, N')}$ or alternatively $F_{4, 4}^{(N, N')}$ in Eqs.~\eqref{SDMel^2},~\eqref{F_44}, related to the $\Op^N_4$ NR operator in \Eq{ONR SD} (see also \QnA{Q&A form factors}). In contrast, the number density of protons and neutrons can be arguably approximated as equal (aside from the actual number of protons and neutrons in the nucleus), implying that the squared form factors $F_{1, 1}^{(N, N')}$ entering the SI interaction cross section have approximately the same $q$ dependence and thus can be parametrized in terms of a single form factor $F_\text{SI}^2$ (see \Eq{F_11} and related discussion). This allows to factor the $q$ dependence of the SI DM-nucleus differential cross section from its dependence on the DM-nucleon couplings, an aspect that is instrumental in the analysis leading to Eqs.~\eqref{barsigma},~\eqref{D factor}. In other words, varying the ratio of the SI DM-nucleon couplings $f_p$ and $f_n$ does not substantially change the $q$ dependence of the DM-nucleus differential cross section (it only changes its overall size). On the contrary, for the SD interaction, varying the ratio of the pure SD couplings $a_p$ and $a_n$ does change the $q$ dependence of the DM-nucleus differential cross section, which leads to a different $\ER$ dependence of the rate and thus to the impossibility of a rigorous recasting of the SD constraints.
\item If the $q$ or $v^\perp_N$ dependence of the DM-nucleon scattering amplitude is different from that of the SI and SD interactions, experimental constraints on the SI and SD interactions can likely not be recast. For instance, if the interaction is described by a $q$- or $v^\perp_N$-dependent NR operator~\eqref{general NR Op}, SI constraints cannot be recast (since the NR operator~\eqref{ONR SI} describing the SI interaction is independent of $q$ and $v^\perp_N$). The DM-nucleon scattering amplitude depends on $q$ or $v^\perp_N$ \eg when the scattering process occurs through a light $t$-channel mediator with mass comparable to $q$ or smaller, see \eg \Sec{Pheno}; most of the electromagnetic DM interactions listed in \Tab{tab: EM interactions} also fall in this category, see \eg \Sec{Electromagnetic interactions} and \Sec{Rate spectrum}.
\end{itemize}

\subsubsection{Is there a way to recast the experimental constraints for other DM velocity distributions?}
The differential scattering rate depends non-trivially on the DM velocity distribution, see \Eq{diffrate}. For instance, it becomes extremely sensitive to the particular shape of its high-speed tail for light DM particles. Assuming a different DM velocity distribution with respect to that used to compute a certain constraint would thus most probably entail that the scattering rate $\ud R_T / \ud \ER$ in \Eq{diffrate} has a different $\ER$ dependence, and therefore that the detection rate $\ud R / \ud \Ed$ in \Eq{dRdEd} has a different $\Ed$ dependence. This in turn implies that a recast is likely impossible, see \eg \QnA{Q&A recast}.

\subsubsection{How well is the Standard Halo Model justified?}
We still do not know the DM spatial distribution close to Earth, nor its local velocity distribution. Experimental constraints on DM properties are commonly computed adopting the SHM local DM velocity distribution, see \QnA{Q&A experiments}. While not completely self-consistent from a theoretical standpoint, the SHM may be thought of as a first approximation to a more realistic description of the DM halo, which should also ideally include local DM substructures and take into account the impact of the baryonic feedback on the distribution of galactic DM. Nevertheless, the SHM has the practical advantage that the related DM velocity distribution and the $\eta_n$ velocity integrals in \Eq{eta} have conveniently analytic forms, see \Sec{SHM}.

\subsubsection{How does the velocity dependence of the differential cross section affect the rate?}
\label{Q&A velocity}
The velocity dependence of the differential cross section determines the exact shape of the velocity integral(s), see \Sec{Scattering rate}. This in turn affects both the recoil-energy and time dependence of the rate, as discussed in the following. As a visual reference, Figs.~\ref{fig: eta},~\ref{fig: etabartilde} illustrate the effect that different velocity dependences of the differential cross section have on the $\vmin$ and time dependence of the velocity integral within the SHM; \Fig{fig: eta(ER)} then shows the effect when the velocity integral is expressed in terms of $\ER$.

The velocity integral contributes to the rate $\ER$ dependence through the $\ER$--$\vmin$ mapping. In fact, as discussed in \Sec{Rate spectrum} (see also \QnA{Q&A rate dependence}), the rate dependence on $\ER$ is due to the inherent momentum-transfer dependence of the DM-nucleon scattering amplitude, to the nuclear form factors, and to the velocity integrals. A good understanding of the properties of the velocity integrals and of the $\ER$--$\vmin$ mapping, which is directly linked to the scattering kinematics, is instrumental in understanding the rate dependence on the nuclear target and on the model parameters such as the DM mass, and consequently the shape of parameter-space constraints, see Secs.~\ref{Rate spectrum},~\ref{Constraints} (see also \QnA{Q&A rate dependence}).

Furthermore, the velocity integral is the sole source of time dependence of the rate. As the DM distribution is not expected to change significantly over the timescale of an experiment (years), this time dependence is prominently due to Earth's motion around the Sun, which causes the DM flux experienced on Earth, and thus the scattering rate, to be annually modulated. More precisely, neglecting the small effect of Earth's rotation around its own axis, the DM velocity distribution at Earth's location in the detector's rest frame $\fE(\bol{v}, t)$ is obtained from that in the rest frame of the galaxy through a boost by $\bol{v}_\text{E}(t)$, Earth's velocity with respect to the galactic rest frame (see \Sec{DM velocity distribution} and in particular \Eq{vec vE}). Therefore, $\fE(\bol{v}, t)$ (and thus the velocity integrals) depends on time exclusively through $\bol{v}_\text{E}(t)$. As explained in \Sec{Modulation}, moreover, for a locally isotropic DM velocity distribution in the galactic rest frame the velocity integral depends on $\vE(t)$ but not on $\hat{\bol{v}}_\text{E}(t)$, implying that $\vE(t)$ in \Eq{vE} (see also \Eq{Taylor vE}) is the only source of time dependence of the velocity integral.

\subsubsection{What information can be obtained from the time dependence of the rate?}
As mentioned in \Sec{Modulation}, the time dependence of the rate has distinctive (though model-dependent) features that can help telling a putative DM signal from mismodeled or unaccounted for backgrounds. The time dependence of known backgrounds is in fact different from what is expected from a DM signal. Moreover, the analysis of the time dependence of the rate can also help discriminating among different models of DM interactions, as commented upon in Secs.~\ref{Differential cross section},~\ref{Pheno}, and among different models of DM velocity distribution. The strict relationship between DM velocity distribution and time dependence of the rate is testified by the fact that the latter arises from the time-dependent boost of the local DM velocity distribution from the galactic frame to the detector's rest frame (see \Sec{DM velocity distribution} and \QnA{Q&A velocity}).

\subsubsection{How does the scattering rate depend on $\ER$, the target material, and the model parameters?}
\label{Q&A rate dependence}
As explained in \Sec{Rate spectrum}, the momentum-transfer dependence of the rate has three sources: the nuclear form factors, the inherent $q$ dependence of the DM-nucleon interaction, and the velocity integrals.

The nuclear form factors are expression of the nuclear compositeness, and their $q$ dependence embodies the loss of coherence in the scattering at finite values of momentum transfer (see \Sec{Differential cross section}). For this reason, when normalized to a finite value at zero momentum transfer, the nuclear form factors decrease with momentum transfer at the small $q$ values of interest to direct DM detection, thus reducing the rate. Their $q$ dependence is specific to each interaction, nucleon type, and nuclear target, but in general they get severely suppressed over momentum-transfer scales comparable with the nuclear radius, see discussion in \Sec{Rate preliminaries} (see also \QnA{Q&A nuc prop} and \QnA{Q&A form factors}).

The $q$ dependence of the DM-nucleon scattering amplitude $\Mel_N$ (and thus of the NR operator $\Op_\NR^N$~\eqref{general NR Op} describing the interaction), is not there for the SI interaction but can arise for other interactions, \eg from mediator propagators or from the NR expansion if its zeroth order vanishes; the SD interaction has an inherent $q$ dependence through the induced pseudo-scalar interaction due to light-meson exchange, see \Eq{ONR SD}. The $q$ dependence of $\Mel_N$ (and $\Op_\NR^N$) is specific to each model and may induce in the DM-nucleus differential cross section (and thus in the rate) a particular dependence on the target mass and on the model parameters (such as $\mDM$ and $\delta$), different from that of other models.

Also the velocity dependence of the DM-nucleon interaction contributes to the $q$ dependence of the rate: in fact, it determines the $v$ dependence of the DM-nucleus differential scattering cross section (see \Sec{Form factors}) which then shapes the velocity integral involved in the rate. For instance, a leading $v^{-2}$ or $v^0$ dependence of $\ud \sigma_T / \ud \ER$ (corresponding respectively to a $v^0$ or $v^2$ dependence of $\ud \sigma_T / \ud \cos\theta$, see \Sec{Differential cross section}) induces $\eta_0$ or $\eta_1$ in the rate (see \Eq{eta}), the two velocity integrals having different dependence on $\vmin$ and thus on $\ER$ (and on $m_T$, $\mDM$, $\delta$). An example model featuring $\eta_1$ (alongside with $\eta_0$) is described in \Sec{Magnetic-dipole DM} (see also \Eq{Rate magnetic}); its rate spectral shape is depicted in \Fig{fig: spectra}, together with that of other qualitatively different models.

Despite their differences, however, all halo-DM velocity integrals share some key properties. They (their annual average) vanish for speeds larger than $\vmax^+$ in \Eq{vmax^+-} ($\vmax$ in \Eq{vmax}), as particles faster than $\vmax^+$ in Earth's frame are not gravitationally bound to our galaxy. Also, the velocity integrals in \Eq{eta} are functions of $\vmin$ that are uniquely determined by the local DM velocity distribution (see Figs.~\ref{fig: eta},~\ref{fig: etabartilde} for the SHM); their dependence on $m_T$, $\mDM$, and $\delta$ is then ascribed exclusively to the $\vmin(\ER)$ function (see Secs.~\ref{Elastic scattering},~\ref{Inelastic scattering} and in particular Eqs.~\eqref{vmin},~\eqref{vmin_inelastic}), as would be the case of any other velocity integral whose integrand does not depend on these parameters. In this sense it is useful to think of the velocity integrals as functions of $\vmin$ that are stretched onto the $\ER$ axis in a $m_T$- and $\mDM$- (and $\ER$-) dependent way.

From these universal properties of the velocity integral follow some important features of the rate, see discussion in \Sec{Rate spectrum}, which in turn shape the parameter-space constraints, see \Sec{Constraints}:
\begin{itemize}
\item The differential rate vanishes for recoil energies above a certain value (that for which $\vmin(\ER)$ exceeds the maximum speed), and this $\ER$ value is smaller for lighter DM, see \eg Figs.~\ref{fig: SI spectra},~\ref{fig: SI modulated spectra} for the SI interaction. This in turn implies that experiments with a finite threshold cannot detect sufficiently light DM, see \eg Figs.~\ref{fig: bounds},~\ref{fig: boundslightmed} where the constraints quickly lose sensitivity to small enough $\mDM$.
\item The $\ER$--$\vmin$ mapping of the velocity integrals, which only depends on $\mDM$ through the DM-nucleus reduced mass $\mu_T$, becomes approximately independent of the DM mass for $\mDM \gg m_T$ (see \Sec{kinematics notation}), thus greatly simplifying the rate dependence on $\mDM$ in this regime. For models where the only other DM-mass dependence of the rate stems from the $\rho / \mDM$ factor from the DM flux, see \eg Eqs.~\eqref{Rate SI},~\eqref{Rate light med},~\eqref{Rate millicharged} and \Sec{Scattering rate}, this implies that any constraint on parameters controlling the overall size of the rate (\eg $\sigma_p$ for the SI interaction with isosinglet couplings, see \Eq{isosinglet sigma}) scales as $\mDM^{-1}$ for $\mDM \gg m_T$.
\end{itemize}

\section*{Acknowledgements}
Thanks to Graciela Gelmini, Anne Green, Ji-Haeng Huh, Paolo Panci, and, more in general, to all from whom I've learnt. Also thanks to Marco Cirelli and Lorenza Scarparo. This work was partially supported by STFC Grant No.~ST/P000703/1.

\end{document}